\documentclass{tcibook}
\usepackage{fancyhea}
\usepackage{work}
\usepackage{amsmath,amssymb}
\usepackage{psfrag}
\usepackage{color}
\usepackage{epsfig}  
\usepackage{graphicx}               
\usepackage{url}
\usepackage{hyperref}
\usepackage{epstopdf}
\usepackage{multirow}
\usepackage{xspace}

\newcommand{\nc}{\newcommand}  

\nc{\beq}{\begin{equation}}  
\nc{\eeq}{\end{equation}}  
\nc{\beqa}{\begin{eqnarray}}  
\nc{\eeqa}{\end{eqnarray}}  
\nc{\bea}{\begin{eqnarray}}  
\nc{\eea}{\end{eqnarray}}  
\nc{\ra}{\rightarrow}  
\nc{\lsim}{\begin{array}{c}\,\sim\vspace{-21pt}\\< \end{array}}  
\nc{\gsim}{\begin{array}{c}\sim\vspace{-21pt}\\> \end{array}}  
\nc{\slsh}{\slash\hspace*{-0.22cm}}

\def\to{\rightarrow}
\def\Re{{\cal R \mskip-4mu \lower.1ex \hbox{\it e}\,}}
\def\Im{{\cal I \mskip-5mu \lower.1ex \hbox{\it m}\,}}
\def\be{\begin{equation}}
\def\ee{\end{equation}}
\def\bea{\begin{eqnarray}}
\def\eea{\end{eqnarray}}
\def\bit{\begin{itemize}}
\def\eit{\end{itemize}}
\nc{\eref}[1]{(\ref{#1})}
\nc{\Eref}[1]{Eq.~(\ref{#1})}

\nc{\vev}[1]{ \left\langle {#1} \right\rangle }
\nc{\bra}[1]{ \langle {#1} | }
\nc{\ket}[1]{ | {#1} \rangle }
\nc{\fb}{\,{\rm fb}^{-1}}
\nc{\ev}{{\rm eV}}
\nc{\kev}{{\rm keV}}
\nc{\Mev}{{\rm MeV}}
\nc{\gev}{{\rm GeV}}
\nc{\tev}{{\rm TeV}}
\nc{\mev}{{\rm MeV}}

\def\D{{\cal D}}

\def\O{{\cal O}}

\def\BR{\mbox{\rm BR}}
\def\ee{e^+e^-}

\def\msb{{\bar{\ssstyle M \kern -1pt S}}}

\begin{document}

\def\bibname{References}
\bibliographystyle{plain}

\raggedbottom
\parindent=0pt
\parskip=8pt
\setlength{\evensidemargin}{0pt}
\setlength{\oddsidemargin}{0pt}
\setlength{\marginparsep}{0.0in}
\setlength{\marginparwidth}{0.0in}
\marginparpush=0pt


\renewcommand{\chapname}{chap:intro_}
\renewcommand{\chapterdir}{.}
\renewcommand{\arraystretch}{1.25}
\addtolength{\arraycolsep}{-3pt}

\thispagestyle{empty}

\rightline{\vbox{\halign{&#\hfil\cr
&ANL-HEP-TR-12-25\cr
&SLAC-R-991\cr}}}

\begin{centering}
\vspace*{3cm}

{\Huge\bf Fundamental Physics\\ at the Intensity Frontier}

\vspace*{14cm}
{\Large\bf  Report of the
Workshop held December 2011 in Rockville, MD}
\end{centering}

\newpage

\pagenumbering{roman}


\begin{center}

Workshop Chairs:  J.L.~Hewett$^{117}$, and H.~Weerts$^{3}$\\
\vspace{0.5cm}

Conveners:  R.~Brock$^{76}$, J.N.~Butler$^{38}$, B.C.K.~Casey$^{38}$, J.~Collar$^{23}$, A.~de Gov\^ea$^{93}$, 
R.~Essig$^{122}$, Y.~Grossman$^{30}$, W.~Haxton$^{11}$, J.A.~Jaros$^{117}$, C.K.~Jung$^{122}$
% ...

\end{center}

\begin{center}

% ...
$^{3}$Argonne National Laboratory, Argonne, IL 60439, USA\\
% ...
$^{11}$University of California, Berkeley, Berkeley, CA 94720, USA\\
% ...
$^{23}$University of Chicago, Enrico Fermi Institute, Chicago, IL 60637, USA\\
% ...
$^{30}$Cornell University, Laboratory for Elementary Particle Physics, Ithaca, NY 14853, USA\\
% ...
$^{38}$Fermi National Accelerator Laboratory, Batavia, IL 60510, USA\\
% ...
$^{76}$Michigan State University, East Lansing, MI 48824-1045, USA\\
% ...
$^{93}$Northwestern University, Evanston, IL  60208 USA\\
% ...
$^{117}$SLAC National Accelerator Laboratory, Menlo Park CA 94025, USA\\
% ...
$^{122}$Stony Brook University, Stony Brook, NY 11794, USA\\
% ...

\end{center}

\newpage

\begin{center}
{\Huge\bf Executive Summary}
\end{center}

Particle physics aims to understand the universe around us.  The Standard Model of particle physics describes the basic structure of matter 
and forces, to the extent we have been able to probe thus far.  However, it leaves some big questions unanswered. Some are within 
the Standard Model itself, such as why there are so many fundamental particles and why they have different masses. In other cases, the 
Standard Model simply fails to explain some phenomena, such as the observed matter-antimatter asymmetry in the universe, the 
existence of dark matter and dark energy, and  the mechanism that reconciles gravity with quantum mechanics. These gaps lead us to conclude 
that the universe must contain new and unexplored elements of Nature.  Most of particle and nuclear physics is directed towards discovering 
and understanding these new laws of physics. 
 
These questions are best pursued with a variety of approaches, rather than with a single experiment or technique.  Particle physics uses 
three basic approaches, often characterized as exploration along the cosmic, energy, and intensity frontiers.  Each employs different tools and 
techniques, but they ultimately address the same fundamental questions.  This allows a multi-pronged approach where attacking basic 
questions from different angles furthers knowledge and provides deeper answers, so that the whole is more than a sum of the parts.   A 
coherent picture or underlying theoretical model can more easily emerge, to be proven correct or not.  

The intensity frontier explores fundamental physics with intense sources and ultra-sensitive, sometimes massive detectors.  
It encompasses searches for extremely rare processes and for tiny deviations from Standard Model expectations.  Intensity
frontier experiments use precision 
measurements to probe quantum effects.   They typically investigate new laws of physics, at energies higher than the kinematic reach of 
high energy particle accelerators. The science addresses basic questions, such as: Are there new sources of $CP$ violation?  Is there 
$CP$ violation in the leptonic sector? Are neutrinos their own antiparticles?  Do the forces unify? Is there a weakly coupled hidden 
sector that is related to dark matter?  Are there undiscovered symmetries?

To identify the most compelling science opportunities in this area, the workshop {\it Fundamental Physics at the Intensity Frontier} 
was held in December 2011,  sponsored by the Office of High Energy Physics in  the US Department of Energy Office of Science.    
Participants investigated the most promising experiments to exploit these opportunities and described the knowledge that can be gained 
from such a program.   The 
workshop generated much interest in the community, as witnessed by the large and energetic participation by a broad spectrum of scientists.
This document chronicles the activities of the workshop, with contributions by more than 450 authors.

The workshop organized the intensity frontier science program along six topics that formed the basis
for working groups:  experiments that probe ($i$) heavy quarks, ($ii$) charged leptons, ($iii$) 
neutrinos, ($iv$) proton decay, ($v$) light, weakly interacting particles, and ($vi$) nucleons, nuclei, and atoms.    The 
conveners for each working group included an experimenter and a theorist working in the field and an observer from the community at large.
The working groups began their efforts well in advance of the workshop, holding
regular meetings and soliciting written contributions.

Specific avenues of exploration were identified by each working group.  Experiments that study rare 
strange, charm, and bottom meson decays provide a broad program of measurements that are sensitive to new laws of physics.  Charged leptons, 
particularly muons and taus, provide a precise probe for new physics because the Standard Model predictions for their properties are very 
accurate.  Research at the intensity frontier can reveal  $CP$ violation in the lepton sector,  and elucidate whether neutrinos are their 
own antiparticles. A very weakly coupled hidden-sector that may comprise the dark matter in the universe could be
discovered.  The search for 
proton decay can probe the unification of the forces with unprecedented reach and test sacrosanct symmetries.  Detecting 
an electric dipole moment for the neutron, or neutral atoms, could establish a clear signal for new physics, while limits on such a 
measurement would place severe constraints on many new theories.

This workshop marked the first instance where discussion of these diverse programs was held under one roof.   As a result, it was realized that 
this broad effort has many connections; a large degree of synergy exists between the different areas and they address similar questions.
Results from one area were found to be pertinent to experiments in another domain. For example, 
participants identified several avenues for exploring new physics that could account for the matter-antimatter asymmetry of the universe, 
and it is the combination of results that could provide the ultimate explanation.  The workshop also revealed 
many synergies between the intensity frontier and the energy and cosmic frontiers.  The three frontiers 
are closely linked, with much overlap in the physics they can discover and interpret.  A clear conclusion of the workshop is that 
intensity frontier experiments constitute an integral and crucial part of a balanced experimental program in particle and nuclear physics.  
It is rich with outstanding science opportunities that have the potential to change paradigms.

The workshop identified the facilities required to do intensity frontier science.
Experiments that can be performed this decade with current or planned facilities were described, as well as those 
proposed for the next decade that require new facilities or technology to reach their full potential.  Some emphasis was placed on 
experiments that could produce data by the end of this decade.  It is clear that the intensity frontier will move the field forward substantially 
this decade. For example, the recent determination that the remaining mixing angle in the neutrino sector is large, opens the possibility 
that  detectors planned to start construction soon can make measurements that address fundamental questions about 
neutrinos and the universe.  Likewise, several experiments searching for the neutron electric dipole moment will start this year and are 
expected to advance the field.  Muon experiments planned for later this decade will allow for an improvement over current results by roughly 
four orders of magnitude. this decade, new 
heavy quark facilities have the potential to search beyond the reach of the Large Hadron Collider this decade.
For the next decade, proposals for new experiments and facilities with more intense beams and more sensitive 
detectors are already under consideration, and will provide a large step forward.  This document compares the science reach of these 
facilities to the reach of others around the globe, new and old, that propose to house similar programs.  The result is a challenging, 
interconnected, and essential science program covered by the intensity frontier.

The vibrancy of the workshop reflected the enthusiasm of the community for intensity frontier physics.  Since the workshop addressed all 
aspects of the intensity frontier, participants were forced to look across traditional boundaries and become aware of the intensity science 
program as a whole.  It was realized that there is a strong intensity frontier community that has common physics interests and is addressing 
fundamental questions.  It is important for the future of the field that broad workshops such as this continue.  In this way members of the 
intensity frontier community can continue to learn from each other and see how their program contributes to our understanding of the universe.

We hope this report provides valuable input into strategic 
decision-making processes in the United States and elsewhere, and contributes to establishing a worldwide intensity frontier program to 
which the US contributes in a major and competitive way. The workshop confirmed that the proposed facilities and experiments will position 
the US as a global leader in intensity frontier science.

\newpage

\tableofcontents

\newpage

\pagenumbering{arabic}
\chapter{Introduction}

The mission of particle physics is to understand the universe around us.  Our field has achieved much success in discovering elementary 
constituents of matter and the forces that govern them.  Nonetheless, unsolved mysteries of Nature remain and must be investigated to 
advance our understanding. The optimal way to probe further is to embark on several avenues of exploration that test Nature from different 
directions.  The three basic avenues of particle physics, known as the cosmic, energy, and intensity frontiers, address the same fundamental  
questions, but are distinguished by the tools they employ in the quest for deeper knowledge.   This report focuses on the intensity frontier 
and constitutes the proceedings of the workshop, {\it Fundamental Physics at the Intensity Frontier}, held in December 2011.  Experiments at 
the intensity frontier search for very rare processes and tiny deviations from Standard Model predictions. Throughout the history of particle 
physics, discoveries in studies of rare processes have led to a deeper understanding of Nature. These precise measurements require 
intense beams of particles and ultra-sensitive, sometimes massive, 
detectors.  This report identifies the science opportunities at the intensity frontier and 
the facilities required to enable that science.  It defines the role of the intensity frontier in exploring the universe, and describes the 
synergy between this approach and the two other frontiers. This workshop report exhibits the diversity and breadth of 
the intensity frontier physics program and illustrates the potential for paradigm-changing discoveries.

Over the last few decades, a series of experiments combined with theoretical insights has led to a description of the elementary constituents 
of matter and their interactions that is known as the Standard Model (SM) of particle physics.  The SM contains many fundamental particles:  
the lepton and quark fermions, which  come in three families; the  gluon, photon, $W$ and $Z$ vector bosons that carry forces; and the Higgs 
scalar boson.  It is expected that the Higgs boson will soon be discovered at the CERN Large Hadron Collider (LHC).  These particles interact via  
electromagnetism and the weak and strong forces.  The strong force is responsible for the formation of protons and neutrons; it is mediated 
by the massless gluons and confines the quarks into hadrons.  Its dynamics are well described by the theory of Quantum Chromodynamics.   The 
weak force mediates radioactive decay via the exchange of the $W$ and $Z$ bosons, which are roughly 100 times heavier than the proton.  All 
fermions carry a weak charge and participate in the weak interaction.  The electromagnetic force sees electric charge, is mediated by the 
massless photon, and is described by the theory of Quantum Electrodynamics.  Gravity has yet to be reconciled with quantum mechanics and is 
not considered part of the SM.  At the distance scales of normal life, these forces act independently.  However, at smaller distances (or 
higher energies) it has been experimentally verified that the electromagnetic and weak force unify into the electroweak force, which is 
described by a single theory.  It is hoped that at even higher energies, the electroweak theory will unify with the strong force. The reverse 
process, {\it i.e.}, going from smaller to larger distance scales, results in electroweak symmetry breaking which we believe is carried out 
via the Higgs mechanism.

Despite the success of the SM in describing Nature, it leaves many questions unanswered.  Why are there so many fundamental particles?   Why 
do they have their intrinsic masses, and why do the values of those masses span 14 orders of magnitude?  Do all the forces unify?  How does 
gravity fit into the picture?  Are there undiscovered laws of Nature?  Does the Higgs mechanism break the electroweak symmetry and, if so, 
how is the electroweak scale stabilized?
 
Astrophysical observations have revealed that most of the energy and matter in the universe $-$ on the order of 96\% of it $-$ takes the 
form of dark energy or  dark matter.   Only the remaining 4\% consists of SM particles, {\it i.e.}, matter like us.  Dark energy is a 
mysterious energy density that permeates space.  Dark matter is a new form of matter that interacts through gravity and possibly through 
very weak couplings to the SM fields; hence the term ``dark.''  In addition, the observed asymmetry of matter and antimatter in the universe 
remains unexplained by the SM.

These considerations point to the existence of new physics, defined as laws and symmetries of Nature that lie beyond the SM.  Currently, 
numerous imaginative theories for new physics have been proposed, but experiments have yet to provide guidance pointing to the correct 
fundamental theory.  Much of the worldwide effort in particle and nuclear physics is driven by searches for evidence of new particles and 
interactions.

\begin{figure}[b!]
\centerline{\includegraphics*[height=7.35cm]{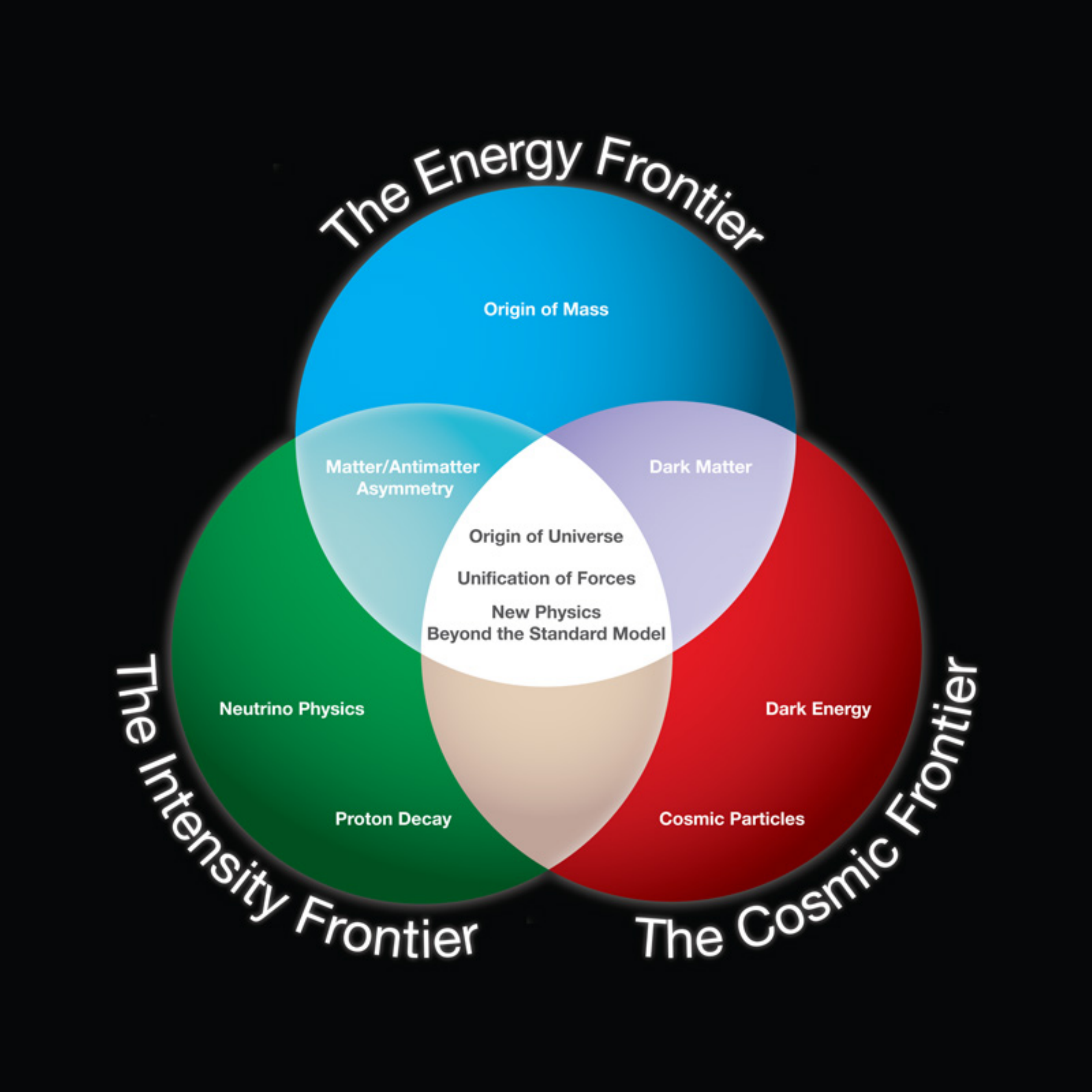}} 
\caption{Illustration of the three frontiers of particle physics from \cite{intro:P5}.}
\label{P5fig}
\end{figure}

The three-frontier model of particle physics was defined by the Particle Physics Project Prioritization Panel in its 2008 report \cite{intro:P5} 
and is often represented by the Venn diagram in Fig. \ref{P5fig}.  It has proven beneficial for various levels of communication and is now 
widely used and recognized. Each frontier employs different tools and techniques, but all frontiers work together to address the
same fundamental questions.

At the cosmic frontier, physicists use the universe as an experimental laboratory and observatory, taking advantage of naturally occurring 
events to observe indications of new interactions.  Research focuses on understanding dark energy and dark matter, employing a variety of 
instruments to measure particles on or close to Earth.  This program is pursued worldwide with a leading component in the United States. 

At the energy frontier, experiments explore the highest possible energies reachable with accelerators, directly looking for new physics via 
the production and identification of new states of matter.  This has the advantage of direct observation in a laboratory setting, but is 
limited by the kinematical reach of high energy colliders.   This work is now being carried out at the LHC at CERN, 
which collides protons at a center of mass energy of 7-8 TeV, increasing to 14 TeV in the next few years. 

At the intensity frontier, experiments use intense sources of particles from accelerators, reactors, the sun and the atmosphere to explore 
new interactions.  This involves ultra-precise measurements to search for quantum effects of new particles in rare processes or  effects 
that give rise to tiny deviations from SM predictions. This technique has the asset of exploring very high energy scales, although 
pinpointing the correct underlying theory is more complex.  This program is currently pursued worldwide. 

It is clear that a multi-pronged approach is best and 
science must advance along all three frontiers to optimize the discovery of the ultimate laws of Nature. 

The US has the opportunity to be a world leader at the intensity frontier.  Several intensity frontier experiments have been 
proposed, approved, or are currently under construction.  These are located at laboratories across the US, with Fermilab playing the 
leading role.  The Office of High Energy Physics  in the DOE   Office of Science  charged this workshop \cite{intro:DOEcharge} with  identifying 
the science opportunities at the intensity frontier, exploring what can be learned  from such experiments, and describing the facilities  
required for such a program to reach its full potential.  The workshop was held in December 2011 in Rockville, MD with more than 500 
participants \cite{intro:IFW}.  This report fulfills the workshop's charge and delivers a cohesive description of the broad, diverse, and 
exciting science potential  at the intensity frontier. 

The intensity frontier physics program may be broken into six areas, each centered on the particles used as probes for new physics and 
represented at the workshop by a separate working group.  Research within these areas focuses on precision measurements of processes involving 
a specific type of particle  that can be performed with increasing accuracy.   This automatically includes nucleons, nuclei, and atoms, and 
therefore the nuclear physics community participated in this workshop.  Searches for new particles and forces that only interact very weakly 
with Standard Model particles and fields were also included, as these searches require very intense beams.  The resulting six working groups 
are listed in Table \ref{WGTable} along with a brief description of their physics programs.  The science represented by these working groups 
provides the definition of the intensity frontier.   The working groups themselves were ultimately responsible for the material included in 
the workshop and in this final report. 

\begin{table}[tb]
\centerline{\begin{tabular}{c|c}
\hline\hline
Working group name & Description\\ \hline
Heavy Quarks &  Study of rare processes and $CP$ violation in\\
 & strange, charm, and bottom quark systems\\ \hline
Charged Leptons & Study of rare processes and precision measurements\\
 & of the properties of the muon and tau leptons\\ \hline
Neutrinos & Physics opportunities associated with neutrino\\
 & oscillations and neutrinoless double beta decay\\ \hline
Proton Decay & Proton lifetime\\ \hline
New Light, Weakly Coupled Particles & Searches for new weakly coupled \\
  &forces and associated light particles\\ \hline
Nucleons, Nuclei, and Atoms & Searches for new physics utilizing precision\\
 & measurements of the properties of
nucleons, nuclei and atoms\\ \hline
\hline\hline
\end{tabular}}
\caption{List of working groups and their physics programs.}
\label{WGTable}
\end{table}

The intensity frontier is not simply all the areas of  particle physics outside  the energy or  cosmic frontiers. For example, experiments 
that study nucleon or nuclear structure with lepton scattering  are an important part of the high energy physics and nuclear physics portfolio, 
but were not included in the workshop.  Rather, the intensity frontier is defined by its focus on programs directed at discovering signatures 
of new physics.  While broad and diverse, these  programs are firmly  connected.  Multiple working groups address the same fundamental 
questions, such as:  Are there new sources of $CP$ violation?  Is there $CP$ violation in the leptonic sector?  What is the nature of 
neutrinos?  Do the forces unify?  Is there a weakly coupled hidden-sector related to dark matter?  Are apparent symmetries violated at 
very high energies?  In addition to these synergies within intensity frontier programs, the  science they pursue is also strongly linked 
to that of the energy and cosmic frontiers.

Each working group was charged with organizing their part of the workshop, engaging the appropriate community (both international and 
domestic), and summarizing their findings in this report. Each group was convened by an experimentalist and a theorist, as well as an 
observer.  The role of the observer was to provide independent, non-expert input and to engage scientists who were not necessarily part 
of the intensity frontier community.  The groups began their work during the fall of 2011 in preparation for the workshop.  Some working 
groups organized meetings before the workshop and requested short written contributions from the community.  The information contained in 
this report was gathered before and during the workshop with much input from the community.  Further information about this exercise can be 
obtained from the workshop webpage \cite{intro:IFW}.

Each  chapter of this report contains the findings of one of the six  working groups.  It  outlines the science opportunities  achievable at 
the intensity frontier during this decade, as well as what is possible next decade with existing, planned or proposed facilities.  It also 
describes the motivation for each program and how its measurements would further our knowledge.   The purpose of this report is not to 
advocate a particular facility, but to provide a coherent, cohesive picture of what can be learned from intensity frontier experiments over 
the next two decades. We hope this will provide useful input to any strategic decision making process in the US or elsewhere and contribute 
towards establishing a worldwide intensity frontier program in which  the US is a scientific leader.


\def\OMIT#1{{}}
\def\babar{\mbox{\slshape B\kern-0.1em{\footnotesize A}\kern-0.08em
  B\kern-0.1em{\footnotesize A\kern-0.12em R}}\xspace}
\newcommand{\Bbar}{\,\overline{\!B}{}}
\newcommand{\Dbar}{\,\overline{\!D}{}}
\newcommand{\Kbar}{\,\overline{\!K}{}}
\def\B0bar{\Bbar{}^0}
\def\D0bar{\Dbar{}^0}
\def\K0bar{\Kbar{}^0}
\def\rhobar{\bar\rho}
\def\etabar{\bar\eta}
\newcommand{\nn}{\nonumber}


\chapter{Heavy Quarks}

%
%
%

\bigskip

\begin{center}

Conveners: J.N.~Butler, Z.~Ligeti, J.R.~Patterson, J.L.~Ritchie

N.~Arkani-Hamed, 
D.M.~Asner,
A.J.~Bevan,
M.~Blanke,
G.~Bonvicini, 
R.A.~Briere, 
T.E.~Browder, 
D.A.~Bryman, 
P.~Campana, 
R.~Cenci,
N.H.~Christ, 
D.~Cline, 
J.~Comfort, 
D.~Cronin-Hennessy, 
A.~Datta, 
S.~Dobbs,
M.~Duraisamy, 
J.E.~Fast,
K.T.~Flood,
R.~Forty, 
T.~Gershon,
D.G.~Hitlin,
A.~Jawahery, 
C.P.~Jessop,
A.L.~Kagan, 
D.M.~Kaplan, 
M.~Kohl, 
P.~Krizan,
A.S.~Kronfeld, 
K.~Lee,
L.S.~Littenberg, 
D.B.~MacFarlane,
P.B.~Mackenzie,
B.T.~Meadows,
J.~Olsen, 
M.~Papucci, 
G.~Paz,
G.~Perez, 
K.~Pitts,
M.V.~Purohit,
B.N.~Ratcliff,
D.A.~Roberts,
J.L.~Rosner,
P.~Rubin,
B.~Schmidt,
A.~Schopper, 
A.J.~Schwartz, 
J.~Seeman, 
K.K.~Seth, 
S.R.~Sharpe, 
T.~Skwarnicki,
A.~Soni,  
S.~Stone,
R.~Sundrum, 
R.~Tschirhart, 
A.~Vainshtein,
R.S.~Van de Water, 
Y.W.~Wah,
G.~Wilkinson,
M.B.~Wise, 
J.~Xu, 
T.~Yamanaka, 
J.~Zupan

\end{center}



\section{Quark Flavor as a Tool for Discovery}
\label{sec:one}

An essential feature of flavor physics experiments is their ability to probe
very high mass scales, beyond the energy accessible in collider experiments.  
In addition, flavor physics can teach us about properties of TeV-scale new
physics, which cannot be learned from the direct production of new particles at
the CERN Large Hadron Collider (LHC).  This is because quantum effects allow virtual particles to modify the
results of precision measurements in ways that reveal the underlying physics.
(The determination of the $t - s,d$ couplings in the Standard Model (SM)
exemplifies how direct measurements of some properties of heavy particles may
only be possible in flavor physics.) Even as the LHC
embarks on probing the TeV scale, the ongoing and planned precision flavor
physics experiments are sensitive to beyond Standard Model (BSM) interactions at
mass scales that are higher by several orders of magnitude.  These experiments
will provide essential constraints and complementary information on the
structure of models put forth to explain any discoveries at the LHC, and they have
the potential to reveal new physics that is inaccessible to the LHC.

Throughout the history of particle physics discoveries made in studies of rare
processes have led to new and deeper understanding of Nature.  A classic example
is beta decay, which foretold the electroweak mass scale and the ultimate
observation of the $W$ boson.  A number of results from kaon decay experiments
were crucial for the development of the Standard Model:  the discovery of $CP$
violation in $K_L^0 \to \pi^+ \pi^-$ decay ultimately pointed toward the
three-generation CKM model~\cite{KM, C}, the absence of strangeness-changing
neutral current decays (i.e., the suppression of $K_L^0 \to \mu^+ \mu^-$ with
respect to $K^+ \to \mu^+ \nu$) led to the prediction of the fourth (charm)
quark~\cite{GIM}, and the measured value of the $K_L$\,--\,$K_S$ mass difference
made it possible to predict the charm quark mass~\cite{Gaillard-Lee, Vainshtein}
before charm particles were directly detected.  More recently the larger-than-expected 
$B_H$\,--\,$B_L$ mass difference foretold the high mass of the top
quark.  Precision measurements of time-dependent $CP$-violating asymmetries in
$B$-meson decays in the \babar and Belle experiments firmly established the CKM
phase as the leading source of $CP$ violation observed to date in flavor-changing
processes --- leading to the 2008 Nobel Prize for Kobayashi and Maskawa.  At the
same time, corrections to the SM at the tens of percent level are still allowed,
and many extensions of the SM that were proposed to solve the hierarchy problem
are likely to give rise to changes in flavor physics that may be observed in the
next generation of experiments.

Today, a well-planned program of flavor physics experiments --- using strange,
charm, and bottom quarks --- has the potential to continue this history of
producing paradigm-changing scientific advances. 

This report from the Heavy Quarks working group responds,
in the context of quark-flavor physics,
to the DOE charge to the Intensity Frontier Workshop to identify
experimental opportunities and to explain their potential, as
well as address the needed facilities.  This report is not a review of
quark-flavor physics, and no attempt has been made to
provide complete references to prior work.

\section{Strange, Charm, and Bottom Quarks as Probes of New Physics}
\label{sec:two}

\begin{figure}[b!]
\centerline{\includegraphics*[height=5.35cm]{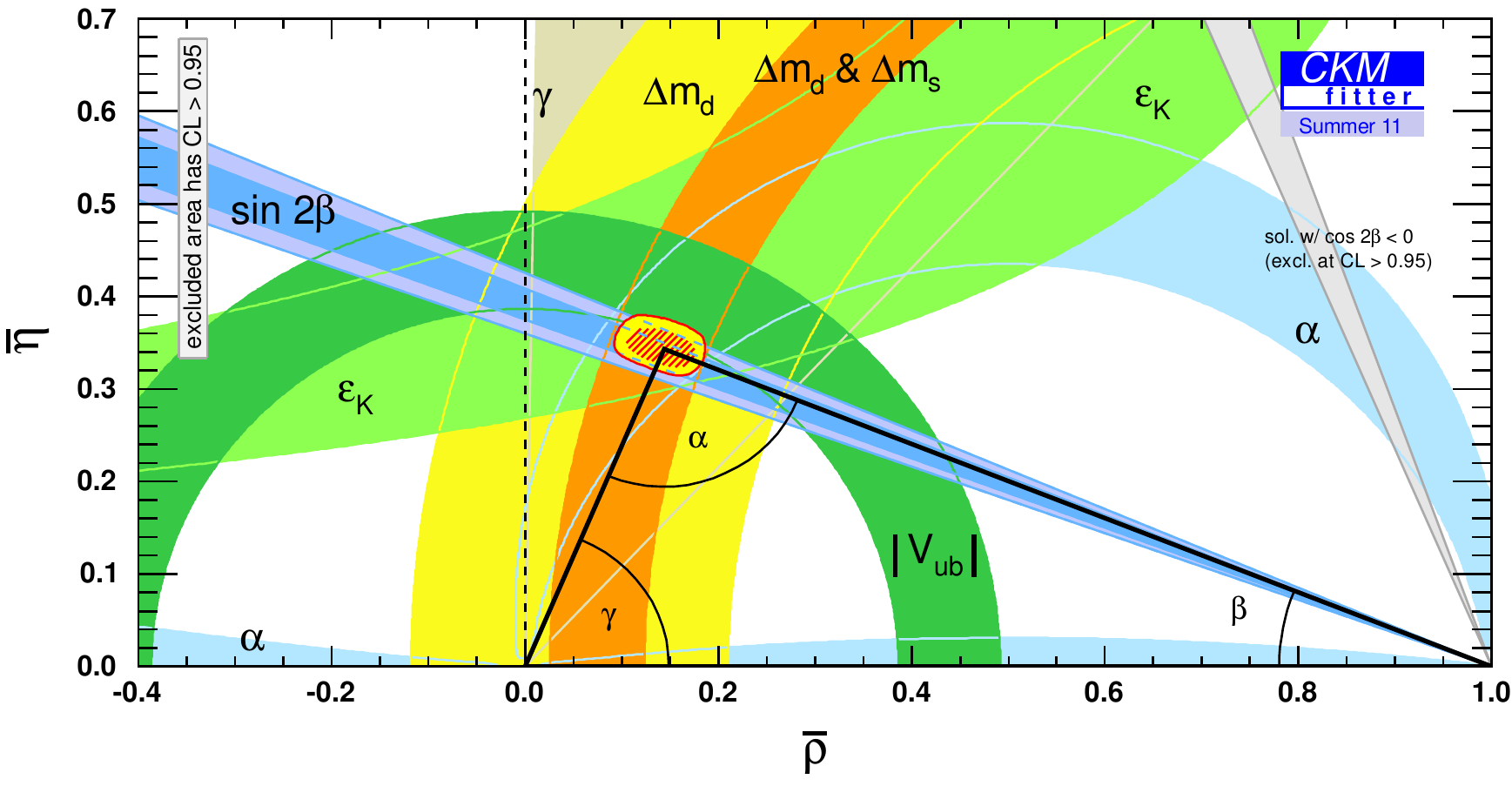} \hfill
\raisebox{5pt}{\includegraphics*[height=5.35cm]{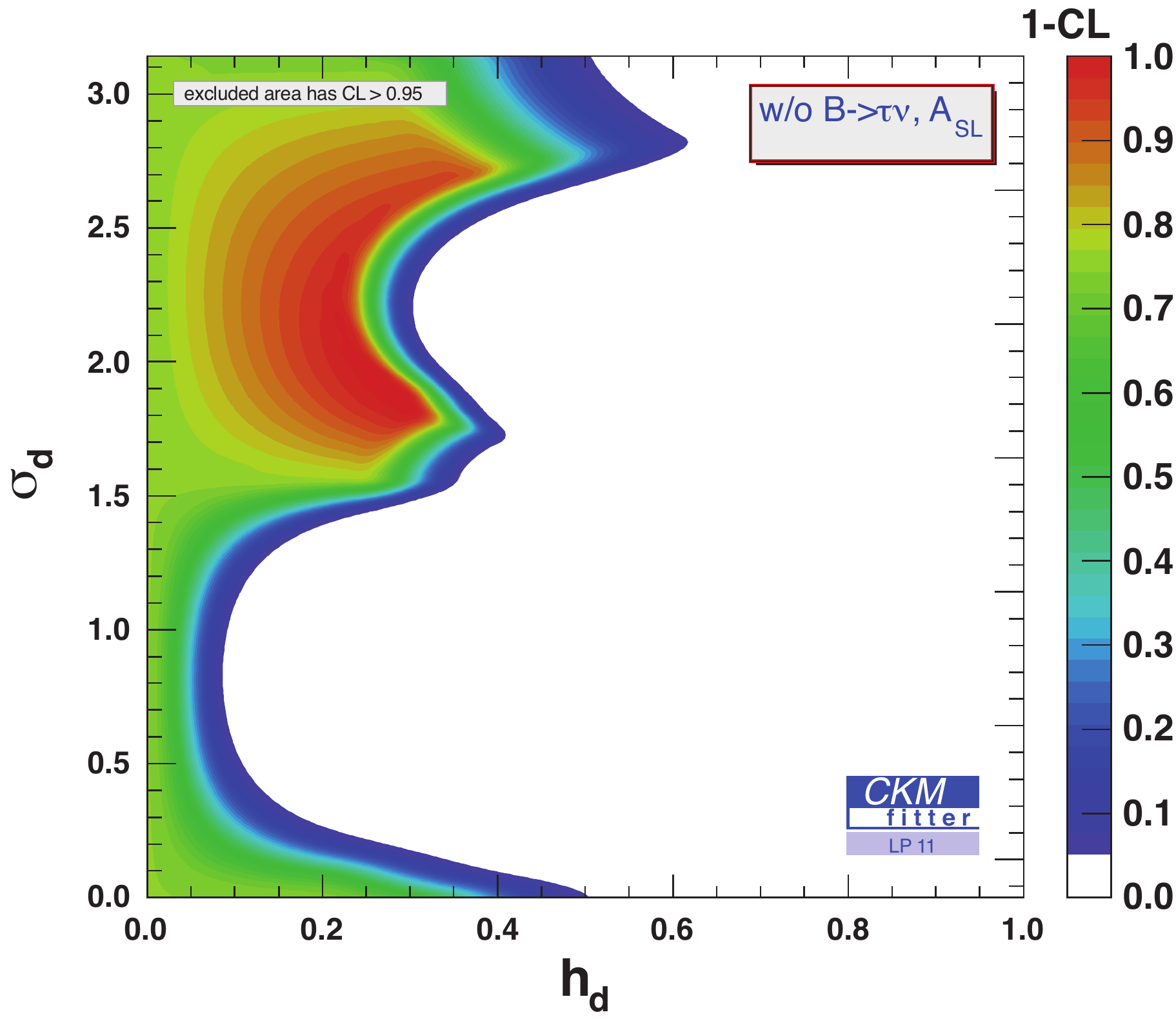}}}
\caption{Left: Constraints on the apex of the unitarity triangle in the
$\rhobar-\etabar$ plane (at 95\% CL). 
Right: the allowed $h_d-\sigma_d$ new physics parameter space (see text) in
$B^0$--$\B0bar$ mixing. (From~\cite{ckmfitter,Charles:2004jd}.)}
\label{fig:hdsd}
\end{figure}

In the past decade our understanding of flavor physics has improved very
significantly due to the $e^+e^-$ $B$ factories - \babar, Belle, and CLEO - and the
Fermilab Tevatron experiments.  While kaon physics was crucial for the development of the
SM, and has provided some of the most stringent constraints on BSM physics since
the 1960s, precision tests of the CKM picture of $CP$ violation in the kaon sector
have been hindered by theoretical uncertainties in calculating direct $CP$
violation in $K$ decay ($\epsilon'$).  The $B$ factories provided many stringent
tests by precisely measuring numerous $CP$-violating and $CP$-conserving quantities,
which in the SM are determined in terms of just a few parameters, but are
sensitive to different possible BSM contributions.  The consistency of the
measurements and their agreement with $CP$ violation in $K^0$--$\K0bar$ mixing,
$\epsilon_K$, and with the SM predictions (shown in the left plot in
Fig.~\ref{fig:hdsd}) strengthened the ``new physics flavor problem."  It is the
tension between the relatively low (TeV) scale required to stabilize the
electroweak scale and the high scale that is seemingly required to suppress BSM
contributions to flavor-changing processes.  This problem arises because the SM
flavor structure is very special, containing small mixing angles, and because of
additional strong suppressions of flavor-changing neutral-current (FCNC)
processes.  Any extension of the SM must preserve these features, which are
crucial for explaining the observed pattern of weak decays.

The motivation for a broad program of precision flavor physics measurements has
gotten even stronger in light of the 2011 LHC data.  With a hint at a particle
that may be a SM-like Higgs boson, but no sign of other high-mass states, the
LHC has begun to test naturalness as a guiding principle of BSM research.  If
the electroweak scale is unnatural, we have little information on the next
energy scale to explore (except for a hint at the TeV scale from dark matter, a
few anomalous experimental results, and neutrinos most likely pointing at a very
high scale).  The flavor physics program will explore much higher scales than
can be directly probed.  However, if the electroweak symmetry breaking scale is
stabilized by a natural mechanism, new particles should be found at the LHC. 
Since the largest quantum correction to the Higgs mass in the SM is due to the
top quark, the new particles will likely share some properties of the SM quarks,
such as symmetries and interactions.  Then they would provide a novel probe of
the flavor sector, and flavor physics and the LHC data would provide
complementary information.  Their combined study is our best chance to learn
more about the origin of both electroweak and flavor symmetry breaking. 

\begin{table}[tb]
\centerline{\begin{tabular}{c|cc|cc|c}
\hline\hline
\multirow{2}{*}{Operator} &
  \multicolumn{2}{c|}{Bounds on $\Lambda$~[TeV]~($C=1$)} &
  \multicolumn{2}{c|}{Bounds on $C$~($\Lambda=1$\,TeV) } & 
  \multirow{2}{*}{Observables}\\
&   Re  & Im  &  Re  &  Im  &  \\
\hline 
$(\bar s_L \gamma^\mu d_L )^2$  &~$9.8 \times 10^{2}$& $1.6 \times 10^{4}$
&$9.0 \times 10^{-7}$& $3.4 \times 10^{-9}$ & $\Delta m_K$; $\epsilon_K$ \\
($\bar s_R\, d_L)(\bar s_L d_R$)   & $1.8 \times 10^{4}$& $3.2 \times 10^{5}$
&$6.9 \times 10^{-9}$& $2.6 \times 10^{-11}$ &  $\Delta m_K$; $\epsilon_K$ \\
\hline
$(\bar c_L \gamma^\mu u_L )^2$  &$1.2 \times 10^{3}$& $2.9 \times 10^{3}$
&$5.6 \times 10^{-7}$& $1.0 \times 10^{-7}$ & $\Delta m_D$; $|q/p|, \phi_D$ \\
($\bar c_R\, u_L)(\bar c_L u_R$)   & $6.2 \times 10^{3}$& $1.5 \times 10^{4}$
&$5.7 \times 10^{-8}$& $1.1 \times 10^{-8}$ &  $\Delta m_D$; $|q/p|, \phi_D$\\
\hline
$(\bar b_L \gamma^\mu d_L )^2$    &  $5.1 \times 10^{2}$ & $9.3
\times 10^{2}$ &  $3.3 \times 10^{-6}$ &
$1.0 \times 10^{-6}$ & $\Delta m_{B_d}$; $S_{\psi K_S}$  \\
($\bar b_R\, d_L)(\bar b_L d_R)$  &   $1.9 \times 10^{3}$ & $3.6
\times 10^{3}$ &  $5.6 \times 10^{-7}$ &   $1.7 \times 10^{-7}$
&   $\Delta m_{B_d}$; $S_{\psi K_S}$ \\
\hline 
$(\bar b_L \gamma^\mu s_L )^2$    &  $1.1 \times 10^2$ & $2.2 \times 10^2$  &
 $7.6\times10^{-5}$  &  $1.7\times10^{-5}$  & $\Delta m_{B_s}$; $S_{\psi\phi}$ \\
($\bar b_R \,s_L)(\bar b_L s_R)$  &  $3.7 \times 10^2$ & $7.4 \times 10^2$ &
 $1.3\times10^{-5}$ &  $3.0\times10^{-6}$  & $\Delta m_{B_s}$; $S_{\psi\phi}$ \\ 
\hline\hline
\end{tabular}}
\caption{Bounds on $\Delta F=2$ operators of the form $(C/\Lambda^2)\, {\cal
O}$, with ${\cal O}$ given in the first column. The bounds on $\Lambda$
assume $C=1$, and the bounds on $C$ assume $\Lambda=1$\,TeV. (From~\cite{Isidori:2010kg}.)}
\label{tab:DF2}
\end{table}

Consider, for example, a model in which the only suppression of new
flavor-changing interactions comes from the large masses of the new particles
that mediate them (at a scale $\Lambda\gg m_W$).  Flavor physics, in particular
measurements of meson mixing and $CP$ violation, put severe lower bounds on
$\Lambda$.  For some of the most important four-quark operators contributing to
the mixing of the neutral $K$, $D$, $B$, and $B_s$ mesons, the bounds on the
coefficients $C/\Lambda^2$ are summarized in Table~\ref{tab:DF2} (for 
$S_{\psi\phi}$ we use the LHCb result).  For $C=1$, they are at the scale
$\Lambda \sim (10^2-10^5)$\,TeV. Conversely, for $\Lambda = 1$\,TeV, the
coefficients have to be extremely small. Therefore, there is a tension.  The
hierarchy problem can be solved with new physics at $\Lambda \sim 1$\,TeV. 
Flavor bounds, however, require much larger scales or tiny couplings.  This
tension implies that TeV-scale new physics must have special flavor structures,
{\it e.g.}, possibly sharing some of the symmetries that shape the SM Yukawa
interactions. The new physics flavor puzzle is thus the question of why, and in
what way, the flavor structure of the new physics is non-generic. As a specific
example, in a supersymmetric extension of the SM, there are box diagrams with
winos and squarks in the loops.  The size of such contributions depends
crucially on the mechanism of Supersymmetry (SUSY) breaking, which we would like to probe.  

To be sensitive to BSM contributions to FCNC processes (where the SM is
suppressed, but not absent), many measurements need to be done, and it is only
their combination that can reveal a signal.  (There are some exceptions, mainly
processes forbidden in the SM, but   considering only those would reduce the
sensitivity of the program to BSM physics.)  To visualize the constraints from
many measurements, it is convenient to use the Wolfenstein
parameterization~\cite{Wolfenstein:1983yz} of the CKM matrix (for a review,
see~\cite{Hocker:2006xb}),
\beq\label{ckmdef}
V_{\rm CKM} = \left( \begin{array}{ccc}
  V_{ud} & V_{us} & V_{ub} \\
  V_{cd} & V_{cs} & V_{cb} \\
  V_{td} & V_{ts} & V_{tb} \end{array} \right)
= \left( \begin{array}{ccc}
  1-\frac{1}{2}\lambda^2 & \lambda  &  A\lambda^3(\rhobar-i\etabar) \cr
  -\lambda  &  1-\frac{1}{2}\lambda^2  &  A\lambda^2 \cr
  A\lambda^3(1-\rhobar-i\etabar)  &  -A\lambda^2  &  1 \end{array} \right)
  + {\cal O}(\lambda^4) \,.
\eeq
It exhibits the hierarchical structure of the CKM matrix by expanding in a small
parameter, $\lambda \simeq 0.23$.  The unitarity of this matrix in the SM
implies many relations, such as that defining the ``unitarity triangle" shown in
Fig.~\ref{fig:hdsd}, which arises from rescaling the $V_{ud}\, V_{ub}^* +
V_{cd}\, V_{cb}^* + V_{td}\, V_{tb}^* = 0$ relation by $V_{cd}\,V_{cb}^*$ and
choosing two vertices of the resulting triangle to be $(0,0)$ and $(1,0)$.
(We use definitions of the $\lambda,\, A,\, \rhobar$ and $\etabar$ parameters
that obey unitarity and ensure that the apex of the unitarity triangle is
$(\rhobar,\etabar)$ exactly~\cite{Charles:2004jd}.)

As a result of second-order weak interaction processes, there are transitions
between the neutral meson flavor eigenstates, so the physical mass eigenstates
are their linear combinations, denoted as $|B_{H,L}\rangle = p |B^0\rangle \mp q
|\Bbar^0\rangle$.  (The $p$ and $q$ parameters differ for the four neutral
mesons, but the same notation is commonly used without distinguishing indices.) 
In a large class of models, the BSM physics modifies the mixing amplitude of
neutral mesons, and leaves tree-level decays unaffected.  This effect can be
parameterized by just two real parameters for each mixing amplitude.  For $B^0 -
\B0bar$ mixing, defining $M_{12} = M_{12}^{\rm SM}\, \big(1 + h_d\,
e^{2i\sigma_d}\big)$, the constraints on $h_d$ and $\sigma_d$ are shown in the
right plot in Fig.~\ref{fig:hdsd}.  (Evidence for $h_d \neq 0$ would rule out
the SM.)  Only in 2004, after the first significant constraints on $\gamma$ and
$\alpha$ became available from \babar and Belle, did we learn that the BSM
contribution to $B$--$\Bbar$ mixing must be less than the SM
amplitude~\cite{Ligeti:2004ak, Charles:2004jd}.  The right plot in
Fig.~\ref{fig:hdsd} shows that order $10-20\%$ corrections to $|M_{12}|$ are
still allowed for (almost) any value of the phase of the new physics
contribution, and if this phase is aligned with the SM ($2\sigma_d = 0$ mod
$\pi$), then the new physics contribution does not yet have to be much smaller
than the SM one.  Similar conclusions apply to other neutral meson
mixings~\cite{Bona:2007vi, Lenz:2010gu}, as well as many other $\Delta F=1$ FCNC
transition amplitudes.

The fact that such large deviations from the SM are not yet excluded gives very
strong motivations to continue flavor physics measurements in order to observe
deviations from the SM predictions or establish a stronger hierarchy between the
SM and new physics contributions.

In considering the future program, the following issues~\cite{Grossman:2009dw}
are of key importance:

\begin{enumerate}\vspace*{-14pt}\itemsep -2pt

\item
What are the expected deviations from the SM predictions induced by
new physics at the TeV scale? \\
As explained above, TeV-scale new physics with generic flavor structure is ruled
out by many orders of magnitude.  However, sizeable deviations from the SM are
still allowed by the current bounds, and in many scenarios 
observable effects are expected.

\item
What are the theoretical uncertainties? \\
These are highly process dependent.  Some measurements are limited by
theoretical uncertainties (due to hadronic, strong interaction, effects), but in
many key processes the theoretical uncertainties are very small, below the expected
sensitivity of future experiments.

\item
What can 
be expected in terms of experimental precision?\\
The useful data sets can increase by a factor on the order of 100 (in most cases
10--1000), and will probe effects predicted by fairly generic BSM scenarios.

\item
What will the measurements 
reveal,
if deviations from the SM are, or are not, seen?\\
The flavor physics data will be complementary with the high-$p_T$  part of
the LHC program. The synergy of measurements can 
reveal a lot about what the
new physics at the TeV scale is, and what it is not.

\end{enumerate}\vspace*{-10pt}

This report concentrates on the physics and prospects of a subset of measurements
for which the answers to these questions are the clearest, both in terms of
theoretical cleanliness and experimental feasibility.  The experiments will
enable many additional measurements that are not discussed here, some due to
lack of space and some because they will be more important than 
can now be anticipated.
(Recall that the best measurements of the CKM angles $\alpha$ and
$\gamma$ at \babar and Belle were not in formerly expected decay modes.)
Theoretical research, including lattice QCD, 
is also an important part of this program because 
future progress will add to the list of observables that can be used as probes for
BSM physics.  Nevertheless, the value of more sensitive experiments is not
contingent upon theoretical progress.

\boldmath
\subsection{$K$ Decays}
\label{sec:two.K}
\unboldmath

As can be seen from Table~\ref{tab:DF2}, some of the strongest constraints on
BSM physics come from the measurements of the $K_L-K_S$ mass difference $\Delta
m_K$ and the $CP$-violating quantities $\epsilon_K$ and $\epsilon'$.  This is
because the SM suppressions are the strongest in the kaon sector, since the $u$
and $c$ contributions to FCNC processes are very strongly suppressed by the
Glashow-Iliopoulos-Maiani (GIM) mechanism, while
that of the $t$ is strongly CKM suppressed.  Hence the agreement of the
measurements with the SM implies that new physics must mimic the SM
suppressions.  While $\epsilon_K$ can be calculated precisely, the hadronic
uncertainties in the SM calculation of $\epsilon'$ are particularly large,
because two terms with comparable magnitude and opposite sign contribute. 
Progress in lattice QCD may make $\epsilon'$ tractable in the future; at
present, however, we can neither rule out nor prove that it receives a
substantial new physics contribution.

In several rare FCNC kaon decays, such as those containing a charged lepton pair
in the final state, a challenge in learning about short distance physics is due to
long distance contributions via one or two photons converting into the
$\ell^+\ell^-$ pair.  However, the decays involving a $\nu\bar\nu$ pair in the
final state are theoretically clean, providing very interesting channels in which to
search for BSM physics.  The $K^+\to \pi^+\nu\bar\nu$ and $K^0_L\to
\pi^0\nu\bar\nu$ decays are determined by short distance physics, and there is a
single operator (both in the SM and in most BSM scenarios) that determines the
decay rates, ${\cal O} = X\, (\bar s d)_V\, (\bar\nu\nu)_{V-A}$. Moreover, the
form factor that parameterizes the matrix element of this operator is the same
as the one measured in $K\to \pi\ell\nu$ decay, in the limit of isospin
symmetry.  The decay rate ${\cal B}(K^+\to \pi^+\nu\bar\nu)$ is proportional to
$|X|^2$, and ${\rm Re}(X)$ gets a contribution from a penguin diagram with a
charm loop.  This contribution has been calculated to next-to-leading order, and
is responsible for the slightly larger irreducible theoretical uncertainty in the
charged mode than in the neutral mode.  The $K^0_L \to \pi^0 \nu \bar\nu$ rate is
even cleaner theoretically, because the final state is almost completely
$CP$-even~\cite{Littenberg:1989ix}, so the decay proceeds dominantly through $CP$
violation in the interference of decay with and without
mixing~\cite{Grossman:1997sk,Buchalla:1998ux}.  The rate is determined by ${\rm
Im}(X) \propto {\rm Im}[(V_{td}V_{ts}^*) / (V_{cd}V_{cs}^*)]\}$.  Both decay
rates are proportional to $(A\lambda^2)^4$, which could, however, be canceled by
taking a ratio of rates, substantially reducing the uncertainties.  The
constraint from a future measurement of ${\cal B}(K^0_L \to \pi^0 \nu\bar\nu)$
would be two horizontal bands centered at a certain value of
$\pm|\overline\eta|$.  At present, the experimental uncertainty of ${\cal
B}(K^+\to \pi^+\nu\bar\nu)$ is ${\cal O}(1)$, while the bound on ${\cal
B}(K^0_L\to \pi^0\nu\bar\nu)$ is $10^3$ times the SM prediction, leaving a lot
of room for future experiments to find unambiguous signs of BSM physics.

An important synergy with $B$-decay measurements is due to the fact that all
three observables $\epsilon_K$, ${\cal B}(K^+\to \pi^+\nu\bar\nu)$, and ${\cal
B}(K^0_L\to \pi^0\nu\bar\nu)$ depend on $|V_{td} V_{ts}|^2$, which is
proportional to $A^4$, which in turn is determined by $|V_{cb}|^4$.  This
provides a strong motivation to improve the determination of $|V_{cb}|$ at the
super flavor factories, combining the much larger data sets with theoretical
improvements.

Lattice QCD is also important for the kaon program.  For $\epsilon_K$, the
determination of the bag parameter, $B_K$, has improved remarkably in the last 
decade~\cite{Bae:2011ff}, and it is hoped that $\epsilon'$ may also become
tractable in the future.  A lattice QCD determination of the charm loop
contribution to $K^+\to \pi^+\nu\bar\nu$ would also be worth pursuing.  And, of
course, lattice QCD is important for determining $|V_{cb}|$ from semileptonic
$B$ decays.

The next generation of kaon experiments will not only measure $K\to
\pi\nu\bar\nu$, but also perform a much broader program, which includes $K\to
\pi\ell^+\ell^-$, $K\to \ell\bar\nu$, $CP$-violating triple products, and many
other interesting measurements sensitive to BSM physics.

\boldmath
\subsection{$B$ and $B_s$ Decays}
\label{sec:two.B}
\unboldmath

The $B$ physics program is remarkably broad, with many measurements sensitive to
complementary ways of extending the SM (its Higgs sector, gauge sector, or
fermion sector).  Here the focus is on a subset of measurements that can be
improved by an order of magnitude or more, and whose interpretations are
definitely not limited by hadronic uncertainties.  Particularly promising
channels to search for new physics are in mixing and in FCNC decays, where the SM
contributions are suppressed, so BSM contributions originating at a higher scale
may compete.  We saw that BSM contributions of order 20\% of the SM ones are
still allowed in most FCNC processes, and improving these constraints will be
important for interpreting the LHC results.  

In this program, the determinations of $\gamma$ and $|V_{ub}|$ are crucial,
because they are obtained from tree-level processes and hence provide a
``reference" determination of the CKM matrix (i.e., $\rhobar$ and $\etabar$, the
apex of the unitarity triangle) to which other measurements can be compared. 
There is ongoing theoretical work to improve the determination of $|V_{ub}|$,
using both continuum methods and lattice QCD, but it is not yet known if more
than a factor-of-a-few improvement will be possible.  At the same time, the
measurement of $\gamma$ from $B \to D K$ decays is limited only by statistics
(the current world average is $\gamma =
\big(68^{+10}_{-11}\big)^\circ$~\cite{ckmfitter}).  It is arguably the cleanest
measurement in terms of theoretical uncertainties, because the necessary
hadronic quantities can be measured. All $B \to D K$ based analyses consider
decays of the type $B \to D (\overline D)\, K\, (X) \to f_D\, K\, (X)$, where
$f_D$ is a final state accessible in both $D$ and $\overline D$ decay and $X$
denotes possible extra particles in the final state~\cite{Gronau:1990ra}.  The
crucial point is that the flavor of the $D$ or $\overline D$ in the intermediate
state is not measured, so the $b\to c \bar u s$ and $b\to u\bar c s$ decay
amplitudes can interfere.  Using several decay modes, one can perform enough
measurements to determine all relevant hadronic parameters, as well as the weak
phase $\gamma$.  Thus, the theoretical uncertainties are much below the
sensitivity of any foreseeable experiment.  A complementary method available at
LHCb, using the four time-dependent $B_s\to D_s^\pm K^\mp$ rates, has not been
tried yet.

The above tree-dominated measurements will allow improvements in the $CP$
asymmetry in $B\to J/\psi\, K_S$ and related modes, determining the angle
$\beta$, to improve the constraints on BSM physics.  In the $B_s$ system, the SM
prediction for $CP$ violation in the similar $b\to c\bar c s$ dominated decays,
such as $B_s \to J/\psi\, \phi$, is suppressed by $\lambda^2$ compared to
$\beta$, yielding for the corresponding time-dependent $CP$ asymmetry
$\beta_s^{\rm (SM)} = 0.0182 \pm 0.0008$.  While the Tevatron measurements
hinted at a possibly large value, the LHCb result, $\beta_s = -0.075\pm
0.095$~\cite{lhcb_psi_phi}, did not confirm those.  The key point is that the
uncertainty is still much larger than that of the SM prediction.  

An important search for new physics in penguin amplitudes comes from the
comparison of $CP$ asymmetries measured in tree-level $b\to c\bar cs$ dominated
decays with those in loop-dominated $b\to q\bar q s$ decays. The specific
measurements that probe such effects include the difference of $CP$ asymmetries
$S_{\psi K_S} - S_{\phi K_S}$ or related modes in $B_d$ decay, and $S_{\psi\phi}
- S_{\phi\phi}$ in $B_s$ decay.

There are some intriguing hints of deviations from the SM in the current data. 
$CP$ violation in neutral meson mixing, the mismatch of the $CP$ and mass
eigenstates, measured by the deviation of $|q/p|$ from 1, is simply $1-|q/p| =
2\,{\rm Re}(\epsilon_K)$ in the $K$ system.  It is sensitive to BSM
contributions in $B$ mesons, since $1-|q/p|$ is model-independently suppressed
by $m_b^2/m_W^2$, and there is an additional $m_c^2/m_b^2$ suppression in the
SM, which new physics may violate.  In $B_d$ mixing, the SM expectation for
$1-|q/p|$ is at the few times $10^{-4}$ level~\cite{Beneke:2003az}, while in
$B_s$ mixing it is suppressed in addition by $|V_{td}/V_{ts}|^2$ to $10^{-5}$. 
Thus, it is remarkable that the D\O\ experiment at the Tevatron 
measured the $CP$-violating dilepton asymmetry
for a mixture of $B_d$ and $B_s$ mesons at the $4\sigma$ level, $A_{\rm SL}^b =
(7.87 \pm 1.96) \times 10^{-3} \approx 0.6\, A_{\rm SL}^d + 0.4\, A_{\rm
SL}^s$~\cite{Abazov:2011yk}, where in each system $A_{\rm SL} \simeq
2(1-|q/p|)$.  It will be important at LHCb and at the super flavor factories to
clarify this situation by more precise measurements.  Since the hint of the
signal is much above the SM, there is a lot of room to find BSM contributions.

Another interesting tension in the current data is from the measurement of the
${\cal B}(B\to \tau\bar\nu)$ rate, which is about $2.5\sigma$ above the
prediction from the SM fit.  This comparison relies on a lattice QCD calculation
of the $B$ meson decay constant.  The simplest BSM explanation would be a
charged Higgs contribution, which in the type-II 2HDM is proportional to $m_b\,
m_\tau \tan^2\beta/m_H^2$.  It will require larger data sets at the future
$e^+e^-$ $B$ factories (and measuring the $B\to\mu\bar\nu$ mode as well) to
clarify the situation.

There is a nearly endless list of interesting measurements.  Many are in rare
decays involving leptons.  LHCb will be able to search for $B_s\to \ell^+\ell^-$
down to the SM level, at a few times $10^{-9}$.  This process received a lot of
attention in the last decade, after it was noticed that a SUSY contribution is
enhanced by $\tan^6\beta$.  With the LHCb upgrade and many years of super flavor
factory running, the search for $B_d\to \ell^+\ell^-$ may also get near the SM
level.  Rare decays involving a $\nu\bar\nu$ pair are theoretically very clean,
and the next generation of $e^+e^-$ machines should reach the SM level in $B\to
K^{(*)}\nu\bar\nu$; the current constraints are an order of magnitude weaker. 
There is also a long list of interesting measurements in $b\to s \gamma$ and
$b\to s \ell^+\ell^-$ mediated inclusive and exclusive decays, $CP$ asymmetries,
angular distributions, triple product correlations, etc., which will be probed
much better in the future.  And the $s\leftrightarrow d$ processes, with lower
SM rates, will provide many other challenging measurements and opportunities to
find new physics.  Furthermore, there is synergy and complementarity with
``hidden sector" particle searches, as explained in that working group report.

While any one of the above measurements could reveal new physics, the strongest
complementary information to the LHC will come not from one measurement, but from the
pattern in which they do or do not show deviations from the SM. In addition, the
experiments that carry out this program will also be able to search for charged
lepton flavor violation at an unprecedented level, {\it e.g.}, $\tau\to \mu \gamma$
and $\tau\to 3\mu$, discussed in the Charged Leptons working group report.
There is also a set of measurements for which our understanding of hadronic
physics is not yet good enough, but it could improve in the next decade.  A high-profile 
example is the difference of direct $CP$ asymmetries, $a_{K^+\pi^0} -
a_{K^+\pi^-} = 0.148 \pm 0.028$, which is expected to be small if corrections to
the heavy quark limit were under control.  Precise measurements at the super
flavor factories of other decay modes related by $SU(3)$ flavor symmetry will
help to clarify this situation and also teach us about hadronic physics.

\boldmath
\subsection{$D$ Decays}
\label{sec:two.D}
\unboldmath

\begin{figure}[t!]
\centerline{\includegraphics[width=0.4\textwidth,clip]{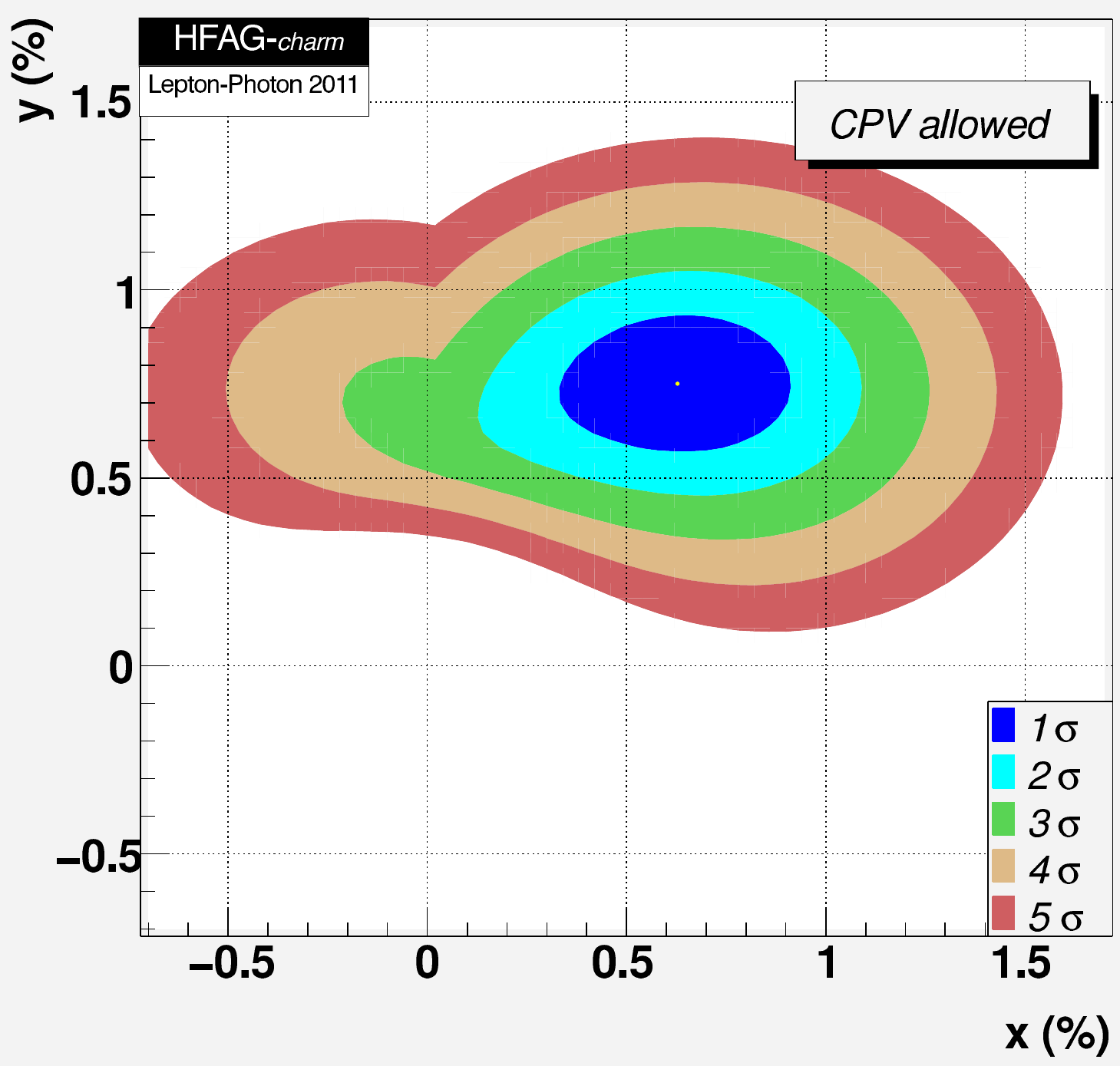}\hfil\hfil
\includegraphics[width=0.4\textwidth,clip]{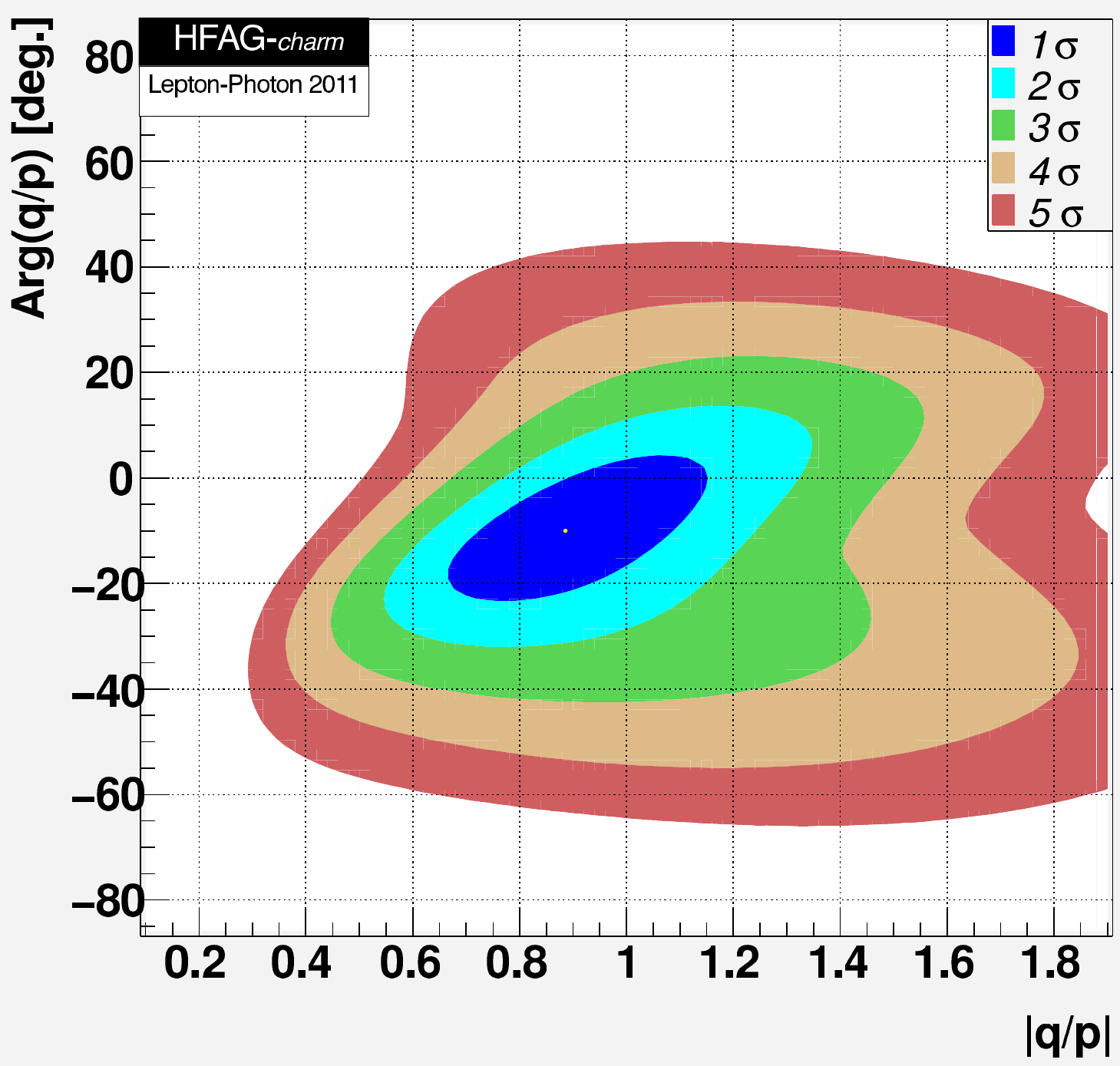}}
\caption{Results on charm mixing parameters $x$ and $y$, showing clear evidence
for mixing (deviation from $x=y=0$) (left); and the magnitude and
phase of $q/p$ (right).  (From~\cite{HFAG}.)}
\label{dmix}
\end{figure}

The $D$ meson system is complementary to $K$ and $B$ mesons because it is the
only neutral-meson system in which mixing and rare FCNC decays are generated by
down-type quarks in the SM loop diagrams.  This complementary sensitivity is
also present for new physics models.  For example, in supersymmetric theories
the gluino and neutralino contributions to
FCNC $K$ and $B$ transitions involve down-type squarks, whereas the $D$ system
is sensitive to the mixing of the up-type squarks in loop diagrams.  In the SM,
since the down-type quarks are much lighter than $m_W$ and the $2\times 2$
Cabibbo matrix is almost unitary, FCNC charm transitions and $CP$ violation in
charm decays are expected to be strongly suppressed.  Only since 2007 has there been
unambiguous evidence for $D^0$--$\D0bar$ mixing, and both $x = \Delta m/\Gamma$
and $y = \Delta\Gamma/(2\Gamma)$ are at or below the 0.01 level~\cite{HFAG}
(left plot in Figure~\ref{dmix}).  

The values of the mixing parameters can be accommodated in the
SM~\cite{Falk:2001hx}, and imply that long-distance physics is important.  
Nevertheless, the measurement of $\Delta m$ (the upper bound on it) already had
important implications for BSM.  For example, in supersymmetry, it was possible
to suppress FCNC $K$ and $B$ processes by aligning the quark and squark mixing
matrices~\cite{Nir:1993mx}, which predicted $x \sim \lambda^2 \sim 0.04$.  The
measurement of $\Delta m$ implies that if the first two squark doublets are
within the reach of the LHC, then they must be degenerate to some extent, since
quark-squark alignment alone cannot provide enough
suppression~\cite{Nir:2007ac}.

$CP$ violation in mixing, the deviation of $|q/p|$ from 1, is very sensitive to
BSM contributions in charm mixing as well.  The SM expectation is below the 0.01
level, while the current uncertainty of $|q/p|-1$ is about 0.2 (right plot in
Figure~\ref{dmix}, where ${\rm arg}(q/p)$ is ${\rm arg}\,\lambda_{K^+K^-}$). 
Thus, future measurements can improve the sensitivity to BSM contributions by an
order of magnitude before becoming limited by hadronic uncertainties.

Direct $CP$ violation has been observed in $K$ and $B$ decays, and was expected to
be at or below the $10^{-3}$ level in charm decays.  Recently LHCb announced a
$3.5\sigma$ evidence for direct $CP$ violation, a nonzero value of $\Delta a_{\rm
CP} \equiv a_{K^+ K^-} - a_{\pi^+ \pi^-} = -(8.2\pm 2.1\pm 1.1)\times
10^{-3}$~\cite{Aaij:2011in}, giving a world average $\Delta a_{\rm CP} =
-(6.5\pm 1.8)\times 10^{-3}$.  In the SM, $\Delta a_{\rm CP}$ is suppressed by
$|V_{cb} V_{ub}|/|V_{cs} V_{us}| \simeq 7 \times 10^{-4}$, so an order of
magnitude enhancement from hadronic physics or new physics is needed to explain
this central value~\cite{Isidori:2011qw, Golden:1989qx}.  To clarify the
situation, precise measurement in many modes, accessible in different
experiments, will be necessary~\cite{Isidori:2011qw, Pirtskhalava:2011va}.

There are many other measurements in charm decays that are
sensitive to new physics and important for the rest of the program.  These
include leptonic and semileptonic rates with much improved precision, testing
lattice QCD calculations, and learning about hadronic physics from charm
spectroscopy and glueball searches.  Experiments producing charm at threshold
can collect large samples of $CP$-tagged $D^0$ decays, which will be very useful
for high-precision measurements of the CKM angle $\gamma$.


\subsection{Effective Theories, Hadronic Physics, and Exotic States}
\label{sec:two.qcd}

Developments in improving the sensitivity to BSM physics and understanding QCD
are strongly connected.  Past experience shows that whenever an order of
magnitude more data becomes available, it always leads to renewed theoretical
activity aimed at understanding the strong dynamics, which often results in improvements
that increase the sensitivity of the measurements to new physics.  The history
of the field is full of unanticipated surprises that enriched this line of
research.

Model-independent tools to tackle some strong interaction phenomena are provided
by effective field theories, such as chiral perturbation theory (CHPT), heavy
quark effective theory (HQET), and soft-collinear effective theory (SCET). 
Lattice QCD allows first-principles calculations of certain hadronic matrix
elements (often using some effective theory methods as well), and 
improvements in lattice QCD calculations that are projected over the next few years
will substantially improve the discovery potential of future flavor
physics experiments.  By now, there are two or more realistic three-flavor
lattice-QCD calculations of all hadronic matrix elements needed for the SM CKM
fit (see, {\it e.g.},~\cite{Laiho:2009eu, Colangelo:2010et}).  The development
of all of these methods was motivated to a large extent by the desire to better
calculate $K$ and $B$ decay matrix elements, and they have provided fundamental
insights into the dynamics of QCD.  These techniques are important in other
parts of particle physics as well; {\it e.g.}, SCET is being applied to jet physics to
aid discoveries at the LHC, and lattice QCD is being applied to the
determination of matrix elements for dark-matter detection cross sections.

The spectrum of states containing heavy quarks has provided some of the most
important insights into the dynamics of QCD.  After decades in which heavy quark
spectroscopy was thought to amount to finding previously unobserved states
anticipated from the quark model picture, \babar and Belle discovered a large
number of unexpected states, as well as states with unexpected masses.  An
important open question is whether states other than mesons composed of
$q\overline{q}$ and baryons composed of $qqq$ are realized in nature.  Possible
``unconventional" combinations include  four-quark mesons,
$q\overline{q}q\overline{q}$ (tetraquarks), five-quark baryons,
$qqqq\overline{q}$ (pentaquarks), ``hybrids" consisting of ``valence" quarks and
gluons, ``glueballs" that are composed of gluons (with no quarks), and hadronic
``molecules."  Some of these states can have exotic quantum numbers, {\it e.g.},
isospins, hypercharges, or values of $J^{PC}$ that cannot be produced in the
quark model by $q\overline{q}$ or $qqq$ constituents.  Lattice QCD calculations
reproduce the spectrum of charmonium and bottomonium states and predict the
glueball spectrum. Many phenomenological models have also been developed to
explain various aspects of these states, and the recent experimental results
triggered lots of new theoretical research.

\section{A World-wide Program of Quark Flavor Experiments}
\label{sec:three}

\subsection{Kaon Experiments}
\label{sec:three.one}

As accelerators and detectors have advanced, the sensitivity of 
rare kaon decay experiments has 
also improved.  In the past the United States led in this arena.  
Kaon experiments that took data more than a decade ago at the Brookhaven National Laboratory
Alternating Gradient Synchrotron (AGS) approached 
the $10^{-11}$ level
of branching fraction sensitivity~\cite{E949,E865}, 
and in one case the $10^{-12}$ level~\cite{E871}.  With current and future accelerators, 
substantial improvements will be possible.  Experiments are now under way in Europe
at CERN and INFN Frascati, and in Japan at J-PARC.  While no experiments are under way in the
US, existing facilities at Fermilab can support
world-leading experiments today, and Project X has the potential to
make further significant improvements possible.
A summary of the forseeable experimental progress is given in
Table~\ref{tab:exptKsummary}, while the individual experimental initiatives are discussed below.

\begin{table}[tb]
\centerline{\begin{tabular}{c|c|c|l}
\hline\hline
{Observable} &
  {SM Theory} &
  {Current Expt.} & 
  {Future Experiments}\\
\hline 
${\cal B}(K^+ \to \pi^+ \nu \overline{\nu})$ 
& $7.8  \times 10^{-11}$
& $1.73^{+1.15}_{-1.05} \times 10^{-10}$ 
& $\sim$10\% measurement from NA62 \\
& & & $\sim$5\% measurement from ORKA \\
& &  & $\sim$2\%  with Project X \\
\hline
${\cal B}(K^0_L \to \pi^0 \nu \overline{\nu})$ 
& $2.43 \times 10^{-11}$  &$<2.6 \times 10^{-8}$&$ 1^{\rm st}$ observation from KOTO \\
&   &  & $\sim$5\% measurement with Project X \\
\hline
${\cal B}(K^0_L \to \pi^0 e^+ e^-)_{SD}$ 
& $1.4 \times 10^{-11}$  & $<2.8 \times 10^{-10}$& $\sim$10\% measurement with Project X\\
\hline
${\cal B}(K^0_L \to \pi^0 \mu^+ \mu^-)_{SD}$ 
& $3.5 \times 10^{-11}$  & $<3.8 \times 10^{-10}$ &  $\sim$10\% measurement with Project X \\
\hline
$|P_T|$ in $K^+ \to \pi^0 \mu^+ \nu$
& $ \sim 10^{-7}$ & $<0.0050$  & $<0.0003$ from TREK  \\
 &   &  & $<0.0001$ with Project X  \\
\hline 
$R_K=\Gamma(K_{e2})/\Gamma(K_{\mu2})$  
& $2.477 \times 10^{-5} $
& $(2.488 \pm 0.080)\times 10^{-5}$   &  $\pm 0.054 \times 10^{-5}$ from TREK \\
   &  &  & $\pm 0.025 \times 10^{-5}$  with Project X \\ 
\hline
${\cal B}(K^0_L \to \mu^\pm e^\mp)$ 
& $< 10^{-25}$  &$< 4.7 \times 10^{-12}$ & $< 2 \times 10^{-13}$ with Project X \\
\hline\hline
\end{tabular}}
\caption{A summary of the reach of current and proposed experiments for some key rare kaon decay measurements, in comparison to Standard Model theory and the current best experimental results.}
\label{tab:exptKsummary}
\end{table}

\subsubsection{KLOE-2}

The KLOE-2 experiment~\cite{KLOE-2} 
will run at the upgraded DA$\Phi$NE $e^+e^-$ storage ring at
INFN Frascati, and it
will extend the results of the earlier KLOE experiment.  The upgraded DA$\Phi$NE will
achieve a factor-of-three increase in instantaneous luminosity with a crab waist at
the interaction point, one of the innovations that will also be used to achieve
large luminosity gains for the super flavor factories.  A number of detector improvements
are being made for KLOE-2, including a new $\gamma\gamma$ tagging system, a new inner
tracker, new small-angle calorimeters, 
improved front-end electronics, and updated computing and software.
Ultimately KLOE-2 aims to collect an integrated luminosity of  $25 \, {\rm fb^{-1}}$,
an order of magnitude more than KLOE.

The KLOE-2 physics program exploits the correlated production of $K$ and $\overline{K}$
mesons in a $J^{PC} = 1^{--}$ state from $\phi$ decays, rather than achieving high 
sensitivity to rare decays (which is the domain of
experiments using kaon beams at proton accelerators).  KLOE-2 will be able
to improve neutral kaon interference measurements, leading to improved tests of $CPT$ and
quantum mechanics and refined measurements of mass and mixing parameters ($\Gamma_L$,
$\Gamma_S$, $\Delta m$) and $CP$ violation parameters.  It will also make a wide range of
measurements of non-leptonic 
and radiative $K$ and $\eta$/$\eta^\prime$ meson decays.

\subsubsection{NA62}

The NA62 experiment~\cite{NA62pro} has the goal of making a 
measurement of the $K^+ \to \pi^+ \nu \overline{\nu}$
branching fraction with uncertainty approaching 10\%.
It will run in the CERN SPS north area extraction line 
that housed the NA48 detector array, some components of which (in particular, 
the liquid krypton calorimeter) are being reused.
NA62 will utilize a high-intensity
(750~MHz)
unseparated charged beam (about 6\% $K^+$'s) to 
search for $K^+ \to \pi^+ \nu \overline{\nu}$
decays in flight.  It will be the first decay-in-flight experiment to search for this mode.
The projected sensitivity of the experiment would allow 
about 55 $K^+ \to \pi^+ \nu \overline{\nu}$
events to be collected per year at the SM branching fraction, 
with signal/noise of about 7/1.

Background rejection in this experiment requires precise measurements of
the incoming $K^+$ and outgoing $\pi^+$.  The former measurement is challenging
in a high-intensity beam, so that NA62 is developing a so-called gigatracker using silicon
pixel detectors.  The latter measurement will be performed by straw tracking chambers operated
in vacuum, in order to minimize multiple scattering in the decay region.  
High efficiency for vetoing
photons from $\pi^0$ decays is assisted by the relatively high beam energy 
and will be accomplished using a combination of different calorimeter 
technologies in different regions.
Very good $\pi$:$\mu$ separation is also required and will be achieved with 
a RICH counter
in combination with an instrumented hadron absorber.

Construction of the NA62 detector systems~\cite{NA62status} 
has been under way for about three years, and an engineering run of 
representative elements is scheduled for the second half of 2012.  Data taking 
is expected to begin in 2014, depending on the LHC upgrade schedule (since
the SPS will run only when the LHC is operating).

\subsubsection{KOTO}

The KOTO experiment~\cite{KOTOpro} will search for the  
$K_L^0 \to \pi^0 \nu \overline{\nu}$ at J-PARC.  It will reuse
parts of the detector of the
E391a experiment that ran at the KEK PS, along with significant modifications.  
E391a set the best
upper limit~\cite{E391a}
so far for this decay ($2.6 \times 10^{-8}$), which is three orders
of magnitude larger than the SM branching fraction.  The goal of KOTO is to
close that gap and to make the first observation of  
$K_L^0 \to \pi^0 \nu \overline{\nu}$.

The $K_L^0 \to \pi^0 \nu \overline{\nu}$ mode is particularly challenging because
the only observable particles are the two photons from the $\pi^0$ decay, and 
there are copious other sources of photons.  To obtain a kinematic constraint, KOTO will
have a tightly collimated neutral beam (a ``pencil beam") 
so that the reconstructed $\pi^0$ momentum component transverse to the beam 
direction can be used as a constraint. 
Imposing a requirement that the transverse momentum be relatively large
forces missing photons from background sources to have higher energies,
which makes them easier to detect.
Neutron interactions must be
suppressed, so most beam and detector volumes are evacuated.  Excellent efficiencies for detecting photons and charged particles from background decays are
achieved by surrounding the entire decay volume with active photon veto counters.

The KOTO experiment has many improvements over the E391a experiment.
At J-PARC the $K_L^0$ flux will be higher by a factor of up to 40,
while the $n/K_L^0$ ratio is expected to be lower by a factor of at least three due to
an improved neutral beamline.  
The CsI calorimeter has been replaced with smaller and longer CsI crystals from the
Fermilab KTeV experiment, to suppress backgrounds, and the data acquisition system is being
upgraded.
An engineering run and first physics running are planned for 2012.  Running with
100~kW beam power will not take place before 2014; subsequently
annual runs of approximately four months duration are expected.  To achieve sufficient sensitivity to observe a few events at the SM branching fraction, it will be 
necessary for KOTO to run for several years.

\subsubsection{TREK}

The TREK experiment~\cite{TREKpro} will run at J-PARC.  
The primary goal of TREK is a search for $T$-violation in
the decay $K^+ \to \pi^0 \mu^+ \nu$ via observation of muon polarization in the direction transverse 
to the $\pi$\,--\,$\mu$ decay plane with 20 times better precision than the 
prior best limit ($|P_T| < 0.005$)~\cite{E-246}, which is from KEK-PS experiment E-246.  
TREK will use the E-246 spectrometer after both detector and data acquisition upgrades.  
The experiment
will use stopped-$K^+$'s (i.e., a low-energy $K^+$ beam enters the detector and
a fraction of the $K^+$'s are brought to rest via $dE/dx$ at the center of the detector
in a scintillating fiber target).  
Charged decay products of the $K^+$ are subsequently detected in a toroidal spectrometer, combined with a calorimeter with large solid angle
to detect photons.  
Muons from $K^+ \to \pi^0 \mu^+ \nu$ stop inside a
muon polarimeter, which detects the direction of the positron in the 
$\mu^+ \to e^+ \nu_e \overline{\nu}_{\mu}$.  

The TREK design calls for a beam power of 270~kW for 30~GeV protons, which will not be 
available for several years.  Other measurements are possible with less beam.  The
ratio of decay rates $R_K = \Gamma(K^+ \to e^+ \nu )/\Gamma(K^+ \to \mu^+ \nu)$ tests
lepton-flavor universality.  The SM ratio depends only on kinematics (i.e., masses)
and small radiative corrections.  The current world average result for $R_K$ (from
NA62 and KLOE) agrees
with the SM expectation with an uncertainty of 0.4\%.  TREK expects to improve this
comparison to the 0.2\% level.  TREK also has the ability to search for a 
heavy sterile neutrino (N) in the decay $K^+ \to \mu^+ N$ down to a
branching ratio $10^{-8}$.  

TREK requires slow extraction from J-PARC and is expected to begin data taking in 2014
with beam power of 50~kW, which is adequate for the $R_K$ measurement and the
heavy neutrino search.

\subsubsection{ORKA}

The $K^+ \to \pi^+ \nu \overline{\nu}$ decay has only been observed so far
in Brookhaven experiments E787 and E949, which used stopped~$K^+$'s.  E949 was an upgrade of E787.  
These experiments
ran at the AGS in several short runs  between 1988 and 2002
(usually 10 to 16 weeks of running in a given year, which
was typical of AGS operations).   
Ultimately these
experiments observed seven signal events~\cite{E949} (with background
$0.93\pm0.17$ events).  In the end E949 did not reach its goal,
since it was terminated early due to lack of funding.  
Nonetheless, E949 demonstrated background 
rejection at the $2 \times 10^{-11}$ level, which is 
sufficient for a high-statistics measurement of ${\cal B }(K^+ \to \pi^+ \nu \overline{\nu})$.

The ORKA experiment~\cite{ORKApro}
at Fermilab would apply the same technique demonstrated in E787/E949, while
taking advantage of the longer running time per year, the higher beam flux possible 
with the Main Injector and
the large acceptance gains that are made 
possible by using updated detector technologies and modern
data acquisition systems.  The ORKA detector will be a completely modernized 
version of the original E949 detector and will benefit from several improvements.
These include increasing the length of
the detector to increase geometrical acceptance, a larger magnetic field to improve
tracking resolution, a new and improved range stack scintillator with higher
light yields, a thicker photon veto system to improve photon detection efficiency,
deadtime-less 
electronics, and a modern high-throughput data acquisition system.
Estimates of ORKA's sensitivity are based on extrapolations
from E949's measured performance, rather than simulations.  
Background rejection does not need to
be better than in E949 for ORKA to reach its goal.

The ORKA proposal received Stage~I approval at Fermilab
in December 2011.  The time scale for receiving final approval is not 
yet known.
If it is approved and funded soon, 
it should be possible to complete detector construction
and begin first data taking by the end of 2016.  The projected sensitivity would
allow ORKA to collect about 200 $K^+ \to \pi^+ \nu \overline{\nu}$ events
per year (at the SM level), enabling a branching fraction measurement with
5\% uncertainty after five years of running.  This would be a strong test
for new physics, since the theory uncertainty in the SM branching fraction will
be at the same level of uncertainty.

\subsubsection{Opportunities with Project X}

Project X at Fermilab~\cite{ProjectX} could provide extremely high intensity kaon beams with
a very well-controlled time structure.  The beam power available to produce
kaons (1500~kW) will be higher by an order of magnitude than any other kaon 
source in
the world. 
Since the proton kinetic energy would be
around 3~GeV, the kaon energy will be low.  While this may not be well-matched
to all experiments, for some it will be nearly optimal~\cite{xKphys}.  
In particular, Project~X
provides the most promising opportunity advanced so far to make a high-statistics
measurement of the $K_L^0 \to \pi^0 \nu \overline{\nu}$
branching fraction.

A challenge for $K_L^0 \to \pi^0 \nu \overline{\nu}$
is the unknown momentum of the incident $K_L^0$.  As discussed in the context
of KOTO, some compensation for this can be achieved by limiting the beam aperture
so that at least the $K_L^0$ direction of flight is known to good precision.
In addition, the precisely controlled beam pulses that can be delivered
by a continuous-wave (CW) linac make it possible to measure the $K_L^0$ momentum using time-of-flight
information.  The 500~MeV $K_L^0$'s typical of Project~X energies are ideal for
this measurement.  This provides a strong kinematic constraint that significantly
improves background rejection while maintaining larger acceptance than the
pure pencil-beam technique.  Initial estimates indicate that it may be possible
to collect as many as 200 $K_L^0 \to \pi^0 \nu \overline{\nu}$ events per year
in a Project~X experiment, making possible a measurement at the 5\% level after
about five years of data taking.

The existence of Project~X will surely stimulate initiatives focusing on other rare 
modes,
such as the lepton-flavor violating decays $K_L^0 \to \mu e$ and
$K^+ \to \pi^+ \mu e$.
Several rare $K$ decay measurements would be possible, some involving subtle effects,
such as interference measurements versus proper time of  $K^0_L$ and $K^0_S$  decaying
into a common $\pi^0 e^+ e^-$ final state.
Such interference measurements can possibly isolate the directly $CP$-violating 
component of the decay amplitude and provide complementary handles to interpret 
new physics that may be observed in  $K^0_L \to \pi^0 \nu \overline{\nu}$ decays.
The unprecedented intensity of a stopped $K^+$ beam from Project X
can be exploited to extend the TREK research program now at J-PARC 
to a sensitivity limited by theoretical uncertainties.   
This ultra-bright stopped $K^+$ source can also enable 
other precision measurements sensitive to new physics, 
such as anomalous polarization of muons in  $K^+ \to  \pi^+ \mu^+ \mu^-$ decays 
and more precise studies of $K^+ \to \pi^+ \nu \overline{\nu}$ decays,
including the measurement of the $\pi^+$ spectrum which can be 
used to test the underlying matrix
element.

Exploiting the opportunities provided by Project~X will also require detector
improvements, so R\&D is needed.  Some areas of importance are: 
ultra-low-mass
tracking detectors that can operate at high rates and in vacuum;
fine-grained, fast, scintillator-based shower counters read-out with 
high quantum efficiency photodetectors that can operate in high
magnetic fields and in vacuum; 
large-scale system time-of-flight resolution better than 20~ps;
high-rate $\gamma$-pointing calorimetry; and fully streaming ``triggerless"
data-acquisition technologies.

\boldmath
\subsection{$B$-meson Experiments}
\label{sec:three.two}
\unboldmath


\subsubsection{Super Flavor Factories}

When Kobayashi and Maskawa shared the Nobel Prize in 2008 for
``the discovery of the origin of the broken symmetry which
predicts the existence of at least three families of quarks in nature,"
it was widely acknowledged that the $B$-factory experiments ---
\babar at SLAC and Belle at KEK --- had provided the essential
experimental
confirmation.
The spectacular 
successes of the $B$-factories KEKB and PEP-II rested on two
important features of these accelerators:  
unprecedented high luminosities which allowed the experiments to
collect data samples on the $\Upsilon(4S)$ resonance
consisting of several hundred million 
$B \Bbar$ pairs, and asymmetric beam energies which made
it possible to measure rate asymmetries in
$B$ and $\Bbar$ decays as a function of
the proper decay time difference.
In addition, $e^+e^-$ collisions provide a relatively
clean environment
so that complex final states  can be reconstructed
(including
those with several daughters, $\pi^0$'s, $K_L^0$'s,
and even $\nu$'s), thereby enabling a broad program of
measurements.

KEKB achieved 
peak luminosity
of
$2.1 \times 10^{34} \, {\rm cm^{-2} s^{-1}}$ 
and an integrated luminosity of $1040 \, {\rm fb^{-1}}$
(i.e., just over 
$1.0 \, {\rm ab^{-1}}$).  
PEP-II achieved a peak of
$1.2 \times 10^{34} \, {\rm cm^{-2} s^{-1}}$ 
and integral of  $550 \, {\rm fb^{-1}}$,
before its running was terminated early due
to a funding crisis in the US\ in 2008.
Achieving these high luminosities
required accelerator advances in a number of areas,
including  
very high-current stored beams ($> 3 \, {\rm A}$),
very large numbers of bunches ($> 1000$),
bunch-by-bunch feedback systems, high-power radio-frequency (RF) systems,
and operational advances such as continuous injection;
KEKB also enhanced its luminosity by using crab cavities
to achieve head-on collisions.

Innovations in the last few years,
in part resulting from linear collider studies and 
light source development, make it possible
to achieve instantaneous luminosity close to 
$1 \times 10^{36} \, {\rm cm^{-2} s^{-1}}$.
This has
led to plans for ``super flavor factories."  
SuperKEKB will be built as an upgrade
to KEKB in Japan.  The new Italian Cabibbo
Laboratory, located near Frascati, 
will host a green field
project to build the Super$B$ collider.  
These machines will achieve dramatic luminosity gains
by making the beams very small at the collision point and, 
in the case of Super$B$,
by implementing a crab-waist crossing.  Beam currents will
be higher than in the $B$ factories, but only by a factor of
about two so that beam-associated backgrounds will not
follow the gains in luminosity.
These machines will  
collect data sets of $50$--$75 \, {\rm ab^{-1}}$.
The cross section on the $\Upsilon(4S)$ is $1.1 \, {\rm nb}$,
so the super flavor factory experiments will have access to
more than $5 \times 10^{10}$ $B \Bbar$ pairs.
This will open the
door to precise measurements of a large number of processes
that have the potential to reveal new physics.

\subsubsection{Physics Reach of Super Flavor Factories}

Complete discussions of the physics programs of the
super flavor factory experiments 
exist~\cite{Belle2physics, SuperBphysics}.  Only a
few highlights are discussed here.

\begin{figure}[tbp]
\centerline{
\includegraphics[width=0.4\textwidth]{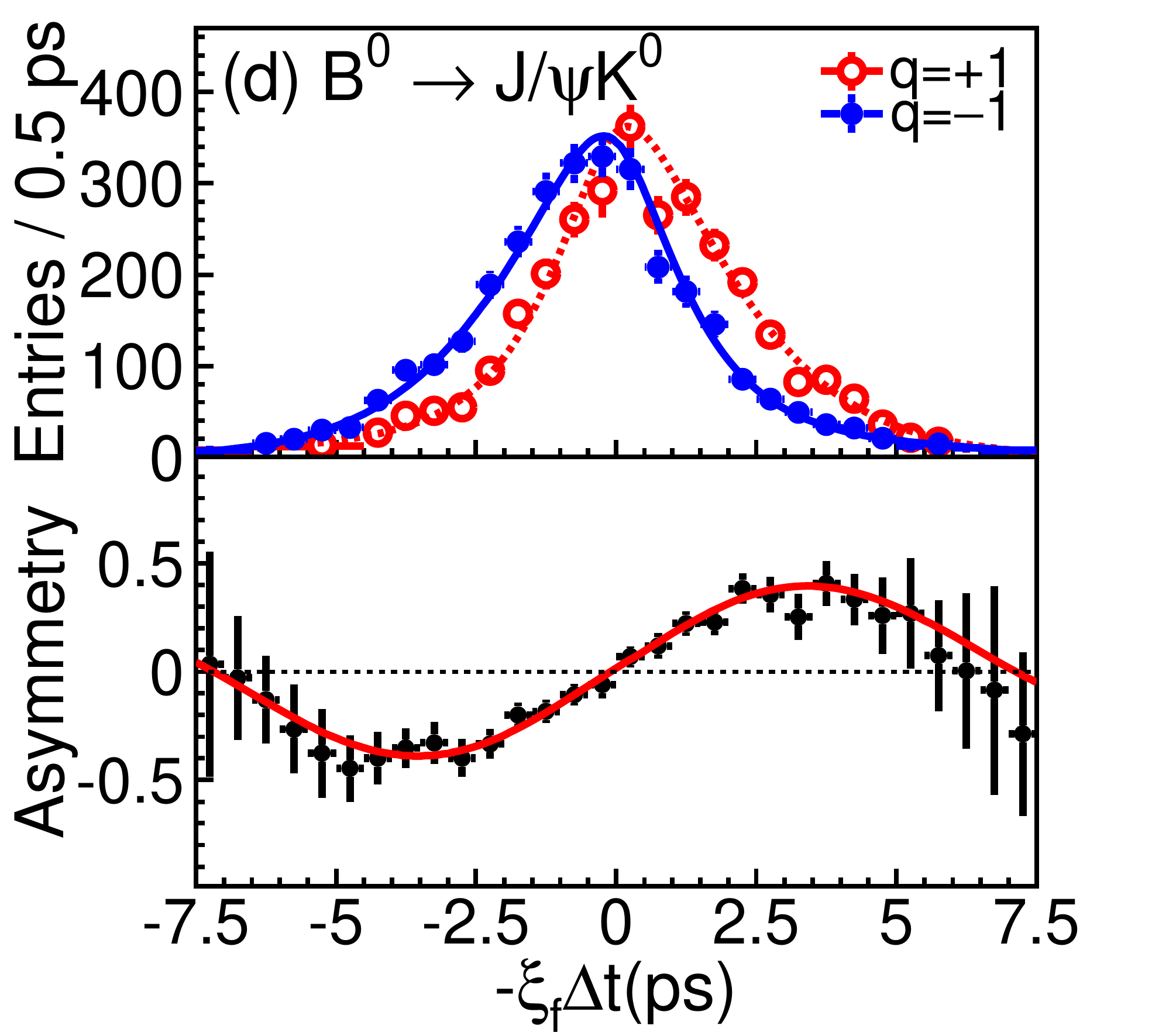}
\hspace{1.0cm}
\includegraphics[width=0.4\textwidth]{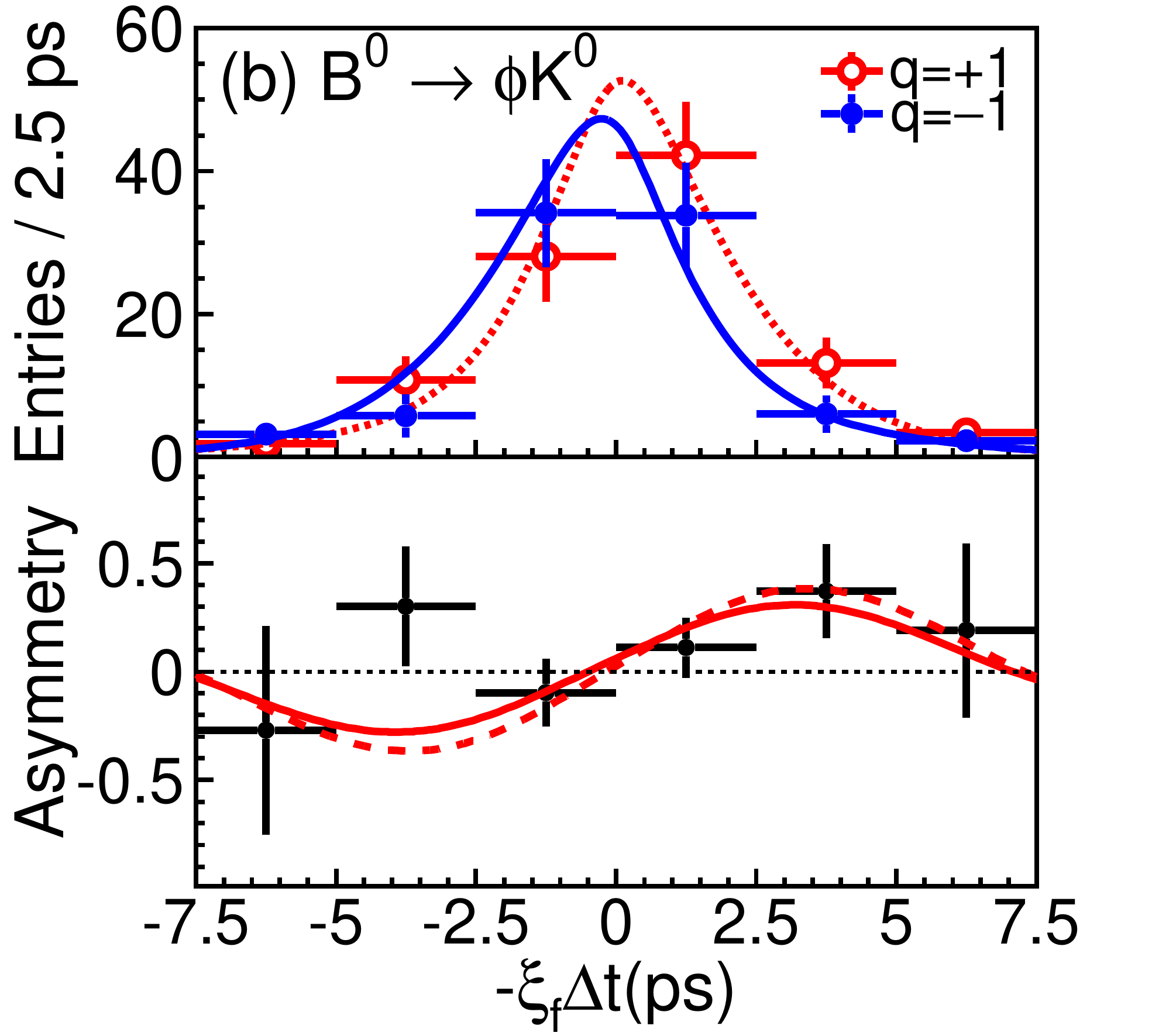}}
\caption{Belle measurements~\cite{Belle-TDCPfigs} of the time-dependent CP asymmetry
versus $\Delta t$ for $B \to J/\psi K^0$ (left) and $B \to \phi K^0$ (right).
$\sin(2\beta)$ is determined from the amplitude of the oscillations evident in the lower plots.
Super flavor factory experiments will obtain statistics for  $B \to \phi K^0$ 
(and other loop-dominated 
modes) as good as those obtained for $B \to J/\psi K^0$ in Belle and \babar. }
\label{fig:TDCPasymm}
\end{figure}

One strength of the super flavor factory experiments will be
their ability to search for non-SM sources of
$CP$~violation.  
The $B \Bbar$ pairs produced at the $\Upsilon(4S)$
are in a coherent quantum state, which allows the decay of one
$B$ to tag the state of the other.   Since
$B^0$ and $\B0bar$ may decay to the same
$CP$ eigenstate, the difference of $B^0$ and 
$\B0bar$ decay rates to a common final state is 
an observable for $CP$~violation.  
When measured versus time, the decay rate asymmetry is sensitive to
$CP$~violation that occurs in the interference between
two amplitudes --- those for $B^0 \to f_{\rm CP}$ and 
$B^0 \to \B0bar \to f_{\rm CP}$, where $f_{\rm CP}$ is the 
$CP$ eigenstate
and in the second instance the $B^0$ ``oscillates" into
$\B0bar$ before decaying.  This interference provides
direct access to underlying CKM parameters, since the
decay rate asymmetry versus time is a simple sine function
whose amplitude is $\sin(2 \beta)$, or equivalently 
$\sin(2 \phi_1)$ in the notation favored in Japan.  
The precision measurement of $\sin(2 \beta)$
is one of the keystone achievements of
the $B$-factory experiments; 
$\sin(2 \beta) = 0.678 \pm 0.020$ is the average~\cite{HFAG} 
of Belle and \babar
from decay modes resulting
from the quark level process $b \to c \overline{c} s$,
such as $B^0 \to J/\psi K^0$.
Since the $b \to c \overline{c} s$ decay is dominantly tree level, this is
effectively the SM value of $\sin(2 \beta)$.  However,
an analogous $\sin(2 \beta)$ measurement 
can be made using 
$b \to s \overline{s} s$ decays, 
such as $B^0 \to \phi K^0$ and $B^0 \to \eta^\prime K^0$,
which occur only through loops (i.e., penguin diagrams).
Loop processes open the door to additional amplitudes (and
complex phases)
from new heavy particles.  Comparison of such measurements
from \babar and Belle to the $b \to c \overline{c} s$ value
is statistically limited and inconclusive,
as illustrated in Figure~\ref{fig:TDCPasymm}.
Belle and \babar averages~\cite{HFAG} 
for $\sin(2 \beta)$ from
$B^0 \to \phi K^0$ and $B^0 \to \eta^\prime K^0$
are
$0.56 \pm 0.17$ and $0.59 \pm 0.07$, respectively.
The super flavor factory experiments 
can reduce the errors on these measurements by
an order of magnitude.  

Dramatically improved tests for direct $CP$~violation
in numerous modes will also be possible at the super flavor
factories.  One example is $B \to X_{s+d} \gamma$,
which results from electromagnetic penguin diagrams
for the quark level processes $b \to s \gamma$ and $b \to d \gamma$;
$X_{s+d}$ represents the hadronic system in these decays.
In a fully inclusive measurement (i.e., one that detects the
$\gamma$ but does not reconstruct the hadronic system 
in order to avoid losing complicated final states), the net flavor of
the $X_{s+d}$ is not determined.  In the Standard Model there
is a robust expectation that direct $CP$~violation is negligible;
that is, the decay rate for $B \to X_{s+d} \gamma$ equals
that for $\Bbar \to X_{s+d} \gamma$ to a very good approximation.
Any detected difference must be an indication of new physics, and
differences of up to 10\% appear in some non-SM  scenarios. 
The best measurement with existing $B$-factory 
data is consistent with no difference and has a 6\% error.
Super flavor factory experiments can reduce the error to below 1\%. 

Many rare $B$ decays that have either not been observed by
Belle or \babar, or that have been observed with only marginal
statistics, will become accessible in super flavor factory experiments.
An example is $B \to \tau \nu$, which in the SM results from
a simple $W$-exchange diagram and has a branching fraction of 
$(1.1 \pm 0.2) \times 10^{-4}$.
This mode is sensitive to supersymmetric models or others that 
predict the existence of a charged Higgs.
The current average branching fraction from \babar and Belle
is $(1.64 \pm 0.34) \times 10^{-4}$, in loose agreement with the
SM expectation.  Super flavor factory experiments can reduce
the error to about $0.04 \times 10^{-4}$.  This mode,
which has multiple neutrinos in the final state, 
is a good example of the power of $e^+ e^-$ experiments 
for $B$ physics.  The technique of reconstructing the ``other'' $B$
in the event can be very effective in reducing backgrounds for
modes in which the signal $B$ is impossible to reconstruct.  
 
Rare decay modes in which the underlying quark-level process is
$b \to s \ell^+ \ell^-$ or $b \to d \ell^+ \ell^-$ (where 
$\ell$ represents $e$ or $\mu$)
provide excellent sensitivity to new
physics because they occur through loop diagrams;
the former have branching fractions of order $10^{-6}$
and the latter of order $10^{-8}$.  Some of these modes,
such as $B^0 \to K^{*0} \mu^+ \mu^-$, can be collected in
very large numbers in hadronic production experiments,
making a good measurement of the
lepton forward-backward asymmetry $A_{\rm FB}$ possible.
However, a full exploration of these decays can be
accomplished only in the $e^+e^-$ environment.  Examples
of important measurements at which the super flavor
factories will excel 
include the inclusive 
decay rates versus
dilepton mass, comparisons of $e^+e^-$ modes to $\mu^+\mu^-$
as tests of universality, and searches for $CP$~violation in
these decays.

The processes discussed above provide only a glimpse of the
rich menu of incisive measurements that will be made by the
super flavor factory experiments running on the $\Upsilon(4S)$.
Running on the $\Upsilon(5S)$ will also give the super flavor factories
access to $B_s$ physics.  This may be important
if LHCb makes a measurement that is inconsistent with the
SM; that is, experimental confirmation
of important $B_s$ results may be needed.  Also, 
some interesting $B_s$ measurements will not be possible
in the hadronic environment, such as $B_s \to \gamma \gamma$
and other decays with neutral particles or neutrinos in the final
state.

\begin{table}[tb]
\centerline{\begin{tabular}{c|c|c|c}
\hline\hline   
{Observable} &
  {SM Theory} &
  {Current Expt.} & 
  {Super Flavor Factories}\\
\hline 
$S(B \to \phi K^0)$  &  $0.68$  &  $0.56 \pm 0.17$ & $ \pm 0.03$ \\
$S(B \to \eta^\prime K^0)$  & $0.68$  & $0.59 \pm 0.07$  & $\pm 0.02$ \\
$\gamma$ from $B \to D K$  &  &  $\pm 11^\circ$ &  $\pm 1.5^\circ$ \\
$A_{\rm SL}$ & $-5 \times 10^{-4}$ & $-0.0049 \pm 0.0038$  & $\pm 0.001$ \\
$S(B \to  K_S \pi^0 \gamma)$ & $< 0.05$  & $-0.15 \pm 0.20$ & $\pm 0.03$ \\
$S(B \to  \rho \gamma)$ & $<0.05$  & $-0.83 \pm 0.65$ & $\pm 0.15$  \\
$A_{\rm CP}(B \to  X_{s+d}  \gamma)$  
  &  $<0.005$ &  $0.06 \pm 0.06$ & $\pm 0.02$  \\
\hline
${\cal B}(B \to  \tau \nu)$ 
& $1.1 \times 10^{-4}$   & $(1.64 \pm 0.34) \times 10^{-4}$ & $\pm 0.05 \times 10^{-4}$ \\
${\cal B}(B \to  \mu \nu)$ 
& $4.7 \times 10^{-7}$   & $< 1.0 \times 10^{-6}$ & $\pm 0.2 \times 10^{-7}$ \\
${\cal B}(B \to  X_s \gamma)$ 
& $3.15 \times 10^{-4}$  & $(3.55 \pm 0.26)\times 10^{-4}$ & $\pm 0.13 \times 10^{-4}$  \\
${\cal B}(B \to X_s  \ell^+ \ell^- )$ 
& $1.6 \times 10^{-6}$   & $(3.66 \pm 0.77) \times 10^{-6}$ & $\pm 0.10 \times 10^{-6}$  \\
${\cal B}(B \to  K \nu \overline{\nu})$ 
& $3.6 \times 10^{-6}$  & $<1.3 \times 10^{-5}$ & $\pm 1 \times 10^{-6}$ \\
$A_{\rm FB}(B \to  K^* \ell^+ \ell^-)_{q^2 < 4.3 \, {\rm GeV^2}}$ 
& $-0.09$ & $0.27 \pm 0.14$  & $\pm 0.04$  \\
\hline\hline
\end{tabular}}
\caption{A summary of the reach of the planned super flavor factory experiments for some key B decay measurements, in comparison to Standard Model theory and the current best experimental results.
Normally Belle~II assumes $50 \, {\rm ab^{-1}}$ for such comparisons, while
Super$B$ assumes $75 \, {\rm ab^{-1}}$.  For this table, $50 \, {\rm ab^{-1}}$ has been assumed.}
\label{tab:eeBsummary}
\end{table}

\subsubsection{Belle II at SuperKEKB}

The SuperKEKB project~\cite{kekbII_machine} in Japan is under construction.  
Commissioning of the accelerator is expected to begin in 2014.
The design luminosity is $8 \times 10^{35} \, {\rm cm^{-2}s^{-1}}$
(40 times larger than KEKB),
which will allow an integrated luminosity of $50 \, {\rm ab^{-1}}$
to be accumulated in five years of running.

The Belle~II detector will be an upgraded version of Belle
that can handle the increased backgrounds associated with
higher luminosity.  The inner vertex detector will
employ DEPleted Field Effect (DEPFET) pixels, inside 
tracking layers that consist of double-sided silicon
strips with high-speed readout.  There will also be a 
new small-cell drift chamber.  The particle identification
system will be a ring-imaging Cherenkov (DIRC-type) detector.  The CsI calorimeter
will be retained, but it will be instrumented with 
waveform sampling readout.  The outer $K^0_L$/$\mu$~detector
will be upgraded to use scintillator to accommodate the 
higher rates.  Belle~II should be ready to roll in by the
end of 2015, after commissioning of SuperKEKB is completed.
The US groups on Belle~II are focusing their efforts on
the particle identification and $K^0_L$/$\mu$ systems.

\boldmath
\subsubsection{Super$B$ in Italy}
\unboldmath

The Super$B$ project~\cite{superb_machine} has been approved by the Italian government
and will be sited at the new Cabibbo Laboratory, which is
at the University of Rome Tor Vergata near Frascati.
The design luminosity will be 
$1 \times 10^{36} \, {\rm cm^{-2}s^{-1}}$.
It is hoped to begin commissioning in 2016.

The Super$B$ detector is based on the \babar detector,
and large parts of \babar will be re-used: the superconducting
coil and steel flux return, the quartz bars from the DIRC,
and the barrel CsI crystals.  New tracking detectors will
be build, including a silicon strip vertex detector whose
inner layer (very close to the beam) will initially be silicon striplets,
which will be upgraded to pixels for high-luminosity running after the
first few years,
and a new central drift chamber.  The DIRC readout will utilize
faster photodetectors, and the CsI barrel calorimeter will be augmented
by a forward calorimeter using LYSO crystals which are much
faster and more radiation hard than CsI.  The flux return
will be augmented with additional absorber to improve the
muon identification.

While large US contributions to Super$B$ are planned in the form of
PEP-II components and \babar components, the status of US
physicist participation is currently unsettled.


\boldmath
\subsubsection{$B$ Physics at Hadron Colliders}
\unboldmath

Hadron colliders have great potential for studying the decays of particles
containing charm and bottom quarks. The production cross sections are quite
large and the machine luminosities are very high, so more than 10 kHz of
$b$-hadrons can be produced per second. This is a much higher production rate
than can be achieved even in the next generation $e^{+}e^{-}$ $B$ factories. 
All species of $b$-flavored hadrons, including $ B_{s} $, $ B_{c} $, and $b$
baryons, are produced.  However, compared to $e^{+}e^{-}$ $b$ and charm
factories, the environment is much more harsh for experiments. At hadron
colliders, the $b$ quarks are accompanied by a very high rate of background
events; they are produced over a very large range of momenta and angles; and
even in $b$-events of interest there is a complicated underlying event. The
overall energy of the center of mass of the hard scatter that produces the $b$
quark, which is usually from the collision of a gluon from each beam particle,
is not known, so the overall energy constraint that is so useful in
$e^{+}e^{-}$ colliders is not available.  These features translate into
difficult challenges in triggering, flavor tagging, and particle identification,
and limit the overall efficiency and background rejection that can be achieved.

The CDF and D\O\ experiments at the Fermilab Tevatron demonstrated that these
problems could be successfully addressed using precision silicon vertex
detectors and specialized triggers.  While these experiments were mainly
designed for high-$p_{T}$ physics, they nevertheless made major contributions to
bottom  and charm physics~\cite{cdf_b_results, d0_b_results}.  Highlights of
their $B$-physics program include the first measurement of $B_{s}$
mixing~\cite{cdf_bs_mixing}; possible deviations from the SM predictions for the
asymmetry between $\mu^{+}\mu^{+}$ and $\mu^{-}\mu^{-}$ from the semileptonic
decays of $B$ mesons from the D\O\ experiment~\cite{d0_like_sign_dimu};
observation of many $B_{s}$ and $b$-baryon decay modes and measurement of the
$B_{s}$ and $\Lambda_{b}$ lifetimes; bottomonium spectroscopy; and the first
observation of the  $B_{c}$ meson and measurement of its
lifetime~\cite{cdf_Bc, d0_Bc}. 

The LHC produced its first collisions at 7 TeV center of mass energy at the end
of March 2010. 
The $b$ cross section at the LHC is a few hundred $\mu$b, a factor of three
higher than at the Tevatron and approximately, 0.5\% of the inelastic cross
section. When the LHC reaches its design center of mass energy of 14 TeV in
2015, the cross section will be a factor of two higher.

\boldmath
\subsubsection{$B$ Physics at LHCb}
\unboldmath

The LHC program features for the first time at a hadron collider a  dedicated
$B$-physics experiment, LHCb~\cite{lhcb_detector}.  LHCb covers the forward
direction from about 10 mr to 300 mr with respect to the beam line. $B$ hadrons
in the forward direction are produced by collisions of gluons of unequal energy,
so that the center of mass of the collision is Lorentz boosted in the direction
of the detector. Because of this, the $b$-hadrons and their decay products are
produced at small angles with respect to the beam and have momenta ranging from
a few GeV/c to more than a hundred GeV/c. Because of the Lorentz boost, even though
the angular range of LHCb is small, its coverage in pseudorapidity is from about
2 to about 5 and both $b$~hadrons travel in the same direction, making $b$
flavor tagging possible.  With the small angular coverage, LHCb can stretch out
over a long distance along the beam without becoming too large transversely. A
silicon microstrip vertex detector (VELO) 
only 8~mm from the beam
provides precision tracking that enables LHCb to separate
weakly decaying particles from particles produced at the interaction vertex.
This allows the measurement of lifetimes and oscillations due to flavor mixing.
A 4 Tm dipole magnet downstream of the collision region, in combination with the
VELO, large area silicon strips (TT) placed downstream of the VELO but upstream
of the dipole,  and a combination of silicon strips (IT) and straw tube chambers
(OT) downstream of the dipole provides a magnetic spectrometer with excellent
mass resolution. There are two  Ring Imaging Cherenkov counters, one upstream of
the dipole and one downstream, that together provide $K$--$\pi$ separation from
2 to 100 GeV/c.  An electromagnetic calorimeter (ECAL) follows the tracking
system and provides electron triggering and $\pi^{0}$ and $\gamma$
reconstruction. This is followed by a hadron calorimeter (HCAL) for triggering
on hadronic final states. A  muon detector at the end of the system provides muon
triggering and identification. 

LHCb has a very sophisticated trigger system that uses hardware at the lowest
level (L0) to process the signals from the ECAL, HCAL and muon systems. The L0
trigger reduces the rate to $\sim$1 MHz followed by  the High Level Trigger
(HLT), a large computer cluster, that reduces the rate to $\sim$3 kHz for
archiving to tape for physics analysis. LHCb is able to run at a luminosity of
4.0$\times$10$^{32}$ cm$^{-2}$s$^{-1}$. This is about 10\% of the current peak
luminosity achieved by the LHC and is about 3\% of the LHC design luminosity. 
The luminosity that LHCb can take efficiently  is currently limited by the 1 MHz
bandwidth between the Level 0 trigger system and the trigger cluster. Therefore,
the physics reach of LHCb is determined by the detector capabilities and not by
the machine luminosity. In fact, the LHC implemented a ``luminosity leveling''
scheme in the LHCb collision region so that LHCb could run at its desired
luminosity throughout the store  while the other experiments, CMS and ATLAS,
could run at higher luminosities.  This mode of running will continue until 2017
when a major upgrade of the LHCb trigger and parts of the detector and front end
electronics will increase the bandwidth to the HLT and permit operation at a
factor of 10 higher luminosity. 

There have been two runs of the LHC. In the first ``pilot" run in 2010, LHCb
recorded 35 pb$^{-1}$, which was enough to allow it to surpass in precision many
existing measurements of $B$ decays. In 2011, the LHC delivered more than 5
fb$^{-1}$  to CMS and ATLAS. Since this luminosity was more than LHCb was
designed to handle, the experiment ran at a maximum luminosity that was 10\% of
the LHC peak luminosity. The total integrated luminosity was about 1 fb$^{-1}$.


The decay $B_{s}\rightarrow J/\psi \phi$ has been used to measure the CKM angle
$\phi_{s} = - 2\beta_{s}$~\cite{lhcb_psi_phi}. The result, using also the decay mode 
$B_{s}\rightarrow J/\psi f_{0}$ first established by LHCb~\cite{lhcb_psi_f0}, is
$\phi_{s} = 0.07 \pm 0.17 \pm 0.06 \ {\rm rad}$~\cite{lhcb_psi_f0_cpv}. The
difference in the width of the $CP$-even and $CP$-odd $B_{s}$ mesons is $\Delta
\Gamma_{s} = (0.123 \pm 0.029 \pm 0.008)$ ps$^{-1}$. These results are
consistent with the SM, resolving a slight tension with earlier measurements
from the Tevatron~\cite{tev_psi_phi}, which deviated somewhat from the SM
predictions. 

\begin{figure}[tb]
\parbox{0.5\textwidth}{
\centerline{\includegraphics[width=6.75cm]{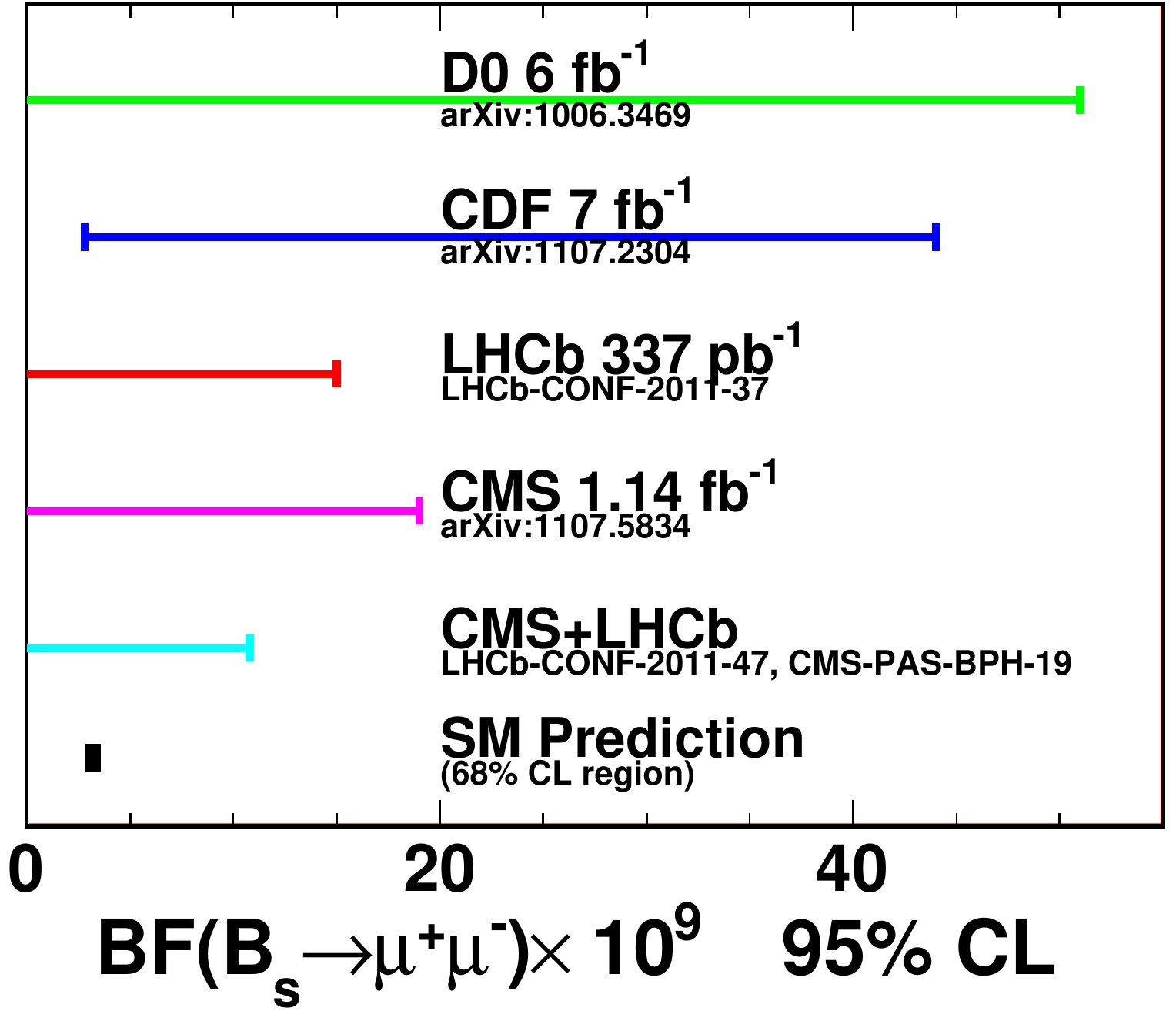}}

\vspace*{4pt}
\caption{The 95\% confidence level upper limits on the $B_{s}\rightarrow
\mu^{+}\mu^{-}$ branching fraction from CDF, D\O, LHCb, CMS, and the
combined value from LHCb and CMS. The SM prediction for this decay is also
shown. The integrated luminosity used in each measurement is shown.  All four
experiments are analyzing additional data.}
\label{hadronB_fig_3}
}
\parbox{0.5\textwidth}{
\centerline{\includegraphics[width=7.25cm, height=5.8cm]{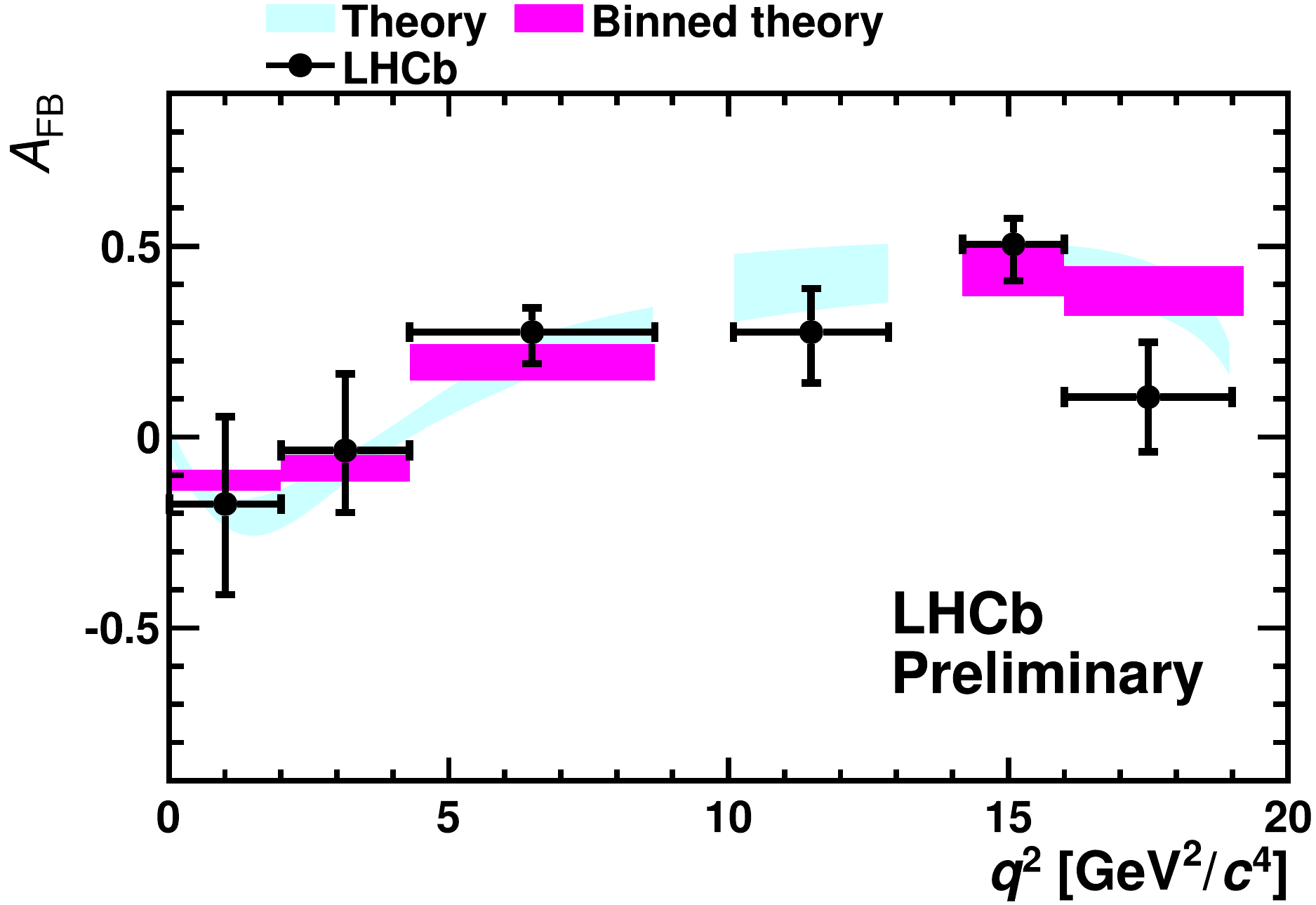}}

\vspace*{4pt}
\caption{$A_{FB}$ for $B^{0}\rightarrow K^{*0}\mu^{+}\mu^{-}$ as a function of
the dimuon mass, $q^{2}$. The SM prediction is given by the cyan (light) band,
and this prediction rate-averaged across the $q^{2}$ bins is indicated by the
purple (dark) regions. No SM predictions are shown in the two mass regions
dominated by J/$\psi$ and $\psi^{\prime}$ dimuon decays.}
\label{hadronB_fig_4}
}
\end{figure}

The rare decay $B_{s} \rightarrow \mu^{+}\mu^{-}$ is predicted in the SM  to
have a branching fraction that is 3$\times$10$^{-9}$.  A higher branching
fraction would be an indicator for new physics beyond the SM. LHCb has now
produced the best limit on this decay mode. While the current upper limit is now
approaching the SM value, there is still room for a substantial contribution
from new physics. The CMS experiment at the LHC is also a contributor to this topic. 
The overall experimental situation and the combined 
limit~\cite{lhcb_bs_mumu, cms_bs_mumu, combined_bs_mumu} from LHCb and CMS is
shown in Fig.~\ref{hadronB_fig_3}. For the LHC experiments, 
this represents about 1/4--1/3 of the data
already taken. Updated results are expected from both experiments soon using
the full 2011 data sets. This measurement will continue and if no new physics
appears, the SM value will be observed sometime between 2015 and 2017, based on
the current LHC midterm schedule and luminosity projections. 

LHCb has also produced results on the key decay $B^{0}\rightarrow
K^{*0}\mu^{+}\mu^{-}$~\cite{lhcb_kstar_mumu} that could reveal evidence for new
physics.  The forward-backward asymmetry of the $\mu^{-}$ relative to the direction
of the parent $B^{0}$ meson in the dimuon center of mass vs the $q^{2}$ (dimuon
invariant mass) is shown in  Fig.~\ref{hadronB_fig_4}. The SM prediction crosses
over through zero in a narrow range of $q^{2}$ due to the interference between
the SM box and electroweak penguin diagrams. New physics can remove the crossover
or displace its location. Indications from low statistics  at Belle, \babar,
and CDF seemed to indicate that this might be happening. The new LHCb results
are the most precise so far and are in good agreement  with the SM. 

\vspace*{-2pt}

Many other decays are being studied, including all-hadronic decays such as $B_{s}\rightarrow \phi\phi $,
$B\rightarrow D\pi$, $B\rightarrow DK$, and states with photons such as $B_{s}\rightarrow \phi \gamma$. 


LHCb will run at a luminosity of 4.0 $\times 10^{32}$cm$^{-2}$s$^{-1}$ for
several years, limited by the bandwidth between the Level 0 trigger and the HLT.
A substantial upgrade~\cite{lhcb_upgrade} that will enable LHCb to run at much
higher rates is being developed. It will be installed in a long shutdown planned
for the LHC in 2018. Between now and then, LHCb will accumulate about 1
fb$^{-1}$ per operating year so a total of about 5 fb$^{-1}$ will be obtained.
The sensitivity will increase by more than this because the LHC will run for at
least three years of this period at 14 TeV, with a correspondingly higher $B$ cross
section.  After the upgrade is installed, LHCb will integrate about 5 fb$^{-1}$
per year so that about 50 fb$^{-1}$ will be obtained over the  decade following
2018. The expected sensitivity to selected important decays during each phase of
LHCb running is shown in Table~\ref{hadronB_tab_1}.

\begin{table}[tbp]
\centering
\begin{tabular}{c|c|c|c}   \hline\hline
\multirow{2}{*}{Observable} & Precision & LHCb & Upgrade  \\
            & as of 2011 & (5 fb$^{-1}$) & (50 fb$^{-1}$) \\ \hline
$\phi_{s}(B_{s}\rightarrow J/\psi \phi)$ & 0.16 & 0.019 & 0.006 \\
$S(B_{s}\rightarrow \phi\phi)$ & --- & 0.08 & 0.02 \\
$S(B_{s}\rightarrow K^{*0}\bar{K}^{*0})$  & --- & 0.07 & 0.02  \\ \hline
$\beta(B^{0}\rightarrow J/\psi K^{0})$ &  $1^\circ$ & 0.5$^\circ$ &
0.2$^\circ$  \\ 
$S(B^{0}\rightarrow \phi K^{0}_{S})$ & 0.17 & 0.15& 0.03 \\ \hline 
$\gamma(B\rightarrow D^{(*)}K^{(*)})$ & $\sim20^\circ$ & $\sim4^\circ$ &
0.9$^\circ$  \\
$\gamma(B\rightarrow D_{s}K)$ & --- &  $\sim7^\circ$ & 1.5$^\circ$ \\ \hline 
$B(B_{s}\rightarrow \mu^{+}\mu^{-}) $ & --- & 30\% & 8\% \\
{\small $B(B^{0}\rightarrow \mu^{+}\mu^{-}) / B(B_{s}\to \mu^{+}\mu^{-})$} 
  & --- & --- &  $\sim$35\% \\ \hline
$S(B_{s}\rightarrow \phi\gamma)$ & --- & 0.07 & 0.02 \\
$A^{\Delta \Gamma_{s}}(B_{s}\rightarrow \phi\gamma )$ & --- & 0.14 & 0.03 \\
$A_{T}^{2}(B^{0}\rightarrow K^{*0}\mu^{+}\mu^{-})$ & --- & 0.14 & 0.04 \\
$s_{0}A_{FB}(B^{0}\rightarrow K^{*0}\mu^{+}\mu^{-})$ & --- & 4\% & 1\% \\ 
\hline\hline
\end{tabular}
\caption{ Sensitivity of LHCb to key observables. The current sensitivity is
compared to that expected after 5 fb$^{-1}$ and that achievable with
50 fb$^{-1}$ by the upgraded experiment, all assuming $\sqrt{s} = 14$ TeV. Note
that at the upgraded  LHCb, the yield in fb$^{-1}$ in hadronic $B$ and $D$
decays will be higher on account of the  software trigger. (Adapted
from Table 2-1, p13, of reference~\cite{lhcb_upgrade}.)}
\label{hadronB_tab_1}
\end{table}

\boldmath
\subsubsection{$B$ Physics at CMS and ATLAS}
\unboldmath

Two detectors, CMS and ATLAS at the LHC, are designed to explore high mass and
high-$p_{T}$ phenomena to look for new physics at the LHC. 
They must operate at luminosities of up to $10^{34}\,$cm$^{-2}$s$^{-1}$, which
implies the need to handle an average event pileup of $\sim$20. 
This demands that the  detectors  cover a large area with very high granularity. 
They both have tracking coverage up to  $|\eta| \, = \, 2.5$ and have silicon
pixel detectors capable of separating $B$ decays from the main interaction vertex. 
The detectors also have other features that are needed to do $B$ physics, including
precise mass reconstruction, electron, photon, and muon identification,
and sophisticated triggering capability. However, they lack some important
characteristics that are necessary to carry out a broad program of $B$ physics.
They have only limited charged hadron identification. 
In addition, as general-purpose detectors they must operate in high-luminosity conditions with trigger configurations that are more oriented to high transverse momentum events. Consequently, they have only 
limited ability to trigger on low-$p_{T}$ $B$ hadrons, which comprise most of the $B$ cross section.

Both experiments can implement muon triggers with relatively low thresholds of a few GeV/c.
However, the rate of low-$p_{T}$ muons from $B$ decays competes for scarce resources with the many other trigger signatures that could contain direct evidence of new physics. The high muon rate at all levels of the trigger system in the LHC environment must be controlled as the luminosity increases, by requiring double muon signatures for low $p_{T}$ muons, raising thresholds, applying invariant mass and lifetime cuts in high level triggers, restricting signatures to certain regions of the detector, and, as a last resort, prescaling. These measures limit the ability to collect large statistics of certain $B$ decays and means that those which are collected have a higher transverse momentum overall than those collected by LHCb.

On the other hand, both experiments can successfully record certain $B$ decays
with reasonable efficiency. These typically involve final states that contain dimuons. 
One example of this, discussed above, is the rare
decay $B_{d,s}\rightarrow \mu^{+}\mu^{-}$, where CMS and ATLAS can be
competitive because they employ clever triggering and can compensate for
lower efficiency by running at more than an order of magnitude  higher
luminosity than LHCb.  If ATLAS and CMS can maintain their
triggering efficiency as the LHC luminosity and eventually its energy increase,
they can continue to be competitive in this study. The decay $B^{0}\rightarrow
K^{*}\mu^{+}\mu^{-}$ presents more problems. The muons are softer and more
difficult to trigger on and the limited $K$--$\pi$ separation increases the
background to the $K^{*}$. It is still hoped that these two experiments can play a
confirming role to LHCb in this study. Despite their limitations, these two experiments
will collect large numbers of $B$ decays and should be able to observe many new
decay modes and new particles containing $b$ and charmed
quarks~\cite{cms_6_pager, atlas_6_pager}.

\subsection{Charm Experiments}
\label{sec:three.three}


In the SM, many charm decay modes involving loop diagrams are suppressed. Therefore, $CP$-violating 
and rare decays of charmed particles are promising places to look for new physics, since new phenomena could make observable contributions to such decays.  In the future, information on charm decays will come from:
\begin{itemize}\vspace*{-10pt}\itemsep -2pt
\item the BES experiment at the Beijing Electron Positron Collider~\cite{besIII_machine}, which is dedicated to the study
of systems containing charmed quarks;
\item The two super flavor factories; and
\item LHCb, the dedicated heavy quark experiment at the LHC, which is described above, with perhaps some
additional results in a few favorable decay modes from CMS and ATLAS. 
\end{itemize}\vspace*{-8pt}

A fourth source of information on charm could come from fixed-target experiments,
of which the only currently approved example is PANDA~\cite{panda_fair} at the
FAIR facility at Darmstadt,  which will collide antiprotons in a storage ring
with gas, solid, or liquid targets. The ability of that experiment to contribute
will depend on the cross section for charm production by low energy antiprotons - 
a quantity that has not been measured and whose theoretical estimates vary from
1$\mu$b to 10$\mu$b and the amount of time dedicated to the charm program,
which competes with other aspects of the program that require the machine to
operate below or close to the bare charm production threshold.

For these experiments, the challenge will be to observe small effects. For theory,
the task will be to pin down the size of the long-range contributions so 
observations can be correctly identified as new physics or conventional
physics.

\subsubsection{Charm Physics at Charm Factories}

The BES program carried out a major upgrade to a two-ring machine optimized for
running at center of mass energies of 3--4 GeV. The accelerator/storage ring,
now called BEPCII, is designed  for a peak luminosity of
1$\times$10$^{33}$~cm$^{-2}$s$^{-1}$. An all-new and improved detector,
BESIII~\cite{besIII_detector}, has been built to exploit the opportunity
afforded by the higher luminosity. The upgraded machine  began to run in July of
2008 and has achieved so far about 2/3 of the design luminosity. BESIII  has now
collected data at the center of mass energy of significant $c\overline{c}$
resonances, including the $\psi^{\prime}$, the J/$\psi$,and  the $\psi$(3770), and
also at center of mass energy of 4010 MeV (0.3 nb of $D_{s}^{+}D_{s}^{-}$). 
BESIII has now integrated about 3.5 times more data than CLEO-c on the
$\psi$(3770).  By studying charm particle properties on the  $\psi$(3770)
resonance, which is very near  $D$--$\overline{D}$ threshold, BESIII has large and
relatively clean source  of $D$ mesons with tightly constrained kinematics. This
provides powerful flavor-tagging capability and unique access to leptonic and
semileptonic decay modes, and enables the study of decays that include
neutrinos. The two $D$ mesons are produced in the $CP$-odd state. This quantum
correlation can be used to study $CP$ violation and strong phases and will extend
the work carried out by CLEO-c.  BESIII also collected 400 pb$^{-1}$ in 2011 at
4010 MeV and in 2013 will take data at 4170 MeV, which is above the $D_{s}^{*}D_{s}$
threshold (0.9 nb cross section), and which has been previously studied by
CLEO-c.  These data, taken all together, will represent a significant advance in
our understanding of the $D_{s}$ meson. 

With these exposures, BESIII could well be the leader in the use of the charm
system as a QCD laboratory.  BESIII should excel in the measurements of leptonic
and semileptonic $D$ and $D_s$ decays. These determine $|V_{cd}|$ and $|V_{cs}|$
(without assuming CKM unitarity) when combined with lattice-QCD calculations,
and can also check lattice QCD with high precision, so that similar calculations
can be trusted when applied to the $B$ system to extract CKM matrix elements
and look for BSM physics.  These results, many of them from data already in
hand, should precede by a few years any data that could come from a super flavor
factory running on a boosted $\psi$(3770), as discussed below.

\boldmath
\subsubsection{Charm Physics at $e^{+}e^{-}$ Super Flavor Factories}
\unboldmath

The major effort at the upgraded $B$ factories, SuperKEKB and Super$B$, is to learn
about new physics by carrying out precision measurements of mixing and $CP$
violation and searching for rare decays of $B_{d}$ and $B_{u}$  mesons,
primarily by running on the $\Upsilon$(4S). However,  massive statistics on
charm decays will be gathered from the charm meson and baryon daughters of the
$B$-decays, as well as from direct charm production from the continuum background
under the resonance.  Most of the charm sensitivity will be obtained from this
mode of running.

A new possibility is being studied
by Super$B$. They are considering a run of 1 ab$^{-1}$ on the $\psi$(3770). The
energies of the two rings will be chosen so that $\beta \gamma$ is between 0.24 and 0.6. This choice provides good acceptance and precision measurements of the time dependence of the decays. The results will occur after the BES results but will exceed them by a factor of 100 in integrated luminosity. 
 For Super$B$ this might
make sense in the early phase of running when the luminosity is still low. This would allow them to carry out
charm studies that take advantage of the production at threshold and quantum coherence with the added advantage that they would be able to study the time dependence of the decays.

\subsubsection{Charm Physics at the LHC}

LHCb, the dedicated $B$ physics experiment at the LHC,  also has significant
capability to study charm decays. The $B$ decays recorded by LHCb are themselves
a copious source of charmed particles. Direct production of charm at the LHC is
a few percent of the total cross section, so the direct charm rate is enormous
and actually has to be suppressed since it competes with $B$ physics for
precious resources, such as output bandwidth between the Level 0 trigger and the
higher-level trigger.  Even with this suppression, LHCb records a very high rate
of directly produced charm. LHCb should be a leader in the spectroscopy and
decay properties of charmed baryons and in the study of rare, lepton flavor-
and lepton number-violating decays. It should be able to carry out a large
number of detailed decay studies including Dalitz plot analyses and
time-dependent Dalitz plot analyses. It does not have an overall energy
constraint so the study of many decays that involve neutrinos in the final state
will be difficult to do. LHCb's ability to observe states with photons and
$\pi^{0}$'s efficiently is still to be demonstrated. 

After LHCb is upgraded, with more events reaching the HLT, it will be possible
to record a much more targeted
selection of events. This should benefit the LHCb
charm program and permit it to improve or at least maintain its efficiency for
charm as the luminosity of the LHC increases.

The basic $CP$-violating parameters in charm can be measured by LHCb 
and the super flavor 
factories. A summary of the  sensitivity for these
quantities is given in Table~\ref{allCexpt_tab_1}. These measurements may reveal
new physics beyond the Standard Model and will help in discriminating among
the various models of new physics that might be employed to explain results from
the LHC.

\begin{table}[tbp]
\centering

\begin{tabular}{c|c|c|c|c}   \hline \hline
\multirow{2}{*}{Observable} & \multirow{2}{*}{Current Expt.}
  & LHCb	& Super Flavor		&  LHCb Upgrade  \\
		&         & (5 fb$^{-1}$) & Factories (50 ab$^{-1}$) &    (50 fb$^{-1}$) \\ \hline
$x$		& (0.63 $\pm$ 0.20)\%  & $\pm$0.06\% & $\pm$0.02\% &  $\pm$0.02\%  \\
$y$		& (0.75 $\pm$ 0.12)\%  & $\pm$0.03\% & $\pm$0.01\% &  $\pm$0.01\%   \\
$y_{\rm CP}$	& (1.11 $\pm$ 0.22)\%  & $\pm$0.02\% & $\pm$0.03\%  & $\pm$0.01\%   \\  
$|q/p|$ & 0.91 $\pm$ 0.17  &  $\pm$0.085 & $\pm$0.03  &   $\pm0.03$     \\ 
${\rm arg}(q/p)$ & ($-10.2 \pm 9.2$)$^{\circ}$   &  $\pm$4.4$^{\circ}$ & 
$\pm$1.4$^{\circ}$  & $\pm$2.0$^{\circ}$   \\ \hline \hline
\end{tabular}
\caption{Sensitivities of super flavor factories and LHCb to charm mixing
related parameters, along with the current results for these measurements (here
${\rm arg}(q/p)$ means ${\rm arg}\,\lambda_{K^+K^-}$).  The second column gives
the 2011 world averages.  The remaining columns give the expected accuracy at
the indicated integrated luminosities. For this table, $50 \, {\rm ab^{-1}}$ has
been assumed for super flavor factory experiments (usually Super$B$ assumes $75 \,
{\rm ab^{-1}}$,  while Belle~II assumes $50 \, {\rm ab^{-1}}$). (Adapted from
Table 1, p13, of reference~\cite{charm_table}.)
}
\label{allCexpt_tab_1}
\end{table}

\subsection{Exotic States}
\label{sec:two.three}


Recently, there has been an explosion of new results on heavy meson 
spectroscopy. The \babar and Belle experiments, in addition to advancing the 
field of bottomonium spectroscopy by observing the $b \overline{b}$ ground state
$\eta_b$ and other missing $b \overline{b}$ states, have observed 18 states in
the mass range 3872~MeV to 4700~MeV. These so-called ``XYZ" states do not easily
fit into the expected spectrum of charmonium states.  An example is the very
narrow $X(3872)$, first observed by Belle and confirmed by \babar, CDF, D\O,
and now also by CMS and LHCb.  Many models have been proposed to explain this
state, including the possibility that it may be a $\D0bar D^{*0}$ molecule. 

In addition to searching for additional states, the experimental agenda includes
the measurement  of masses and widths, branching fractions, and quantum numbers
of the  observed states. 

The super flavor factories study charmonium states in the decay of $B$ mesons.
They may also directly produce charmonium and bottomonium states that have
$1^{--}$ quantum numbers. The $e^{+}e^{-}$ charm factories can study $1^{--}$ 
charmonium resonances. The LHC experiments may produce charmonium states 
directly or observe them in $B$-meson decays. They can also study bottomonium
states. The PANDA experiment at the new $\overline{p}$ facility, FAIR in
Darmstadt, can study charmonium. The $\overline{p}$ experiments can  produce
charmonium states exclusively by annihilation or in association with other
particles. Particularly for narrow-width meson resonances that can be produced
by annihilation in $p\overline{p}$ collisions at FAIR, the measurement of the
mass and width ($\Gamma  \simeq 50$ KeV) can be obtained very accurately from
machine scans across the resonances.    

These studies complement the ability of these experiments to probe high mass 
scales. They provide an opportunity to study one of Nature's fundamental
interactions, QCD,   in a regime where it is poorly understood. A large
community of  theorists and experimentalists is focused on these topics.

\section{The Need for New Experiments and Facilities}
\label{sec:four}

Before looking forward,  
it provides useful context to briefly review some history.
In the 1990's, the US was the leader both on the energy frontier and in
quark flavor-physics experiments at the intensity frontier.  $B$ physics
was still dominated by the CLEO experiment for most of that decade.  
The most sensitive rare $K$
decay experiments performed to date were then under way at the Brookhaven AGS,
and direct $CP$ violation in $K_L^0 \to \pi \pi$ decays was the focus of
a fixed-target experiment using the Tevatron at Fermilab.
Toward the end of that decade, the asymmetric $e^+e^-$ 
$B$ factories began running at SLAC and KEK, leading to increases in  the
size of $B$ meson data sets by two orders of magnitude and also opening
the door to measurements of time-dependent $CP$ asymmetries,
which provided the experimental basis for the 2008 Nobel Prize.
In the midst of this success,  a number of new and aggressive
quark-flavor initiatives were put forward in the US.  These included the BTeV
proposal, which would have used the Tevatron collider for 
$B$ physics; the CKM proposal, which would have made the first high-statistics
measurement of $K^+ \to \pi^+ \nu \overline{\nu}$ using the Fermilab Main Injector;
and the RSVP proposal, which included an experiment (KOPIO) to measure
$K_L^0 \to \pi^0 \nu \overline{\nu}$ at the Brookhaven AGS.
After lengthy consideration in an environment characterized by flat budgets and 
a predilection for a fast start on the International Linear Collider
on US soil, all of these initiatives were ultimately terminated.
Also, as accelerator breakthroughs capable of increasing $B$-factory luminosity
by more than another order of magnitude were made,
the opportunity to upgrade the PEP-II $B$ factory at SLAC was not pursued;
subsequently, the proponents coalesced around what is now the Italian
super flavor factory planned to be built at the new Cabibbo Lab near Rome.  
This history is relevant in order to stress that the US has been a leader in 
flavor-physics experiments --- involving a vigorous community ---
until very recently.  Nonetheless, 
this sequence of events  inevitably encouraged many in the 
flavor-physics community in the US to migrate elsewhere, 
most often to ATLAS or CMS at the LHC.

In spite of these developments in the US, a rich heavy-quark flavor physics
program is flourishing around the world.  Kaon experiments, $B$-physics
experiments, and charm experiments are running and under construction in Asia
and Europe. Indeed, CERN --- the laboratory that now owns the energy
frontier~--- is also the home of a running $B$-physics experiment (LHCb) that
has a clear upgrade path, and a rare $K$ decay experiment (NA62) that is under
construction.  This reflects the world-wide consensus that  flavor-physics
experiments are critical to progress in  particle physics, as described in
Section~\ref{sec:two}.


Looking forward, it is clear --- based on this workshop --- 
that there continues to be strong
interest and a potentially substantial community in the US for an intensity
frontier flavor physics program.  Indeed, US physicists participate in
almost all the offshore experiments, although on a modest scale. 
Two conclusions are
obvious: US participation in offshore intensity frontier experiments should
be supported on a scale sufficient to make a significant impact on those 
experiments, and facilities are needed in the US that can support
a leadership role at the quark-flavor intensity frontier.

The basic motivation for this program can be described very simply.
If the LHC observes
new high-mass states, it will be necessary to distinguish between models
proposed to explain them.  This will require tighter constraints from the
flavor sector, which can come from more precise experiments using
strange, charm, and bottom quark systems.  
If the LHC does not make such discoveries, then the ability of 
precision flavor-physics experiments to probe mass scales far above LHC, 
through virtual effects, is the best hope to see signals that may point toward
the next energy scale to explore.
Therefore, a healthy US particle physics program
must include a vigorous flavor-physics component.

A few conclusions from this working group can be summarized briefly:

\begin{itemize}\vspace*{-12pt}\itemsep 2pt

\item Intensity Frontier experiments using strange, charm, and bottom quark
systems are an essential component of a balanced world-wide particle physics
program. The US, which led in this area only a few years ago, should endeavor
to  be among the leaders in the future.

\item The compelling case for flavor-physics experiments, as described in this
report, is not predicated on future theoretical progress.  Nonetheless,
theoretical progress, including improvements in lattice QCD calculations, will 
strengthen the program by increasing the set of observables that can be used to
search for new physics.  Continued support of the theory community that is engaged in
this research is important.

\item Several intensity frontier experiments using strange, charm, and bottom
quark systems are under way or planned at laboratories around the world
(including KEK and J-PARC in Japan, BESIII in China, and CERN and
Frascati/Cabibbo laboratories in Europe).  The US needs to be involved in
these experiments on a significant scale in order to exploit the expertise
gained over the many years that US facilities led in these areas and to share
in possible new discoveries.

\item At the present time, no intensity frontier experiments  using strange,
charm, or bottom quark systems are under way in the US, in spite of the fact
that existing facilities at Fermilab provide powerful capabilities.   In
particular, world-leading rare kaon decay experiments can be mounted at
Fermilab, using the Main Injector, with relatively modest investment. The ORKA
experiment, if it proceeds, would exploit this opportunity.

\item Kaon beams from Project~X can provide a singular opportunity for intensity
frontier flavor physics experiments.   These experiments comprise an important
element within the world-wide flavor-physics program, and their physics case is
compelling.

\item To exploit the potential that Project~X can provide, improved detectors will be needed.  Therefore, an active program of detector R\&D focused on the key issues is 
critical.

\end{itemize}\vspace*{-10pt}

A well-planned program of flavor physics experiments --- using strange,
charm, and bottom quarks --- has the potential to produce new paradigm-changing scientific advances.

\def\Discussion{\setlength{\parskip}{0.3cm}\setlength{\parindent}{0.0cm}
     \bigskip\bigskip      {\Large {\bf Discussion}} \bigskip}\def\speaker#1{{\bf #1:}\ }
\def\endDiscussion{}

\def\babar{\mbox{\slshape B\kern-0.1em{\small A}\kern-0.1em
    B\kern-0.1em{\small A\kern-0.2em R}}}

\chapter{Charged Leptons}
\label{chap:charged-leptons}


\begin{center}

Conveners: B.C.K.~Casey, Y.~Grossman, A.~Roodman

K.~Agashe, M.~Aoki, K.L.~Babu, N.~Berger, R.~Bernstein, T.~Blum, B.~Dutta, M.~Eads, R.C.~Group, F.A.~Harris, D.~Hertzog, D.H.~Hitlin, K.S.~Kumar, Y.~Kuno, M.~Lancaster, J.P.~Miller, Y.~Okada, C.C.~Polly, M.~Pospelov, B.L.~Roberts, M.~Rominsky, N.~Saito, A.~Schwartz, D.~Stoeckinger, P.~Tanedo, Y.~Tsai, I.~Yavin
\end{center}

\section{Overview}\label{sec:cl:over}

When the muon was discovered, I.~I.~Rabi famously asked, ``Who ordered
that?''  The question could have been asked again with the discovery of the tau.  For all
the successes of the Standard Model (SM), we still have no insight into the origin of the three
generations of leptons.  However, the study of charged leptons has been an integral part of the advances in our
understanding of particle physics. The journey is not over yet
and the charged lepton sector still has significant potential to
teach us much more about the fundamental principles of Nature.

This working group focused on how precision measurements of charged
lepton interactions can be used to test the Standard Model and look
for signs of new physics. Charged leptons are unique in the following ways:
\begin{itemize}
\item
They directly probe the couplings of new particles to leptons.  This is unique in that the current energy frontier machine, the 
CERN Large Hadron Collider (LHC), is a hadron
collider. It is very effective for probing the quark sector, but
is significantly more limited in the lepton sector.
\item
They are relatively easy to produce and detect. Thus, very
precise measurements can be made at a level that is difficult to achieve in
other sectors.
\item
They can be studied using a diverse set of independent
processes. The combination of these studies can provide additional
insights into the structure of the lepton sector.
\item
Hadronic  uncertainties in the Standard Model predictions are either insignificant, or in the case of muon $g-2$, are controlled using independent data sets.
\item
There are  many cases, particularly charged lepton flavor violation (CLFV), where any signal would indicate an indisputable discovery of new physics.
\end{itemize}

There are many important charged lepton observables that are best studied using electrons, most notably the search for an electron electric dipole moment (EDM).  In most cases, these experiments are performed using outer-shell or shared electrons in either atoms or molecules.  These topics are therefore covered in detail by the Nucleons/Nuclei/Atoms working group and we refer the interested reader to the corresponding chapter of this document.

The objective of the experimental program in the charged lepton sector is discovery of new physics.  The program is very diverse, containing multi-purpose experiments that take advantage of the large tau-pair production cross section at $B$ factories and highly optimized, single-purpose experiments that focus on near-forbidden interactions of muons.
A very encouraging fact about this experimental program is that huge
improvements in sensitivity are possible in the immediate future.  There are exciting new facilities currently being designed that will enable this program, and a few examples are highlighted below.  New
experiments can probe rare processes at rates that are five orders
of magnitude more sensitive than current bounds. These  enormous
improvements will be a significant part of the program to understand new short-distance dynamics or new ultra-weak interactions.

The charged lepton sector is an integral part of the broader intensity
frontier program and provides a vital link to the energy frontier. In the same way that the results of each charged
lepton experiment are much more sensitive to new physics when taken together, charged lepton sector results as a whole are more
powerful in concert with other intensity and energy frontier experiments. In
particular, there are three domains where such a combination is a
crucial probe of new physics.   First, since neutrinos and charged leptons form a 
natural doublet, one would expect any new physics effects
 in neutrinos to also be seen in charged leptons.  
 Second, any complete theory of 
 flavor generation and the observed matter-antimatter asymmetry in the universe must 
 relate flavor and $C\!P$ (or $T$) violation in the heavy quark, neutrino, and charged lepton sectors.  
 Third,  any theory that predicts new particles or interactions at the LHC must also account for the virtual effects of those particles on decays and interactions of charged leptons and heavy quarks.
Thus, the major expansion in the study of charged leptons now under way in the international particle physics community is a natural extension of the successful heavy quark,  neutrino, and energy frontier programs of the previous decades.

A fourth domain is the probe of new ultra-weak, low energy interactions, referred to collectively  as hidden or dark sectors.  Here, charged lepton experiments overlap  with a wide variety of
experiments at the intensity, cosmic, and energy frontiers.  A large
experimental program is now under way to directly probe for new
hidden sectors, particularly in regions of parameter space
consistent with the muon $g-2$ anomaly.  This program is covered
in detail in the ``New, Light, Weakly-Coupled Particles'' chapter of this document.

\begin{figure}[t!]
\begin{center}
\includegraphics[width=10cm]{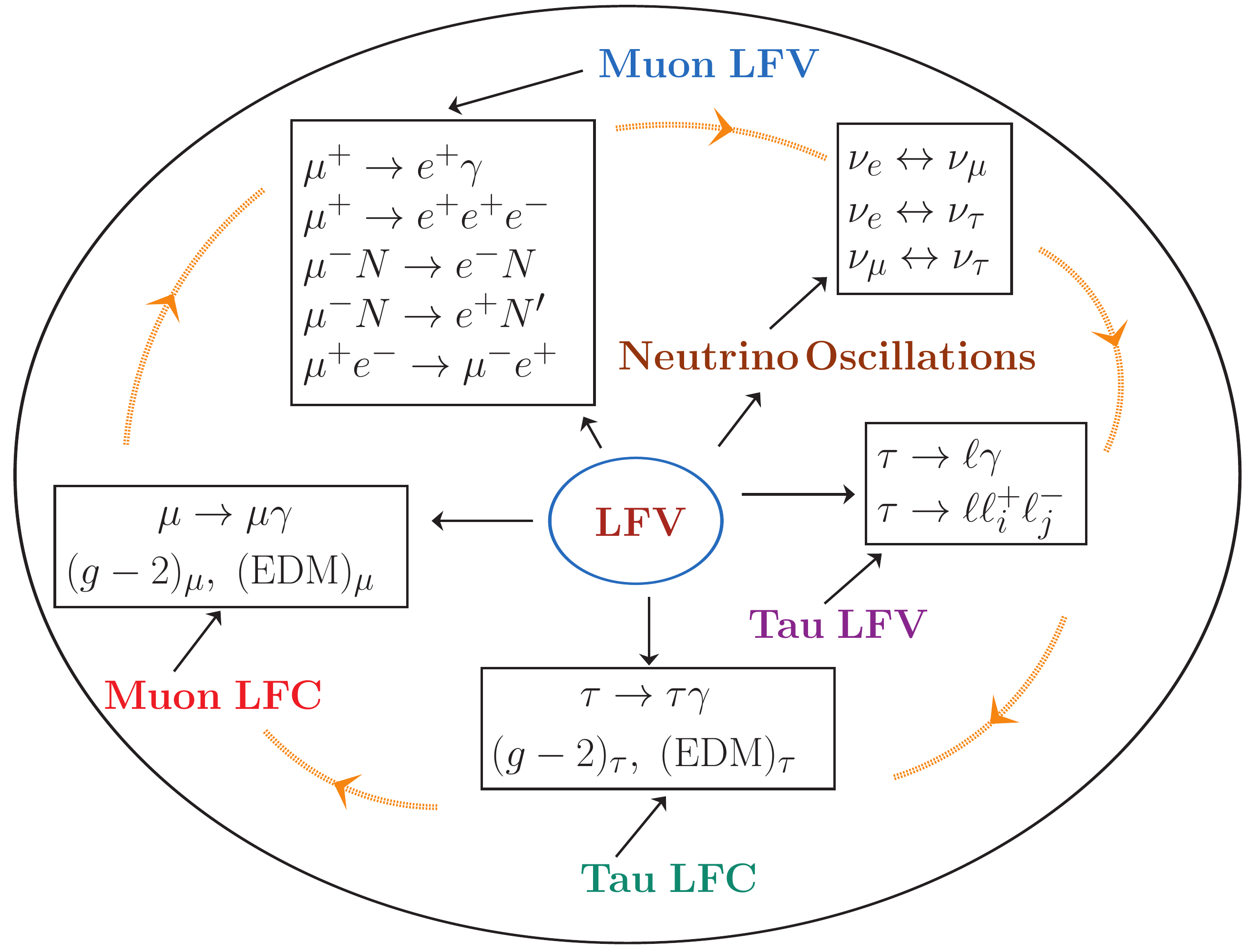}
\caption{\label{CL:chart}Interconnection between various lepton flavor violating and  lepton flavor conserving processes.}
\end{center}
\end{figure}

Fig. \ref{CL:chart} schematically depicts the interconnection between
various flavor conserving and violating processes in the lepton
sector.  In an underlying theory, neutrino flavor oscillations,
charged lepton flavor violation,  the anomalous magnetic moments, and permanent electric
dipole moments are related.  It should be emphasized that each experimental avenue we pursue allows us to uncover more aspects of the underlying theory. 

There are many important physical observables potentially sensitive to
new physics effects in charged lepton processes. Below, they are  split into  flavor violating observables and flavor conserving observables  such as $g-2$, EDMs, and parity violation measurements. Tau decays offer a unique opportunity to simultaneously study flavor
 conserving, flavor violating, $C\!P$ violating,  and $T$ violating effects and are discussed in
 their own section below.


\section{Muon Flavor Violating Processes}\label{sec:cl:spec}

\subsection{Theory Overview}

Neutrino flavor oscillations are well established. This requires charged lepton flavor violation at some level as well.  However flavor violation in charged lepton interactions has never been observed.
If neutrino mass is the only source of new physics, and if
the mass generation occurs at a very high energy scale, CLFV processes are highly suppressed.  For example, if neutrinos are 
Dirac particles, the branching ratio for $\mu\rightarrow e\gamma$ is
\begin{equation}
BR(\mu \rightarrow e \gamma)=\frac{3\alpha}{32\pi}\left|\sum_i U_{\mu
i}^* U_{e i}\frac{m_{\nu_{i}}^2}{m_{W}^2}\right|^2 \sim 10^{-52} \,,
\end{equation}
where $U_{ei}$ are the leptonic mixing matrix elements. This value,
which suffers from extreme suppression from the small neutrino masses,
is experimentally inaccessible. In many extensions of the Standard Model,
however, there are much larger contributions to CLFV, and current experimental bounds set strict limits on the parameter space available for new physics models.

\subsubsection{CLFV Decays and Interactions}

The effective Lagrangian relevant for the $\mu \rightarrow e\gamma$ and $\mu^+ \rightarrow e^+e^-e^+$ decays can be parametrized,
regardless of the origin of CLFV, as
\begin{eqnarray}
{\cal L}_{\mu\rightarrow e\gamma, eee} &=& -{4G_{F}\over\sqrt{2}}
\left[  {m_{\mu }}{A_R}\overline{\mu_{R}}
        {{\sigma }^{\mu \nu}{e_L}{F_{\mu \nu}}}
       + {m_{\mu }}{A_L}\overline{\mu_{L}}
        {{\sigma }^{\mu \nu}{e_R}{F_{\mu \nu}}} \right. \nonumber \\
    && + {g_1}(\overline{{{\mu }_R}}{e_L})
              (\overline{{e_R}}{e_L})
       + {g_2}(\overline{{{\mu }_L}}{e_R})
              (\overline{{e_L}}{e_R}) \nonumber \\
    &&   +{g_3}(\overline{{{\mu }_R}}{{\gamma }^{\mu }}{e_R})
              (\overline{{e_R}}{{\gamma }_{\mu }}{e_R})
       + {g_4}(\overline{{{\mu }_L}}{{\gamma }^{\mu }}{e_L})
              (\overline{{e_L}}{{\gamma }_{\mu }}{e_L})  \nonumber \\
    && \left.  +{g_5}(\overline{{{\mu }_R}}{{\gamma }^{\mu }}{e_R})
              (\overline{{e_L}}{{\gamma }_{\mu }}{e_L})
       + {g_6}(\overline{{{\mu }_L}}{{\gamma }^{\mu }}{e_L})
              (\overline{{e_R}}{{\gamma }_{\mu }}{e_R})
       +  h.c. \right].
       \label{CL:int}
\end{eqnarray}
The decay $\mu \rightarrow e\gamma$ is mediated by the first two terms 
of Eq. (\ref{CL:int}), the dipole terms.  These
terms, as well as the remaining contact terms, all contribute to the 
decay $\mu^+ \rightarrow e^+e^-e^+$.  The relative strength of
these decay rates depends on the relative strength of the dipole and contact terms. 
 Turning this around, searches for
these two decays reveal much about the underlying flavor structure.  
In some models the dipole contribution 
dominates both decays.  In this case, a simple relation
exists for the relative branching ratio:
\begin{equation}
\frac{B(\mu^{+}\rightarrow e^{+}e^{-}e^{+})}{B(\mu^{+} \rightarrow
e^{+} \gamma)} \simeq
\frac{\alpha}{3\pi}\left(\ln(\frac{m_{\mu}^{2}}{m_{e}^{2}})-\frac{11}{4}\right)
= 0.006.
\label{CL:branching}
\end{equation}
However, contact terms arise
frequently in popular models where the relation
Eq.~(\ref{CL:branching}) does not hold.  A good example is the type II seesaw mechanism for small neutrino masses.  Here, one
does not add right-handed neutrinos to the spectrum; rather one
includes an iso--triplet scalar $\Delta = (\Delta^{++},\,\Delta^+,\,\Delta^0)$ with quantum numbers
$(1,3,+2)$ under $SU(3)_c \times SU(2)_L \times U(1)_Y$.  Neutrino
masses are generated via the Yukawa coupling $\frac{f_{ij}}{2}
\ell_i^T C \ell_j \Delta$, once a nonzero $\langle
\Delta^0\rangle$ develops.  The doubly charged scalar
$\Delta^{++}$ could mediate the decay $\mu^+
\rightarrow e^+e^-e^+$ at tree level.  In this case, the branching ratios for $\mu
\rightarrow e\gamma$ and $\mu^+ \rightarrow e^+e^-e^+$ become comparable.


\begin{figure}[t]
\begin{center}
\includegraphics[width=5cm]{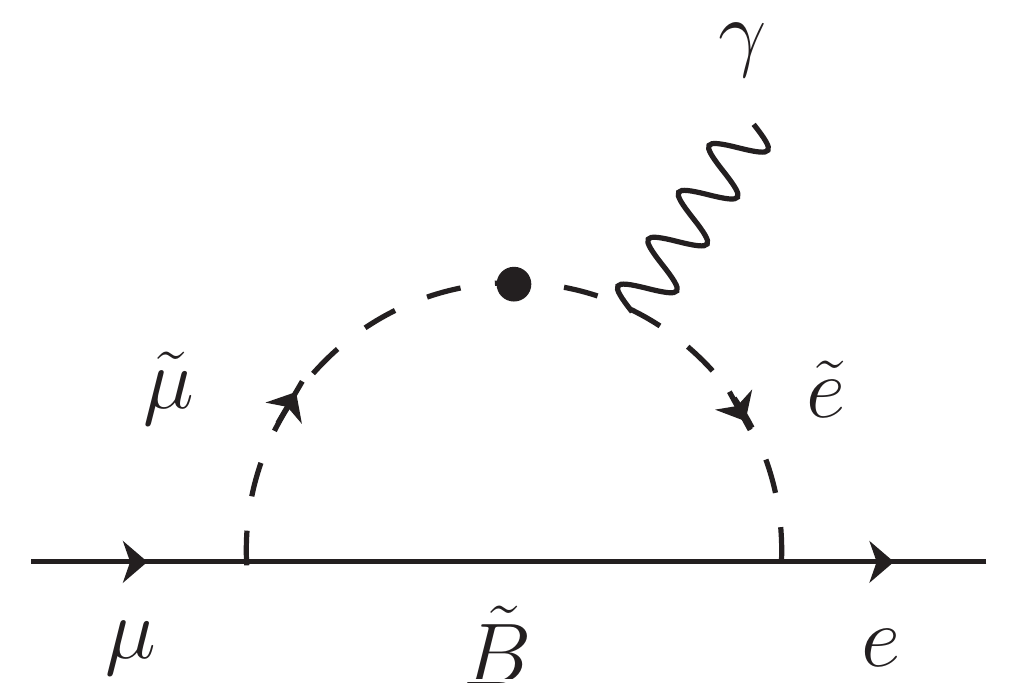} \hspace*{2cm}
\includegraphics[width=5cm]{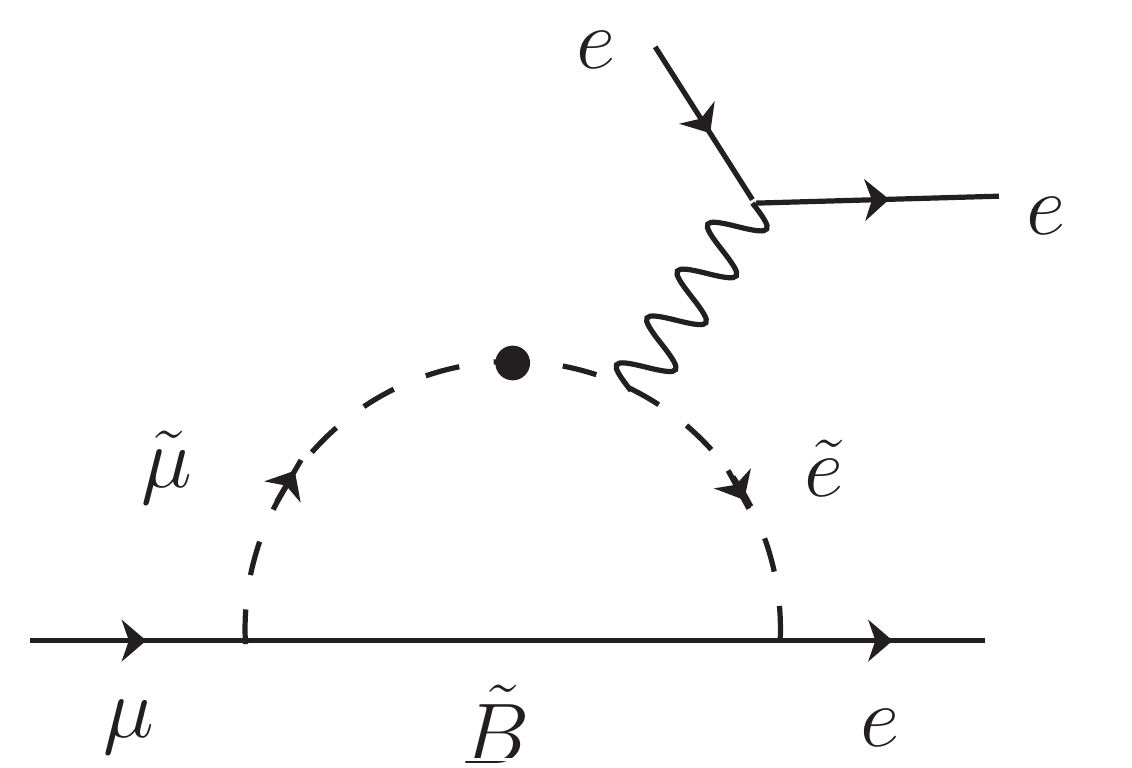}
\caption{\label{CL:mutoegamma}$\mu \rightarrow e\gamma$ decay (left panel), and $\mu \rightarrow 3e$ decay (right panel)
mediated by supersymmetric particles.}
\end{center}
\end{figure}

Equally important as the decays $\mu \rightarrow e\gamma$ and $\mu^+
\rightarrow e^+e^-e^+$ is the coherent $\mu^-N \rightarrow e^-N$ conversion
process in nuclei.  Muonic atoms are formed
when negative muons are stopped in matter.  In the ground state of these
atoms, the muon can decay in orbit or be captured with the emission of
a neutrino via the process $\mu^{-} + (A,Z) \rightarrow
\nu_{\mu} + (A,Z-1)$.  If there are new sources of CLFV, muon capture
without the emission of a neutrino can occur: $\mu^{-} + (A,Z)
\rightarrow e^{-} + (A,Z)$.   This would occur in Supersymmetry (SUSY)  via
the diagram of Fig.~\ref{CL:mutoegamma}, when the photon is attached to
a quark line.  Like $\mu^+\rightarrow e^+e^-e^+$, this process can occur through dipole
interactions or through contact interactions.  Such contact
interactions arise naturally in leptoquark models at the tree level,
while in  SUSY the dipole interactions dominate.  If only the
dipole couplings are important, one can obtain a relation for the ratio of rates
\begin{equation}
\frac{B(\mu^{+}\rightarrow e^{+}\gamma)}
{B(\mu^{-}N\rightarrow e^{-}N)}
=\frac{96\pi^{3} \alpha}{G_F^2 m_{\mu}^4}\cdot
{1\over{3\cdot 10^{12}B(A,Z)}}\simeq \frac{428}{B(A,Z)},
\end{equation}
where $B(A,Z)$ is a function of the atomic number and atomic weight,
with its value ranging from 1.1 to 2.2 for Al, Ti and Pb atoms.  
The best limits on these processes are $B(\mu^- + {\rm Ti}
\rightarrow e^- + {\rm Ti}) < 4.3 \times 10^{-12}$~\cite{Dohmen:1993mp} and $B(\mu^- + {\rm Au}
\rightarrow e^- + {\rm Au}) < 7.0 \times 10^{-13}$~\cite{Bertl:2006up} from experiments conducted at the Paul Scherrer Institute (PSI) in 
Switzerland.
For these searches the limits quoted are with respect to the muon capture process $\mu^{-} + (A,Z) \rightarrow \nu_{\mu} + (A,Z-1)$.
Future
experiments can improve tremendously on these limits down to  $10^{-18}$.

A related process is the incoherent, lepton number violating
process $\mu^{-} + (A,Z) \rightarrow e^{+} + (A,Z-2)^{*}$, which
occurs in left--right symmetric models via the exchange of
right--handed neutrinos and $W_R^\pm$ gauge bosons.  The best limit
presently on this process is $B(\mu^- + {\rm Ti} \rightarrow
e^+ + {\rm Ca}) < 1.7 \times 10^{-12}$, also from PSI.  TeV scale
left--right symmetry predicts observable rates for this
transition.

CLFV could also be seen in other muonic systems.  Muonium is a  ($\mu^+ e^-$) bound state   analogous to the hydrogen atom, which
in the presence of a CLFV interaction can oscillate into antimuonium
($\mu^-e^+$).  The doubly charged scalar of the seesaw model, the left--right
symmetric model, and the radiative neutrino mass model would all lead to
this process.  If
the Lagrangian for this process is parametrized as
\begin{equation}
H_{\rm Mu\overline{Mu}} = \left({G_{\rm Mu\overline{Mu}} \over
\sqrt{2}} \right)
\overline{\mu}\gamma_{\lambda}(1-\gamma_5){e}
\overline{\mu}\gamma^{\lambda}(1-\gamma_5){e} + h.c.,
\end{equation}
the current limit from PSI experiments is $G_{\rm Mu\overline{Mu}} < 0.003
\,G_F$, with room for improvement by several orders of magnitude in
the near future.

\subsubsection{Specific New Physics Models}

If we assume that neutrino mass is generated by a seesaw mechanism \cite{Minkowski:1977sc}, we can see effects in CLFV if the seesaw scale is low~\cite{de
Gouvea:2007uz}. It is perhaps more natural that the seesaw mechanism
is realized at a very high energy scale, $M_R \sim 10^{10} - 10^{14}$
GeV. In this case there can be significant CLFV provided that there is
some new physics at the TeV scale. Below we briefly mention two such
scenarios, SUSY and Randall-Sundrum (RS) warped extra dimensions.

Implementing the seesaw mechanism within the context of SUSY leads to
a new source of CLFV. In a momentum range  between $M_R$ and  $M_{\rm
Pl}$, where $M_{\rm Pl}$ is the fundamental Planck scale, the
right-handed neutrinos are active and their Dirac
Yukawa couplings with the lepton doublets induce flavor
violation among the sleptons.  The sleptons must have masses of order
TeV or less, if SUSY is to solve the hierarchy problem, and they carry
information on flavor violation originating from the seesaw.
Specifically, the squared masses of the sleptons would receive flavor
violating contributions given by
\begin{equation}
(m_{\tilde{l}_{L}}^{2})_{ij} \simeq -\frac{1}{8\pi^{2}}
(Y_{\nu}^\dagger Y_{\nu})_{ij} (3 m_{0}^{2} + |A_{0}|^{2})\ln\left({M_{\rm
Pl}\over M_{R}}\right)\,,
\label{CL:SUSY-LFV}
\end{equation}
where $Y_\nu$ is the Dirac Yukawa coupling of the neutrinos, and $m_0$
and $A_0$ are SUSY-breaking mass parameters of order $100$ GeV.  In
SUSY Grand Unified Theories (GUTs), even without neutrino masses, there is an independent
contribution to CLFV, originating from the grouping of quarks and
leptons in the same GUT multiplet.   The squared masses of
the right-handed sleptons would receive contributions to CLFV in this momentum range from the GUT
scale particles that are active, given by
\begin{equation}
(m_{\tilde{e}_{R}}^{2})_{ij} \simeq -\frac{3}{8\pi^{2}}
V_{3i}V_{3j}^{*}|Y_t|^{2}(3 m_{0}^{2}  +
|A_{0}|^{2})\ln\left(\frac{M_{\rm Pl}}{M_{{\rm GUT}}}\right).
\label{CL:GUT_LFV}
\end{equation}
Here $V_{ij}$ denote the known CKM quark mixing matrix elements, and
$Y_t$ is the top quark Yukawa coupling.  Unlike
Eq.~(\ref{CL:SUSY-LFV}), which has some ambiguity since $Y_\nu$ is not
fully known, the CLFV contribution from Eq.~(\ref{CL:GUT_LFV}) is
experimentally determined, apart from the SUSY parameters. 

We next consider RS models with
bulk gauge fields and fermions.   In these models, our universe
 is localized on one (ultraviolet) membrane of a multidimensional 
 space while the Higgs field is localized on a different (infrared) membrane. 
  Each particle has a wave function that is localized near the Higgs 
  membrane for heavy particles or near our membrane 
  for light particles. Thus localization of different wave functions between
   the membranes  generates flavor.
For a given fermion mass spectrum, there are only two free parameters,
 an energy scale to set the Yukawa couplings and a length scale of 
 compactification that sets the level of Kaluza-Klein  (KK) excitations.  
The two scales can be accessed using a combination of tree-induced CLFV 
processes that occur in $\mu N\rightarrow e N$ or $\mu\rightarrow 3e$ and 
loop-induced interactions such as $\mu\rightarrow e\gamma$ \cite{hep-ph/0606021,arXiv:1004.2037}. The amplitude of loop-induced
flavor-changing decays, such as $\mu\to e \gamma$, is given by a
positive power of the Yukawa and a negative power of the KK
scale.  Tree-level flavor-changing diagrams, on the other hand, come
from four-fermion interactions whose flavor-changing vertices come
from the non-universal profile of an intermediate KK gauge boson. This
non-universality is an effect of electroweak symmetry breaking so that
the flavor-changing part of the KK gauge boson profile is localized
near the IR brane and the size of flavor-changing effects depend on
the size of the zero-mode fermion profile towards the IR
brane. However, in order to maintain the Standard Model fermion
spectrum the zero-mode fermion profiles must be pushed away from the
Higgs vacuum expectation value on the IR brane as the anarchic Yukawa scale is
increased. Thus the tree-level flavor changing amplitudes go like a
negative power of the anarchic Yukawa scale. For a given KK scale,
experimental constraints on lepton flavor-changing processes at tree
and loop level thus set lower and upper bounds on the Yukawa scale,
respectively.

A version of minimal flavor violation exists in RS models where the new
 scales have a very small effect on low energy flavor changing processes.  
 It was noted in \cite{hep-ph/0606021} and \cite{Agashe:2004cp} that
certain flavor changing diagrams are suppressed in the RS scenario
because the particular structure of zero-mode wave functions and
Yukawa matrices is the same as the zero-mode mass terms induced by
electroweak symmetry breaking. When passing to the physical basis
of light fermions, these processes are also nearly diagonalized, or
aligned, and off-diagonal elements of these transitions are
suppressed. These flavor-changing processes are not completely zero 
since the fermion bulk masses are an additional flavor spurion
in these theories. In other words, the $U(3)^3$ lepton flavor 
symmetry is not restored in the limit where the Yukawa terms vanish. 
The full one-loop calculation of $\mu\to e \gamma$ in Randall-Sundrum
models including these misalignment effects and a proof of finiteness 
was performed in \cite{arXiv:1004.2037}.

\subsection{Muon Experimental Overview}\label{sec:cl:exp}

Experimental searches for lepton flavor violation in the muon sector
began in Canada in 1947~\cite{cl:1948_muegamma} and over 64 years the
limits have improved by 10 orders of magnitude. Recent advances in
accelerator and detector technology mean it is possible to extend the
search considerably, by up to four orders of magnitude in some
processes, in the next 5-10 years. The region such searches would be
probing is an extremely rich one theoretically. The predicted rate for
muon flavor violation is ${\cal{O}}(10^{-52})$ in the Standard Model
and therefore any positive signal is unambiguous evidence for new
physics beyond the Standard Model. 

\begin{table}[!b]
  \caption{
    \label{tab:muon-lfv}
   Evolution of the 95\% CL limits on the main CLFV observables with initial state muons.  The expected limits in the 5-to-10 year range are based on running or proposed experiments at existing facilities.  The expected bounds in the 10-to-20 year range are based on sensitivity studies using muon rates available at proposed new facilities. The numbers quoted for  $\mu^{+} \rightarrow e^{+}\gamma$ and  $\mu^{+} \rightarrow e^{+}e^-e^+$ are limits on the branching fraction.  The numbers quoted for $\mu^{-}N \rightarrow e^{-}N$ are limits on the rate with respect to the muon capture process $\mu^{-}N \rightarrow \nu_\mu N^\prime$.  Below the numbers are the corresponding experiments or facilities and the year the current limit was set.
  }
  \begin{center}
    \begin{tabular}{lllll}
      \hline \hline
Process & Current limit &\multicolumn{2}{c}{ Expected limit} & Expected limit\\
 & & \multicolumn{2}{c}{5-10 years }& 10-20 years  \\
      \hline
      $\mu^{+} \rightarrow e^{+}\gamma$ & $2.4 \times 10^{-12}$& \multicolumn{2}{l}{$1 \times 10^{-13}$}& $1 \times 10^{-14}$ \\
      & PSI/MEG (2011) &\multicolumn{2}{l}{ PSI/MEG }& PSI,  Project X  \\
& & & \\
 \hline
      $\mu^+ \rightarrow e^+e^-e^+$ & $1 \times 10^{-12}$ & $1\times 10^{-15}$ & $1 \times 10^{-16}$ & $1\times 10^{-17}$ \\
           & PSI/SINDRUM-I (1988) & Osaka/MuSIC & PSI/$\mu3e$ & PSI, Project X  \\
& & & \\
\hline
$\mu^{-}N \rightarrow e^{-}N$ & $7\times 10^{-13}$ & $1\times 10^{-14} $ & $6\times 10^{-17}$  & $1\times 10^{-18} $\\	
     & PSI/SINDRUM-II (2006) & J-PARC/DeeMee &  FNAL/Mu2e & J-PARC, Project X  \\
     & & & \\
     \hline \hline
    \end{tabular}
  \end{center}
\end{table}

The theoretical models predict a
variety of rates for muon lepton flavor violating processes, the
majority of which are comfortably within reach of proposed programs
over the next decade. The predicted rates for the three experimental
processes with the greatest theoretical sensitivity: $\mu^{-}N
\rightarrow e^{-}N$, $\mu^{+} \rightarrow e^{+}\gamma$, $\mu^{+}
\rightarrow e^{+}e^{-}e^{+}$ (and corresponding $\tau$ decays) vary
significantly between the models. It is thus imperative that all three
processes are measured with commensurate sensitivity.  
 Additionally the rates for $\mu^{-}N \rightarrow e^{-}N$, which
is a coherent process amongst all nuclei, have a strong dependence on
the atomic number, and this dependence is again model dependent. In the
event of a signal being observed, a comparison of the $\mu^{-}N
\rightarrow e^{-}N$ rates between a low-$Z$ and a high-$Z$ target is
very important.
Table~\ref{tab:muon-lfv} lists the expected evolution in limits on the main three processes over the next two decades.  Details of each experiment are given in the following sections.

\subsubsection{Muon Flavor Violation Experiments in the Next Decade}

The upcoming decade will see the launch of several new CLFV
experiments that promise substantial improvements over the
sensitivities of previous experiments.  First, the field has a running
experiment in MEG at PSI, searching for $\mu\rightarrow e \gamma$,
 that is refining its techniques after its
most recent run.  In $\mu^+ \rightarrow e^+e^-e^+$, the $\mu 3 e$ collaboration
at PSI is preparing a proposal to improve that limit from $1.0 \times
10^{-12}$ to $10^{-16}$; at Osaka, MuSIC hopes to reach $10^{-15}$.
The DeeMe experiment at J-PARC could improve the existing
muon-to-electron conversion limit by two orders of magnitude.
 Later
in the decade, two muon-to-electron conversion experiments, COMET at
J-PARC and Mu2e at Fermilab, will start taking data and improve the
limits on $\mu N \rightarrow eN$ by four orders of magnitude.

The only running experiment is MEG measuring the $\mu^{+}\rightarrow e^{+}\gamma$
process at PSI. In 2011 it placed a 90\% CL upper limit
$BR(\mu^{+}\rightarrow e^{+}\gamma) < 2.4 \times
10^{-12}$~\cite{cl:MEG_2011_limit}. The MEG limit, in the absence of a
signal, will go below $10^{-12}$ by the end of 2012 owing to a
doubling in the number of stopped muons per annum compared to 2010.
In the longer term MEG plans to improve its positron detection
efficiency and resolutions to achieve a sensitivity of $\approx
10^{-13}$ in 2016. The dominant background to the search is from a
coincident positron and a photon from different stopped muons which
can only be countered by tight cuts on the energy, angle and time of
the $e^+$ and photon candidates. Experimental resolution, along with a
sufficient rate of stopped muons, is therefore the main driver in
determining the ultimate sensitivity. The critical resolutions are the
photon energy and angle, which come in squared in the sensitivity. A
sensitivity below $10^{-13}$ will likely only be achieved by measuring
the $e^+e^-$ conversion products of the photon in a tracking detector,
where superior angle and energy measurements are possible compared to
a calorimetric measurement as presently used in MEG. The thickness of
the material prior to the tracker for the conversion will need to be
optimized to balance rate against resolution degradation. Such a
track-only detector in conjunction with a stopped muon rate of
$10^{9}~{\rm s^{-1}}$ over a four-year period may be sufficient to
attain a sensitivity of $10^{-14}$ provided a reasonable detection
efficiency can be achieved in an environment of significant event
pile-up.

The $\mu^+ \rightarrow e^+e^-e^+$ experiments are studying how to
improve the existing limits from SINDRUM, more than a decade old, by up to
$10^4$.  There are two significant backgrounds in a $\mu^+ \rightarrow e^+e^-e^+$ 
search: the intrinsic physics background of radiative $\mu^+
\rightarrow e^+e^+e^- \nu \nu$ decays with small neutrino energy, and
backgrounds from multiple events where electrons and positrons from
different stopped muons combine to form accidental backgrounds.
 MuSIC will likely use a time projection chamber (TPC) to track the $e^{\pm}$ whereas
$\mu 3e$, which is ultimately seeking a sensitivity beyond $10^{-16}$,
will use HV-MAPS~\cite{cl:HV_MAPS} silicon pixels in conjunction with
a fiber tracker. The HV-MAPS sensors are very thin ($\sim 30~\mu{\rm
m})$ ensuring the resolution contribution from multiple scattering is
$\sim$~5~mrad and that precision tracking in a high occupancy
environment can be achieved. The $\mu 3e$ proposal will initially use the
MEG ($\pi{E5}$) beamline but ultimately would move to a new beamline,
such as the neutron spallation SINQ beamline~\cite{cl:SINQ}, where
stopped muon rates as high as $10^{10}~{\rm s}^{-1}$ are
envisaged.

The $\mu \rightarrow e$ conversion experiments COMET and Mu2e make two
major changes from prior experiments.  First, they use a pulsed beam.
The advantage of a pulsed beam is that the experiment can wait for the
activity from the initial proton pulse to decay or pass through the
detector and examine events only after this activity has ended, both
eliminating backgrounds and reducing rates.  Second, they use a
three-stage solenoidal beamline system: (1) a production target solenoid, which
makes pions that decay into muons; (2) a magnetic solenoid system that
transports the muons; (3) a stopping target solenoid that captures the muons
and allows them to interact with nuclei and convert into detectable
electrons. The DeeMe proposal simplifies this system by eliminating
the stopping target and looking for conversions in the production
target, and aims to improve sensitivity by two orders of magnitude at a lower 
cost than Mu2e or COMET.

Mu2e and COMET will probe with single-event sensitivities of $\approx
2 \times 10^{-17}$. If there is no signal, these experiments will set
new limits on CLFV-mediating particles near $10^4$ TeV.  If there is a
signal, the new physics responsible must be pinned down by varying  the
$Z$ of the stopping target.  The biggest discriminating power is found
for high-$Z$ materials such as Au.  However, the lifetime of a
captured muon drops with $Z$ so that new beamlines and
new detectors will be required.

\subsubsection{Muon Flavor Violation: The Next Generation}

The experiments described above will establish new techniques and
extend the measurements several orders of magnitude beyond the
state of the art.  Since the Standard Model backgrounds are ${\cal O}(
10^{-52} )$, the experiments can continue to improve without
theoretical uncertainties clouding a discovery. The results of MEG, or
information from the proposed $\mu  3e$ or DeeMe
experiments, will inform future choices for any of the experiments
under discussion.  Other $\Delta F=2$ processes, such as
muonium-antimuonium oscillations, can be greatly improved as well.
One common theme for improving all these experiments is flexibility in
the time structure and new muon sources of unprecedented intensity.

As described above, Mu2e and COMET use a pulsed beam. In
high-$Z$ elements the lifetime is so short ({\it e.g.}, 72 nsec for Au) that
the technique does not work. The initial activity from the proton
pulse will overlap with an electron conversion signal, blinding the
detectors and making the experiments susceptible to backgrounds.  Also,
a potentially limiting background of current $\mu N \rightarrow eN$
searches is radiative pion capture, where remnant pions in the muon
beam line are captured on nuclei in the stopping target and emit an
energetic photon. That photon can then convert in the same material
used to capture the muon and produce an electron in the right energy
range, which is indistinguishable from the signal.  This background
will not limit the Mu2e or COMET experiments since by waiting for the pions
to decay, the background is reduced to an acceptable level.
However, in going beyond the first round, this background becomes
a serious issue. Therefore, in order to progress, the mechanism
of muon transport must change.  Particularly one would
like to run with a single thin stopping target to reduce energy loss,
which is only possible if the muon beam has a very low kinetic energy (about
2~MeV) with a small momentum and time spread. This could potentially
be achieved using a ``dipole wedge'' arrangement in conjunction with
helical-cooling channels~\cite{cl:dipole_wedge} as proposed for
Project X or a  fixed-field alternating gradient accelerator 
(FFAG) as envisaged in the PRISM/PRIME~\cite{cl:PRISM}
experiment. The use of an FFAG would also ensure all the pions have
decayed prior to the stopping target. In any such experiment the
number of stopped muons per second would have to exceed $10^{13}$.

The
challenges in achieving this high rate in conjunction with the low kinetic energy,
small $\Delta{p}$, and $\Delta{t}$ have much in common with those facing
the design of a Neutrino Factory/Muon Collider. Thus the same research and 
development in accelerator technology is required to advance CLFV, 
the intensity frontier neutrino program, and the energy frontier lepton collider program. 

The processes considered so far violate lepton flavor by one unit. The
muonium ($\mu^{+}e^{-}$) to anti-muonium ($\mu^{-}e^{+}$) transition
violates lepton flavor by two units. The signature of a 13.5 eV
$e^{+}$ in conjunction with an electron from a $\mu^{-}$ decay has
been sought in several experiments, and an upper limit on the
probability of a spontaneous transition of $8.3 \times 10^{-11}$ has
been determined at 90\% CL by the PSI MACS
experiment~\cite{Willmann:1998gd}. Recent advances in detector
technology should allow far better resolutions to be achieved and a
reduction in this limit by at least two orders of magnitude to be
readily obtained.

Many of the requirements of muon experiments in the next decade can be met by Project X.
Project X is a US led accelerator initiative with strong international participation
that aims to realize a next-generation proton source that will dramatically extend
the reach of intensity frontier research.
The state of the art in super-conducting RF has advanced to a point where it
can be considered and implemented as the core enabling technology for a next
generation multi-megawatt proton source--reliably delivering unprecedented
beam power at duty factors ranging from $10^{-5}$ to $100\%$.
The base super-conducting radio-frequency (RF) technology also supports flexible
beam-timing configurations among simultaneous experiments, allowing a
broad range of experiments to develop and operate in parallel.
The DOE Office of High Energy Physics and its advisory bodies have
recognized this potential and are supporting R$\&$D for Project X
that could lead to a construction start as early as 2016.

The high-power rare-decay campus at Project X is driven by 3000 kW of
beam power at 3 GeV (proton kinetic energy) that can be distributed
with very low losses between multiple experiments using splitter
technology developed and refined by Jefferson Lab.  The front end of
the Project X accelerator includes an innovative beam chopper system
that supports essentially any beam pulse train with frequency
components between 1 MHz and 160 MHz for experiments.  The Project X
muon program could expect about one third of the total rare-decay
campus beam power, about 1000 kW.  Muon capture initiatives such as
muon-to-electron conversion experiments can receive pulse trains
matched to the lifetime of muonic atoms ranging from 1 MHz (low-Z
atoms with 1-$\mu$ second lifetime) to 10 MHz (high-Z atoms with 0.1
$\mu$ -second lifetime).  Experiments searching for multi-body final
states such as $\mu \rightarrow e \gamma$ and $\mu^+ \rightarrow e^+e^-e^+$
would benefit from the highest frequency beam (160 MHz) to minimize
accidental rate effects.  Recent studies of muon production yield on
practical high-power carbon targets running at 1000 kW have shown good
scaling of the yield compared to lower power high-Z targets
({\it e.g.}, COMET at J-PARC and Mu2e at Fermilab), supporting the development
of muon-to-electron concepts at Project X with two orders of magnitude
beyond the sensitivity of experiments planned this decade.

\section{Muon Flavor-Conserving Processes}\label{sec:cl:fcp}

\subsection{Muon $g-2$ and EDM}

The muon provides a unique opportunity to explore the properties of a
second-generation particle  with great precision. Several muon properties make
these measurements possible:  it has a long
lifetime of $\simeq 2.2~\mu$s;  it is produced in the weak decay
$\pi^- \rightarrow \mu^- \bar \nu_\mu$, providing copious numbers of
polarized muons; and the weak decay
$\mu^- \rightarrow e^- \nu_\mu \bar \nu_e $ is self-analyzing, providing information on the muon spin direction at the time of decay.

In his famous paper on the relativistic theory of the electron,
Dirac\cite{Dirac28} obtained the correct magnetic moment for the
electron, and he also mentioned the possibility of an electric
 dipole moment, which like the magnetic dipole moment (MDM)
would be directed along the electron spin direction.
  The magnetic dipole and electric dipole moments
are given by
\begin{equation}
\vec \mu = g \left( \frac{Qe}{ 2m}\right) \vec s\, , \qquad
 \vec d = \eta  \left(\frac {Qe  }{ 2
     mc}\right)
\vec s \, ,
\label{eq:MDM-EDMdef}
\end{equation}
where  $Q =  \pm 1$ and $e>0$. Dirac theory predicts $g \equiv 2$,
but radiative corrections dominated by the
lowest-order (mass-independent) Schwinger contribution $a_{e,\mu,\tau} =
 \alpha/(2\pi)$~\cite{Schwinger48} make it necessary to
write the magnetic moment as
\be
\mu = \left(1 + a\right)\frac{Qe \hbar }{ 2m}\quad {\rm with} \quad
a = \frac{{g - 2} }{ 2}.
\label{eq:muon-MDM}
\eeq

The muon played an important role in our discovery of the generation
structure of the SM when
 experiments at the Nevis
cyclotron
showed  that $g_\mu$ was consistent with 2~\cite{Garwin57}.
Subsequent experiments at Nevis and CERN showed  that
$a_\mu \simeq \alpha/(2\pi)$~\cite{Garwin60,Charpak61},
implying that in a magnetic field, the muon
behaves like a heavy electron.
The SM value of the muon anomaly is now known
to better than half a part per million (ppm), and has
been measured to a similar precision~\cite{Bennett06}.

The quantity $\eta$ in Eq.~\ref{eq:MDM-EDMdef}
 is analogous to the $g$-value for the magnetic dipole
moment. An EDM violates both {\sl P} and {\sl T}
symmetries~\cite{Purcell50,Landau57,Ramsey58}, and assuming the
 $C\!PT$ theorem, {\sl C\!P} as well.  Thus
searches for EDMs provide an important tool in our quest to
find non-Standard Model {\sl C\!P} violation.

The measured value of the muon anomalous magnetic moment is in apparent
disagreement with the expected value based on the
SM.  The Brookhaven National Laboratory (BNL) E821 experiment finds~\cite{hep-ex/0602035}
\begin{equation} \hspace*{-23pt}
    a_\mu(\textrm{Expt}) = 116\,592\,089(54)(33)\times10^{-11},
    \label{eq:e821}
\end{equation}
where $a_\mu=(g-2)/2$ is the muon anomaly, and the uncertainties are
statistical and systematic, respectively.  This can be compared with
the SM prediction~\cite{arXiv:1010.4180,931465}
\begin{equation}
    a_\mu(\textrm{SM})   = 116\,591\,802(42)(26)(02)\times10^{-11},
    \label{eq:SM}
\end{equation}
where the uncertainties are from the $\mathrm{O}(\alpha^2)$ hadronic vacuum
polarization (HVP) contribution, $\mathrm{O}(\alpha^3)$ hadronic
contributions (including hadronic light-by-light (HLbL) scattering),
and all others (pure QED, including a 5-loop
estimate~\cite{arXiv:1110.2826}, and electroweak, including
2-loops~\cite{hep-ph/0212229}). The hadronic contributions dominate
the uncertainty in $a_\mu(\rm SM)$.  The discrepancy between the
measurement and the SM stands at
\begin{equation}
\Delta a_\mu=287(80)\times 10^{-11}
\end{equation}
(3.6 standard deviations ($\sigma$)), when based on the $e^+e^-\to\rm
hadrons$ analysis for the HVP
contribution~\cite{arXiv:1010.4180}. When the HVP analysis is
complemented by $\tau\to\rm hadrons$, the discrepancy is reduced to
2.4$\sigma$~\cite{arXiv:1010.4180}. However, a recent re-analysis,
employing effective field theory techniques, of the $\tau$
data~\cite{arXiv:1101.2872} shows virtual agreement with the
$e^+e^-$-based analysis, which would solidify the current discrepancy
at the 3.6$\sigma$ level. $\Delta a_\mu$ is large, roughly two times
the electroweak contribution~\cite{hep-ph/0212229}, indicating potentially
large new physics contributions.

\subsection{Muon $g-2$: Standard Model Contributions}\label{sec:cl:g-2the}

\begin{figure}[bp]
    \centering
    \includegraphics[width=0.3\columnwidth]{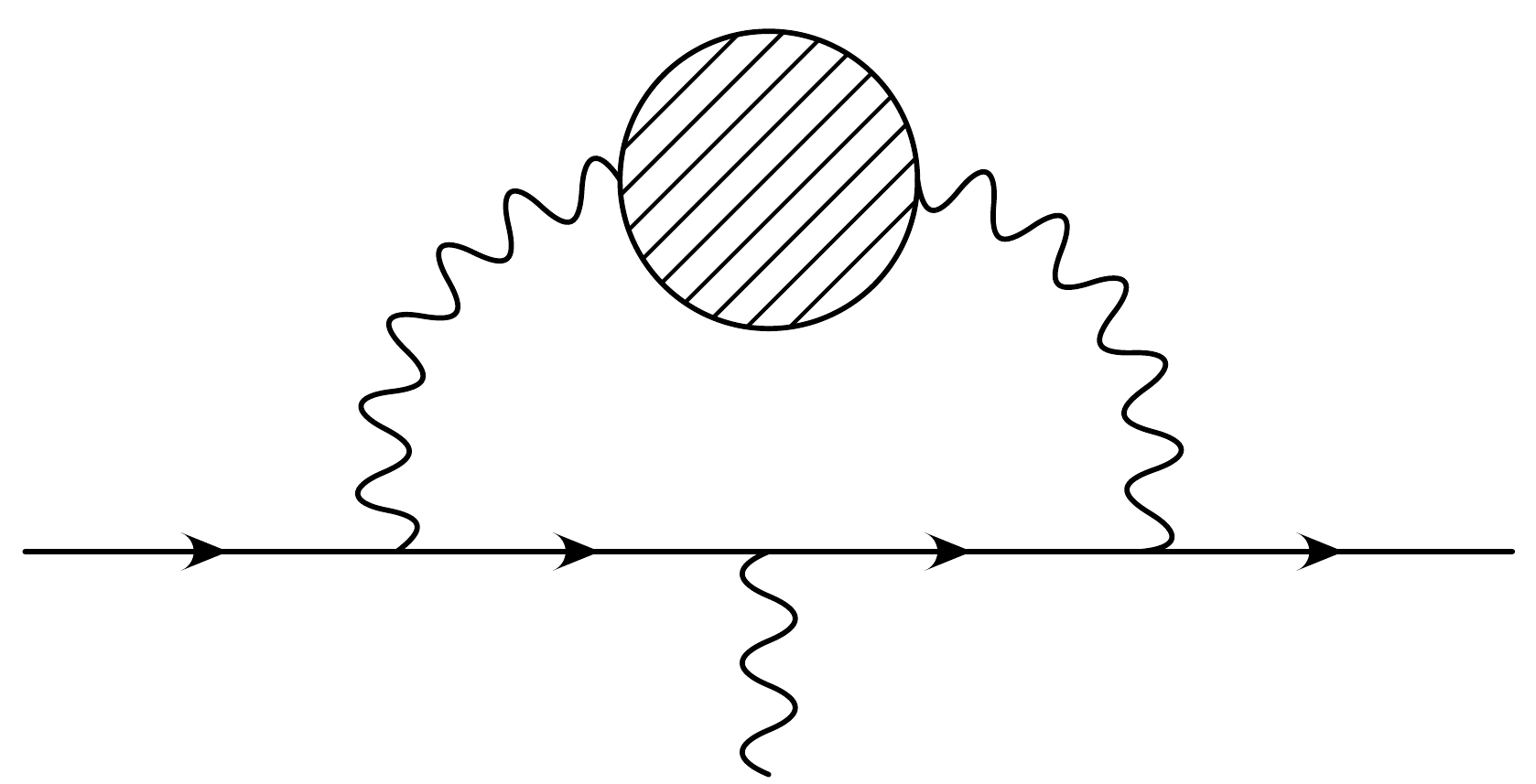}\hskip 1cm
    \includegraphics[width=0.3\columnwidth]{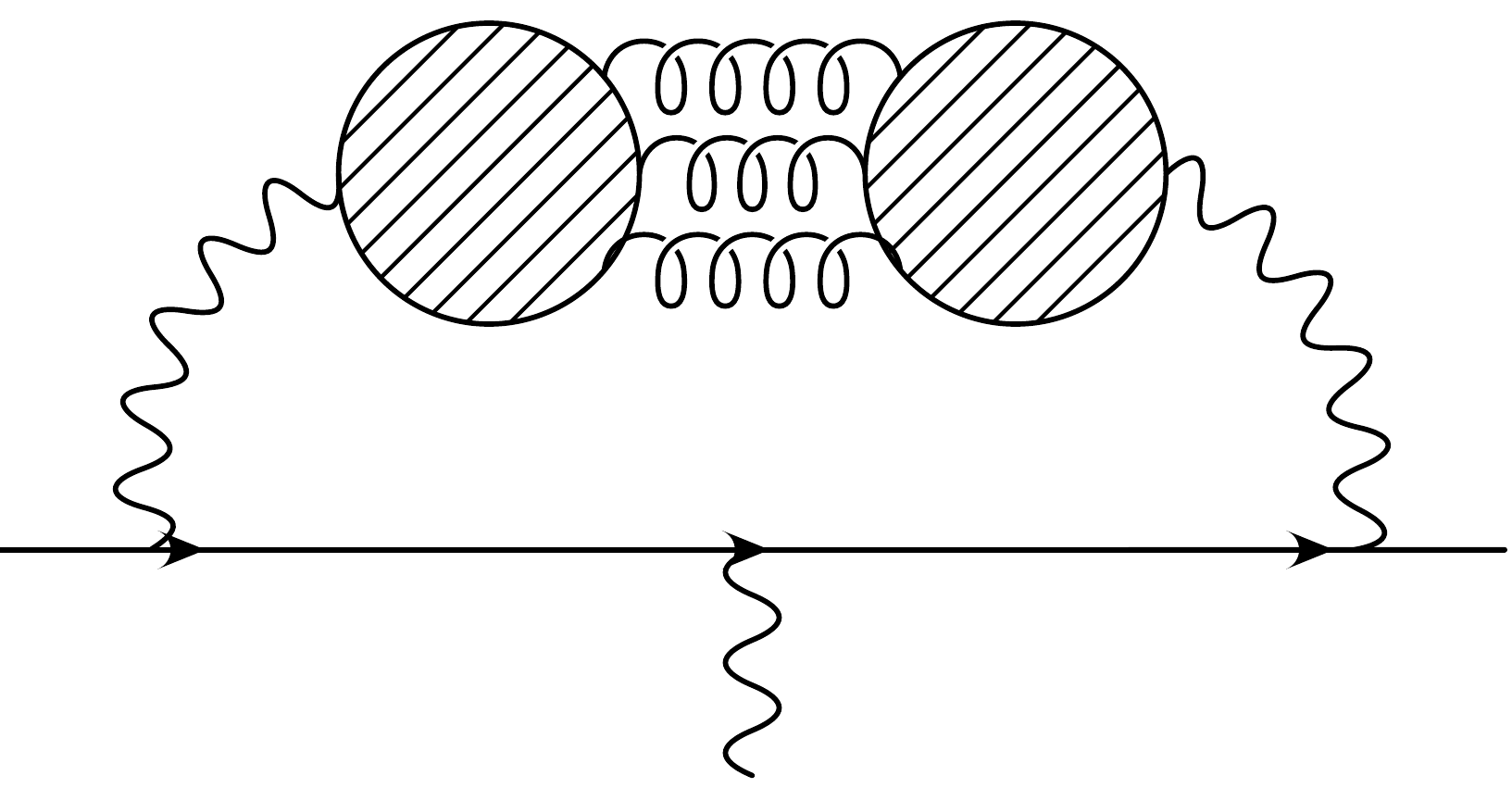}
\caption{Hadronic vacuum polarization diagrams contributing to the SM muon anomaly. The horizontal lines represent the muon. (Left panel) The blob formed by the quark-antiquark loop represents all possible hadronic intermediate states.  (Right panel) Disconnected quark line contribution. The quark loops are connected by gluons.}

    \label{fig:hvp}
\end{figure}

The HVP contribution to $a_\mu$ can be determined from the
cross-section for $e^+e^-\to\rm hadrons$ (and over a certain energy
range, by $\tau\to\rm hadrons$) and a dispersion relation. It can also
be computed from purely first principles using lattice QCD to
calculate the HVP directly~\cite{hep-lat/0212018}. The two methods are
complementary and can be used to check each other. The current best
uncertainty comes from the first method,
\begin{equation}
a_\mu(\rm HVP)=(692.3\pm4.2)\times 10^{-10},
\end{equation}
or about 0.61\%~\cite{arXiv:1010.4180} when only $e^+e^-$ data are used. 
If $\tau$ data are included, $a_\mu=701.5\pm4.7\times 10^{-10}$, or 0.67\% 
(but see~\cite{arXiv:1101.2872} for the analysis that brings the $\tau$ 
into good agreement with $e^+e^-$). In the next 3-5 years the uncertainty on 
$a_\mu(\rm HVP)$ is expected to drop by roughly a factor of two, relying 
on new results from {\babar}, Belle, BES, and VEPP2000.
The lattice calculations presently have an uncertainty of 
about 5\%~\cite{hep-lat/0608011, arXiv:1103.4818, Boyle:2011hu,  DellaMorte:2011aa}, which is 
expected to decrease to 1-2\% in the next 3-5 years~\cite{USQCD}. At the 
1\%~level, contributions from the charm and so-called disconnected 
diagrams (right panel, Fig.~\ref{fig:hvp}) enter. Both are currently under investigation.

\begin{figure}[bp]
    \centering
\vspace*{-2pt}
    \includegraphics[width=0.3\columnwidth]{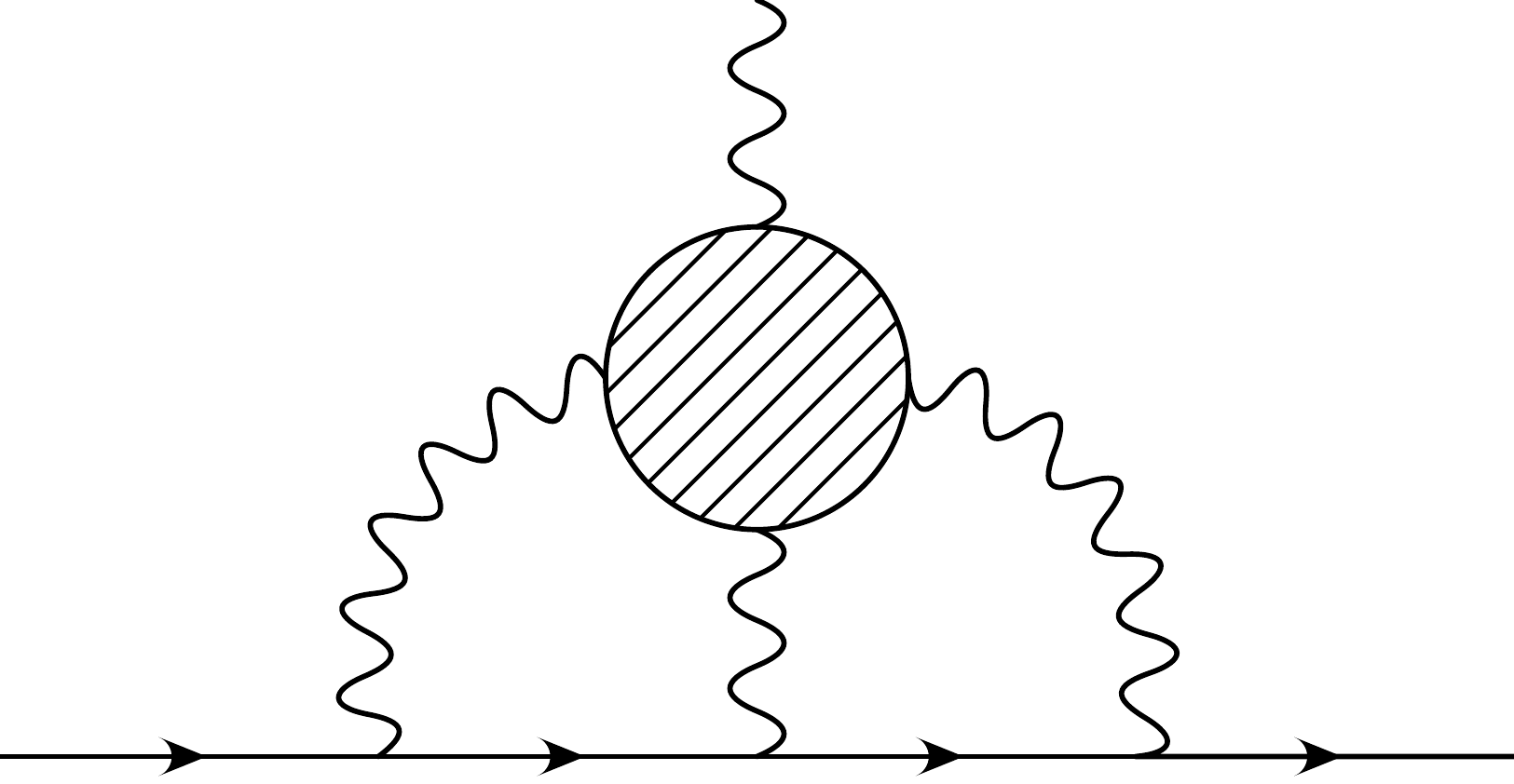}\hskip 1cm
    \includegraphics[width=0.3\columnwidth]{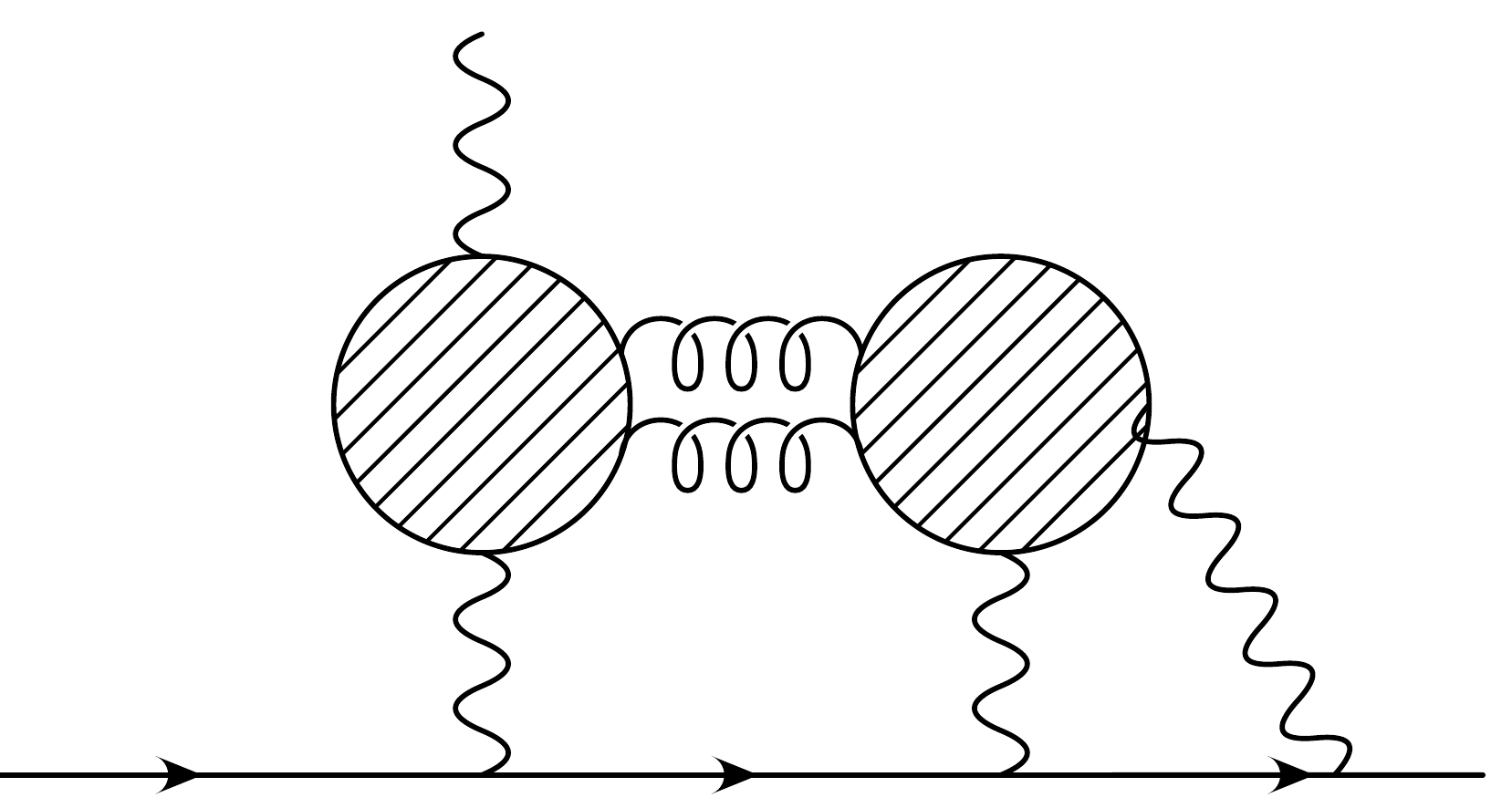}
    \caption{Hadronic light-by-light scattering diagrams contributing to the SM muon anomaly. The horizontal lines represent the muon. (Left panel) The blob formed by the quark loop represents all possible hadronic intermediate states.  (Right panel) One of the disconnected quark line contributions. The quark loops are connected by gluons.}
    \label{fig:hlbl}
\end{figure}

The hadronic light-by-light scattering amplitude shown in
Fig.~\ref{fig:hlbl} is much more challenging.  The contribution to
$g-2$,
\begin{equation}
    a_\mu(\textrm{HLbL}) = 105(26) \times 10^{-11},
    \label{eq:PRV}
\end{equation}
is not well known. It is based on the size of various hadronic contributions estimated in
several different models~\cite{arXiv:0901.0306}.
Its uncertainty, though less than that in $a_\mu(\textrm{HVP})$ by about a factor of two, seems harder to reduce and is expected to be the dominant uncertainty as the HVP uncertainty is reduced.
Finding a new approach, such as lattice QCD, in which uncertainties are systematically improvable,
is crucial for making greatest use of the next round of experiments.
With this in mind, a workshop was recently convened at the Institute for Nuclear Theory~\cite{INTws}.
Workshop participants discussed how models, lattice QCD, and data-driven methods could be exploited to reduce the
uncertainty on $a_\mu(\textrm{HLbL})$.
The outcome of this workshop is that a SM calculation of the HLbL contribution with a total
uncertainty of 10\% or less can be achieved within five years. A detailed discussion of the computation of $a_\mu(\rm HLbL)$ in lattice QCD is given in the USQCD Collaboration white paper on $g-2$~\cite{USQCD}.

There are two methods, using the lattice framework, under investigation. The conventional one, analogous to the HVP calculation, is to calculate the correlation function of four electromagnetic currents for the quarks in pure QCD, one for each possible, independent, momentum configuration (there are $V^2$), fit the resulting function of discrete momenta to a smooth function and insert it into the two-loop QED integrals. The resulting four-Lorentz-index hadronic tensor has 32 independent contractions. For these reasons, the calculation is computationally demanding. An intermediate but useful step is to calculate the four-point correlation function at well chosen values of the vertex momenta to partially check model calculations.

A second method is to compute the entire amplitude on the lattice, including the muon, in a  combined QED+QCD gauge field~\cite{hep-lat/9602005,hep-lat/0509016,827504}. The method has passed several non-trivial tests. First, it has been successfully checked against perturbation theory in pure QED. Large finite volume effects (the photons are long range) appear manageable. Preliminary calculations in full QED+QCD, at unphysical quark and muon mass and momentum transfer $q^2$, show a statistically significant result. The method requires a non-perturbative subtraction of leading order in $\alpha$ contributions that has been checked by varying the strength of the electric charge in the calculations and observing the expected scaling, before and after the subtraction. Disconnected contributions like the one shown in the right panel of Fig.~\ref{fig:hlbl} have not been included yet, but will be once the simpler first diagram (left panel, same figure) is fully under control. Calculations on a larger volume with smaller masses are in progress.

In addition to these direct approaches, there is other ongoing work on lattice QCD calculations that check or
supplement the model calculations.
For example, it is well known that the pion pole (namely, $\gamma\gamma^*\to\pi^0\to\gamma^*\gamma^*$)
provides the largest contribution to the QCD blob in Fig.~\ref{fig:hlbl}.
Just as experiments are being mounted to examine this physics ({\it e.g.}, PrimEx at Jefferson Laboratory and KLOE at INFN Frascati
National Laboratory (LNF)),
several groups~\cite{arXiv:0810.5550,arXiv:0912.0253,XFeng} are using lattice QCD to compute the amplitudes for
$\pi^0\to\gamma\gamma^*$ and $\pi^0\to\gamma^*\gamma^*$ (with one or two virtual photons).

If the SM and experiment central values do not change while both experimental and theoretical uncertainties are reduced, the discrepancy between the two becomes irresistible. The improvement expected from E989 (0.14 ppm) by itself improves $\Delta a_\mu$ to 5$\sigma$. A simultaneous decrease in the HLbL uncertainty to 10\% from the current 25\% pushes it to 6$\sigma$, and finally, reducing the uncertainty on the HVP contribution by a factor of two increases it to 9 $\sigma$. Such a large and clear difference between experiment and the Standard Model for the muon $g-2$ will be extremely  discriminating between new physics scenarios responsible for this discrepancy and will significantly leverage results from the energy frontier being explored at the LHC.

\subsection{Muon $g-2$ Beyond the Standard Model Contributions}

The importance of $a_\mu$ as a constraint on physics beyond the Standard Model
(BSM) is due to two facts. First, different types of BSM physics can
contribute to $a_\mu$ in very different amounts, so $a_\mu$
constitutes a meaningful benchmark  and discriminator between BSM
models. Second, the
constraints from $a_\mu$ on BSM models are different and
complementary to constraints from other observables from the intensity  and energy frontiers.

The role of $a_\mu$ as a discriminator between very different Standard
Model extensions is well illustrated by a general relation discussed by
Czarnecki and Marciano~\cite{czmar}. If a new
physics model with a mass scale $\Lambda$
contributes to the muon mass by an amount $\delta m_\mu$, it also
contributes to $a_\mu$, and the two contributions are related as
\begin{equation}
\label{CzMbound} a_\mu|_{\mbox{NewPhysics}}={\cal O}(1)\times
\left(\frac{m_\mu}{\Lambda}\right)^2 \times \left(\frac{\delta
m_\mu}{m_\mu}\right).
\end{equation}
 The ratio
$C\equiv\delta m_\mu/{m_\mu}$ spans a wide
 range.   For models with new weakly interacting particles ({\it e.g.}, $Z'$,
 $W'$, little Higgs or universal extra dimension models
 \cite{Blanke:2007db,AppelqDob}) one typically
 obtains perturbative contributions to the muon mass ${\cal
  O}(\alpha/4\pi)$ and hence very small contributions to $a_\mu$.
For supersymmetric  models one obtains an additional factor
$\tan\beta$, which can naturally bring the contributions to $a_\mu$
into the region of the currently observed $\sim 3\sigma$ deviation,
depending on the precise values of the SUSY-breaking
parameters \cite{dsreview}. For models with radiative muon mass
generation one expects
$C\simeq1$ and contributions to $a_\mu$ can be quite
significant \cite{czmar}.

The complementarity between $a_\mu$ and other observables can be easily
seen. $a_\mu$ corresponds to a flavor- and
$C\!P$-conserving interaction that is sensitive to, and potentially
enhanced by, chirality flips.
Many other intensity frontier observables are
chirality flipping but flavor violating ($b$- or $K$-decays, $\mu\to
e$ conversion, etc.), or $C\!P$-violating (electric dipole
moments). Furthermore, while $a_\mu$ is sensitive to leptonic couplings,
$b$- or $K$-physics more naturally probe the hadronic couplings of new
physics. If CLFV exists, observables such
as $\mu\to e$ conversion can only determine a combination of the
strength of lepton-flavor violation and the mass scale of new
physics. Many
high energy collider observables are insensitive to chirality
flips. Particularly the LHC experiments are mainly sensitive to new
strongly interacting particles, while $a_\mu$ is sensitive to weakly
interacting new particles that interact with muons.

\begin{figure}[t]
\begin{center}
\includegraphics[width=14cm]{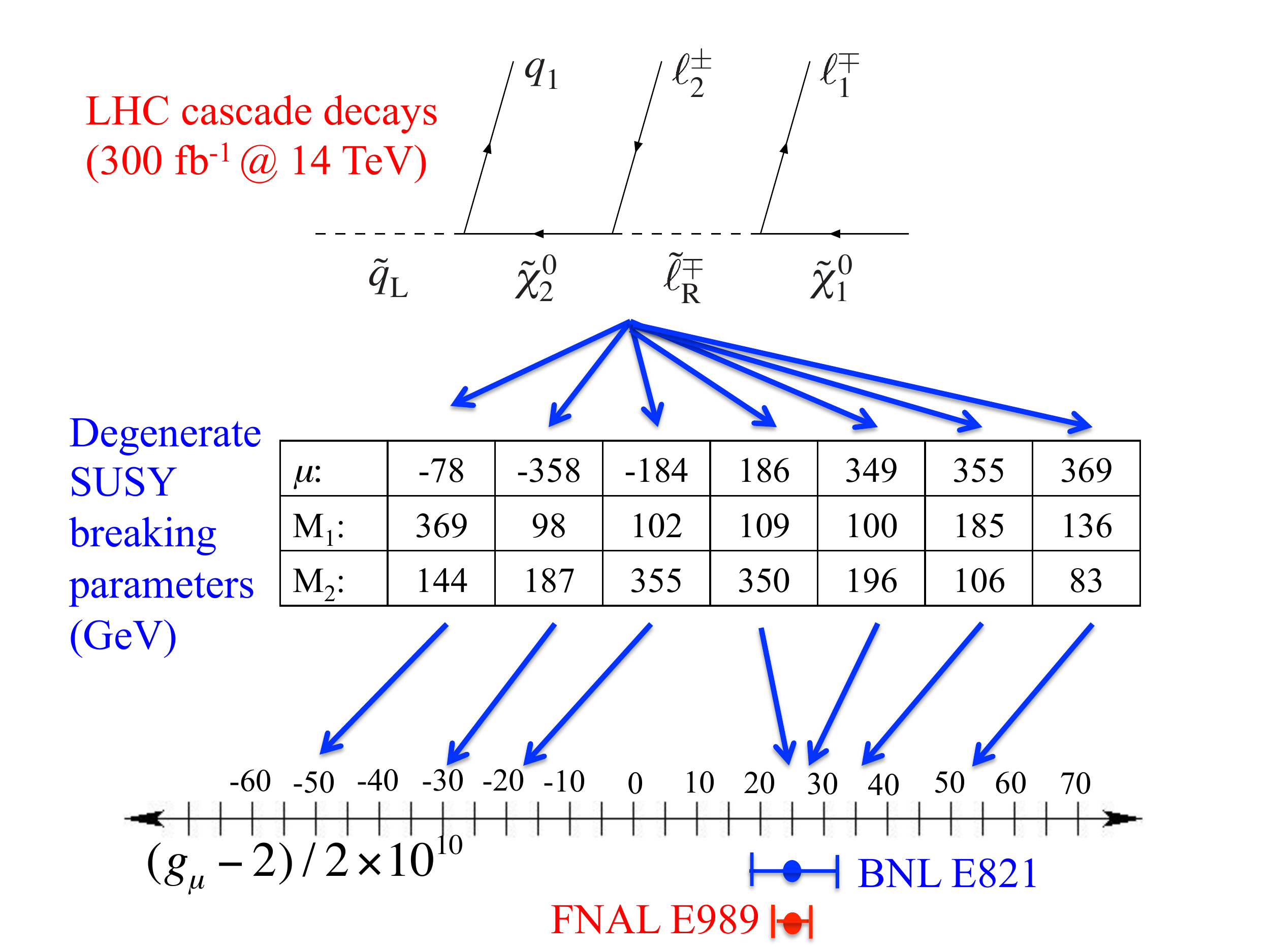} 
\caption{\label{fig:gm2susy} Graphical depiction of the muon $g-2$ solution to the LHC inverse problem. Depicted is a typical cascade decay of heavy SUSY particles into the lightest SUSY particle at the LHC.  A study indicates 12 degenerate solutions of GUT scale SUSY breaking parameters, six of which are shown.  The degeneracy is broken since the scenarios predict different values for the muon $g-2$ value.  The value from E821 and the expected result from E989, assuming the same central value, are also shown.}
\end{center}
\end{figure}

The complementarity between LHC and $a_\mu$
has been made quantitative in \cite{PhysicsCaseWP,PlehnRauchNew}
for a prospective $\tan\beta$ measurement within
SUSY. Similarly, $a_\mu$ can disentangle between the many
distinct (``degenerate'') SUSY scenarios of
\cite{Adam:2010uz} that can't be distinguished by LHC data
alone, as shown schematically in Fig.~\ref{fig:gm2susy}.  Multiple 
final states that would be produced at the LHC for a fixed point in the 
multidimensional SUSY parameter space were studied assuming 
300 fb$^{-1}$ at 14 TeV.  A 12-fold degeneracy was uncovered in 
the mapping back to the GUT scale SUSY breaking parameters, 
the Higgs mass parameter $\mu$ and the gaugino mass parameters M$_1$ and M$_2$.  
However, these solutions lead to distinct predictions for the $g-2$ anomaly.
 For discussions that
combine $a_\mu$ with LHC data see
{\it e.g.}, \cite{Bechtle:2011dm,Buchmueller:2011sw}.

\subsection{Muon $g-2$ and EDM: Experiment}\label{sec:cl:g-2exp}

Measurements of the magnetic and electric dipole moments make use of the
torque on a dipole in an external field, $\vec \tau = \vec\mu \times
\vec B + \vec d \times \vec E$. All muon MDM experiments except the original
Nevis ones used polarized muons in flight, and
 measured the rate at which the spin turns relative to the momentum,
$\vec \omega_a =\vec \omega_S - \vec \omega_C$, when a
 beam of polarized muons is injected into a magnetic field.
The resulting frequency, assuming that $\vec \beta \cdot \vec B = 0$,
 is given
by~\cite{Thomas26,Bargmann59}
\begin{equation}
\vec{\omega}_{a\eta}= \vec \omega_a + \vec \omega_\eta = -
 \frac {Qe}  {m}
\left[
a \vec{B}
+ \left( a - \left(  \frac {m} {p} \right)^2  \right)
 \frac {\vec{\beta} \times \vec{E}} {c} \right] -  \eta \frac {Qe}{2m}
 \left[ \frac {\vec{E}} {c}  +  \vec{\beta} \times \vec{B} \right] .
 \label{eq:omegaa-edm1}
\end{equation}
Important features of this equation are the motional magnetic and
electric fields:
$\vec \beta \times \vec E$ and $\vec \beta \times \vec B$.

The E821 Collaboration, working at the
Brookhaven Alternating Gradient Synchrotron (AGS), used an electric quadrupole field
to provide vertical focusing in the storage ring, and shimmed the magnetic
field to 1 ppm uniformity on average.  The storage ring was operated
 at the ``magic'' momentum, $p_{magic} = 3.094$~GeV/c,
($\gamma_{magic}= 29.3$),
so that $a_\mu = (m/p)^2$ and the electric field did not
contribute to $\omega_a$.
They obtained\cite{Bennett06}
\begin{equation}
  a_\mu^{(\mathrm{E821})} = 116\,592\,089(63) \times
  10^{-11}~~\mbox{(0.54\,ppm)}
\end{equation}
 The final uncertainty of
0.54~ppm consists of a 0.46~ppm statistical component and a 0.28~ppm
systematic
component.  

The present limit on the EDM also comes from E821~\cite{Bennett08-edm}
\begin{equation}
d_\mu = (0.1 \pm 0.9) \times 10^{-19} e  \cdot {\rm cm}; \
 \vert d_\mu \vert < 1.9 \times 10^{-19}  e  \cdot {\rm cm}\ (95\%\ {\rm C.L.})\, ,
\label{muedm-result}
\end{equation}
so the EDM contribution to the precession is very small.  In the muon $g-2$
experiments, the motional electric field dominates the $\omega_\eta$ term,
which means that $\vec \omega_a$ and $\vec \omega_\eta$ are orthogonal.
The presence of an EDM in the magic momentum experiments has two effects:
the measured frequency is the quadrature sum of the two frequencies,
 $\omega = \sqrt{\omega_a^2 + \omega_\eta^2}$, and the EDM causes a tipping of
 the plane of precession, by an angle
 $\delta = \tan ^{-1}[ \eta \beta/(2a_\mu)]$. This tipping results 
 in an up-down oscillation of the decay
 positrons relative to the midplane of the storage ring with frequency
$\omega_a$ out of phase by $\pi/2$ with the $a_\mu$ precession.

\subsubsection{ Future $g-2$ Experiments}

The E989 collaboration
at Fermilab will move the E821 muon storage ring to Fermilab, and
will use the magic momentum technique to measure $a_{\mu^+}$.
 New detectors and electronics, and a
beam handling scheme that increases the stored muon rate per hour
 by a factor of six
over E821, will be implemented.  The goal is at least 21 times the
statistics of E821 and a factor of four overall uncertainty reduction, with
equal systematic and statistical uncertainties of $\pm 0.1$ ppm.

The scope of Project X includes 50-200kW of beam power at 8 GeV,
about three to 15 times the beam power of E989.  This large step in beam
power could be used to measure $g-2$ for negative muons
and provide muon beams with lower emittance, thereby reducing
experimental systematics. 

Given the high impact of the E821 result and the 
crucial role the value of $g-2$ plays in interpreting energy frontier results, 
it is imperative to have a second measurement with at least equal 
precision but with a complementary approach to the measurement. 
An alternate approach planned for J-PARC~\cite{JPARC-Lg2} uses a much lower muon
energy, and does not use the magic momentum technique. A surface muon beam
produced by  the low energy Booster is brought to rest in an aerogel
target, where muonium (the $\mu^+ e^-$ atom) is formed.  The muonium
is ionized by a powerful laser, which produces a very slow muon beam with
extremely small emittance. This low-emittance beam is then accelerated by a
linac to 300 MeV, and injected into an~$\sim 1$~m diameter
solenoidal magnet with point-to-point
uniformity of $\pm 1$ ppm, approximately 100 times better than at the Brookhaven experiment.  The average uniformity is expected to be 
known to better than 0.1 ppm.  The decays are detected by a full volume tracker consisting of an array of silicon
detectors.  This provides time, energy, and decay angle information for every positron, maximizing the sensitivity to separate the  $g-2$ and EDM precession frequencies.  The expected   $g-2$ sensitivity is comparable to the
 Fermilab experiment but will have very different systematic uncertainties, and the combined results from the two experiments should bring the precision to below the 100 ppb level.

\subsubsection{ Future Muon EDM Experiments}

 A number of the E989 detector stations will
be instrumented with straw chambers to measure the decay positron
tracks. With this instrumentation, a simultaneous EDM
measurement can be made during the $a_\mu$ data collection,
improving on the  E821 muon EDM~\cite{Bennett08-edm}
 limit by up to two orders of magnitude down to
$\sim 10^{-21}\,  e \cdot {\rm cm}$.   The J-PARC muon $g-2$ proposal also
 will have decay angle information for all tracks and expects a similar improvement.

To go beyond this level will require a dedicated EDM experiment that
uses
the ``frozen spin'' method~\cite{Farley04,Roberts2010}.
 The idea is to operate a
muon storage ring off of the magic momentum and to use a radial electric field
to cancel the $\omega_a$ term in Eq.~\ref{eq:omegaa-edm1},
 the $g-2$ precession.  The  electric field needed to freeze the spin is
$E \simeq aBc\beta\gamma^2$.
Once the spin is frozen, the EDM will cause a steadily increasing
out-of-plane motion of the spin vector. One stores polarized muons in a ring
with detectors above and below the storage region and forms the asymmetry
(up - down)/(up + down).  Reaching a sensitivity of $10^{-24}e \cdot {\rm cm}$ would
require $\sim 4 \times 10^{16}$ recorded events~\cite{Farley04}.
 Preliminary discussions have begun on a frozen spin experiment
using the ~1000 kW beam power available at the Project X 3 GeV rare process campus.

\subsection{Parity-Violating Experiments}

The MOLLER experiment proposes to measure the parity-violating asymmetry  $A_{PV}$ in the scattering
of longitudinally polarized electrons off unpolarized target electrons using the upgraded 11~GeV beam in Hall~A at Jefferson
Laboratory (JLab), to an overall fractional accuracy of 2.3\%. Such a
measurement would constitute more than a factor of five improvement in
fractional precision over the only other measurement of the same
quantity by the E158 experiment at SLAC~\cite{Anthony:2005pm}.  The
electron beam energy, luminosity, and stability at Jefferson Laboratory
are uniquely suited to carry out such a measurement.  After the energy
upgrade, a 11 GeV JLab beam provides a compelling new opportunity to
achieve a new benchmark in sensitivity. The physics motivation has two
important aspects:
\begin{enumerate}
\item The proposed $A_{PV}$ measurement is sensitive to interaction amplitudes as small as $1.5\times 10^{-3}\times G_F$. This corresponds to a sensitivity of $\Lambda/g =  7.5$~TeV, where $g$ characterizes the strength and $\Lambda$ is the scale of the new dynamics. This would be the most sensitive probe of new flavor and $C\!P$-conserving
neutral current interactions in the leptonic sector until the advent
of a linear collider or a neutrino factory.
\item Within the Standard Model, weak neutral current amplitudes are functions of the weak mixing angle $\sin^2\theta_W$.
The proposed $A_{PV}$ measurement  would achieve a sensitivity of
$\delta(\sin^2\theta_W) = \pm 0.00029$ compared to the current best low energy measurements with uncertainties of $\pm 0.0014$~\cite{Bennett:1999pd} and $\pm 0.0015$~\cite{Anthony:2005pm}.
\end{enumerate}

$A_{PV}$ in M\o ller scattering measures the weak charge of the
electron $Q^e_W$, which is proportional to the product of the
electron's vector and axial-vector couplings to the $Z^0$ boson.  At
low energy, $Q^e_W$ is predicted to be $0.0469\pm 0.0006$, a $\sim
40$\%\ change from its tree-level value of $\sim 0.075$ (when evaluated
at $M_Z$).  The prediction for $A_{PV}$ for the proposed experimental
design is $\approx 35$~ppb and the goal is to
measure this quantity with a statistical precision of 0.73 ppb.  The
reduction in the numerical value of $Q^e_W$ due to radiative
corrections leads to increased fractional accuracy in the
determination of the weak mixing angle, $\sim 0.1$\%, comparable to
the two best such determinations from measurements of asymmetries in
$Z^0$ decays in the $\mathrm{e}^+\mathrm{e}^-$ colliders Large Electron Positron (LEP) at CERN and 
Stanford Linear Collider (SLC) at SLAC.

At the level of sensitivity probed, the proposed measurement could be
influenced by radiative loop effects of new particles predicted in Supersymmetry~\cite{RamseyMusolf:2006vr}. Fractional $Q^e_W$ deviations as
large as $+8$\%\ are allowed.  If the assumption of R-parity
conservation is relaxed, tree-level interactions could generate
deviations as large as -18\%, a shift of almost 8~$\sigma$.  A
comprehensive analysis of the MOLLER sensitivity to TeV-scale
$Z^\prime$ bosons has recently been carried out~\cite{Erler:2011iw} for a
fairly large class of family-universal models contained in the $E_6$
gauge group.  The MOLLER reach is found to be significant and
complementary to collider searches for the 1-2 TeV range.  

An $e^+e^-$ super flavor factory with a polarized electron beam can make a
measurement of $\sin^2\theta^{\rm eff}_{\rm W}$ using $e^+ e^- \to
\mu^+\mu^-$ scattering using $A_{\rm LR}$ to a precision equal to that
of the SLD measurement. This would then resolve the issue caused by
the disagreement of the NuTeV result\cite{ref:nutev} with Standard
Model running of $\sin^2\theta^{\rm eff}_{\rm W}$ since it would provide a data point between the $Z$-pole measurements and the neutrino scattering measurement. A less precise
measurement of $\sin^2\theta^{\rm eff}_{\rm W}$ using $A_{\rm LR}$ in
$e^+ e^- \to \tau^+\tau^-$ will then provide a very sensitive test of lepton universality.

Further discussion of parity-violating experiments can be found in the chapter below on ``Nucleons, Nuclei, and Atoms.''

\section{Tau Decays}\label{sec:cl:taus}

\subsection{Theory}\label{sec:cl:tauthe}

$\tau$ decays provide a variety of opportunities to probe new
interactions and new sources of $C\!P$ violation. Unlike $\mu$,
decays which are sensitive only to new leptonic
interactions, the $\tau$ can decay into many final states involving
hadrons. $\tau$ processes are theoretically clean
compared to charm and bottom decay processes, where searches for new
interactions are often limited by hadronic uncertainties.
As a member of the third generation, $\tau$ may play a special role
in the search for new physics. Although the similarity of the three generations
is a basic property of gauge interactions that have been tested up to now,
the current most important issue of particle physics is understanding the
mass-generation mechanism; sensitive studies of the heaviest
lepton could contribute in a unique way.

There are three important physical observables potentially sensitive
to new physics effects in charged lepton processes, namely the anomalous
magnetic moment $g-2$, the EDM, and CLFV. It is instructive to compare the naive
lepton mass dependence of new physics effects in these quantities.
The new physics effects on $g-2$ scale roughly as the square of the lepton mass,
because it is defined relative to the Bohr magneton. Since
current precisions are at the ppb level for the electron and the ppm level for
the muon, the muon $g-2$ is most sensitive to new physics effects. On the
other hand, the experimental precision for the tau $g-2$ has not yet
reached the lowest-order prediction. EDMs naively scale as the lepton mass.
Since the current experimental upper bounds are $O(10^{-27})$ $e\cdot$cm,
$O(10^{-19})$ $e\cdot$cm, $O(10^{-17})$ $e\cdot$cm for the electron EDM, the muon EDM and
the tau EDM respectively, the electron EDM nominally provides the most important constraint
on a new $C\!P$-violating phase, unless a large flavor dependence is involved.
CLFV branching ratios do not
have apparent lepton mass scaling. Current experimental upper bounds on various
CLFV processes are in the range of $O(10^{-13})-O(10^{-11})$ for muon processes
and $O(10^{-8})-O(10^{-7})$ for tau processes. Since the flavor
dependence is essential for CLFV processes, whether the $\mu$
or the $\tau$ process is more important depends on the details of the particular new physics
mechanism. There are also many examples of exceptions to the naive
lepton mass dependence.

An important feature of $\tau$ CLFV searches is the range of processes that can be studied. Compared to
the $\mu$ CLFV case, where $\mu^+ \to e \gamma$, $\mu^+ \to e^+e^+e^-$ and
the $\mu$ to $e$ conversion in muonic atoms are the three major processes, there
are many possible $\tau$ CLFV decay modes in which searches can be carried out
at $e^+e^-$ $B$ factories. For example, there are six different flavor
combinations in $\tau$ to three-charged-lepton decay processes, to be compared in the muon case
 to only $\mu^+ \to e^+e^+e^-$, and there are many CLFV $\tau$ decay modes with hadrons
in the final state that have been searched for experimentally. These distinct searches
are useful in looking for different CLFV interactions and distinguishing
new physics models. If the photonic dipole moment operator is dominant,
which is the case in most supersymmetric models,
the $\tau^+ \to \mu^+\mu^-\mu^+$ branching ratio should be about 500 times smaller than
that of $\tau \to \mu \gamma$. Combined with similar relations in $\mu$ CLFV
processes, such information is crucial in determining the detailed nature of CLFV interactions.

There are many well-motivated models of new physics that predict large
$\tau$ CLFV branching ratios. In the supersymmetric seesaw neutrino model
and supersymmetric grand unified theories, Yukawa coupling constants related
to neutrino mass generation become sources of CLFV, and various CLFV
processes are generated by flavor mixings in slepton mass matrices.
The relationship between $\mu$ CLFV and $\tau$ CLFV branching ratios depends
on heavy neutrino parameters as well as on light neutrino mixing parameters
and the mass hierarchy, and it is possible that $\tau$ CLFV branching ratios
can be close to the present experimental upper bounds, without violating current $\mu$ CLFV constraints. In the little Higgs model
with $T$-parity, new flavor mixing matrices are introduced for the $T$-odd
fermion sector, which is independent of both the CKM matrix and
the neutrino mixing matrix.
In this case, $\tau  \to \mu \gamma$ and $\tau^+ \to  \mu^+\mu^-\mu^+$ branching ratios
are similar in magnitude, in contrast to supersymmetric models, in which
$\tau  \to \mu \gamma$ is much larger. In models with doubly charged Higgs
bosons that couple to a pair of charged leptons, such as the left-right
symmetric model,  the $\tau$ to three-lepton modes are more important
than the $\tau \to \mu (e) \gamma$ process, because the former arise
at tree level.  In this manner CLFV decays can discriminate among new 
physics models that would yield similar signatures at the LHC.

Polarization of the $\tau$ is useful in determining the chiral structure and $C\!P$ nature
of new interactions. At $e^+e^-$ colliders, $\tau$ polarization information can
be extracted by measuring the angular correlation of decay products of a $\tau$ pair. This is in
contrast to polarized muon decays, where a dedicated experimental setup is needed.
If the initial electron beam is polarized, an additional observable quantity, defined by reconstructing a single $\tau$, can be used to obtain
polarized $\tau$ decay information. These measurements are important for
both $\tau$ EDM and CLFV searches. In the case of supersymmetric models,
the angular distribution of the final photon with respect to the $\tau$
polarization depends on whether flavor mixing exists in the left-handed
and/or the right-handed slepton sector, and therefore can provide
important information on the origin of the CLFV interaction. Many asymmetries,
including $T$-odd asymmetries, can be defined for $\tau$ to three-lepton processes.
Direct $C\!P$ violation in $\tau$ decays provides a method to search for $C\!P$-violating
phases. It is known that a sizable SM $C\!P$ asymmetry is induced in $\tau \to K_S \pi \nu$ decay due to
$K^0 -\bar{K^0}$ mixing. On the other hand,
the $\tau \to K^{\pm} \pi \nu$ decay mode is sensitive to a new $C\!P$ phase
that can be present, for example, in the charged-Higgs-boson exchange
interaction.

 \subsection{Experiment}\label{sec:cl:tauexp}

CLFV searches using $\tau$ leptons are excellent probes since one can use a single experiment to search in many different decays.  Tests with the heavy $\tau$ can be more powerful on an event-by-event basis than those using the lighter muon, since the large $\tau$ mass greatly decreases 
Glashow-Iliopoulos-Maiani (GIM) suppression, correspondingly increasing new physics partial widths (typically $\geq \times 500$ more sensitive in $\tau \rightarrow \mu \gamma$ or $e \gamma$ vs.\ $\mu \rightarrow e \gamma$).  The difficulty is that one can make $10^{11}$ muons per second,  and the current event samples from \babar\ and Belle are much smaller.  The proposed new super flavor factories, \cite{ref:superb,ref:superkekb} promise to extend
experimental sensitivities in $\tau$ decays to levels that sensitively
probe new physics in the lepton sector. Since CLFV is severely suppressed in the Standard Model, CLFV $\tau$ decays
are especially clean and unambiguous probes for new physics
effects. Super flavor factories can search for CLFV decays at a
sensitivity that directly confronts many models of new physics.  The
super flavor factories can access $\tau$ CLFV decay rates two orders of magnitude smaller than current limits for the cleanest channels
({\it e.g.}, $\tau\to 3\ell$), and one order of magnitude smaller for other
modes that have irreducible backgrounds, such as $\tau\to \ell\gamma$.

A polarized beam at an $e^+e^-$ collider can provide further
experimental advantages.   It allows  a reduction of backgrounds in certain
CLFV decay modes, as well as providing sensitive new
observables that increase precision in other important measurements.  These include
searches for $C\!P$ violation in $\tau$ decay, the measurement
of $g-2$ of the $\tau$, and the search for a $\tau$ EDM.  Most studies indicate that polarization improves the sensitivity on these quantities by a factor of two to three.

The
provision of polarization requires a polarized electron gun, 
a lattice that supports transverse
polarization at an energy that allows collisions at the $\Upsilon(4S)$
resonance, a means of interchanging transverse polarization in the
ring and longitudinal polarization at the interaction point (two superconducting
solenoid insertions in the machine lattice), and a means of monitoring
the polarization, typically a Compton polarimeter to monitor the
backscattering of circularly polarized laser light. Provision of a
polarized positron beam is difficult and expensive; it is generally
also regarded as unnecessary, as typically 80\% to 90\% of the
advantages of polarization for the measurements cited above can be
accomplished with a single polarized beam. Super$B$, the super flavor factory to be built at the Nicola Cabibbo Laboratory at Tor Vergata,
Italy, intends to have a polarized beam making use of the SLC gun,
 providing a longitudinal polarization of
80\% for electrons at the interaction region\cite{ref:pol}.

The sensitivity of $\tau$ CLFV searches at super flavor factories is
conservatively estimated by extrapolating from current \babar\ and Belle
limits. The optimization of search sensitivities depends on the size
of the sample as well as on the size of the expected backgrounds. For Super$B$, the
experimental reach for the (background-free) $\tau\to\ell\ell\ell$
modes assumes $1/{\mathcal L}$ scaling; that for $\tau\to\ell\gamma$
modes uses $1/{\sqrt{\mathcal L}}$. Further improvement is expected
due to smaller beam size, better solid angle coverage, and electron
beam polarization.  For Belle II, similar scaling is performed with the exception 
that all modes are assumed to have background beyond 5 ab$^{-1}$.
The expected sensitivities for several modes are shown for the Belle II experiment in Table~\ref{tab:LFVExptSensitivities-BelleII}~\cite{Abe:2010sj}  and for the Super$B$ experiment in Table~\ref{tab:LFVExptSensitivities}~\cite{Bona:2007qt}. 
The sensitivities shown in
Table~\ref{tab:LFVExptSensitivities} are expressed in terms of the
expected 90\% CL upper limit assuming no signal, as well as in terms
of a 3$\sigma$ evidence branching fraction in the presence of
projected backgrounds\cite{ref:superb}. In the case of Super$B$,
exploitation of the angular correlations in polarized $\tau$ decay
produces a significant  improvement of the sensitivity for the $\tau\to\ell\gamma$
modes, using $\rho\nu_\tau$ and $\pi\nu_\tau$ modes assuming left-handed couplings.

\begin{table}[!b]
  \caption{
    \label{tab:LFVExptSensitivities-BelleII}
    Expected $90\%$ CL upper limits
    on $\tau\to\mu\gamma$, $\tau\to \mu\mu\mu$, and $\tau\to \mu\eta$
    with $5 \ {\rm ab}^{-1}$ and  $50 \ {\rm ab}^{-1}$ data sets from Belle II and  Super KEKB.
  }
  \begin{center}
    \begin{tabular}{lll}
      \hline \hline
Process & $5\ {\rm ab}^{-1}$ & $50\ {\rm ab}^{-1}$  \\
      \hline
      $\BR(\tau \to \mu\,\gamma) \rule{0pt}{2.6ex}$ &  $10 \times 10^{-9}$ &  $3 \times 10^{-9}$  \\
      $\BR(\tau \to \mu\, \mu\, \mu)$ &  $3 \times 10^{-9}$ & $1 \times 10^{-9}$  \\
      $\BR(\tau \to \mu \eta)$             &  $5 \times 10^{-9}$ & $2\times 10^{-9}$    \\
 \hline\hline
    \end{tabular}
  \end{center}
\end{table}

\begin{table}[!b]
  \caption{
    \label{tab:LFVExptSensitivities}
    Expected $90\%$ CL upper limits and 3$\sigma$ discovery reach
    on $\tau\to\mu\gamma$ and $\tau\to \mu\mu\mu$ and other CLFV decays with $75 \ {\
\rm ab}^{-1}$
at Super$B$    with a polarized electron beam.
  }
  \begin{center}
    \begin{tabular}{lll}
      \hline \hline
    Process &  \rule{0pt}{2.6ex} Expected 90\%CL & 3$\sigma$ Evidence \\
              &  upper limit ($75$ ab$^{-1}$)   &  Reach   ($75$ ab$^{-1}$) {\rule[-1.2ex]{0pt}{0pt}} \\
      \hline
      $\BR(\tau \to \mu\,\gamma) \rule{0pt}{2.6ex}$          &  $1.8 \times 10^{-9}\
$ &  $4.1 \times 10^{-9}$  \\
      $\BR(\tau \to e\,\gamma)$            &  $2.3 \times 10^{-9}$  &  $5.1 \times \
10^{-9}$  \\
      $\BR(\tau \to \mu\, \mu\, \mu)$ &  $2 \times 10^{-10}$ & $8.8 \times 10^{-10}$  \\
      $\BR(\tau \to e e e )$               &  $2 \times 10^{-10}$  \\
      $\BR(\tau \to \mu \eta)$             &  $4 \times 10^{-10}$    \\
      $\BR(\tau \to e \eta)$               &  $6 \times 10^{-10}$    \\
      $\BR(\tau \to \ell K^0_S)$   {\rule[-1.2ex]{0pt}{0pt} }       &  $2 \times 10\
^{-10}$    \\
      \hline\hline
    \end{tabular}
  \end{center}
\end{table}

These CLFV sensitivities directly confront a large variety of new
physics models. Of particular interest is the correlation between
$\tau$ CLFV branching ratios such as $\tau\to \mu\gamma$ and $\tau\to e
\gamma$, as well as the correlation with $\mu\to e \gamma$ and the
$\mu\to e$ conversion rate, all of which are diagnostic of particular
models.  A polarized electron beam potentially allows the possibility of determining the helicity structure of CLFV couplings from Dalitz plot analyses of, for example, $\tau \to
3\ell$ decays.

The experimental discrepancy with the Standard Model prediction for
the muon anomalous magnetic moment heightens interest in the
possibility of measuring $g-2$ of the $\tau$ lepton using angular
distributions in $\tau$-pair production. This can be done at a super flavor factory, with or without electron polarization.  With polarized taus one can access new observables that are estimated by
Bernab\'eu {\it et al.}\cite{ref:b1} to increase the sensitivity to
$g-2$ by a factor of three, to $\sim2\times 10^{-6}$ with 80\%
electron polarization, which could allow a measurement of the Standard
Model moment to a precision of several percent with a data sample of
75 ab$^{-1}$.

Observation of a $\tau$ EDM would be evidence of $T$ violation.  $T$-odd observables can be isolated by the study of $\tau$ angular
distributions using unpolarized beams. Having a polarized electron
beam allows these investigations to be done using the decay products
of individual polarized taus.  The upper-limit sensitivity for the real part of the
$\tau$ EDM has been estimated to be to be $|\Re{d_\gamma}|\simeq 3\times 10^{-19} \ e \cdot {\rm cm}$
with 50 ab$^{-1 }$ at Belle II and $|\Re{d_\gamma}|\simeq 7\times 10^{-20} \ e \cdot {\rm cm}$
with 75 ab$^{-1 }$ at Super$B$\cite{ref:b2}.

A $C\!P$-violating asymmetry in $\tau$ decay would be manifest evidence
for physics beyond the Standard Model. \babar\ has recently published
a 3$\sigma$ asymmetry in $\tau\to\pi K_S^0(\ge 0\pi^0)$
decay\cite{ref:taucp}. The super flavor factories have the sensitivity to
definitely confirm or refute this measurement, and, further, provide
access to new $C\!P$-odd observables that increase the sensitivity in the
search for a $C\!P$ asymmetry to the level of $\sim 10^{-3}$.

\section{Summary}\label{sec:cl:gm2edmdisc}

The enormous physics potential of the charged
lepton experimental program was very much in evidence at this workshop. There are discovery opportunities both in experiments that will be conducted over the coming decade using existing facilities and in more sensitive experiments possible with future facilities such as Project X.
Sensitive searches for rare decays of muons and tau leptons, together with precision measurements of their properties, will either elucidate the scale and dynamics of flavor generation or
limit the scale of flavor generation to well above $10^4$ TeV.  This information
will be vital to understanding the underlying physics responsible for new particles discovered at the LHC.

The crown jewel of the program is the discovery potential of muon and tau decay experiments searching for charged lepton flavor violation with several orders of magnitude improvement in sensitivity in
multiple processes.  This is an
international program, with experiments recently completed, currently running, and
soon to be constructed in the United States, Japan, and Europe.  The program is very interesting over the near term, with the completion of the now-running MEG experiment at PSI, and with the construction and completion of the Mu2e and COMET experiments at Fermilab and J-PARC, respectively.  It will also be substantially improved by new facilities such as the super flavor factories for taus and Project X for muons.

Over the next decade,
gains of up to four orders of magnitude are feasible in
muon-to-electron conversion and in the $\mu^+ \to e^+e^-e^+$
searches.  Over a longer time scale, at higher intensity machines, another two orders of magnitude
are possible for muon-to-electron conversion.  
Two orders of magnitude are possible in $\mu \to e\gamma$ and $\tau \to 3\ell$ decay, and more than one order of magnitude in $\tau \to \ell\gamma$ CLFV
searches.  Existing searches already place strong constraints on
many models of physics beyond the Standard Model; the contemplated improvements increase these constraints significantly, covering substantial regions of the parameter space of many new physics models.
These improvements are important regardless of the outcome of new particle searches at the
LHC; the next generation of CLFV searches are an essential
component of the particle physics road map going forward.  If the LHC finds new
physics, then CLFV searches will confront the lepton sector in ways
that are not possible at the LHC, while if the LHC uncovers no sign of
new physics, CLFV may provide the path to discovery.

In general, muon measurements have the best
sensitivity over the largest range of the parameter space of many new
physics models. There are, however, models
in which  rare tau decays could provide the discovery
channel. It was clear from the discussion that as many different
CLFV searches as feasible should be conducted, since our guess at the best discovery
channel is model-dependent and the model is not yet known.  Should a
signal be observed in any channel, searches and measurements in as
many CLFV channels as possible will be crucial to determining the nature
of the underlying physics, since correlations between the rates
expected in different channels provide a powerful discriminator between
physics models.

The next generation muon $g-2$ experiments will measure the anomaly to close to 100 ppb precision using rather different experimental techniques. Many models that predict new physics at the LHC also predict effects much larger than the future $g-2$ precision.  Indeed, some models are already strongly disfavored because they predict corrections to $g-2$ much larger than the current discrepancy.  The predictions of all models will be confronted by the 
next-generation quark and lepton flavor programs.  The combined information from these programs,  $g-2$, and the LHC should allow a clean interpretation of the dynamics of electroweak symmetry breaking.  This has been most extensively studied in the context of supersymmetry. The next-generation muon $g-2$ experiments will greatly improve our knowledge of fundamental parameters of the Higgs sector. A particular example is the measured ratio of vacuum expectation values between the up- and down-type Higgs ($\tan\beta$) and whether there is constructive or destructive interference among the Higgs bosons (sign$(\mu)$).  This is expected to be the case even after $100$ fb$^{-1}$ of data has been recorded at $14$ TeV. 

Within this decade, every relevant term in the prediction for the Standard Model contributions to the
 muon $g-2$ value will be data
driven.  The hadronic vacuum polarization is already taken directly from $e^+e^-$ data.
There will soon  be several independent experimental measurements with comparable precision, but with different systematic uncertainties, adding redundancy and illuminating the cause of any discrepancies among measurements.  There will also
be an enormous increase in tau data, which can be used to understand any remaining discrepancies between tau and
$e^+e^-$ data.  The models used to estimate the hadronic light-by-light scattering
will be confronted by new two-photon data very close to $q^2=0$.  This
data, along with lattice calculations, will demonstrate that the uncertainties
quoted are reasonable.

High luminosity $e^+e^-$ colliders will provide data sets that could allow one to measure, for the first time, the anomalous magnetic moment of the tau lepton.  This is done using angular correlations in tau-pair production, and also observables involving tau polarization.

EDMs also play a quintessential role in new physics
searches. The achievable limit on the electron EDM is the most stringent, but searches for muon and tau EDMs are nonetheless of interest, particularly to limit models with non-trivial mass scaling among the leptons. These will be
important: if an electron EDM were to be found, the value of second and third generation EDMs would be of great interest.  Parasitic measurements with the next round of $g-2$ experiments will improve the $\mu$ EDM limit by two
orders of magnitude. Improvement of this limit would also help to rule out
the possibility that the muon EDM is the cause of the current discrepancy in the
$g-2$ measurement. New dedicated experiments now being discussed
could bring the limit down to the $10^{-24}$ to $10^{-25}$ $e \cdot $cm level.   In the same vein, super flavor factories can limit the tau EDM.  The ultimate sensitivity will make use of beam polarization and improve existing limits by two orders of magnitude.
Additional symmetry tests will also be possible, including sensitive searches for $C\!P$ violation in $\tau$ decay and 
tests of electroweak parity violation using electron scattering and $e^+e^-$ collisions. 

An extraordinary program of sensitive searches for new physics using the large samples of $\mu$ and $\tau$ decays in experiments at the intensity frontier awaits us. These experiments will likely be central to our understanding of physics beyond the Standard Model.

\def\Discussion{\setlength{\parskip}{0.3cm}\setlength{\parindent}{0.0cm}
     \bigskip\bigskip      {\Large {\bf Discussion}} \bigskip}\def\speaker#1{{\bf #1:}\ }
\def\endDiscussion{}


\chapter{Neutrinos}
\label{chap:neutrinos}

\begin{center}\begin{boldmath}
%

\normalsize
Conveners: A.~de Gouv\^ea, K.~Pitts, K.~Scholberg, G.P.~Zeller

C.~Albright, J.~Alonso, C.~Ankenbrandt, C.~A.~Arguelles, E.~Arrieta Diaz, J.~Asaadi
% ...
\end{boldmath}\end{center}



\section{Introduction}\label{sec:intro}



Neutrinos are the most elusive of the known fundamental particles. They are color- and charge- neutral spin one-half fermions, and, to the best of our knowledge, only interact with charged fermions and massive gauge bosons, through the weak interactions. For this reason, neutrinos can only be observed and studied because there are very intense neutrino sources (natural and artificial) and only if one is willing to work with large detectors. The existence of neutrinos was postulated in the early 1930s, but they were only first observed in the 1950s. The third neutrino flavor eigenstate, the tau-type neutrino $\nu_{\tau}$, was the last of the fundamental particles to be observed \cite{Kodama:2000mp}, eluding direct observation six years longer than the top quark \cite{Abe:1995hr,Abachi:1995iq}.  More relevant to this chapter, in the late 1990s the discovery of non-zero neutrino masses moved the study of neutrino properties to the forefront of experimental and theoretical particle physics.

Experiments with solar \cite{Cleveland:1998nv,Hampel:1998xg,Ahmad:2002jz,Abdurashitov:2002nt,Fukuda:2001nj,Ahmed:2003kj}, atmospheric \cite{Fukuda:1998mi,Ashie:2004mr}, reactor \cite{Eguchi:2002dm,Araki:2004mb} and accelerator \cite{Ahn:2002up,Michael:2006rx} neutrinos  have established, beyond reasonable doubt, that a neutrino produced in a well-defined flavor state (say, a muon-type neutrino $\nu_{\mu}$) has a non-zero probability of being detected in a different flavor state  (say, an electron-type neutrino $\nu_e$). This flavor-changing probability depends on the neutrino energy and the distance traversed between the source and the detector. The simplest and only consistent explanation of all neutrino data collected over the last two decades is that neutrinos have mass and neutrino mass eigenstates are different from neutrino weak eigenstates, {\it i.e.}, leptons mix. The neutrino flavor-change is a consequence of so-called neutrino oscillations. 


Massive neutrinos imply that a neutrino produced as a coherent superposition of mass-eigenstates, such as a neutrino $\nu_{\alpha}$ with a well-defined flavor, has a non-zero probability to be measured as a neutrino $\nu_{\beta}$ of a different flavor ($\alpha,\beta=e,\mu,\tau$). This oscillation probability $P_{\alpha\beta}$ depends on the neutrino energy $E$, the propagation distance $L$, and on the neutrino mass-squared differences, $\Delta m^2_{ij}\equiv m_i^2-m_j^2$, $i,j=1,2,3,\ldots$, and the elements of the leptonic mixing matrix,\footnote{Often referred to as the Maki-Nakagawa-Sakata (MNS) Matrix, or the Pontecorvo-Maki-Nakagawa-Sakata (PMNS) Matrix.} $U$, which relates neutrinos with a well-defined flavor ($\nu_e,\nu_{\mu},\nu_{\tau}$) and neutrinos with a well-defined mass ($\nu_{1},\nu_2,\nu_3,\ldots$). For three neutrino flavors, the elements of $U$ are defined by
\begin{equation}
\left(\begin{array}{c}\nu_e \\ \nu_{\mu} \\ \nu_{\tau} \end{array} \right) =\left(\begin{array}{ccc} U_{e1} & U_{e2} & U_{e3} \\ U_{\mu1} & U_{\mu2} & U_{\mu3} \\ U_{\tau1} & U_{e\tau2} & U_{\tau3}\end{array}\right)
\left(\begin{array}{c}\nu_1 \\ \nu_2 \\ \nu_3 \end{array} \right).
\label{UMNS}
\end{equation}
Almost all neutrino data to date can be explained assuming that neutrinos interact as prescribed by the Standard Model, there are only three neutrino mass eigenstates, and $U$ is unitary.
Under these circumstances, it is customary to parameterize $U$ in Eq.~(\ref{UMNS}) with three mixing angles $\theta_{12},\theta_{13},\theta_{23}$ and three complex phases, $\delta,\xi,\zeta$, defined by
\begin{equation}
\frac{|U_{e2}|^2}{|U_{e1}|^2}\equiv \tan^2\theta_{12};
~~~~\frac{|U_{\mu3}|^2}{|U_{\tau3}|^2}\equiv \tan^2\theta_{23};~~~~
U_{e3}\equiv\sin\theta_{13}e^{-i\delta},
\end{equation}
with the exception of $\xi$ and $\zeta$, the so-called Majorana $CP$-odd phases. These are only physical if the neutrinos are Majorana fermions, and have essentially no effect in flavor-changing phenomena.

\begin{figure}[ht]
\centerline{\includegraphics[width=0.45\textwidth]{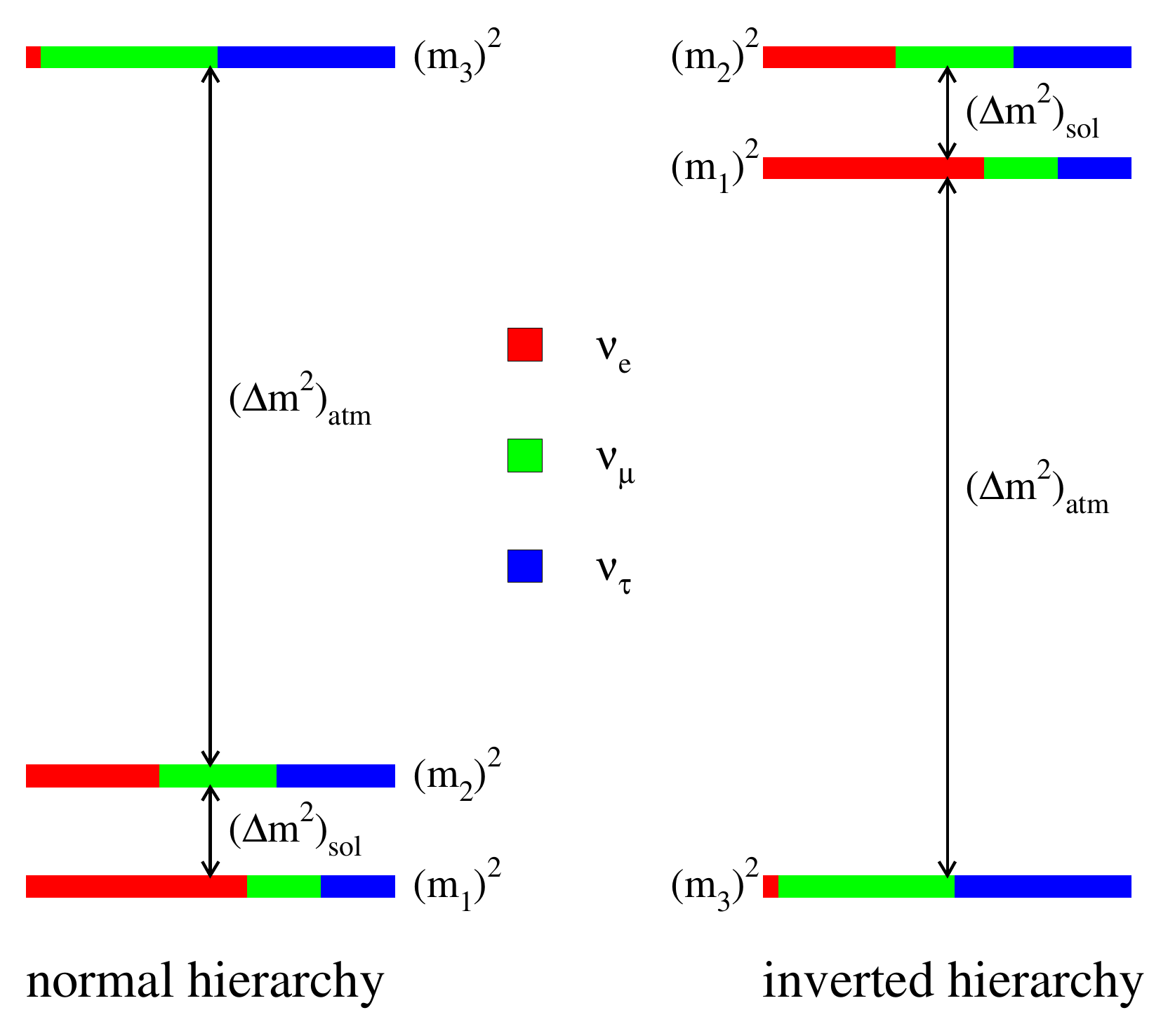}}
\caption{Cartoon of the two distinct neutrino mass hierarchies that fit all of the current neutrino data, for fixed values of all mixing angles and mass-squared differences. The color coding (shading) indicates the fraction $|U_{\alpha i}|^2$ of each distinct flavor $\nu_{\alpha}$, $\alpha=e,\mu,\tau$ contained in each mass eigenstate $\nu_i$, $i=1,2,3$. For example, $|U_{e2}|^2$ is equal to the fraction of the $(m_2)^2$ ``bar'' that is painted red (shading labeled as ``$\nu_e$'').
\label{3nus_pic}}
\end{figure}

In order to relate the mixing elements to experimental observables, it is necessary to define the neutrino mass eigenstates, {\it i.e.}, to ``order'' the neutrino masses. This is done in the following way: $m_2^2>m_1^2$ and $\Delta m^2_{21}<|\Delta m^2_{31}|$. In this case, there are three mass-related oscillation observables: $\Delta m^2_{21}$ (positive-definite), $|\Delta m^2_{31}|$, and the sign of $\Delta m^2_{31}$. A positive (negative) sign for $\Delta m^2_{31}$ implies $m_3^2>m_2^2$  ($m_3^2<m_1^2$) and characterizes a so-called normal (inverted) neutrino mass hierarchy. 
The two mass hierarchies are depicted in Fig.~\ref{3nus_pic}.

Our knowledge of neutrino oscillation parameters can be summarized as~\cite{Schwetz:2011zk,Machado:2011ar,An:2012eh}
\begin{eqnarray}
\Delta m^2_{21} = 7.59^{+0.20}_{-0.18}\times 10^{-5}~{\rm eV}^2, & \Delta m^2_{31}=2.50^{+0.09}_{-0.16}\times 10^{-3}~{\rm eV}^2 & (-2.40^{+0.08}_{-0.09}\times 10^{-3}~{\rm eV}^2), \\
\sin^2\theta_{12}=0.312^{+0.017}_{-0.015}, & \sin^2\theta_{23}=0.52\pm0.06, & \sin^2\theta_{13}=0.023\pm0.004.
\end{eqnarray}
We have virtually no information concerning $\delta$ (and, for that matter, $\xi$ and $\zeta$) or the sign of $\Delta m^2_{31}$. Indications by T2K~\cite{Abe:2011sj}, MINOS~\cite{Adamson:2011qu} and Double Chooz~\cite{Abe:2011fz} in 2011 pointed to $\sin^22\theta_{13}\simeq0.08$. In combination, these results excluded $\sin^22\theta_{13}=0$ at more than three sigma without direct reference to solar or atmospheric data, as was required prior to the result from Double Chooz~\cite{Fogli:2011qn}. Very recently, the Daya Bay collaboration, after analyzing 55~days of data, claimed 5 sigma evidence that $\sin^2\theta_{13}$ is not zero. Their best fit result is $\sin^22\theta_{13}=0.092\pm0.017$ \cite{An:2012eh}. It is clear that our knowledge of the smallest of the lepton mixing angles is quickly evolving. In the immediate future, reactor neutrino experiments~\cite{Ardellier:2006mn,Guo:2007ug,Ahn:2010vy} will soon provide improved results, T2K will resume full operation, and MINOS and NO$\nu$A are expected to add very useful information. 

The main goal of next-generation neutrino oscillation experiments is to test whether the scenario outlined above, the standard three-massive-neutrinos paradigm, is correct and complete. This is to be achieved not simply by determining all of the parameters above, but by ``over-constraining'' the parameter space in order to identify potential inconsistencies. This is not a simple task, and the data collected thus far, albeit invaluable, allow for only the simplest consistency checks. In the future, precision measurements, as will be discussed in Sec.~\ref{sec:3nus}, will be required.

Large, qualitative modifications to the standard paradigm are allowed. Furthermore, there are several, none too significant, hints in the world neutrino data that point to a neutrino sector that is more complex than the one outlined above. These will be discussed in Sec.~\ref{sec:sbl}. Possible surprises include new, gauge singlet fermion states that manifest themselves only by mixing with the known neutrinos, and new weaker-than-weak interactions.

In the Standard Model, neutrinos were predicted to be exactly massless. The discovery of neutrino masses hence qualifies as the first instance where the Standard Model failed. This is true even if the three-massive-neutrino paradigm described above turns out to be the whole story. More important is the fact that all modifications to the Standard Model that lead to massive neutrinos change it qualitatively. For a more detailed discussion of this point see, for example,  \cite{DeGouvea:2005gd}.

\begin{figure}[ht]
\centerline{\includegraphics[width=0.65\textwidth]{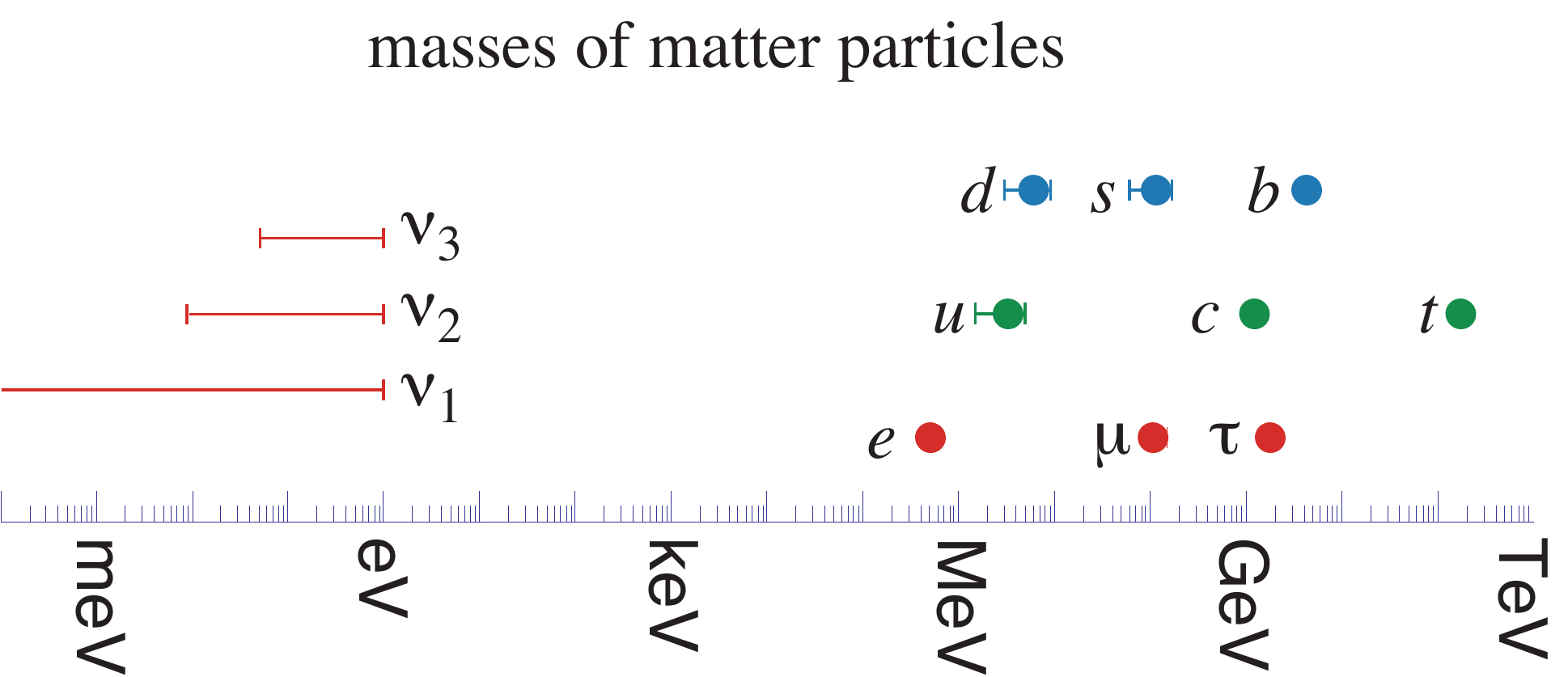}}
\caption{Standard Model fermion masses. For the neutrino masses, the normal mass hierarchy was assumed, and a loose upper bound $m_i<1$~eV, for all $i=1,2,3$ was imposed.}
\label{fig:fermionmasses}
\end{figure}

Neutrino masses, while non-zero, are tiny when compared to all other known mass scales in the Standard Model,\footnote{Except, perhaps, for the mysterious cosmological constant.} as depicted in Fig.~\ref{fig:fermionmasses}. Two features readily stand out: (i) neutrino masses are at least six orders of magnitude smaller than the electron mass, and 
(ii) there is, to the best of our knowledge, a ``gap'' between the largest allowed neutrino mass and the electron mass.
We don't know why neutrino masses are so small or why there is such a large gap between the neutrino and the charged fermion masses. We suspect, however, that this may be Nature's way of telling us that neutrino masses are ``different.''

This suspicion is only magnified by the possibility that massive neutrinos, unlike all other fermions in the Standard Model, may be Majorana fermions. The reason is simple: neutrinos are the only electrically neutral fundamental fermions and hence need not be distinct from their antiparticles. Determining the nature of the neutrino -- Majorana or Dirac -- would not only help guide theoretical work related to uncovering the origin of neutrino masses, but could also reveal that the conservation of lepton number is not a fundamental law of Nature. The most promising avenue for learning the fate of lepton number, as will be discussed in Sec.~\ref{sec:majorana},  is to look for neutrinoless double-beta decay, a lepton-number violating nuclear process. The observation of a non-zero rate for this hypothetical process would easily rival, as far as its implications for our understanding of nature are concerned, the first observations of parity violation and $CP$-invariance violation in the mid-twentieth century.

It is natural to ask what augmented, ``new'' Standard Model ($\nu$SM) leads to non-zero neutrino masses. The answer is that we are not sure. There are many different ways to modify the Standard Model in order to accommodate neutrino masses. While these can differ greatly from one another, all succeed -- by design -- in explaining small neutrino masses and all are allowed by the current particle physics experimental data. The most appropriate question, therefore, is not what are the candidate $\nu$SM's, but how can one identify the ``correct'' $\nu$SM? The answer lies in next-generation experiments, which will be described throughout this chapter.


For concreteness we discuss one generic mechanism in more detail.  The effect of heavy new degrees of freedom in low-energy phenomena can often be captured by adding to the Standard Model higher-dimensional operators. As first pointed out in~\cite{Weinberg:1979sa}, given the Standard Model particle content and gauge symmetries, one is allowed to write only one type of dimension-five operator -- all others are dimension-six or higher:
\begin{equation}
 {1 \over \Lambda}\ (L H) (L H) + h.c.\quad  \Rightarrow \quad {v^{2} \over \Lambda}\nu\nu + h.c.,
\label{eq:seesaw}
\end{equation}
where $L$ and $H$ are the lepton and Higgs boson $SU(2)_L$ doublets, and the arrow indicates one of the components of the operator after electroweak symmetry is broken. $v$ is the vacuum expectation value of the neutral component of $H$, and $\Lambda$ is the effective new physics scale. If this operator is indeed generated by some new physics, neutrinos obtain Majorana masses $m_\nu \sim v^2/\Lambda$. For $\Lambda\sim 10^{15}$~GeV, $m_{\nu}\sim 10^{-1}$~eV, in agreement with the current neutrino data. This formalism explains the small neutrino masses via a seesaw mechanism: $m_{\nu}\ll v$ because $\Lambda\gg v$.

$\Lambda$ is an upper bound for the masses of the new particles that lead to Eq.~(\ref{eq:seesaw}). If the new physics is strongly coupled and Eq.~(\ref{eq:seesaw}) is generated at the tree-level, the new degrees of freedom are super-heavy: $M_{\rm new}\sim 10^{15}$~GeV. If, however,  the new physics is weakly coupled or Eq.~(\ref{eq:seesaw}) is generated at the loop level, virtually any value for $M_{\rm new}\gtrsim 1$~eV is allowed. In summary, if Eq.~(\ref{eq:seesaw}) is correct, we expect new physics to show up at a new mass scale $M_{\rm new}$ which lies somewhere between $10^{-9}$~GeV and $10^{15}$~GeV. Clearly, more experimental information is required.

At the tree-level, there are only three renormalizable extensions of the Standard Model that lead to Eq.~(\ref{eq:seesaw}). They are referred to as the three types of seesaw mechanisms, and are summarized as follows. For more details, see, for example, \cite{Mohapatra:2005wg,Langacker:2011bi}.
\begin{itemize}
\item  \textit{Type I}~\cite{neu:Minkowski:1977sc,Yanagida:1979as,GellMann:1980vs,Glashow1980,Mohapatra:1979ia}:
The fermion sector of the Standard Model is augmented by at least two gauge singlets $N_i$, which couple to the lepton and Higgs scalar doublets via a new Yukawa coupling $y_{\nu}$. These so-called right-handed neutrinos are allowed to have Majorana masses $M_N$. After electroweak symmetry breaking, assuming $M_N\gg y_{\nu}v$, one generates Eq.~(\ref{eq:seesaw}). Here $\Lambda=M_N/y_{\nu}^2$.
\item \textit{Type II}~\cite{Konetschny:1977bn,Cheng:1980qt,Lazarides:1980nt,Schechter:1980gr,Mohapatra:1980yp}:
The Higgs sector of the Standard Model is extended by one $SU(2)_L$ Higgs triplet $\Delta$. The neutrino masses are $m_\nu \approx Y_\nu v_{\Delta}$, where  $v_{\Delta}$ is the
vacuum expectation value (vev) of the neutral component of the triplet and $Y_\nu$ is the Yukawa coupling that describes the strength of the $\Delta$ coupling to two lepton doublets. If the doublet and triplet mix via a dimensionful parameter $\mu$, electroweak symmetry breaking can translate into $v_{\Delta} \sim \mu v^2/M_{\Delta}^2$,
where $M_{\Delta}$ is the mass of the triplet. In this case, after one integrates out the $\Delta$ states Eq.~(\ref{eq:seesaw}) is generated, and $\Lambda=M_{\Delta}^2/(\mu Y_{\nu})$. Small neutrino masses require either $M_{\Delta}\gg v$ or $\mu\ll v$.
%
\item \textit{Type III}~\cite{Foot:1988aq}:
The fermion sector of the Standard Model is augmented by at least two $SU(2)_L$ triplets $T_i$ with zero hypercharge. As in the Type I case, if these triplets couple to the lepton and Higgs scalar doublets via a new Yukawa coupling $y_{T}$, and are endowed with Majorana masses $M_T$, after electroweak symmetry breaking, assuming $M_T\gg y_{T}v$, one generates Eq.~(\ref{eq:seesaw}). Here $\Lambda=M_T/y_{T}^2$.
\end{itemize}
We will refer to different manifestations of these scenarios throughout this chapter.
Some predict new physics at scales that can be probed at the energy frontier or elsewhere in the intensity frontier, while others predict new physics scales that are way beyond the reach of laboratory experiments. If that turns out to be the case, we will only be able to access the new physics indirectly through neutrino experiments and the study of relics in the cosmic frontier. The synergy of neutrino physics with other fundamental physics is discussed in Sec.~\ref{sec:synergy}.

Neutrino data also provide a new piece to the flavor puzzle: the pattern of neutrino mixing. The absolute value of the entries of the CKM quark mixing matrix are, qualitatively, given by
\begin{equation}
|V_{\rm CKM}|\sim \left(\begin{array}{ccc} 1 & 0.2 & 0.004\\ 0.2 & 1 & 0.04 \\ 0.008 & 0.04 & 1\end{array} \right),
\end{equation}
while those of the entries of the PMNS matrix are given by
\begin{equation}
|U_{\rm PMNS}|\sim \left(\begin{array}{ccc} 0.8 & 0.5 & 0.2 \\ 0.4 & 0.6 & 0.7 \\ 0.4 & 0.6 & 0.7\end{array} \right).
\end{equation}
It is clear that the two matrices look very different. While the CKM matrix is almost proportional to the identity matrix plus hierarchically ordered off-diagonal elements, the PMNS matrix is far from diagonal and, with the possible exception of the $U_{e3}$ element, all elements are ${\cal O}(1)$. Significant research efforts are concentrated on understanding what, if any, is the relationship between the quark and lepton mixing matrices and what, if any, is the ``organizing principle'' responsible for the observed pattern of neutrino masses and lepton mixing. There are several different theoretical ideas in the market  (for summaries, overviews and more references see, for example, \cite{Mohapatra:2005wg,Albright:2006cw}). Typical results include predictions for the currently unknown neutrino mass and mixing parameters ($\sin^2\theta_{13}$, $\cos2\theta_{23}$, the mass hierarchy, {\it etc.}) and the establishment of sum rules involving different parameters.

From the flavor physics perspective,  the goals of precision neutrino oscillation physics can be broken down into the following
questions: Is the atmospheric mixing angle maximal ($\theta_{23}=\pi/4$)? Is there leptonic $CP$ violation (or what is the value of $\delta$)? What is the ordering of mass eigenstates (sign of
$\Delta m^2_{31}$)? What is the value of the ``reactor'' angle, $\theta_{13}$?

Precision neutrino oscillation measurements are required to address the flavor questions above. That can only be achieved as the result of significant investments in intense, well-characterized neutrino sources and massive high-precision detectors. Some of these are discussed in Sec.~\ref{sec:facilities}. Excellent understanding of neutrino interactions -- beyond the current state of the art --  is also mandatory. This will require a comprehensive experimental program on neutrino scattering, as summarized in Sec.~\ref{sec:scattering}. These, of course, are not only ancillary to neutrino oscillation experiments, but are also interesting in their own right. Neutrinos, since they interact only weakly, serve as a unique probe of nucleon and nuclear properties, and may reveal new physics phenomena at the electroweak scale, including some that are virtually invisible to the Tevatron and the LHC.


\section{Testing the Standard Oscillation Paradigm}\label{sec:3nus}


Physical effects of non-zero neutrino masses, to date, have been observed only in neutrino oscillation experiments. Those are expected to remain, for the foreseeable future, the most powerful tools available for exploring the new physics revealed by solar and atmospheric neutrino experiments at the end of the twentieth century.

\subsection{Overview of Neutrino Oscillations}

The standard setup of a neutrino oscillation experiment is as follows. A detector is located a distance $L$ away from a source, which emits ultra-relativistic neutrinos or antineutrinos with, most often, a continuous spectrum of energies $E$, and flavor $\alpha=e,\mu$, or $\tau$. According to the Standard Model, the neutrinos interact with matter either via $W$-boson exchange charged-current interactions where a neutrino with a well-defined flavor $\nu_{\alpha}$ gets converted into a charged lepton of the same flavor ($\nu_e X\to eX'$, {\it etc.}) or via $Z$-boson exchange neutral-current interactions, which preserve the neutrino flavor ($\nu_{\mu}X\to\nu_{\mu}X'$). The occurrence of a neutral-current process is tagged by observing the system against which the neutrinos are recoiling. The detector hence is capable of measuring the flux of neutrinos or antineutrinos with flavor $\beta=e,\mu$, or $\tau$, or combinations thereof, often as a function of the neutrino energy. By comparing measurements in the detector with expectations from the source, one can infer $P_{\alpha\beta}(L,E)$ or $\bar{P}_{\alpha\beta}(L,E)$, the probability that a(n)  (anti)neutrino with energy $E$ produced in a flavor eigenstate $\nu_{\alpha}$ is measured in a flavor $\nu_{\beta}$ after it propagates a distance $L$.  In practice, it is often preferable to make multiple measurements of neutrinos at different distances from the source, which can be helpful for cancellation of systematic uncertainties.

\begin{figure}[ht]
\begin{center}
\includegraphics[width=0.6\textwidth]{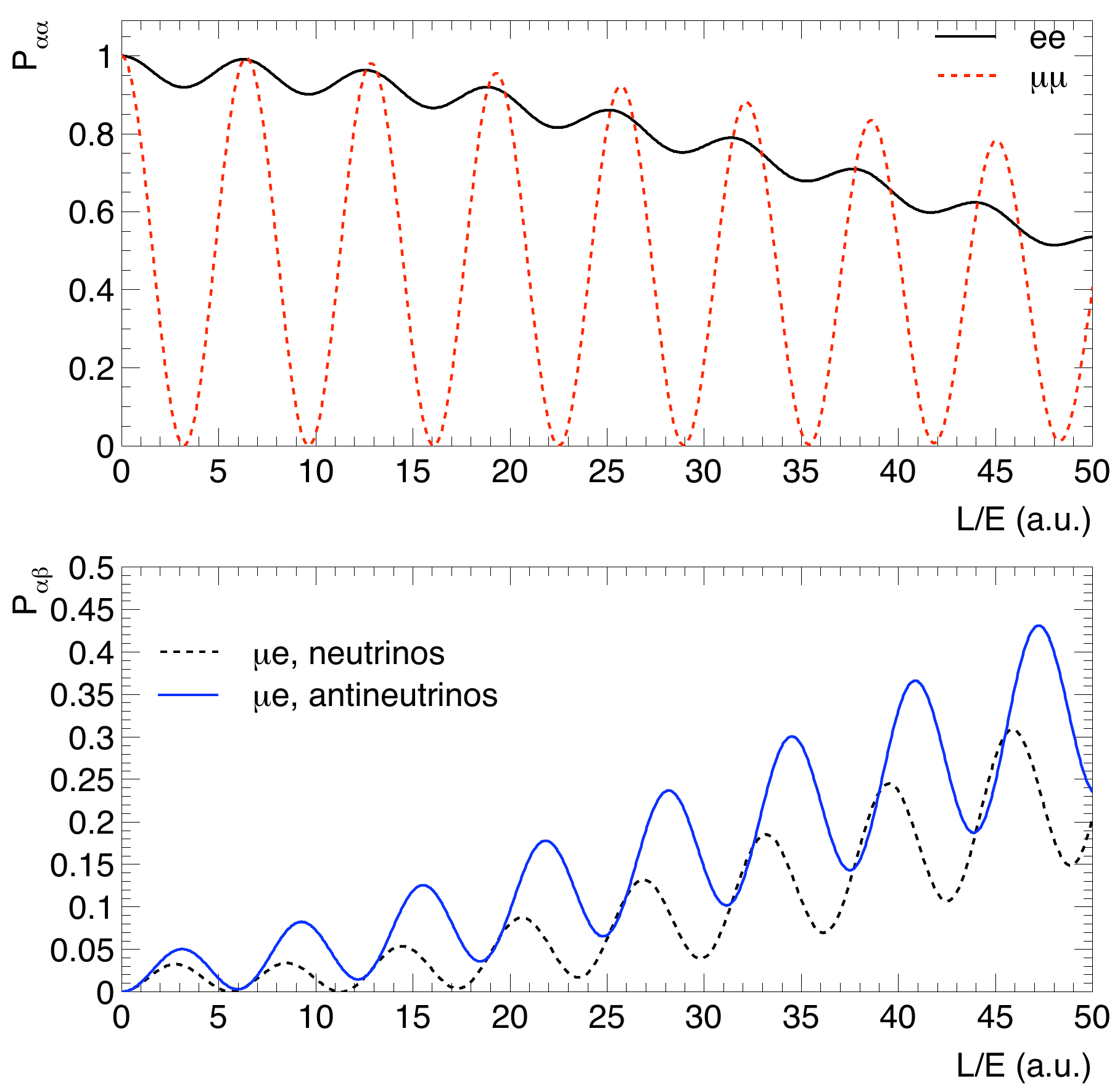}
\end{center}
\caption{Top: $P_{ee}$ and $P_{\mu\mu}$ in vacuum as a function of $L/E$ (in arbitrary units), for representative values of the neutrino oscillation parameters, including a non-zero value of $\delta$. Bottom: $P_{\mu e}$ and $\bar{P}_{\mu e}$ in vacuum as a function of $L/E$, for representative values of the neutrino oscillation parameters.}
\label{fig:oscillations}
\end{figure}

In the standard three-flavor paradigm, $P_{\alpha\beta}$ is a function of the mixing angles $\theta_{12,13,23}$, the Dirac $CP$-odd phase $\delta$, and the two independent neutrino mass-squared differences $\Delta m^2_{21,31}$, defined in the Introduction.   Assuming the neutrinos propagate in vacuum, and making explicit use of the unitarity of $U$, one can express $P_{\alpha\beta}(L,E)=|A_{\alpha\beta}|^2$, where

\begin{eqnarray}
A_{\alpha\beta}=\delta_{\alpha\beta}+U_{\alpha2}U^*_{\beta2}\left(\exp\left(-i\frac{\Delta m^2_{21}L}{2E}\right)-1\right)+U_{\alpha3}U^*_{\beta3}\left(\exp\left(-i\frac{\Delta m^2_{31}L}{2E}\right)-1\right), \label{eq:Aab}
\\
\bar{A}_{\alpha\beta}=\delta_{\alpha\beta}+U^*_{\alpha2}U_{\beta2}\left(\exp\left(-i\frac{\Delta m^2_{21}L}{2E}\right)-1\right)+U^*_{\alpha3}U_{\beta3}\left(\exp\left(-i\frac{\Delta m^2_{31}L}{2E}\right)-1\right),
\label{eq:barAab}
\end{eqnarray}
up to an unphysical overall phase. $A$ ($\bar{A}$) is the amplitude for (anti)neutrino oscillations. It is easy to see that $P_{\alpha\beta}$ are oscillatory functions of $L/E$ with, in general, three distinct, two independent oscillation lengths proportional to  $\Delta m^2_{21}$, $\Delta m^2_{31}$ and $\Delta m^2_{32}\equiv\Delta m^2_{31}-\Delta m^2_{21}$, as depicted in Figure~\ref{fig:oscillations}. Ideally, measurements of some $P_{\alpha\beta}$ as a function of $L/E$ would suffice to determine all neutrino oscillation parameters. These would also allow one to determine whether the standard paradigm is correct, {\it i.e.}, whether Eqs.~(\ref{eq:Aab},\ref{eq:barAab}) properly describe neutrino flavor-changing phenomena.

For example, if one could measure both $P_{ee}$ and $P_{\mu\mu}$ as a function of $L/E$, one should be able to determine not only $\Delta m^2_{21}$ and $|\Delta m^2_{31}|$, but also $|U_{e2}|^2$,  $|U_{e3}|^2$, $|U_{\mu2}|^2$ and $|U_{\mu3}|^2$, and the sign of $\Delta m^2_{31}$. This in turn would translate into measurements of all mixing parameters, including the $CP$-odd phase $\delta$. One would also be able to determine, for example, whether there are other oscillation lengths, which would indicate there are new, yet-to-be-observed, neutrino states, or whether $P_{ee,\mu\mu}\neq 1$ in the limit $L\to 0$, which would indicate, for example, the existence of new, weaker-than-weak, charged-current type interactions.

In the real world, such measurements are, to say the least, very hard to perform, for several reasons. $\Delta m^2_{21}$ is much smaller than the magnitude of $\Delta m^2_{31,32}$, which in turn makes it challenging to observe two independent oscillation frequencies in the same experimental setup. For this reason all measurements of $P_{\mu\mu}$ performed to date are, effectively, only sensitive to $|\Delta m^2_{31}|$ and $|U_{\mu3}|$ -- the $L/E$ factors probed are too small to ``see'' the $\Delta m^2_{21}$-driven oscillations or distinguish $\Delta m^2_{31}$ from $\Delta m^2_{32}$. On the other hand, the magnitude of $|U_{e3}|$ turns out to be much smaller than that of the other entries of $U$. For this reason, measurements of $\bar{P}_{ee}$ have only been precise enough to definitively observe $\Delta m^2_{21}$-driven oscillations and hence determine its magnitude, along with that of $U_{e2}$. The same is true of solar neutrino experiments.
The current generation of reactor neutrino experiments is expected to measure $|U_{e3}|$ via precision measurements of $\bar{P}_{ee}$ governed by $\Delta m^2_{31}$. These results are expected to be insensitive to all other oscillation parameters.

Another real-world issue is that, for any setup, it is not possible to measure any $P_{\alpha\beta}$ with perfect $L/E$ resolution. Furthermore, the available $L/E$ ranges are, in most cases, narrow. More realistically, one expects to measure, with decent statistics and small systematic errors, $P_{\alpha\beta}$ integrated over a few finite-sized $L/E$ bins. This discreteness of the data leads to ambiguities when it comes to measuring the different mixing parameters. For example, different pairs of $\theta_{13},\delta$ values lead to identical values for $P_{\alpha\beta}$ integrated over a fixed $L/E$. The same is true for pairs of $\theta_{13},\theta_{23}$, and so on. A so-called eight-fold degeneracy has been identified and studied in great detail in the neutrino literature (see, for example, \cite{Cervera:2000kp,BurguetCastell:2001ez,Barger:2001yr}). The solution to this challenge is to perform several measurements of different $P_{\alpha\beta}$ at different values of $L$ and $E$ (and $L/E$). This is especially true if one is interested in not only measuring the three-flavor neutrino mixing parameters but also, much more importantly, over-constraining the standard paradigm and hence testing its validity. For example, one would like to precisely measure $\theta_{13}$ in different channels, for different values of $L$ and $E$, to find out if all of them agree.

Measurements of vacuum survival probabilities, $P_{\alpha\alpha}$ or $\bar{P}_{\alpha\alpha}$ do not violate $CP$ invariance: $P_{\alpha\alpha}=\bar{P}_{\alpha\alpha}$  is guaranteed by $CPT$-invariance. In order to directly observe $CP$-invariance violation, one needs to measure an appearance probability, say $P_{\mu e}$. $P_{\mu e}$ is different from $\bar{P}_{\mu e}$,\footnote{Note that T-invariance violation, $P_{e\mu}\neq P_{\mu e}$, is also present under the same conditions.} as depicted in Fig.~\ref{fig:oscillations} (bottom), if the following conditions are met, as one can readily confirm by studying Eqs.~(\ref{eq:Aab},\ref{eq:barAab}): (i) all $U_{\alpha i}$ have non-zero magnitude, (ii) $U_{\alpha2}U^*_{\beta2}$ and $U_{\alpha3}U^*_{\beta3}$ are relatively complex, (iii) $L/E$ is large enough that both $\Delta m^2_{21,31}\times L/E$ are significantly different from zero. Given what is known about the oscillation parameters, condition (iii) can be met for any given neutrino source by choosing a large enough value for $L$. This, in turn, translates into the need for a very intense source and a very large, yet high-precision, detector, given that for all known neutrino sources the neutrino flux falls off like $1/L^2$ for any meaningful value of $L$. Whether conditions (i) and (ii) are met lies outside the control of the experimental setups. Given our current understanding, including the newly acquired knowledge that $|U_{e3}|\neq0$,  condition (i) holds. That being the case, condition (ii) is equivalent to $\delta \neq 0,\pi$. In the standard paradigm, the existence of $CP$-invariance violation is entirely at the mercy of the value of $CP$-odd phase $\delta$, currently unconstrained.

The fact that neutrino oscillation experiments might be sensitive to a directly $CP$-violating effect cannot be overemphasized. To date, all 
observed $CP$-invariance violating effects have occurred in experiments involving strange and $B$-mesons \cite{Nakamura:2010zzi}, along with tantalizing new hints from the charm sector (see, for example, \cite{Aaltonen:2011se,neu:Aaij:2011in}). Furthermore, in spite of several decades of experimental searches, all of these are explained by the CKM paradigm -- the three-flavor mixing paradigm in the quark sector -- and all are functions of a unique $CP$-odd parameter in the Standard Model Lagrangian: the phase $\delta_{CKM}$ in the quark mixing matrix. Neutrino oscillations provide a unique opportunity to probe a brand new $CP$-violating sector of Nature.

All neutrino data accumulated so far provide only hints for non-zero $P_{\mu\tau}$ \cite{Abe:2006fu,Agafonova:2010dc} and $P_{\mu e}$ \cite{Abe:2011sj,Adamson:2011qu}.\footnote{Solar data translate into overwhelming evidence for $P_{e\mu}+P_{e\tau}\neq 0$. In the standard paradigm, this is indistinguishable from $1-P_{ee}\neq 1$ and hence cannot, even in principle, provide more information than a disappearance result.} Both results are only sensitive to one mass-square difference ($|\Delta m^2_{31}|$) and to $|U_{\mu3}U_{\tau3}|$ and $|U_{\mu3}U_{e3}|$, respectively. The goal of the current neutrino oscillation experiments NO$\nu$A and T2K is to observe and study $P_{\mu e}$ and $\bar{P}_{\mu e}$ governed by $\Delta m^2_{31}$, aiming at measuring $U_{e3}$ and, perhaps, determining the sign of $\Delta m^2_{31}$ through matter effects, as will be discussed promptly.

Eqs.~(\ref{eq:Aab},\ref{eq:barAab}) are valid only when the neutrinos propagate in a vacuum. When neutrinos propagate through a medium, the oscillation physics is modified by so-called matter effects \cite{Wolfenstein:1977ue}. These are due to the coherent forward scattering of neutrinos with the electrons present in the medium, and they create an additional contribution to the phase differences.  Notably, this additional contribution distinguishes between neutrinos and antineutrinos, since there are no positrons present in the Earth.\footnote{In fact, the electron background explicitly violates $CPT$ symmetry. For neutrinos oscillating in matter, it is no longer true, for example, that $P_{\alpha\alpha}=\bar{P}_{\alpha\alpha}$.} Matter effects also depend on whether the electron neutrino is predominantly made out of the heaviest or lightest mass eigenstates, thus allowing one to address the ordering of the neutrino mass eigenstates. For one mass hierarchy, the oscillation of neutrinos for a certain range of $L/E$ values can be enhanced with respect to that of antineutrinos, while for the other mass hierarchy the effect is reversed. On the flip side, if the mass hierarchy is not known, matter effects lead to ambiguities in determining the oscillation parameters, as discussed briefly earlier. Matter effects have already allowed the determination of one ``mass hierarchy,'' that of $\nu_1$ and $\nu_2$. Thanks to matter effects in the sun, we know that $\nu_1$, which is lighter than $\nu_2$, has the larger electron component: $|U_{e1}|^2>|U_{e2}|^2$. A similar phenomenon may be observable in the $\Delta m^2_{31}$ sector, as long as $|U_{e3}|$ is not zero. Quantitatively, the importance of matter effects will depend on the density of the medium being traversed, which determines the so-called matter potential $A\equiv \sqrt{2}G_FN_e$, where $G_F$ is the Fermi constant and $N_e$ is the electron number-density of the medium, and on the value of $\Delta m^2_{21,31}/E$. Matter effects are irrelevant when  $A\ll\Delta m^2_{21,31}/E$. For $\Delta m^2_{31(21)}$ matter effects in the Earth's crust are significant for $E\gtrsim 1$~GeV (20~MeV).

\subsection{Neutrino Experiments: Sources and Detectors}

\begin{figure}[ht]
\begin{center}
\includegraphics[width=0.74\textwidth]{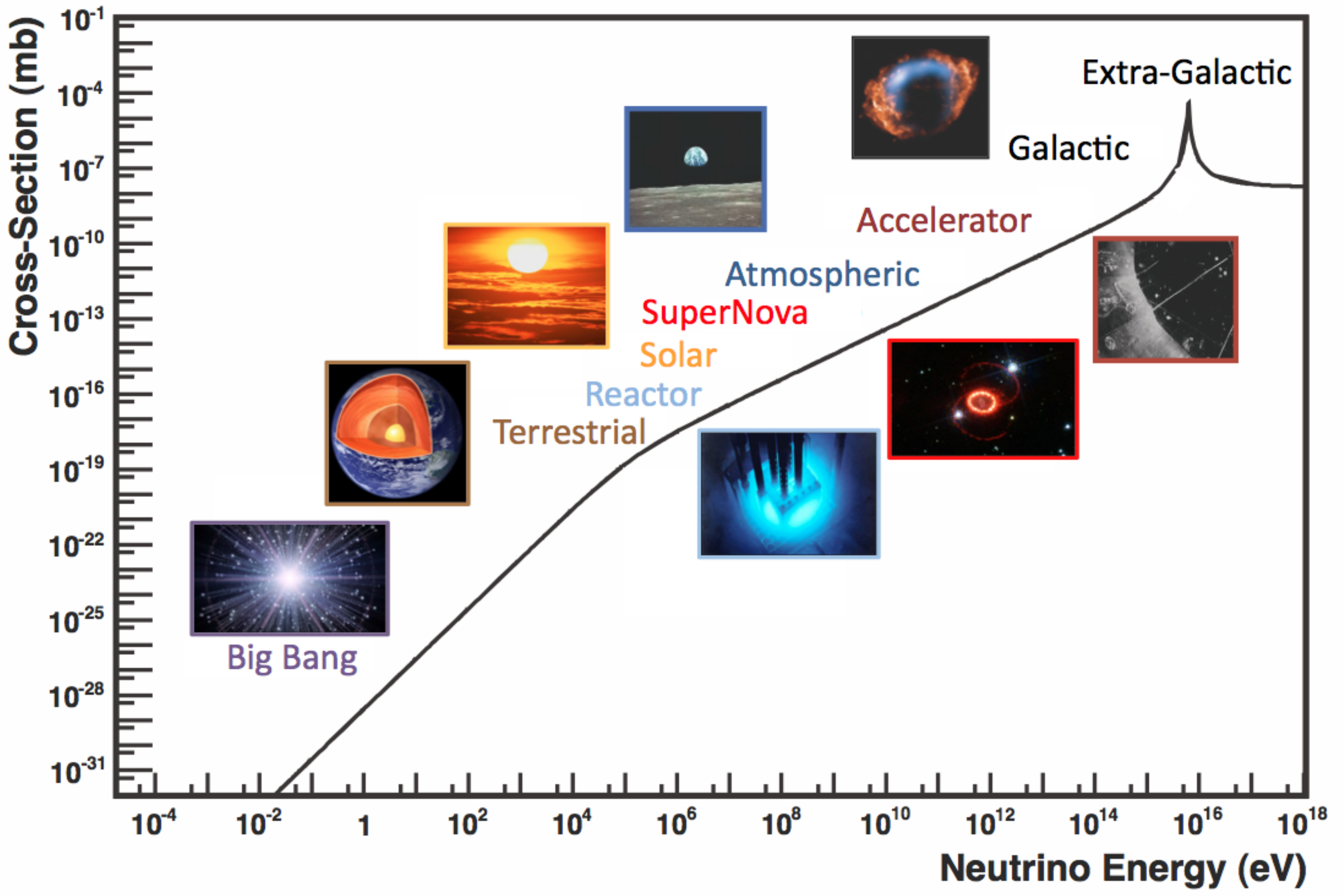}
\end{center}
\caption{Neutrino interaction cross section as a function of energy, showing typical energy regimes for different sources. The scattering cross section for $\bar{\nu}_e \, e^- \rightarrow e^- \, \bar{\nu}_e$ on free electrons is shown for comparison. Plot is reproduced from~\cite{FormaggioZeller:2012}.}
\label{fig:sources}
\end{figure}

Next-generation experiments have at their disposal a handful of neutrino sources, which we describe qualitatively here, concentrating on their prospects for neutrino oscillation searches.   The sources span many orders of magnitude in energy: see Fig.~\ref{fig:sources}.   Associated with each experiment is an appropriate detector. The natures and the capabilities of the detectors depend on the neutrino source.   

The sun is a very intense source of $\nu_e$ with energies between 100~keV and 10~MeV. Precision measurements of the low-energy component of the solar neutrino flux (the so-called $pp$-neutrinos) may provide an unique opportunity to improve on the precision with which $\sin^2\theta_{12}$ is known \cite{Bahcall:2003ce}. The detection of very low-energy solar neutrinos is very challenging, but R\&D related to building such detectors profits from significant synergy with efforts to look for dark matter and observe neutrinoless double-beta decay. Solar neutrinos in the few-MeV range are very sensitive to solar matter effects, and provide a unique opportunity to test the Standard Model through the Mikheev-Smirnov-Wolfenstein (MSW) matter effect \cite{Wolfenstein:1977ue,Mikheev:1986gs}. Indeed, data from the SNO experiment seem to hint at potential deviations from Standard Model expectations~\cite{Aharmim:2011vm}. During this decade, more (neutrino) light is expected to shine on this potentially very important matter, from the Borexino \cite{Alimonti:2008gc} and the SNO+ \cite{Kraus:2010zzb} experiments.

Nuclear reactors are an intense, very pure source of $\bar{\nu}_e$ with energies between a few and several MeV. Due to the low neutrino energies, only $\bar{\nu}_e$ can be detected in the final state, which is done via inverse $\beta$-decay, $\bar{\nu}_e+p \rightarrow e^+ + n$.  The current generation of reactor experiments aims at percent-level measurements of the $\bar{\nu}_e$ spectrum, one or two kilometers away from the source. At these distances and energies one is sensitive only to $\Delta m^2_{31}$-driven oscillations. The necessary precision is expected to be achieved through the comparison of data obtained at near and far detectors. In a nutshell, the near detector measures the neutrino flux before oscillations have had time to act, while the far detector measures the effects of the oscillations \cite{Anderson:2004pk}. Reactor neutrino experiments with much longer baselines (say, 50~km) have been considered: see, for example, \cite{Minakata:2004jt,Petcov:2006gy}. These would be sensitive to both $\Delta m^2_{31}$ and $\Delta m^2_{21}$-driven oscillations, and, in principle, would allow much more precise measurements of $\Delta m^2_{21}$ and $|U_{e2}|$. A detector with exquisite energy resolution may also be sensitive to the neutrino mass hierarchy (see, for example, \cite{Ghoshal:2010wt}).

Meson decays are a very good source of $\nu_{\mu}$ and $\nu_{\tau}$ and their antiparticles. The heavy $\tau$-lepton mass, however, prevents any realistic means of producing anything that would qualify as a $\nu_{\tau}$-beam, so we will only discuss $\nu_{\mu}$ beams. Pions and, to a lesser extent, kaons are produced in large numbers through proton--nucleus interactions. These, in turn, can be sign-selected in a variety of ways to yield a mostly pure $\nu_{\mu}$ or $\bar{\nu}_{\mu}$ beam. The neutrino energy is directly related to the pion energy.

The lowest energy $\nu_{\mu}$ ``beams'' (really, isotropic sources) are achieved from pion decay at rest. A large sample of mostly $\pi^+$ at rest yields a very well-characterized flux of mono-energetic $\nu_{\mu}$ (from the $\pi^+$ decay), along with $\bar{\nu}_{\mu}$ and $\nu_e$ from the subsequent daughter muon decay. All neutrino energies are below the muon production threshold, so only $\nu_e$ and $\bar{\nu}_e$ can be detected via charged-current interactions. An interesting experimental strategy is to search for $\bar{\nu}_e$ via inverse $\beta$-decay, a very well understood physics process, and hence measure with good precision $\bar{P}_{\mu e}$ \cite{Conrad:2009mh}. Matter effects play an insignificant role for the decay-at-rest beams, rendering oscillation results less ambiguous. On the other hand, even very precise measurements of $\bar{P}_{\mu e}$ from pion decay at rest are insensitive to the neutrino mass hierarchy.

Boosted pion-decay beams are the gold standard of readily accessible neutrino oscillation experiments. A pion beam is readily produced by shooting protons on a target. These can be charge- and energy-selected, yielding a beam of either mostly $\nu_{\mu}$ or $\bar{\nu}_{\mu}$. Larger neutrino energies allow one to look for $\nu_e$, $\nu_{\mu}$ and, for energies above a few GeV, $\nu_{\tau}$ in the far detector. Large neutrino energies, in turn, require very long baselines\footnote{The oscillation phase scales like $L/E$. For a 1~GeV beam, one aims at $L$ values close to 1000~km.} and hence very intense neutrino sources and very large detectors. Intense neutrino sources, in turn, require very intense proton sources, of the type described in Sec.~\ref{sec:facilities}. For this reason, these pion-decay-in-flight beams are often referred to as superbeams. Larger neutrino energies and longer baselines also imply nontrivial matter effects even for $\Delta m^2_{31}$-driven oscillations. A neutrino beam with energies around 1~GeV and baselines around 1000~km will allow the study of $P_{\mu\mu}$ and $P_{\mu e}$ (and, in principle, the equivalent oscillation probabilities for antineutrinos) as long as the far detector is sensitive to both $\nu_{\mu}$ and $\nu_e$ charged-current interactions.  One may choose to observe the neutrino flux a few degrees
off the central beam axis,  where the pion decay kinematics result in a narrowly peaked neutrino spectrum. This is beneficial for optimizing sensitivity at the oscillation maximum and for reducing backgrounds outside the energy regime of interest.

The constant collision of cosmic rays with the atmosphere produces mesons (mostly pions and kaons) and, upon their decays, $\nu_{\mu}$, $\bar{\nu}_{\mu}$,  $\nu_e$, $\bar{\nu}_e$. These atmospheric neutrinos cover a very wide energy range (100~MeV to 100~GeV and beyond) and many different distances (15~km to 13000~km), some going through the core of the Earth and hence probing matter densities not available for Earth-skimming neutrino beams. This is, by far, the broadest (in terms of $L/E$ range) neutrino ``beam.'' As far as challenges are concerned, uncertainties in the atmospheric neutrino flux are not small, and the incoming neutrino energy and direction must be reconstructed only with information from the neutrino detector.

In the past, atmospheric neutrinos have provided the first concrete evidence for  neutrino oscillations, and at present they are still a major contributor to the global fits to neutrino oscillation parameters. They will continue to be important in the future. They are also ubiquitous and unavoidable. IceCube DeepCore is already taking data and will accumulate close to a million events with energies above about 10 GeV over the next decade \cite{FernandezMartinez:2010am}. Any other very large detector associated with the intensity frontier program will also collect a large number of atmospheric neutrino events in various energy ranges, through different types of signatures. While atmospheric neutrino data suffer from larger systematic uncertainties, some of these can be greatly  reduced by studying angular and energy distributions of the very high statistics data. Their study can  complement that of the high precision measurements from fixed baseline experiments. For example, non-standard  interactions of neutrinos, additional neutrino flavors and other new physics phenomena  affecting neutrinos could be present, and their effects are likely to be more important at higher energies or in the presence of matter, thus making atmospheric neutrinos an ideal testing ground (see, for example, \cite{GonzalezGarcia:2011my}). Furthermore, a precise, very high statistics measurement of the atmospheric neutrino flux itself over a very  large range of energies will also contribute to a better understanding of cosmic ray propagation through the atmosphere \cite{Gaisser:2002jj,GonzalezGarcia:2006ay,Abbasi:2010ie}.

Muon decays are also an excellent source of neutrinos. The physics and the kinematics of muon decay are very well known and yield two well-characterized neutrino beams for the price of one: $\nu_{\mu}+\bar{\nu}_e$ in case of $\mu^-$ decays,  $\bar{\nu}_{\mu}+\nu_e$ in the case of $\mu^+$. A neutrino factory is a storage ring for muons with a well-defined energy.  Depending on the muon energy, one can measure, with great precision, $P_{\mu\mu}$ and $P_{e\mu}$, assuming the far detector can tell positive from negative muons, potentially along with $P_{\mu e}$ and $P_{ee}$, if the far detector is sensitive to electron charged-current events and can deal with the $\pi^0$ backgrounds, or $P_{\mu\tau}$ and $P_{e\tau}$, if the muon energy is large enough and if the far detector has the ability to identify $\tau$-leptons with enough efficiency.
Neutrino factories are widely considered the ultimate sources for neutrino oscillation experiments \cite{neu:NF:2011aa}, and probably allow for the most comprehensive tests of the standard three-neutrino paradigm.

Finally, nuclei that undergo $\beta$-decay serve as a very well-characterized source of $\nu_e$ or $\bar{\nu}_e$. An intense, highly boosted beam of $\beta$-decaying nuclei would allow for the study of $P_{e\mu}$. 
Such sources are known as ``$\beta$-beams'' \cite{Zucchelli:2002sa}.

\subsection{Neutrino Oscillation Experiments: Achievements and Opportunities}

The experimental achievements of the past 15 years that have filled in the three-flavor neutrino picture have been astonishing.   In both the atmospheric and solar sectors, an effect first observed with natural neutrinos was explored by several experiments and 
confirmed independently with artificial sources of neutrinos.  
In the atmospheric case, the $\nu_\mu$ disappearance was confirmed by long-baseline beam experiments, K2K and MINOS.  Solar neutrino oscillations measured by radiochemical experiments, Super-Kamiokande and SNO,  were confirmed by, and the parameters constrained by, reactor $\bar{\nu}_e$ with KamLAND.   A decade ago, the space of allowed oscillation parameters spanned many orders of magnitude;
allowed regions have now shrunk to better than the 10\% precision level.

The current generation of detectors on the hunt to measure $\theta_{13}$ are employing both reactor and boosted-pion-decay beam sources of neutrinos.
The known natural sources of neutrinos used for oscillation experiments-- solar and atmospheric neutrinos-- still have information to be gained from them, both for understanding of neutrino properties and for studies of the sources themselves; 
furthermore, a core-collapse supernova will produce an enormous burst of neutrinos of all flavors that can be exploited for both
neutrino oscillation physics and core-collapse astrophysics. 
By the end of this decade, we anticipate new invaluable information from the current generation of neutrino oscillation experiments, namely the long-baseline beam experiments 
MINOS,  T2K, and NO$\nu$A and the reactor experiments Double Chooz, Daya Bay and RENO.
 In the language of the standard paradigm, these are very likely to measure $\theta_{13}$, $\theta_{23}$, and $|\Delta m^2_{31}|$ much more precisely, and may provide nontrivial hints regarding the neutrino mass hierarchy. While the possibility of surprises cannot, and certainly should not, be discarded, it is expected that the neutrino data accumulated until the end of the decade will be unable to definitively test the standard three-neutrino paradigm or observe $CP$-invariance violation in the lepton sector. That will be the task of next-generation experiments.

\begin{figure}[!h]
\centerline{
\includegraphics[width=0.5\textwidth]{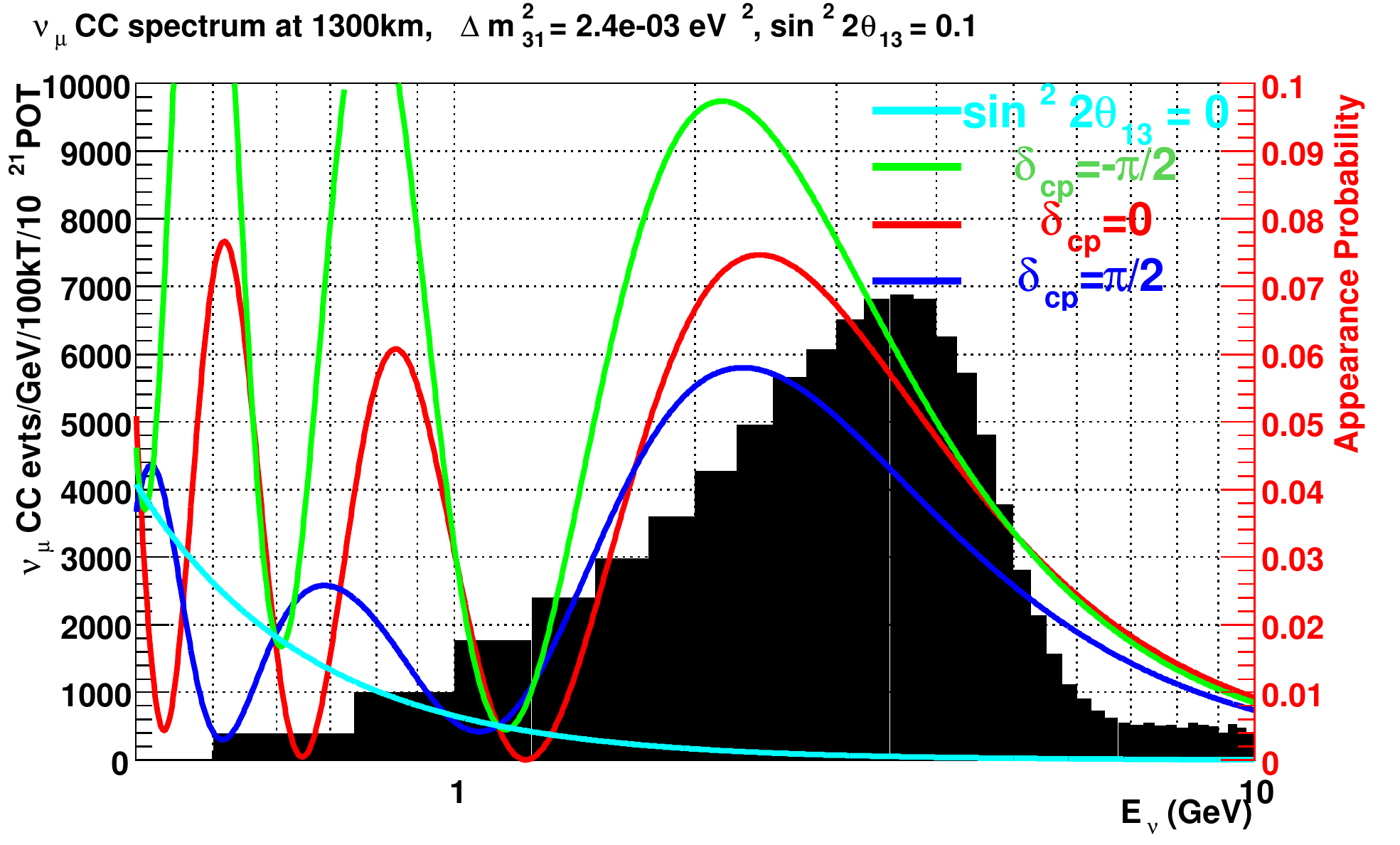}
\includegraphics[width=0.5\textwidth]{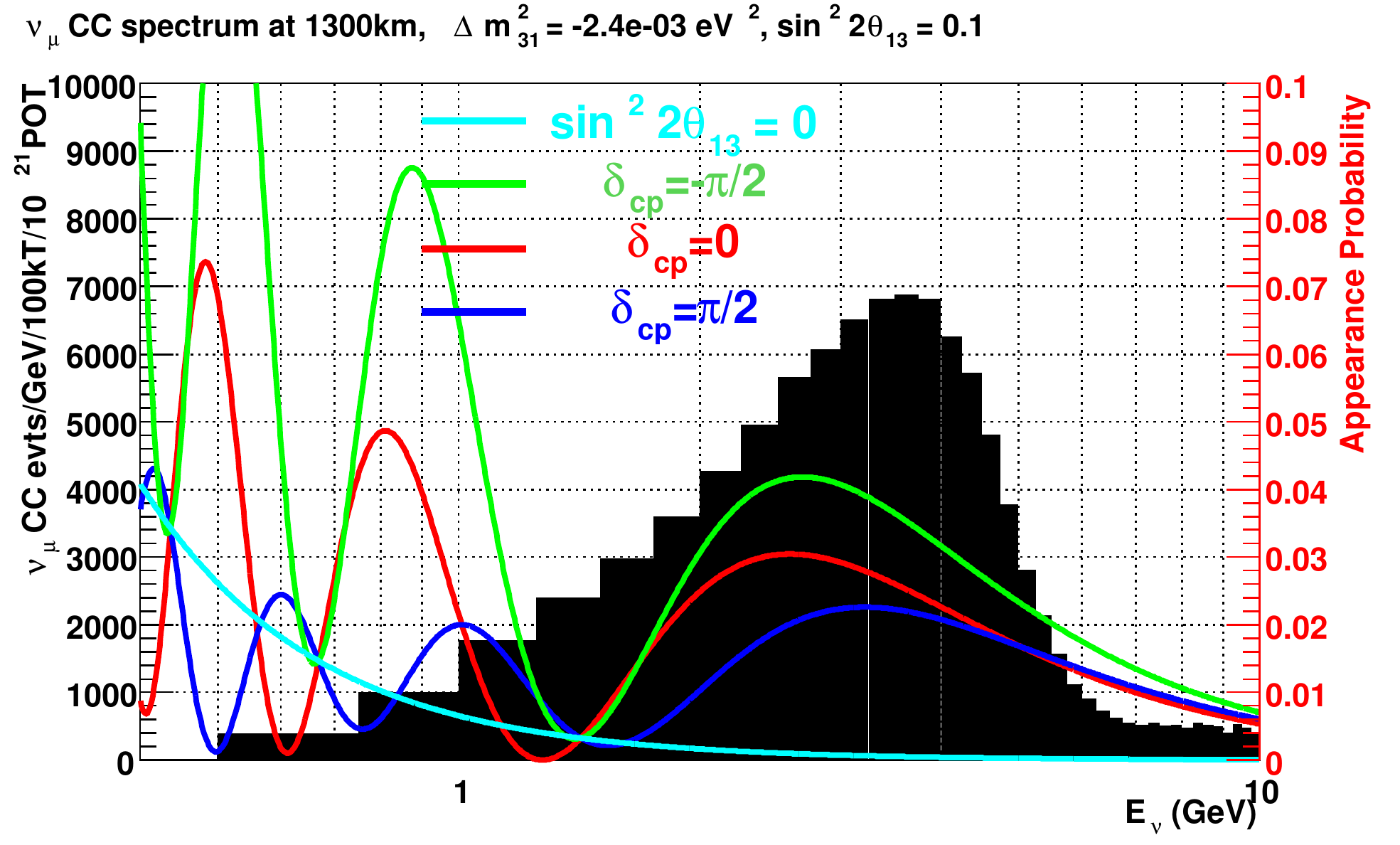}
}
\caption{The $\nu_\mu \rightarrow \nu_e$ oscillation probability for the LBNE
to Homestake baseline of 1300~km for different mixing parameters with normal
hierarchy (left) and inverted hierarchy (right) is shown as colored curves.
The unoscillated charged-current $\nu_\mu$ spectrum from an LBNE candidate beam is shown as
the solid black histogram.  From study in reference~\cite{Akiri:2011dv}.}
\label{fig:lbnebeam}
\end{figure}

Future opportunities for testing the paradigm and probing new physics for next-generation neutrino oscillation experiments are broad and exciting.  We will outline in detail the challenges in Section~\ref{sec:facilities}, and describe there both the global context and opportunities for the United States, but we briefly mention a few highlights here.  The next focus for the US is the Long-Baseline Neutrino Experiment (LBNE), which will employ a new beam from Fermilab and a large liquid argon time projection chamber (TPC) at the Homestake mine in South Dakota, 1300~km away. LBNE will make use of a new beam from Fermilab of 700~kW power; the detector will be on the beam axis. Figure~\ref{fig:lbnebeam} shows the expected event spectrum with oscillation probabilities for a representative value of the mixing parameters superimposed.

\begin{figure}[htb]
\centering\includegraphics[width=.47\textwidth]{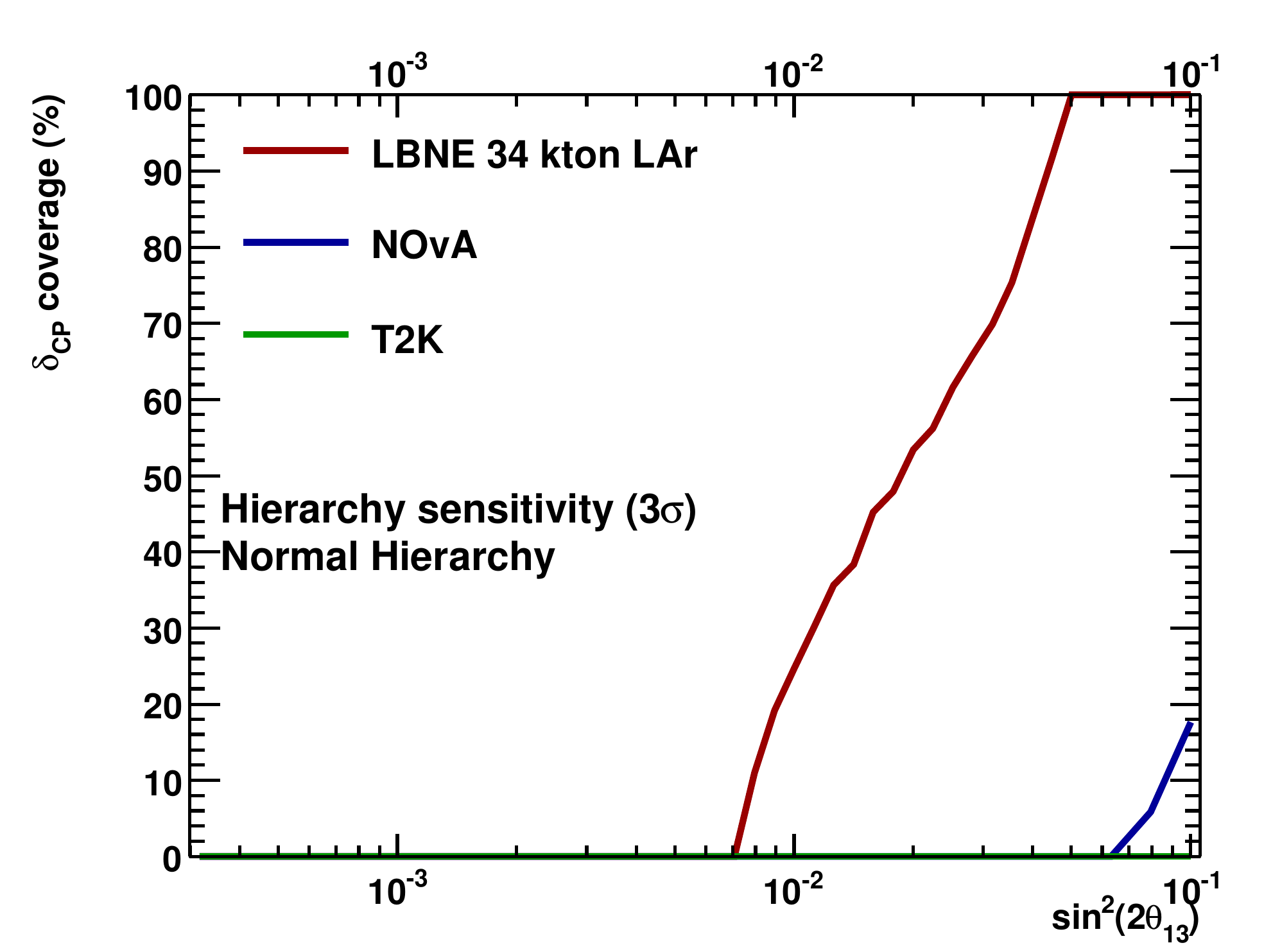}
\centering\includegraphics[width=.47\textwidth]{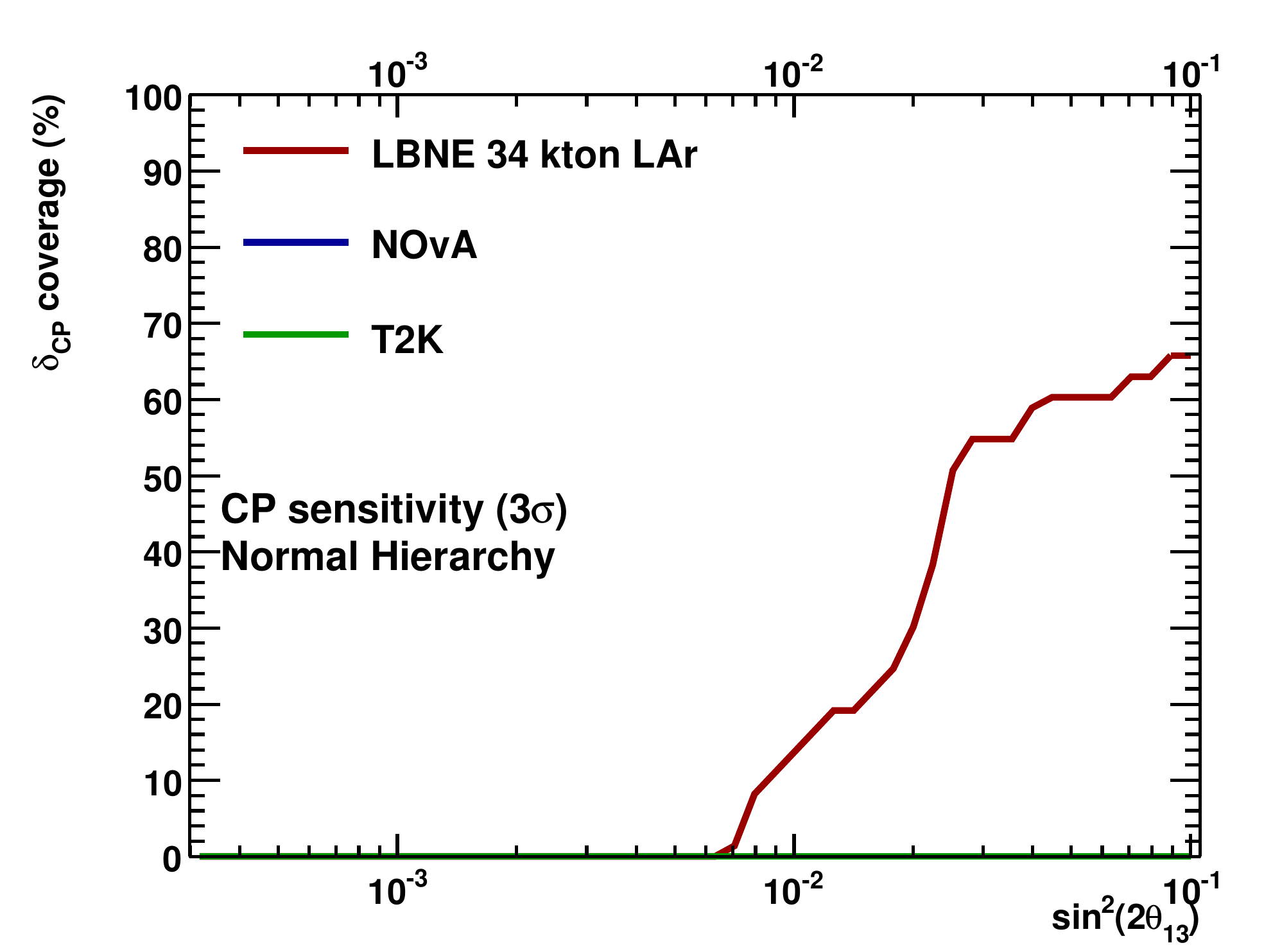}
 \caption{$3\sigma$ discovery potential of LBNE for determining
  the mass hierarchy (left), and $CP$
  violation (right) as function of $\sin^22\theta_{13}$ and the fraction
  of $\delta_{CP}$ coverage. Here the fraction of $\delta_{\mathrm CP}$ reflects
  the fraction of all true values of $\delta_{CP}$ for which the corresponding
  quantity can be measured. Sensitivities are shown for normal mass hierarchy. 
Results for   5+5 years of $\nu$+$\bar{\nu}$ running in a 700~kW beam
for LBNE 34~kt LAr, NO$\nu$A (3+3 years of $\nu+\bar{\nu}$ running in a 700 kW beam), and T2K (3+3 years of $\nu+\bar{\nu}$ running in a 770 kW beam) are shown. Note that NO$\nu$A and T2K have no sensitivity to $CP$ violation, and T2K has no sensitivity to hierarchy at 3$\sigma$ for this range of $\sin^2 2\theta_{13}$ using GLoBES model projections.
 From~\cite{casestudy}.}
  \label{fig:lbnesens}
\end{figure}

Figure~\ref{fig:lbnesens} shows the expected sensitivity of NO$\nu$A, T2K, and LBNE~\cite{Akiri:2011dv} to oscillation parameters of relevance.  These plots are given in terms of ``fraction of $\delta_{CP}$'': as a function of $\sin^22\theta_{13}$, they show the fraction of $\delta_{CP}$ values from 0 to $2\pi$ for which a 3$\sigma$ discovery could be made (measurement of non-zero value for the cases of $\theta_{13}$ and $\delta_{CP}$, or determination of the hierarchy).  
If $\sin^22\theta_{13}$  is indeed very close to $\sim 0.1$, as current data seem to prefer, 
these plots show that the 
prospects for determining the mass
hierarchy are excellent, and chances for measurement of $\delta_{CP}$ are also very good.
We note that the sensitivity of LBNE 
will be enhanced with the addition of precision $\sin^22\theta_{13}$ measurements
from other experiments (notably Daya Bay).

Beyond LBNE, Project X  will sharply extend the reach for neutrino physics, in concert with diverse intensity frontier physics programs.  The community's ideas for future physics measurements do not stop with these; there is an abundance of creative concepts for sources and detectors, many currently being explored. We describe some of those in more detail in Section~\ref{sec:facilities}. Tables~\ref{experimentstable} and \ref{experimentstable2} summarize the capabilities of current and future neutrino oscillation experiments.
\begin{table}
\footnotesize
\caption{Types of current or proposed neutrino oscillation experiments, with some current and future examples (not exhaustive), along with their accessibility to different oscillation channels. $\surd\surd$ indicates the most important  oscillation channel(s) while $\surd$ indicates other accessible channels. `$\nu_{e,\mu}$ disapp' refers to the disappearance of $\nu_{e}$ or $\nu_{\mu}$, which are related to $P_{ee}$ and $P_{\mu\mu}$, respectively. `$\nu_{\mu}\leftrightarrow\nu_e$' refers to the appearance of $\nu_e$ in a $\nu_{\mu}$ beam or vice versa, related to $P_{e\mu}$ or $P_{\mu e}$. `$\nu_{\tau}$ app' refers to the appearance of $\nu_{\tau}$ from an initial state $\nu_e$ or $\nu_{\mu}$, related to $P_{(e,\mu)\tau}$. `Pion DAR/DIF' refers to neutrinos from pion decay at rest or in flight. `$\mu$ DAR/DIF' and `$\beta$ Beam' refer to neutrinos from muon decay and nuclear decay in flight, respectively. In particular Pion DIF stands for a so-called conventional neutrino beam. `Coherent $\nu$-A' stands for very low-energy neutrino experiments, usually from spallation sources, aiming at measuring coherent neutrino--nucleus scattering. See text for more details.}
\label{experimentstable}
\begin{center}
\begin{tabular}{|c|c|c|c|c|c|c|c|c|c|c|c|c|c|c|} \hline
Expt. Type & $\nu_{e}$~disapp & $\nu_{\mu}$~disapp & $\nu_{\mu}\leftrightarrow\nu_{e}$  & $\nu_{\tau}$~app$^1$ & Examples \\ \hline \hline 
Reactor & $\surd\surd$ & -- & -- & -- & KamLAND, Daya Bay, Double Chooz, RENO \\ \hline
Solar$^2$ & $\surd\surd$ & -- & $\surd$ & -- & Super-K, Borexino, SNO+, LENS, Hyper-K (prop)\\ \hline 
Supernova$^3$ & $\surd\surd$ & $\surd$ & $\surd\surd$ & -- & Super-K, KamLAND, Borexino, IceCube, \\ 
 &  &  & &  & LBNE (prop), Hyper-K (prop) \\ \hline  
Atmospheric & $\surd$ & $\surd\surd$ & $\surd$ & $\surd$ & Super-K, LBNE (prop), INO (prop), IceCube, Hyper-K (prop)\\ \hline  
Pion DAR & $\surd$ & -- & $\surd\surd$ & -- & DAE$\delta$ALUS \\ \hline
Pion DIF & -- & $\surd\surd $ & $\surd\surd$ & $\surd$ & MiniBooNE, MINER$\nu$A$^4$, MINOS(+, prop), T2K \\ 
 &  &   &  &  & NO$\nu$A, MicroBooNE, LBNE (prop), Hyper-K (prop) \\ \hline 
Coherent $\nu-$A$^5$ & -- & -- & -- & -- & CLEAR (prop), Ricochet (prop) \\ \hline
$\mu$ DIF$^6$ & $\surd$ & $\surd\surd$ & $\surd\surd$ & $\surd$ & VLENF, NuFact \\ \hline
$\beta$ Beam & $\surd$ & -- & $\surd\surd$ & -- & \\ \hline
\end{tabular}
\end{center}
$^1$In order to observe $\nu_{\tau}$ appearance, a dedicated detector or analysis is required, along with a high-enough neutrino energy. $^2$Solar neutrino experiments are sensitive, at most, to the $\nu_e$ and the $\nu_{e}+\nu_{\mu}+\nu_{\tau}$ components of the solar neutrino flux. $^3$Signatures of neutrino oscillation occurring both in the collapsed star matter and in the Earth will be present in the spectra of observed fluxes of different flavors, and do not strictly fall in these categories; detectors are sensitive to $\nu_e$ and $\bar{\nu}_e$ fluxes, and to all other flavors by NC interactions.
$^4$MINER$\nu$A measures neutrino cross sections with the aim of reducing systematics for oscillation experiments.   $^5$Coherent elastic neutrino-nucleus scattering is purely NC and not sensitive to oscillation between active flavors. $^6$The ``standard'' high-energy neutrino factory setups are not sensitive to electron appearance or disappearance.   
\end{table}
\begin{table}
\footnotesize
\caption{Types of current or proposed neutrino oscillation experiments and their ability to address some of the outstanding issues in neutrino physics. `NSI' stands for non-standard neutrino interactions, while $\nu_s$ ($s$ for sterile neutrino) stands for the sensitivity to new neutrino mass eigenstates. `$\star\star\star$' indicates a very significant contribution from the current or proposed version of these experimental efforts, `$\star\star$' indicates an interesting contribution from current or proposed experiments, or a significant contribution from a next-next generation type experiment, `$\star$' indicates a marginal contribution from the current or proposed experiments, or an interesting contribution from a next-next generation type experiment. See Table \ref{experimentstable} and text for more details.}
\label{experimentstable2}
\begin{center}
\begin{tabular}{|c|c|c|c|c|c|c|c|c|c|c|c|c|c|c|} \hline
Expt. Type & 
$\sin^2\theta_{13}$ & sign($\Delta m^2_{31}$) & $\delta$ & $\sin^2\theta_{23}$ & $\left|\Delta m^2_{31}\right|$ &  $\sin^2\theta_{12}$ & $\Delta m^2_{21}$ & NSI & $\nu_{s}$\\ \hline \hline 
Reactor &  $\star\star\star$ & $\star$& -- & -- & $\star$ & $\star\star$ & $\star\star$ & -- & $\star\star$ \\ \hline
Solar &  $\star$ & -- & -- & -- & -- & $\star\star\star$ & $\star$ & $\star\star$ & $\star\star$ \\ \hline  
Supernova &  $\star$ & $\star \star \star$ & -- & -- & -- & $\star$ & $\star$ & $\star\star$ & $\star\star$ \\ \hline  
Atmospheric & $\star\star$ & $\star\star$ & $\star\star$ & $\star\star$ & $\star\star$ & -- & -- & $\star\star\star$ & $\star\star$ \\ \hline  
Pion DAR &  $\star\star\star$ & -- & $\star\star\star$ & $\star$ & $\star\star$ & $\star$ & $\star$ & -- & $\star\star$ \\ \hline
Pion DIF &  $\star\star\star$ & $\star\star\star$ & $\star\star\star$ & $\star\star$ & $\star\star$ & $\star$ & $\star$ & $\star\star$ & $\star\star$ \\ \hline
Coherent $\nu-$A & -- & -- & -- & -- & -- & -- & -- & $\star\star\star$ & $\star\star$ \\ \hline
$\mu$ DIF & $\star\star\star$ & $\star\star\star$ & $\star\star\star$ & $\star\star\star$ & $\star\star\star$ & $\star$ & $\star$ & $\star\star$ & $\star\star$ \\ \hline
$\beta$ Beam & $\star\star\star$ & -- & $\star\star\star$ & $\star\star$ & $\star\star$ & $\star$ & $\star$ & -- & $\star\star$ \\ \hline
\end{tabular}
\end{center}
\end{table}

\section{The Nature of the Neutrino -- Majorana versus Dirac}
\label{sec:majorana}




With the realization that neutrinos are massive, there is an increased
interest in investigating their intrinsic properties.  Understanding
the neutrino mass generation mechanism, the absolute neutrino mass
scale, and the neutrino mass spectrum are some of the main focuses
of future neutrino experiments.   Whether neutrinos are Dirac fermions (\textit{i.e.}, exist as separate massive neutrino and 
antineutrino states) or Majorana fermions (neutrino and antineutrino states are equivalent) is a key experimental
question, the answer to which will guide the theoretical description of neutrinos.

All observations involving leptons are consistent with their appearance and
disappearance in particle anti-particle pairs. This property is expressed in the
form of lepton
number, $L$, being conserved by all fundamental forces.
We know of no fundamental symmetry relating to this empirical conservation law.
Neutrinoless double-beta decay, a weak nuclear decay process in which a nucleus decays to a different nucleus
emitting two beta-rays and no neutrinos,
 violates lepton number conservation by two units and thus,
if observed, requires a revision of our current understanding of particle physics.
In terms of field theories, such as the
Standard Model, neutrinos are assumed
to be massless and there is no chirally right-handed neutrino field.
The guiding principles
for extending the Standard Model are the conservation of electroweak isospin and renormalizability,
which do not preclude each neutrino mass eigenstate $\nu_i$ to be identical to its
antiparticle $\overline{\nu}_i$, or a Majorana particle.
However, $L$ is no longer conserved if $\nu = \overline{\nu}$.
Theoretical models, such as the seesaw mechanism that can explain the smallness of neutrino
mass, favor this scenario. Therefore, the discovery of Majorana neutrinos would
have profound theoretical implications in the formulation of a new Standard Model while
yielding insights into the origin of mass itself. If neutrinos are Majorana particles, they
may fit into the leptogenesis scenario for creating the baryon asymmetry, and hence ordinary
matter, of the universe.

As of yet, there is no firm experimental evidence to confirm
or refute this theoretical prejudice. Experimental evidence of neutrinoless double-beta
($0\nu\beta\beta$) decay would establish the Majorana nature of neutrinos.
It is clear that $0\nu\beta\beta$ experiments sensitive at least
to the mass scale indicated by the atmospheric neutrino oscillation results are needed.
\begin{figure}[ht]
\begin{center}
\includegraphics[width=0.65\textwidth]{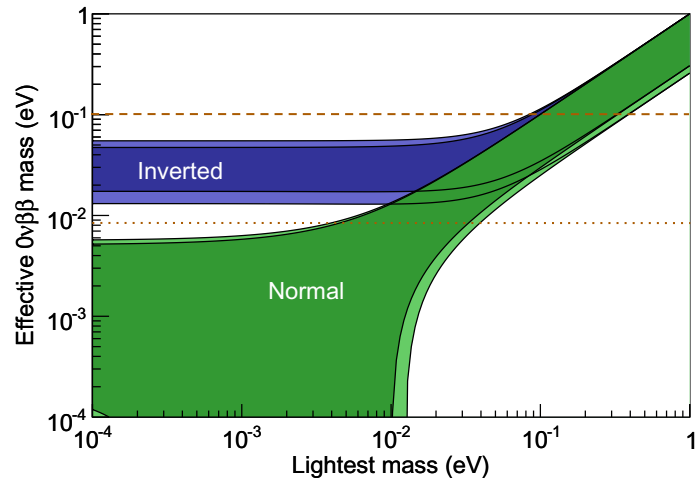}
\end{center}
\caption{Allowed values of $\langle m_{\beta \beta} \rangle$ as a function of the lightest neutrino mass for the inverted and normal hierarchies. The dark shaded regions correspond
to the best-fit neutrino mixing parameters from \cite{Amsler:2008zzb}  and account for the degeneracy
due to the unknown Majorana phases. The lighter shading corresponds to the maximal allowed regions
including mixing parameter uncertainties as evaluated in \cite{Amsler:2008zzb}.
The dashed line shows expected sensitivity of next-generation $\sim$100~kg class experiments and the dotted line shows potential reach of multi-ton scale future experiments.}
\label{fig:SensitivityExposure}
\end{figure}
%

For $0\nu\beta\beta$ decay the summed energy of the emitted electrons is mono-energetic.
Observation of a sharp peak at the $\beta\beta$ endpoint would thus quantify the $0\nu\beta\beta$
decay rate,
demonstrate that neutrinos are Majorana particles, indicate that lepton number is not
conserved, and, paired with nuclear structure calculations, provide a measure of an
effective Majorana mass, $\langle m_{\beta\beta} \rangle$. 
There is consensus within the neutrino physics community that such a decay peak
would have to be observed for at least two different decaying isotopes at two different
energies to make a credible claim for $0\nu\beta\beta$ decay.

In more detail, the observed half-life can be related to an effective Majorana mass
according to 
$(T_{1/2,0\nu\beta\beta})^{-1} 
= G_{0\nu}|M_{0\nu}|^2\langle m_{ \beta \beta} \rangle^2$,
where $\langle m_{ \beta \beta} \rangle^2 \equiv | \sum_i U_{ei}^2 m_i|^2$.
$G_{0\nu}$ is a phase space factor, $m_i$ is the mass
of neutrino mass eigenstate $\nu_i$, and $M_{0\nu}$ is the transition nuclear matrix
element.  The matrix element has significant nuclear theoretical uncertainties,
dependent on the nuclide under consideration.  

In the standard three-massive-neutrinos paradigm,  
\begin{equation}
\langle m_{\beta\beta}\rangle= |\cos^2\theta_{12}\cos^2\theta_{13}e^{-2i\xi}m_1+\sin^2\theta_{12}\cos^2\theta_{13}e^{-2i\zeta}m_2+\sin^2\theta_{13}e^{-2i\delta}m_3|.
\end{equation}
If none of the neutrino masses vanish, $\langle m_{\beta\beta} \rangle$ is a function of not only the oscillation parameters $\theta_{12,13},\delta$ and the neutrino masses $m_{1,2,3}$ but also the two Majorana phases $\xi,\zeta$.
Neutrino oscillation experiments indicate that at least one
neutrino has a mass of $\sim 45$~meV or more. As a result and as shown in Fig.~\ref{fig:SensitivityExposure}, in the
inverted hierarchy mass spectrum with $m_3 = 0$~meV, $\langle m_{\beta\beta} \rangle$ is
between 10 and 55 meV
depending on the values of the Majorana phases. This is sometimes referred to as the
atmospheric mass scale. Exploring this region requires a sensitivity to half-lives exceeding
$10^{27}$ years. This is a challenging goal requiring several ton-years of exposure
and very low backgrounds. The accomplishment of this goal requires a detector at the ton
scale of enriched material and a background level below 1 count/(ton y) in the spectral
region of interest (ROI).  Very good energy resolution is also required.


There is one controversial result from a subset of collaborators
of the Heidelberg-Moscow experiment, who claim a measurement
of the process in $^{76}$Ge, with 70 kg-years of data~\cite{KlapdorKleingrothaus:2001ke}.
These authors interpret the observation as giving an $\langle m_{\beta\beta}\rangle$ of
440~meV.   Recent limits from NEMO-3 and Cuoricino (see below) are impinging
on this $\langle m_{\beta\beta}\rangle$ regime, for $^{100}$Mo and $^{130}$Te respectively.

There is a large number of current neutrinoless double-beta decay search efforts, employing very different techniques; a recent review is~\cite{Avignone:2007fu}.  Here we will highlight some for which there is a component
of effort from physicists based in the US.  These represent different kinds of detectors and experimental approaches.

The MAJORANA \cite{Aguayo:2011sr,Phillips:2011db,Schubert:2011nm} experiment employs the germanium isotope $^{76}$Ge, to be enriched.  The current phase of the experiment is the ``Demonstrator'', which will employ 30~kg of Ge enriched to 86\% $^{76}$Ge and 10~kg of Ge P-type point contact detectors, with an aim of being underground at the Sanford Underground Research Facility (SURF) in 2013.    The MAJORANA collaboration is planning a ton-scale effort in collaboration with its European counterpart GERDA.  

The ``bolometric'' CUORE experiment~\cite{Alessandria:2011rc}, located at Gran Sasso National Laboratory in Italy, employs $^{130}$Te in the form of TeO$_2$ crystals.  This is a cryogenic setup that determines energy loss via temperature rise measured with thermistors.  The first phase of this experiment, Cuoricino, ran from 2003-2008 with 11.3~kg of $^{130}$Te mass.  The current version of the experiment, CUORE-0, has 11~kg, and the plan for full CUORE starting in 2014 will have 206~kg.

The EXO experiment makes use of $^{136}$Xe, which double-beta
decays as $^{136}\rm{Xe} \rightarrow ^{136}{}\rm{Ba}^{++} + e^- + e^-$.
The first version of EXO, EXO-200, is currently taking data at the Waste Isolation Pilot
Plant in New Mexico  with 175~kg of xenon enriched to 80\% in the isotope 136.
Both scintillation light from the interaction and ionization energy deposited by the electrons is detected in the xenon, which is used in the liquid phase.
The EXO collaboration's novel idea for an upgrade
is the use of barium tagging: the principle is to reduce backgrounds 
by identifying the resulting nucleus by laser spectroscopy~\cite{Danilov:2000pp}.
This ambitious plan-- to tag a single ion in as much as 10 tons of xenon --
is currently under development, and there are several schemes under investigation,
including gaseous versions of EXO.    EXO-200 has recently reported the first observation of two-neutrino double-beta decay \cite{Ackerman:2011gz} in $^{136}$Xe.

Another ambitious idea for a double-beta decay experiment is SNO+~\cite{Kraus:2010zzb}.
SNO+ is an experiment at SNOLAB in Canada which plans to refill
the acrylic vessel of SNO with liquid scintillator.  This experiment would
in addition provide a rich physics program of solar neutrino, geoneutrino and supernova
neutrino physics.
It may also be possible to add 0.1\% Nd
(possibly enriched with $^{150}$Nd, the $0\nu\beta\beta$ decay
isotope of interest) to the scintillator.  Although typically
the energy resolution of such a large detector would not naively be expected
to meet the stringent standards of a neutrinoless double-beta decay search,
the quantity of dissolved Nd would be so large that the neutrinoless signal
would be visible as a clear feature in the spectrum. 

KamLAND-Zen~\cite{Gando2012} (Kamioka Liquid Anti-Neutrino Detector, ZEro Neutrino double-beta decay) is an extension of the KamLAND\cite{Abe08} experiment. KamLAND is a 6.5-m radius balloon filled with 1000 tons of liquid scintillator, submerged inside a 9-m radius stainless-steel sphere filled with 3000 tons of mineral oil with PMTs mounted on the wall. The cavity outside this sphere is filled with water also instrumented with PMTs. KamLAND was built to search for reactor anti-neutrinos and the extension is intended as a search for neutrinoless double-beta decay. The collaboration added an additional low-background miniballoon into the inner sphere that contains 13 tons of liquid scintillator loaded with 330 kg of dissolved Xe gas enriched to 91\% in $^{136}${Xe}. This detector at the Kamioka mine in Japan began operation in September 2011,
and initial results include an improved limit on neutrinoless double-beta decay for $^{136}${Xe} and a measurement of two-neutrino double-beta decay that agrees with the recent EXO-200 result~\cite{kz:2012fc}.

NEXT~\cite{Gom11,Yah10} (Neutrino Experiment with Xenon TPC) intends to use $>$100 kg of Xe enriched to $\sim$90\% in $^{136}${Xe}. The detector will be a moderate-density gas TPC $\sim$0.08 g/cm$^3$ that will also detect scintillation light. By operating at low pressures ($\sim$15 bar), the design should not only provide good energy resolution, but also permit tracking that allows fairly detailed track reconstruction to confirm that candidate events involve two electrons moving in opposite directions. The collaboration has recently demonstrated impressive 1\%  resolution in a limited fiducial volume device. Construction is scheduled to start in 2012 with commissioning to start in 2014. It will operate at the Laboratorio Subterr\'{a}neo 
de Canfranc (LSC) in Spain.

The SuperNEMO~\cite{Arn10} proposal builds on the great success of the NEMO-3 (Neutrino Ettore Majorana Observatory) experiment, which measured two-neutrino double-beta decay rates in seven isotopes~\cite{Bon11}. NEMO-3 has provided the best two-neutrino double-beta decay data to date, including information on single-electron energy distributions and opening angles. The design uses calorimetry to measure energies and tracking to gather kinematical information about the individual electrons. SuperNEMO will improve on NEMO-3 by using a larger mass of isotope, lowering backgrounds, and improving the energy resolution. The present design is for 100 kg of $^{82}${Se}, but other isotopes are being considered. It will have a modular design of 20 thin-source planes of 40 mg/cm$^2$ thickness. Each source will be contained within a Geiger-mode drift chamber enclosed by scintillator and phototubes. Timing measurements from digitization of the scintillator and drift chamber signals will provide topological information such as the event vertex and particle directionality. The modules will be surrounded by water and passive shielding. A one-module demonstrator with 7~kg of $^{82}${Se} is planned to be in operation by 2014. This Demonstrator will have only passive shielding. The complete experiment will be ready by the end of the decade in an extension of the LSM Modane in the Fr\'{e}jus Tunnel in France.

The current and next-generation experiments are of 10-100~kg masses;
these have sensitivities down to about 100~meV.  Further ton-scale
experiments are planned for the generation beyond that: these should
have sensitivities reaching the 10 meV or smaller scale.  Reaching
this regime will be very interesting in its complementarity with
oscillation experiments: if independent oscillation experiments (or
data from supernovae or colliders) determine the mass hierarchy to be inverted,
and there is no $0\nu\beta\beta$ decay signal at the 10 meV scale,
then neutrinos must be Dirac (assuming Nature is not too diabolical).
If a signal is observed at the few meV scale, then not only will we
know that neutrinos are Majorana, but we will also know that the
hierarchy must be normal, even in the absence of an independent
determination.

\begin{table}[h]
\hspace{-1cm}
\begin{tabular}{|c|c|c|c|c|c|}
\hline
Experiment &   Isotope
& Mass &  Technique
& Status  & Location  \\ \hline
AMoRE\cite{Kim11,Lee11}
&$^{100}${Mo}  & 50 kg &CaMoO$_4$ scint. bolometer crystals
& Devel.  & Yangyang \\
CANDLES\cite{Kis09}  &  $^{48}${Ca} & 0.35 kg
& CaF$_2$ scint. crystals
& Prototype &  Kamioka \\
CARVEL\cite{Zde05}  &$^{48}${Ca} & 1 ton
& CaF$_2$ scint. crystals
& Devel.  & Solotvina        \\
COBRA\cite{Zub01}  &  $^{116}${Cd} & 183 kg
& $^{enr}${Cd} CZT semicond. det.
& Prototype             &  Gran Sasso   \\
CUORE-0\cite{Alessandria:2011rc}  &  $^{130}${Te} & 11 kg
& TeO$_2$ bolometers  & Constr. (2012)  &  Gran Sasso            \\
CUORE\cite{Alessandria:2011rc}  &  $^{130}${Te} & 203 kg
& TeO$_2$ bolometers  & Constr. (2013)  &  Gran Sasso            \\
DCBA\cite{ish00}  & $^{150}${Ne} & 20 kg
&$^{enr}${Nd} foils and tracking
& Devel.  & Kamioka  \\
EXO-200\cite{Ackerman:2011gz}  &  $^{136}${Xe} & 160 kg
&Liq. $^{enr}${Xe} TPC/scint.
& Op. (2011)  & WIPP             \\
EXO\cite{Danilov:2000pp}  &  $^{136}${Xe} & 1-10 t
&Liq.  $^{enr}${Xe} TPC/scint.
& Proposal              &  SURF             \\
GERDA\cite{sch05}  &  $^{76}${Ge} & $\approx$35 kg
&$^{enr}${Ge} semicond. det.  
& Op. (2011)  &  Gran Sasso             \\
GSO\cite{dane01}  & $^{160}${Gd} & 2 t
&Gd$_2$SiO$_5$:Ce crys. scint. in liq. scint.
& Devel. &  \\
KamLAND-Zen\cite{Efr11}
& $^{136}${Xe}  & 400 kg & $^{enr}${Xe} dissolved in liq. scint.          
& Op. (2011)    & Kamioka \\
LUCIFER\cite{Giu10,Arnaboldi11}
& $^{82}${Se} & 18 kg
& ZnSe scint. bolometer crystals
& Devel.  & Gran Sasso \\
MAJORANA \cite{Aguayo:2011sr,Phillips:2011db,Schubert:2011nm}
&  $^{76}${Ge}  & 26 kg &$^{enr}${Ge} semicond. det. & Constr. (2013)
&  SURF             \\
MOON \cite{eji07}
&  $^{100}${Mo}  & 1 t &$^{enr}${Mo} foils/scint.
& Devel.  &               \\
SuperNEMO-Dem\cite{Arn10}
&  $^{82}${Se}  & 7 kg & $^{enr}${Se} foils/tracking
& Constr. (2014)       &  Fr\'{e}jus             \\
SuperNEMO\cite{Arn10}  &  $^{82}${Se} & 100 kg
& $^{enr}${Se} foils/tracking
& Proposal (2019) &  Fr\'{e}jus             \\
NEXT \cite{Gom11,Yah10}
&  $^{136}${Xe} & 100 kg
& gas TPC                                    
& Devel. (2014)     &   Canfranc            \\
SNO+\cite{che05,Chen2008,Kraus:2010zzb}
&  $^{150}${Nd} & 55 kg
& Nd loaded liq. scint.                       
& Constr. (2013)    &  SNOLab             \\
\hline
\end{tabular}\label{tab:FutureExperiments} 
\caption{A summary list of neutrinoless double-beta decay proposals and experiments.}
\end{table}%

It is important to understand that several experiments using different isotopes are in order, at each step of sensitivity.  This is because different isotopes involve different matrix elements with their uncertainties.  In addition, unknown small-probability gamma transitions may occur at or near the end point of a particular isotope, but it is very unlikely that they occur for {\it every} double-beta decay emitter.    Finally, and maybe most importantly, different isotopes generally correspond to radically different techniques, and since neutrinoless double-beta decay searches require exceedingly low backgrounds, it is virtually impossible to decide {\it a priori} which technique will truly produce a background-free measurement.
The long-term future for double-beta decay experiments will depend on what is observed: if no experiments, or only some experiments, see a signal at the 100~kg scale, then ton-scale experiments are in order.  If a signal is confirmed, the next generation of detectors may be low-energy trackers, in order to better investigate the $0\nu\beta\beta$ mechanism by separately measuring the energies of each electron as well as their angular correlations.

\section{Weighing Neutrinos}
\label{sec:mass}


The neutrino's absolute mass cannot be determined by oscillation experiments, which give information only on mass differences.
The neutrino's rest mass has a small but potentially measurable
effect on its kinematics, in particular on the phase space available
in low-energy nuclear beta decay.  The effect is indifferent to the
distinction between Majorana and Dirac masses, and hence its
observation would provide information complementary to neutrinoless
double-beta decay.  

Two nuclides are of major importance to current experiments: tritium
($^3$H or T) and $^{187}$Re.  The particle physics is the same in both
cases, but the experiments differ greatly.  Consider the superallowed decay
$^3\mathrm{H}\rightarrow \mathrm{^3He} + e^- + \bar{\nu}_e$.  The
electron energy spectrum has the form:

\begin{equation}
dN/dE \propto F(Z,E) p_e (E+m_e) (E_0 - E)\sqrt{(E_0-E)^2 - m_\nu^2}
\end{equation}

where $E,p_e$ are the electron energy and momentum, $E_0$ the Q-value,
and $F(Z,E)$ the Fermi function.   If the neutrino is massless, the
spectrum near the endpoint is approximately parabolic around $E_0$.   A finite neutrino mass makes
the parabola ``steeper'', then cuts it off $m_\nu$ before
the zero-mass endpoint.  $m_\nu$ can be extracted from the shape without knowing $E_0$ precisely, and without resolving the cutoff.

The flavor state $\nu_e$ is an admixture of three mass states
$\nu_1$, $\nu_2$, and $\nu_3$.  Beta decay yields a superposition of three
spectra, with three different endpoint shapes and cutoffs,
whose relative weights depend on the magnitude of elements of the mixing matrix.   Unless the three endpoint steps are fully resolved, the spectrum is
well approximated by the single-neutrino spectrum with an effective
mass $m_\beta^2 = \Sigma_{i}U_{ei}^2m_i^2$.  Past tritium experiments have determined $m_\beta < 2.0$ eV.

To measure this spectrum distortion, any experiment must have
the following properties:  \begin{itemize}\item High energy resolution---in particular, a resolution function
lacking high-energy tails---to isolate the near-endpoint electrons
from the more numerous low-energy electrons.  \item An extremely
well-known spectrometer resolution.  The neutrino mass
parameter covaries very strongly with the detector resolution.
\item The ability to observe
a very large number of decays, with high-acceptance
spectrometers and/or ultra-intense sources, in order to collect adequate
statistics in the extreme tail of a rapidly falling spectrum. \end{itemize}

 

The KATRIN experiment~\cite{osi2001,Robertson:2007xx}, now under construction, will attempt to extract
the neutrino mass from decays of gaseous T$_2$.  KATRIN achieves 
high energy resolution using a MAC-E (Magnetic Adiabatic
Collimation-Electrostatic) filter.  In this technique, the T$_2$
source is held at high magnetic field. Beta-decay
electrons within a broad acceptance cone are magnetically guided towards a
low-field region; the guiding is adiabatic and forces the electrons nearly parallel to $B$ field lines.  In the parallel region, an electrostatic field serves as a sharp energy filter.  Only the highest-energy electrons can pass
the filter and reach the detector, so MAC-E filters can tolerate huge
low energy decay rates without encountering detector rate problems.
In order to achieve high statistics, KATRIN
needs a very strong source, supplying $10^{11}$ e$^-/s$ to the spectrometer acceptance.   This cannot be done by increasing the source thickness, which is
limited by self-scattering, so the cross-sectional area of the
source and spectrometer must be made very large, 53 cm$^2$ and 65 m$^2$
respectively.   KATRIN anticipates achieving a neutrino mass exclusion limit
down to $0.2$ eV at 95\% confidence, or $0.35$ eV for a 3-sigma
discovery.


The MARE~\cite{Andreotti:2007eq,Nucciotti:2010tx} experiment attempts to extract the neutrino endpoint from the
endpoint of $^{187}$Re.  Rhenium's extremely low endpoint, 2.6 keV,
is seven times lower than tritium's; all else being
equal, a $^{187}$Re endpoint measurement has $7^3$ times as much statistical
power as a tritium measurement.   However, because the $^{187}$Re half-life is so
long, it is impossible to make a strong transparent source; the decay energy
is always self-absorbed.   MARE attempts to capture this energy in a
microcalorimeter with 1--3 eV energy resolution.  To amass high statistics without pileup, MARE needs a large number of individual counters.
MARE is made possible by microbolometer-array technology pioneered in the x-ray astronomy community.  MARE's arrays might include, on each of thousands of pixels: a rhenium source/absorber/calorimeter, a transition-edge sensor including readout wiring, and a weak thermal link to a cold support, all
fabricated using lithographic techniques.  A future implementation of
MARE might include 10$^5$--10$^6$ microcalorimeters and achieve
neutrino mass sensitivity comparable to KATRIN, with independent systematics.


Project 8 is a new technology for pursuing the tritium endpoint \cite{Monreal:2009za}; it is
currently running proof-of-concept experiments, but anticipates providing a
roadmap towards a large tritium experiment with new neutrino mass
sensitivity. In Project 8, a low-pressure gaseous tritium source is
stored in a magnetic bottle.  Magnetically trapped
decay electrons undergo cyclotron motion for $\sim 10^6$ orbits.  This motion emits microwave radiation at frequency $\omega = qB/\gamma m$, where $\gamma$ is the Lorentz factor.  A measurement of the frequency can be translated into an energy.  A prototype, now operating at the University of
Washington, is attempting to detect and characterize single conversion
electrons from a $^{83m}$Kr conversion electron calibration source.
If this is successful, Project 8 offers a tritium measurement strategy
with very different scaling laws and systematics than KATRIN.

Another way of addressing the question of absolute neutrino masses connects to the cosmic frontier. The field of observational cosmology now
has a wealth of data.   Global fits to the data -- large-scale
structure, high-redshift supernovae, cosmic microwave background, and Lyman
$\alpha$ forest measurements -- yield limits on the sum of the three
neutrino masses of less than about
0.3-0.6~eV, although specific results depend on assumptions.  Future
cosmological measurements will further constrain the absolute mass
scale. References~\cite{Lesgourgues:2006nd,Abazajian:2011dt,Wong:2011ip} are recent
reviews, and see also Sec.~\ref{cosmology}.

\section{Neutrino Scattering}
\label{sec:scattering}

\begin{figure}[ht]
\begin{center}
       \includegraphics[width=0.48\textwidth]{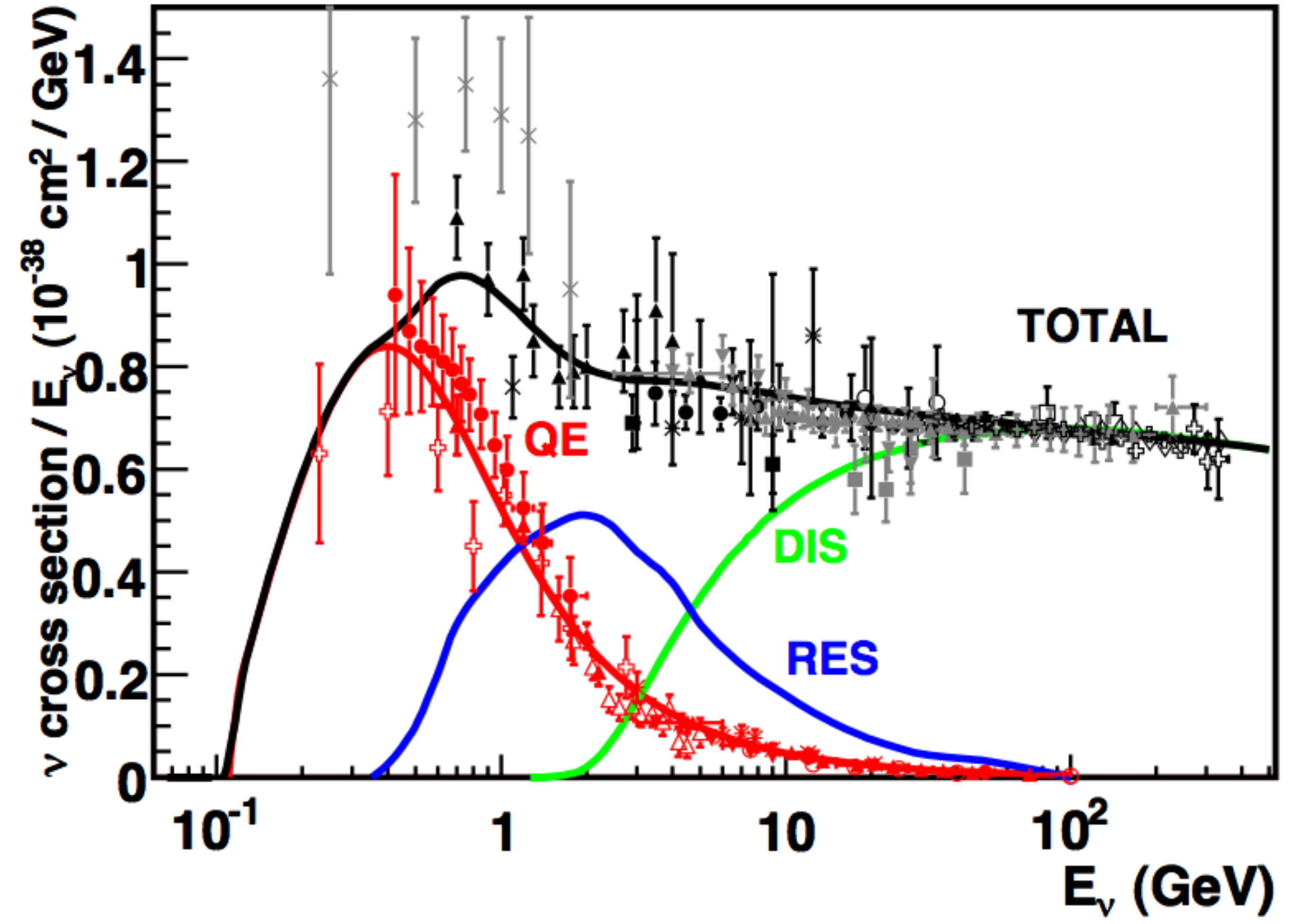}
       \includegraphics[width=0.48\textwidth]{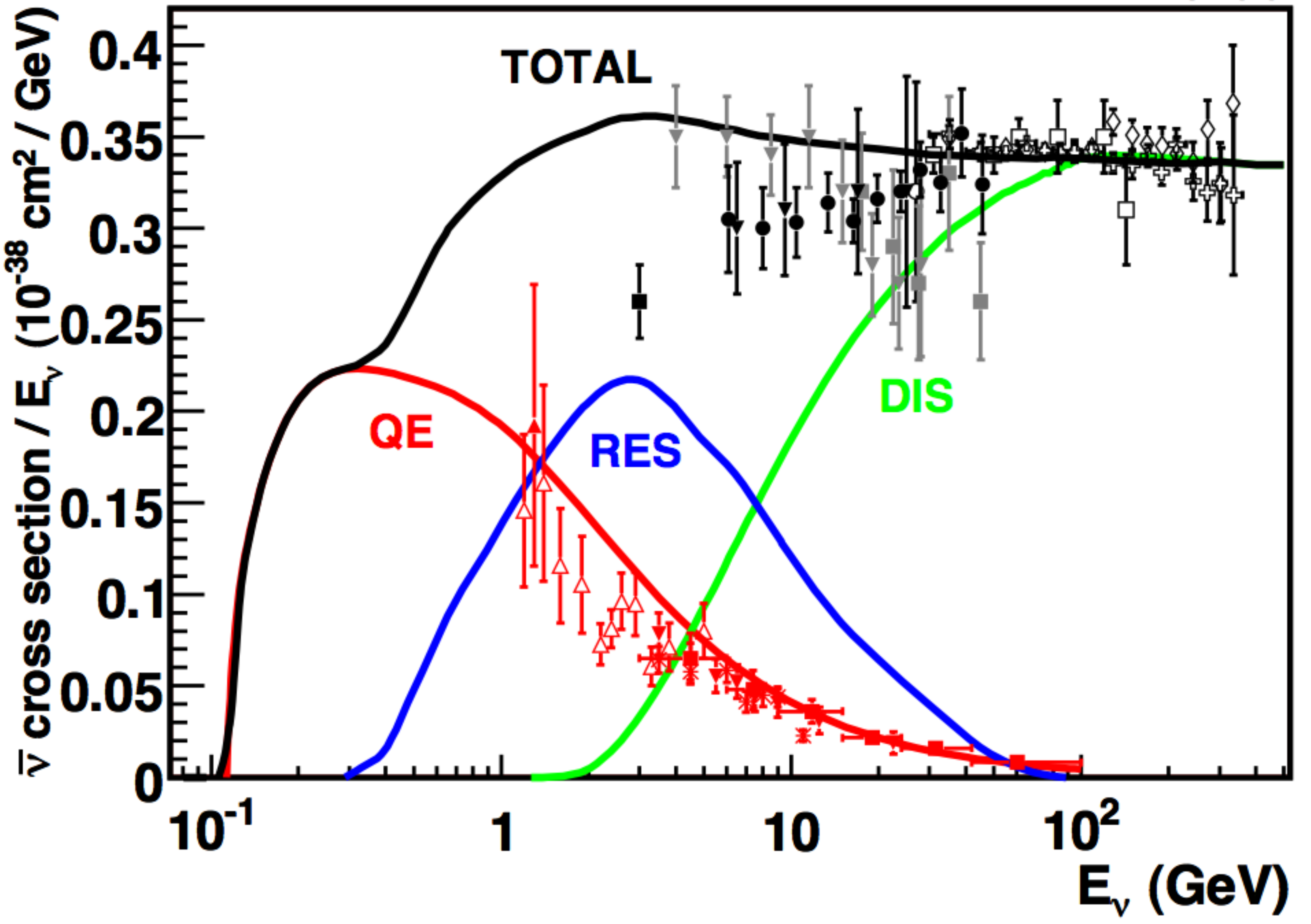}
\end{center}
\caption{Existing muon neutrino (left) and antineutrino (right) charged-current 
         cross section measurements~\cite{FormaggioZeller:2012}
         and predictions~\cite{Casper:2002sd} as a function of neutrino energy.
         The contributing processes in this energy region include 
         quasi-elastic (QE) scattering, resonance production (RES), 
         and deep inelastic scattering (DIS). The error bars in the 
         intermediate energy range reflect the uncertainties in these
         cross sections (typically $10-40\%$, depending on the 
         channel).}
\label{fig:neutrino-xsecs}
\end{figure}

While the initial discovery 
of neutrino oscillations was established using natural (solar and atmospheric)
neutrino sources, many of the high precision investigations in the future will 
be performed with artificial neutrinos. In particular, long-baseline 
accelerator neutrino beams will play a fundamental role. Such long-baseline 
experiments will rely on intense neutrino sources to reduce statistical 
uncertainties, and on very careful control of systematic errors. As such, 
these efforts will require detailed understanding of the interaction of 
few-GeV neutrinos to complete their experimental programs (the energy 
region being dictated by the baseline). One of the main sources of systematic 
uncertainty has and will continue to be poor knowledge of the underlying 
neutrino interaction cross sections. Figure~\ref{fig:neutrino-xsecs} shows 
existing measurements of charged-current neutrino cross sections in the 
relevant energy range. Such measurements form the foundation of our knowledge 
of neutrino interactions and provide the basis for simulations in present use. 

In this energy regime, neutrino interactions are a complex combination 
of quasi-elastic scattering, resonance production, and deep inelastic 
scattering processes, each of which has its own model and associated 
uncertainties. While solar and reactor experiments operating at very 
low neutrino energies (10's of MeV) and scattering experiments at very 
high energies (100's of GeV) have enjoyed very precise knowledge of their
respective neutrino interaction cross sections (at the few-$\%$ level), 
the same is not true for this intermediate energy regime. In this region, 
the cross sections are not very well known (at the $10-40\%$ level) 
and the data are in frequent conflict with theoretical predictions.

Neutrino cross section uncertainties are already becoming a limiting factor in 
the determination of neutrino oscillation parameters in many experiments. 
Understanding the underlying neutrino processes directly affects how well 
one can separate signal from background. Furthermore, experiments using 
heavier nuclear targets to increase their signal yields have to deal with 
the presence of significant nuclear effects impacting both the interaction 
cross sections and final state topologies. Such nuclear effects also impact 
one's ability to reconstruct the incoming neutrino energy, a key parameter 
in the determination of neutrino oscillation parameters. Uncertainties in 
both the neutrino interaction cross sections and associated nuclear effects 
must be understood to maximize the sensitivity of an experiment to 
neutrino oscillations. Of course, depending on the detector, the scientific 
question being asked, and the oscillation parameters, different cross section
uncertainties can have different levels of importance. For example, careful 
control of neutrino/antineutrino cross section differences will be 
particularly important in establishing $CP$ violation in the neutrino 
sector~\cite{jorge}. 
In fact, if $|U_{e3}|$ is large, such systematic uncertainties become 
even more important because the expected neutrino/antineutrino asymmetry 
becomes increasingly smaller for larger $|U_{e3}|$.

Interest in neutrino interaction physics has recently surged due, in large 
part, to the demand for accurately predicting signal and background rates 
in such neutrino oscillation searches. Despite the presence of existing 
measurements from past experiments (which were pioneering at the time), 
these data sets are decades old, were not collected on the type of targets 
relevant for modern oscillation experiments, and generally are not of the 
precision needed for neutrino oscillation physics. Taking advantage of new 
intense sources of neutrinos, modern experiments have begun to remeasure 
these neutrino interaction cross sections, most importantly on nuclear 
targets relevant to the neutrino oscillation program. This includes 
programs at the ArgoNeuT~\cite{Anderson:2011ce}, 
K2K~\cite{Mariani:2010ez,Rodriguez:2008aa,Gran:2006jn,Hasegawa:2005td,Nakayama:2004dp}, MINER$\nu$A~\cite{Harris:2011zz}, 
MiniBooNE~\cite{Grange:2011zi,Dharmapalan:2011sa,AguilarArevalo:2010bm,AguilarArevalo:2010xt,AguilarArevalo:2010cx,AguilarArevalo:2010zc,AguilarArevalo:2009ww,AguilarArevalo:2009eb,AguilarArevalo:2008xs,AguilarArevalo:2007ab},
MINOS~\cite{Mayer:2011zz,Cherdack:2011zz,Dorman:2009zz,Adamson:2009ju}, 
NOMAD~\cite{Kullenberg:2011rd,Kullenberg:2009pu,Lyubushkin:2008pe,wu:2007ab}, 
and SciBooNE~\cite{Nakajima:2010fp,Kurimoto:2010rc,Kurimoto:2009wq,Hiraide:2008eu} experiments. One of several intriguing results from these new data 
comes from recent measurements of quasi-elastic (QE) scattering. QE scattering
is a simple reaction historically thought to have a well-known cross section;
this is one reason why it is chosen as the signal channel in many neutrino 
oscillation experiments. Interestingly, the QE cross section recently measured
on carbon at low energy is about $30\%$ higher than the most widely used 
predictions~\cite{Smith:1972xh} and is even larger than the free nucleon 
scattering cross section in some energy regions~\cite{AguilarArevalo:2010zc}. 
This is surprising because nuclear effects have always been expected to 
reduce the cross section, not enhance it. A recent QE cross section 
measurement at higher energies does not exhibit such an 
enhancement~\cite{Lyubushkin:2008pe}. A possible reconciliation between 
the two classes of measurements has suggested that previously neglected 
nuclear effects could in fact significantly increase the QE cross section 
on nuclei at low energy~\cite{Martini:2009uj}. A similar enhancement has 
been observed in electron scattering~\cite{Carlson:2001mp}. Significant 
discrepancies have also been noted in non-QE neutrino data 
sets~\cite{AguilarArevalo:2010bm,AguilarArevalo:2010xt,AguilarArevalo:2009ww}. 
If true, this radically changes our thinking on nuclear effects and their 
impact on low energy neutrino interactions. This revelation has been the 
subject of intense theoretical scrutiny and experimental investigation 
over the past year (for some examples, 
see~\cite{Sobczyk:2012ms,Amaro:2011aa,Mosel:2011vf,Meucci:2011ce,Benhar:2011ef,Nieves:2011yp,Martini:2011wp,Sobczyk:2012ah,Bhattacharya:2011ah,Bodek:2011ps,Amaro:2011qb,Benhar:2011wy,Ankowski:2011dc,Nieves:2011pp,AlvarezRuso:2010ia,Benhar:2010jw,Amaro:2010sd,Juszczak:2010ve,Benhar:2010nx,Butkevich:2010cr,Martini:2010ex}.
These recent discoveries emphasize
that neutrino interactions on nuclei are quite complex, especially in the 
energy regime where we are conducting our neutrino oscillation measurements. 
Modern data are uncovering new and unexpected phenomena, but more data are 
surely needed to understand them.

In the near future, significant advances in our understanding of neutrino 
interactions are expected as new data arrive and are analyzed. New information will include
neutrino and antineutrino data expected on a variety of nuclear targets and
across a broad energy range from MINER$\nu$A, MicroBooNE, and the T2K and 
NO$\nu$A near detectors in the upcoming years. Beyond this, the proposed LBNE 
near detector complex will be add to this wealth of knowledge, as 
well as supply the constraints needed for the long-baseline oscillation 
program in that experiment. Given the increasing importance of neutrino 
interaction physics and some surprising discrepancies being unearthed 
by recent experimental results, people have already started to think beyond 
this planned program and how we might add to it in both the short and long 
term. Such ideas include instrumenting the existing MINER$\nu$A detector with 
either hydrogen or deuterium targets~\cite{minerva-d2}, construction of a 
fine-grained detector in the NO$\nu$A narrow-band beam 
(SciNO$\nu$A)~\cite{scinova}, design of a very low energy neutrino factory 
(VLENF) to provide precision $\nu_e$ and $\overline{\nu}_e$ cross section 
measurements~\cite{vlenf}, and installation of a magnetized, high resolution 
neutrino detector (HiResM$\nu$) to explore a host of neutrino 
interaction physics in the Project X neutrino beams \cite{hiresmnu}.

We note also that for neutrino interactions for which cross sections are very well predicted in the 
context of the Standard Model, precision measurements provide a means of determining Standard Model
parameters such as the weak mixing angle, and enable searches for new physics.   Included in this
category are neutrino-electron elastic scattering and neutrino-nucleus coherent elastic scattering~\cite{orrell}.  
Stopped-pion sources are particularly promising for this kind of measurement.  Large detectors
are needed for $\nu$-e scattering experiments, for which cross section is quite small; however,
for very high-rate neutrino-nucleus coherent  scattering,
relatively small but very low-threshold dark matter WIMP-style detectors are suitable. (\textit{e.g.}~\cite{Scholberg:2009ha,Formaggio:2011jt}). 
A recent proposal is to 
use the far off-axis Fermilab Booster neutrino beam as a source of neutrinos for this measurement~\cite{yoo}.

\section{Beyond the Standard Paradigm -- Anomalies and New Physics}
\label{sec:sbl}



Data from a variety of short-baseline experiments as well as astrophysical 
observations and cosmology hint at the existence of additional neutrino mass 
states beyond the three active species in the Standard Model. The possible 
implications of additional sterile neutrino states would be profound, and would change 
the paradigm of the Standard Model of particle physics. As a result, great 
interest has developed in testing the hypothesis of sterile neutrinos and 
providing a definitive resolution to the question: do sterile neutrinos 
exist? 

Recently, a number of tantalizing results (anomalies) have emerged from 
short-baseline neutrino experiments that cannot be explained by the current 
three-neutrino paradigm. These anomalies are not directly ruled out by other 
experiments and include the excess of electron-antineutrino events 
($3.8\sigma$) observed by the LSND experiment~\cite{Aguilar:2001ty}, 
the excess of electron-neutrino events ($3.0\sigma$) observed by the 
MiniBooNE experiment in neutrino mode~\cite{AguilarArevalo:2008rc}, 
the excess of electron-antineutrino events ($2.3\sigma$) observed by 
the MiniBooNE experiment in antineutrino mode~\cite{AguilarArevalo:2010wv}, 
the deficit of electron-antineutrino events ($0.937\pm0.027$) observed 
by reactor neutrino experiments~\cite{Mention:2011rk}, and the deficit 
of electron-neutrino events ($0.86\pm0.05$) observed by the SAGE and 
GALLEX gallium calibration experiments (see \cite{Giunti:2010zu}). 

How can we explain these anomalies? Although there are several possibilities 
(e.g., Lorentz invariance violation), one of the simplest explanations 
is the $3+N$ sterile neutrino model, in which there are three light, mostly 
active neutrinos and $N$ heavy, mostly sterile neutrinos. For $N>1$, 
these models allow for $CP$ violation in short-baseline experiments. 
These $3+N$ models fit the world's neutrino and antineutrino data 
fairly well~\cite{Giunti:2011cp,Kopp:2011qd}, albeit it in a  
not-too-convincing fashion.  One key test of these $3+N$ models is the 
existence of muon-neutrino disappearance ($\sin^22\theta>10\%$) 
at a $\Delta m^2 \sim 1$ eV$^2$. Several  workshops have been held 
over the past year to critically review the evidence for and against 
sterile neutrinos and the need to pursue new experiments and strategies 
to address the experimental observations~\cite{sbl-workshop,snac-workshop}.

In order to determine whether these short-baseline anomalies are due to neutrino
oscillations in a $3+N$ sterile neutrino model and not to some other process or 
background, future short-baseline experiments with good electron and muon 
identification will need to measure (with precision) the $L/E$ dependence of 
neutrino appearance and disappearance at $L/E$ values of order 1. Various ways of 
measuring the $L/E$ dependence have been proposed. These include: (1) positioning
two or more detectors at different distances in an accelerator-induced neutrino
beam to reduce systematic errors, (2) placing a large detector close 
to a source of low-energy neutrinos from a reactor or intense radioactive 
source and measuring the $L/E$ distribution of neutrino events in the single 
detector, and (3) measuring the $L/E$ distribution of high energy (TeV) 
atmospheric-induced neutrinos, where strong matter effects are expected at 
particular values of $L/E$.

\begin{figure}[ht]
\begin{center}
     \includegraphics[width=0.6\textwidth]{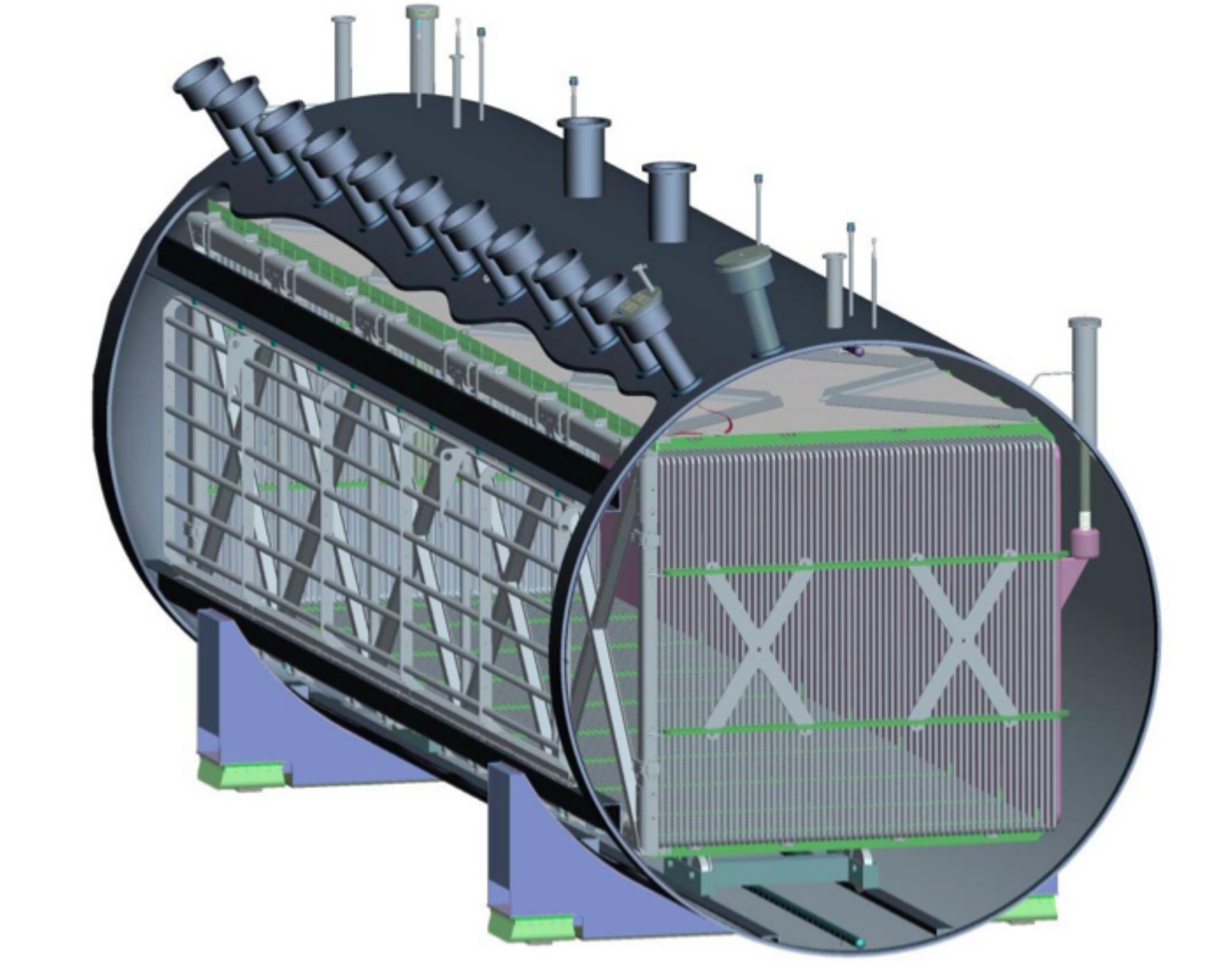}
\end{center}
\caption{\footnotesize Cutaway view of the MicroBooNE detector. }
\label{fig:microboone}
\end{figure}

Diverse experiments, spanning vastly different energy scales, have 
been proposed or are being built to test the $3+N$ models and resolve the present 
anomalies. The MicroBooNE experiment is building a liquid-argon (LAr) TPC just 
upstream of the MiniBooNE detector that will be able to determine whether the event 
excesses observed by MiniBooNE are due to electron events, as expected from $3+N$ 
models, or are simply due to unmodeled photon backgrounds (see Fig.~\ref{fig:microboone}). 
Another LAr TPC proposal is to move the ICARUS detector, now taking data in the 
Gran Sasso National Laboratory, to the PS neutrino beamline at CERN and to build a second, 
smaller LAr TPC~\cite{Calligarich:2010zz}. Similar options also exist in the
Booster neutrino beamline at Fermilab. With two detectors at different distances, 
many of the associated systematic errors cancel, which will allow a definitive 
test of the LSND neutrino oscillation signal. Other accelerator neutrino experiments 
at Fermilab include the MINOS+ experiment~\cite{Tzanankos:2011zz}, which will 
search with high sensitivity for muon neutrino to sterile disappearance, and the BooNE 
experiment~\cite{Stancu:2009zz}, which proposes the construction of a second 
MiniBooNE-like detector at a different distance (200m) than the original 
MiniBooNE detector (541m). BooNE would have the potential to measure electron 
neutrino and electron antineutrino appearance, muon neutrino and muon antineutrino 
disappearance, and $CP$ violation in the lepton sector, as well as demonstrate the 
existence of sterile neutrinos by comparing neutral current $\pi^0$ scattering 
at different distances. Fermilab already has world-class neutrino beams (the 
Booster neutrino beamline and NuMI); however, future facilities could significantly 
enhance these capabilities. These future facilities include Project-X, which would 
increase present proton intensities by an order of magnitude or more, and a muon 
storage ring, which would enable an extremely precise search for electron neutrino 
and electron antineutrino disappearance. 

Besides those at Fermilab and CERN, there are also several other opportunities for pursuing 
short-baseline neutrino physics. The Spallation Neutron Source (SNS) facility at 
Oak Ridge National Laboratory produces an intense and well-understood flux of neutrinos from $\pi^+$ and 
$\mu^+$ decay at rest.  An idea has been put forward, OscSNS~\cite{oscsns}, for 
building a MiniBooNE-like detector approximately 60m from the SNS beam dump. 
OscSNS would be capable of making precision measurements of electron antineutrino 
appearance and muon neutrino disappearance. Also, the Southern California Reactor 
Antineutrino Anomaly Monitor (SCRAAM)~\cite{scraam} experiment could be built at 
the San Onofre nuclear generating station in California or at the Advanced Test 
Reactor, a research reactor at the Idaho National Laboratory. SCRAAM would have 
less baseline spread than previous reactor neutrino experiments and would be able 
to measure oscillations by looking for a spectral distortion in the reactor neutrino 
energy spectrum. In addition, neutrino radioactive source experiments could be 
mounted in either the Borexino, Daya Bay, KamLAND, or SNO+ 
detectors~\cite{Cribier:2011fv,Dwyer:2011xs}. The advantage of radioactive source 
experiments is that due to the low neutrino energies, oscillations could be 
observed in a single detector or in several closely separated detectors. There
are also possibilities for performing sterile neutrino measurements in neutral
current coherent neutrino-nucleon scattering using cryogenic solid state 
bolometers~\cite{Formaggio:2011jt}. A final 
opportunity for measuring short-baseline oscillations is to search for atmospheric 
muon antineutrino disappearance with the IceCube experiment at the South Pole~\cite{Barger:2011rc}. 
With a typical atmospheric neutrino energy of a few TeV and a typical distance 
of a few thousand kilometer, IceCube is very sensitive to oscillations at the roughly 
1 eV mass scale, especially because these oscillations would be matter-enhanced 
via the MSW mechanism.

Finally, we emphasize that satisfactorily resolving these short-baseline anomalies 
is very important for carrying out the neutrino oscillation program described earlier. 
The two to three sigma effects reported, even if unrelated to sterile neutrinos, 
are at the sub-percent to the several-percent level, similar to, for example, the 
$|U_{e3}|$ and $CP$-violating signals being pursued in long-baseline experiments. 


Other than new light neutrino degrees of freedom -- sterile neutrinos -- neutrino experiments 
are sensitive to several other manifestations of new physics. For example, many 
proposals for new physics beyond the Standard Model predict novel, weakly interacting, 
light scalar or vector particles. Classical examples of such particles include 
Majorons, axions, Kaluza-Klein modes in the Randall-Sundrum scenarios with extra 
dimensions, and many others. As discussed over the years, novel light particles 
could be responsible, among other things, for solving the strong $CP$ problem in QCD, 
giving neutrinos their mass, or even explaining the origin of dark energy. These 
new particles can be produced by proton bremsstrahlung and detected, assuming 
they are long-lived, by particle decays or scatters in the center of neutrino 
detectors, if the proton beam is on-axis.  

Neutrino experiments in general, and neutrino oscillation experiments in particular, are also very sensitive to new, heavy degrees of freedom that mediate new ``weaker-than-weak'' neutral current interactions. These so-called non-standard interactions (NSI) between neutrinos and charged fermions modify not only neutrino production and detection, but also neutrino propagation through matter effects. In a little more detail, NSI are described by effective operators proportional to, for example, $G_F\epsilon^f_{\alpha\beta}\nu_{\alpha}\gamma_{\mu}\nu_{\beta}\bar{f}\gamma^{\mu}f$, where $\nu_{\alpha,\beta}=\nu_{e,\mu,\tau}$, $f$ are charged fermions ($e, u, d, \mu, s$, \ldots), $G_F$ is the Fermi constant, and $\epsilon$ are dimensionless couplings.\footnote{$\epsilon\sim 1$ ($\ll 1$) implies that the new physics effects are on the order of (much weaker than) those of the weak interactions.} When $f$ is a first-generation fermion, the NSI contribute to neutrino detection and production at order $\epsilon^2$ (ignoring potential interference effects between the Standard Model and the NSI). On the other hand, the NSI also contribute to the forward-scattering amplitude for neutrinos propagating in matter, modifying the neutrino dispersion relation and hence its oscillation length and mixing parameters. These modified matter effects are of order $\epsilon^1$ and potentially more important than the NSI effects at production or detection. Furthermore, for $\alpha\neq\beta$, the NSI-related matter effects lead to $P_{\alpha\beta}\neq\delta_{\alpha\beta}$ in the very short baseline limit ($L\to 0$), which are not present in the Standard Model case. More information -- including relations to charged-lepton processes -- current bounds, and prospects are discussed in detail in, for example, \cite{Gavela:2008ra,Biggio:2009nt}, and references therein. 

Very recently, the OPERA Collaboration observed that $\nu_\mu$ produced at CERN and detected in Gran Sasso arrive at the detector before expected~\cite{Adam:2011zb}. Interpreted as an anomalous velocity, the neutrinos travel faster than $c$ with fractional difference $(v-c)/c = (2.37 \pm 0.32 {\rm (stat.)}^{+0.34}_{-0.24} {\rm(sys.)} ) \times 10^{-5}$. This result has caused much discussion and needs to be tested, although recent reports indicate some problems with the measurement.  Similar long-baseline beam experiments will repeat the measurement, and similar measurements can be done by future long-baseline experiments discussed in this document.
If the result holds up, independent experiments using different methods will be critical; for example, an experiment to measure neutrinos in coincidence with their ``siblings'' from the same parent mesons could be done at Fermilab~\cite{Schmitt}.
Such experiments may be part of the intensity frontier.

\section{Synergy with Other Fundamental Physics Efforts}\label{sec:synergy}

The study of neutrino properties and interactions is truly interdisciplinary. For example, neutrino scattering provides information to, and requires input from, nuclear physics, low-energy strong interactions, and perturbative QCD, as discussed in Sec.~\ref{sec:scattering}. The very large detectors required for detecting beam neutrinos are also ideal for searching for nucleon decay, and the very intense proton sources required to produce neutrino beams are also necessary when it comes to searching for rare muon or kaon properties. The measurement of astrophysical neutrinos reveals a great deal about their sources, and different neutrino properties shape the observable universe in different ways. Conversely, we may be able to infer elusive neutrino properties by observing astrophysical neutrinos or precisely understanding the energy budget and distribution of the universe.

Theoretically, one of the key unanswered questions is the nature of the mechanism behind neutrino masses and lepton mixing. Regardless of the mechanism, one generically expects new degrees of freedom. We are not sure what those might be but, depending on the energy scale of this new physics, we expect to produce and detect new particles and phenomena through a variety of fundamental physics observables.

\subsection{Neutrinos in Cosmology and Astrophysics}

Neutrinos are copiously produced in astrophysical objects, including the Earth, the sun, and supernova explosions. Neutrinos are also predicted to be relics from the big bang. Finally, the physics behind neutrino masses may provide the answer to one of the most ambitious fundamental physics questions we are allowed to ask: why is there so much matter in the universe?

\subsubsection{Geo- and solar neutrinos}


The importance of solar neutrinos for understanding neutrino properties has been emphasized in Sec.~\ref{sec:3nus}. One of the original motivations for studying solar neutrinos, however, was to understand how the sun works. Neutrinos are produced deep inside the sun and exit effortlessly, carrying information that is not accessible through the solar photons. Neutrinos are still revealing details concerning the inner workings of the sun, and future measurements of the $pep$ and CNO neutrinos appear very promising \cite{Bahcall:2003ce,Serenelli:2011py}. For example, the Borexino Collaboration recently announced the first positive measurement of $pep$ neutrinos \cite{Collaboration:2011ng}, along with a nontrivial upper bound on neutrinos from the CNO cycle, which are yet to be observed.  Real-time $pp$ neutrino observation is still a future goal, which may be within reach with next-generation, very low energy threshold detectors.

Closer to home, the Earth is also a potent source of low-energy antineutrinos and neutrinos, thought to be mostly produced in the decay of uranium, thorium and potassium in the crust. Geoneutrinos were first observed in KamLAND \cite{Araki:2005qa} and Borexino \cite{Bellini:2010hy}, experiments designed to study reactor antineutrinos and solar neutrinos. More detailed studies of geoneutrinos, which can be made, for example, with next-generation liquid scintillator detectors, are expected to shine more light on the Earth as a heat source and on the Earth's composition (see, for example, \cite{Fiorentini:2007te}).

The center of the Earth, and especially that of the sun, is expected to act as a gravitational attractor for dark matter. The self-annihilation of dark matter inside these astrophysical bodies may yield high energy ($E\gtrsim 1$~GeV) neutrinos. Those, in turn, might be seen in large neutrino detectors, including IceCube (especially with the DeepCore ``infill'')~\cite{rott}. The absence of a flux of high energy neutrinos from the sun provides the most stringent constraint on a class of dark matter models. In the future, the search for high energy solar and terrestrial neutrinos is expected to be among the most promising dark matter search channels \cite{Bertone:2004pz}.


\subsubsection{Neutrinos from Core-Collapse Supernovae}

Approximately 99\% of the energy released in the explosion of a core-collapse supernova is emitted in the form of neutrinos. While these events are somewhat rare in our corner of the universe, the large neutrino detectors of the next generation can operate for decades.
On this time scale, there is a significant likelihood of a core-collapse supernova exploding in our galaxy. 

Compared to the 1987A event, when only two dozen neutrinos were observed, future detectors may register tens -- or even hundreds -- of thousands of neutrino interactions.  Furthermore,
flavor sensitivity -- not only interaction rate but the ability to tag different interaction channels-- 
is critical for maximizing the science harvest from a burst observation.  Current-generation large detectors made of water and scintillator are primarily sensitive to $\bar{\nu}_e$; however, next-generation detectors will expand worldwide flavor sensitivity.  For example, liquid argon detectors are primarily sensitive to $\nu_e$~\cite{Scholberg:2007nu,Akiri:2011dv}.
If future Mton-scale detectors are built, prospects are excellent for a vast yield of information from a nearby burst, and we even hope for a reach extending well beyond the Milky Way~\cite{Ando:2005ka}.
With such tremendous rates, it will be possible to precisely measure not only the time-integrated spectra, but also their second-by-second evolution. 

We quickly summarize what physics might be gleaned from this data. The first item is the mechanism of the explosion, which has been an unsolved issue in astrophysics for more than half a century.
Supernova neutrinos record the information about the physical processes in the center of the explosion during the first several seconds, as it happens.
Next, the neutrino oscillation physics in a supernova is incredibly rich. As neutrinos stream out of the collapse core, their number densities are so large that their flavor states become coupled due to the mutual coherent scattering. This âself-MSWâ phenomenon results in non-linear, many-body flavor evolution and has been under active exploration for the last five years, as supercomputers caught up with the physics demands of the problem (see, for example~\cite{Duan:2005cp,Fogli:2007bk,Raffelt:2007cb,Raffelt:2007xt,EstebanPretel:2008ni,Duan:2009cd,Dasgupta:2009mg,Duan:2010bg,Duan:2010bf}.)
 While the full picture is yet to be established, it is already clear that the spectra of neutrinos arriving on Earth will have spectacular nonthermal features. Neutrino flavor evolution is also affected by the moving front shock and by stochastic density fluctuations behind it, which may also imprint unique signatures on the signal. All of this will give new large detectors a chance to observe neutrino oscillations in qualitatively new regimes, inaccessible on Earth, and will very likely yield information on the neutrino mass hierarchy.
Last but not least, the future data will allow us to place significant constraints on many extensions of particle physics beyond the Standard Model. This includes scenarios with weakly interacting particles, such as axions, Majorons, Kaluza-Klein gravitons, and others (see, for example~\cite{Raffelt:1999tx,Hannestad:2001jv}). These new particles could be produced in the extreme conditions in the core of the star and could modify how it evolves and cools.
The problem thus is very rich and truly multidisciplinary, with neutrino physics and astrophysics going hand in hand. 


Looking even farther out for sources of neutrinos, one can imagine measuring the flux of neutrinos from all the supernovae in cosmic history.  This ``diffuse supernova neutrino background'' (DSNB) is sometimes referred to as the ``relic'' supernova neutrino flux.  The physics of the DSNB is reviewed in~\cite{Beacom:2010kk,Lunardini:2010ab}.  The DSNB flux depends on the historical rate of core collapse, average neutrino production, cosmological redshift effects and neutrino oscillation effects.  For neutrino energies above about 19~MeV, estimates for the $\bar{\nu}_e$ component of the DSNB range from about 0.1 to 1 cm$^{-2}$s$^{-1}$.  The detection interactions remain the same as for burst neutrinos; however, the experimental issues become entirely background-dominated, as there is no external trigger and events will be measured singly.  Large gadolinium-doped water and scintillator detectors are especially promising for measuring this flux.

\subsubsection{Ultra-High Energy Neutrinos from Cosmic Sources}\label{cosmicnus}

The topic of ultra-high energy cosmic neutrinos is also at the intersection  of the intensity and cosmic frontiers.  Thanks to their weak interactions, neutrinos are able to traverse very long distances free of impediment, unlike photons or charged particles.  They bring us unique information about exotic astrophysical objects such as active galactic nuclei and gamma ray bursters, and in addition test neutrino oscillations and fundamental physics.  Because the flux of neutrinos from such cosmic sources is expected to be very small, detectors must be extremely large. The current detectors searching for them are the cubic kilometer scale IceCube and ANTARES; upgraded versions of these are also planned.  These detectors are also sensitive to atmospheric and supernova neutrinos as discussed elsewhere in this document.
The physics and astrophysics reach of such searches is reviewed in~\cite{Gaisser:1994yf,Halzen:2000km,Anchordoqui:2009nf}.


\subsubsection{Neutrinos and the Early Universe}\label{cosmology}


The Concordance Cosmological Model predicts the existence of a relic neutrino background, currently somewhat colder than the cosmic microwave background, $T_{\nu}=1.95$~K. While relic neutrinos have never been directly observed, their presence is corroborated by several cosmological observables that are sensitive to the amount of radiation in the universe at different epochs. For example, precision measurements of the cosmic microwave background and measurements of the relic abundances of light elements independently require relativistic degrees of freedom other than photons, and are safely consistent with the three known light neutrino species; see, for example, \cite{Hamann:2011ge}. It is curious that the current data are also consistent with four light neutrinos. Future cosmological data are expected to provide precision measurements of the number of relativistic species, and have the potential to not only probe the Concordance Cosmological Model but also help reveal whether there are more very light degrees of freedom (such as more neutrinos).

Neutrino properties directly impact the dynamics of the relic neutrinos, which in turn impact cosmological observables. Non-zero neutrino masses, for example, modify the dynamics of structure formation in the early universe. In a nutshell, massive neutrinos constitute a small part of the dark matter. However, given what is known about neutrino masses, neutrinos are relativistic at the time of decoupling -- neutrinos are hot dark matter -- and their presence dampens the formation of structure at small distance scales. The heavier the neutrinos, the more they influence structure formation, and the less structure is expected at small scales. They are consistent with 100\% cold dark matter and allow one to place an upper bound on the neutrino masses.  Current data constrain, assuming the Concordance Cosmological Model, the sum of the neutrino masses to be less than around 0.5~eV. In the future, it is expected that cosmological observables will be sensitive to the sum of neutrino masses if it is larger than 0.05~eV, perhaps smaller \cite{Abazajian:2011dt}. It is worthwhile pointing out that the current neutrino oscillation data require $\sum m_i>0.05$~eV, meaning that, assuming the Concordance Cosmological Model, next-generation cosmological observations are expected to observe effects of the relic neutrino masses.

Deviations from the Concordance Cosmological Model or new physics beyond the Standard Model of fundamental particles can dramatically modify the 
relationship between cosmological observables and neutrino properties. The extraction of neutrino properties from cosmological observables is, in some sense, complementary to that from terrestrial experiments. By comparing the results from these two classes of experimental efforts, we can not only determine properties of the massive neutrinos, including exotic ones, but also hope to test and, perhaps, move beyond the Concordance Cosmological Model.

The ``holy grail'' of neutrino astrophysics/cosmology is the direct detection of the relic neutrino background. This is extremely cold ($1.95~{\rm K}=1.7\times 10^{-5}$~eV) and today, at least two of the neutrino species are nonrelativistic. Several ideas have been pursued, and a clear path towards successfully measuring relic neutrinos is yet to emerge. Recently, the idea, first discussed in \cite{Weinberg:1962zz}, of detecting relic neutrinos through threshold-less inverse-beta decay -- e.g., $\nu_e+^3{\rm\hspace*{-1mm} H}\to ^3{\rm\hspace*{-1mm} He}+e^-$ -- has received some attention \cite{Cocco:2007za}. In a nutshell, the $\beta$-rays produced by the relic neutrino capture have energies above the end point of the $\beta$-rays produced by the ordinary nuclear decay. The expected number of interactions turns out to be accessible for intense enough nuclear samples, such that with enough energy resolution, one can aim at directly determining the existence of the relic neutrinos.


\subsubsection{Leptogenesis}

The right-handed neutrinos in the Type I seesaw mechanism are Majorana fermions and hence do not have a well-defined lepton number. This means that right-handed neutrinos, assuming these are heavy enough, can decay into final states with positive lepton number and final states with negative lepton number. If these decays are $CP$-violating, right-handed neutrino decays in the early universe can, under the right conditions, generate a lepton-number asymmetry. As the universe cools, part of this lepton asymmetry gets converted into a baryon asymmetry, which translates into a universe with more matter than antimatter and hence a large matter density -- like the universe we find ourselves in. This process of generating a baryon asymmetry through a lepton asymmetry is referred to as leptogenesis \cite{Fukugita:1986hr}.\footnote{Similar mechanisms can also be realized in several other neutrino mass models, including the Type II and III seesaws.} In straightforward versions of leptogenesis, these heavy neutrinos must have masses of $10^9$ GeV or more, so that, at least for a long time to  come, we will not be able to confirm their existence directly by producing them at an energy frontier collider. Instead, the hypothesis of leptogenesis must be explored indirectly through intensity frontier experiments with the light neutrinos $\nu$ related to the heavy neutrinos $N$ through the seesaw mechanism.

A major motivation to look for $CP$ violation in neutrino oscillations is that  its observation would make it more plausible that the baryon-antibaryon asymmetry of the universe arose, at least in part, through leptogenesis.
To be sure, it can be shown that if all the heavy neutrino masses exceed $10^{12}$ GeV, then the phases that drive leptogenesis are independent of  those in the neutrino mixing matrix \cite{Casas:2001sr,Abada:2006fw,Nardi:2006fx,Abada:2006ea}. However, there is no need  for the heavy neutrino masses to be this large. Indeed, supersymmetry
suggests that the mass of the lightest $N$ must be $\sim10^9$~GeV \cite{Kohri:2005wn}. It has been shown that when the smallest $N$ mass is below  $10^{12}$ GeV, $CP$-violating phases in the neutrino mixing matrix, which produce $CP$ violation in light-neutrino oscillation and influence the rate for neutrinoless double-beta decay, also lead
to a baryon-antibaryon asymmetry \cite{Pascoli:2006ie}, barring accidental cancellations. Assuming the seesaw picture, leptogenesis and light-neutrino $CP$ violation generically do imply each other.


\subsection{Neutrinos and the Energy Frontier}

The new degrees of freedom associated with non-zero but small neutrino masses can be directly produced and detected in fundamental physics experiments as long as $M_{\rm new}$ is not exceedingly large. In the Type I seesaw, for example, the right-handed neutrinos can be produced and detected in neutrino oscillation experiments if $M_{\rm new}\lesssim 10$~eV (see, for example, \cite{deGouvea:2011zz}). They look like sterile neutrinos. If $M_{\rm new}\gg 10$~eV, these may still manifest themselves in medium ($M_{\rm new}\lesssim 1$~GeV) and high energy experiments ($M_{\rm new}\lesssim 1000$~GeV). We note, however, that, realistically, any signal of a heavyish $N$ would indicate a more subtle manifestation of the Type I seesaw \cite{deGouvea:2006gz,neu:deGouvea:2007uz}, for the following reason. A generic prediction of the Type I seesaw is that the induced mixing  between active and sterile states, which governs right-handed neutrino production, is $U_{N\ell}^2 \sim m_\nu/M_N$ and hence unobservably small for $M_N\gg 100$~keV.


\newcommand{\pslash}{\mbox{${p\!\!\!/ }$}}

In the case of the Type II and III versions of the seesaw mechanism, the situation is markedly different. We comment on those in more detail. Most important, here the new degrees of freedom -- extra Higgs scalars or extra charged fermions -- can be produced via electroweak interactions at, for example, the LHC, as long as the new states are not too heavy. Furthermore, by studying the production and decay properties of the new states one can both verify that they are connected with the neutrino mass generating mechanism and, perhaps, measure properties of the neutrino oscillation parameters, like the neutrino mass hierarchy.

\begin{figure}[ht]
\includegraphics[scale=1,width=7.4cm]{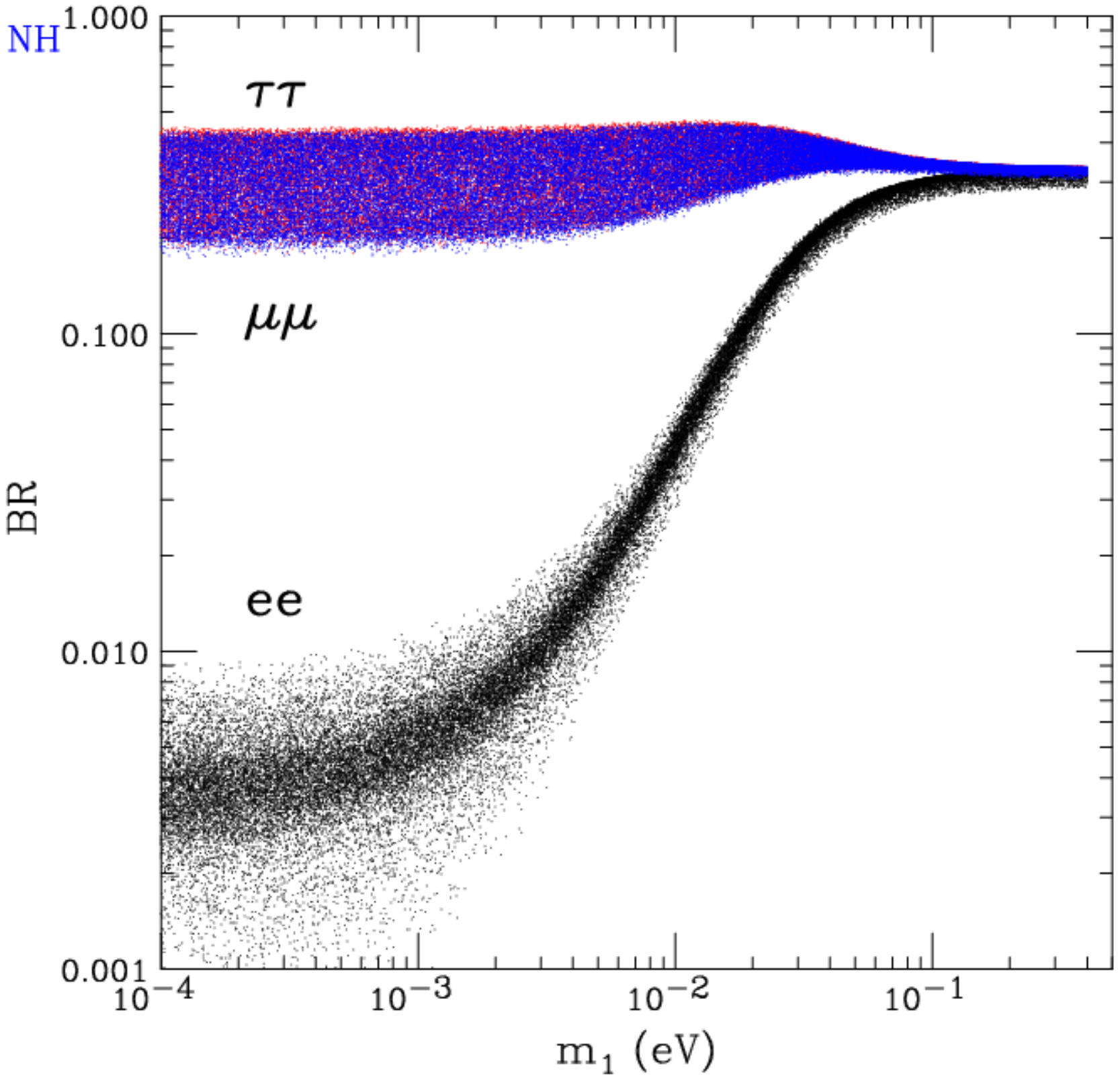}
\includegraphics[scale=1,width=7.4cm]{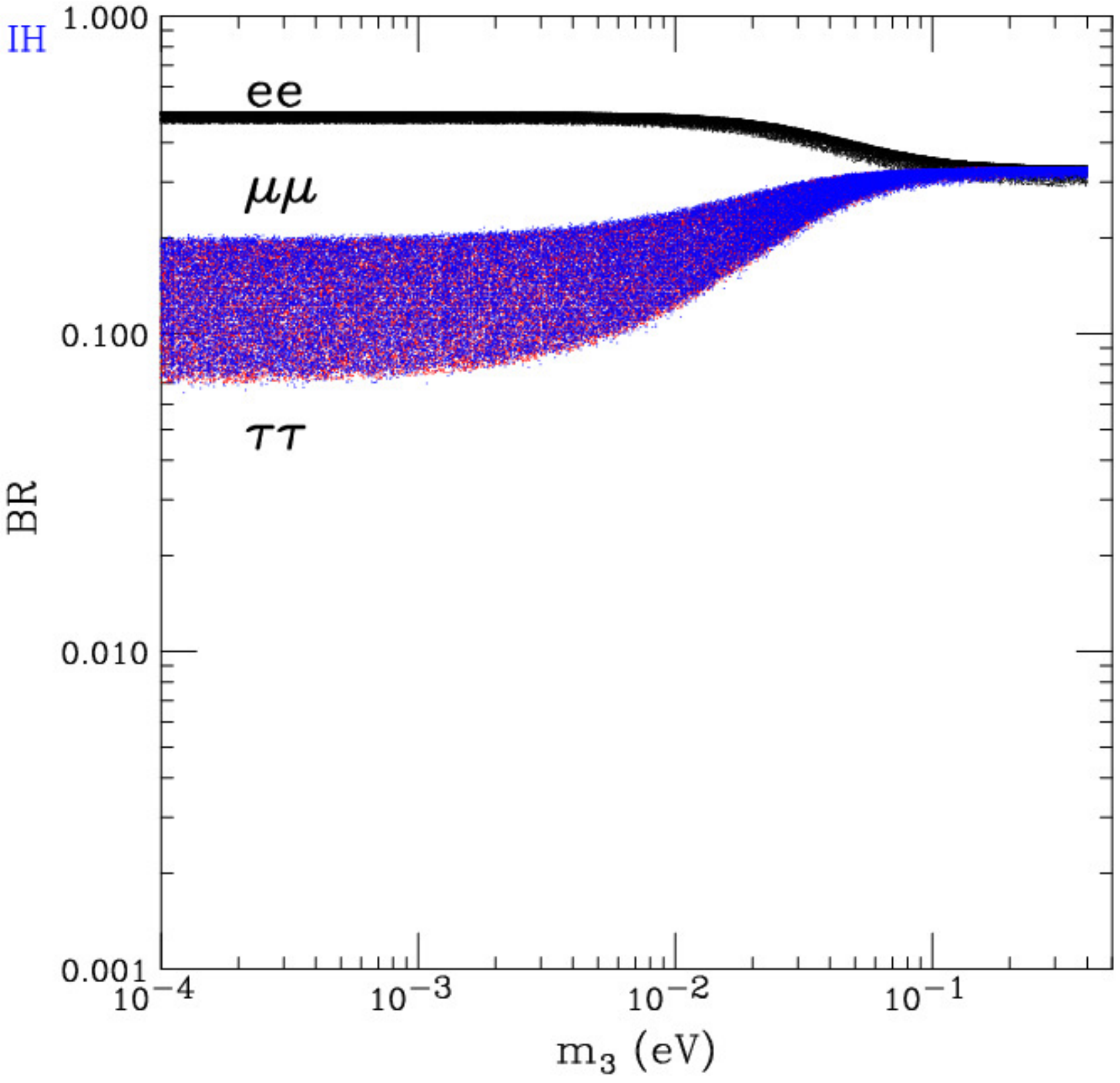}
\caption{$H^{\pm\pm}$ decay branching fractions as a function of the lightest neutrino mass, assuming a normal (left) or inverted (right) neutrino mass hierarchy. See \cite{Perez:2008ha} for details.}
\label{HBRs}
\end{figure}

Several earlier studies for certain aspects of the Type II seesaw model
at the LHC exist \cite{Chun:2003ej,Han:2005nk,Hektor:2007uu,Han:2007bk,Kadastik:2007yd,Akeroyd:2007zv}. The couplings of the new states in the Higgs sector to the Standard Model fermions are directly proportional to the neutrino masses and the lepton mixing angles. According to the current understanding of neutrino oscillation parameters, Fig.~\ref{HBRs} depicts the different branching ratios of the $H^{\pm\pm}$ states into leptonic final states  \cite{Perez:2008ha}.
The synergy is apparent: by identifying the flavor structure of the lepton number-violating decays of the charged Higgs bosons at the LHC, one can establish the neutrino mass pattern -- normal mass hierarchy, inverted mass hierarchy, or a quasi-degenerate neutrino mass spectrum. 

In the case of the Type III seesaw, the components of the lepton triplet can be produced via electroweak interactions at the LHC. Furthermore, similar to the Type II seesaw model, some versions of the Type III seesaw model (e.g., \cite{Bajc:2006ia}) also predict a clear correlation \cite{Arhrib:2009mz}  between the neutrino mass matrix and the couplings of the new heavy lepton triplet $T^{\pm}$ and $T^{0}$,
\bea
m_\nu^{ij} \sim -v^2 \frac{y_T^iy_T^j}{M_T},\quad BR(T^{\pm,0} \to W^{\pm}\ell,\ Z\ell) \sim y^2_T \sim U^2\
{M_T m_\nu\over v^2},
\eea
where $U$ is the familiar neutrino mixing matrix. 
The collider search for the production of, say, $T^+ + T^0 \to \ell^+Z(h) + \ell^+ W^- \to \ell^+ jj (b\bar b) + \ell^+ jj $ directly probes, as long as $M_T$ is within reach, the underlying neutrino mass generation mechanism via the lepton flavor correlations governed by the $\nu$ mass pattern. 

\subsection{Neutrinos and the ``Other'' Intensity Frontier}


Our understanding of neutrinos stands to increase significantly through the efforts of intensity frontier research outside of those discussed in this chapter. Moreover, as already mentioned in passing, several intensity frontier efforts share facility and detector needs with neutrino physics efforts. The most transparent examples include searches for nucleon decay, which also require very massive detectors capable of detecting {$\cal O$}(1~GeV) muons and electrons, and searches for rare meson and muon processes, which also require very intense proton sources. There is also strong experimental synergy between neutrinoless double-beta decay and dark matter direct-search detectors.

\subsubsection{Charged Lepton Properties}

Experiments related to determining the properties and interactions of charged leptons, including precision measurements of the muon magnetic dipole moment, searches for a non-zero muon or electron electric dipole moment, and searches for rare processes that violate charged-lepton flavor ($\mu\to e$ conversion in nuclei, $\ell\to\ell' \gamma$, {\it etc.}), are potentially very sensitive to the physics responsible for neutrino masses. Conversely, precision measurements of muon properties and other rare charged-lepton processes may not only help reveal the mechanism behind neutrino masses, but also play a role in determining neutrino oscillation parameters. The general argument is as follows. The new physics that has manifested itself in the neutrino sector in the form of non-zero neutrino masses is likely to leave its largest imprint in the charged-lepton sector, given the stringent connection between charged leptons and neutrinos. Charged-lepton flavor violation searches have the ability to help determine the flavor structure of the new physics, while searches for electric dipole moments may help establish the relationship between the new physics and $CP$ invariance.

As a more concrete example, neutrino oscillation data already reveal that charged-lepton flavor-violating processes must happen. The reason is simple. Neutrino oscillation experiments reveal that individual lepton-flavor numbers (electron number, muon number, tau number) are not conserved. On the other hand, individual lepton-flavor number conservation was the only symmetry preventing, {\it e.g.}, $\mu\to e\gamma$ decays from happening. Because the mechanism behind neutrino masses is unknown, the expected ``neutrino induced'' contribution is unknown. If the neutrinos are Dirac fermions, for example, or if neutrino masses are a consequence of the Type-I seesaw and the new physics scale is very high (say, $M_N\sim10^{14}$~GeV), the expected rates for charged-lepton flavor-violating processes  are exceedingly small.\footnote{For example, if the neutrinos are Dirac fermions and there is no other new physics, one can compute $B(\mu\to e\gamma)\lesssim 10^{-54}$.} If the new physics scale $M_{\rm new}$ is closer by, however, the situation can be markedly different.

\begin{figure}[ht]
\centerline{\includegraphics[width=0.5\textwidth]{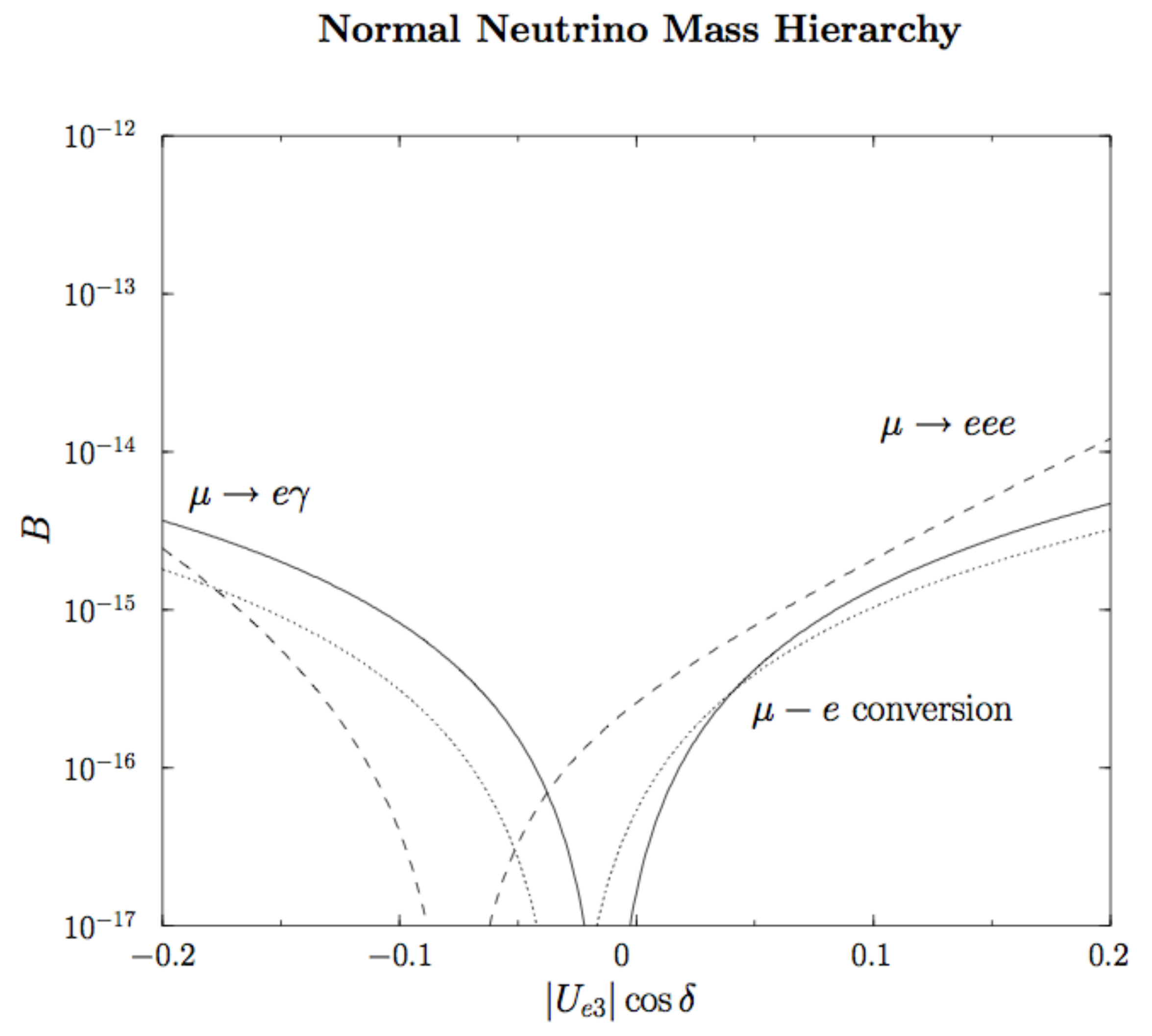}
\includegraphics[width=0.5\textwidth]{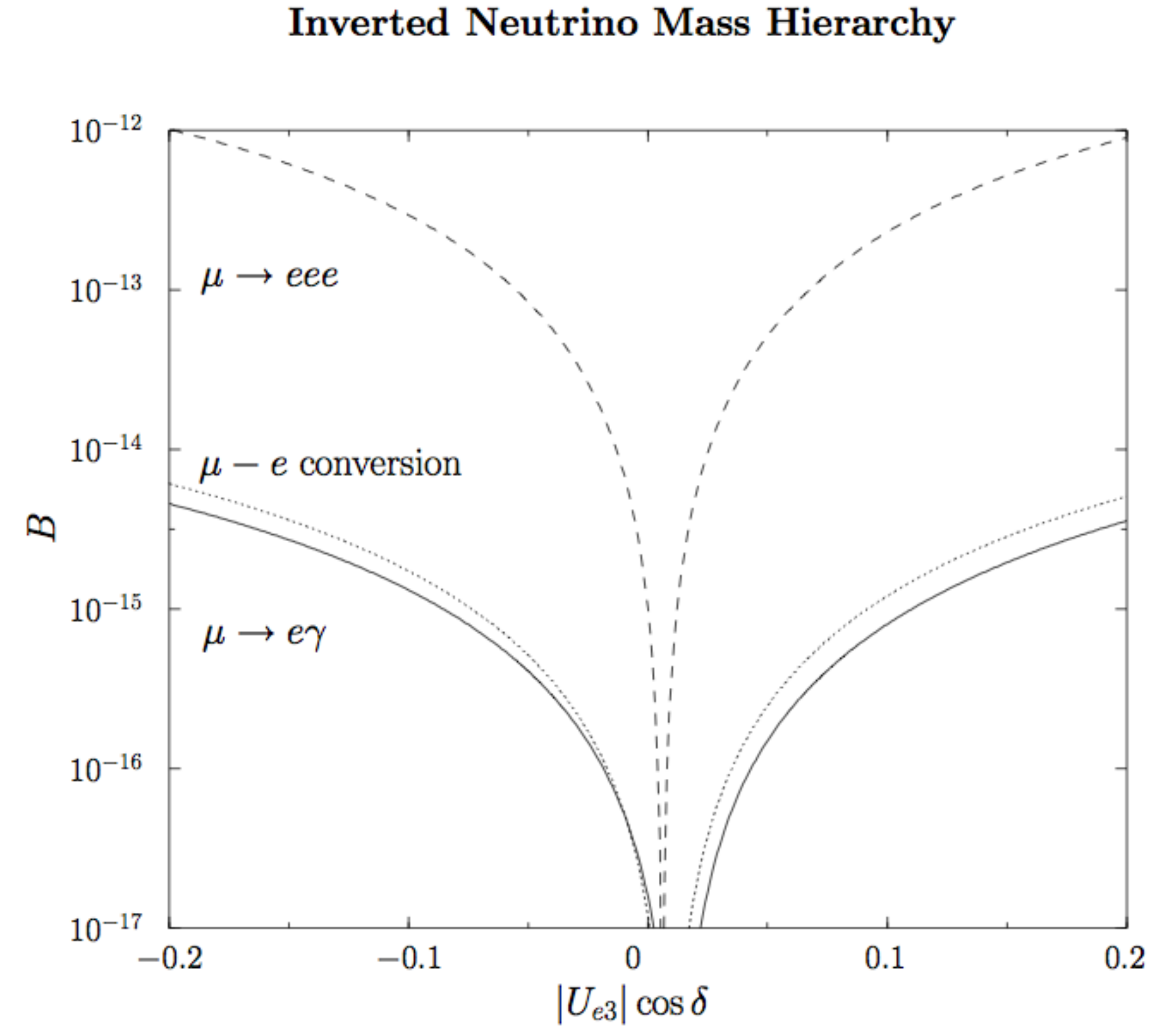}}
\caption{The branching ratios $B$ for $\mu \to e \gamma$ (solid line) and $\mu \to eee$ (dashed line), and the normalized capture rate $B$ for $\mu\to e$-conversion in Ti (dotted line) as a function of $|U_{e3}|\cos\delta$ in a scenario where neutrino masses arise as a consequence of the presence of a triplet Higgs field with a small vacuum expectation value. The lightest neutrino mass is assumed to be negligible, while the neutrino mass hierarchy is assumed to be normal (left) and inverted (right). See \cite{Kakizaki:2003jk} for details.\label{triplet}}
\end{figure}

There are abundant examples in the literature. Fig.~\ref{triplet} depicts the rate for different muon-flavor violating processes as a function of $|U_{e3}|\cos\delta$, in the case of the Type-II seesaw \cite{Kakizaki:2003jk}. The overall expectation for the transition rates depends on parameters external to the neutrino mass matrix, like the triplet mass and vacuum expectation value. The combination of data from neutrino oscillation experiments, the energy frontier (say, the LHC) and charged lepton flavor-violating searches in the intensity frontier should ultimately allow one to thoroughly test particular Higgs triplet models and, if these turn out to be correct, unambiguously reveal the physics behind neutrino masses.

Another important example of the potential interplay between the energy, cosmic, and intensity frontiers, via experiments with charged leptons and neutrinos, is the possibility of testing the leptogenesis hypothesis. If all the new physics associated with generating the baryon asymmetry of the universe is very heavy, it won't be directly tested at the energy frontier. If, however, the LHC uncovers new particles at the TeV scale, it is likely that these new degrees of freedom will also be imprinted, just like the neutrinos, with information from very high energy scales. This information, most likely, can only be revealed through searches for charged lepton flavor violation, which are closely related to the flavor properties of the few TeV scale degrees of freedom. All this information, combined with precision measurements of neutrino oscillation parameters -- including $CP$-invariance violating observables --  and the confirmation that neutrinos are Majorana fermions, should allow a very powerful, albeit still indirect, test of the leptogenesis hypothesis. For concrete realizations, see, {\it e.g.}, \cite{Pascoli:2003uh,Buckley:2006nv}. If the seed for the baryon asymmetry was planted when the universe was $10^{-30}$ seconds old, an indirect test of the mechanism may be as good as it gets.


\subsubsection{Lepton-Number Violation in Rare Meson Decays}

The relation between neutrinos and quarks is not as straightforward as the one between neutrinos and charged leptons, but the synergy between neutrino physics and meson properties and processes is also significant. As an example, rare meson decays offer several channels for the exploration of lepton-number violation and, as already emphasized, whether lepton number is an exact symmetry of Nature is among the most important anticipated pieces of the neutrino mass puzzle. Some of the most stringent constraints on lepton-number violation, other than searches for neutrinoless double-beta decay, come from the meson decay processes
\begin{equation}
M^{\pm}\to \ell^{\pm}\ell^{\prime\pm} M^{\mp},
\end{equation}
where $M^{\pm}$ are charged mesons, e.g., $\pi^{\pm}$, $K^{\pm}$, $D^{\pm}$, or $B^{\pm}$, while $\ell,\ell'=e,\mu,\tau$. The searches for these very rare decay processes can be carried out at a variety of intensity frontier facilities, from dedicated kaon experiments to $B$-factories to the LHCb detector. They currently provide the strongest constraints on MeV-scale new, mostly sterile, Majorana neutrinos. For details see, for example, \cite{Atre:2009rg}. These states, if discovered, would reveal that lepton number is not a fundamental symmetry of Nature and would provide precious information regarding the origin of neutrino masses.


\section{Facilities and Instrumentation Challenges}
\label{sec:facilities}

In this section we will describe the basic approaches for future experimental steps, and then describe specific programs globally and in the United States.

\subsection{Experimental Approaches}

As described in Section~\ref{sec:3nus}, the basic approach for measuring oscillation parameters and testing the three-flavor paradigm is to observe either appearance or disappearance of flavors in well-understood neutrino fluxes, as a function of energy and baseline.   Well-understood neutrino sources can also be used for cross section measurements and other physics.

\subsubsection{Neutrino Source Development Challenges}

Sources of neutrinos for oscillation experiments can be either natural or artificial.  As described in the previous sections, experiments will continue to exploit natural sources of neutrinos -- solar, geo-, supernova, and atmospheric --  for both particle physics and astrophysics.
Of course, natural neutrinos require no source development; the experimental challenges in these cases are in the detection technology.

In contrast, development of artificial sources of neutrinos present significant challenges for experimentalists.
For the foreseeable future, nuclear power
reactors~\cite{Ardellier:2006mn,Guo:2007ug,Ahn:2010vy}, which will be exploited over the 
next 10 years for various experiments, are not really candidates for neutrino source R\&D:  high-power 
reactors are of course
typically designed to optimize electrical power generation, and neutrino experimentalists benefit from the
inevitable neutrino by-products.  However, radioactive sources of low-energy neutrinos have been proposed for some experiments, and creation of high-intensity sources of radioactive isotopes (in reactors or with cyclotrons) requires technical development.

Stopped pion (and muon) neutrino beams (really, isotropic sources) have been employed in the past~\cite{Athanassopoulos:1996ds,Armbruster:2002mp} and can be employed for oscillation studies, cross section
measurements and Standard Model tests.  High-intensity sources such as the Spallation Neutron Source exist but as yet have not been used for neutrino experiments.
Recently, there have
 been proposals to use this type of neutrino source alone or in
 combination with conventional beams to study $CP$ violation; see~\cite{Alonso:2010fs,Agarwalla:2010nn} and Section~\ref{sec:us}.   A high-intensity stopped-pion source employing novel cyclotrons requires some significant work, but there are connections with applications in industry. 

For the ``standard'' boosted-pion-decay beam of the type used at KEK, FNAL and J-PARC, in which
pions are produced by proton irradiation of a thick target, focused forward with a ``magnetic horn,''
and allowed to decay in a long evacuated pipe, design issues are relatively well understood.
There are indeed challenges to achieve the proton intensities required for MW ``superbeam''-scale
beams; however the difficulties seem relatively straightforward to surmount and
there are several future proposals for superbeams.
There are several challenges associated with measuring neutrino oscillations in superbeam experiments. The beam flavor composition is not simple. Kaon decays and the decays of daughter muons lead to a non-zero flux of $\nu_{e}$, while a subdominant population of ``wrong sign'' $\nu_{\mu}$ survives the pion charge-selection mechanism. Furthermore, the energy dependence of the neutrino flux is not very well characterized, and near detectors are necessary in order to ``measure'' the different neutrino fluxes. Furthermore, the charged-current scattering cross section for GeV-scale neutrinos is very hard to model and is yet to be measured with the required precision. This fundamentally important issue is the subject of ongoing research and is expected to play a central role in next-generation neutrino experiments.   This issue was discussed in some detail in Sec.~\ref{sec:scattering}. Finally, the measurement of $\nu_e$ charged-current events is challenging due to beam-induced backgrounds, mostly from neutral pions produced in neutral current $\nu_{\mu}$ events. Ultimately, it is fair to say that sub-percent measurements of oscillation probabilities using neutrino superbeams appear to be, at the very best, very challenging.

Farther-future oscillation neutrino sources include neutrino factories and $\beta$-beams.  The potential
high intensity, boost tunability, and lepton sign selection are highly desirable.  The technical challenges,
however, are large.  In particular, for a neutrino factory, methods for ``cooling'' the muon transverse
momentum require further development.
An international design study for a neutrino factory is currently under way~\cite{neu:NF:2011aa}. It aims at identifying in detail the steps necessary to demonstrate that a neutrino factory can be built, and at laying out a plan for doing it. 
The challenges associated with the production of $\beta$-beams are also many, but well known. The international design study for a neutrino factory is also charged with assessing the challenges and the virtues of $\beta$-beams.


\subsubsection{Large Detector Development Challenges}

Because of the tiny neutrino interaction cross section, detectors for neutrino experiments must typically
be very large.  Although for some short-baseline oscillation (and other neutrino) experiments relatively modest scale
(kTon or less) detectors can be used, for long-baseline oscillation experiments and for astrophysical neutrinos 
detector masses of multi-kTon and upwards scale are required.  

\begin{figure}[ht]
\includegraphics[width=0.4\textwidth]{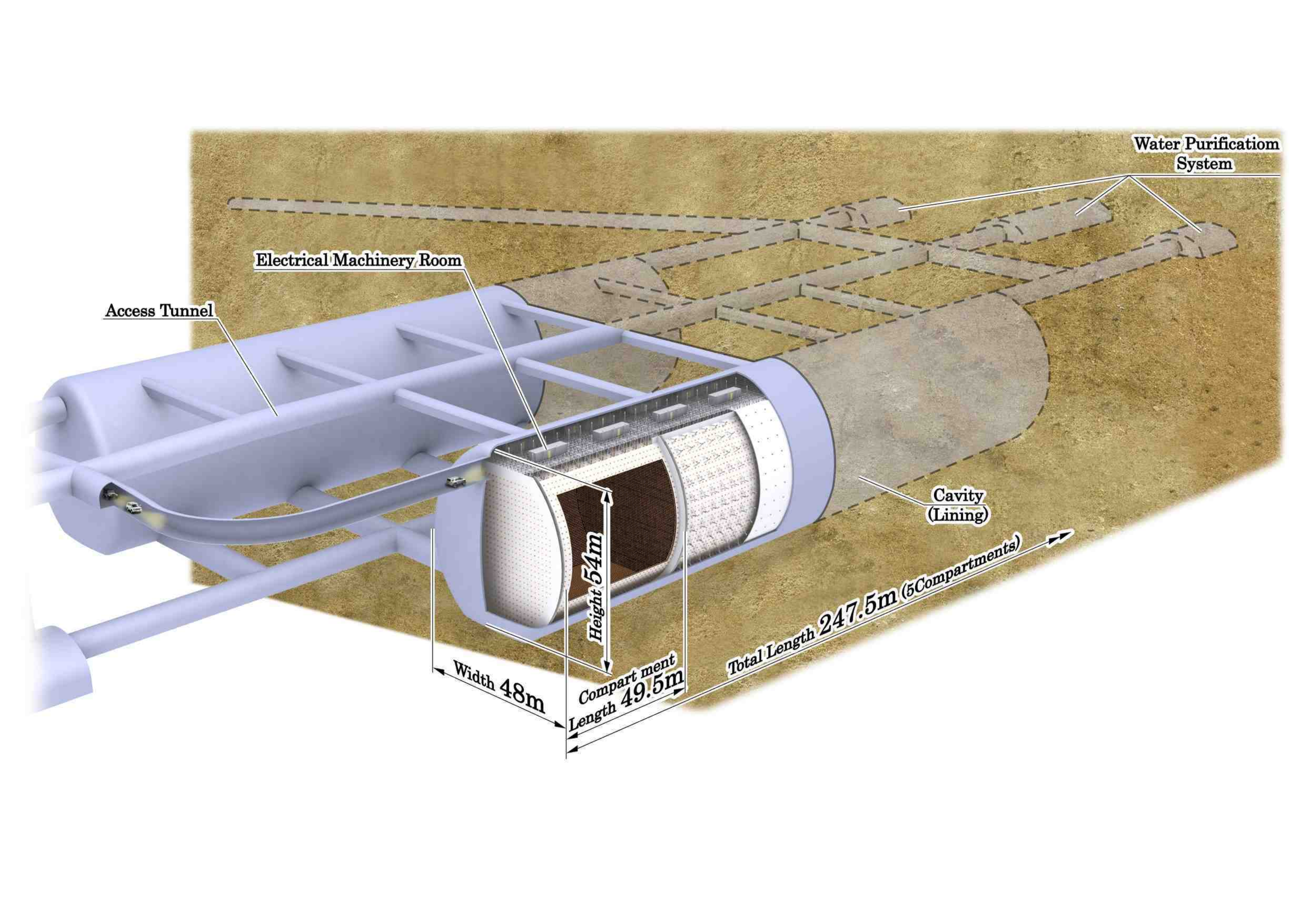}
\includegraphics[width=0.3\textwidth]{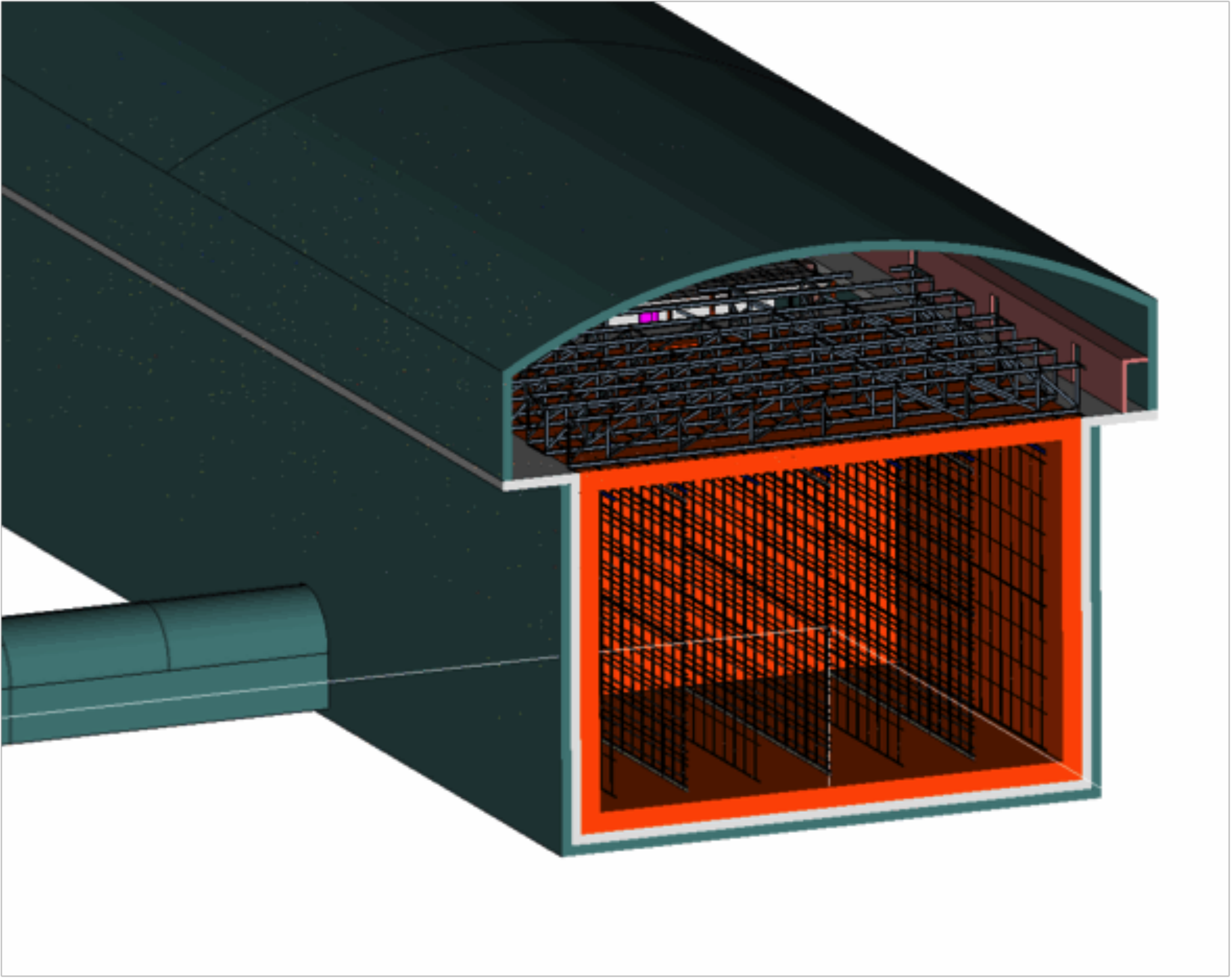}
\includegraphics[width=0.2\textwidth]{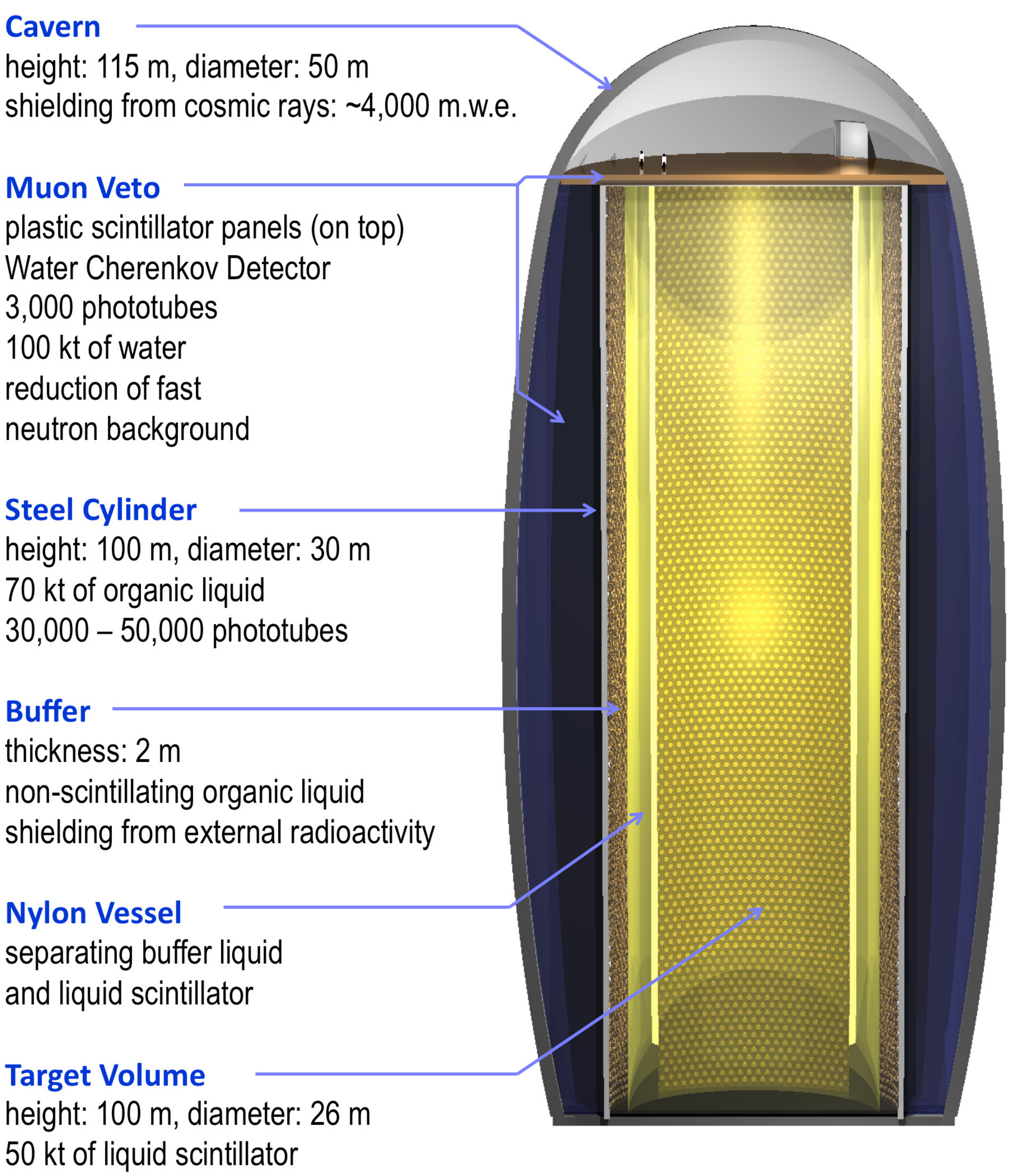}
\caption{Left: example of a large water detector concept (Hyper-Kamiokande). Center: example of a large liquid argon detector concept (LBNE).  Right: example of a large scintillator detector concept (LENA).}\label{fig:large_detectors}
\end{figure}

There are three main kinds of detector technology under consideration
for the next generation of multi-kTon detectors: water Cherenkov,
liquid argon and liquid scintillator.  See Figure~\ref{fig:large_detectors}.    Of these, water and liquid argon are the most suitable for long-baseline oscillation experiments, due to better high-energy event reconstruction capabilities.

\paragraph{Water Cherenkov Detectors}:
Water Cherenkov detectors, in the form of large volumes of ultra-pure
water surrounded by photomultiplier tubes (PMTs), are sensitive to the
charged particles produced by interactions of neutrinos with energies
greater than a few MeV.  Charged particles
moving faster than the speed of light in a medium produce Cherenkov
photons if $\beta>1/n$, where $n$ is the refractive index of the
medium.  Water is a convenient and cheap detector material, suitable
for neutrino detection because very large volumes can be deployed
cheaply, even though light yields are typically much lower than for
scintillator. 
However, due to the Cherenkov threshold, heavy
particles are invisible, and signals from low energy
electrons, positrons and gammas (which are detected via
Compton-scattered electrons) may be lost.
Both low (few to few tens of MeV, relevant for solar, supernova and stopped-pion neutrinos) and high ($\gsim$ GeV, relevant for beam and atmospheric neutrinos as well as proton decay) energy neutrino
detection are possible in large water Cherenkov detectors.
In Cherenkov detectors, particle direction can be determined
using the angular information from the
Cherenkov ring.   This is helpful both for reconstructing multiple
particles in high energy interactions, and for reconstructing
neutrino-electron elastic scattering events at low energy.   Particle
type can be determined by evaluating the ``fuzziness'' of a track: electrons
and gammas scatter and shower, whereas muons and pions have sharp
tracks; the Cherenkov angle can also be of use for particle identification.
Low energy event detection may potentially be
enhanced in water using Gd doping~\cite{Beacom:2003nk},
for which neutrons are captured on gadolinium nuclei, producing a
cascade of gammas with $\sim$4~MeV visible energy; this allows tagging
of interactions that produce neutrons, such as inverse beta decay.
There are also ideas for developing a large area picosecond photosensor-based
detector filled with water-based liquid scintillator to expand the capabilities
of these detectors to low energy particles and for improved particle 
identification \cite{lappd}. 

Past water Cherenkov detectors include IMB~\cite{BeckerSzendy:1992hr}
and Kamiokande~\cite{Hirata:1991ub}.
The successful use of the technology for a wide range of physics
topics is well proven
at the few-tens-of-kTon scale:  Super-Kamiokande~\cite{Fukuda:2002uc} has been running for more than 15 years
and has demonstrated a broad range of physics capabilities over several orders
of magnitude in energy.  Proposed next-generation water Cherenkov detectors include
Hyper-Kamiokande~\cite{Abe:2011ts} in Japan
and MEMPHYS~\cite{Borne:2011zz} in Europe.

Phototubes can also be embedded in ice or suspended in water in long-string configurations. This kind of configuration is primarily sensitive to high energy (greater than multi-GeV) neutrinos, although there is 
also some sensitivity to supernova neutrinos~\cite{Abbasi:2011ss}.    The  science goals of existing long-string detectors are primarily to study astrophysical objects, although there is oscillation sensitivity with atmospheric (or cosmic) neutrinos as well.   Examples are IceCube, ANTARES, and Lake Baikal; planned next-generation efforts are KM3NET and ideas for denser infill of Antarctic ice, possibly with enhanced photosensors (PINGU~\cite{Tyce2011}).
In the much more distant future, such an enhanced detector could conceivably serve as a target for a long-baseline beam~\cite{Tang:2011wn}.

\paragraph{Liquid Argon Time Projection Chambers}:
Liquid argon (LAr) time projection chambers do not suffer from the Cherenkov
threshold issue, and in principle extremely high-quality particle reconstruction is
possible.  The ionization charge from the passage of particles through
argon is drifted with an electric field and collected on readout wire planes; a 3D track can be
reconstructed using charge arrival time information.  Furthermore,
scintillation light signals in argon detected by photomultiplier tubes
can allow fast timing of signals and enhance event localization.  Very high-purity
cryogenic argon is required.   Track granularity is determined by
wire spacing, and in principle very fine-grained tracking can be
achieved.  Particle identification is possible by measuring ionization energy loss
along a track~\cite{Palamara:2011zz}.

Because of the excellent, full-particle tracking capability of liquid
argon, very high-efficiency particle reconstruction allows a smaller
LAr detector to match the efficiency of a water detector of a given mass, with an approximate
ratio of $1:6$.  In principle, low
energy physics ($<$ 100 MeV, \textit{e.g.} for supernova neutrinos) is
possible in LAr as well, assuming adequate triggering capability.

The current largest liquid argon detector is 
ICARUS~\cite{Amerio:2004ze} in Europe.   In the US, a dedicated program of
development towards large liquid argon detectors has begun at Fermilab with
ArgoNeuT~\cite{Palamara:2011zz}, including a material test stand program, liquid argon
purity demonstrator, and reconstruction software development. This path forward
in the US will continue with MicroBooNE~\cite{Ignarra:2011yq}, 
a 35 ton membrane cryostat, testbeam studies, and an 800 ton LAr TPC 
(LAr1)~\cite{integrated-plan}. Farther future possibilities will include a 
future LAr detector for LBNE~\cite{Akiri:2011dv}, GLACIER~\cite{Curioni:2011tz}
in Europe, and detectors in Japan~\cite{Hasegawa:2011zz}.The LAr technique is 
as yet unproven at multi-kTon scale, but R\&D has so far yielded promising 
results.

\paragraph{Liquid Scintillator Detectors}:
Liquid scintillator detectors consist of large volumes of clear
hydrocarbon, in a homogeneous or segmented volume viewed by
photomultiplier tubes.  Light yield can be very high -- typically 50
times more light per energy loss than Cherenkov detectors.  This
enables both low energy thresholds and good energy resolution.
However, in order to detect neutrinos at low energy, extremely good
radioactive purity is also required.  Particle energy loss is
proportional to number of photoelectrons detected, and particle
interaction vertices can be reconstructed by timing; to a lesser
extent direction and other properties can be
reconstructed.  Unfortunately because of the
isotropy of scintillation light, directionality and tracking
capabilities are relatively weak.  Nevertheless, some high energy
particle
reconstruction may be possible (\textit{e.g.}~\cite{Mollenberg:2011zz})
using photon timing.  
This kind of detector excels for detection of low energy (tens of MeV) signals, such 
as reactor, geo- and solar neutrinos; furthermore $0\nu\beta\beta$ candidate isotopes can be
added for high-mass searches (see Section~\ref{sec:majorana}).
There is a long history of successful kTon-scale scintillation detectors,
starting with the segmented Baksan~\cite{Pomansky:1986jv},
MACRO~\cite{Ambrosio:2002mb} and LVD~\cite{Bari:1989nw} detectors, and
followed by KamLAND~\cite{Eguchi:2002dm}
and Borexino~\cite{Alimonti:2008gc}.  The near-future
SNO+\cite{Kraus:2010zzb} will be next.  Proposed large future
detectors include HanoHano~\cite{Learned:2007zz} and LENA~\cite{Wurm:2011zn}.

\subsubsection{Underground Facilities}

Some neutrino experiments can be sited on or near the surface by employing beams with sharply pulsed timing to separate signal events from cosmogenic background,  or by using other strong background rejection techniques; and some are not especially sensitive to cosmogenic background (\textit{e.g.} kinematic experiments for absolute neutrino mass).
However  it is highly desirable, and in some cases mandatory, to site next-generation detectors in deep
underground laboratories.
For neutrinoless double-beta decay searches, low rates of cosmic rays and very deep locations are absolutely essential.
These detectors, furthermore, have very stringent requirements on rates of natural radioactivity from both surroundings and detector materials, so very clean underground facilities are also important.
Such requirements are shared with dark matter WIMP search detectors as well.

It is also desirable to site large detectors designed to be targets for long-baseline beams at deep locations, as these detectors can address a much broader range of physics topics when protected from cosmogenic background.
The depth required depends both on physics topic and on detector technology, as the specific nature of the background will vary according to the particular signal.  For natural
sources of neutrinos, the depth requirements may be fairly stringent.
Reference~\cite{Bernstein:2009ms} explores depth requirements for physics sensitivity to different topics.

For these reasons, plans for next-generation experimental programs focus primarily on deep underground laboratory facilities.  Infrastructure at a common site can furthermore be shared between different experiments.

\subsection{Opportunities for Experimental Programs}

\subsubsection{Global Context}

Worldwide there are multiple existing and planned programs of beams and detectors at surface and underground sites.  We present a brief survey of international projects here.

In Canada, the premier underground laboratory is SNOLAB, in Sudbury, Ontario, at a great 2070~m depth.  This laboratory hosted the SNO heavy water experiment, soon to be converted to the SNO+ scintillator experiment, which will be doped with $^{150}$Nd for a neutrinoless double-beta decay search.  SNOLAB hosts, or is preparing to host, a variety of dark matter and neutrinoless double-beta decay search experiments.
Some underground space remains unallocated at this time, although perhaps not for long.

In Japan, the Super-Kamiokande experiment near Kamioka, Japan, has been operational since 1996.  It has produced a very broad range of results using solar and atmospheric neutrinos, and has the world's best limits on
proton decay.  It also served as the target for the
first long-baseline neutrino experiment K2K (KEK to Kamioka), and is currently the target for the T2K (Tokai to Kamioka) off-axis beam from J-PARC near Tokai.   The beam will be upgraded over the next several years.
Japan also hosts the KamLAND experiment, which measured
long-baseline reactor $\bar{\nu}_e$ disappearance and is now focusing on a neutrinoless double-beta decay search using dissolved xenon.  Other underground experiments are sited in the Kamioka mine.

A prominent proposed new experiment in Japan is Hyper-Kamiokande, to be sited 
near the Super-K detector~\cite{Abe:2011ts}.  Two sites in the Tochibora mine (1500-1750
meters water equivalent depth) are under study.  The proposed detector has a fiducial mass
of 560 kTon and 10-20\% Super-K-equivalent photomultiplier coverage.  The plan is for an
eventual upgrade of the T2K beam to 1.7~MW.  There are also ideas for
Japan-based
liquid argon detectors, including a 100~kTon facility at Okinoshima
island halfway between Japan and Korea~\cite{Hasegawa:2011zz}.

A future underground large detector optimized for atmospheric neutrinos
is the planned 50-kTon ICAL iron calorimeter detector for the India-based 
Neutrino Observatory~\cite{Athar:2006yb}.
This detector has lepton sign-selection capability using a magnetic field 
to enable separation of neutrinos and antineutrinos. If $\theta_{13}$ is
large, ICAL will have reasonable sensitivity to the neutrino mass hierarchy
via the $P_{\mu\mu}$ channel~\cite{Gandhi:2007td}. At the same laboratory 
site, other experiments are also planned.

Europe currently hosts the world's highest-power conventional boosted meson-decay beam, 
the 510~kW CNGS (CERN Neutrinos to Gran Sasso) beam, sent 730~km from CERN to the experiments at the Gran
Sasso laboratory, including OPERA and Icarus.  There is some potential for upgrade of CNGS to 750~kW, but
no practical near detector location for this beam.
A number of future beam, detector and siting possibilities are being
explored by the LAGUNA study~\cite{Rubbia:2010zz}.    The envisioned EUROnu superbeam from CERN would
be a 4~MW beam to Fr\'ejus.
Detectors under consideration for LAGUNA include a 0.5
Mt water Cherenkov detector (MEMPHYS~\cite{Borne:2011zz}), a 100 kt LAr detector (GLACIER~\cite{Curioni:2011tz})
and a 50~kTon scintillator detector (LENA~\cite{Wurm:2011zn}).  
Beam options from CERN sending neutrinos to these sites also include beta beams or a
neutrino factory in the farther future.
Top sites under consideration
are Pyh\"asalmi in Finland (2300 km from CERN), Fr\'ejus (130 km from
CERN) and Umbria (665 km from CERN, off-axis from the existing CNGS beam).
A possible staged program is described in~\cite{Agarwalla:2011hh}.
Europe is also a major participant in the international neutrino factory and $\beta$-beam design studies
and associated R\&D experiments (MERIT, MICE and EMMA).

This survey of proposed programs globally is far from exhaustive: a number of smaller-scale laboratory facilities exist or are planned at various European locations~\cite{Bettini:2007xc,Bettini:2011zza}.  Efforts in other parts of the world include the JinPing underground laboratory in China, and the proposed ANDES facility in South America.

\subsubsection{US Contributions and Facilities}\label{sec:us}

\paragraph{Current and Near-Future Programs}:
Existing facilities in the US dedicated to neutrinos are centered at Fermilab.  The NuMI (Neutrinos at the Main Injector) beam facility~\cite{Hylen:1997xd} 
employs 120 GeV/c protons impinging on a graphite target, and two magnetic horns and a 675-m decay pipe; beam power is approximately 400~kW.   Neutrinos are sent to MINOS, a 5~kTon iron tracker detector at the Soudan mine 734 km away in Minnesota, located on the beam axis.  The MINOS experiment includes a near detector facility at Fermilab.  Since January 2005, NuMI has delivered more than 10$^{21}$ protons on target.
This beam also serves the MINER$\nu$A experiment.
The MiniBooNE experiment employs a separate Booster beam, which delivers 8~GeV protons on target to a secondary beamline with a focusing horn and 50-m decay region.
In both MINOS and MiniBooNE cases, the polarity of the horns can be reversed to produce a flux that is primarily
$\nu_\mu$ or $\bar{\nu}_\mu$.

Highlights of the US neutrino program of the past $\sim$15 years include the most precise  measurements of $\nu_\mu$ disappearance $|\Delta m^2|$ from MINOS~\cite{Adamson:2011ig,Adamson:2012rm}, $|U_{e3}|$ results~\cite{Adamson:2011qu},  as well as numerous searches for new physics.  MINOS
has plans for extended running (MINOS+~\cite{Tzanankos:2011zz}.)
The MiniBooNE program has also produced several important neutrino oscillation searches (Sec.~\ref{sec:sbl}) in addition to multiple neutrino interaction measurements as described in Sec.~\ref{sec:scattering}. 

\begin{figure}[ht]
\begin{center}
     \includegraphics[width=0.46\textwidth]{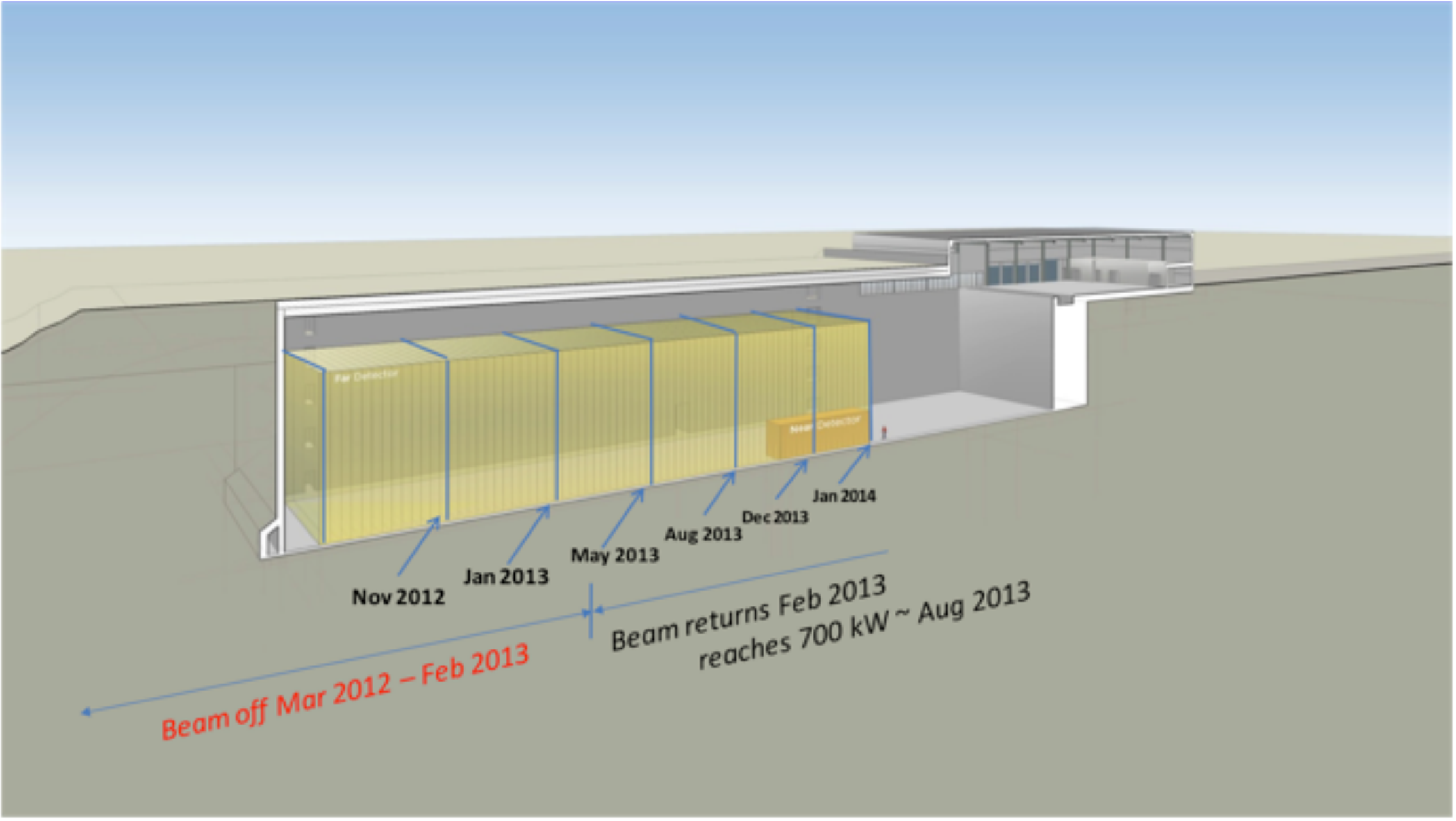}
     \includegraphics[width=0.41\textwidth]{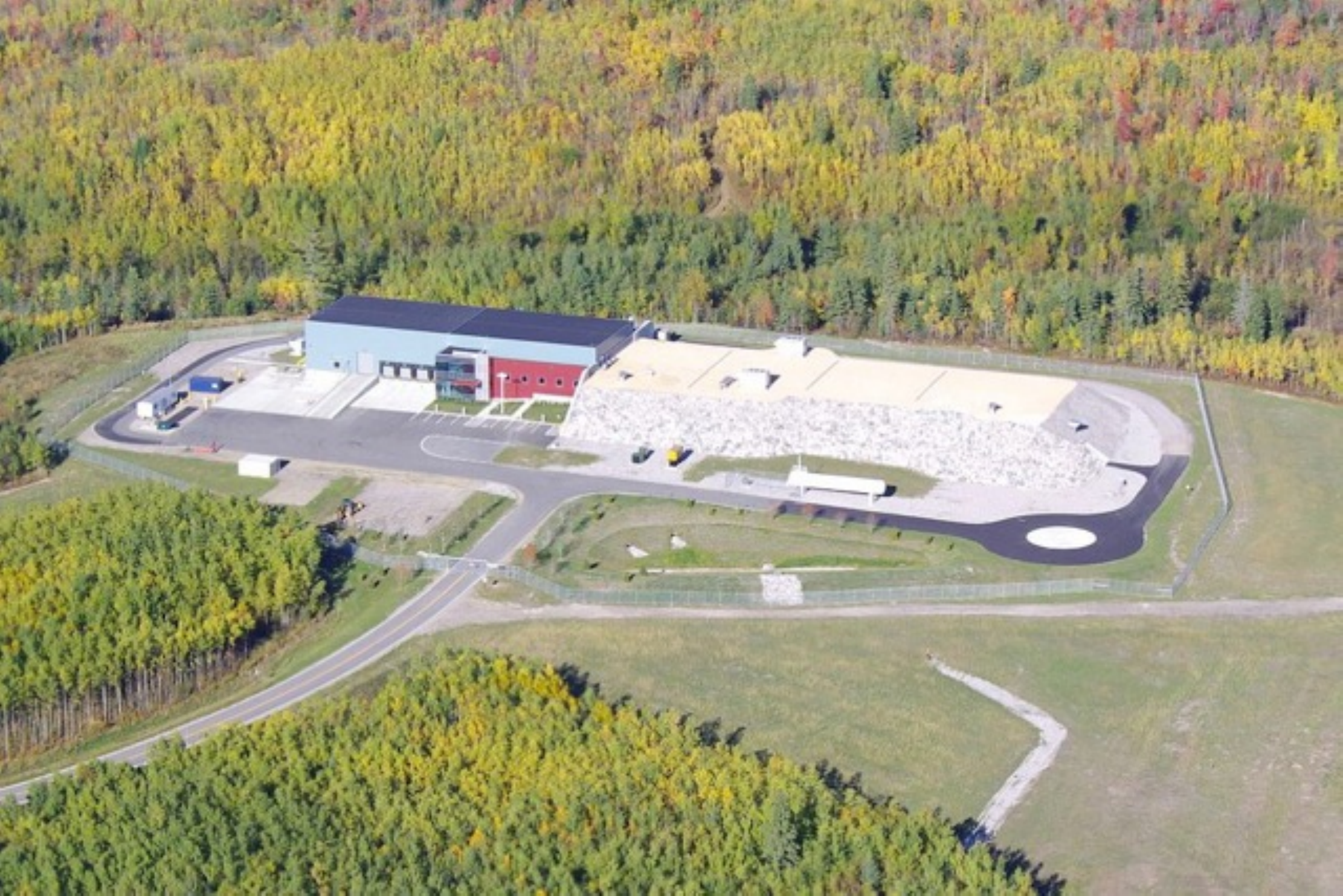}
\end{center}
\caption{\footnotesize NO$\nu$A schematic with timeline (left) and site (right). }
\label{fig:nova}
\end{figure}

The NO$\nu$A experiment will employ an upgraded 700~kW off-axis NuMI beam. The 14 kTon liquid scintillator detector is under construction at Ash River (see Fig.~\ref{fig:nova}), at an 810-km baseline; this location is off the beam axis, where the beam neutrino spectrum is narrower.
The NO$\nu$A program will improve the measurements of $|U_{e3}|$ and parameters related to $\nu_\mu$ disappearance for both neutrinos and antineutrinos.   In particular, NO$\nu$A will have a chance to determine the mass hierarchy, thanks to its long baseline. 

A number of proposals for small experiments and extensions to existing experiments could enhance the 
US program \textit{e.g.}~\cite{Thomas}.
Several of these are described in various sections of this Chapter.

US physicists have also made significant contributions to offshore neutrino experiments.
These include longstanding contributions to the Super-K and T2K programs, which will continue to run through this decade.  
The US has been a major participant in the Double Chooz (in France) and Daya Bay (in China) reactor $|U_{e3}|$ experiments, both of which have recently started production running.  The physics runs of these experiments planned over the next several years will have high impact.  

\begin{figure}[ht]
\begin{center}
      \includegraphics[width=0.6\textwidth]{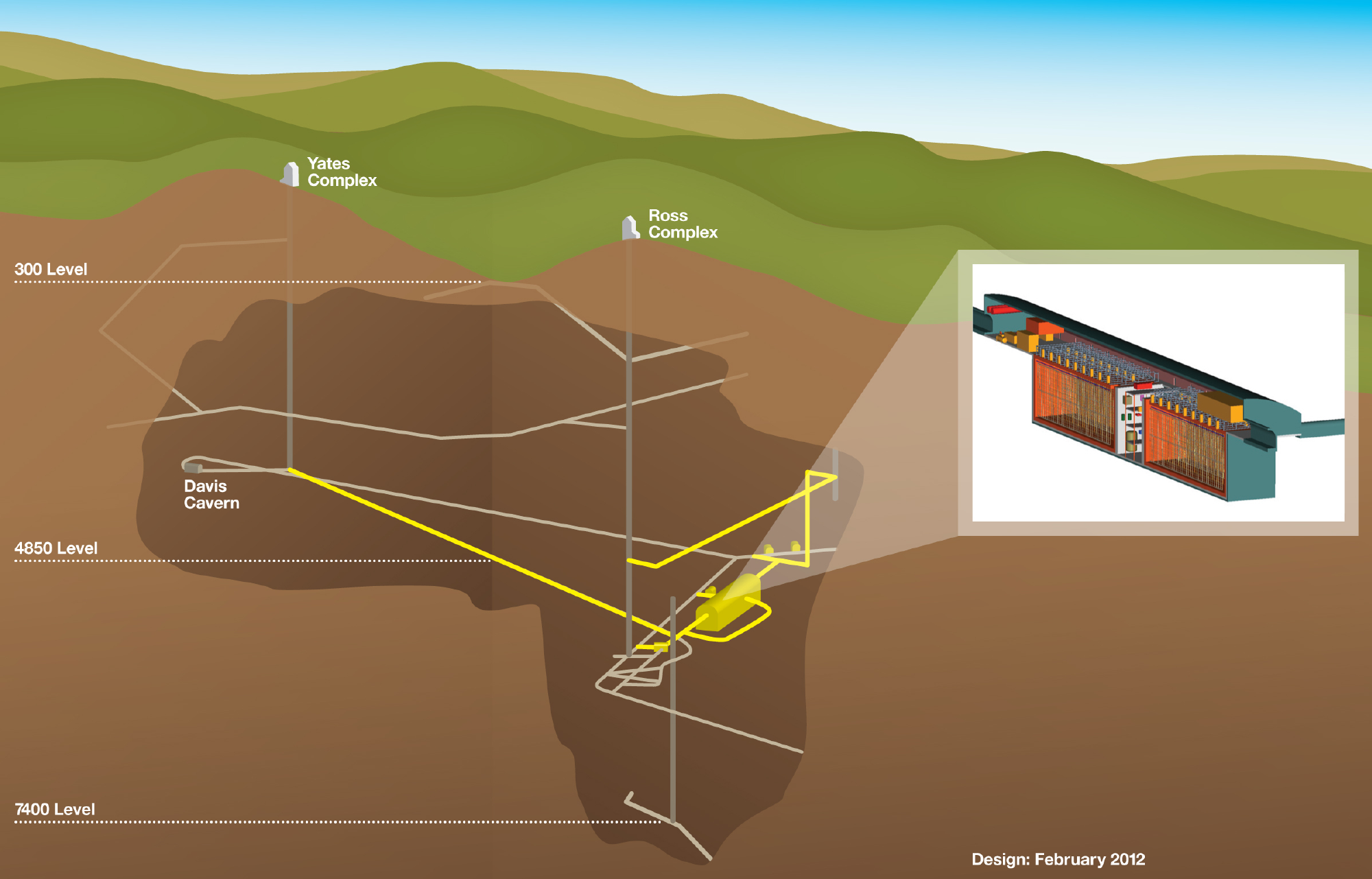}
\end{center}
\caption{\footnotesize The proposed LBNE LAr detector at the 4850 ft level of the Homestake mine. }
\label{fig:lardet}
\end{figure}

\paragraph{LBNE}:
The next major planned neutrino program in the US is the Long-Baseline Neutrino Experiment (LBNE).  The experiment as currently envisioned comprises a new 700~kW beam at Fermilab, a near detector complex, and a large far detector at the Homestake mine in South Dakota, at a baseline of 1300~km.  
Extensive design work and physics sensitivity studies were done over the past few years for two detector options for LBNE: a 200-kTon single-module water Cherenkov detector and a 34-kTon dual-module liquid argon TPC
~\cite{Akiri:2011dv}. Although a configuration with both technologies would be preferable for physics, the cost was prohibitive.  After an exhaustive  decision-making process, the LAr detector option was selected
(see Fig.~\ref{fig:lardet}).  The deep site at 4850 ft is strongly favored for this program, thanks to improved cosmogenic background rejection for astrophysical neutrino and proton decay studies, as well as the possibility for shared infrastructure with a broader underground program.

The sensitivity of LBNE was discussed in Sec.~\ref{sec:3nus} (see Fig.~\ref{fig:lbnesens}). At the proposed deep site, the LBNE program will be enriched by additional sensitivity to proton decay and atmospheric and supernova neutrino physics.  

\paragraph{Homestake}:
The site currently under consideration for the LBNE far detector is the Homestake mine in Lead, South Dakota, the former site of the Davis chlorine solar neutrino experiment.  This site could also host a suite of other underground experiments with both intensity and cosmic frontier impacts.
The facility currently at this site is known as the Sanford Underground Research Facility (SURF).
The facility will host its first dark matter (LUX) and neutrinoless double-beta decay (Majorana Demonstrator) experiments early in 2012.  The state of South Dakota established the South Dakota Science Technology Authority (SDSTA) to oversee the conversion of the former mine into a research facility.
The SDSTA has accomplished rehabilitation of the facility, assuring safe redundant access from the surface to the 4850 level (4850 feet below ground).  The pumping system has been restored and the accumulated water pool has been lowered below the 5800 foot level.  The 7400 foot level is anticipated 
to be reached by early 2014.  
Two shafts require upgrading to enable experimental activity underground.
The Ross shaft is being refurbished to restore rock hoisting capacity necessary for major construction activities.  It will serve as the primary operations and maintenance access for facility operations personnel and materials.  The Yates shaft is being upgraded to primary access capability and will serve as the principal scientific access underground. This upgraded Yates shaft is anticipated to be completed early in 2012 to support the new physics experiments.  
Conceptual designs for the excavations and outfitting necessary 
for the LBNE 34 kT liquid argon detector either at the 800L or the 4850L are well under way.
The plans for using the 4850L maximize the synergies of co-locating the experiments and sharing facility infrastructure and access. 

\paragraph{Project X}:
   Project~X~\cite{projectx} is a US-led accelerator initiative with strong international participation that aims to realize a next-generation proton source that will dramatically extend the reach of intensity frontier research.   The state of the art in superconducting radio frequency (RF) has advanced to a point where it can be considered and implemented as the core enabling technology for a next-generation multi-megawatt proton source--reliably delivering unprecedented beam power at duty factors ranging from $10^{-5}$ to 100\%.   The base super-conducting RF technology also supports flexible beam-timing configurations among simultaneous experiments, allowing a broad range of experiments to develop and operate in parallel.   The DOE Office of High Energy Physics and its advisory bodies have recognized this potential and are supporting R\&D for Project X that could lead to a construction start as early as 2016.  
Project X has leadership potential in the future landscape of proton sources, and can enable new lines of research including
neutrino interaction and oscillation experiments. 
Project X is a multi-megawatt proton source with proton kinetic energies of 3, 8, and 120 GeV that can drive intense simultaneous neutrino beams directed toward near detectors on the Fermilab site and massive detectors at distant underground laboratories.  In addition there are possibilities for kaon, muon, nuclear, and neutron precision experiments, as described elsewhere in this document.
Furthermore, 
neutrino factory concepts, and beyond that, muon-collider concepts depend critically on developing multi-megawatt proton source technologies.  A technology roadmap has been developed that will directly leverage the Project X infrastructure into a platform for these future concepts.  
Initial review of the potential research program has identified 25+ world-class and world-leading experimental programs that can be driven by Project X.   Many of these programs are under way now and can serve as cost-effective day-one beneficiaries of Project X beam power.  

     Notable in the Project X program is the deep reach in neutrino physics.  
The direct scope of Project X includes 2000-2400 kW of beam power at 60-120 
GeV and 50-190 kW at 8 GeV, corresponding to three times the initial beam 
power of the Long Baseline Neutrino Experiment (LBNE) and three to 12 
times the beam power delivered to the MiniBooNE experiment. This extraordinary 
beam power is particularly important to long-baseline experiments in which the 
sensitivity is ruled by the product of beam power and detector mass, where 
detectors are pushed to the limit of massive scale. The benefits of Project X
to neutrino physics can be expressed both in terms of beam power and in terms
of its ability to bring new physics within reach more quickly. Project X beam 
intensities allow the long-baseline oscillation physics program to be 
accomplished three times faster ({\it i.e.}, oscillation measurements made in 10 
years with Project X would take more than 30 years without). More importantly, 
searches for $CP$ violation in the neutrino sector quickly become potential 
discovery level measurements with Project X (a $3\sigma$ measurement of 
$CP$ violation in a 10 year exposure of LBNE in a 700 kW beam becomes $5\sigma$ 
in the same time with Project X). Figure~\ref{fig:pxcomp} gives an example of 
the extended reach that will result from Project X's beam power.  With the 
increased beam intensity, a larger fraction of phase space can be probed 
at high significance, thus ensuring more complete coverage of whatever possible
value of $\delta_{CP}$ Nature has chosen. In addition, Project X uniquely 
enables high intensity running at lower neutrino energies. For example, a step 
beyond the direct scope of Project X is an upgrade of the 8 GeV pulsed beam 
power to 4000 kW simultaneous with 2000 kW at 60 GeV. This joint beam power 
can enable precise tuning of the neutrino energy spectrum to simultaneously 
illuminate both the first and second interference maxima, which greatly 
enhances the sensitivity of LBNE to matter-antimatter asymmetries among 
neutrinos (\textit{e.g.},~\cite{Bishai:2012ss}).  

\begin{figure}[ht]
\begin{center}
      \includegraphics[width=0.5\textwidth]{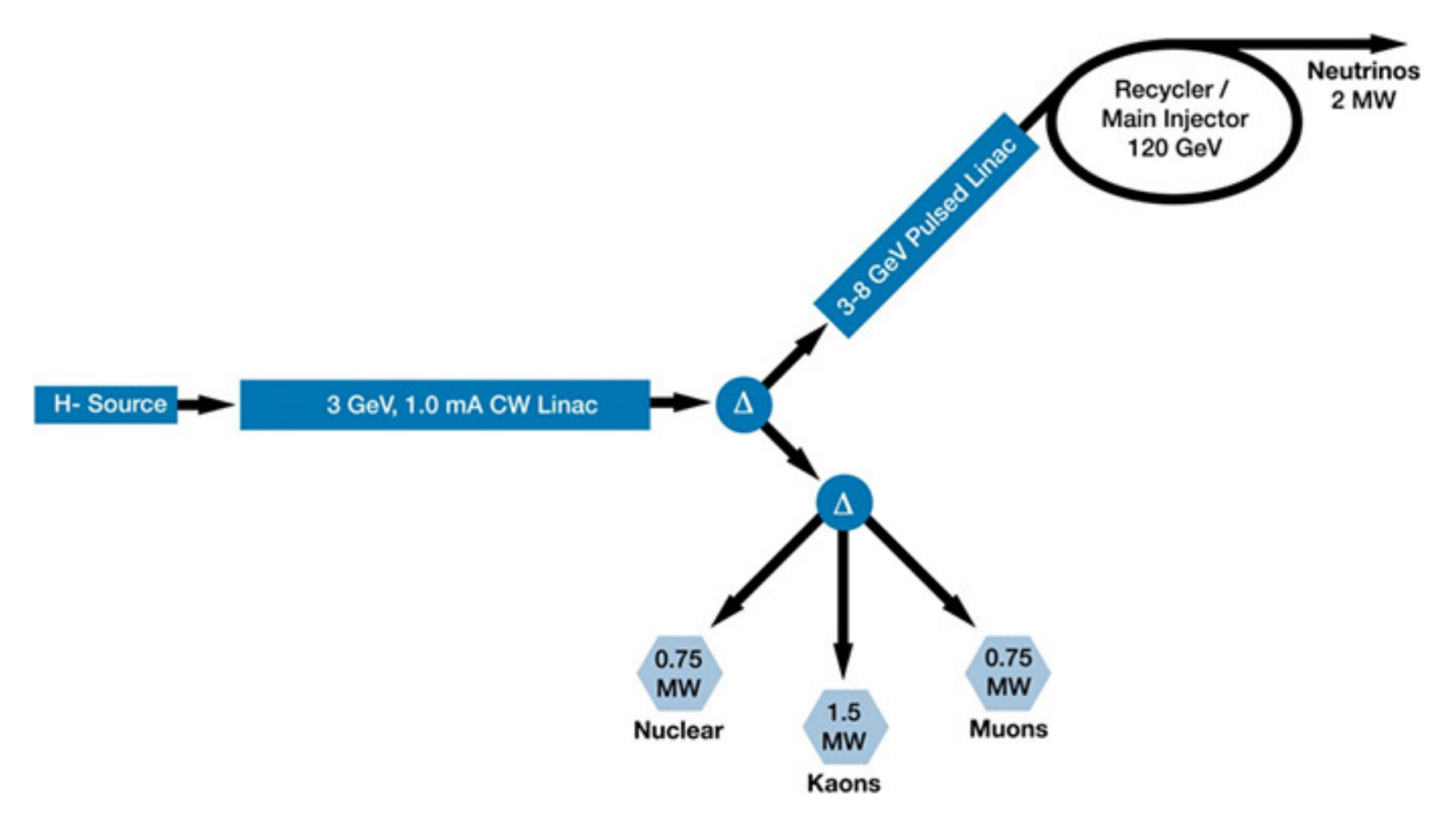}
      \includegraphics[width=0.3\textwidth]{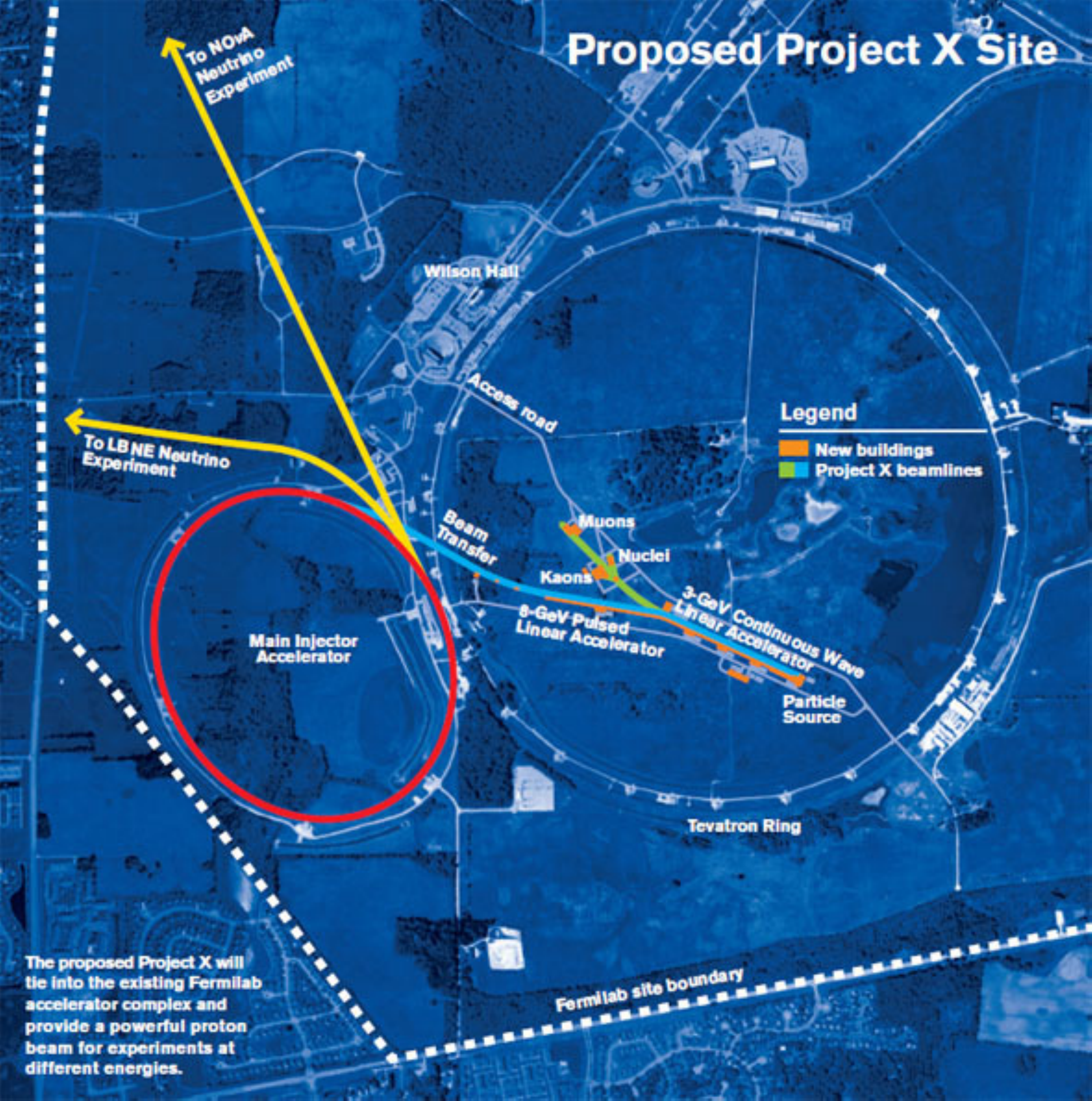}
\end{center}
\caption{\footnotesize Left: Project X reference design.  Right: Project X proposed site at Fermilab. }
\label{fig:projectx}
\end{figure}

\begin{figure}[ht]
\begin{center}
     \includegraphics[width=0.6\textwidth]{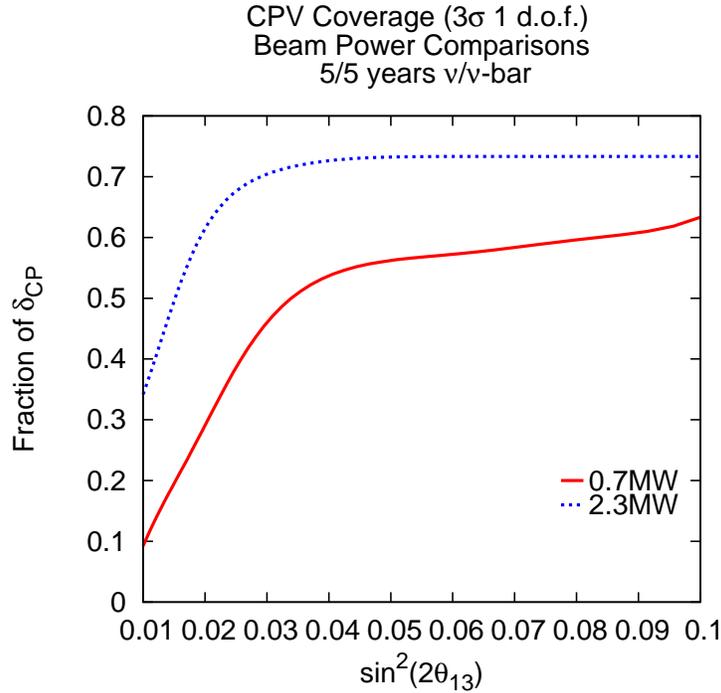}
\end{center}
\caption{\footnotesize Comparison of the fraction of $\delta_{CP}$ values for which  a 3$\sigma$ discovery of $CP$ violation would be possible, for 10 years of LBNE running with 0.7~MW and with the beam power enhancement to 2.3~MW that would be enabled by Project X. Plot is from M. Bass.}
\label{fig:pxcomp}
\end{figure}

\paragraph{Other Ideas for Future Experiments}:
Neutrino factories remain an interesting potential source of neutrinos for the farther future.
 The International Design Study for a Neutrino Factory (the IDS-NF)\cite{neu:NF:2011aa}
has a baseline neutrino factory design, 
involving a high-energy ($E_\mu$=25 GeV) two-baseline facility.
It remains the best facility to accurately measure the remaining parameters in the 3 $\nu$ mixing parameter space, if it turns out that the value of $\sin^22\theta_{13}$ is actually 3$\sigma$ below the central value of the latest global fit ($\sin^2 2\theta_{13} > 0.005$). In the case of large $\sin^22\theta_{13}$ ($\sin^2 2\theta_{13} > 0.01$),
another possibility could be more effective:
a single baseline (L = 2000 km), lower energy facility ($E_\mu$=10 GeV), termed the LENF, that uses a 100 kT magnetized iron detector (MIND) as the far detector. 
A recently developed strategy (again in the context of $\sin^2\theta_{13}$ being large) is for a phased or staged approach for the LENF $-$ the low-luminosity-low-energy Neutrino Factory, L$^3$ENF.  This facility would require neither a proton driver to begin
the physics program, nor muon ionization cooling.  It could enable an
earlier start and presents an obvious upgrade path.

The Very-Low-Energy Neutrino Factory (VLENF)~\cite{vlenf} is another idea for a near-term, relatively inexpensive facility that could address large $\Delta m^2$ neutrino oscillations, make precision neutrino cross section measurements in the few GeV range with a well-known beam, and also serve as a technology test demonstration.  The facility is very simple and consists of a conventional target station, a capture and transport section, and 
an injector feeding into a race track ring with a straight length of between 50-75~m.

An additional idea to improve $CP$ violation sensitivity is DAE$\delta$ALUS~\cite{Alonso:2010fs}, which is complementary to LBNE and which may be 
a possibility for an inverse-beta-decay-sensitive
far detector.
DAE$\delta$ALUS combines multiple pion decay-at-rest neutrino sources with a water Cherenkov detector doped with gadolinium, or possibly a scintillator detector.   The experimental configuration includes multiple cyclotron
complexes at baselines ranging from 1-20~km.  The experiment uses only $\bar{\nu}_e$ appearance,
comparing oscillation probabilities at different $L/E$ and exploiting the $L$-dependence of the $CP$-violating interference terms in the three-neutrino oscillation probability.
This experimental setup allows for a powerful search for $CP$ violation in three-neutrino mixing.  Furthermore, there are further improvements in combination with a conventional long-baseline measurement~\cite{Alonso:2010fy}.  This kind of stopped-pion source is also valuable for other kinds of physics, including sterile neutrino searches.
DAE$\delta$ALUS requires development of new high-power cyclotrons (which we note may have industrial applications).
A promising concept~\cite{Calabretta:2011nr} 
involves acceleration of H$_2^+$ ions, which offers excellent prospects for meeting the ambitious requirements. 

\section{Conclusions}

The Standard Model has been one of the most successful theoretical descriptions of Nature in the history of humankind.  Decades of precision tests have revealed only one concrete violation of the Standard Model:  the existence of non-zero neutrino mass.     While many experiments continue to look for other Standard-Model-violating processes, it is clear that continued study of the neutrino sector is of the upmost importance.    
 
Compared to the other fermions, the elusive nature of the neutrino has made it extremely difficult to study in detail.    While the field of neutrino physics has been making continuous progress over many decades, the rate of progress in recent years has been impressive.   The current generation of neutrino experiments is producing important results that help us to better understand the neutrino sector.    In some cases, these experiments have uncovered intriguing anomalies that require additional study and will prompt future experiments.  Furthermore, the current generation of neutrino experiments is providing advances in detector technology and analytical techniques needed for the next generation of neutrino experiments.   
 
This synergy $-$ the physics of the neutrino as a key to understanding the fundamental nature of the physical world, along with technological advances in experimental techniques $-$ make this an exciting time for neutrino physics.   The coming decade will provide us with an opportunity to answer some of the most fundamental and important questions of our time:  Are neutrinos Majorana or Dirac particles?    Is there $CP$ violation in the lepton sector?   Does the small, but non-zero neutrino mass couple to a mass scale that is far beyond what we can hope to reach in colliders?    Although these questions have been asked for many years, we now have opportunities to finally answer some of them.  
 
The coming decade promises significant experimental progress around the world.  In the search for neutrinoless double-beta decay, a number of experiments rely on complementary isotopes and experimental techniques.  The next generation of $\sim$100~kg-class $0\nu\beta\beta$ experiments should reach effective masses in the 100 meV range; beyond that, there are opportunities for multi-ton-class experiments that will reach $<$10~meV effective mass sensitivity, pushing below the inverted hierarchy region.  The next-generation tritium beta decay kinematic experiment, KATRIN, will push limits a factor of 10 beyond the current best ones; innovative new ideas may help to go beyond.  Long-baseline neutrino oscillation experiments will clarify the neutrino mass hierarchy and search for $CP$ violation; these require new high-power beams and large underground detectors.  Both T2K and MINOS are currently running, with NO$\nu$A expected to begin in 2014.  Reactor experiments will also continue to take data this decade. 
There is vigorous worldwide activity towards planning for large-scale next-generation long-baseline efforts.  There are exciting opportunities for the US to take leadership in this arena with LBNE, and beyond that, Project X, for increased neutrino intensity at several beam energies.  Given the challenges associated with precision measurements in the neutrino sector, complementary baselines, sources and detector techniques will be needed to help further understand the nature of $CP$ violation in the neutrino sector.  Smaller experiments will also help address some of the remaining anomalies and hints for new physics beyond the three-flavor paradigm.
 
The diversity of physics topics that can be probed through the neutrino sector is significant, and the interplay between neutrino physics and other fields is vast.   Neutrinos can and will provide important information on structure formation in the early universe; Earth, solar and supernova physics; nuclear properties; and rare decays of charged leptons and hadrons.   In other words, the neutrino sector sits at the nexus of the worldwide effort in energy, intensity and cosmic frontier physics.
 
Finally, the unique physics potential and technological advancements have conspired to produce a fertile environment for new ideas for improved measurements and new techniques.  This provides an important training ground for the next generation of scientists and engineers, motivated and excited about ground-breaking experiments that can benefit from their contributions.


\def\Discussion{\setlength{\parskip}{0.3cm}\setlength{\parindent}{0.0cm}
     \bigskip\bigskip      {\Large {\bf Discussion}} \bigskip}\def\speaker#1{{\bf #1:}\ }
\def\endDiscussion{} 



\chapter{Proton Decay}
\label{chap:pdk}

\begin{center}\begin{boldmath}


\begin{center}

Conveners: R.~Brock, C.K.~Jung, C.E.M.~Wagner

K.S.~Babu,
R.~Dermisek,
B.~Dutta,
P.~Fileviez Perez,
K.S.~Ganezer,
I.~Gogoladze,
Y.~Kamyshkov,
T.~Lachenmaier,
P.~Langacker,
W.J.~Marciano,
R.N.~Mohapatra,
P.~Nath,
Y.~Obayashi,
J.C.~Pati,
S.~Raby,
J.L.~Raaf,
Q.~Shafi,
G.C.~Stavenga,
R.~Svoboda,
R.J.~Wilson

\end{center}




\end{boldmath}\end{center}


\section{Theoretical Perspectives on Proton Decay}\label{sec:pdk_theory}

\newcommand{\ka}{\kappa}
\newcommand{\mcA}{{\mathcal A}}
\newcommand{\mcD}{{\mathcal D}}
\newcommand{\mcF}{{\mathcal F}}
\newcommand{\mcL}{{\mathcal L}}
\newcommand{\mcN}{{\mathcal N}}
\newcommand{\mcO}{{\mathcal O}}
\newcommand{\mcV}{{\mathcal V}}
\newcommand{\mcW}{{\mathcal W}}
\def\Dbarhat{\hat{\makebox[0pt][c]{\raisebox{0.5pt}[0pt][0pt]{$\not$}}\mcD}}
\def\Fbarhat{\hat{\makebox[0pt][c]{\raisebox{0.5pt}[0pt][0pt]{$\not$}}F}}

\newcommand{\PRD}[3]{Phys. Rev. {\bf D#1}, #2 (#3)}
\newcommand{\PLB}[3]{Phys. Lett. {\bf B#1}, #2 (#3)}
\newcommand{\PRL}[3]{Phys. Rev. Lett. {\bf#1}, #2 (#3)}
\newcommand{\NPB}[3]{Nucl. Phys. {\bf B#1}, #2 (#3)}
\newcommand{\vn}{{\vec{n}}}
\newcommand{\vm}{{\vec{m}}}
\newcommand{\si}{\sigma}
\newcommand{\hmu}{{\hat\mu}}
\newcommand{\hnu}{{\hat\nu}}
\newcommand{\hrho}{{\hat\rho}}
\newcommand{\hh}{{\hat{h}}}
\newcommand{\hg}{{\hat{g}}}
\newcommand{\hk}{{\hat\kappa}}

\def\F{{\bf F}}
\def\A{{\bf A}}
\def\J{{\bf J}}
\def\af{{\bf \alpha}}
\def\beqn{\begin{eqnarray}}
\def\eeqn{\end{eqnarray}}

\def\dspace{\baselineskip = .30in}
\def\beq{\begin{equation}}
\def\eeq{\end{equation}}
\def\bea{\begin{equation}}
\def\eea{\end{equation}}
\def\pl{\partial}
\def\na{\nabla}
\def\al{\alpha}
\def\bt{\beta}
\def\Ga{\Gamma}
\def\ga{\gamma}
\def\de{\delta}
\def\De{\Delta}
\def\da{\dagger}
\def\ka{\kappa}
\def\si{\sigma}
\def\Si{\Sigma}
\def\te{\theta}
\def\La{\Lambda}
\def\lam{\lambda}
\def\Om{\Omega}
\def\om{\omega}
\def\ep{\epsilon}
\def\non{\nonumber}
\def\sq{\sqrt}
\def\sqg{\sqrt{G}}
\def\sp{\supset}
\def\sb{\subset}
\def\l{\left (}
\def\r{\right )}
\def\lq{\left [}
\def\rq{\right ]}
\def\fr{\frac}
\def\la{\label}
\def\hs{\hspace}
\def\vs{\vspace}
\def\inf{\infty}
\def\ran{\rangle}
\def\lan{\langle}
\def\ov{\overline}
\def\tl{\tilde}
\def\tm{\times}
\def\lrar{\leftrightarrow}
\def\orvec{\overrightarrow}


The search for proton decay addresses one of the  most important open questions in high energy physics.  Its discovery  would provide a wealth of insight into physics at the deepest level, bearing on the question of a unification of the fundamental forces of nature.
Within the Standard Model, proton stability is associated with a conservation law, the conservation of baryon number. Protons and neutrons carry one unit of baryon charge, and there is no other lighter particle that is charged under baryon number in Nature.  The proton being lighter than the neutron,  conservation of baryon number forbids the decay of the proton. But baryon number is just an accidental symmetry of the Standard Model, 
which is already violated  at the quantum level. Since it is not fundamental, it is expected to be violated in any extension of the Standard Model.   If baryon charge conservation were violated, then protons might decay in a variety of ways, for example via the processes $p \rightarrow e^+ \pi^0$ or $p\rightarrow  e^+ \gamma$. If protons were to decay rapidly, then all chemistry and life as we know it would come to an abrupt end. 

Current limits on proton decay suggest that if baryon number 
is violated, the natural physics scale associated with the baryon
number violating processes must be larger than about $10^{15}$~GeV. 
The search for proton decay at a large underground detector can therefore
dramatically shed light on  fundamental aspects of the laws of
nature:
\begin{itemize}
\item Improved studies of proton decay would enable us to probe
nature at the highest energy scale of order $10^{16}$ GeV, or equivalently 
$10^{-30}$ cm---something that would not be possible by any
other means.
\item  The discovery of proton decay would have profound significance
for the idea of grand unification, which   
proposes
to unify the basic constituents of matter
and also the three basic forces---the strong, weak and
electromagnetic. Grand Unified Theories (GUTs) predict
that the proton must decay,  albeit with a long
lifetime exceeding $10^{30}$ years. While proton decay has yet to be
seen, the grand unification idea has turned out to be spectacularly
successful as regards its other predictions. These include in
particular the phenomena of ``coupling unification," amounting to an
equality of the strengths of the three forces at very high energies,
which has been verified to hold, in the context of low energy supersymmetry,  
at an energy scale of $10^{16}$ GeV.
Furthermore, a class of grand unified models
naturally predicts that the heaviest of the three neutrinos should
have a mass in the range of one hundredth to one electron volt, with
the next-to-heaviest being an order of magnitude lighter, the two being
quantum-mechanical mixtures of what one calls nu-mu and nu-tau. This
is in full accord with the discovery of neutrino oscillation. 
In this sense,
proton decay now remains the missing piece of evidence for grand
unification.
\end{itemize}

One can in fact argue,
within a class of well-motivated ideas on grand unification, that
proton decay should occur at accessible rates, with a lifetime of
about $10^{35}$ years, 
for
protons decaying into a positron plus a neutral pion, and a lifetime of
less than a few $\times 10^{34}$ years  for protons decaying into 
an anti-neutrino and a positively charged $K$-meson. 
The most stringent limits on proton lifetimes now come from 
Super-Kamiokande~\cite{raaf}.
  For  the two important decay modes mentioned above, they  are: 
\begin{equation}
\tau(p \rightarrow e^+\pi^0) > 1.4 \times 10^{34} \; {\rm yrs}, \;\;\;\;\;\;\;\;\;
 \tau(p \rightarrow \bar \nu K^+  )>  4 \times 10^{33} \; {\rm yrs}.
 \end{equation}
These well-motivated models   then predict the observation of proton decay 
if one can improve the current sensitivity (of
Super-Kamiokande) by a factor of five to 10. This is why an improved
search for proton decay, possible only with a large  underground detector,
is now most pressing.


\subsection{Grand Unification and Proton Decay}

\begin{figure}[h!]
\begin{center}
\includegraphics[width=12cm]{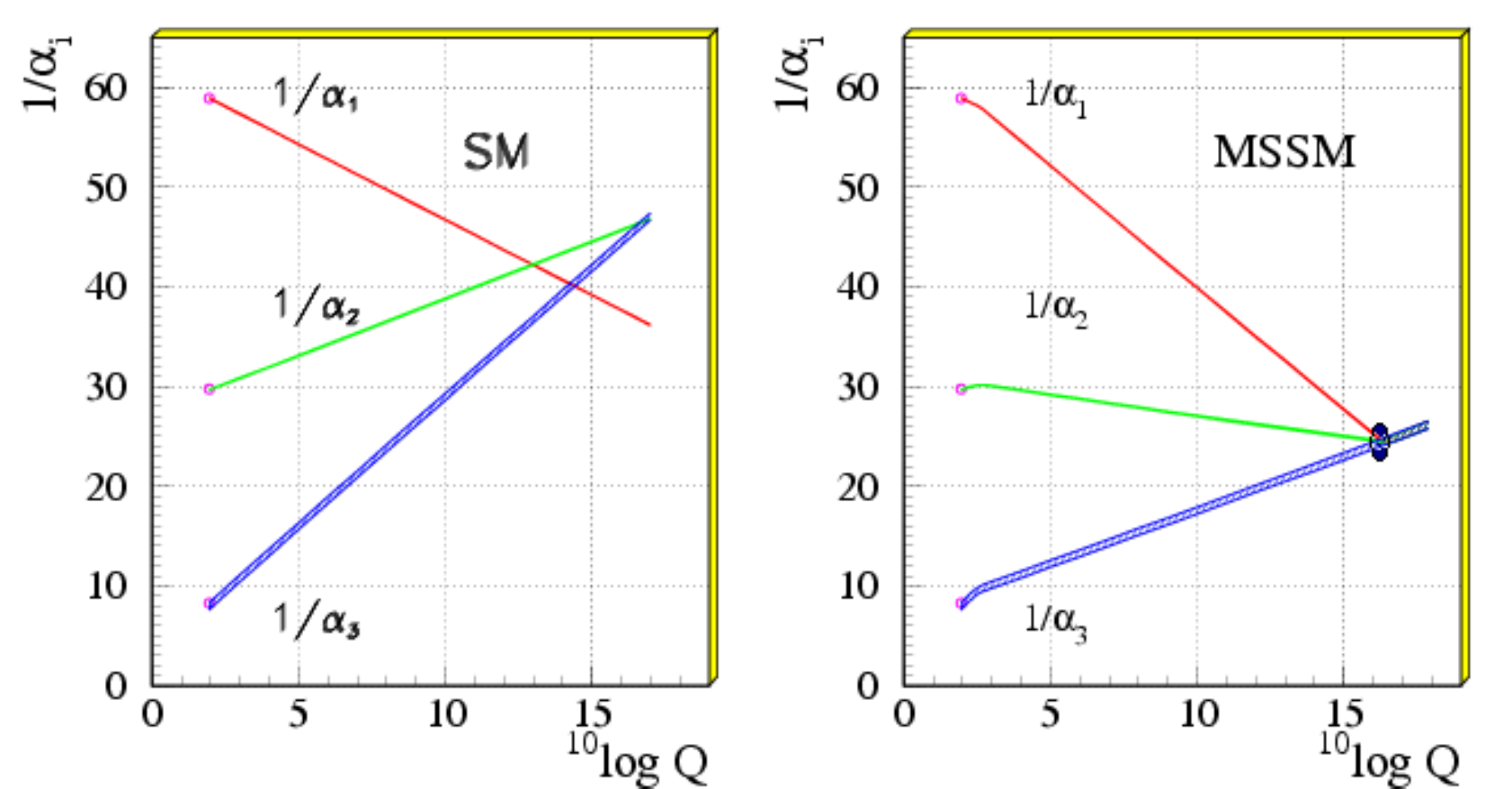}
\caption{\label{unif-diag} Evolution of the three gauge couplings $\alpha_i$ with momentum $Q$:  Standard Model (left panel) and Minimal
Supersymmetric Standard Model (right panel) }\label{unif}
\end{center}
\end{figure}

The decay of the proton~\cite{Pati} is one of the most  exciting predictions of the idea of the unification of matter and of forces at the very highest energy scales~\cite{GG},\cite{Georgi:1974yf}, which is motivated on several grounds (for a review, see \cite{Langacker:1980js}).  For example, the experimental observation that electric charge is quantized, together with  $|Q_{\rm proton}| = |Q_{\rm electron}|$ (to better than 1 part in $10^{21}$), has a natural explanation in GUTs owing to their non--Abelian nature. The miraculous cancellation of chiral anomalies that occurs among each family of quarks and leptons has a symmetry--based explanation in GUTs.  Furthermore, GUTs provide a natural understanding of the quantum numbers of quarks and leptons.  
With the grouping of quarks with leptons, and particles with antiparticles, in a common GUT multiplet, these theories predict that baryon number would be violated and that the proton must decay.  Finally, with the assumption of low energy supersymmetry, motivated by the naturalness of the Higgs boson mass, the strong, weak and electromagnetic gauge couplings are found to unify nicely at a scale $M_X \approx 2 \times 10^{16}$ GeV, the scale of interest for proton decay 
(see right panel of Fig.~\ref{unif-diag}). It should be noted that low energy supersymmetry would allow baryon and lepton number violating interactions of the type $QLD^c$, $U^c D^c D^c$ and $L L E^c$ in the superpotential
($Q$, $L$ etc are the quark and lepton superfields).  However, these operators can be eliminated by imposing R-parity
conservation~\cite{Rparity},\cite{Farrar:1978xj}, which could arise from a gauged $B-L$ 
symmetry~\cite{bminusl}  that  occurs in many unified theories.



$SU(5)$ is the simplest grand unified group, and it turns out to be the most predictive as regards proton lifetime and the unification of the three gauge couplings, owing to small GUT scale threshold effects.  The minimal non-supersymmetric version of $SU(5)$~\cite{GG} has already been excluded by the experimental lower limit on $p \rightarrow e^+ \pi^0$ lifetime and the mismatch of the three gauge couplings when extrapolated to high energies 
(see left panel of Fig.~\ref{unif-diag}).  Yet low energy supersymmetry, which is independently motivated by the naturalness of the Higgs boson mass, provides a simple solution to these problems of $SU(5)$, as it increases the prediction of the lifetime for the decay process $p \rightarrow e^+ \pi^0$ due to the larger value of $M_X$ and also corrects the unification mismatch 
(see right panel of Fig.~\ref{unif-diag})~\cite{Langacker:1980js}. 

\begin{figure}[h!]
\begin{center}
\includegraphics[width=14cm]{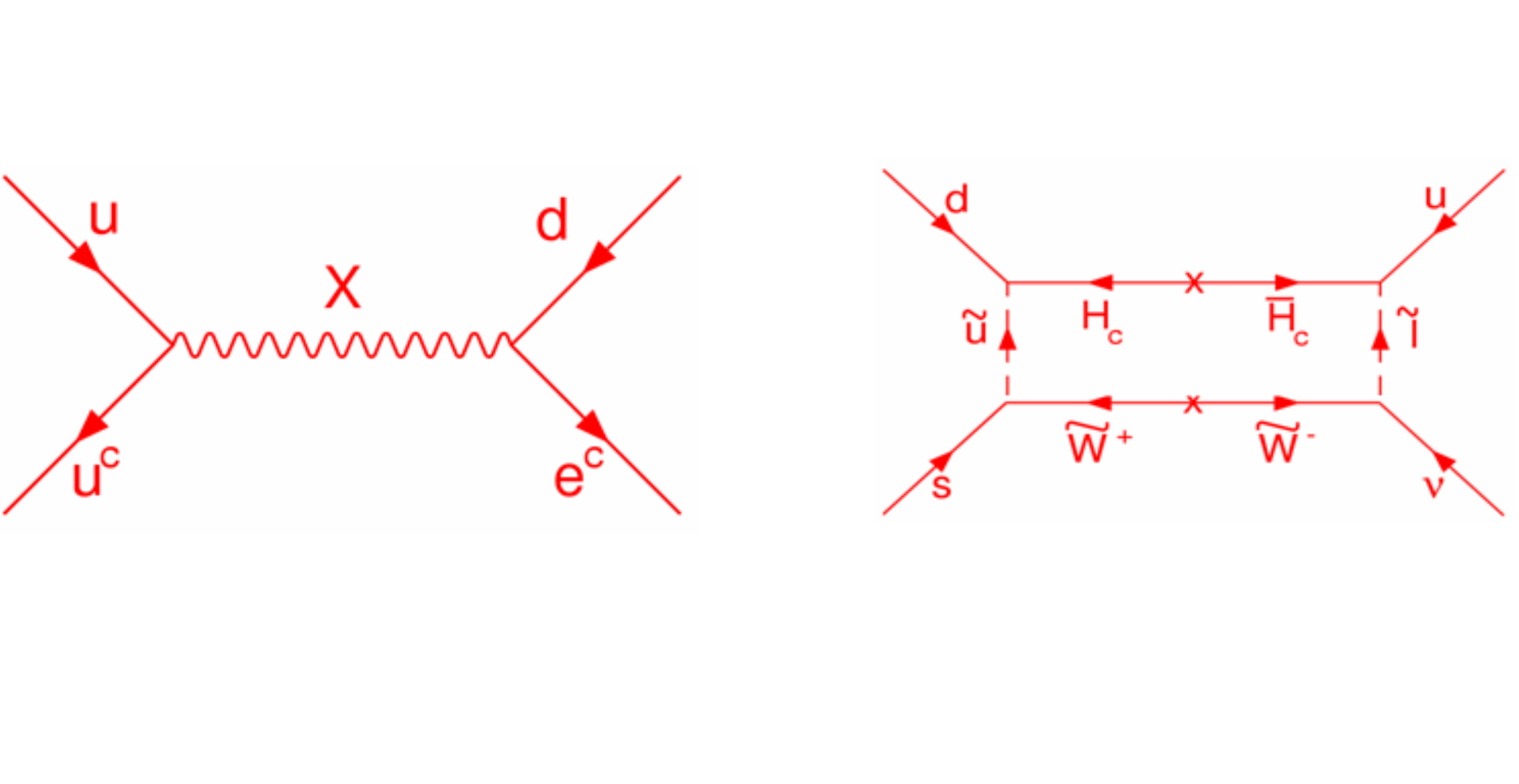}
\vspace*{-0.8in}
\caption{\label{pdk-diag} Diagrams inducing proton decay in SUSY GUTs.  $p \rightarrow e^+ \pi^0$ mediated by $X$ gauge boson (left), and
$p \rightarrow \overline{\nu} K^+$ mediated by colored Higgsino (right).}\label{pdk-diagrams}
\end{center}
\end{figure}

Supersymmetric grand unified theories (SUSY GUTs)~ \cite{Dimopoulos:1981zb},\cite{Dimopoulos:1981dw},\cite{sugragut}--\cite{SUGRA20} are natural extensions of the Standard Model that preserve the attractive features of GUTs such as quantization of electric charge,  and lead to the unification of the three gauge couplings. They also 
explain the  existence of the  weak scale, which is much smaller than the GUT scale,
and provide a dark matter candidate in the lightest SUSY particle. Low energy SUSY brings in a new twist to proton decay, however, as it predicts a new decay mode $p \rightarrow \overline{\nu} K^+$ that would be mediated by the colored Higgsino~\cite{Sakai:1981pk},\cite{Weinberg:1981wj}, the GUT/SUSY partner of the Higgs doublets (see Fig. \ref{pdk-diag}, right panel).  Typically, the lifetime for this mode in many models is shorter than the current experimental lower limit.


In order to evaluate the lifetimes for the $p \rightarrow \overline{\nu}K^+$ and $p \rightarrow e^+ \pi^0$ decay modes in SUSY $SU(5)$~\cite{Hisano:1992jj}, a symmetry breaking sector and a consistent Yukawa coupling sector must be specified.  In $SU(5)$, one family of quarks and leptons is organized as $\{10 + \overline{5} + 1\}$, where $10 \supset \{Q, u^c, e^c\},\overline{5} \supset \{d^c, L\}$, and $1 \sim \nu^c$.  $SU(5)$ contains 24 gauge bosons, 12 of which are the gluons, $W^\pm, Z^0$ and the photon, while the remaining 12 are the $(X,Y)$ bosons that transform as $(3,2,-5/6)$ under $SU(3)_c \times SU(2)_L \times U(1)_Y$.  These bosons have both diquark and leptoquark couplings, which lead to baryon number violating processes. The diagram leading to the decay $p \rightarrow e^+ \pi^0$ is shown in Fig. \ref{pdk-diag}, left panel.  $SU(5)$ breaks down to the Standard Model symmetry in the supersymmetric limit by employing a ${24_H}$ Higgs boson.  Additionally, a $\{5_H + \overline{5}_H\}$ pair of Higgs bosons is employed, for electroweak symmetry breaking and the generation of quark and lepton masses.  

The masses of the super-heavy particles of the theory can be related to low energy observables in minimal SUSY $SU(5)$
via the renormalization group evolution of the three gauge couplings, which depends through the threshold correction on  
$M_T$,   the mass of the color
triplet Higgsinos which mediate $p \rightarrow \overline{\nu} K^+$ decay.
In general, agreement with the experimental value of $\alpha_3(M_Z) = 0.1184\pm 0.0007$ demands the color triplet mass to be  lower than the GUT scale. This tends to lead to a rate of proton decay into $ \bar{\nu} K^+$ which is in disagreement with observations~\cite{SU5}. 

It should be noted, however,  that the Yukawa sector of minimal SUSY $SU(5)$ enters in a crucial way in the rate of proton decay into $K^+\bar{\nu}$. Minimal $SU(5)$  leads to the relation $M_d = M_\ell^T$, relating the down quark and charged lepton mass matrices.  Consequently, the asymptotic relations $m_b^0 =m_\tau^0,~m_s^0 = m_\mu^0,~m_d^0 = m_e^0$ follow for the masses of quarks and leptons at the GUT scale.  Although the first of these relations agrees reasonably well with observations once it is extrapolated to low energies, the relations involving the two light family fermions are not in agreement with observations. 
Rather than concluding that the minimal SUSY $SU(5)$ is excluded, it would be beneficial to consider the simplest modification that corrects these bad mass relations, and then explore its prediction for proton decay.  
One possibility is to add higher dimensional operators to the minimal theory. 
An alternative   possibility, which appears to be simple and predictive, is to add a vector-like pair of $\{5+\overline{5}\}$ fermions~\cite{Babutalk}.  The quarks and leptons from these multiplets can mix differently with the usual quarks and leptons, and thereby correct the bad mass relations.  
Optimizing these mixings so as to enhance the dominant $p \rightarrow
\overline{\nu} K^+$ lifetime, approximate upper limits for the various
partial lifetimes are found: $\tau(p \rightarrow \overline{\nu} K^+)
\sim 4 \cdot 10^{33}$ yrs, $\tau(p \rightarrow \mu^+ K^0) 
\sim 6 \cdot 10^{33}$ yrs, and $\tau(p \rightarrow \mu^+ \pi^0) 
\sim 1 \cdot 10^{34}$ yrs.  In obtaining these numbers, the SUSY particles
are assumed to have masses below 2 TeV, and the unification scale is taken to be at
least a factor of 50 below the Planck scale, so that quantum gravity effects
remain negligible. Here lattice calculations for the proton decay matrix
elements have been used~\cite{Aoki}~\footnote{Alternative computational methods, which suggest some level of suppression of these operators~\cite{Martin:2011nd}, are affected by large theoretical uncertainties}.  Since the predicted rates are close
to the present experimental limits, these models can be tested by improving
the current sensitivity for proton lifetime by a factor of ten.

\boldmath  
\noindent{\bf $SO(10)$ Unification and Proton Decay}:\unboldmath~Models
based on $SO(10)$ gauge symmetry are especially attractive since quarks, leptons, anti-quarks, and
anti-leptons of a family are unified  in a single ${\bf 16}$-dimensional spinor representation of the gauge group~\cite{GeorgiSO(10)}.
This explains  the quantum numbers (electric charge, weak charge, color charge)
of fermions, as depicted in Table~1.  $SO(10)$ symmetry contains five independent
internal spins, denoted as $+$ or $-$ signs (for spin--up and spin--down) in Table 1.  Subject to the condition that the number of
down spins must be even, there are 16 combinations for the spin orientations, each corresponding to one fermionic degree. The first three spins denote color charges, while the last two are weak charges.
In addition to the three independent color spins ($r,b,g)$, there is a fourth color (the fourth row), identified as lepton number~\cite{Pati}.
The first and the third columns (and similarly the second and the fourth) are left--right conjugates. Thus $SO(10)$ contains
quark--lepton symmetry as well as parity.  Thus a right--handed neutrino state ($\nu^c$)
is predicted because it is needed to complete the multiplet. Being a singlet of the Standard Model, it naturally acquires a superheavy
Majorana mass and leads in a compelling manner to the generation of light neutrino masses via the seesaw mechanism. 
Hypercharge of each fermion follows from the formula $Y = \frac{1}{3} \Sigma (C) - \frac{1}{2} \Sigma (W)$, where  $\Sigma (C)$ is the summation of color spins (first three entries) and $\Sigma (W)$ is the sum of weak spins (last two entries).
This leads to quantization of hypercharge, and thus of electric charge.
Such a simple organization of matter is remarkably beautiful and can be argued as a hint in favor of GUTs based on $SO(10)$.
\begin{table}[h]
{\footnotesize
\begin{center}
\begin{tabular}{||c|c|c|c||}\hline\hline
$u_r:~\{-++~+-\}$ & $d_r:~ \{-++~-+\}$ & $u^c_r:~\{+--~++\}$ & $d^c_r:~ \{+--~--\}$ \\
$u_b:~\{+-+~+-\}$ &  $d_b:~ \{+-+~-+\}$ & $u^c_b:~\{-+-~++\}$ &  $d^c_b:~ \{-+-~--\}$ \\
$u_g:~\{++-~+-\}$ &  $d_g:~\{++-~-+\}$ & $u^c_g:~\{--+~++\}$ &  $d^c_g:~ \{--+~--\}$ \\
$~\nu:~\{---~+-\}$ &  $~e:~~ \{---~-+\}$ & $~\nu^c:~\{+++~++\}$ &  $~e^c:~ \{+++~--\}$ \\ \hline\hline
\end{tabular}
\end{center}
\caption{Quantum numbers of quarks and leptons. The first three 
signs refer to color charge, and the last two to weak charge.
To obtain hypercharge, use $Y = \frac{1}{3}\Sigma(C)-\frac{1}{2}\Sigma (W)$.}
}
\end{table}

As in the case of $SU(5)$, when embedded with low energy supersymmetry so that the mass of the Higgs boson is stabilized,
the three gauge couplings of the Standard Model (SM) nearly unify at an energy scale of $M_X \approx 2 \cdot 10^{16}$ GeV in $SO(10)$ models.
The light neutrino masses inferred from neutrino oscillation data ($m_{\nu_3} \sim 0.05$ eV) suggest the Majorana mass of the heaviest of the
three $\nu^c$'s to be $M_{\nu^c} \sim 10^{14}$ GeV, which is close to $M_X$. In a class of $SO(10)$ models discussed further here, $M_{\nu^c} \sim M_X^2/M_{\rm Pl} \sim 10^{14}$ GeV quite naturally.  The lepton number violating decays of $\nu^c$ can elegantly explain the observed baryon asymmetry of the universe via leptogenesis. Furthermore, the unified setup of quarks and leptons in $SO(10)$ serves as a powerful framework in realizing predictive schemes for the masses and mixings of all fermions, including the neutrinos, in association with flavor symmetries in many cases.  All these features make SUSY $SO(10)$ models compelling candidates for the study of proton decay.

Even without supersymmetry, $SO(10)$ models are fully consistent with the unification of the three gauge couplings and the experimental
limit on proton lifetime, unlike non--SUSY $SU(5)$.  This is possible since $SO(10)$ can break to the SM via an intermediate symmetry such as $SU(4)_C \times
SU(2)_L \times SU(2)_R$.  Such models would predict that a proton would decay predominantly to $e^+ \pi^0$ with a lifetime in
the range $10^{33} - 10^{36}$ yrs, depending on which intermediate gauge symmetry is realized~\cite{Mohapatra}.

In SUSY $SO(10)$ models, symmetry breaking can occur in two interesting ways.  One type adopts a ${\bf 126}$ of Higgs, a tensor, which couples directly to the $\nu^c$ states and generates large Majorana masses for them.  This class of models has the attractive feature that
the $R$--parity of the Minimal Supersymmetry Standard Model (MSSM), which is 
so crucial for identifying the lightest SUSY particle as the dark matter candidate, is an
automatic symmetry, which is part of $SO(10)$.  
In this category, a class of minimal $SO(10)$ models employing a single $\overline{\bf 126}$ 
and a single ${\bf 10}$ of Higgs bosons that couple to the fermions has been developed~\cite{ProtonSO(10)}.  
Owing to their minimality, 
these models are quite predictive as regards the neutrino  mass spectrum and oscillation angles.  Small quark mixing angles and large neutrino oscillation angles emerge simultaneously in these models, despite their parity at the fundamental level.  The neutrino oscillation angle
$\theta_{13}$ is predicted to be  large in these models, $\sin^2\theta_{13} \geq 0.02$.  Proton decay studies of these models~\cite{SO10Pati} show that at least some of the modes among $p \rightarrow \overline{\nu} \pi^+$, $n \rightarrow \overline{\nu} \pi^0$,
$p \rightarrow \mu^+ \pi^0$ and $p \rightarrow \mu^+ K^0$ have inverse decay rates of order  $10^{34}$ yrs, 
while that for $p \rightarrow e^+ \pi^0$ is of order $10^{35}$ yrs.

The second type of SUSY $SO(10)$ model adopts a set of low-dimensional Higgs fields for symmetry breaking~\cite{Dimopoulos:1981zu}--\cite{Babu:2010ej}.  
This includes
spinors ${\bf 16} + {\bf \overline{16}}$, vectors ${\bf 10}$ and an adjoint ${\bf 45}$ which acquires a vacuum expectation value along
the $B-L$ direction of the form $\lan A\ran ={\rm i}\si_2\otimes {\rm Diag}\l a, ~a,~a,~0,~0\r$.  This has quite an interesting effect~\cite{Dimopoulos:1981zu}, \cite{Babu:1994dq}, 
since it would leave a pair of Higgs doublets from the {\bf 10} naturally light, while giving superheavy mass to the color triplets -- a feature
that is necessary to avoid rapid proton decay -- when the ${\bf 45}$ couples to the vector ${\bf 10}$--plets.  These models predict that
the heaviest of the light neutrinos has a mass that is naturally of order one tenth of an eV, consistent with atmospheric neutrino oscillation data.  This setup
also allows for a predictive system for fermion masses and mixings, in combination with a flavor symmetry.  
Models that appear 
rather different in the fermion mass matrix sector result in very similar predictions for $p \rightarrow \overline{\nu} K^+$ inverse decay rate, which 
has been found to be typically shorter than a few times $10^{34}$ yrs.

Recent work in the same class of SUSY $SO(10)$ models has shown that there is an interesting correlation between the inverse
decay rates for the $p \rightarrow \overline{\nu} K^+$ and $p \rightarrow e^+ \pi^0$ modes.  The amplitude for the former
scales inversely as the three-halves power of that for the latter, with only a mild dependence on the SUSY spectrum in the
constant of proportionality~\cite{Babu:2010ej}.  This intriguing correlation leads to the most interesting result that the empirical lower limit
of the lifetime for $p \rightarrow \overline{\nu} K^+$ decay provides a theoretical upper limit on the lifetime for $p \rightarrow
e^+ \pi^0$ decay, and vice versa, as noted below.
\begin{eqnarray}
\tau(p \rightarrow e^+ \pi^0) & \leq & 5.7 \times 10^{34} ~{\rm yrs}\,, \nonumber \\
\tau(p \rightarrow \overline{\nu} K^+) & \leq & \left(4 \times 10^{34} ~{\rm yrs}\right) \cdot \left(\frac{m_{\tilde{q}}}{1.8~{\rm TeV}}\right)^4
\cdot \left(\frac{m_{\tilde{W}}}{190~{\rm GeV}}\right)^2 
\cdot\left(\frac{3}{\tan\beta} \right)^2~.
\end{eqnarray}
These predictions are accessible to future experiments, with an improvement in current sensitivity by about a factor of 10.
%

Thus, well-motivated supersymmetric GUTs generically predict proton decay rates that can be probed by next-generation experiments. 
One could conceive, however, variations of these predictions by either cancellation of contributions from different $B$- and $L$-violating dimension-five operators~\cite{ns1}, by  suppression of Higgsino couplings with matter, by judicious choice of the flavor structure~\cite{Dutta:2004zh}, or by the largeness of the scalar masses (see, for example,~\cite{chatto}). For further studies see~\cite{pdkth:Lazarides:1980nt}--\cite{Dermisek:2001hp}, and for a connection between the inflation mechanism and the proton decay rate, see, for example,~\cite{Rehman:2009yj}. 

Let us stress in closing that an important prediction of the simple $SU(5)$ and $SO(10)$ GUTs is that proton decay modes obey the selection rule $\Delta (B-L) =0$ and are mediated by effective operators with dim=6.  
A general effective operator analysis of baryon number violation reveals two conclusions~\cite{BMS}.
One is that  $\Delta B\neq 0$ operators with   dim=7 always predict that $|\Delta(B-L)| =2$, which leads to decays such as  $n\to e^-+\pi^+$. 
A similar such class of operators is those with dim=9, which leads 
to processes such as $n-\bar{n}$ oscillations. 
%
 If these higher dimensional operators, involving mass scales that are far below the conventional GUT scale, are found to be relevant,  then  minimal SUSY GUTs would be excluded.  In this case, however, 
the empirical successes of minimal SUSY GUTs  would have to be regarded 
as fortuitous.  Searches for these $(B-L)$-violating processes 
would thus be  helpful in judging the validity of minimal SUSY GUTs.

 \subsection{Proton Decay in Extra Dimensional GUTs}

The issues of Higgs doublet-triplet splitting and GUT symmetry breaking have been addressed within four-dimensional GUTs as discussed above. 
 In higher dimensions it is possible to solve these two problems in an elegant way via 
 boundary conditions in the extra dimensions~\cite{Kawamura:2000ev}.   Moreover, the successes of SUSY GUTs can be maintained~\cite{Hall:2001xr},\cite{Hall:2002ci}.
 String theories 
 that manifest the nice features of 4D SUSY GUTs   
 have been constructed with a discrete $Z_4^R$ symmetry.   This symmetry prevents dimension 3 and 4 and 5 lepton and baryon number violating operators, which are potentially dangerous, to all orders in perturbation theory.   The $\mu$-term also vanishes perturbatively.  However non-perturbative effects will generate a $\mu$-term of order the SUSY breaking scale. 
 On the other hand,  the low energy theory is guaranteed to be invariant under matter parity.  Thus the lightest SUSY particle is stable and is a perfect dark matter candidate.

Nucleon decay in theories with a $Z_4^R$ symmetry~\cite{Gogoladze},\cite{Lee:2010gv} is dominated by dimension 6 operators which lead to the classic decay modes,  $ p \rightarrow e^+ \pi^0, \bar \nu \pi^+$  and $n \rightarrow \bar \nu  \pi^0,  e^+  \pi^-$.   The lifetime for these modes is of order  $\tau \sim \frac{M_C^4}{\alpha_G^2 m_p^5}$  where $M_C$ is the compactification scale of the extra dimension.   This scale is typically less than the 4D GUT scale, {\it i.e.},  $M_C \leq M_G \approx 3 \times 10^{16}$ GeV.   Thus the rate for nucleon decay in these modes is typically also within the reach of the next generation of experiments. Moreover, due to the different dominant decay modes, the observation of proton decay may allow one to distinguish minimal four-dimensional unification models from extra-dimensional ones.


\section{Current and Proposed Proton Decay Search
Experiments}\label{sec:pdk_exp}

\subsection{Current Proton Decay Search Experiments}\label{sec:pdk_current_exp}

The current limits for proton decay searches are dominated by results from 
the Super-Kamiokande experiment. In the past there have been several 
large underground detectors that set the lower limits 
on the partial lifetime of various decay modes. Those were IMB and SOUDAN
in the United States, Kamiokande in Japan, and Frejus in France. 
Among these, IMB and Kamiokande were water Cherenkov 
detectors, and SOUDAN and Frejus were iron tracking detectors.
Since Super-Kamiokande has far surpassed the limits set by previous 
experiments, the challenge for the next-generation detectors will be improving 
the sensitivities beyond the Super-Kamiokande limits.

In the following section we describe the current proton/nucleon decay 
search status with the Super-Kamiokande detector.


\subsubsection{Status of Super-Kamiokande Proton Decay 
Searches}

The Super-Kamiokande water Cherenkov detector has excellent capability to
search for nucleon decay.  The 22,500-ton fiducial mass of the detector has
$7.5\times10^{33}$~protons and $6.0\times10^{33}$~neutrons.  Fully contained
atmospheric neutrino interactions in the $\sim$~GeV range constitute the
background to nucleon decay searches by means of neutral- and charged-current
neutrino-nucleon interactions in the water.  The experiment has been
collecting data since 1996 and there have been four distinct data-taking
periods, called SK-I, -II, -III, and -IV.  During the SK-I,
-III, and -IV periods, $\sim$11,000 inward-facing 20-inch photomultiplier
tubes (PMTs) were distributed evenly on the entire inner detector (ID)
surface to provide 40\% photocathode coverage.  Recovery from an accident
that destroyed roughly half of the PMTs in the inner detector marked the
beginning of the SK-II period, where the remaining functional PMTs were
redistributed evenly across the ID surface, giving $\sim$20\% photocathode
coverage.  This period of reduced coverage is notable for future generation
water Cherenkov detectors because the Super-K nucleon decay and atmospheric
neutrino analyses show that the reduction in photocathode coverage does not
have a large adverse effect on the nucleon decay detection efficiency.

The two most commonly discussed decay modes are $p\rightarrow e^{+}\pi^{0}$
and $p\rightarrow \overline{\nu}K^{+}$. However, there are a large number of
other decay modes also predicted by GUTs.  Different GUTs
predict that different modes will have the dominant branching fraction, making it
critical for experiments to search in every mode that is accessible to their
respective detectors.  Non-observation of nucleon decay places strong
constraints that model-builders must evade; an equally important (but more
exciting) outcome is that the observation of differing rates in more than one
decay channel could provide enough extra information to allow distinction
among the various models of grand unification theories.

\begin{figure}[htbp]
  \begin{center}
    \includegraphics[width=0.40\textwidth]
{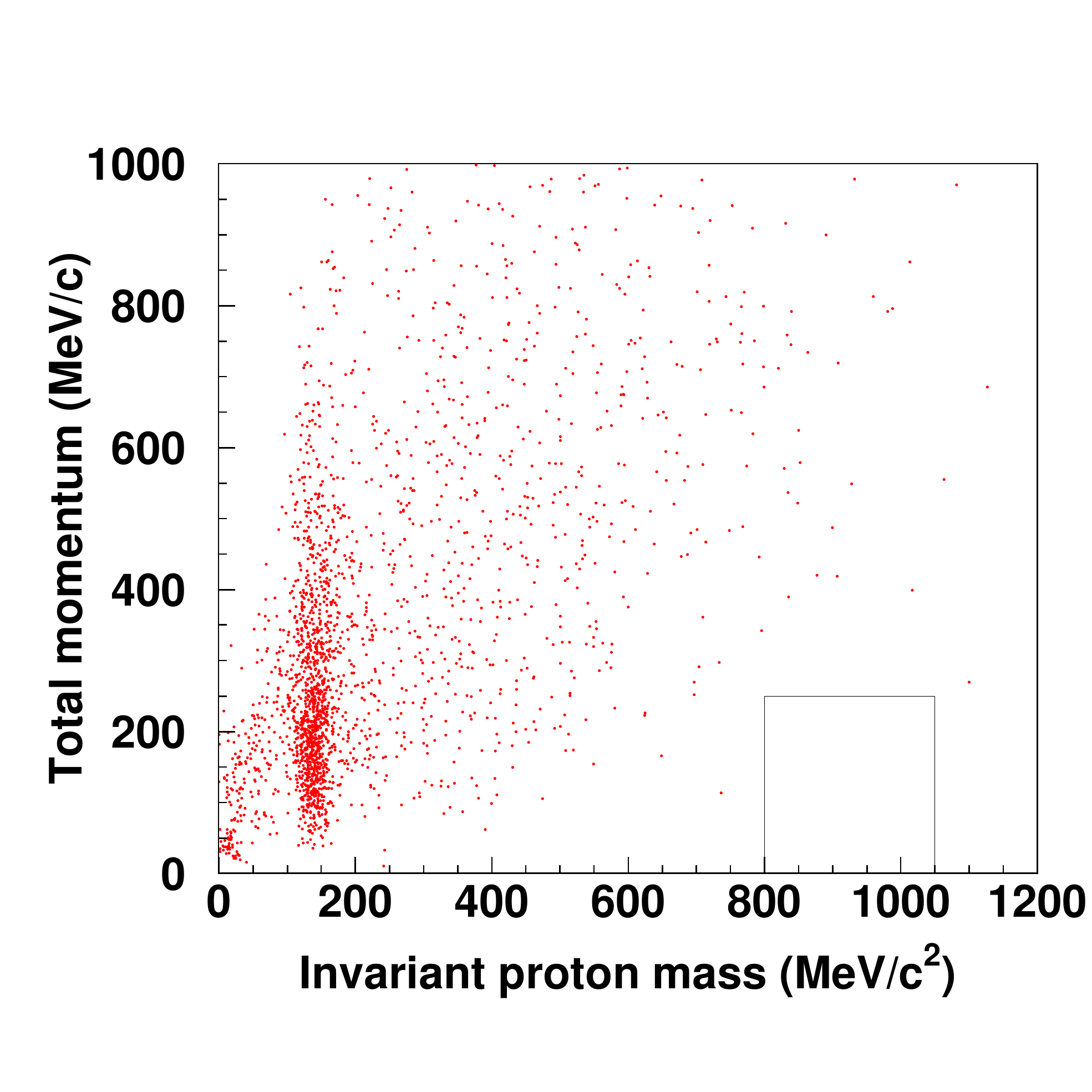}
    \includegraphics[width=0.59\textwidth]
{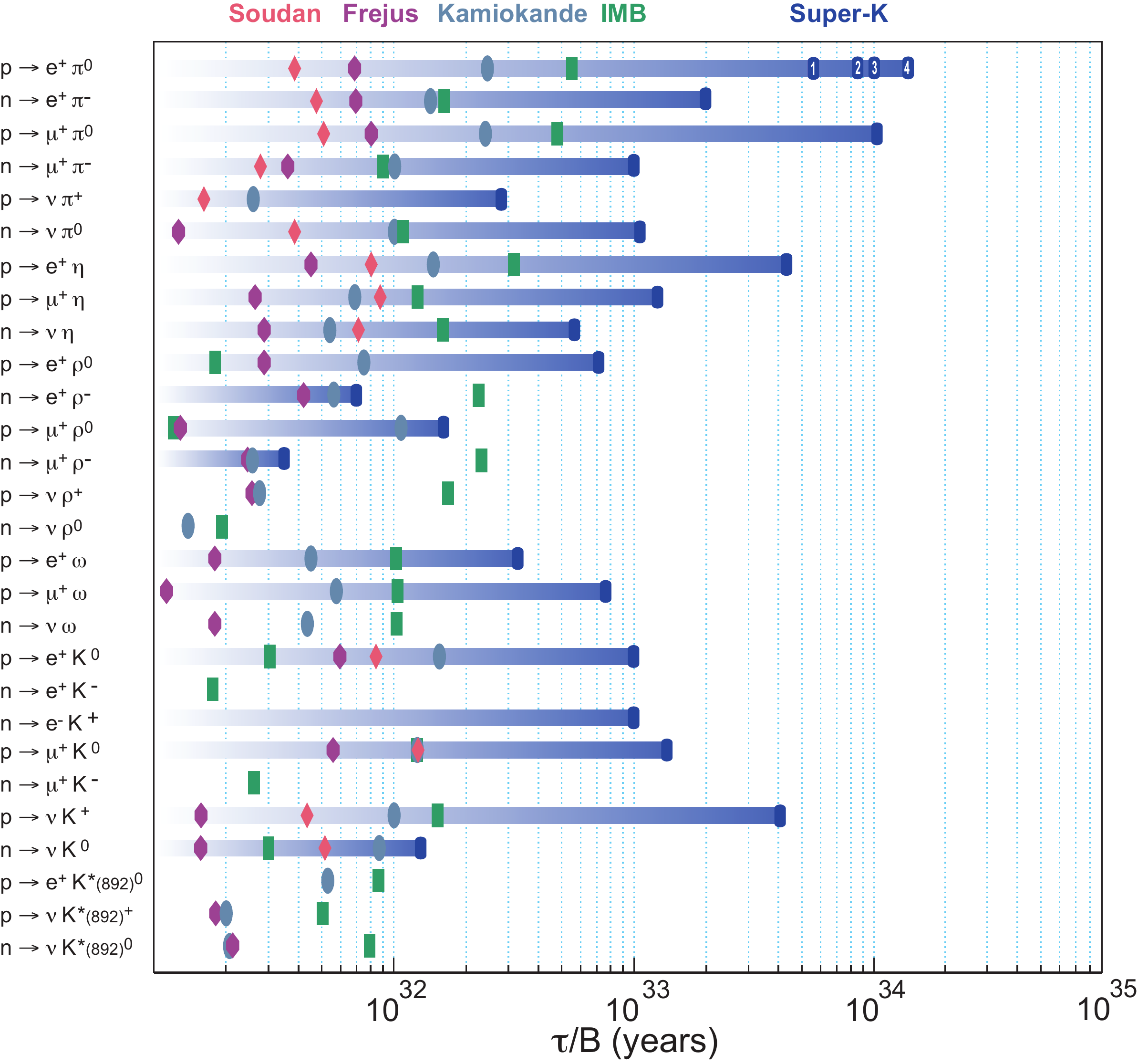}
  \end{center}
  \caption{\label{fig:superk_pdk}
Left Panel: SK-I+II+III+IV fully-contained dataset, with the $p\rightarrow  
e^{+}\pi^{0}$ signal region shown by the black box.
Right Panel: Summary of experimental proton decay searches by Super-K 
(dark blue gradient band with marker) and previous experiments, 
Soudan (pink diamonds), Frejus (purple hexagons), 
Kamiokande (light blue ovals), and IMB (light green rectangles)~\cite{raaf}.}
\end{figure}

There are several methods of searching for nucleon decay in Super-K.  The
most straightforward method is to define a set of selection criteria that
maximize the signal detection efficiency and minimize the background.  The
$p\rightarrow e^{+}\pi^{0}$ mode is a good example of this technique.  As can
be seen in Fig.~\ref{fig:superk_pdk}, the signal region is defined by a box
indicating the expected ranges of total reconstructed momentum and invariant
proton mass.  The background events that pass all other selection cuts
(atmospheric neutrino interactions) do not typically fall into the range of
momentum and invariant mass that one expects for proton decay events, making
this a low-background search mode.  Other searches with a similar technique
have been also performed using Super-K data. 

A second technique is used for some decay modes in which a low background
cannot be achieved. For these modes, a ``bump search'' is done.  An example
of this is the $n\rightarrow \overline{\nu}\pi^{0}$ mode, where one must look
for a mono-energetic peak of single $\pi^{0}$'s on top of a background
consisting mostly of neutral-current atmospheric neutrino events with a
single $\pi^{0}$.  For this type of search, understanding the shape of the
background event spectrum is critical.

A third technique, which is used for the SUSY GUT favored $p \rightarrow
\overline{\nu}K^{+}$ mode, uses a combination of the first two techniques,
and an additional trick that helps to reduce the amount of background by
tagging the mono-energetic low energy photon from the de-excitation of the
excited nucleus that is left after the decay of a proton in $^{16}$O. Using a
combination of these techniques allows the measurement to push the limit on the
proton decay lifetime further than using any of the individual methods.

Finally, decay modes that have a unique event topology can be searched for
in Super-K as well. One example of this is dinucleon decay into two kaons,
$^{16}O(pp)\rightarrow K^{+}K^{+}$. In order to improve sensitivity to the
kaon modes, an improved kaon-like particle identification algorithm was
implemented and a new multiple vertex finder that looks for the displaced
vertices of the two kaon decays was developed.  These improved tools allow
Super-K to set the strongest lower limit on the partial lifetime for
dinucleon decay to two kaons, which in turn can be used to constrain some
supersymmetric models.

The results of all Super-K single nucleon decay searches using the various
techniques described above are shown in Fig~\ref{fig:superk_pdk} compared
with measurements from past experiments. Although no signs of nucleon decay
have been seen yet, such an observation would mark a revolutionary discovery in
particle physics. We need to continue searching in the current generation of
large detectors, and high priority should be placed on nucleon decay searches
in the next generation of neutrino detectors as well.


\subsection{Proposed Proton Decay Search
Experiments}\label{sec:pdk_proposed_exp}


There have been many next-generation large underground/underwater/under-ice 
detectors proposed to search for proton decays and do neutrino 
physics since UNO\cite{UNO} was proposed at the first 
Next generation Nucleon decay and Neutrino detector 
(NNN) workshop at Stony Brook, New York in 1999. Some of these 
proposals are inactive or discontinued, 
while others are being actively discussed 
in various parts of the world. The proposed detectors can be categorized 
broadly in three distinctive technologies: water Cherenkov detectors, 
liquid argon TPCs and scintillator detectors.
Table~\ref{tab:NNN_detectors} shows a summary of these detectors categorized 
by technology and region. 

\begin{table}[htb]
\caption{NNN detector proposals categorized by technology and region.
The detector proposals listed in
parenthesis are considered  inactive or discontinued.
\label{tab:NNN_detectors}}
\begin{center}
\begin{tabular}{c|c|c|c}
\hline        & Water Cherenkov & Liquid Argon & Scintillator \\
\hline
\hline Europe   & MEMPHYS & MODULAr, GLACIER & LENA \\
\hline Japan    & HyperK, Deep-TITAND (T2KK) & GLAO & \\
\hline U.S.     & LBNE-WCh, PINGU (UNO, 3M) & LBNE-LAr (LANDD) 
& TASD (SciPIO) \\
\hline
\hline
\end{tabular}
\end{center}
\end{table}

In the following sections some of the notable proposals  
contributed to this report are presented. 

\subsubsection{Proton Decay Searches with the Long Baseline Neutrino Experiment (LBNE)}


LBNE is planning for two 
types of detectors at an underground site in Homestake, South Dakota: a 
200 kt water Cherenkov (WCh) detector or a 34 kt liquid Argon (LAr) TPC. 
These detectors have complementary strengths in the search for proton decay, 
and either represents a considerable step forward from existing 
facilities. (At the time of this writing the LAr option has 
been chosen for the LBNE far detector technology. However, in order 
to be complete the proton decay sensitivities of both detector options are 
described in this report.) 

\begin{figure}
  \begin{center}
    \includegraphics[width=0.49\textwidth]{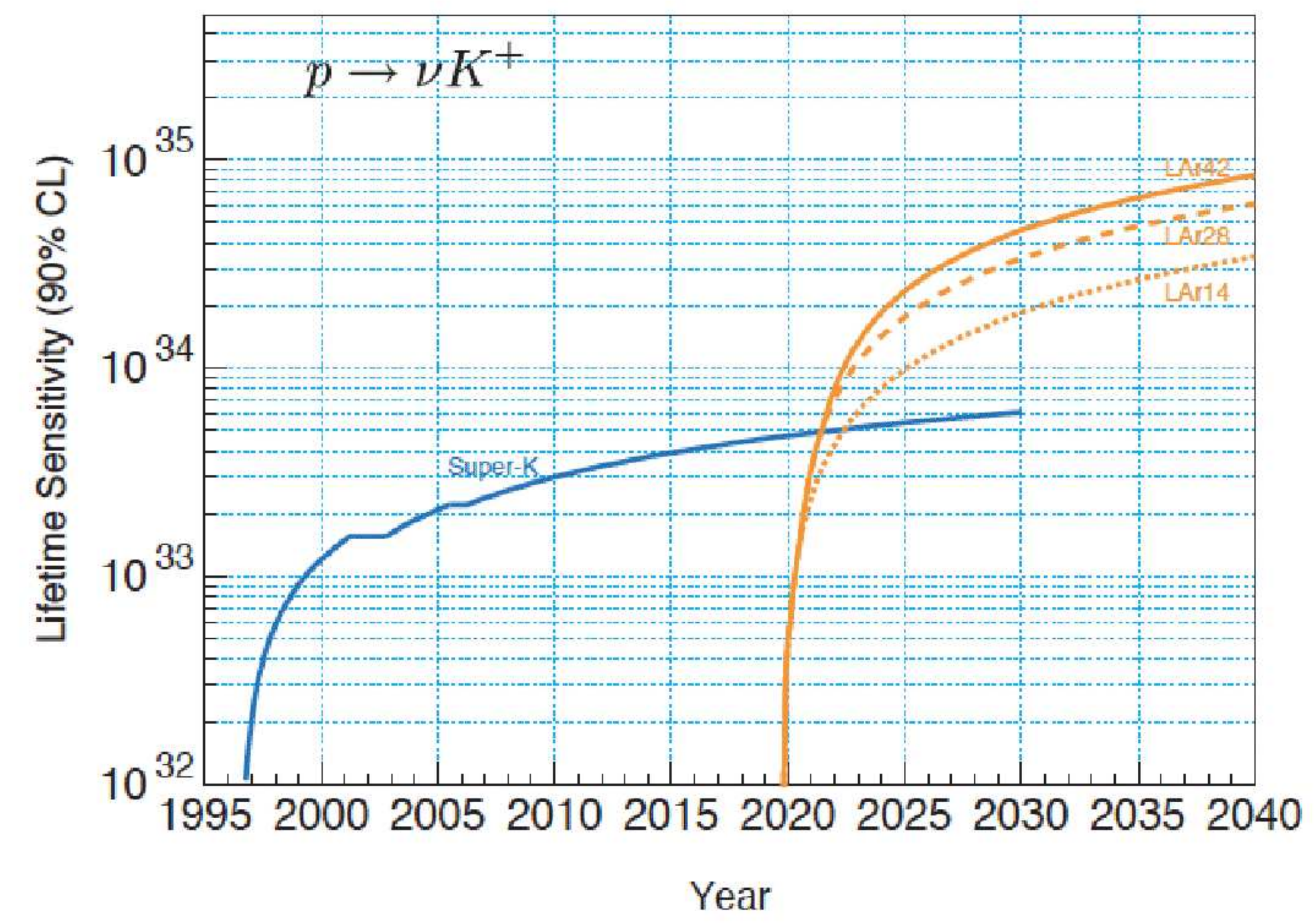}
    \includegraphics[width=0.49\textwidth]{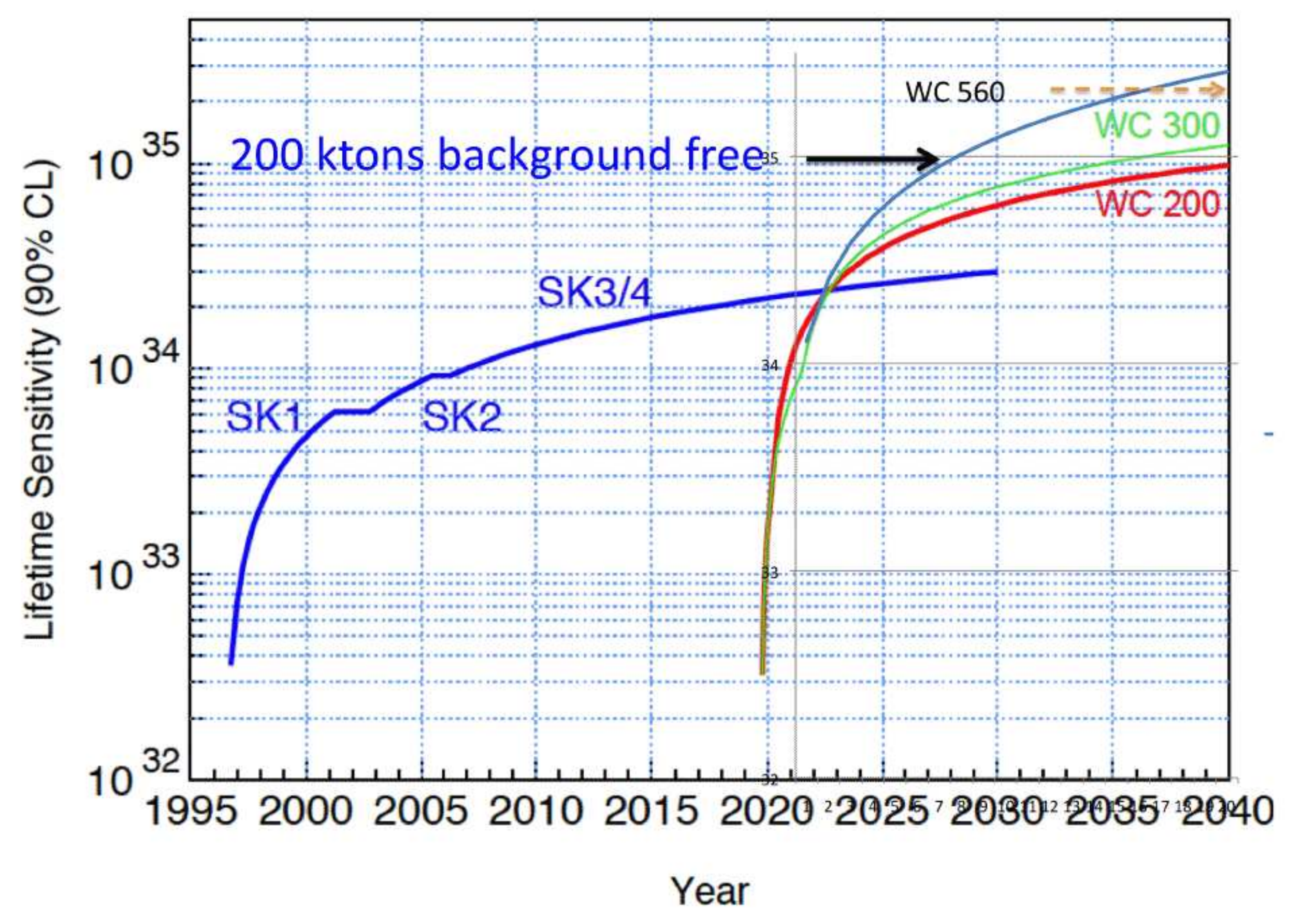}
  \end{center}
  \caption{\label{fig:LBNE_pdk_sensitivities} 
  The LBNE projected sensitivities for proton decay searches as a function 
  of calendar year. 
Left panel: the sensitivity of the liquid argon detector option 
for the $p \rightarrow \bar{\nu} K^+$ mode; 
Right panel: the sensitivity of the water Cherenkov detector option
for the $p \rightarrow e^+ \pi^0$ mode. The dashed arrow marked as 
``WC 560'' is the expected sensitivity by Hyper-K around year 2040, 
assuming, of course, it will be built as proposed.}
\end{figure}

\paragraph{Liquid Argon TPC:} 
This detector would be a significant size scale-up of the current largest 
liquid Argon TPCs, the 300~t ICARUS modules that came into full 
underground operation in 2010. The plan is for two separate 
modules, each of 16.5~kt fiducial mass. While this is less than a 
factor of two larger than Super-K (22 kt), the ability to observe 
charged particle tracks below the Cherenkov threshold in water means 
that some modes poorly observed in Super-K would be much better 
measured in LBNE LAr. For example, there is a significant 
sensitivity increase for the supersymmetric grand unified theories 
motivated mode $p \rightarrow \bar{\nu} K^+$.  
The charged kaon, with a momentum of 340 MeV/c (neglecting nuclear effects), 
has a range of 14~cm in LAr, so ionization energy loss measurements 
are expected to give high particle identification efficiency. 
The $K^+$ will also decay at rest to fully reconstructable final states 
such as a muon with reconstructed momentum of 236~MeV/$c$ and no other 
visible particle, a clear signature for 
$K^+ \rightarrow \mu^+\nu$ (65\% branching fraction). 
Therefore efficiency in excess of 90\% with very low background 
from atmospheric neutrinos is quite plausible. 
In contrast, since the $K^+$ has a relativistic gamma of 1.2, 
below the water Cherenkov threshold of 1.5, in Super-K 
this mode is measured only by the less-efficient gamma tag method. 
The LAr efficiency for this mode is roughly a factor of five higher than 
Super-K, as shown in Fig.~\ref{fig:LBNE_pdk_sensitivities}. 
The top curve in the left panel shows the sensitivity that could be 
reached assuming the full active volume can be used. 
Also shown are the expectations if only one module is built (dotted) or 
a significant fiducial volume cut is needed (dashed) to reduce cosmogenic 
backgrounds.  The most serious background is from neutral kaons 
produced by cosmic ray interactions in the surrounding rock that 
subsequently undergo charge exchange in the sensitive volume. 
These could result in the appearance of a charged kaon that mimics 
proton decay if the $K^+$ has the right momentum. This background 
process will be studied by measuring the rate of such events 
in momentum sidebands, but cutting out candidates near the side walls 
will eliminate such events at a cost of reduced fiducial mass. 
In the Bueno {\it et al.} paper~\cite{Bueno:2007um}, 
several different overburden and active veto scenarios were considered, 
with fiducial cuts as much as 7 meters from the wall, 
resulting in fiducial mass reductions ranging from 66\% to 90\%. 
A 2-meter cut from the sidewalls of the planned LBNE
detector, reduces the volume by 70\% from 
14$\times$15$\times$71~meters$^3$ to 14$\times$11$\times$67~meters$^3$, 
corresponding roughly to the dotted lines in 
Fig.~\ref{fig:LBNE_pdk_sensitivities}. 
The current LBNE reference design locates the LAr at the 4850-ft 
level of Homestake to mitigate this background; 
if the detector is located at the 800-ft level a substantial 
muon tracking veto system will also be required. 

The $p \rightarrow e^+ \pi^0$ mode can be detected in LAr 
with an expected efficiency somewhat lower than the 45\% efficiency 
achieved in WCh, which is set by the irreducible pion absorption rate 
in the nucleus. Although not yet calculated explicitly, 
the expected efficiency reduction should scale roughly with the 
linear size of the nucleus. Thus a naive estimate would be 
$45\% \times (16/39)^{1/3} = 33\%$. Even with a very low background 
rate of 0.1 events per 100~kt-years~\cite{Bueno:2007um}, 
the improvement in sensitivity compared to Super-K is not significant.
 
\paragraph{Water Cherenkov Detector:} The water Cherenkov detector 
planned is roughly nine times larger than Super-K, 
and hence for modes with no background sensitivity will simply 
scale with volume. For modes like $p \rightarrow e^+ \pi^0$ 
some amount of background is expected in the LBNE regime
based on measurements done by the K2K experiment, which also validated 
the neutrino simulations used for the water Cherenkov experiments. 
The level of expected background 
is 2 events/Mton/year, all from atmospheric neutrino interactions.
 
The right panel in Fig.~\ref{fig:LBNE_pdk_sensitivities} 
shows the sensitivity that could be reached in this mode 
assuming no background improvements are made (red line). 
In fact, it may be possible to significantly reduce this background 
if a neutron detection capability (such as addition of Gadolinium) 
is realized. This is due to the fact that while 80\% of proton decays 
in water should not have associated neutrons, atmospheric neutrino 
interactions are likely to produce one or more neutrons. 
These neutrons come from direct production via anti-neutrinos on oxygen, 
final state scattering of hadrons and $\pi^-$ capture on oxygen, 
and nuclear de-excitation. The figure shows the sensitivity reached 
if improvements allow rejection of all atmospheric neutrino backgrounds 
(blue line). The actual efficiency for background rejection will require 
measurements in a neutrino beam, and such an experiment is being 
planned for the FNAL booster neutrino beam.

\subsubsection{Proton Decay Searches with the Hyper-Kamiokande Experiment}

A next-generation underground water Cherenkov detector,
Hyper-Kamiokande (Hyper-K), is proposed in Japan.
If built, it will serve as a far detector of a long baseline 
neutrino oscillation experiment envisioned for the upgraded J-PARC, 
and as a detector capable of observing  $---$ 
far beyond the sensitivity of Super-K $---$ nucleon decays, 
atmospheric neutrinos, and neutrinos from
astronomical origins. The baseline design of Hyper-K is based 
on the highly successful Super-K, taking full
advantage of a well-proven technology. The total (fiducial) mass of 
the detector is 0.99 (0.56) million metric tons, 
which is about 20 (25) times larger than that of Super-K.
The details of the proposed experimental setup are described in the 
earlier sections, and also can be found in the recently published
Hyper-K Letter of  Intent {Abe:2011ts}.

The sensitivity of Hyper-K for nucleon decays 
has been studied with a Monte Carlo (MC) simulation 
based on the Super-Kamiokande analysis.  
An estimate of the atmospheric neutrino background is necessarily
included in the study.

\boldmath
\noindent{\bf Sensitivity study for the $p \rightarrow e^{+} \pi^{0}$ mode}:\unboldmath~Signal
candidates for $p \rightarrow e^{+} \pi^{0}$ mode are 
selected with the same selection criteria used by the
 present Super-K analysis.
The overall proton decay efficiency of $p \rightarrow e^{+} \pi^{0}$ 
is estimated to be 45\%.
The main background source of the proton decay search is the 
atmospheric neutrino events, which can occasionally produce 
an electron and a $\pi^{0}$ in the final state.
The remaining background events are estimated to 
be 1.6~events/Megaton$\cdot$year from the atmospheric neutrino MC simulation.
This result of the MC simulation has been experimentally confirmed by the 
K2K experiment~\cite{Mine:2008pr}. 

\begin{figure}[htbp]
  \begin{center}
    \includegraphics[width=0.45\textwidth]{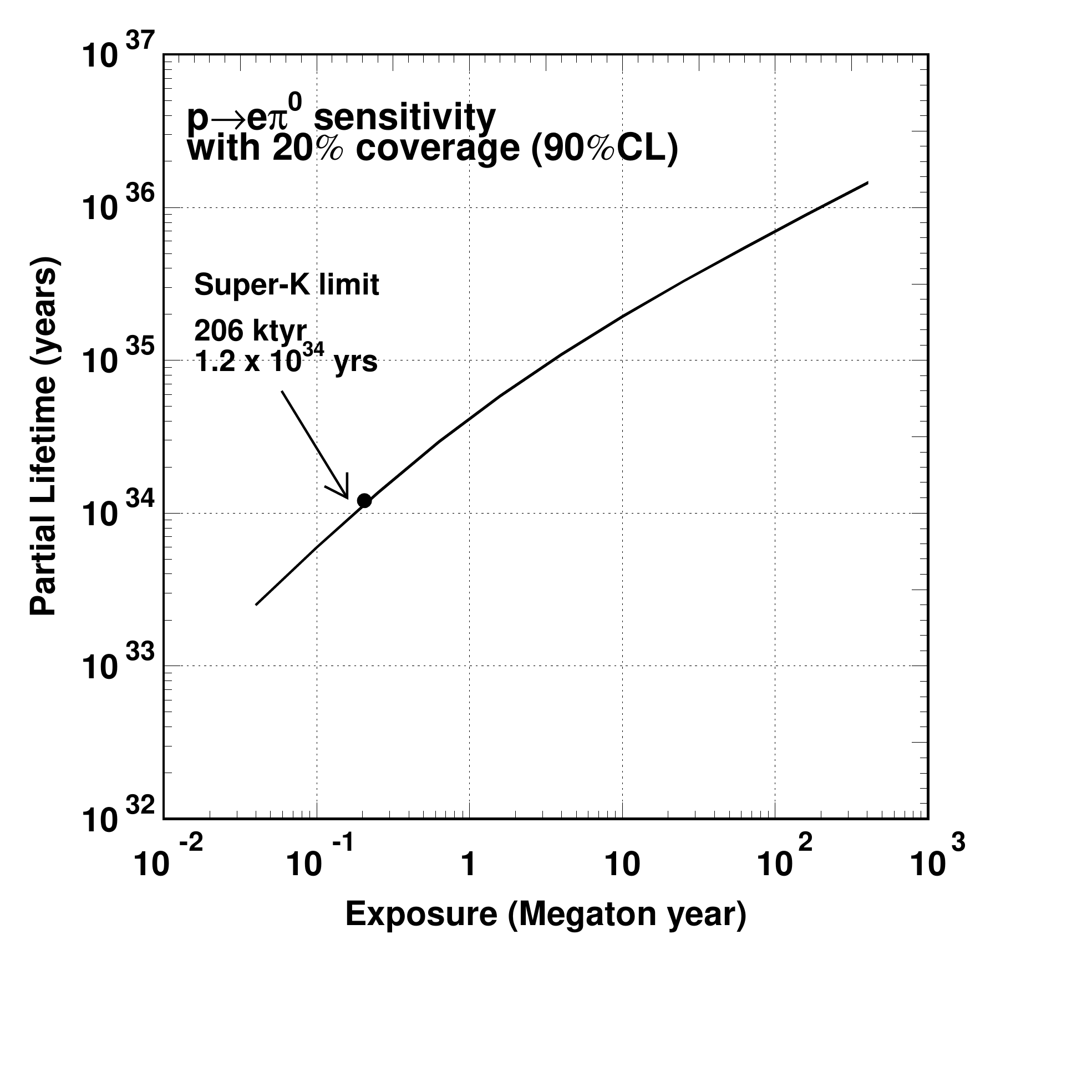}
    \includegraphics[width=0.45\textwidth]{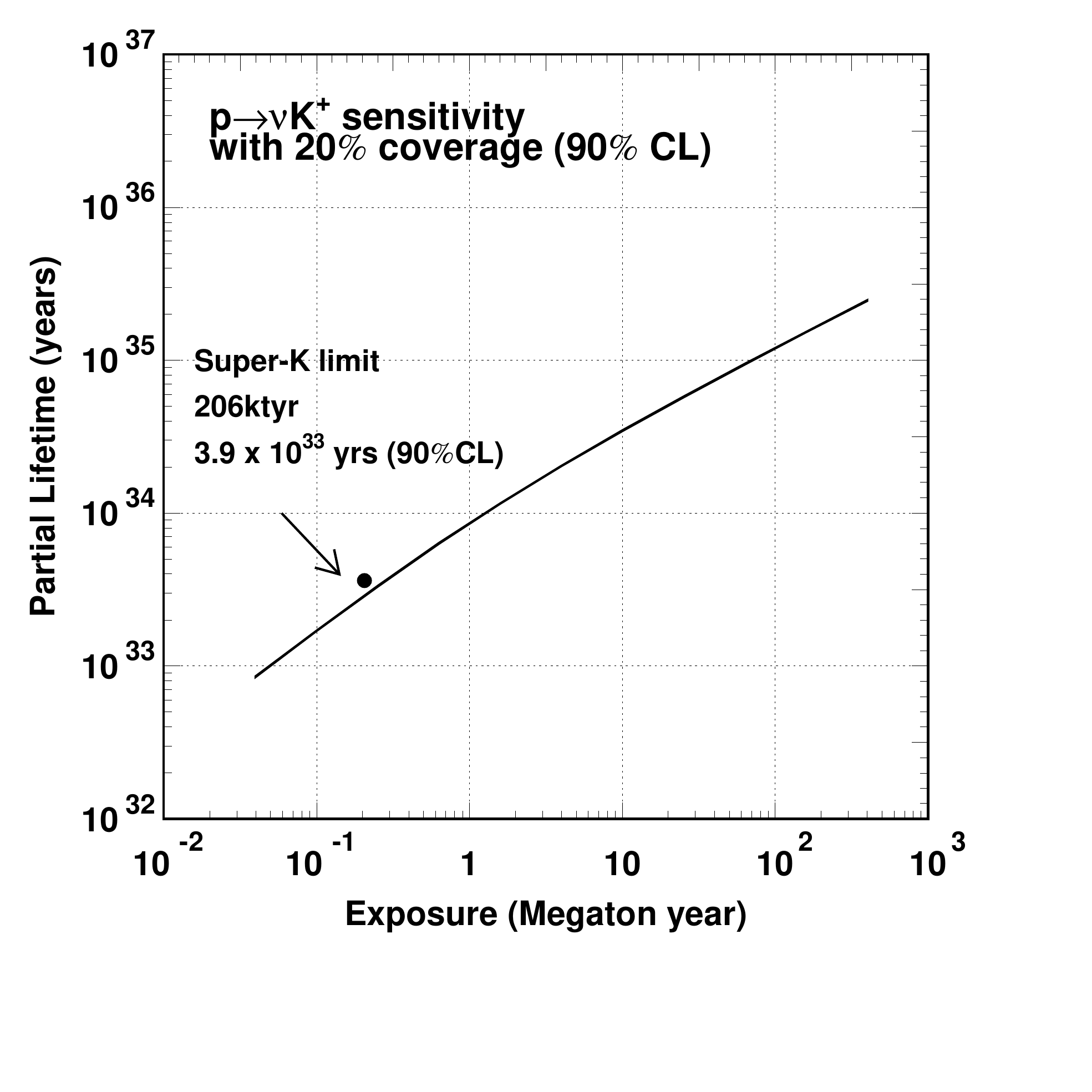}
  \end{center}
\caption {Sensitivities of the Hyper-Kamiokande proton decay search as 
a function of detector exposure.
Left Panel: for the $p \rightarrow e^{+} \pi^{0}$ mode; 
Right Panel: for the $p \rightarrow \overline{\nu} K^{+}$ mode}       
  \label{fig:hyperk_pdk_sens}                                                       \end{figure}

Fig.~\ref{fig:hyperk_pdk_sens} shows the sensitivity for proton decay with 
a 90$\%$ CL as a function of the detector exposure. A  
1.0$\times 10^{35}$ years partial lifetime can be reached 
with a 4~Megaton$\cdot$year 
exposure, which corresponds to eight years' running of Hyper-K; 
by contrast, it would take Super-K 178 years to reach this level.


\boldmath
\noindent{\bf Sensitivity study for the 
$p \rightarrow \overline{\nu} K^{+}$ mode}:\unboldmath~For
the $p \rightarrow \overline{\nu} K^{+}$ mode $K^{+}$ itself is not 
visible in a water Cherenkov detector due to having a low, sub-Cherenkov 
threshold,  momentum. 
However, $K^{+}$ can be identified by the decay products of 
$K^+ \to \mu^{+}+\nu$  (64\% branching fraction) and $K^+ 
\to \pi^{+}+\pi^{0}$  (21\% branching fraction).
The muons and pions from the $K^+$ decays have monochromatic momenta 
due to being produced via two-body decays. 
Furthermore, when a proton in an oxygen nucleus decays, the proton hole is 
filled by de-excitation of another proton, resulting in  $\gamma$ ray 
emission. The probability of a 6~MeV $\gamma$ ray being emitted is about 40\%. 
This 6~MeV $\gamma$ is a characteristic signal used to identify a proton 
decay and to reduce the atmospheric neutrino background. 
There are three established methods for the $p \rightarrow \overline{\nu} 
K^{+}$ mode search~\cite{Kobayashi:2005pe}: 
(1) look for single muon events with a de-excitation $\gamma$ ray just 
before the time of the muon, since the $\gamma$ ray is emitted at the 
time of $K^+$ production; 
(2) search for an excess of muon events with a momentum of 236~MeV/$c$ 
in the momentum distribution; and 
(3) search for  $\pi^{0}$ events with a momentum of 205~MeV/$c$. 
The detection efficiencies are calculated to be 7.1\% for method (1), 
43\% for method (2), and 6.7\% for method (3).
The background rates from atmospheric neutrinos are 1.6, 1940, and 6.7 
events/Megaton$\cdot$year for methods (1), (2), and (3), respectively.  
The number of atmospheric neutrino background events is estimated to be 
9.0 events.
Fig.~\ref{fig:hyperk_pdk_sens} shows the 90\% CL sensitivity curve  for the  
$p \rightarrow \overline{\nu} K^{+}$ mode, by combining all three methods,  
as a function of the detector exposure. 


Table~\ref{tab:pdk-summary} shows the summary of the study for the 
highlighted modes, 
$p \rightarrow e^{+} \pi^{0}$ and $p \rightarrow \overline{\nu} K^{+}$. If the
proton lifetime is shorter than 5.7$\times 10^{34}$ years for the 
$p \rightarrow e^{+} \pi^{0}$ mode, or shorter than 1.0$\times 10^{34}$ 
years for $p \rightarrow \overline{\nu} K^{+}$, 
a proton decay signal over the atmospheric neutrino 
background events could be discovered at a 
3$\sigma$ significance by collecting data corresponding to 
a  5.6~Megaton$\cdot$year exposure. 

\begin{table}[htb]
\caption{Summary of the sensitivity study for a 5.6~Megaton$\cdot$year 
exposure 
for the $p \rightarrow e^{+} \pi^{0}$ and 
$p \rightarrow \overline{\nu} K^{+}$ modes. 
For $p \rightarrow \overline{\nu} K^{+}$,  
method (1) $\mu$ + 6MeV$\gamma$ is labeled ``Meth.1", 
method (2) ($\mu$) as ``Meth.2", and method (3) $\pi^{+}\pi^{0}$ as ``Meth.3".
\label{tab:pdk-summary}}
\begin{center}
\begin{tabular}{l|r|r|r|r}
\hline
        & $p \rightarrow e^{+} \pi^{0}$ & 
        \multicolumn{3}{c}{$p \rightarrow \overline{\nu} K^{+}$} \\ 
\hline
       &  & Meth.1 & Meth.2 & Meth.3 \\
\hline
\hline
Efficiency ($\%$)& 45   & 7.1  & 43  & 6.7  \\ 
\hline
Background (/Mton$\cdot$yr) & 1.6  & 1.6 & 1940 & 6.7 \\
\hline 
90$\%$ Sensitivity ($\times 10^{34}$ yrs)
                   & 13  & \multicolumn{3}{c}{2.5} \\
\hline
3$\sigma$ Discovery potential ($\times 10^{34}$ yrs)
                   & 5.7  & \multicolumn{3}{c}{1.0} \\
\hline 
\end{tabular}
\end{center}
\end{table}

There is no question that in order to explore order of magnitude longer 
lifetime regions than Super-K, a larger detector is absolutely necessary.  
As seen in this sensitivity study, if built, Hyper-K will open 
up a new era in the search for nucleon decay, as would LBNE.

\subsubsection{Proton Decay Searches with the Low Energy Neutrino Astronomy (LENA) Experiment}

LENA (Low Energy Neutrino Astronomy) is a proposed unsegmented 
liquid-scintillator
detector of 50\,kt target mass proposed as a true multi-purpose facility.
LENA at the Pyh\"asalmi mine (Finland) is one of three detector options
discussed within the European LAGUNO-LBNO design study.
The details of the proposed detector design are described 
in the ``Neutrinos'' chapter of this report.



While the emphasis of the LENA physics program is on low-energy neutrinos 
and anti-neutrinos
(E$<$100 MeV), the experiment
can also contribute to several aspects of neutrino
and particle physics at GeV energies. Actually,
the search for proton decay into kaons and antineutrinos
was one of the first items considered to play
an integral part in the LENA concept, since the visibility
of the kaon's energy deposition in the scintillator
substantially increases the detection efficiency
in comparison to water Cherenkov detectors.


Currently, the best limits on proton lifetime are held by 
Super-Kamiokande, and it seems unlikely that LENA will substantially 
improve the limit for $p\rightarrow e^+ \pi^0$. However, 
the sensitivity for the decay mode $p\rightarrow \bar{\nu} K^+$ 
is an order of magnitude larger than in water Cherenkov detectors. 
Moreover, the search in LENA is expected to be background-free for 
about 10 years, allowing  a lifetime limit of 
$\tau_{p}>4$$\times$10$^{34}$\,yrs (90\,\% CL)  if no event is observed.

Within the target volume of LENA, about 1.6$\times$10$^{34}$ protons, 
both from carbon and hydrogen nuclei, are candidates for the decay. 
As all decay particles must be contained inside the active volume, the 
fiducial volume is about 5\,\% smaller.


In the case of protons from hydrogen nuclei ($\sim$0.25$\times$10$^{34}$ 
protons in the fiducial volume of LENA), the proton can be assumed at rest. 
Therefore, the proton decay $p\to \bar{\nu} K^+$ can be considered a 
two-body decay problem, where $K^+$ and $\bar\nu$ always receive the same 
energy (105 and 339 MeV, respectively). The large sensitivity of LENA for this 
decay channel arises from the visibility of the ionization signal generated 
by the kinetic energy deposition of the kaon. A water Cherenkov detector 
is blind to this signal, as the kaon is produced below the Cherenkov 
threshold in water, reducing the detection sensitivity.

In LENA, the prompt signal of the decelerating kaon is followed by the 
signal arising from the decay particle(s):  After $\tau_{K^+}=12.8$\,ns, 
the kaon decays either by $K^+\to\mu^+\nu_\mu$ (63.43\,$\%$) or by 
$K^+\to\pi^+\pi^0$ (21.13\,\%). In 90\% of these cases, the kaon decays 
at rest. If so, the second signal is again monoenergetic, either 
corresponding to the 152\,MeV kinetic energy of the $\mu^+$ or 246\,MeV 
from the kinetic energy of the $\pi^+$ and the rest mass of the $\pi^0$ 
(which decays into two gamma rays, creating electromagnetic showers). 
A third signal arising from the decay of the muon will be observed with 
a comparably large time delay.

If the proton decays inside a carbon nucleus ($\sim$1.2$\times$10$^{34}$ 
protons in the fiducial volume), further nuclear effects have to be 
considered. First of all, since the protons are bound to the nucleus, 
their effective mass will be reduced by the nuclear binding energy 
$E_b$, 37\,MeV and 16\,MeV for protons in s-states and p-states, respectively. 
Secondly, decay kinematics will be altered compared to free protons due to 
the Fermi motion of the proton. The experimental signature of the proton 
decay in LENA is not substantially affected by nuclear effects or the kaon 
decay mode: A coincidence signal arising from the kinetic energy deposited 
by the kaon and from the delayed energy deposit of its decay particles will 
be observed.

The main background source in the energy range of the proton decay is 
atmospheric muon neutrinos $\nu_\mu$. Via weak charge-current interactions, 
these $\nu_\mu$ create muons inside the detector, with a substantial fraction 
in the energy range relevant for the proton decay search. Moreover, 
additional kaons can be produced in deep inelastic scattering reactions 
adding to the $\nu_\mu$ background.

The double signature of kaon energy deposition and decay can be used to 
discriminate atmospheric $\nu_\mu$ events as long as the kaon decay is 
sufficiently delayed to produce a discernible double signal, {\it i.e.}, the 
delay is large compared to the time resolution of the detector. 
MC simulations show (based on 2$\times$10$^4$ proton decay and muon 
events in the relevant energy regime) that an analysis cut can be defined 
which rejects all muons and retains a detection efficiency of 
$\varepsilon_{p}\approx 65$\,\% for proton decay. The sensitivity 
$\varepsilon_{p}$ is an order of magnitude larger than the one obtained 
in the Super-Kamiokande analysis, corresponding to a similar increase 
in the proton lifetime limit. The corresponding background rejection 
efficiency is at least $\varepsilon_\mu \geq 1-5$$\times$10$^{-5}$. 
This results in an upper limit of $\sim$0.05 muon events per year 
that are misidentified as proton decay events.


In the case of charged-current reactions of atmospheric $\nu_\mu$'s at larger 
energies, hadrons can be produced along with the final state muon. 
These events are dangerous if they are able to mimic the double signature 
of the proton decay. While this is not the case in pion and hyperon 
production, interaction modes creating an additional kaon in the final 
state may be mistaken as signal events. In principle, these events can 
be discriminated by the additional decay electron of the muon created in 
the charged-current reaction. However, this signal is sometimes covered by the muon 
signal itself: Monte Carlo simulations return an upper limit of 0.06 
irreducible background events per year for this channel.

Based on the efficiencies of the rise time cut, the sensitivity of LENA 
for the proton decay search can be determined. Combining the expected 
background rates from atmospheric neutrino-induced muon and kaon production, 
a rate of 0.11 background events per year or 1.1 events in 10 years can be 
obtained. In case there is no signal observed in LENA within these 10 
years, the lower limit for the lifetime of the proton will be placed at 
$\tau_{p}>4$$\times$10$^{34}$\,yrs at 90\,\% CL If one candidate is 
detected, the lower limit will be reduced to 
$\tau_{p}>3$$\times$10$^{34}$\,yrs (90\,\% CL). 
LENA might also provide relevant sensitivity levels to other 
nucleon decay channels. While the analysis presented here is 
independent of the tracking capabilities of the detector, 
in others (e.g.~$p\to e^+ \pi^0$) the possibility of reconstructing the 
decay vertex might be necessary to discriminate background signals. 
However, these aspects require further studies.

\boldmath
\section{Neutron Anti-Neutron ($n-\bar{n}$) Oscillation}\label{sec:pdk_nnbar}
\unboldmath

\boldmath
\subsection{Theoretical Motivation for $n-\bar{n}$ 
Oscillation Searches}\label{sec:pdk_nnbar_theory}
\unboldmath

A true understanding of the physics of baryon number
violation would require comprehensive knowledge of the underlying symmetry
principles, with distinct selection rules corresponding to different
complementary scenarios for unification and for the generation of baryon
asymmetry of the universe. Thus, discovery of proton decay with the selection
rule $\Delta B = 1$ would imply the existence of new physics at an
energy scale of  $10^{15}$ GeV,  while discovery of $n-\bar{n}$ oscillation
with the selection rule $\Delta B = 2$ would point to new
physics near and above the TeV scale. There also exist many models,
including those with extra space dimensions at TeV scale, with local
or global $B$ or $B-L$ symmetry that do not allow proton decay, and
where $n-\bar{n}$ oscillation is the only observable baryon number
violating process.  

The discovery of neutrino mass  has  provided the first
direct evidence for physics beyond the Standard Model. A simple way to
understand the small neutrino masses is by the seesaw mechanism, which
predicts that the neutrino is a Majorana fermion,  {\it i.e.}, it breaks
lepton number by two units. Even if the Majorana nature of the
neutrino is established through observation of neutrinoless double
beta decay, we still need to understand at what scale the dynamics
occurs.  Since the true anomaly-free symmetry of the Standard Model
is the combination $B-L$, if $L$ is broken by two units, it is natural
for $B$ to be broken by two units as well.  Indeed, quark-lepton unified
theories that predict Majorana neutrinos also predict $n-\bar{n}$
oscillations.  The search for  $n-\bar{n}$ oscillations will
therefore supplement neutrinoless double beta decay experiments
in establishing a common mechanism of these processes. In particular,
an observation of $n-\bar{n}$ oscillation may indicate that the small neutrino
mass is not a signal of physics at the GUT scale but rather at the
scale not much above a TeV.

Originally it was
thought that proton decay predicted by grand unified theories could
generate the matter-antimatter asymmetry. However, since sphaleron
processes in the Standard Model violate $B+L$ number, any $B-L$ conserving
GUT--scale--induced
baryon asymmetry would be washed out 
at the electroweak phase transition~\cite{bmn}.
As noted before, lepton number violating decays of the right--handed
neutrino within minimal $SO$(10)--type GUTs can naturally explain the observed
matter-antimatter asymmetry via the process of leptogenesis~\cite{fy}.
Observation of $n-\overline{n}$ oscillations at currently achievable
sensitivity would, however, hint at a new mechanism for the generation of
matter-antimatter asymmetry, since the baryon  excess generated by
leptogenesis would be washed away. It has been shown that the same physics
that leads to $n-\overline{n}$ oscillation also provides a mechanism for
baryogenesis at scales below the electroweak phase transition.
Existing theories describing such processes
typically also predict colored scalars within the reach of the LHC, along
with an observable electric dipole moment of the neutron and some
rare $B$-meson decay channels.

The probability of $n-\bar{n}$ transformation in vacuum in the absence
of a magnetic field is $P \cong (t/\tau)^2$, where $t$ is the free neutron
propagation time in vacuum and $\tau$ is a characteristic oscillation time
determined by new physics processes that induce $\Delta B = 2$
transitions. If the scale of the relevant new physics is around
$(10^4-10^6)$ GeV, as predicted by various theoretical models,
the possible range of $n-\bar{n}$ oscillation time is
$\tau \sim (10^9- 10^{11})$ seconds. (See \cite{nnbar_Theory} for 
a review and more detailed discussion of $n-\bar{n}$ oscillation.)

\boldmath
\subsection{Current and Proposed $n-\bar{n}$ 
Oscillation Search Experiments}\label{sec:pdk_nnbar_exp}
\unboldmath

\boldmath
\subsubsection{Current $n-\bar{n}$ Oscillation Search Experiments}\label{sec:pdk_nnbar_current_exp}
\unboldmath

Transformation of neutrons to antineutrons with neutrons bound inside nuclei 
has been sought in large underground proton decay and neutrino detectors: 
inside iron by the Soudan-2 experiment, inside deuterium by the SNO experiment, 
and inside oxygen by the Super-Kamiokande experiment \cite{superk_nnbar}. 
Compared to free neutron transformation in vacuum, the intranuclear   
$n-\bar{n}$ transformation is strongly suppressed by the 
difference of nuclear potential 
for neutrons and antineutrons. This suppression was calculated by 
quantum mechanical nuclear theoretical models~\cite{Dover}\cite{Friedman}. 

An antineutron transformed from the neutron inside the nucleus is expected 
to annihilate quickly with one of the surrounding nucleons and to 
produce multiple secondary hadrons, mainly pions that will be available 
for detection. Experimentally the search for $n-\bar{n}$ transformation in 
all the above-mentioned experiments is limited by atmospheric 
neutrino backgrounds. The best result so far was obtained in 
the Super-Kamiokande experiment, where 24 $n-\bar{n}$ oscillation 
candidate events were observed for 1489 days 
with an estimated atmospheric neutrino background of 24.1 events. 
Based on this observation, the lower limit on the lifetime for 
neutrons bound in ${}^{16}O$  was calculated to be $1.89 \times 10^{32}$ 
years at the 90\% CL~\cite{superk_nnbar}.

The lifetime limit for bound nucleons in an ${}^{16}O$ nucleus ($T$) 
can be converted to the $n-\bar{n}$ oscillation time for 
a free neutron ($\tau$) using the relationship: 
$T(\text{intranuclear}) = R \cdot \tau_{n-\bar{n}}^2 (\text{free})$, 
where $R$ is the nuclear suppression factor \cite{superk_nnbar}. 
Thus, the corresponding limit for the oscillation time of free neutrons 
from the Super-K limit can be calculated as $2.44 \times 10^8$ s 
using $R= 1.0 \times 10^{23} \text{s}^{-1}$ from~\cite{Dover}. 
For a more recent theoretical model\cite{Friedman} 
with $R = 0.52 \times 10^{23} \text{s}^{-1}$, 
one can find a limit for the free neutron oscillation time of 
$3.38 \times 10^8$ s.

\boldmath
\subsubsection{Proposed $n-\bar{n}$ Oscillation Search Experiments}\label{sec:pdk_nnbar_proposed_exp}
\unboldmath

As described above, the  presence of a large atmospheric neutrino background makes 
further improvement beyond the Super-K result on $n-\bar{n}$ oscillations
in next-generation water Chrenkov detectors difficult.
Alternatively, a search for $n-\bar{n}$ oscillations 
with free neutrons possesses excellent background rejection, also  allowing 
the possibility of turning off the signal using a small magnetic 
field, and therefore has enormous potential in exploring the stability 
of matter. Thus, for example, a limit on the free-neutron oscillation 
time $\tau >10^{10}$ s would correspond to the limit on matter stability 
of $\tau_A=1.6-3.1 \times 10^{35}$ years.

The previous experimental 
search for free $n-\bar{n}$ transformations using a cold neutron 
beam from the research reactor at Institute Laune-Langevin (ILL) 
in Grenoble gave a limit on 
$\tau>8.6 \times 10^7$ seconds in 1991 \cite{ILL}. The average 
velocity of the cold neutrons used was $\sim 700$  ${\rm m/s}$ and the 
average neutron observation time was $\sim 0.1 $\,s. Antineutron 
appearance was sought through annihilation in a $\sim 100 \mu$ 
carbon film target, generating a star pattern of several secondary 
pions, viewed by a tracking detector, and an energy deposition of 1-2 GeV 
in the surrounding calorimeter. This detection process strongly 
suppresses backgrounds. In one year of operation this ILL experiment 
saw zero candidate events. 

The figure of merit for a free-neutron $n-\bar{n}$ search experiment 
is $N_n \times t^2$, where $N_n$ is the number of free neutrons observed 
and $t$ is the observation time. A dedicated spallation neutron source coupled
to a high-intensity beam provided by 
Project X could be optimized for the enhanced production of slow ultra-cold 
(UCN) and very-cold (VCN) neutrons (velocities below $\sim 100$ ${\rm m/s}$) 
with the use of modern neutron optics, neutron moderators, and cryogenic 
technologies. Significant progress has been made in the field of neutron 
optics since the time of the ILL experiment, and recent advances in 
neutron source design promise to enhance the brightness of moderators 
as well as reduce their effective spectral temperature. Furthermore, 
the existing 105m vertical shaft at Fermilab could be used to develop 
a vertical layout of the experiment, which provides additional gains by 
employing gravity to increase even further the observation time.  The 
optimization of parameters for the target/moderator/cold-source design, 
neutron optical layout, neutron flight vessel (with vacuum better than 
$10^{-5}$ Pa and magnetic shielding down to $1$  $nT$) and annihilation 
detector will require detailed R\&D studies. An optimized design that 
employs all these advances would support an experiment that can achieve  
major improvements over the ILL sensitivity.

\noindent{\bf Experimental goal}: Assuming beam power of 0.2-0.5 MW 
on the spallation target (e.g. with a 3 GeV proton or deuteron beam from the 
Project X linac), the goal of a new $n-\bar{n}$ search experiment will 
be to improve the sensitivity to $n-\bar{n}$ transformation probability 
by 3-4 orders of magnitude  beyond that in the ILL-based experiment and 
to probe the range of the free neutron oscillation time around $10^{10}$ s. 
An antineutron annihilation  detector with negligible background could 
allow a single observed $n-\bar{n}$  event to be a discovery. The active 
magnetic shielding should be tunable to suppress oscillations if needed 
to confirm a signal. 

\boldmath
\noindent{\bf Timeline of $n-\bar{n}$ experiment}:\unboldmath
An  $n-\bar{n}$ 
experiment can be implemented with the ``Nuclei" beamline of the 3 GeV 
linac of Project X as a 0.2 -- 0.5 MW spallation target, or possibly 
with Main Injector or Booster beams of lower power. Two to three years of R\&D 
research is required for the configuration optimization and the 
conceptual design. Provided that the vertical shaft is available, 
the construction stage of the experiment will take approximately 
3--4 years. The anticipated running time of an $n-\bar{n}$ experiment would be 
3--4 years.

\section{Conclusions}

While yet to be seen, proton decay is an indispensable tool for
probing Nature at truly high energies. It remains the missing piece for evidence 
of grand unification. The dramatic meeting of the three gauge couplings at
a scale of about $2 \times 10^{16}$~GeV, which is found to occur in the
context of low energy supersymmetry, and the tiny neutrino masses as
observed in the neutrino oscillation experiments, lend  strong support
to the idea of supersymmetric grand unification.
Moreover, grand unified theories that are in accord with the observed
masses and mixings of all fermions, including neutrinos, typically predict
proton lifetimes within a factor of five to 10 of current Super-Kamiokande
limits.
This is why an improved search for proton decay is now most  pressing.
This can only be done with a large 
detector built deep underground.
Such a detector, coupled to a long-baseline intense neutrino beam
(as would be available from Fermilab), can simultaneously 
sensitively study neutrino oscillations so as to shed light on
neutrino mixing parameters, mass-ordering, and most importantly
$CP$ violation in the neutrino system. And it can help efficiently study
supernova neutrinos. In short, such a detector would have a unique
multi-purpose value with high discovery potential in all three areas.

Diverse next-generation nucleon decay and neutrino detectors based on a
plausible extrapolation of existing technologies are being discussed in
Europe, Japan and the US. Building such a large underground detector coupled
to a long-baseline neutrino beam in the US, in a timely fashion, would not
only probe a set of fundamental issues in physics, but would
enable the US to assume a leadership position by having a stellar facility that
would be an asset to the world as a whole. Sensitive searches for proton
decay, measurements of neutrino oscillation parameters and 
observation of supernova neutrinos with such a large
underground detector should thus be given high priority in the intensity
frontier program. Proton decay, if found, would no doubt constitute a
landmark discovery for mankind.

\def\Discussion{\setlength{\parskip}{0.3cm}\setlength{\parindent}{0.0cm}
     \bigskip\bigskip      {\Large {\bf Discussion}} \bigskip}\def\speaker#1{{\bf #1:}\ }
\def\endDiscussion{}


\newcommand{\red}[1]{{\bf\textcolor{red}{[#1]}}}
\newcommand{\Aslash}{A\!\!\!/\,}
\newcommand{\Bslash}{X\!\!\!\!/\,}
\def\babar{\mbox{\sl B\hspace{-0.4em} {\small\sl A}\hspace{-0.37em} \sl B\hspace{-0.4em} {\small\sl A\hspace{-0.02em}R}}}

\chapter{New light, weakly-coupled particles}
\label{chap:hspaw}

\begin{center}\begin{boldmath}



\begin{center}

Conveners: J.I.~Collar, R.~Essig, J.A.~Jaros

A.~Afanasev,
J.R.~Armendariz, 
O.~Baker,
B.~Batell,
J.~Beacham, 
F.~Bossi,
J.~Boyce,
M.~Buckley,
G.~Carosi,
R.~Cowan,
A.~Denig,
B.~Echenard,
A.~Freyberger,
A.~Gasparian,
M.~Graham,
P.W.~Graham, 
A.~Haas,
J.~Hartnett,
I.~Irastorza,
J.~Jaeckel,
I.~Jaegle, 
M.~Lamm,
A.~Lindner,
W.C.~Louis,
D.~McKeen,
H.~Merkel,
G.~Mills,
L.A.~Moustakas,
G.~Mueller,
M.~Pivovaroff,
R.~Povey,
S.~Rajendran,
J.~Redondo,
A.~Ringwald,
P.~Schuster,
M.~Schwarz, 
K.~Sigurdson,
P.~Sikivie, 
J.H.~Steffen, 
S.~Stepanyan, 
M.~Strassler,
D.B.~Tanner,
M.~Tobar,
N.~Toro,
A.~Upadhye,
S.~Vahsen, 
R.~Van de Water,
J.~Vogel, 
D.~Walker,
N.~Weiner,
A.~Weltman,
W.~Wester,
G.~Wiedemann,
B.~Wojtsekhowski,
K.~Zioutas

\end{center}




\end{boldmath}\end{center}

\section{Overview}\label{sec:hspaw-overview}

New light particles that couple only weakly to ordinary matter are ubiquitous in new physics extensions of the 
Standard Model.  Their existence is motivated by several theoretical and observational puzzles, 
many of which are central in our quest to obtain a comprehensive understanding of the constituents of our 
universe and their interactions.  These include the nature of dark matter and dark energy, the 
strong CP problem, and a variety of astrophysical puzzles and dark matter-related anomalies.   

Our working group examines axions, axion-like particles, hidden-sector photons, milli-charged particles, chameleons, 
and related particles (see~\cite{arXiv:1002.0329} for a recent review).  Their masses can range anywhere 
from sub-femto-eV to the weak scale ($\sim 100$ GeV), and they are characterized by their small coupling or mixing 
with the photon.  
This allows them to be produced with intense beams of photons, electrons, or protons and detected 
with sensitive equipment.  This makes them, by definition, targets for the intensity frontier.  
We will sometimes refer to these weakly interacting sub-eV (or ``slim'') particles as ``WISPs''.  

Axions are pseudo-scalar particles that solve the strong $CP$ problem.  
They have extremely small masses, because they arise as pseudo-Nambu-Goldstone bosons of an 
almost exact symmetry, the ``Peccei-Quinn'' symmetry, which is spontaneously broken at a very high energy scale.
The spontaneous breaking of other, non-Peccei-Quinn global symmetries is common in new physics models 
(including string theory) and can give rise to light axion-like scalar or pseudo-scalar particles, called ALPs. 
Axions and ALPs can constitute the dark matter of our universe and can explain a variety of astrophysical 
observations. 
The axion couples to gluons, photons, and Standard Model fermions, and it is the latter two that are 
the most easily detected.  ALPs also naturally couple to photons, 
although this coupling is not guaranteed.

Hidden-sector photons, called $A'$ bosons, are massive vector bosons that can mix with 
the ordinary photon via ``kinetic-mixing.''  
This mixing can permit photon-$A'$ oscillations (observable for sub-eV $A'$ bosons) 
and produces a small coupling of the $A'$ to electrically charged matter.  
A sub-eV mass $A'$ could be the dark matter particle or contribute to the observed number of 
relativistic degrees of freedom in the early universe. 
A MeV-GeV mass $A'$ could explain the discrepancy between the measured and 
calculated muon anomalous magnetic moment in the Standard Model.  

Intriguingly, an $A'$ allows ordinary matter to have a small coupling to new particles in a ``hidden-sector'' 
that do not interact with the Standard Model's strong, weak, or 
electromagnetic forces.  These hidden-sectors can have a rich structure and have thus far remained undetected.  
They appear in many extensions of the Standard Model, and are often required for consistency or phenomenological reasons.
Dark matter could be part of such a hidden-sector and couple to an $A'$, producing new dark matter 
interactions with ordinary matter that might explain some of the data of 
cosmic-ray, balloon-borne, or terrestrial dark matter experiments.  
A hidden-sector coupled to a massless $A'$ would include particles that are milli-charged under 
the ordinary photon.  

Chameleons are scalar particles that might be responsible for the ``dark energy'' that is accelerating our universe's expansion.  They are one of only a few known ways to hide dark energy from local gravity 
(equivalence principle and inverse-square law) tests.  
Chameleons remain hidden because their mass depends on the local matter density.  
In high-density regions, such as on Earth where these tests are done, their mass is large, so the new 
force generated by the chameleons has a range too short to be detectable.  
In low-density regions, such as in the universe at large, their mass is small and they are a candidate for dark energy.  
They can couple to photons, allowing photons to oscillate to chameleons and vice versa.  

Searches for these new particles are naturally part of the intensity frontier, since 
intense beams of photons, electrons, or protons are required to produce them in sufficient quantities to compensate 
for their weak coupling to ordinary matter.  
Several experimental proposals exist to search for these particles.  Many existing facilities have not yet been 
fully exploited in this search.  Technology and accelerator developments, in many cases very modest by 
current standards in both cost and effort, can allow exploration of even more of the motivated parameter space.  
Large-scale experiments are also under consideration, well-motivated, and require more substantial investments.  
In many instances, simple configuration changes can switch their sensitivity from one type of particle to another.  
All these experiments provide an amazing opportunity for major fundamental discoveries.  
The discovery of a new force of Nature, or a whole new sector of particles is within their reach.
There is world-wide interest and competition in this physics, and the US needs to pursue it aggressively. 
The strong motivation for the existence of new light, weakly coupled particles combined with a very modest 
cost-to-benefit ratio implies that any sensible future intensity frontier program should include this physics 
as an important component.  

The outline of the remainder of this section is as follows.  \S\ref{sec:hspaw-motivation} discusses the theory and physics 
motivation for axions, axion-like particles, hidden-sector photons, milli-charged particles, and chameleons.  
\S\ref{subsec:hspaw-searches} provides an overview of the experimental searches for these 
various particles, including current and future planned experiments.  
In \S\ref{sec:hspaw-future-techno} and \S\ref{sec:hspaw-future-accelerators}, we discuss the 
technologies and accelerators, respectively, needed for future progress.  
\S \ref{sec:hspaw-conclusions} contains our conclusions.

\section{Physics Motivation}\label{sec:hspaw-motivation}

\subsection{Axions and Axion-Like Particles}\label{subsec:hspaw-motivation-ALPs} 

One of the biggest unresolved puzzles in the Standard Model is the lack of any observed $CP$ violation in the strong nuclear interactions described by Quantum Chromodynamics (QCD).  While the weak interactions are known to violate $CP$, 
the strong interactions also contain a $CP$-violating term in the Lagrangian, $\frac{\Theta}{32 \pi^2} G_{\mu\nu} \tilde G^{\mu\nu}$, where $G^{\mu\nu}$ is the gluon field strength.  
For non-zero quark masses, this term leads to (unobserved) $CP$-violating effects of the strong interactions.  
This so-called ``strong $CP$ problem'' is often exemplified by the lack of observation of a neutron dipole moment down 
to a present experimental upper limit 10 orders of magnitude smaller than what is expected from a $CP$-violating QCD.

Solutions to this problem are scarce.  Perhaps the most popular suggestion is the so-called Peccei-Quinn (PQ) $U(1)$ approximate global symmetry, which is spontaneously broken at a scale $f_a$. The axion is a hypothetical particle that arises as the pseudo-Nambu-Goldstone boson (PNGB) of this symmetry~\cite{Peccei:1977hh,Weinberg:1977ma,Wilczek:1977pj}.

The axion mass is $m_a \sim 6~{\rm meV}~(10^9~{\rm GeV}/f_a)$. Its coupling to ordinary matter is proportional to $1/f_a$ and can be calculated in specific models.  It couples to leptons and to photons,
the latter being of the form 
$\mathcal{L}\supset -\frac{1}{4} \,g_{a\gamma} \,a \,F_{\mu\nu}\, \tilde{F}^{\mu\nu}$, where 
$g_{a\gamma} \sim 10^{-13}~{\rm GeV}~(10^{10}~{\rm GeV}/f_a)$~\cite{hpsaw:Nakamura:2010zzi} is a coupling that is model-dependent up to an $\mathcal{O}(1)$ factor.
Moreover, since $m_a\ll \Lambda_{QCD}$, the axion's coupling to quarks should be described through its coupling to hadrons, which occurs through small mixing with the $\pi^0$ and $\eta$ mesons.  All of these interactions can play a role in searches for the axion, and allows the axion to be produced or detected in the laboratory and emitted by the sun or other stars.  

\begin{figure}[!t]
\centering
\includegraphics[width=0.74\textwidth]{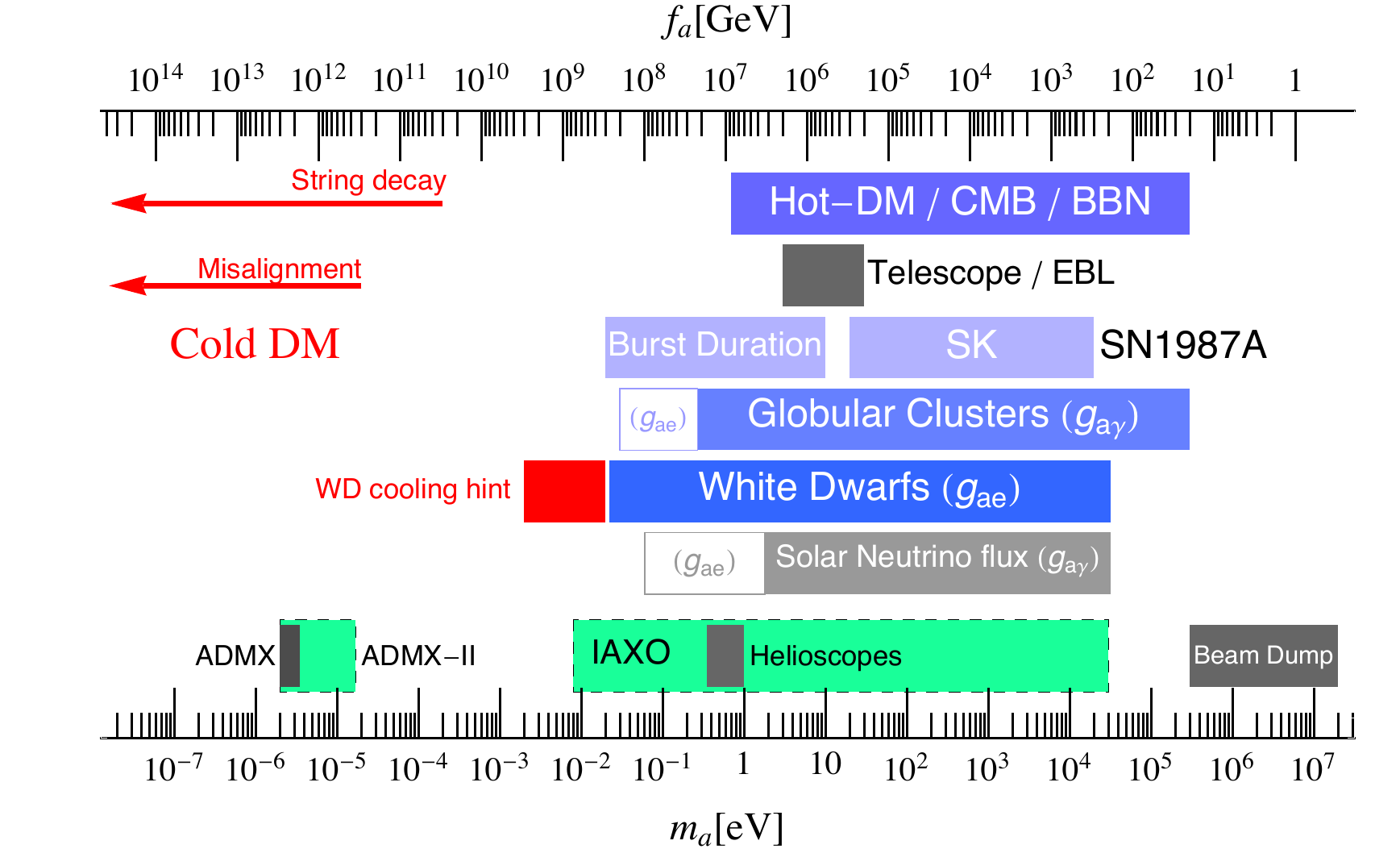} \\
\vskip 2mm
\includegraphics[width=0.74\textwidth]{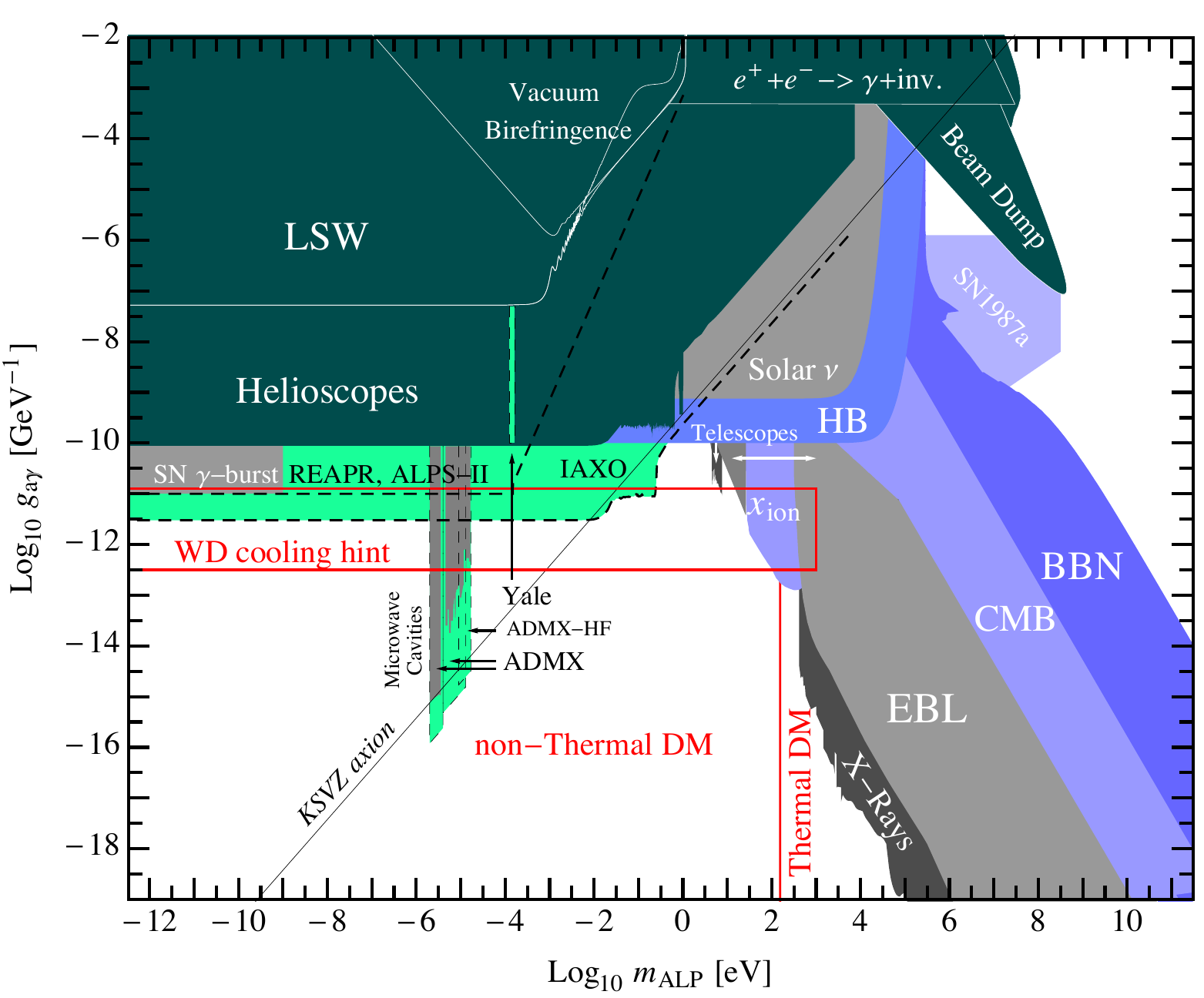} 
\caption{
Parameter space for axions (top) and axion-like particles (ALPs) (bottom).  In the bottom plot, the 
QCD axion models lie within an order of magnitude from the explicitly shown ``KSVZ'' axion line. 
Colored regions are: experimentally excluded regions (dark green), constraints from 
astronomical observations (gray) or from astrophysical or cosmological arguments (blue), and 
sensitivity of planned experiments (light green). Shown in red are boundaries where ALPs can account for all the dark 
matter produced either thermally in the big bang or non-thermally by the misalignment mechanism.
}\label{fig:hspaw-ALPs}
\end{figure}

Fig.~\ref{fig:hspaw-ALPs} (top) shows the allowed axion parameter space as a function of $f_a$ or, equivalently, $m_a$.  
Direct searches for such particles and calculations of their effect on the cooling of stars and on the supernova SN1987A 
excludes most values of $f_a\lesssim 10^9$~GeV.  Some of these constrain only the axion coupling to 
photons ($g_{a\gamma}$), while others constrain the axion coupling to electrons ($g_{ae}$).  
Recent and future laboratory tests (the latter shown in light green) can probe $f_a\sim 10^9-10^{12}$ GeV, or even higher $f_a$.  

The basic physical mechanism that leads to the axion --- the spontaneous breaking at a high energy scale of a $U(1)$ approximate global symmetry, generating a light PNGB --- also allows for other axion-like particles (ALPs).  Unlike axions, which are linked to the strong interactions and whose masses and couplings are determined by a single new parameter $f_a$, ALPs are much less constrained, and their masses and couplings to photons are independent parameters.  
Searches for ALPs should not therefore be limited to the parameter space of the axion itself.
Both ALPs and axions are generic in string theory~\cite{Print-84-0838 (PRINCETON),hep-th/0602233,hep-th/0605206,arXiv:0905.4720}, with the natural size of 
their decay constant $f_a$ being the string scale, varying typically between $10^9$ and
$10^{17}$~GeV. 

The parameter space for ALPs is shown in Fig.~\ref{fig:hspaw-ALPs} (bottom).  The axion parameter space lies within 
an order of magnitude from the line labelled ``KSVZ axion,'' which represents a particular QCD axion model.  
Experimentally excluded regions (dark green), constraints 
from astronomical observations (gray) or from astrophysical, or cosmological arguments (blue) are shown.  Sensitivity of 
a few planned experiments are shown in light green.  

Axions and ALPs can naturally serve as the universe's dark matter, meaning that the galactic halo may be formed partly or entirely from these particles.  They can be produced thermally or non-thermally in the early universe.  Thermally produced axions are disfavored by observations of the universe's large scale structure \cite{Hannestad:2010yi}, but thermally produced 
ALP dark matter is still allowed in parts of the parameter space.  Non-thermal production can occur through the ``vacuum misalignment mechanism,'' and axions could provide all the observed dark matter within this mechanism for $f_a\sim 3\times 10^{11}$ GeV ($m_a\sim 2\times 10^{-5}$ eV)~\cite{Wantz:2009it}.  This presents a clear experimental target.  The  Axion Dark Matter eXperiment (ADMX) will soon probe part of this preferred parameter space.  

More generally, however, a large number of non-thermal axion and ALP production mechanisms exist, including the decay 
of strings.  
Together with a whole range of possible initial conditions, a much larger parameter space needs to be 
explored as indicated in Fig.~\ref{fig:hspaw-ALPs}; see also {\it e.g.},~\cite{Arias:2012mb}.  

The presence of an axion or ALP during inflation generates isocurvature temperature fluctuations that can be searched for by the Planck satellite~\cite{Hamann:2009yf}. 
The axion dark matter may also form a Bose-Einstein condensate~\cite{Erken:2011dz}, which may lead to  
caustic rings in spiral galaxies, which may already have been observed.

There are some astrophysical puzzles that may be solved by the existence of axions or 
ALPs, namely the apparent non-standard energy loss of white dwarf stars, {\it e.g.},~\cite{arXiv:0806.2807,arXiv:0812.3043}, 
and the anomalous transparency of the universe for TeV gamma rays, {\it e.g.},~\cite{Horns:2012fx,arXiv:0707.4312,arXiv:0712.2825,arXiv:0905.3270,arXiv:1106.1132}. 
The required coupling strengths seem within reach in controlled laboratory experiments at the intensity frontier, 
and can serve as useful benchmarks.  

\subsection{Hidden-Sector Photons}\label{subsec:hspaw-motivation-A's} 

This section describes the theory and motivation for new forces mediated by new abelian $U(1)$ 
gauge bosons $A'$ --- also called  ``U-bosons,'' or ``hidden-sector,'' ``heavy,'' ``dark,'' ``para-,'' and ``secluded'' photons --- that couple very weakly to electrically charged particles through 
``kinetic mixing'' with the photon~\cite{Holdom:1985ag,Galison:1983pa}.  

\begin{figure}[!t]
\centering
\vspace*{-5mm}
\includegraphics[width=0.75\textwidth]{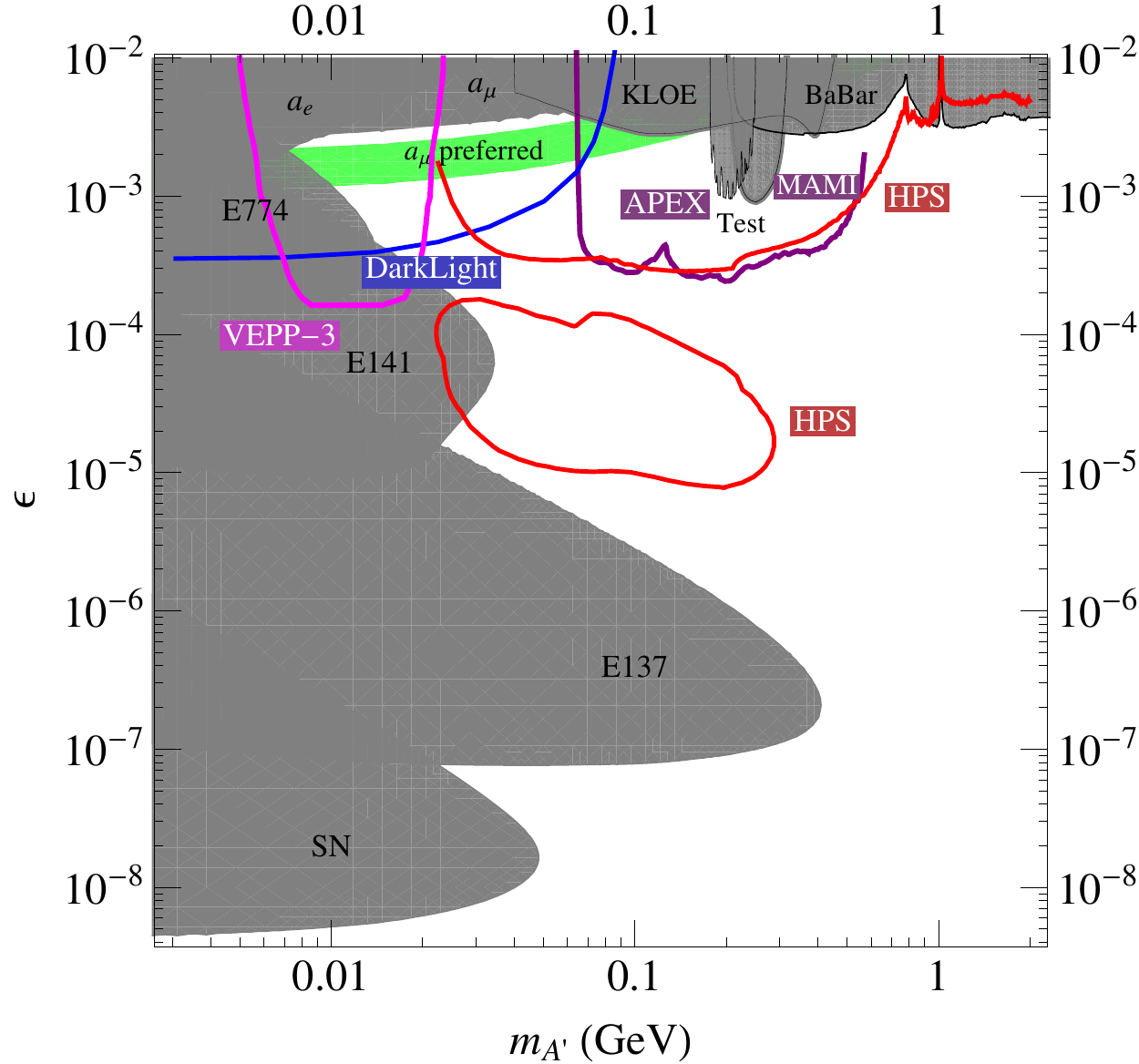} 
\caption{
Parameter space for hidden-photons ($A'$) with mass $m_{A'}> 1$~MeV 
(see Fig.~\ref{fig:hspaw-light-A'} for $m_{A'}\lesssim 1$~MeV).
Shown are existing 90\% confidence level limits from the SLAC and Fermilab beam dump experiments 
E137, E141, and E774~\cite{Bjorken:1988as,Riordan:1987aw,Bross:1989mp,Bjorken:2009mm} 
the muon anomalous magnetic moment $a_\mu$~\cite{Pospelov:2008zw},  
KLOE~\cite{Collaboration:2011zc}, 
the test run results reported by APEX~\cite{Abrahamyan:2011gv} and MAMI~\cite{Merkel:2011ze}, 
an estimate using a BaBar result~\cite{Bjorken:2009mm,Reece:2009un,Aubert:2009cp}, and 
a constraint from supernova cooling~\cite{Bjorken:2009mm} (see also~\cite{Dent:2012mx}).
In the green band, the $A'$ can explain the observed discrepancy between the
calculated and measured muon anomalous magnetic moment~\cite{Pospelov:2008zw} 
at 90\% confidence level.
Projected sensitivities are shown for the full APEX run~\cite{Essig:2010xa}, 
HPS~\cite{HPS}, DarkLight~\cite{Freytsis:2009bh}, and VEPP-3~\cite{Wojtsekhowski:2009vz}.  
MAMI has plans (not shown) to probe similar parameter regions as these experiments. 
Existing and future $e^+e^-$ colliders such as \babar, BELLE, KLOE, Super$B$, BELLE-2, and KLOE-2  can also probe large 
parts of the parameter space for $\epsilon\gtrsim 10^{-4}-10^{-3}$; their reach is also not explicitly shown.  
}
\label{fig:hspaw-heavy-A'}
\end{figure}

Kinetic mixing 
produces an effective parity-conserving interaction
$\epsilon e A'_\mu J^\mu_{\rm EM}$ of the $A'$ to the 
electromagnetic current $J^\mu_{EM}$,  suppressed relative to the electron charge 
$e$ by the parameter $\epsilon$, which can be naturally small (we often write the coupling strength as 
$\alpha' \equiv  \epsilon^2 \alpha$  where $\alpha= e^2/4\pi\simeq 1/137$). 
In particular, if the value of $\epsilon$ at very high energies is zero, 
then $\epsilon$ can be generated by perturbative or non-perturbative effects.  Perturbative contributions can include heavy messengers that carry both hypercharge and the new $U(1)$ charge, and quantum loops of various order can 
generate $\epsilon \sim 10^{-8} -10^{-2}$ \cite{Essig:2009nc}.   
Non-perturbative and large-volume effects common in string theory constructions can generate much smaller 
$\epsilon$.  
While there is no clear minimum for $\epsilon$, values in the $10^{-12}-10^{-3}$ range have been predicted in the 
literature~\cite{Goodsell:2009xc,Cicoli:2011yh,Goodsell:2011wn}. 

A hidden-sector consisting of particles that do not couple to any of the known forces and containing an $A'$ is generic 
in many new physics scenarios.  Hidden-sectors can have a rich structure, consisting of, for example, fermions 
and many other gauge bosons.  The photon coupling to the $A'$ could provide the only non-gravitational window into 
their existence.   
Hidden-sectors are generic, for example, in string theory 
constructions~\cite{Goodsell:2010ie,NSF-ITP-84-170,PRINT-86-0084 (PRINCETON),Andreas:2011in}. 
Several other ``portals'' (connections between a visible and hidden-sector) beyond the kinetic mixing portal are 
possible, many of which can be investigated at the intensity frontier. 

\begin{figure}[!t]
\centering
\includegraphics[width=\textwidth]{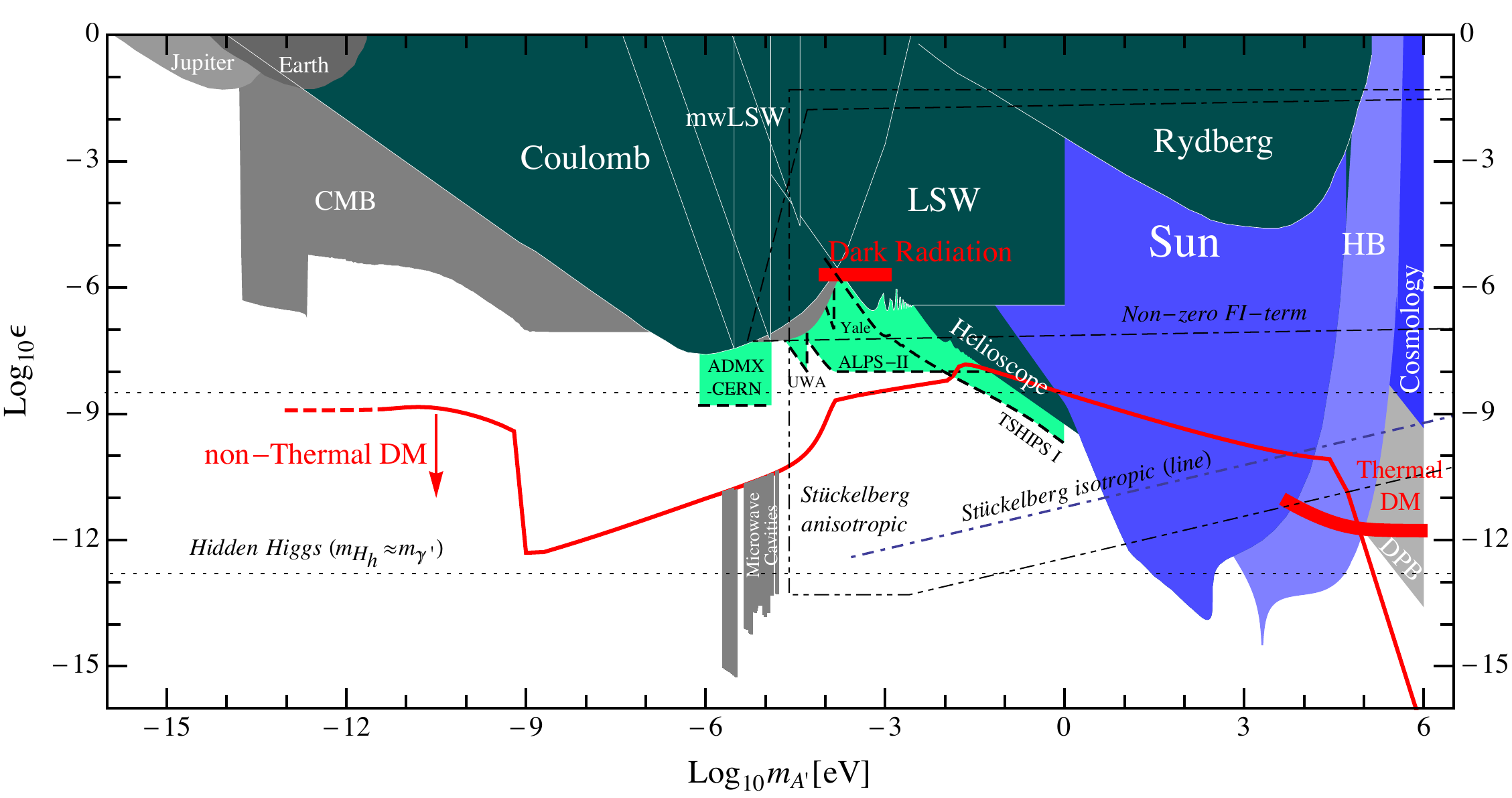} 
\caption{
Parameter space for hidden-photons ($A'$) with mass $m_{A'}\lesssim 1$~MeV (see Fig.~\ref{fig:hspaw-heavy-A'} for 
$m_{A'}>1$~MeV). 
Colored regions are: experimentally excluded regions (dark green), constraints from astronomical 
observations (gray) or from astrophysical or cosmological arguments (blue), and 
sensitivity of planned experiments (light green). Shown in red are boundaries where the $A'$ would account for all the dark matter produced either thermally in the big bang or non-thermally by the misalignment mechanism (the corresponding line is an upper bound).  The regions bounded by dotted lines show predictions from string theory 
corresponding to different possibilities for the nature of the $A'$ mass: Hidden-Higgs, a Fayet-Iliopoulos term, or the St\"uckelberg mechanism. In general, predictions are uncertain by factors of order 1. }
\label{fig:hspaw-light-A'}
\end{figure}

Masses for the $A'$ can arise via the Higgs mechanism and can take on a large range of values. 
$A'$ masses in the MeV--GeV range arise in the models 
of~\cite{Fayet:2007ua,Cheung:2009qd,ArkaniHamed:2008qp,Morrissey:2009ur} (these models often involve supersymmetry).  
However, much smaller (sub-eV) masses are also possible.  
Masses can also be generated via the St\"uckelberg mechanism, which is 
especially relevant in the case of large volume string compactifications with branes. 
In this case, the mass and size of the kinetic mixing are typically linked through one scale, the string
scale $M_s$, and therefore related to each other.  In Fig.~\ref{fig:hspaw-light-A'}, various theoretically motivated 
regions are shown~\cite{Goodsell:2009xc,Cicoli:2011yh}.  
The $A'$ mass can be as small as $M_s^2/M_{\rm Pl}$, i.e.~$m_{A'} \sim$~meV (GeV) for $M_s\sim$~TeV ($10^{10}$~GeV).  
Note that particles charged under a \emph{massive} $A'$ do not have an electromagnetic millicharge, but a \emph{massless} 
$A'$ can lead to millicharged particles (see \S\ref{subsec:hspaw-motivation-MCP}).  

A natural dividing line is $m_{A'} \sim 2 m_e \sim 1$~MeV.  For $m_{A'}>1$~MeV, 
an $A'$ can decay to electrically charged particles ({\it e.g.}, $e^+e^-$, $\mu^+\mu^-$, or $\pi^+ \pi^-$) or to light hidden-sector particles (if available), which can in turn decay to ordinary matter.   
Such an $A'$ can be efficiently produced in 
electron or proton fixed-target experiments \cite{Bjorken:2009mm,Freytsis:2009bh,Essig:2010xa,Abrahamyan:2011gv,HPS,Merkel:2011ze,Batell:2009di,Essig:2010gu,Wojtsekhowski:2009vz} 
and at $e^+e^-$ and hadron colliders  
\cite{Essig:2009nc,Cheung:2009qd,Strassler:2006im,Reece:2009un,AmelinoCamelia:2010me,:2009cp,Batell:2009yf,Aubert:2009pw,Abazov:2009hn,Abazov:2010uc,Collaboration:2011zc,Lees:2012ra,Baumgart:2009tn}.  
Hidden-sector particles could be directly produced through an off-shell $A'$ and decay to ordinary matter.  
An $A'$ in this mass range is motivated by the theoretical considerations discussed above, by anomalies related 
to dark matter~\cite{ArkaniHamed:2008qn,Pospelov:2008jd}, 
and by the discrepancy between the measured and calculated value of the anomalous magnetic moment of the 
muon~\cite{Pospelov:2008zw}.  

Fig.~\ref{fig:hspaw-heavy-A'} shows existing constraints for $m_{A'}>1$~MeV~\cite{Bjorken:2009mm} and the sensitivity of 
several planned experiments that will explore part of the remaining allowed parameter space. 
These include the future fixed-target experiments APEX~\cite{Essig:2010xa,Abrahamyan:2011gv}, HPS~\cite{HPS}, 
DarkLight~\cite{Freytsis:2009bh} at Jefferson Laboratory, 
experiments at MAMI~\cite{Merkel:2011ze} at the University of Mainz (whose reach are not shown, but which may probe similar parameter regions as other experiments), and another at VEPP-3~\cite{Wojtsekhowski:2009vz}.  
Existing and future $e^+e^-$ colliders can also probe large parts of the parameter space for $\epsilon\gtrsim 10^{-4}-10^{-3}$, and include \babar, Belle, KLOE, Super$B$, Belle-2, and KLOE-2 
(the figure only shows existing constraints, and no future sensitivity).  
Proton colliders such as the LHC and Tevatron can also see remarkable signatures for light hidden-sectors~\cite{Strassler:2006im}.  
This rich experimental program is discussed in more detail in \S\ref{subsec:hspaw-searches}.  

For $m_{A'}<1$ MeV, the $A'$ decay to $e^+e^-$ is kinematically forbidden, and only a much slower decay to 
three photons is allowed.  Fig.~\ref{fig:hspaw-light-A'} shows the constraints, theoretically and phenomenologically motivated regions, and some soon-to-be-probed parameter space.  
At very low masses, the most prominent implication of kinetic mixing is that,
similar to neutrino mixing, the propagation and the interaction eigenstates are misaligned,
giving rise to the phenomenon of photon $\leftrightarrow$ $A'$ oscillations~\cite{ITEP-48-1982}.
In the early universe, these oscillations convert thermal photons into $A'$ bosons, generating 
a ``hidden Cosmic Microwave Background'' (hCMB)~\cite{arXiv:0804.4157}.
For $\sim$~meV masses and $\epsilon \sim 10^{-6}$, they 
occur resonantly after big bang nucleosynthesis and before the decoupling of the CMB, and
the corresponding hCMB could lead to an apparent increase in the effective number of 
relativistic degrees of freedom, consistent with some recent  global cosmological analyses~\cite{arXiv:1001.4538,arXiv:1009.0866,arXiv:1105.3182,arXiv:0804.4157}. 
These observations will soon be tested by the Planck satellite. 
At the same time, an $A'$ in the parameter range of interest can also be searched for in the laboratory by 
light-shining-through-wall experiments and helioscopes (see \S\ref{subsec:hspaw-searches}).

The photon $\leftrightarrow$ $A'$ oscillation mechanism can also generate the required $A'$ energy density for them 
to account for all the dark matter for $m_{A'} \sim 100$ keV and $\epsilon \sim 10^{-12}$~\cite{Redondo:2008ec}. This hypothesis can be tested in direct dark matter detection experiments or indirectly through the $A'$ decay into three photons, which could be observed above the astrophysical diffuse X-ray backgrounds~\cite{Pospelov:2008jk}. 

As axions or ALPs, $A'$ bosons can also be dark matter through the realignment mechanism~\cite{Nelson:2011sf}. This intriguing possibility can be realized in a wide range of values for $m_{A'}$ and $\epsilon$~\cite{Arias:2012mb}, see Fig.~\ref{fig:hspaw-light-A'}. It appears that experiments such as ADMX, looking for axion dark matter, can be very sensitive to $A'$ bosons as well, but in this case the use of magnetic fields to trigger the $A' \to$photon conversion is not required. As we will see 
in \S\ref{subsec:hspaw-searches}, the same experimental apparatus can often look for several kinds of particles.  

Other existing constraints, theoretically and phenomenologically motivated parameter regions, and future experimental searches 
for $A'$ bosons with $m_{A'}<1$~MeV are shown in Fig.~\ref{fig:hspaw-light-A'}.  A few planned experimental searches are discussed in 
\S\ref{subsec:hspaw-searches}, although a large parameter space still remains to be experimentally explored.  

\subsection{Milli-charged particles}\label{subsec:hspaw-motivation-MCP}

\begin{figure}[!t]
\centering
\includegraphics[width=0.7\textwidth]{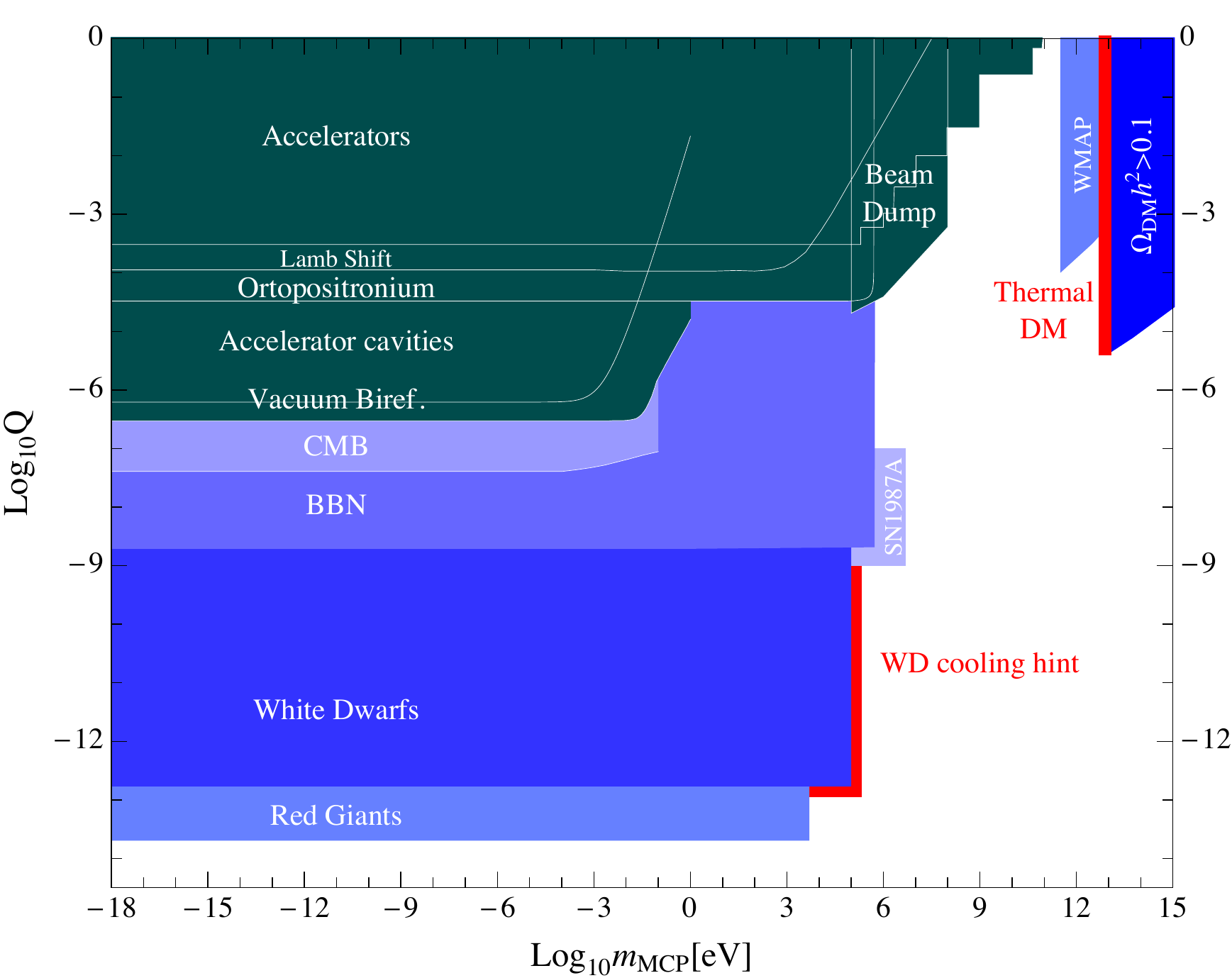} 
\caption{Parameter space for milli-charged particles, where $Q$ is the fractional charge 
in units of the electron charge.  
Colored regions are: experimentally excluded regions (dark green) and constraints from 
astrophysical or cosmological arguments (blue). Shown in red are regions where milli-charged particles could 
account for the dark matter in our universe or explain the energy loss in white dwarf stars; 
see~\S\ref{subsec:hspaw-motivation-ALPs}.}
\label{fig:hspaw-MCP}
\end{figure}

Particles with small un-quantized electric charge, often called mini- or milli-charged particles (MCPs), also arise 
naturally in many extensions of the Standard Model.
MCPs are a natural consequence of extra $U(1)$s and the kinetic mixing discussed in 
\S\ref{subsec:hspaw-motivation-A's} for massless $A'$ fields. 
In this case any matter charged (solely) under the hidden $U(1)$ obtains a small electric charge.
MCPs can also arise in extra-dimensional scenarios or as hidden magnetic monopoles receiving their mass from a magnetic mixing effect~\cite{Batell:2005wa,Brummer:2009cs,Bruemmer:2009ky}.
Milli-charged fermions are particularly attractive because chiral symmetry protects their mass against quantum corrections, making it more natural to have small or even vanishing masses.
MCPs have also been suggested as dark matter candidates~\cite{Goldberg:1986nk,Cheung:2007ut,Feldman:2007wj}. 

Experiments as well as astrophysical and cosmological observation provide interesting constraints on MCPs. These are 
summarized in Fig.~\ref{fig:hspaw-MCP}.  It needs to be investigated whether currently planned experiments can probe any of 
the remaining parameter space.

\subsection{Chameleons}\label{subsec:hspaw-motivation-chameleons}

\begin{figure}[!t]
\centering
\vspace*{-20mm}
\includegraphics[width=0.85\textwidth]{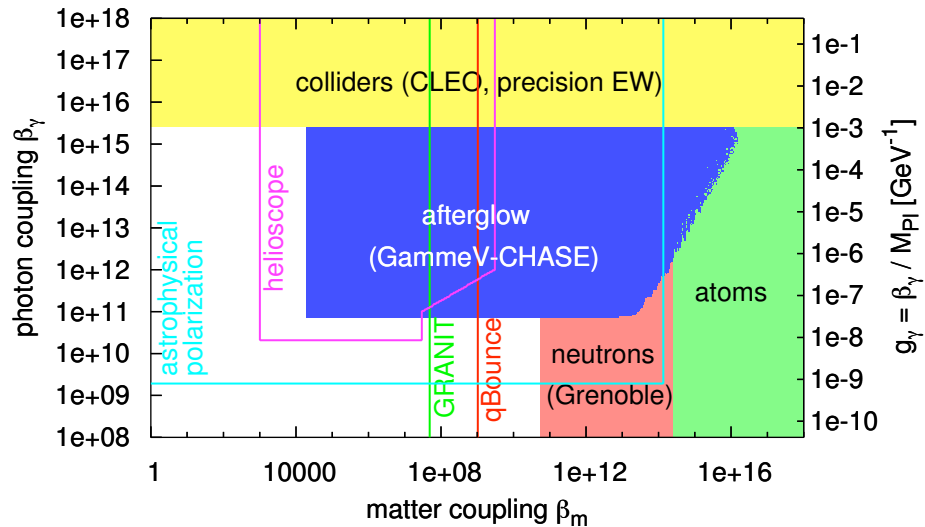} 
\vspace*{-10mm}
\caption{Constraints on a specific chameleon dark energy model (see text for details).  Solid regions labelled by horizontal text represent current constraints; curves labelled by vertical text represent forecasts.}
\label{fig:hspaw-chameleon}
\end{figure}

The acceleration of the expansion of the universe, the discovery of which has won the 2011 Nobel Prize in Physics, remains the greatest mystery in modern cosmology. Any field responsible for this acceleration may couple directly to Standard Model particles. Chameleon fields are a compelling dark energy candidate, as they would do exactly this without violating any known laws or experiments of physics. Importantly, these fields are testable in ways entirely complementary to the standard observational cosmology techniques, and thus provide a new window into dark energy through an array of possible laboratory and astrophysical tests and space tests of gravity. Such a coupling, if detected, could reveal the nature of dark energy and may help lead the way to the development of a quantum theory of gravity.

A canonical scalar field is the simplest dynamical extension of the Standard Model that could explain dark energy.  In the absence of a self interaction, this field's couplings to matter --- which we would expect to exist unless a symmetry forbids them --- would lead to a new, fifth fundamental force whose effects have yet to be observed. However, scalar field dark energy models typically require a self interaction, resulting in a nonlinear equation of motion~\cite{Peebles_Ratra_1988,Ratra_Peebles_1988}.  Such a self interaction, in conjunction with a matter coupling, gives the scalar field a large effective mass in regions of high matter density~\cite{Khoury_Weltman_2004a,Khoury_Weltman_2004b}. A scalar field that is massive locally mediates a short-range fifth force that is difficult to detect, earning it the name ``chameleon field.''  Furthermore, the massive chameleon field is sourced only by the thin shell of matter on the outer surface of a dense extended object.  These nonlinear effects serve to screen fifth forces, making them more difficult to detect in certain environments.

Current best theories treat chameleon dark energy as an effective field theory~\cite{Brax_etal_2004,Hui_Nicolis_2010} describing new particles and forces that might be seen in upcoming experiments, and whose detection would point the way to a more fundamental theory. The ultraviolet (UV) behavior of such theories and their connection to fundamental physics are not yet understood, although progress is being made~\cite{Hinterbichler_Khoury_Nastase_2011}.  

A chameleon field couples to dark matter and all matter types, in principle with independent strengths. At the classical level, a chameleon field is not required to couple to photons, though such a coupling is not forbidden.
However, when quantum corrections are included, a photon coupling about three orders of magnitude smaller than the matter coupling is typically generated~\cite{Brax_etal_2010}.  The lowest order chameleon-photon interaction couples the chameleon field to the square of the photon field strength tensor, implying that in a background electromagnetic field, photons and chameleon particles can interconvert through oscillations.  
The mass of chameleon fields produced will depend on the environmental energy density as well as the electromagnetic field strength. This opens the vista to an array of different tests for these fields on Earth, in space, and through astrophysical observations.  Several astrophysical puzzles could also be explained by chameleons, {\it e.g.},~\cite{Hui:2009kc}. 

The chameleon dark energy parameter space is considerably more complicated than that of axions, but constraints can be provided under some assumptions. With the caveat that all matter couplings are the same but not equal to the photon coupling,  and the assumption of a specific chameleon potential,  $V(\phi) = M_\Lambda^4(1 + M_\Lambda^n/\phi^n)$ in which we set 
the scale $M_\Lambda = 2.4\times 10^{-3}$~eV to the observed dark energy density and, for concreteness, $n=1$, our constraints and forecasts are provided by Figure~\ref{fig:hspaw-chameleon}. This choice prevents us from including Casimir force constraints~\cite{Brax_etal_2007c}, which are most powerful for negative $n$, as well as torsion pendulum constraints~\cite{Kapner_etal_2007,Adelberger_etal_2007}, which have been computed only for $n = -4$. Current constraints (solid regions) and forecasts (curves) are discussed in \S\ref{subsec:hspaw-searches} with an emphasis on afterglow experiments and 
astrophysics expectations.

\section{Experimental Searches}\label{subsec:hspaw-searches} 

This section discusses various experimental searches for axions, axion-like particles (ALPs), hidden-photons, chameleons, 
and milli-charged particles.   Minor modifications to a given experimental technique often allows sensitivity to more than 
one type of particle, so rather than organizing the discussion according to particle type, we here organize it according to 
experimental technique.  The proposed sensitivities of several currently planned experiments are shown in 
Figs.~\ref{fig:hspaw-ALPs} to \ref{fig:hspaw-chameleon}, but \emph{several other experiments have been proposed and many more are possible, and not all proposed experiments discussed below appear in the figures}.

\subsection{Microwave Cavities}\label{subsec:hspaw-searches-MC}

Soon after the axion was realized to be a natural dark matter candidate, a detection concept was proposed that relies on the resonant conversion of dark matter axions into photons via the Primakoff effect \cite{Sikivie_1}. Though the axion mass is unknown, various production mechanisms in the early universe point to a mass scale of a few to tens of $\mu$eV if the axion is the dominant form of dark matter. The detection concept relies on dark matter axions passing through a microwave cavity in the presence of a strong magnetic field where they can resonantly convert into photons when the cavity frequency matches the axion mass. A 4.13 $\mu$eV axion would convert into a 1 GHz photon, which can be detected with an ultra-sensitive receiver. Axions in the dark matter halo are predicted to have virial velocities of $10^{-3} c$, leading to a spread in axion energies 
of $\Delta E_a/E_a \sim 10^{-6}$ (or 1 kHz for our 1 GHz axion example).

Initial experiments run at Brookhaven National Laboratory~\cite{BRF} and the University of Florida~\cite{UofF} came within an order of magnitude of the sensitivity needed to reach plausible axion couplings. 
ADMX~\cite{ADMX_1} was assembled at Lawrence Livermore National Laboratory and consists of a large, 8~T superconducting solenoid magnet with a 0.5 m diameter, 1 m long, open bore. Copper-plated stainless steel microwave cavities are used and have $Q_C \sim 10^5$, low enough to be insensitive to the expected spread
in axion energies. The TM$_{010}$ mode has the largest cavity form factor and is moved to scan axion masses by translating vertical copper or dielectric tuning rods inside the cavity from the edge to the center. TE and TEM modes do not couple to the pseudoscalar axion. 

Using the ADMX setup and an estimated local dark matter density of $\rho_{DM} = 0.45$ GeV/cm$^3$ \cite{Local_DM}, an axion conversion power $P_a \sim 10^{-24}$ W is expected for plausible dark matter axions, with the possibility of scanning an appreciable frequency space (hundreds of MHz) in just a few years. Initial data runs were cooled with pumped LHe to achieve physical temperatures of $<$ 2 K and used SQUID amplifiers to reach plausible dark matter axion couplings \cite{SQUID_results}. Recently the ADMX experiment has been moved to the University of Washington where it will be outfitted with a dilution refrigerator that will increase sensitivity and scan rate. A second ADMX site, dubbed ADMX-HF, is being constructed at Yale and will allow access to $>$ 2 GHz while ADMX scans from 0.4 - 2 GHz. To achieve a greater mass reach, near-quantum limited X-band amplifiers and large volume resonant cavities will have to be developed.

As shown in Fig.~\ref{fig:hspaw-ALPs}, ADMX and ADMX-HF are uniquely sensitive to axion and ALP dark matter in the range of a few to tens of $\mu$eV. The experiments also have exceptional sensitivity to hidden-photons in the same mass region, as shown in Fig.~\ref{fig:hspaw-light-A'}.   
\subsection{Light Shining Through A Wall}\label{subsec:hspaw-searches-LSW} 

The ``Light Shining through a Wall'' (LSW) technique
is a
method where photons of light, $\gamma$'s, are injected against an opaque barrier, and 
a search is made for photons on the other side of this wall.
A positive result would suggest that incoming photons transform into
weakly interacting particles that traverse the barrier, and then
reconvert back into detectable photons.
This technique was originally envisioned as a method suitable for searching
for axion-like particles that couple to two 
photons \cite{Anselm:1986gz,VanBibber:1987rq}, with previous mention in the context of paraphoton searches~\cite{ITEP-48-1982}. A laser beam was shone into
the bore of a
dipole magnet where a laser photon and magnetic field photon
could interact and produce an ALP that could traverse the wall. In
the magnetic field region on the other side of the wall, 
the ALP could regenerate back into a detectable photon.

A first experimental search using spare
accelerator magnets \cite{Cameron:1993mr} in the early 1990s
excluded ALPs with a mass below 1~meV and a
coupling to photons $g_{a \gamma}> 7 \times 10^{-7}$~GeV$^{-1}$. 
Motivated by a possible positive ALP signal 
reported in 2006 \cite{pvlas}, recent LSW experiments have not found evidence for ALPs with 
$g_{a \gamma} > 7 \times 10^{-8}$~GeV$^{-1}$~\cite{worldwide}. This
limit amounts to one rare reconversion process in 10$^{25}$ interactions.

The LSW technique can also be used to search
for processes where photons might transform into 
very light ($<1$~eV) weakly interacting
slim particles (WISPs)
such as light hidden-sector photons or milli-charged particles
\cite{Ahlers:2007qf}. In this case, kinetic mixing would allow the process
to take place even in the absence of a magnetic field. Experiments
\cite{worldwide} have excluded important
regions of WISP parameter space.  

In the short term, LSW experiments will continue to explore open
regions of WISP parameter space. In the mid term,
optimized future LSW experiments will
make use of advanced techniques
such as matched Fabry-Perot optical cavities on both sides of the wall to 
increase the incident 
laser power and to build up the regenerated signal resonantly.
Such efforts can reach sensitivities of $g_{a \gamma} < 1 \times 10^{-11}$~GeV$^{-1}$ \cite{futurelsw}. A comprehensive review of LSW experiments can be found in~\cite{Redondo:2010dp}.
The reach of early LSW experiments and REAPR and ALPS-II are shown in Fig.~\ref{fig:hspaw-ALPs}. LSW sensitivity to low mass hidden-photons is given in Fig.~\ref{fig:hspaw-light-A'}.

\subsection{Axion Helioscopes}\label{subsec:hspaw-searches-helio} 

Axions can be created inside the sun via the Primakoff effect \cite{Primakoff51}. Due to their extraordinarily weak coupling to matter, these particles can escape from the solar interior. An axion helioscope relies on the inverse Primakoff effect \cite{Cavity_idea,Cavity_idea_2,vmmr89}, i.e., reconversion into $0.1-10$~keV photons in an intense transverse magnetic field to detect them. The minimum requirements for a helioscope are therefore a powerful magnet and an X-ray sensor.  Sensitivity can be enhanced by providing a mechanical system to allow for solar tracking, and by use of X-ray optics to reduce detector size and with it the backgrounds.

The first axion helioscope search was carried out at Brookhaven National Lab in 1992 with a stationary dipole magnet \cite{Lazarus92}.
A second-generation experiment, the Tokyo Axion Helioscope, used a
more powerful magnet and dynamic tracking of the sun \cite{Moriyama1998147,Inoue200218,Inoue200893}.
A third-generation experiment, the CERN Axion Solar Telescope (CAST), began data collection in 2003. It employs an LHC dipole test
magnet \cite{cast99} with an elaborate elevation and azimuth drive to track the sun. CAST is also the first
helioscope to use an X-ray optic to focus the photon signal \cite{cast_xrt}.  Each generation of axion helioscope has achieved 
a six-fold improvement in sensitivity to the axion coupling constant over its predecessors.
For $m_{a} <  0.02$~eV, CAST has set an upper limit of $g_{a \gamma} < 8.8\times 10^{-11}$~GeV$^{-1}$ and a larger value of $g_{a \gamma}$ for higher axion masses \cite{CAST_PRL05,cast_jcap07,cast_jcap09,CAST_PRL11}. These sensitivities are shown in Fig.~\ref{fig:hspaw-ALPs} and Fig.~\ref{fig:hspaw-light-A'}.  

To date, all axion helioscopes have ``recycled'' magnets built for other purposes.  A significantly improved sensitivity  is to be expected from a custom-built magnet.
The fourth-generation axion helioscope proposed in \cite{ngah11}, dubbed the International Axion Observatory (IAXO), envisions an
ATLAS-like magnet, a detection system consisting of large X-ray telescopes coupled to ultra-low background X-ray detectors, and a large, robust tracking system.  IAXO would enable the search for solar axions and ALPs in a broad mass range, with sensitivity down to a few $10^{-12}$ GeV$^{-1}$, with sensitivity to QCD axion models down to the meV scale. Lowering X-ray detector thresholds to 0.1~keV would allow IAXO to test whether solar processes can create chameleons \cite{BraxKon10}.  More speculative, but of tremendous potential scientific gain, would be the operation of microwave cavities inside IAXO's magnet, to allow a simultaneous search for solar and dark matter axions [{\it e.g.}, \cite{hrelics}]. Searches for solar axions and chameleons which exploit naturally occurring magnetic fields are described in \cite{hcham,hrelics,hnjp} and reviewed in \cite{hpatras}.

\subsection{Chameleon Searches: Afterglow Experiments + Others}\label{subsec:hspaw-searches-afterglow} 

Afterglow experiments, similar in design to axion searches, produce chameleon particles through oscillations inside a vacuum chamber, which traps them in the chamber walls if their energies are less than their effective mass.  The trapped chameleon particles oscillate back into photons and emerge from the vacuum chamber as an ``afterglow'' after the photon source has been turned off.
The GammeV experiment~\cite{Chou_etal_2009} constrained chameleon fields by modifying the earlier GammeV axion search~\cite{Chou_2008}.  The recent afterglow search GammeV-CHASE~\cite{Steffen_etal_2010,Upadhye_Steffen_Weltman_2010} implemented better control of systematic effects, significantly improving on GammeV constraints.  
A proposed afterglow experiment based upon a modification of the LIPSS axion search can further improve sensitivity to chameleon-photon interactions by use of a higher-intensity photon beam~\cite{Afanasev_etal_2010}.  

Chameleon particles produced by astrophysical sources may also be detectable.  The sun is a well-understood source whose large magnetic field could produce chameleon particles capable of reaching the Earth.  These could regenerate detectable photons in the high magnetic field of a helioscope~\cite{Brax_Zioutas_2010,Brax_Lindner_Zioutas_2011}.  Yet more distant sources could be dimmed or polarized due to photon-chameleon oscillation~\cite{Burrage_Davis_Shaw_2009,Burrage_Davis_Shaw_2009b}.  Since these provide preliminary constraints subject to systematic uncertainties, we include them in Fig.~\ref{fig:hspaw-chameleon} as forecasts.

Colliders exclude very large chameleon-photon couplings $\beta_\gamma$~\cite{Brax_etal_2009,Brax_etal_2010,Kleban_Rabadan_2005,Balest_etal_1995} (see Fig.~\ref{fig:hspaw-chameleon}).  At large chameleon-matter couplings $\beta_m$, limits are provided by measurements of the energy levels of atoms~\cite{Brax_Burrage_2011}, as well as of neutrons bouncing in a gravitational field~\cite{Brax_Pignol_2011,Nesvizhevsky_etal_2002}.  Upcoming neutron bouncing experiments are expected to improve constraints on $\beta_m$ at low $\beta_\gamma$ by orders of magnitude~\cite{Kreuz_etal_2009}. 

It is worth noting that these constraints are model dependent, and providing completely model independent constraints is very challenging. Present constraints cluster in the very large coupling regions for both matter and photon couplings. Order-one couplings can be probed by space tests of gravity that   are proposed but yet to launch. 

\subsection{Electron Fixed-Target Experiments}\label{subsec:hspaw-searches-electron-FT} 

Fixed-target experiments using high-current electron beams are an excellent place to search for $A^\prime$s with 
masses $2m_e<m_{A'} \lesssim$~GeV and couplings down to $\epsilon^2 \equiv \alpha'/\alpha \gtrsim 10^{-10}$.   In these experiments,
the $A^\prime$ is radiated off electrons that scatter on target nuclei. Radiative and Bethe-Heitler trident production give rise to large backgrounds.   
Generally speaking, three experimental approaches have been proposed:   
dual-arm spectrometers, forward vertexing spectrometers, and full final-state reconstruction.  
In most cases, the detectors are optimized to detect the $e^+e^-$ daughters of the $A'$. The complementary approaches
map out different regions in the mass-coupling parameter space. General strategies for $A'$ searches 
with electron fixed-target experiments were laid out in \cite{Bjorken:2009mm}. The reach for recently proposed heavy photon searches
is shown in Fig.~\ref{fig:hspaw-heavy-A'}.

Existing dual-arm spectrometers at Hall A at Jefferson Lab (JLab) and MAMI at Mainz can be used to 
search for heavy photons.  These experiments use  high-current beams ($\sim 100~\mu$A) on relatively thick targets 
(radiation length $X_{0} \sim$ 1-10\%)  to overcome the low geometric acceptance of the detectors ($\sim 10^{-3}$).  
Beam energy and spectrometer angles are varied to cover overlapping regions of invariant mass.  Searches for $A'$ involve looking for 
a bump in the $e^+e^-$ invariant mass distribution over the large trident background, which requires an excellent mass 
resolution.    
Two groups, APEX at JLab and the A1 collaboration at Mainz, have recently performed short test runs and published search 
results with sensitivity down to  $\alpha'/\alpha>10^{-6}$ over narrow mass ranges~\cite{Merkel:2011ze,Abrahamyan:2011gv}.  
Both groups have plans for more extensive searches in the near future~\cite{Essig:2010xa}. 

The HPS collaboration~\cite{HPS} has proposed an experiment to take place in Hall B at JLAB which would use a Si-strip based vertex tracker 
inside a magnet to measure the invariant mass and decay point of  $e^+e^-$ pairs. It uses lower beam currents and thinner targets than the dual arm spectrometers, but has much greater acceptance.  Because it can discriminate $A'$ decays displaced more 
than a few millimeters from the large, prompt, trident background, HPS is sensitive to roughly $10^{-7}\gtrsim \alpha'/\alpha\gtrsim 10^{-10}$ 
for masses $30<m_{A'}<500$~MeV.  Without requiring a displaced vertex, HPS will also explore 
couplings $\alpha^\prime/\alpha>10^{-7}$ over a similar mass range.  HPS has been approved by JLab for a test run with a scaled-down detector in the spring of 2012,
and plans to take data with the full detector after the 12 GeV upgrade at the lab's CEBAF accelerator.

Finally, the experiment DarkLight is proposing to run at the JLab FEL~\cite{Freytsis:2009bh}.  DarkLight proposes to collide the FEL's 10mA, 
100MeV beam on a hydrogen gas jet target surrounded by a high-acceptance detector in a magnetic field.  A key component to the 
experiment is a proton-recoil detector, which would allow for a full reconstruction of the event and should greatly reduce 
accidental background.  This experiment will explore the low-mass region and coupling $\alpha'/\alpha>10^{-7}$, which 
covers almost all of the parameter space needed to explain the measured anomalous magnetic moment of the muon.  DarkLight plans to submit a full proposal to JLab in 2012.  

Future experiments will need to fill the gaps in $m_{A'}-\alpha'/\alpha$ parameter space. To explore the gap region at medium 
mass with medium coupling strengths, they will need to handle more intense beams and have improved vertexing and invariant mass resolution.
To search for higher-mass $A'$ decays, higher beam energies, particle ID, and very high luminosities are needed.

\subsection{Proton Fixed-Target}\label{subsec:hspaw-searches-proton-FT} 

Proton beam fixed-target experiments, especially the neutrino experiments CHARM~\cite{Bergsma:1985qz}, 
LSND \cite{hpsaw:Athanassopoulos:1996ds}, 
MiniBooNE~\cite{AguilarArevalo:2008yp}, 
MINOS~\cite{:1998zzb}, MINERvA~\cite{Osmanov:2011ig}, 
T2K~\cite{Abe:2011ks}, and a future Project X facility~\cite{Holmes:2010zza}, provide an opportunity to search for new light, 
weakly coupled states from a hidden-sector. The basic experimental setup begins with an intense proton beam impinging on a target, producing 
large numbers of secondary hadrons, most of which subsequently decay into neutrinos and other particles. After passage through a shield or earth, 
all the other particles are absorbed, leaving a neutrino beam. If light, weakly coupled 
hidden-sector particles exist, a beam of hidden-sector particles might be produced in similar fashion, and leave distinctive signatures in a downstream detector.

To illustrate the essential experimental principle, consider the case of a heavy photon $A'$ that kinetically 
mixes with the photon $\gamma$, with mixing angle $\epsilon$. Due to the kinetic mixing, neutral pions produced in primary 
collisions will decay a small fraction of the time into a $\gamma A'$ pair. The heavy photon $A'$ may travel to the detector 
and decay into, or interact to produce, an $e^+ e^-$ pair. The neutrino experiments listed above are sensitive to this signature and thus have the 
potential to probe a large part of the $m_{A'}-\alpha'/\alpha$ parameter space of the heavy photon, 
including regions that cannot be tested by any other experiment~\cite{Batell:2009di,Essig:2010gu}.   

Proton fixed-target experiments are unique among particle physics experiments in being sensitive
to some hidden-sector particles, such as axions and
ALPs~\cite{Essig:2010gu}. In addition they are sensitive to heavy photons and hidden
sector scalars and Higgs bosons~\cite{Batell:2009di} and light dark matter~\cite{Batell:2009di,deNiverville:2011it}.  An example 
of a specific search at proton accelerators is for a paraphoton (which is similar to the $A'$, but with slightly different couplings)~\cite{Nelson:2007yq}.  This can be produced in the extreme forward direction (less than a few milliradians w.r.t. the 
proton beam) by bremsstrahlung 
off the incident protons. 
In the case of the
MiniBooNE experiment at FNAL, with $\sim 1.8 \times 10^{21}$
protons on target, the paraphoton can decay or shower in the detector. The signature is thus
electromagnetic-like events directly in line with the incident proton beam. Furthermore, if
these events scatter off electrons, then the recoil electron would point in the very forward direction and could thus be easily
disentangled from the neutrino background.  This is only one of many
examples of the unique event signatures that hidden-sector particles may
have in neutrino experiments, which enhance the overall search
sensitivity relative to simple event counting.

Besides distinct event signatures, the other driving factor in sensitivity is the number of  protons on target.   With the possibility of Project X, the protons on target
intensities could easily be increased by two orders of magnitude at
8 GeV and about an order of magnitude at 120 GeV
over current FNAL rates.  These searches could form
an important part of the motivation for Project X.

\subsection{Electron-Positron Colliders}\label{subsec:hspaw-searches-e+e-} 

Due to their large luminosities and low background environments, low-energy $e^+e^-$ colliders 
are ideal for searching for a light hidden-sector and probing its structure. 

A hidden-sector (dark) photon can be produced in the reaction $e^+e^- \rightarrow \gamma A'$, and decay subsequently 
into a lepton pair. This signature is similar to that of light $CP$-odd Higgs production, $A^0$, in 
$\Upsilon(2S,3S) \rightarrow \gamma A^0, A^0 \rightarrow l^+l^-$. Searches for narrow di-muon 
resonances in $\Upsilon(3S)$ and $\Upsilon(2S)$ decays at \babar\ based on $\sim 40 \rm~fb^{-1}$ of data 
\cite{Aubert:2009cp} can be readily reinterpreted as constraints on dark photon production. 
Limits on the ratio $\alpha'/\alpha$ at the level of $10^{-5}$ have been set.  
Future analyses based on the full \babar\ and Belle data sets are expected to increase this sensitivity by an 
order of magnitude and extend the coverage down to $\sim 10~\Mev$. In particular, the region favored by 
an explanation of the muon anomalous magnetic moment~\cite{hpsaw:arXiv:1010.4180} can be almost entirely probed, 
see Fig.~\ref{fig:hspaw-heavy-A'}.

The simplest extension to a non-Abelian model contains four gauge bosons, one dark photon, and three 
additional dark vector bosons, generically denoted $W'$. A search for di-boson production has been performed 
at \babar\ in the four lepton final state, $e^+e^- \rightarrow W' W'', W'(W'') \rightarrow l^+l^-$ ($l=e,\mu$), 
assuming both the $W'$ and $W''$ have similar masses~\cite{arXiv:0908.2821}. No significant signal is observed 
and limits on the product $\alpha_D \epsilon^2$ at the level of $10^{-10} - 10^{-7}$ have been 
set, assuming equal branching fractions of a dark gauge boson to $e^+e^-$ and $\mu^+\mu^-$ 
(here $\alpha_D = g_D^2/4\pi$, where $g_D$ is the hidden $U(1)$ gauge coupling). 

The dark boson masses are usually generated via the Higgs mechanism, adding a dark Higgs ($h'$) to the theory. 
\babar\  has searched for dark Higgs boson production in 
$e^+e^- \rightarrow A'^* \rightarrow A' h', h' \rightarrow A' A'$, with the constraint $m_{h'} > 2 m_{A'}$~\cite{Lees:2012ra}. 
No signal was seen, and limits on the product $\alpha_D \epsilon^2$ have been set at 
the level of $10^{-10} - 10^{-8}$, depending on the dark photon and dark Higgs boson masses.  A search at Belle is in progress.  

The large number of photons produced by light meson decays offers another gateway to the dark sector. KLOE has 
performed a search for dark photon production in 
$\phi \rightarrow \eta A', \eta \rightarrow \pi^+ \pi^- \pi^0, A' \rightarrow e^+e^-$ 
decays using $1.5 \rm~fb^{-1}$ of data collected at Da$\Phi$Ne \cite{Collaboration:2011zc}. No signal has been seen, and 
limits on $\alpha'/\alpha$ at the level of $10^{-5}$ have been derived in the range $80-400~\Mev$ 
(Fig.~\ref{fig:hspaw-heavy-A'}). 

The existence of a dark scalar or pseudo-scalar particle can also be investigated in $B \rightarrow K^{(*)} l^+l^-$ 
decays. The sensitivity of \babar\ and Belle searches to the mixing angle between the Standard Model Higgs and the dark scalar, as
well as the pseudo-scalar 
coupling constants, is projected to be $\sim 10^{-4} - 10^{-3}$ and $10^3~\tev$, respectively \cite{Batell:2009jf}.

The next generation of flavor factories, Super$B$ and SuperKEKB, are expected to collect an integrated luminosity 
of $50-75 \rm~ab^{-1}$, increasing the current \babar\ and Belle datasets by two orders of magnitude. The sensitivity 
of the searches described above is therefore expected to improve by a factor of $10-100$, depending on the level of 
background. In particular, super flavor factories could probe values of $\alpha'/\alpha$ down to a level comparable 
to dedicated fixed-target experiments like APEX, with a significantly larger mass coverage.

\subsection{Proton Colliders}\label{subsec:hspaw-searches-proton} 

Proton colliders have the ability to reach high center-of-mass energy, making it possible to produce $Z$ bosons, Higgs bosons, 
and perhaps other new, heavy particles (such as supersymmetric particles, $W'$/$Z'$ states, or hidden-sector particles) 
directly. As pointed out in many theoretical studies \cite{ArkaniHamed:2008qp,Baumgart:2009tn,Ruderman:2009tj,Cheung:2009su}, if new states are produced ({\it e.g.},~supersymmetric particles), they could decay to $A'$ bosons and other hidden-sector states, sometimes with very large branching ratios. 
For GeV-scale $A'$ masses, they would be highly boosted when produced in such decays and their 
decay products would form collimated jets, mostly composed of leptons (``lepton-jets'' \cite{ArkaniHamed:2008qp}).

The existing general-purpose proton collider experiments at the Tevatron and LHC have all presented first searches for 
lepton-jets in heavy-particle decays \cite{Abazov:2009hn,Abazov:2010uc,cdflj,atlaslj,Chatrchyan:2011hr}. 
The searches usually employ a specialized lepton-jet identification algorithm to distinguish them from the large multi-jet background. 
Events with additional large missing transverse energy (from other escaping hidden-sector particles) or  a particular di-lepton 
mass (corresponding to the $A'$ mass) have also been searched for. Results have often been interpreted in supersymmetric 
scenarios, and typically exclude di-squark production with a squark mass of 500 GeV decaying through cascades to two lepton-jets.

With large datasets now available at the LHC (20~fb$^{-1}$ expected by 2012) and even 10--100 times larger data sets expected in the 
future, good sensitivity to new, heavy particles decaying to lepton jets will be achievable. Current searches have mostly focused 
on $A'$ bosons heavy enough to decay to muon pairs, since this offers a cleaner signal than electron pairs, but good sensitivity 
is expected in the future down to $\sim 20$~MeV (limited by photon conversions to $e^+e^-$ pairs). Large datasets will contain 
billions of $Z$ and millions of Higgs bosons, allowing branching ratios to lepton-jets as low as $10^{-6}$ (or $\epsilon\simeq 10^{-3}$) 
to be probed for $Z$ decays and $10^{-3}$ for Higgs decays. Searches have also mostly focused so far on prompt decays of dark photons, 
but studies are under way to perform searches for longer-lived decays in the tracking chambers and even muon chambers, allowing 
sensitivity down to $\epsilon \simeq 10^{-6}$, if a new, heavy particle with a large branching ratio into $A'$ bosons exists 
and can be produced at a high enough rate.

\subsection{Satellite Based Experiments}\label{subsec:hspaw-searches-satellite}

The OMEGA Explorer Project~\cite{Moustakas:2008ib}, which has been proposed to NASA, uses strong gravitational lensing to search for light dark matter as well as additional hidden-sector particles that interact with the dark matter~\cite{Moustakas:2008ib,walker:2012,Kaplinghat:2012}.  Dark matter clumps generate large gravitational potentials that lense the light from distant sources.  The observed multiple images (and time-delayed light) from the original source allow precise measurements of the dark matter halo and ``clumpiness."  The potential for observing light dark matter, such as axions, is described in~\cite{Kaplinghat:2012}.  The impact of mediator particles, such as dark photons, is described in~\cite{walker:2012}.  For brevity, we focus on the latter in order to complement the accelerator-based searches discussed above.

The light dark matter, described above, gravitates into clumps with a maximum mass that depends sensitively~\cite{astro-ph/0504112,astro-ph/0607319} 
on the dark matter kinematic decoupling temperature 
$M_{cut} \simeq 10^{-4} (T_d/10\,\,\mathrm{MeV})^{-3} M_\odot$
where $M_\odot$ is the sun's mass.  The dark matter decoupling temperature in turn sensitively depends on the dark matter particle physics as
$T_d =  (M^4_\sigma \, M/M_\mathrm{pl})^{1/4}$,  
where $M$ is the dark matter mass and $M_\sigma$ is the mass scale associated with the dark matter--dark photon interaction cross section.  The dark matter elastically Compton scatters  with the dark photon and is parametrized by 
$\sigma \approx T^2/M_\sigma^4$.
For $M_\sigma \sim 1$ GeV (which is common for Compton scattering), $M_\mathrm{cut} \sim 10^4$ which is within OMEGA's sensitivity in \cite{walker:2012}.  
Thus depending on the dark matter mass, the OMEGA project is potentially sensitive to $m_{A'}\lesssim \mathcal{O}(10)$ GeV. 


\section{Technologies Needed for Future Progress}\label{sec:hspaw-future-techno} 

Searches for axions, ALPs, and WISPs need the highest possible beam intensities to generate these weakly interacting particles.  On the other hand, detection 
requires the ultimate in sensitivity to extract extremely weak signals from noise. Improved technologies will significantly improve detection sensitivity.

Increased magnetic field strength would have the greatest impact on detector sensitivity. Current magnet technology 
used in these searches typically employs NbTi wire, giving fields of 8 T in solenoids 
and 5 T in dipole magnets. A number of steps may increase the maximum field strength. Nb$_3$Sn magnets would be a first step; 
current technology points to a rough doubling of the field with this material, to 16 T (solenoid) and 10 T (dipole) by a change 
in metal. Cooling to 1.7 K from 4.2 K can produce another 1-2 T. Note that the sensitivity goes as $B^2$, where $B$ is the 
field strength, so the resulting gain would be substantial. 

In the longer term, development of magnets from one of the ``new'' superconductors, such as MgB$_2$ or one of the cuprate 
high-$T_c$ materials, could yield fields in the 40 T range. To develop such a magnet will require substantial R\&D, but
could impact many fields in physics. Magnets with fields above 40 T already exist at the National High Magnetic Field Laboratory, 
but have small field volumes (few tens of cm$^2$) and consume huge amounts of power ($\sim 25$~MW).

Increasing the magnetic field volume will also boost sensitivity, since the full figure of 
merit is $B^2V$, with $V$ the volume.  (For laser experiments, the factor is usually written as $B^2L^2$, with $L$ the length 
of the magnetic region, but the laser beam spreads in proportion to its length, so an increase in length requires a 
corresponding increase in diameter to avoid clipping.)  Current experiments have magnetic volumes in the range of 0.035 (OSQAR) 
to 0.2 m$^3$ (ADMX). Laser experiments on the drawing board use similar volumes, but one may imagine
5--10 times increases in length and diameter without reaching unreasonable sizes, giving magnetic volumes of several cubic meters. IAXO requires a large aperature magnet of 3--5 T, with a total length of 17--23 m to provide great sensitivity to axions and ALPs.   

Many searches for hidden-sector photons, axions, and WISPs employ resonators, used either to increase intensity in generation 
of these particles or for resonantly enhanced conversion to photons. There are performance gains to be had by improving the designs 
of these cavities. Microwave experiments may improve cavity quality factor $Q$ by plating a thin layer of type-II superconductor 
on the portions of the cavity wall where the magnetic field is parallel to the wall. Experiments to develop this idea are under way. 
Laser resonators are typically limited by the damage threshold in the mirror coatings. Mirrors with improved damage thresholds would translate 
immediately to higher stored power in the cavities and increased production rates.

Improved detection of extremely weak signals would also improve the sensitivity of these experiments. This has already 
happened in some areas. ADMX has moved from semiconductor-based to superconductor-based (SQUID) radio-frequency amplifiers, reaching nearly the 
quantum limit. There is a need for technology development to push the operating frequencies of such amplifiers to the next decade. 
Similarly, there are conceptual designs for heterodyne detection schemes for laser-based detectors that offer shot-noise limited 
measurements, {\it i.e.}, the standard quantum limit. The next step in both cases would be to apply squeezing or quantum non-demolition schemes to the detector.

An approach suggested to search for axions with $f_a$ between the GUT and Planck scale uses molecular interferometry to detect time-varying energy shifts in $CP$-odd nuclear moments~\cite{Graham:2011qk}. These effects are enhanced in the light Actinides~\cite{Auerbach:1996zd,Spevak:1996tu} and in molecules in background electric fields, which show spontaneous parity violation. Technology developments in producing and manipulating molecules containing light Actinides~\cite{Scielzo:2006xc} and improving molecular interferometry~\cite{Cold Molecules} are needed to realize these experiments.

Finally, large diameter reflective X-ray optics and low background x-ray detectors are needed for future helioscopes to improve sensitivity over present experiments. Fabrication techniques developed for X-ray telescopes deployed on current satellites would be one cost-effective approach for building the required x-ray optics. Further reductions in the background levels in MicroMEGAS detectors would make them satisfy  requirements for low background X-ray detectors.

Advances in traditional particle physics detector technology will benefit searches for more massive hidden-sector particles. Very fast, very radiation hard,
very thin pixel detectors will benefit vertex searches for $\sim 100$~MeV hidden-sector photons by allowing more intense beams and improving vertex resolution. 
Developments in very high rate data acquisition, including faster high-level triggering, will enable experiments to run with more intense 
beams and thereby probe smaller couplings.

\section{Accelerators Needed for Future Progress}\label{sec:hspaw-future-accelerators}

Accelerator-based searches for hidden-sector photons with  $2m_{e}
< m_{A'} \lesssim 1$~GeV  have been performed at $e^+ e^-$ colliders 
\cite{Collaboration:2011zc}, 
$e^-$ fixed-target facilities~\cite{Bjorken:1988as,
Riordan:1987aw,Bross:1989mp}, and proton colliders~\cite{Abazov:2009hn,Abazov:2010uc,cdflj,atlaslj,Chatrchyan:2011hr}. 
New searches with increased
luminosity and improved detector sensitivities will
significantly increase the coverage of the allowed 
parameter space.   This section briefly describes the accelerator
facilities that are envisioned for these searches.

Existing $e^+e^-$ collider data sets have not yet been fully exploited in searches for hidden-sector particles.
Improved limits should be forthcoming. Proposed $e^+e^-$ colliders~\cite{Venanzoni:2006cr,Albert:2005tg,
Kageyama:2006zd} will provide an increased coverage in the $A'$ parameter
space due to the improved luminosity (factor of 10- to 100-fold
increase).  
This increase in luminosity will result in 
an increased coupling strength sensitivity
($\alpha'/\alpha = \epsilon^2$) by more than an order of magnitude:
$\epsilon^2 \approx  10^{-7}$.  
The mass range explored by searches at
$e^+e^-$ colliders will remain in the range $m_{A'} > 10$ MeV due to
backgrounds at low $m_{e^+e^-}$.  

\begin{table}
\begin{center}
\begin{tabular}{llccc}
Facility & Type &Operations &  Energy  & $\mathcal{L}$   \\
         &      &           &  (GeV)   & ($\textrm{cm}^{-2}\textrm{s}^{-1}$)  \\
         \hline
CEBAF (6~GeV) & $e^-$ fixed target   & now-2012  & 0.5-6  & $10^{39}$\\
JLab FEL   & $e^+$ energy recovery linac & now- & 0.14  & $10^{35}$ \\
MAMI        & $e^-$ fixed target   & now-  & 0.855 \& 1.55  &$10^{39}$\\
KLOE-2      & $e^+e^-$ & now-      & $m_{\Phi}$ &   $5\times 10^{32}$     \\
VEPP-3     & $e^+$ storage ring & now- & 0.5 - 2  & $10^{32}$ \\
CEBAF (12~GeV) & $e^-$ fixed target   & 2014-  & 1-12  &  $10^{39}$\\
MESA@Mainz & $e^-$ energy recovery linac   & 2016-  &  0.1 & $10^{35} $ \\
Belle-II & $e^+e^-$ & 2016-      & $m_{\Upsilon(4S)}$ &       $10^{36}$  \\
Super-B  & $e^+e^-$ & 2017-      & $m_{\Upsilon(4S)}$ &    $10^{36}$   \\ \hline

\end{tabular}
\caption{
  List of present and proposed electron collider and fixed-target facilities with their operational dates,  energies and
  luminosity. Fixed-target luminosities assume a 0.1 radiation
  length Tungsten target and 100 $\mu$A beam current. Searches at  the VEPP-3, JLab FEL and MESA
  facilities propose to use a
  internal gaseous hydrogen target. 
  \label{table:electron_machines}
} 
\end{center}
\end{table}

The high luminosity ($\mathcal{O}(\textrm{1 ab}^{-1}/\textrm{day})$) presently available
at  electron accelerators, CEBAF~\cite{Leemann:2001dg},
MAMI~\cite{Dehn:2011zz} and the JLab FEL~\cite{Benson:2011zz}, and their beam
characteristics make them well suited for $A'$ searches. This potential is
just beginning to be exploited.  The beam 
parameters of the fixed-target machines (see
Table~\ref{table:electron_machines}) are well suited for exploring 
$\alpha'/\alpha$ values as low as $10^{-11}$ with a mass coverage of $10 < m_{A'} <
1000$ MeV.   In addition to well-matched beam
parameters, both MAMI
and CEBAF
have double-arm spectrometers that are
well suited for $A'$ searches with little or no modifications.  Short
test runs at both these facilities have recently led to published
results~\cite{Merkel:2011ze,  Abrahamyan:2011gv}  

CEBAF and MAMI  with their flexible configurations, multiple
end-stations and large luminosities,  can explore a large fraction of
the  available $A'$ mass and
coupling range (see Figure~\ref{fig:hspaw-heavy-A'}).
However, the coverage is not complete, and new proposals are needed to fill in the gaps.
The wide range of beam energies and currents available at CEBAF and MAMI allows flexibility in experimental approaches.
If an $A'$ is
discovered within their kinematic reach, they would be in a good position to explore the $A'$ properties in detail.

In addition to the GeV range continuous wave (CW) electron experiments, coverage in range
$m_{e} < m_{A'} < 10$~MeV  has been proposed by using electron beams (at
the JLab FEL and the proposed MESA accelerator at Mainz)
 or
positron beams (at the VEPP-3 storage ring)  with energy of
order 100 MeV (see Table~\ref{table:electron_machines}).   These
energies are available from the VEPP-3 storage ring, JLab FEL, and the
proposed MESA facility at Mainz.    

Searches at proton accelerators, both collider and fixed-target experiments, are
sensitive to heavy photons produced in decays of new particles or secondary particles.
Exploring the possibility of performing $A'$
searches in neutrino experiments should be
encouraged. 
Current data sets need to be fully exploited, and
next-generation neutrino experiments, with their very high luminosities and sensitive
detectors, may be ideal hunting grounds for heavy photons.    

In summary, much of preferred parameter space for heavy hidden-photons
can be explored with existing and proposed accelerator
facilities.   Specifically, the luminosity and flexibility 
of the electron fixed-target facilities are well suited for $A'$
searches in the mass range $2m_{e} < m_{A'} < 1$ GeV. More generally,
exploring the capabilities of all facilities and further exploiting current data sets to
search as much of the allowed region as possible
is a priority. A great deal can be done with existing facilities and planned upgrades. 

\section{Conclusions}\label{sec:hspaw-conclusions}

Searches for new light, weakly coupled particles are motivated by some of the most important questions in particle physics. They are dependent on the tools and techniques of the intensity frontier,  {\it i.e.}, intense beams 
of photons and charged particles, and on extremely sensitive detection techniques.  These searches 
reside at the nexus of the cosmic, energy, and intensity frontiers: new light particles may constitute the dark matter, or be the force carriers responsible for its interactions; they may even help explain the origin of dark energy, or be low-energy 
remnants of physics at the highest energy scales.  
These searches expand our notion of the energy frontier to the very high and the very low, and explore new and 
exciting territory.  

The Intensity Frontier Workshop brought together two related, but largely independent communities, 
both exploring particles that either couple directly to the photon or mix with it to couple to electrically charged particles. 
One is focusing on axions, axion-like particles, and low-mass 
hidden-sector photons. The other is searching for  higher-mass particles residing in hidden-sectors, 
heavy ``photons,'' hidden-Higgs bosons, and the like. 
The resulting  interchange was exciting and will likely  provoke some new directions, experiments, and insights. 
Both communities are exploring vast parameter spaces in particle mass and coupling, holding the potential for truly momentous discoveries. However, this is done utilizing 
what are really modest experimental efforts, compared to the general trend in particle physics. 

A great deal remains to be done with existing tools and techniques, especially in 
searching for heavy photons. Existing facilities can support many new experiments, extending experimental reach significantly. A great deal more territory can be explored for the low-mass hidden-sector and axion-like particles, including theoretically favored regions, with relatively 
modest advances in superconducting magnets, microwave detection, and resonant optical cavities.
Large-scale experiments are also under consideration, well-motivated, and require more substantial investments for 
their implementation.  

Searches for new, light, weakly interacting particles are ongoing around the world, but essentially all the experimental efforts have strong US participation or strong US leadership. Continued support in the US  is essential if the present US role is to continue, and the potential for truly fundamental discoveries preserved. 

The cutting-edge physics experiments searching for new low-mass, weakly interacting particles provide 
ideal educational opportunities for undergraduate, graduate, and post-graduate students. 
These experiments are small, and so demand and deliver the full breadth of experimental opportunity:
design, hardware construction and commissioning, software implementation, 
data taking, and analysis. These experiments have also joined experimentalists and theorists 
into common enterprises to an uncommon degree, providing a considerable benefit for the field 
and pleasure for the participants.

\def\Discussion{\setlength{\parskip}{0.3cm}\setlength{\parindent}{0.0cm}
     \bigskip\bigskip      {\Large {\bf Discussion}} \bigskip}\def\speaker#1{{\bf #1:}\ }
\def\endDiscussion{}


%

\newcommand{\mrm}{\mathrm}
\newcommand{\sub}[1]{\mathrm{\scriptscriptstyle{#1}}}

\newcommand{\lampHg}{\ensuremath{{}^{204}\mathrm{Hg}}}
\newcommand{\magHg}{\ensuremath{{}^{199}\mathrm{Hg}}}

\newcommand{\parafield}{\ensuremath{\uparrow\!\uparrow}}
\newcommand{\aparafield}{\ensuremath{\uparrow\!\downarrow}}

\newcommand{\nedm}{\ensuremath{d_{\sub{n}}}}
\newcommand{\hgedm}{\ensuremath{d_{\sub{Hg}}}}
\newcommand{\ecm}{\ensuremath{e\!\cdot\!\mathrm{cm}}}
\newcommand{\ntwoedm}{\ensuremath{\mathrm{n^2EDM}}}
\newcommand{\fTHz}{\ensuremath{\mathrm{fT/\sqrt{Hz}}}}

\newcommand{\tsups}[1]{\textsuperscript{#1}}
\newcommand{\trinat}{{\scshape Trinat}}
\newcommand{\triumf}{{\scshape Triumf}}
\newcommand{\trex}{T{\small REX}}
\newcommand{\zerotozero}{\mbox{$0^+\!\!\rightarrow 0^+\ $}}
\newcommand{\tamutrap}[0]{{\scshape Tamutrap}}


\chapter{Nucleons, Nuclei, and Atoms}
\label{chap:chapx}

\begin{center}\begin{boldmath}
Conveners:  W.~Haxton, Z.T.~Lu, M.~Ramsey Musolf

K.~Bailey, 
J.~Behr, 
S.~Brodsky, 
D.~Bryman, 
R.~Carlini, 
T.E.~Chupp, 
D.P.~DeMille, 
A.~Deshpande,  
M.S.~Dewey,  
M.R.~Dietrich, 
J.~Fajans,  
C.~Faroughy, 
B.~Feinberg, 
P.~Fierlinger, 
B.~Filippone, 
A.~Garcia, 
S.~Gardner,   
M.~Gonderinger,  
H.~Gould, 
J.P.~Green, 
V.~Gudkov, 
J.C.~Hardy,  
P.G.~Harris,  
B.R.~Heckel, 
E.A.~Hinds, 
R.J.~Holt, 
M.~Kalita, 
Y.~Kamyshkov,  
K.~Kirch, 
W.~Korsch, 
A.D.~Krisch, 
A.~Kronfeld, 
K.~Kumar,  
Y.~Li,  
C.Y.~Liu,  
W.~Lorenzon,  
W.~Marciano,  
J.W.~Martin, 
D.~Melconian, 
K.~Minamisono,  
R.~Mohapatra, 
P.~Mueller,
H.P.~Mumm,  
C.~Munger, 
O.~Naviliat-Cuncic, 
J.S.~Nico,  
T.P.~O'Connor, 
L.~Orozco, 
R.H.~Parker, 
P.~Reimer, 
F.~Ringer, 
S.~Riordan,  
F.~Robicheaux, 
Y.K.~Semertzidis, 
J.~Singh, 
D.~Sivers,  
Y.H.~Song, 
P.~Souder, 
I.A.~Sulai, 
S.~Taneja, 
I.S.~Towner, 
W.~Vogelsang, 
E.~Widmann, 
J.~Wurtele, 
A.~Young

\end{boldmath}\end{center}

\section{Overview}\label{sec:NNAintro}

Despite the success of the Standard Model in explaining so many subatomic physics phenomena, we recognize that
the model is incomplete and must eventually give way to a more fundamental description of Nature.  We have discovered
massive neutrinos and associated flavor violation, which require the introduction of new mass terms in the Standard Model.
We have an excess of baryons over antibaryons in our universe, indicating baryon-number-violating interactions
and likely new sources of $CP$ violation.  We know from the inventory of matter and energy in our universe that the
portion associated with Standard Model physics is only about 5\% of the total.  The rest remains unidentified and quite
mysterious.

These ``big questions" -- the origin of matter, the nature of neutrino mass, the identification of dark matter and dark
energy - have driven the two major thrusts of subatomic physics.  One is the effort to probe ever shorter distance scales
by advancing the energy frontier.  Today that frontier in represented by the CERN Large Hadron Collider.   The alternative is to seek 
signals of the new physics in subtle violations of symmetry in our low-energy world -- ultra-weak interactions that
might mediate lepton- or baryon-number violation, generate electric dipole moments, or lead to unexpected flavor
physics.  This second approach is the theme of this chapter:  ultra-sensitive techniques in atomic, nuclear, and
particle physics that might reveal the nature of our ``next Standard Model."  Like particle astrophysics
and cosmology, the third leg of our
efforts to find new physics, this second approach uses our world as a laboratory, and depends on precision to
identify the new physics.

This field has a long and quite successful history. Tests of fundamental symmetries in experiments involving nucleons, nuclei, and atoms have played an essential role in developing and testing the Standard Model. The observation of parity-violation in the radioactive decay of $^{60}$Co, shortly preceding the observation of parity violation in pion decay, provided the first experimental evidence that the weak interaction does not respect this symmetry, ultimately leading to the Standard Model description of charged weak currents as being purely left-handed. Similarly, the measurements of the parity violating asymmetry in polarized deep-inelastic electron-deuteron scattering in the 1970s  singled out the Standard Model structure for the neutral weak current from among competing alternatives, well in advance of the discovery of the electroweak gauge bosons at CERN. And the non-observation of a permanent electric dipole moment (EDM) of the neutron and $^{199}$Hg atoms has placed stringent bounds on the possibility of combined parity and time-reversal violation (or $CP$ violation) in the strong interaction, motivating the idea of the spontaneously broken Peccei-Quinn symmetry and the associated axion that remains a viable candidate for the cosmic dark matter. 

Present and prospective fundamental symmetry tests with nucleons, nuclei, and atoms are now poised to probe for possible new physics at the Terascale and beyond, making them a vital component of the intensity frontier. At the same time, these tests provide increasingly sophisticated probes of poorly understood features of long-distance strong interactions that are responsible for the structure of nucleons and nuclei. The potential for both discovery and insight has motivated the nuclear physics community to identify studies of fundamental symmetries and neutrino properties as one of the top four priorities for the field in the 2007 Nuclear Science Advisory Committee (NSAC) Long Range Plan \cite{lrp2007}, perhaps anticipating the present broader interest in the intensity frontier that underlies this document. Below, some of the most compelling opportunities with nucleons, nuclei, and atoms are summarized, drawing largely on input received from the nuclear and atomic physics communities. 

Fundamental symmetry tests with nucleons, nuclei, and atoms is remarkably diverse. In preparation for this report, the working group conveners received more than 30 two-page written contributions outlining the progress and opportunities for the field. Given the limitations of space, it is not feasible in this chapter to include all of the detailed information received, so we refer the reader to the website where this input is available \cite{ifwref}. For similar reasons, we do not provide a comprehensive theoretical framework here, again referring the reader to recent review papers \cite{Erler:2004cx,nna:RamseyMusolf:2006vr}. Some theoretical context and basic terminology is included in each of the subsections below, though the primary focus falls on the experimental opportunities. 

With this context in mind, it is useful to delineate three broad classes of studies with nucleons, nuclei, and atoms:
\begin{itemize}
\item[(i)] Rare or forbidden processes: observables that one expects -- based on the Standard Model -- to be either zero or far too suppressed to be observed. The observation of a non-zero signal in such a process would constitute ``smoking gun" evidence for physics beyond the Standard Model. From the standpoint of this report, the permanent electric dipole moment of the neutron or a neutral atom represents the flagship example of such an observable. Other examples include tests of Lorentz symmetry or $CPT$ (defined below). 
\item[(ii)] Precision tests: such studies seek to measure, with high precision, observables that are not suppressed within the Standard Model. Any significant deviation from the Standard Model prediction would again point to physics beyond the Standard Model, whereas agreement can imply severe constraints on various model possibilities. For this class of studies, obtaining robust theoretical Standard Model predictions is vital to the interpretation in terms of ``new physics," as one has already seen earlier in the discussion of the muon anomalous magnetic moment. As discussed below, the primary precision tests further break down into two classifications: those involving the charged-current weak interaction (primarily weak decays) and weak neutral current processes, such as parity violation in electron scattering. 
\item[(iii)] Electroweak probes of the strong interaction: the motivation for this set of studies is to exploit the unique sensitivity of electroweak observables to aspects of nucleon and nuclear structure that are not readily accessible with a purely electromagnetic probe. During the last two decades, for example, measurements of parity-violating asymmetries in fixed-target, polarized electron-proton and electron-nucleus scattering have been performed at MIT-Bates, Jefferson Laboratory, and the University of Mainz  with the aim of determining the strange quark contribution to the electric and magnetic properties of the strongly interacting targets. The interpretation of these experiments treated the Standard Model weak neutral current interaction as sufficiently well known that it could be used to probe this interesting question in hadronic structure. Since the focus of this report is on physics beyond the Standard Model, we touch only briefly on this third class of studies, but emphasize that it remains an area of considerable interest and high priority within the nuclear physics community. 
\end{itemize}

Before proceeding, it is important to emphasize several additional points. First, there exists considerable overlap between the efforts of the nuclear and atomic physics communities described below and the studies described elsewhere in this report. In particular, the 2007 NSAC Long Range Plan identifies both fundamental symmetry tests and neutrino studies as one of the top priorities for the field in the coming decade. Indeed, a substantial fraction of the nuclear physics community is playing a leading role in searches for neutrinoless double beta decay, neutrino mass measurements, and neutrino oscillation studies. Similar leadership roles continue to be filled by members of the nuclear physics community in muon physics and ``dark photon" searches. Conversely, members of the high-energy physics community are in some instances leading the experiments described below -- perhaps most notably in searches for pion leptonic decays. One should not conclude from the organization of this chapter that it reflects the self-organization of the various communities or the primary sources of federal research support.

Second, the array of opportunities described below largely reflects the outcome of a process of ``self-reporting," wherein the conveners have attempted to incorporate information provided by members of the community on a voluntary basis in response to a call for input on a very short timescale. The presence or absence of various topics does not, therefore, reflect any consensus on the part of the broader nuclear and atomic physics communities as to  the top priorities for the future -- except as they make contact with the NSAC Long Range Plan that emerged from a year-long process of town meetings and working group discussions. In some cases, what appears below constitutes a natural continuation of the Long Range Plan content, while other content reflects new ideas that may not have been fully vetted by the community or a peer-review process. Moreover, some areas of study -- such as parity violation in purely hadronic and nuclear systems -- have not been included even though they are the focus of considerable present effort. Such omissions do not imply any relative prioritization by the conveners but rather the ability of the relevant investigators to respond to the call for input on a short timescale.

Finally, one should bear in mind the international context for fundamental symmetry tests. The working group conveners have made efforts to reach out to the international community in order to provide this international context, and several investigators have responded quickly and graciously. However, significant omissions remain, including the highly successful program of neutron decay studies at the Institut Laue-Langevin (ILL) in Grenoble, France and the ambitious future program involving investigators in Heidelberg, Vienna, and Munich. Similarly, a new high-intensity, low-energy electron beam has been proposed by colleagues in Mainz, providing an opportunity to carry out a measurement of the proton's weak charge that is also the subject of a possible Jefferson Lab free electron laser (FEL) concept described below. Again, the omission of these and other international efforts merely reflects the timescale for preparation of this document, the breadth of studies being carried out, and the limited availability of international colleagues to contribute to this process while carrying on their research programs locally. Indeed, fundamental symmetry tests with nucleons, nuclei, and atoms is a worldwide effort, and significant partnerships and sharing of scientific expertise between the community in North America and elsewhere in the globe is vital to the overall scientific impact of this field. 

With these caveats in mind, we turn to an overview of the exciting opportunities to utilize nucleons, nuclei, and atoms as ``laboratories" for tests of fundamental interactions and to uncover new clues about what may lie beyond the Standard Model.

\section{Electric Dipole Moments}
\label{sec:EDM}

\begin{figure}[htbp]
\centering
\vspace*{3mm}
\includegraphics[width=0.4\textwidth]{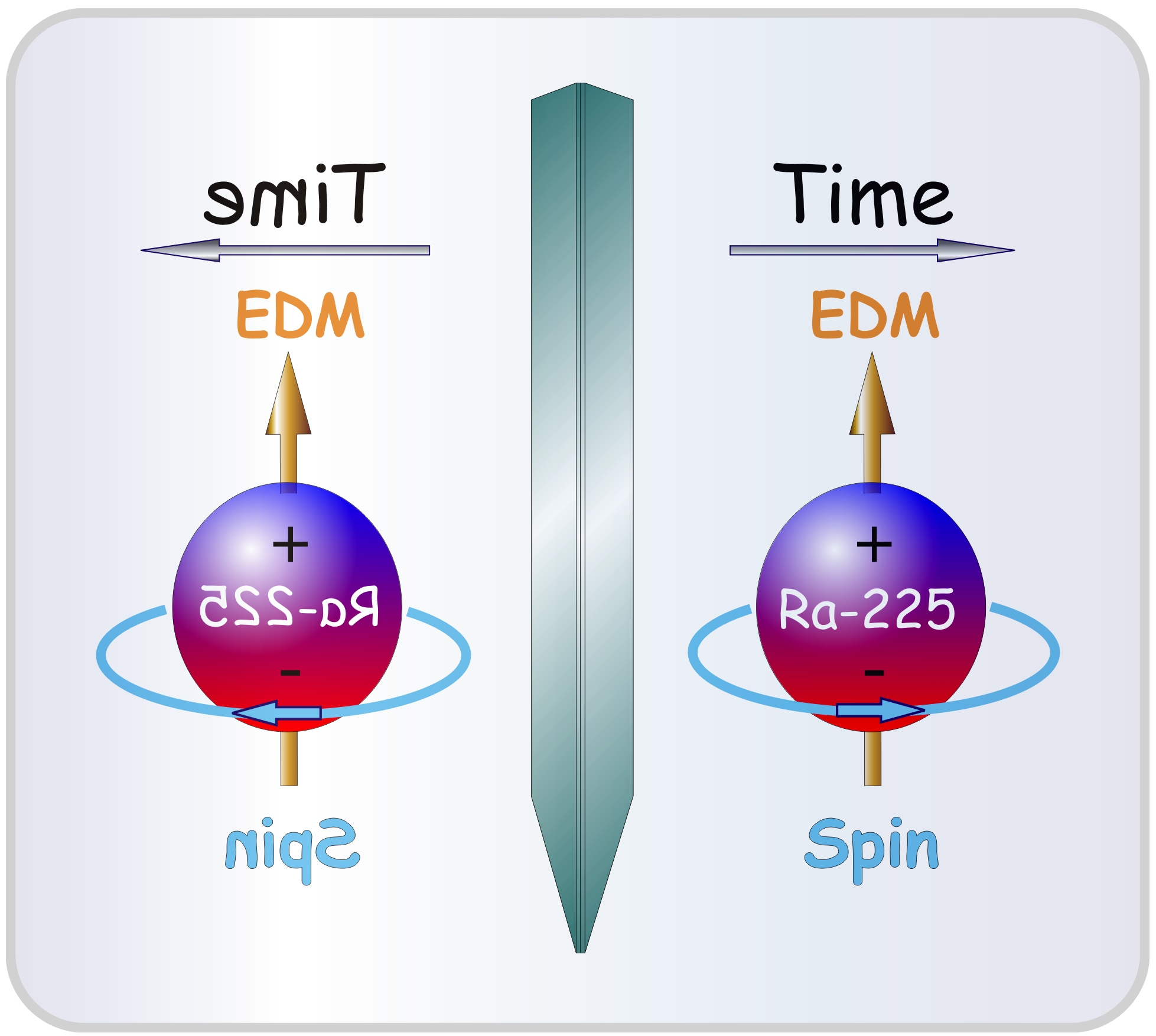}
\caption{EDM violates time-reversal symmetry. When time is reversed, the spin is reversed, but the EDM is not. Therefore, there is a measurable difference between the particle and its image in the time-reversed world.}
\label{EDM1}
\end{figure}

At a classical level, the permanent electric dipole moment (EDM) of a particle arises from the spatial separation of opposite charges along the axis of the particle's angular momentum. The existence of an EDM would be a direct signature of the violation of both parity ($P$) and time-reversal symmetry ($T$) (Fig. \ref{EDM1}). It would also probe physics of $CP$ violation ($C$ stands for charge conjugation) which necessarily accompanies $T$ violation under the assumption of the $CPT$ theorem. EDM measurements conducted in many laboratories around the world employing a variety of techniques have made tremendous progress, and all have so far obtained results consistent with zero EDM. For example, in the past six decades, the search sensitivity of the neutron EDM has improved by six orders of magnitude to reach the current upper limit of $2.9 \times 10^{-26} \ecm$ \cite{Bak06}.

$CP$ violation in flavor-changing decays of $K$- and $B$-mesons have been observed. The results can be explained, and indeed in many cases predicted, by the CKM mechanism within the framework of the Standard Model, in which all of the observed $CP$ violation phenomena originate from a single complex phase in the CKM matrix that governs the mixing of quark flavors. While this elegant solution has been well established following a string of precise measurements, additional sources of $CP$ violation are generally anticipated in extensions of the Standard Model. For example, in Supersymmetry (SUSY), the supersymmetric partners of quarks would naturally allow additional complex phases in the expanded mixing matrix and induce new $CP$-violating phenomena. Within the Standard Model, a $CP$-violating term is known to be allowed in the general form of the QCD Lagrangian, and would have $CP$-violating consequences specifically in the strong interaction. Additional $CP$-violating mechanisms are also called for by the observation that the baryon-to-photon ratio in the universe is as much as nine orders of magnitude higher than the level that can be accommodated by the Standard Model. A much more significant matter-antimatter asymmetry is likely to have been present in the early universe, and provided the favorable conditions for the survival of matter that we observe today.

A permanent EDM is a sensitive probe for new $CP$-violating mechanisms, and is generally considered to be one of the most promising paths towards new physics beyond the Standard Model. The CKM mechanism in the Standard Model can generate EDMs only at the three- and four- loop level, leading to values many orders of magnitude lower than the current experimental limits. For example, the Standard Model value for the neutron EDM is expected to be approximately $10^{-32}$ \ecm, or six orders of magnitude below the current upper limit. Any non-zero EDM observed in the foreseeable future would have to require either $CP$ violation in the strong interaction or physics beyond the Standard Model. Perhaps unsurprisingly, extensions of the Standard Model generally allow a range of EDM values that are within the reach of ongoing experiments, including scenarios that would generate the cosmic baryon asymmetry with new Terascale $CP$-violating interactions (see, {\em e.g.},~\cite{ref2} and references therein). 
The scientific importance and discovery potential of EDM searches are strongly endorsed by the communities of both particle physics and nuclear physics. The Nuclear Science Advisory Committee (NSAC) proclaimed in the 2007 Long Range Plan~\cite{lrp2007} that ``a non-zero EDM would constitute a truly revolutionary discovery.'' The negative findings so far are also valuable. As was pointed out in the 2006 P5 report~\cite{p5}, \emph{The Particle Physics Roadmap}, ``the non-observation of EDMs to date, thus provides tight restrictions to building theories beyond the Standard Model.'' Specifically, the upper limits on EDM provide insight into the scale of the next energy frontier.

The most sensitive EDM searches have so far been conducted on the neutron, neutral atoms ($^{199}$Hg), and the electron (Table 1). On the one hand, experiments in these three categories all compete for the prize of being the first to observe a non-zero EDM. On the other hand, they complement each other, as each category is most sensitive to different sources of $CP$ violation. For example, the neutron is relatively more sensitive to the EDMs of its constituent quarks; heavy nuclei are more sensitive to the quark ``chromo-EDM" (the $CP$-violating quark-gluon interaction) and other $CP$ violation mechanisms in the nuclear force. The recently proposed storage ring EDM experiments of the proton and deuteron aim to probe combinations of $CP$-violating contributions that differ from the neutron EDM. Experiments with paramagnetic atoms or molecules are sensitive to the EDM of the electron and a possible new $CP$-violating electron-quark interaction. In the future, if a non-zero EDM is discovered in one particular system, it would still be necessary to measure EDMs in other categories to help resolve the underlying $CP$ violation mechanisms. 

At the intensity frontier, a new generation of sources for cold neutrons and ultra-cold neutrons (UCNs) is becoming available. Their higher output in neutron flux will enable searches for the neutron EDM projected at a sensitivity level of 10$^{-28}$ \ecm, or two orders of magnitude below the current best limit. A survey of neutron EDM experiments is presented in this section. Also at the intensity frontier, future isotope production facilities such as the Facility for Rare Isotope Beams (FRIB) at Michigan State University (after upgrade) or Project X at Fermilab, will produce prolific amounts of selected isotopes that possess enhanced sensitivities to the EDMs of the nuclei or the electron. Included in this section are the cases for the radium, radon, and francium isotopes.

\begin{table}
\centering
\caption{Upper limits on EDMs in three different categories.}
\label{table1}
\begin{tabular}{|l|c|c|c|}
\hline
 Category & EDM Limit (\ecm)  & Experiment &  Standard Model value (\ecm)   \\
\hline
 Electron             & $1.0\times10^{-27}$    &   YbF molecules in a beam \cite{Hudson:2011}  &   10$^{-38}$       \\
 Neutron             & $2.9\times10^{-26}$    &   ultra-cold neutrons in a bottle \cite{Bak06}   &   10$^{-31}$       \\
 Nucleus              & $3.1\times10^{-29}$    &  $^{199}$Hg atoms in a vapor cell  \cite{Griffith:2009zz}   &   10$^{-33}$      \\
\hline
\end{tabular}
\end{table}

\subsection{PSI Neutron EDM}
\label{sec:PSIEDM}

The search for the EDM of the neutron is considered one of the most important
particle physics experiments at the low energy, high precision, high
intensity frontier \cite{Rai08, Pos05}. The non-observation so far, with the most stringent limit
of $2.9\times10^{-26}$ \ecm~(90\% C.L.) set by the Sussex-RAL-ILL collaboration~\cite{Bak06},
has far-reaching consequences:  The extreme smallness of
$CP$ violation in QCD, apparent in the smallness of the neutron EDM,
is not understood at all and has led to the strong $CP$ problem. 

A nEDM experiment is being developed in steps at the Paul Scherrer Institute (PSI)~\cite{Alt09a}. The collaboration is 
pursuing a considerable technical R\&D effort
but also exploiting the complementary physics potential of the nEDM
apparatus with respect to exotic interactions~\cite{Ban07}.
The experiment is located at the new source for ultra-cold neutrons  at
PSI~\cite{lauss,blau}. This source uses neutron production via
proton-induced spallation on lead, moderation in heavy water and
solid deuterium, and downscattering to UCN. Through an intermediate
storage volume UCN can be distributed to three experimental beam ports. 
The performance of the source is continuously being optimized.
Besides nEDM, the UCN source can also serve other experiments.

The collaboration is presently using the original but upgraded
Sussex-RAL-ILL spectrometer~\cite{Bak06}. In its configuration at
the PSI UCN source, it is estimated to yield a factor of 25 higher
statistics as compared to the earlier ILL setup. This increased statistical
sensitivity needs to be accompanied by a comparable reduction of systematic uncertainties. The
following improvements have been implemented: offline scanning for magnetic contaminations,
surrounding magnetic field compensation, magnetic field correction coils, demagnetization,
optically pumped cesium atomic vapor magnetometers, and a mercury cohabiting magnetometer. 
The goal is to accumulate enough data in 2012 and 2013
to reach a sensitivity of $\sigma(\nedm)=2.6\times10^{-27}\,\ecm$, which
corresponds to an upper limit of $d_{\sub{n}}<
5\times10^{-27}\,\ecm$~(95\,\% C.L.) in case of a null result.

The next-generation neutron EDM
experiment at PSI, named n2EDM, is being designed and will be constructed and
offline tested in parallel to operating nEDM.  It will be operated at
room temperature and in vacuum, aiming at a sensitivity of
$d_{\sub{n}}< 5\times10^{-28}\,\ecm$~\cite{Alt09a} (95\% C.L. limit
in case of no signal). The setup is built around two stacked neutron
precession chambers for simultaneous measurements of both E-field
orientations. Precise control and measurement of the magnetic
environment inside the apparatus is possible via laser
read-out Hg co-magnetometers, multiple Cs magnetometers as
gradiometers surrounding the neutron precession chamber, and
additional $^3$He magnetometer cells both above and below the neutron
chambers. A multi-layer mu-metal shield will provide a passive
shielding factor approaching $10^5$ and will be surrounded by a
multi-coil, multi-sensor, active compensation coil system. At present
the setup area for the n2EDM apparatus is being prepared.
Construction is scheduled to start in mid-2012. Transport to the
beam position, depending on nEDM measurements and n2EDM progress, is
foreseen for 2014. EDM data taking could start in 2015 with first
results in 2016.

\subsection{ILL Neutron EDM}
\label{sec:ILLnEDM}

\paragraph{\bf CryoEDM:} The CryoEDM experiment at ILL uses resonant downscattering of 
{\bf fix tex error}
neutrons in a bath of superfluid $^4$He as a source of UCN.  The UCNs are transported to magnetically shielded storage cells where, as in the previous generation of this experiment carried out at room temperature, the Ramsey technique of separated oscillatory fields is used to measure the precession frequency of the neutron in parallel and antiparallel electric and magnetic fields.  There are two Ramsey chambers; one has no electric field applied, and serves as a control.  Magnetic field fluctuations are monitored with SQUIDs.  The neutrons are counted using detectors situated within the liquid helium~\cite{Bak03}.  

The experiment is in its commissioning stage.  It is anticipated that by 2013 the sensitivity will reach that of the room-temperature experiment~\cite{Bak06}, after which time it will be moved to a new beamline, where upgrades to various components of the apparatus should lead to an improvement of about an order of magnitude in sensitivity.

\paragraph{\bf PNPI/ILL nEDM:} Also at ILL, a  Petersburg Nuclear Physics Institute (PNPI)/ILL experiment~\cite{Ser09} to measure nEDM is currently being prepared at the UCN facility PF2. To enable an improvement of sensitivity, one of the PF2's beam positions has been equipped with new components for UCN transport, polarization, and beam characterization, comprised of a superconducting solenoid polarizer with a 4 Tesla magnetic field, a neutron guide system with a diameter of 136 mm prepared in replica technology, and a novel beam chopper for time-of-flight analysis. The whole EDM apparatus is set up on a non-magnetic platform. A higher density of polarized UCNs at the experimental position, at approximately 5 $cm^{-3}$, shall lead to an EDM measurement with a counting statistical accuracy of  $1.5\times10^{-26}$ \ecm~during 200 days of operation at PF2.

\subsection{SNS Neutron EDM}
\label{sec:USnEDM}

The goal of the SNS nEDM experiment, to be carried out at the 
Spallation Neutron Source (SNS) at Oak Ridge National Laboratory, is to achieve a sensitivity of $< 3 \times
10^{-28}$ \ecm. A value (or limit) for the neutron EDM will be extracted from the difference
between neutron spin precession frequencies for parallel and anti-parallel
magnetic ($\sim 30$ mGauss) and electric ($\sim 70$ kV/cm) fields.
This experiment, based on~\cite{ref3}, uses a novel 
polarized $^3$He co-magnetometer and will 
detect the neutron precession via the spin-dependent neutron capture on 
$^3$He. The capture reaction produces energetic protons and tritons, which
ionize liquid helium and generate scintillation light that can be detected.
Since the EDM of $^3$He is strongly suppressed by electron screening in the
atom, it can be used as a sensitive magnetic field monitor. 
High densities of trapped UCNs are produced via
phonon production in superfluid $^4$He which can also support large 
electric fields. This technique allows for a number of independent 
checks on systematics, including: 

\begin{enumerate}
\item Studies of the temperature dependence of false EDM signals in the $^3$He.
\item Measurement of the $^3$He precession frequency using SQUIDs.
\item Cancellation of magnetic field fluctuations by matching the 
effective gyromagnetic ratios of neutrons and $^3$He with the ``spin dressing'' technique~\cite{ref3}). 
\end{enumerate}

The collaboration is continuing to address critical R\&D developments in 
preparation for construction of a full experiment. Key issues being 
addressed include:

\begin{enumerate}
\item Maximum electric field strength for large-scale electrodes made of
appropriate materials in superfluid helium below a temperature of 1 K. 
\item Magnetic field uniformity for a large-scale magnetic coil and a
superconducting Pb magnetic shield.

\item Development of coated measurement cells that preserve both neutron and 
$^3$He polarization along with neutron storage time. 

\item Understanding of polarized $^3$He injection and transport in the superfluid.

\item Estimation of the detected light signal from the scintillation
in superfluid helium.

\end{enumerate}

The experiment will
be installed at the FNPB (Fundamental Neutron Physics Beamline) at the
SNS and construction is likely to take at least five years, followed by
hardware commissioning and data taking. Thus first results could 
be anticipated by the end of the decade.

\subsection{TRIUMF Neutron EDM}
\label{sec:TRIUMFnEDM}

The basic design of this experiment calls for a room-temperature EDM
experiment to be connected to a cryogenic UCN source \cite{masuda}.  
Neutrons will be moderated and converted into
UCNs via down-scattering in superfluid He. The source will be operated at the Research
Center for Nuclear Physics (RCNP, Osaka) and then moved to TRIUMF
(Canada's National Laboratory for Particle and Nuclear Physics,
Vancouver).  The goal is to achieve $>$ 5000 UCN/cm$^3$ in an nEDM measurement cell.
A prototype nEDM apparatus has been characterized in beam tests
at RCNP Osaka.  Using this apparatus the collaboration has already demonstrated long
UCN storage lifetimes, polarization lifetimes, and transverse spin
relaxation times.

The EDM apparatus has a few unique features: A spherical
coil within a cylindrical magnetic shield is used to generate the DC magnetic
field;  a $^{129}$Xe comagnetometer is used to address false EDMs due to a geometric phase
effect; and due to the expected higher UCN density, 
the measurement cell size is designed to be considerably smaller than the previous ILL
apparatus.  While having a negative impact on statistics, the reduced
cell size limits systematic effects, particularly from the geometric phase effect.

In 2012-13, the collaboration will develop an improved EDM experiment,
including a new superconducting polarizer system for the UCN, and a
demonstration of precision Xe comagnetometry.  In 2013-14 the collaboration intends to
complete an nEDM experiment at RCNP, with a targeted precision of $d_n<1\times 10^{-26}$ \ecm, a factor of three
better than the present limit. The experiment and source will then be moved to TRIUMF and
recommissioned (on a new proton beamline currently under development)
in 2015-16.  Further improvements to the magnetic shielding,
comagnetometry, EDM cell, and detectors will be made, resulting in a
precision of $d_n<1\times 10^{-27}$~\ecm. The long-term goal, to be reached in 2018 and beyond, is $d_n<1\times
10^{-28}$~\ecm. Experiments on the neutron lifetime and
on neutron interferometry are also considered as candidates for the
long-term physics program.

\subsection{Munich Neutron EDM}
\label{sec:MunichnEDM}

At the new UCN source of FRM-II in Garching, Germany, a next-generation neutron EDM experiment aims to achieve a statistical limit of $d_n < 5\cdot 10^{-28}$~\ecm~at $3 \sigma$ and a corresponding control of systematic effects of $\sigma_{d,syst} < 2\cdot 10^{-28}$~\ecm ($1 \sigma$). 
The source of UCN is placed in a tangential beam tube inside the reactor with a thermal neutron flux of $10^{14}$~s$^{-1}$. 
Solid deuterium is used as a super-thermal converter for the production of UCN~\cite{Frei07}.
Operation of the source at the reactor is expected in 2013. A beamline made from specially prepared replica foil tubes with a relative transmission of $> 0.99$ per meter guides the UCNs to the nEDM spectrometer, which is placed outside the reactor building in a new experiment hall 27 m from the solid deuterium source. 
Taking into account production, volumes and losses of all components, and the EDM chambers, the projected polarized UCN density is $>$ 3000 cm$^{-3}$ in the EDM experiment.

This experiment is based on UCN stored in two vertically aligned cylindrical vessels at room temperature and a vertical magnetic field $B_0$.  In between the cells a high voltage electrode is placed to enable measurements with an electric field parallel and anti-parallel to $B_0$ simultaneously.
For EDM measurements, Ramsey's method of separated oscillatory fields is applied to these trapped UCN.
With a precession time of $T = 250$~s and  an electric field $E = 18$~kV$/$cm, the statistical sensitivity goal can be achieved in 200 days.
In addition, a co-magnetometer based on polarized $^{199}$Hg vapor with a laser-based optical system is placed in these cells~\cite{Griffith:2009zz}.
In addition, external magnetometers are used to measure the field distribution online.
Buffer gases can be added to all magnetometers to investigate various systematic effects and to eventually increase the high-voltage behavior.

The construction work for the beam position, as well as the installation of clean rooms, compensation system and outer magnetic shielding is ongoing. Subsequently, the installation of magnetometry systems and the inner magnetic environment is scheduled for 2012, after finalizing ongoing tests of a small scale prototype.

\subsection{Proton Storage Ring EDM}
\label{sec:pEDM}

The storage ring EDM collaboration has submitted  a proposal to DOE for a proton EDM experiment sensitive
to $10^{-29} \ecm$~\cite{edmweb}. This experiment can be done at Brookhaven National Laboratory (BNL) or another facility 
that can provide highly polarized protons with an intensity of more than $10^{10}$ particles per cycle of 15 minutes. 
The method utilizes polarized protons at the so-called magic 
momentum of 0.7 GeV/{\it c} in an all-electric storage ring with a radius of $\sim 4$0 m.  At this momentum, the proton spin and 
momentum vectors precess at the same rate in any transverse electric field.  When the spin is kept along the momentum
direction, the radial electric field acts on the EDM vector, causing the proton spin to precess vertically. The vertical component 
of the proton spin builds up for the duration of the storage time, which is limited to $10^{3}$~$s$
by the estimated horizontal spin coherence time (hSCT) of the beam within the admittance of the ring.

The strength of the storage ring EDM method comes from the fact that a large number of highly polarized particles can be stored
for a long time, a large hSCT can be achieved, and the transverse spin components can be probed as a function of time with a 
high-sensitivity polarimeter.  The polarimeter uses elastic nuclear scattering off a solid carbon target placed in a 
straight section of the ring; this serves as a limiting aperture. The collaboration has over the years developed the method and 
improved their understanding and confidence in it. Some notable accomplishments are listed below:

\begin{enumerate}

\item Systematic errors and the efficiency and analyzing power of the polarimeter have been studied.  The polarimeter systematic 
errors, caused by possible beam drifting, are found to be much lower than the statistical sensitivity.

\item A tracking program has been developed to accurately simulate the spin and beam dynamics of the 
stored particles in the all-electric ring~\cite{sateesh}. The required ring parameters are readily available at BNL with current capabilities.  

\item E-field can be measured at BNL using the technology developed as part of  the International
Linear Collider (ILC) and energy recovery linacs (ERL) R\&D efforts \cite{dunham}. 
Tests indicate that more than 100 kV/cm across a 3 cm plate separation can be achieved.

\item The geometrical phase effect can be reduced to a level comparable to the statistical sensitivity based on
a position tolerance of commonly achievable $\sim 25 \mu {\rm m}$ in 
the relative positioning of the E-field plates around the ring.

\end {enumerate}

\subsection{Mercury-199 Atomic EDM}
\label{sec:HgEDM}
The mercury atom provides a rich hunting ground for sources of $CP$ violation. An EDM in $^{199}$Hg could be generated by EDMs of the neutrons, protons, or electrons, by chromo-EDMs of the quarks, by $CP$-odd electron-nucleon couplings, or by $\theta_{QCD}$, the $CP$-odd term in the strong interaction Lagrangian. The current upper limit on the Hg EDM~\cite{Griffith:2009zz},  $d(\mathrm{^{199}Hg})<3.1\times10^{-29}\ecm$, places the tightest of all limits on chromo-EDMs, the proton EDM, and $CP$-odd electron-nucleon couplings.

The statistical sensitivity of the current upper limit on $d(^{199}Hg)$ was limited by two noise sources: light shift noise and magnetic Johnson noise. The light shift noise was due to a combination of residual circular polarization of the probe light and a small projection of the probe light axis, along the main magnetic field axis. This noise was subsequently reduced by a factor of 10 by better alignment of the probe light axis and will be further reduced by letting the atoms precess in the dark. The next data runs will be taken with the probe light on only at the start and end of the precession period. The magnetic Johnson noise was generated by thermally excited currents in the aluminum cylinder that held the windings of the main magnetic field coil. The aluminum coil form has been replaced by an insulating coil form, leaving magnetic field noise from the innermost magnetic shield as the dominant remaining noise source.  If the dominant noise source in the next data runs is indeed noise from the magnetic shield, then a factor of 10 improvement in statistical sensitivity can be achieved with the existing Hg EDM apparatus.

An increase in statistical sensitivity requires a corresponding increase in the control of systematic errors. The dominant systematic error has been imperfect knowledge about the magnetic fields produced by leakage currents across the Hg vapor cells when high voltage is applied across the cells. Recently, it was found that most of the leakage current flows along electric field lines in the dry nitrogen gas exterior to the cells; these gas currents can be amplified and have been shown to not produce measurable systematic errors. Roughly 10\% of the total current flows along the cell walls, and will be a source for concern. However, by maintaining these cell wall leakage currents below 0.01 pA, as has been achieved in earlier EDM measurements, a 10-fold improvement in the leakage current systematic error can be achieved.

In summary, unless unforeseen problems emerge, the existing Hg EDM apparatus can provide a 10-fold increase in sensitivity to an Hg atom EDM. This would still be roughly a factor of 10 larger than the shot noise limit of the current apparatus. If warranted, a new apparatus could be developed to go further. A larger diameter and thicker-walled innermost magnetic shield would reduce the magnetic field noise, and additional magnetometers could be installed to provide further information about the field stability. Redesigned vapor cells could reduce the leakage currents and better direct their paths ({\it e.g.}, rectangular cells with a reduced electric field gap). An additional factor of five increase in sensitivity would be feasible.

\subsection{Radon-221,223 Atomic EDM}
\label{sec:RnEDM}

In a heavy atom of a rare isotope, for which the nucleus has octupole strength or permanent deformation, the dipole charge distribution in the nucleus, characterized by the Schiff moment, may be significantly enhanced compared to $^{199}$Hg. 
This enhancement is due to the parity-odd moment arising from quadrupole-octupole interference, and the enhanced E1 polarizability effected by closely spaced levels of the same $J$ and opposite parity.   The strongest octupole correlations occur near $Z=88$ and $N=134$, and isotopes $^{221/223}$Rn and $^{225}$Ra are promising both for practical experimental reasons and as candidates for octupole-enhanced Schiff moments. Enhancements of the nuclear Schiff moment by a factor of 100 or more compared to $^{199}$Hg have been predicted by models using Skyrme-Hartree-Fock for $^{225}$Ra~\cite{rf:deJesus2005} and Woods-Saxon and Nilsson potentials in the case of $^{223}$Rn~\cite{rf:Spevak1997}. However, the uncertainties on the size of enhancements are quite large, in part due to uncertainty in the $^{199}$Hg Schiff moment, and, in the case of $^{221/223}$Rn isotopes, the absence of nuclear structure data. 

The RadonEDM collaboration are focusing on potential EDM measurements with radon isotopes for several reasons.  Most importantly, precision measurements with  polarized noble gases in cells  have demonstrated the feasibility of an EDM experiment. For $^{129}$Xe, it was measured that $d=0.7\pm 3.4\times 10^{-27}$ \ecm~\cite{rf:Rosenberry2001}. A number of techniques have been developed including spin-exchange-optical-pumping (SEOP) using rubidium and construction of EDM cells and wall coatings that reduce wall interactions, in particular for spin greater than 1/2. The half-lives of $^{221/223}$Rn are 
on the order of 20-30 minutes, so an on-line experiment at an isotope production facility is essential. The proposed experiment (S-929) at TRIUMF's ISAC, an on-line isotope separator facility, has been approved with high priority. The experimental program includes development of on-line techniques, including collection of rare-gas isotopes and transfer to a cell, optical pumping, and  techniques for detection of spin precession based on gamma-ray anisotropy, beta asymmetry and laser techniques.

For polarized rare-isotope nuclei, the excited states of the daughter nucleus populated by beta decay are generally aligned, leading to a $P_2(\cos\theta)$ distribution of gamma-ray emission. The gamma anisotropy effect has been used to detect nuclear polarization in $^{209}$Rn and $^{223}$Rn \cite{rf:Tardiff2008, rf:Kitano1988}. At TRIUMF, the large-coverage HPGe gamma-detector array TIGRESS or the new GRIFFIN array may be used. Alternatively, beta asymmetry can be used to detect nuclear polarization with a higher efficiency. Both the gamma-anisotropy and beta-asymmetry detection techniques have analyzing power expected to be limited to 0.1-0.2. The sensitivity of the EDM measurement is proportional to analyzing power; thus laser-based techniques are also under investigation. The collaboration is currently developing two-photon magnetometry for $^{129}$Xe that may also be useful as a co-magnetometer in neutron-EDM measurements. The analyzing power for two-photon transitions can be close to unity as long as the density is sufficient.

EDM measurements in radon isotopes will ultimately be limited by production rates. Fragmentation can produce useful quantities of these isotopes for development, and the beam dump at FRIB may be a source for harvesting large quantities for an EDM measurement. Isotope-separator techniques, such as those used at TRIUMF and ISOLDE, have direct yields that are much higher, and would be a great advantage for the future of the RadonEDM program.

\subsection{Radium-225 Atomic EDM}
\label{sec:RaEDM}

The primary advantage of $^{225}$Ra is the large enhancement \cite{rf:Spevak1997, Ban:2010, Doba:2005}, approximately a factor of 1000, of the atomic EDM  over $^{199}$Hg that arises from both the octupole deformation of the nucleus and the highly relativistic atomic electrons. This favorable case is being studied at both Argonne National Laboratory \cite{Guest:2007} and Kernfysisch Versneller Instituut (KVI) \cite{De:2009}. The scheme at Argonne is to measure the EDM of $^{225}$Ra atoms in an optical dipole trap (ODT) as first suggested in \cite{Romalis:2001}. The ODT offers the following advantages:  $\vec{v}\times\vec{E}$ and geometric phase effects are suppressed, collisions are suppressed between cold fermionic atoms, vector light shifts and parity 
mixing-induced shifts are small.  The systematic limit from an EDM measurement in an ODT can be controlled at the level of 10$^{-30}$\ecm~\cite{Romalis:2001}.

The Argonne collaboration demonstrated the first magneto-optical trap (MOT) of Ra atoms~\cite{Guest:2007}, the transfer of atoms from the MOT to the ODT with an efficiency exceeding 80\%, and the transport of atoms to an ODT in a measurement chamber 0.5 m from the MOT.  In the near future, they plan a vacuum upgrade that should permit the lifetime of atoms in the ODT to improve from 6 $s$ to 60 $s$, and begin the first phase of the EDM measurement at the sensitivity level of 10$^{-26}\ecm$, which should be competitive with 10$^{-29}\ecm$ for $^{199}$Hg in terms of sensitivity to $T$-violating physics. For phase 2 of this experiment, the collaboration plans to upgrade the optical trap.  In the present MOT, the slower and trap laser operate at 714 nm where there is a relatively weak atomic transition rate.  In phase 2, they would upgrade the trap to operate at 483 nm where a strong transition can be exploited for slowing and trapping.  

In Phases 1\&2, a typical experimental run will use 1-10 mCi of $^{225}$Ra presently available.  The next-generation isotope facility, such as FRIB after upgrade or Project X, is expected to produce more than 10$^{13}$ $^{225}$Ra atoms/s \cite{px2009}.  In this case it should be possible to extract more than 1 Ci of $^{225}$Ra for use in the EDM apparatus.  This would lead to a projected sensitivity of $10^{-28}-10^{-29}\ecm$ for $^{225}$Ra, competitive with $10^{-31}-10^{-32}\ecm$ for $^{199}$Hg. Table \ref{table2} summarizes the projected sensitivities.

\begin{table}[htbp]
\centering
\caption{Projected sensitivities for $^{225}$Ra and $^{199}$Hg equivalent for three scenarios}
\label{table2}
\begin{tabular}{|l|c|c|c|}
\hline
 Phase & Phase 1  & Phase 2 (upgrade) &  FRIB after upgrade, Project X   \\
\hline
 Ra (mCi)                           & 1-10    &   10   &   $>$ 1000    \\
$d(^{225}\mathrm{Ra})$ ($10^{-28}\ecm$)  &   100   &  10    &  0.1-1     \\
equiv. $d(^{199}\mathrm{Hg})$ ($10^{-30}\ecm$)   &   10    &  1     & 0.01-0.1  \\
\hline
\end{tabular}
\end{table}

\subsection{Electron EDM with polar molecules}
\label{sec:moleculeEDM}
\paragraph{\bf YbF:} Although the Standard Model predicts that the EDM of the electron is far too small to detect, being some 11 orders of magnitude smaller than the current experimental sensitivity, many extensions of the Standard Model naturally predict much larger values of eEDM that should be detectable. This makes the search for eEDM a powerful way to search for new physics and constrain the possible extensions. Cold polar molecules of YbF have been used to measure eEDM at the highest level of precision reported so far, setting the upper limit at $d_{\sub{e}}<1.05\times10^{-27}\,\ecm$~(90\,\% C.L.) \cite{Hudson:2011}.

Previous eEDM measurements were performed on neutral heavy atoms such Tl~\cite{Regan:2002}. Dipolar molecules have two great advantages over atoms. First, at a modest operating electric field the interaction energy of YbF due to eEDM is 220 times larger than that obtained using Tl in a much larger electric field. Second, the motional magnetic field, a source of systematic error that plagued the Tl experiment, has a negligible effect on YbF. Because of these advantages, it is possible to improve on the Tl experiment by using YbF molecules, even though the molecules are produced in much smaller numbers. The collaboration is developing a cryogenic source of YbF that yields a higher flux of molecules at three times slower velocity. With this new source, the eEDM sensitivity is likely to be pushed down to $10^{-28}\,\ecm$. The long-term plan aims to reach $10^{-30}\,\ecm$ with the development of a molecular fountain based on laser cooling of YbF.

\paragraph{\bf ThO:} The Advanced Cold Molecule EDM (ACME) collaboration uses a newly developed cryogenic technique for creating molecular beams of unprecedented brightness~\cite{Hutzler:2011}, hence allowing large improvements in statistical sensitivity to an eEDM. ACME studies thorium monoxide (ThO), which combines the most favorable features of species used in other experiments~\cite{Vutha:2010}. In particular, the measurement takes place in the metastable H $^{3}\Delta_{1}$ state of ThO; here the effective electric field acting on the eEDM is the largest known (104 GV/cm). This state has $\Omega$-doublet substructure, which makes it possible to spectroscopically reverse the internal E-field within the molecule; this in turn enables powerful methods for rejecting most anticipated systematic errors. Finally, in the H $^{3}\Delta_{1}$ state there is a near-perfect cancellation of magnetic moments due to spin and orbital angular momenta; the resulting small magnetic moment ($< 0.01$ Bohr magnetons) makes the experiment insensitive to systematic errors and noise due to uncontrolled magnetic fields.

The initial phases of apparatus construction are complete, and the entire apparatus is working robustly.  Based on the recent data, the collaboration projects that the statistical sensitivity will be at least at the level of $1\times10^{-28}$ \ecm~by 2013.   Quantitative projections for systematic error limits are difficult in the absence of extensive data, but the collaboration is hopeful that the many built-in features for identifying and rejecting systematics will allow them to make a statistics-limited measurement. In the longer term, the collaboration has identified a host of methods to improve the molecular beam flux and the efficiency of state preparation and detection.  Upgrades to signal size will be incorporated into the experiment after the initial measurement with the current apparatus.  Overall, the collaboration projects a sensitivity that could ultimately reach $3\times10^{-31}$ \ecm.

\subsection{Electron EDM with Francium}
\label{sec:FrEDM}

An eEDM experiment using francium atoms can challenge SUSY with unambiguous results. There are no hadronic effects that need to be subtracted out. The relation between an EDM of an alkali atom and that of an electron is the simplest and most reliably calculated EDM effect in any multi-electron system. The calculations have been performed using different techniques and by different authors with numerical differences typically less than $20\%$. Moreover, the calculations are similar to those used for calculating parity violating effects in atoms and so have indirectly been validated by experiments. 

With francium comes a higher sensitivity to an electron EDM than any atom previously used. The large nuclear spin and magnetic dipole moment of $^{211}$Fr, when combined with laser cooling, bring the potential benefit of the most complete systematic rejection of any eEDM experiment yet attempted. Magnetic fields that change synchronously with the electric field can mimic an EDM. Even in experiments where there is no net motion, the Lorentz transform due to the atom's motion through the electric field gives rise to a motional magnetic field $\bf{B}_{\mathrm{mot}} = \bf{v} \times \bf{E}/c^2$
that can lead to first-order systematic effects. These effects can be removed in first order if the atom is quantized in the electric field, no external magnetic fields are applied, and motional and remnant magnetic fields are made small.  The remaining systematic effects scale as inverse powers of the electric field, allowing one to quickly distinguish between a true EDM (linear in $E$) and the systematic effect (proportional to $1/E^3$). The ratio of systematic effect sensitivity to eEDM sensitivity in $^{211}$Fr is two orders of magnitude smaller than in any other alkali atom.

What is presently lacking is a source of francium intense enough to make measurements sensitive enough to lower the electron EDM limit by three orders of magnitude and to test for systematics, both false positives and false negatives. The proposed Joint Nuclear Facility at Project X will have proton beam currents about two orders of magnitude larger than TRIUMF and ISOLDE, and may produce $10^{13}$  $^{211}$Fr/s $-$ sufficient to lower the electron EDM upper limit by a factor of $10^3$.

\section{Weak Decays}
\label{sec:Weak}

The study of weak decays of hadrons built from light quarks continues to provide some of the most precise input for the Standard Model as well as stringent constraints of various Standard Model extensions. In particular, the most precisely determined element of the Cabibbo-Kobayashi-Maskawa matrix, $V_\mathrm{ud}$, is obtained from ``superallowed" Fermi nuclear transitions, while tests of the universality of the charged-current weak interactions of leptons obtained with pion decays are approaching similar levels of precision. At present, there exists considerable effort in three directions involving semileptonic charged-current weak interactions of hadrons: studies of pion and kaon leptonic decays; ongoing measurements of weak interaction nuclear transitions; and studies of neutron decays and interactions. Here we survey some of the present activity and future opportunities. 

\subsection{Pion and Kaon Leptonic Decay}
\label{sec:Pion}
The absence of an explanation for the generation puzzle -- why do we have exactly three
versions of each quark and lepton? -- is a major flaw in the highly successful Standard Model.
Electron-muon universality, within the context of the SM, is the hypothesis
that charged leptons have identical electroweak gauge interactions and differ only in
their masses and coupling to the Higgs. However, there could be additional new physics
effects, such as non-universal gauge interactions or scalar or pseudoscalar bosons with
couplings not simply proportional to the lepton masses, that would
violate universality.

One of the most sensitive approaches to seeking such new interactions is the study of the
ratio of decay rates of pions \cite{Bryman2011}
\begin{equation}
R^\pi_{e/\mu} \equiv {\Gamma(\pi \rightarrow e \nu (\gamma)) \over \Gamma(\pi \rightarrow \mu \nu (\gamma))}
\end{equation}
which in the SM is predicted to be 1.2351(2) with an uncertainty of  $\pm$ 0.02\% \cite{Kinoshita1959,
Marciano1976,Cirigliano2007}.  New physics at scales as high as 1000 TeV can be
constrained or conceivably unveiled by improved measurements of this ratio. One example \cite{Bryman2011}
is a charged physical Higgs boson with couplings $g \lambda_{ud}/2 \sqrt{2}$ to the 
pseudoscalar current $\bar{u} \gamma_5 d$ and $g \lambda_{l \nu}/2 \sqrt{2}$ to the current
$\bar{l}(1-\gamma_5)\nu_l$, $l=e,\mu$, where $g$ is the $SU(2)_L$ gauge
coupling and $\lambda$ represents chiral breaking suppression factors. One finds
\begin{equation}
m_{H^\pm} \sim 200 \mathrm{~TeV}~\sqrt{\lambda_{ud} (\lambda_{e \nu}-{m_e \over m_\mu} \lambda_{\mu \nu})}
\end{equation}
If $\lambda_{e \nu}/\lambda_{\mu \nu} = m_e/m_\mu$, as in the minimal two-Higgs doublet model, 
there is no sensitivity to new physics. However, in more general multi-Higgs models
such a chiral relationship is not necessary and $\lambda$ may not be too suppressed. For example,
in the case of loop-induced charged Higgs couplings $\lambda_{e \nu} \sim \lambda_{\mu \nu} \sim 
\lambda_{ud} \sim \alpha/\pi$, one finds a $R^\pi_{e/\nu}$ determination to $\pm$ 0.1\% is sensitive to
$m_{H^\pm} \sim 400$ GeV.
 If a discrepancy between theory
and experiment is found in $R^\pi_{e/\mu}$, some type of charged Higgs explanation would be quite
natural.  However, such a result could also point to additional charged axial-vector interactions or loop
effects due to new physics \cite{sean}. It could also be interpreted as the damping of one of the $\pi_{l2}$ modes by heavy neutrino
mixing.  Analogous K  decays,
\begin{equation}
R^K_{e/\mu} \equiv {\Gamma(K \rightarrow e \nu (\gamma)) \over \Gamma(K \rightarrow \mu \nu (\gamma))},
\end{equation}
can also probe high scales and have the
added appeal of being particularly sensitive to the lepton flavor-violating decay  $K^+ \rightarrow e^+ + \nu_\tau$,
a process that might be induced through loop effects \cite{Masiero2006}.\\

\subsubsection{Experimental Studies of $\pi \rightarrow e \nu$ and $K \rightarrow e \nu$ Decays}
The most recent $\pi^+ \rightarrow e^+ \nu$ ($\pi_{e2}$) branching ratio measurements and subsequent
determination of the ratio $R^\pi_{e/\mu}$ were done at TRIUMF \cite{Britton1992} and PSI \cite{Czapek1993}
in the 1990s.  The
results from the two experiments are consistent and in  good agreement with the SM
expectation previously discussed,
\begin{equation}
R^{\pi-\mathrm{TRIUMF}}_{e/\mu} = 1.2265(34)(44) \times 10^{-4} ~~~~~R^{\pi-\mathrm{PSI}}_{e/\mu} = 1.2346(35)(36) \times 10^{-4},
\end{equation}
where the first and second uncertainties are due to
statistical and systematic effects. The Particle Data Group (PDG) average value is
$R^{\pi}_{e/\mu} = 1.230(4) \times 10^{-4}$ \cite{Nakamura2010},
which includes results from \cite{Bryman1986}. Two new experiments are under way 
at TRIUMF \cite{Aguilar2009} and PSI \cite{Pocanic2009} which promise to
improve the precision of $R^{\pi}_{e/\mu}$ by a factor of 5 or more, thereby
testing the SM prediction to better than ?0.1\%. At that level of precision, new physics effects could
appear as a deviation; in the absence of a deviation,
strong new
constraints would be imposed on such physics.\\

The PIENU experiment at TRIUMF \cite{Aguilar2009} is based on a refinement of the technique used in the previous TRIUMF
experiment \cite{Britton1992}. The branching ratio will be obtained from the ratio of positron yields from
the $\pi \rightarrow e \nu$ and from the $\pi \rightarrow \mu \rightarrow e$ decay chain. By measuring 
the positrons from these decays
in a non-magnetic spectrometer, many normalization
factors, such as the solid angle of positron detection, cancel to first order, so that only small corrections for 
energy-dependent effects, such as those for multiple Coulomb scattering (MCS) and
positron annihilation, remain. Major improvements in precision stem from
the use of an improved geometry, a superior calorimeter, high-speed digitizing of all pulses,
Si strip tracking, and higher statistics. The improvements lead to an expected precision on
the $R^{\pi}_{e/\mu}$  branching ratio of $<0.06$ \%, which corresponds to a 0.03 \% uncertainty in the ratio
of the gauge boson-lepton coupling constants $g_e/g_\mu$ testing electron-muon universality.\\

At PSI, the PIBETA CsI spectrometer \cite{Pocanic2004}, built for a determination of the $\pi^+ \rightarrow \pi^0e\nu$
branching ratio and other measurements \cite{Frlez2004}, has been upgraded and enhanced for the
PEN \cite{Pocanic2009} measurement of the $\pi \rightarrow e \nu$ branching ratio. The PEN 
technique is similar to that
employed in the previous PSI experiment \cite{Czapek1993}, which used a nearly 4$\pi$-sr BGO spectrometer.
PEN began operations in 2007 and has observed $>10^7$ $\pi \rightarrow e \nu$ decays. PEN completed
data acquisition in 2010 and expects to reach a precision of $<$ 0.05\% in $R^\pi_{e/\mu}$. \\

Recent progress on $R^K_{e/\mu}$ 
has been made by KLOE \cite{Ambrosino2009} and NA62 \cite{Goudzovski2009},  with current NA62 efforts
aimed at reaching a precision of  0.4\%. KLOE collected 3.3 billion $K^+K^-$
pairs, observing decay products in a drift chamber in a 0.52T axial magnetic field
surrounded by an electromagnetic calorimeter. The measurement of $R^K_{e/\mu}$ consisted of comparing the
corrected numbers of decays observed from the $K \rightarrow e \nu(\gamma)$ and $K \rightarrow \mu \nu(\gamma)$ channels. The result,  $R^{K~KLOE}_{e/\mu} = (2.493 \pm 0.025(\mathrm{stat}) \pm 0.019(\mathrm{syst})) \times 10^{-5}$ \cite{Ambrosino2009},
agrees with the SM
prediction at
the 1\% level.\\

NA62 at CERN, using the setup from NA48/2, has embarked on a series of $K_{e2}/K_{\mu 2}$
measurements \cite{Goudzovski2009}. The $K^+$ beam is produced by the 400 GeV/c SPS. Positively charged
particles within a narrow momentum band of (74.0 $\pm$ 1.6) GeV/c are selected by an
achromatic system of four dipole magnets and a muon sweeping system and enter a fiducial
decay volume contained in a 114 m long cylindrical vacuum tank, producing a secondary
beam. The $K \rightarrow e \nu(\gamma)$ and $K \rightarrow \mu \nu (\gamma)$  detection system includes a magnetic
spectrometer, a plastic scintillator hodoscope, and a liquid krypton
electromagnetic calorimeter. As in KLOE, the experimental strategy is based on
counting the numbers of reconstructed $K \rightarrow e \nu(\gamma)$ and $K \rightarrow \mu \nu (\gamma)$ events concurrently,
eliminating dependence on the absolute beam flux and other potential systematic
uncertainties. The result,  $R^{K~NA62}_{e/\mu} = (2.487 \pm 0.011(\mathrm{stat}) \pm 0.007(\mathrm{stat})) \times 10^{-5}$ \cite{Goudzovski2009},
is based on 40\% of the data acquired in 2007 and agrees with the SM prediction.
The full data sample may allow a statistical uncertainty of 0.3\% and a total
uncertainty of 0.4-0.5\% .\\

\subsubsection{Future Prospects}
If PIENU and PEN achieve their sensitivity goals there will still be a considerable window
in which new physics could appear without complications from SM prediction uncertainties. Reaching
the uncertainty ~0.02\%, the level  of current SM calculations, would require
a new generation of experiments capable of controlling
systematic uncertainties at or below 0.01\%. High-precision measurements of
$R^{\pi/K}_{e/\mu}$ will be an important complement to LHC high-energy studies. 
High intensity beams with 100\% duty factors and ultra-high intensities
and purities will be important to ultra-high precision experiments on $\pi/K \rightarrow e/\mu$.
Such experiments potentially could lead to breakthroughs in our understanding of $e-\mu$ universality and
are sensitive to a variety of subtle non-SM effects.
Project X would provide such beams for pions and kaons, while
pion studies could continue with the beams available at PSI or TRIUMF.

\subsection{Superallowed Nuclear Decay and CKM Unitarity}
\label{sec:Super}

\subsubsection{Goals of $\beta$-Decay Studies}
\label{ss:goals}

Studies of weak interactions based on nuclear $\beta$ decay are currently focused on
probing the limits of the Standard Model: Both the conserved vector current (CVC)
hypothesis and the unitarity of the Cabibbo-Kobayashi-Maskawa (CKM) matrix are being
tested with ever-increasing precision. To do so, it is necessary first to isolate the
vector part of the combined vector and axial-vector ($V$$-$$A$) structure of the weak
interaction, a requirement that is satisfied by superallowed $0^+$$\rightarrow 0^+, \Delta T = 0$
nuclear decays, which are pure vector transitions.
These transitions occur in a wide range of nuclei (10$\le A \le$74) and yield
the vector interaction strength in each case.  If CVC is valid, then the
strength of the vector interaction is not renormalized in the nuclear medium but is a
``true'' constant; and, so far, the constancy of the strength is confirmed by these
superallowed decays at the level of $10^{-4}$ \cite{HT09}.  With CVC satisfied, these
results also lead to today's most precisely determined value of the CKM matrix element,
$V_{ud}$: $0.97425 \pm 0.00022$ \cite{HT09}.

When $V_{ud}$ is combined with the $V_{us}$ and $V_{ub}$ values obtained from kaon and $B$-meson decays, 
respectively, the sum of squares of the top-row elements of the CKM matrix provides the
most demanding test available of the unitarity of the matrix.  The result, currently standing
at 0.99990(60) \cite{TH10a}, agrees with unitarity; and its precision,
which can still be improved, sets limits on extensions to the Standard Model
that could break the three-term unitarity of the CKM matrix.  Other tests of the weak
interaction involve asking whether it is pure $V$$-$$A$ as assumed in the Standard Model or
whether there are small components of $S$ (scalar) or $T$ (tensor) interactions.  The
$0^+$$\rightarrow 0^+$ superallowed decays also set a limit on $S$ but not on $T$.  This topic is taken
up further in the report of Savard and Behr \cite{SB11}.

\subsubsection{Superallowed $0^+ \rightarrow 0^+$ Transitions}
\label{ss:00}

The study of $\beta$ decay between ($J^\pi, T$) = ($0^+, 1$) nuclear analog states has
been a fertile means of testing the Standard Model.  Because the axial current cannot
contribute to transitions between spin-0 states, only the vector current is involved in
these transitions.  Thus, according to CVC, the experimental $ft$-value for each of these
superallowed transitions should be simply related to the fundamental weak-interaction
coupling constant, $G_V$.  The $ft$-value itself depends on three measured quantities: the
total transition energy, $Q_{EC}$, the half-life, $t_{1/2}$, of the parent state, and the
branching ratio, $R$, for the particular transition of interest.  The first of these can
be measured in a Penning trap, where a very few ions are sufficient for high precision;
however, both $t_{1/2}$ and $R$ require large numbers of observed decays to achieve the
required statistics.
Currently, the best-known superallowed decays are from nuclei rather close to stability,
which are easily produced.  However, future improvements in precision will need
comparable measurements on transitions from nuclei much farther from stability.  Higher
intensity beams will be required to produce them.

\subsubsection{Superallowed $J \rightarrow J$ Transitions}
\label{ss:JJ}

The vector current also contributes to $J$$\rightarrow$$J$, $T$\,=\,1/2 mirror decays, which can also,
in principle, be used to determine the CKM matrix element, $V_{ud}$.
The difficulty is that these transitions are not pure vector
in character but include a mix of vector and axial-vector contributions.
Thus a correlation experiment is needed to separate one from the other.  Today's
accuracy in correlation experiments is insufficient for this method to be
competitive.  Nevertheless, there are also many examples of $J$$\rightarrow$$J$
transitions, and the uncertainty can be reduced if enough are included in an average.
Naviliat-Cuncic and Severijns \cite{NS09} have studied $J$$\rightarrow$$J$
transitions between odd-mass mirror states in $T$\,=\,1/2 nuclei.
Their average from the five measured transitions leads to a $V_{ud}$ value whose uncertainty
is about 10 times larger than the result from superallowed beta decay.
In future, further experiments on $T$\,=\,1/2 mirror decays, both near and far from stability, can certainly
improve the accuracy of this $V_{ud}$ determination.

\subsubsection{Neutron $\beta$ Decay}
\label{ss:Neutron}

Neutron $\beta$ decay is conceptually the simplest example of a mixed vector and axial-vector
$T = 1/2$ mirror decay.  Unfortunately, the simplicity that this brings to its theoretical
analysis is more than counterbalanced by the experimental difficulties encountered in confining the neutron
long enough to measure its properties.  At present there are conflicting results -- well outside of statistics --
for the neutron mean-life, a situation that has driven the Particle Data Group \cite{PDG10} to ``call upon the
experimenters to clear this up."  The internal agreement among the correlation measurements is better, but
is far from satisfactory since there has been a systematic drift in the measured central value of $\lambda = g_A/g_V$
over the past two decades.  More measurements are clearly required.

\subsubsection{Roadblocks and Opportunities}
\label{ss:blocks}

There are two roadblocks to the experimental quest for a more accurate
$V_{ud}$ value.  First, there is a radiative correction to be included
in the analysis, and a part of this, known as the inner radiative 
correction, is not well determined.  A recent evaluation by Marciano
and Sirlin \cite{MS06} reduced its uncertainty by a factor of two;
but, even so, the new uncertainty remains the largest contributor
to the error budget for $V_{ud}$.  Second, the use of the CVC hypothesis
is valid only in the isospin-symmetry limit.  In nuclei, the presence
of the Coulomb force acting between protons breaks isospin symmetry, as
does charge dependence in the nuclear force to a lesser extent.  So
an isospin-symmetry breaking correction, denoted $\delta_C$, needs
to be evaluated and this involves a nuclear-structure-dependent
calculation.  The estimated uncertainty in the model dependence of
these calculations is the second largest contributor to the error
budget for $V_{ud}$.  There has been a lot of recent activity
in $\delta_C$ calculations for the $0^+$$\rightarrow 0^+$ superallowed
decays \cite{Sa11,Gr10,MS08,Au09,LGM09,TH08}, with considerable
disparity being evident among the results.

Fortunately, there is an opportunity here.  Towner and Hardy \cite{TH10b}
have proposed an experimental test rooted in the requirement that the
$\delta_C$ calculations should yield results consistent with a 
conserved vector current.  To date, shell-model-based calculations \cite{TH08}
satisfy this test the best, and it is these calculations that have been
used in the extraction of $V_{ud}$.  Further precise experiments could
improve the test and reduce the uncertainty on $V_{ud}$.  Calculations
anticipate somewhat larger $\delta_C$ values for the decays of $T_z$\,=\,-1
superallowed emitters with $A \leq 42$, and much larger values for $T_z$\,=\,0
emitters with $A \geq 62$.  If experiment confirms these large calculated
$\delta_C$ values, then it validates the much smaller values for the transitions
now used to obtain $V_{ud}$. In particular, the different calculations give a
range of predictions for mirror pairs of transition such as: $^{26}$Si ($T_z$\,=\,-1)
$\rightarrow$ $^{26}$Al ($T_z$\,=\,0) versus $^{26}$Al ($T_z$\,=\,0) $\rightarrow$
$^{26}$Mg ($T_z$\,=\,1).  The decay of the $T_z = -1$ member of each pair requires a very difficult
branching-ratio experiment to be performed but, if successful, the result would be
very revealing for the $\delta_C$ calculations.  For the other group of emitters with $A \geq 62$,
only $^{62}$Ga has so far been measured with the requisite precision.  Experiments on
lifetimes, branching ratios, and $Q$-values for $^{66}$As, $^{70}$Br and $^{74}$Rb
would be very welcome.  

A recent bellwether experiment \cite{Me11} focused on the beta decay of $^{32}$Cl.
One of its branches is a $J$$\rightarrow$$J$, $\Delta T$\,=\,0 transition for which theory
fortuitously predicts a negligibly small axial-vector component.  Thus it could be
analysed as if it were a pure Fermi transition.  The result it yielded corresponded
to a very large isospin-symmetry-breaking correction of order $5 \%$.  Being a nucleus with
$A = 4n$, large isospin-symmetry breaking could be anticipated since the daughter state
lies very close in energy to a state of the same spin and different isospin.  As a consequence,
the case provides a critical challenge to the $\delta_C$ calculations, a challenge that was successfully
met by the same type of shell-model-based calculations used to analyze the $0^+$$\rightarrow 0^+$
transitions.  In future an examination of other $J \rightarrow J$
transitions from $A = 4n$ nuclei should be undertaken to seek other examples with small
axial-vector contributions accompanied by large isospin-symmetry breaking.


\subsection{Neutron Decay}
\label{sec:nDecay}

Beta-decay studies provide a remarkable tool to probe the helicity structure
and quark-lepton universality of the electroweak interaction, providing model
independent constraints on the effective new physics energy  scale in the
multi-TeV range.  Measurements of neutron beta-decay also provide basic
parameters for the charged weak current of the
nucleon.\cite{Abel08}  In particular, neutron beta-decay measurements are the definitive
source for $g_A$, the axial form factor, and provide a nuclear-structure-independent value
for the CKM element $V_{ud}$.  Although at
present the $0^+ \rightarrow 0^+$ superallowed decays provide the most precise value
for $V_{ud}$, the experimental data for the neutron continue to improve, and should
become directly competitive with $0^+ \rightarrow 0^+$ during the next 10 years.

We note that $g_A$ is important in our understanding of
the spin and flavor structure of the nucleon\cite{Bass05,Clos88}, a central target
for high-precision lattice QCD studies\cite{Yama08,Choi10}, an essential parameter
for effective field theories\cite{Gock05}, and one of a small set of parameters
necessary in establishing high-precision predictions of solar fusion\cite{Adel10}.
The neutron lifetime figures prominently in high precision
predictions for big bang nucleo-synthesis as well\cite{Burl99}.  High precision values
for $g_A$ are also important for the reactor neutrino-anomaly question, one
of the results driving current interest in short baseline oscillations
studies\cite{Ment11}.  The current value of $g_A$ is $g_A$ = 1.2701(25).\cite{Naka10}

High precision neutron beta-decay studies also provide constraints on a large 
variety of extensions to the Standard Model.  The most stringent
constraints come through tests of the quark-lepton universality
of the weak interaction, which within the Standard Model are equivalent to tests
of the unitarity of the CKM quark mixing matrix.  The strongest constraints come
through unitarity tests on the first row:
$V_{ud} ^2 + V_{us} ^2 + V_{ub} ^2$ = 0.9999(6).\cite{Naka10}
For any such test, the diagonal element dominates the sum, and a very
high-precision ($<$ 0.02\%) determination of this diagonal element ($V_{ud}$ for unitarity tests
of the first row of the CKM matrix) is necessary for a competitive unitarity constraint.
Here beta decay also provides the highest-precision measurement
of a diagonal CKM element, with the value extracted from $0^+ \rightarrow 0^+$ decay being
$V_{ud}$ = 0.97425(22), and with the uncertainties from $V_{ud}$ and $V_{us}$ now contributing
equally to the sum.\cite{Naka10}

The resultant constraints on (v,a) current interactions are quite stringent,
with generic limits on the effective scale for new physics at roughly the 10 TeV level.\cite{Ciri10}
A large assortment of extensions to the Standard Model, including new Z' gauge bosons,
generic Kaluza-Klein W* excitations, and charged Higgs bosons, are tightly
constrained by the unitarity sum.\cite{Anto10}  Flavor universality in supersymmetric
extensions of the Standard Model are also constrained by the unitarity sum.\cite{Kury02}
The robustness of these limits and enormous progress made in the kaon-sector
in the precision and reliability with which $V_{us}$ can be determined motivate
continued effort on the experimental extraction of $V_{ud}$.\cite{Anto10}  At present,
the precision of $V_{ud}$ from $0^+ \rightarrow 0^+$ decays is nominally limited by loop-level
electroweak radiative corrections\cite{Hard09,Marc06}; however the nuclear-structur-dependent 
corrections for the $0^+ \rightarrow 0^+$ systems remain an area of active
concern. Neutron beta-decay can provide a structure-independent value for $V_{ud}$, a
significant contribution to the status of the current unitarity test.

The observables in neutron decay include a number of correlations (and
the Fierz term, which influences the energy dependence of the total beta-decay
rate) that provide multiple probes of non v-a interactions generated by
Standard Model extensions.\cite{Abel08,Gudk06}  For example, constraints on ($S,T$) interactions
arise from angular correlation measurements such as the neutrino asymmetry
and the Fierz term.  Because two observables with similar sensitivities to
these terms are available, there is a consistency test within the neutron decay
system itself for these effects. In particular, it is the aim of some beta-decay
experiments (in the planning or construction phase at present) to reach sensitivities
of a few parts in $10^{-4}$.  In this case, the model-independent constraints for
interactions that only couple to electrons and induce scalar and tensor terms can be made
quite stringent with next-generation beta decay experiments. For example,
limits in the 5-10 TeV range which are significantly stronger than expected LHC limits
are expected to be feasible.  In addition, if a new particle resonance is discovered
at the LHC, beta-decay experiments at this level of precision may provide complementary
information on the quantum numbers and weak couplings of such a resonance, as was
recently demonstrated for the case of a scalar resonance\cite{Batt11}. Relevant limits
(complementary to those placed by LHC) can also be placed
on supersymmetric couplings\cite{Prof06} and couplings to leptoquarks\cite{Seve06}. 

$T$-noninvariant angular correlations can also be probed in beta decay.
These experiments can provide constraints on $CP$-violating phases beyond the Standard Model
that are complementary to the ones derived from  EDMs.\cite{Herc01,Tuli11}
In particular, a number of measurements have been performed of angular correlations
proportional to complex, (V,A)couplings (parameterized by the ``D'' coefficient), with
the ongoing work of the emiT\cite{Mumm11} and TRINE\cite{Sold04} collaborations having
established the basis for pushing sensitivities for $T$-violating phases to the final
state effect level ($10^{-5}$ level).

The past 10 years have seen significant growth in the number of physicists
involved in neutron beta-decay measurements. Although it is beyond the scope
of this brief summary to catalog all of the experimental activity in this subfield,
it is characterized by often complementary experiments with cold and ultra-cold neutrons
and has seen the emergence of precision measurements of radiative decay of the neutron
\cite{Nico05} for the first time. A number of experiments are under way that target
precisions near or at the 0.1\% level in the next few
years for the lifetime of the neutron\cite{Dewe09}, the electron-neutrino-asymmetry\cite{Sims09,Wiet09} and the
beta-asymmetry\cite{Maer09,Liu10}.  Taken as a group, they provide a powerful
consistency test for the form factors and Standard Model constraints that can be
extracted at this level of precision\cite{Gard01}.

Ongoing measurements have also set the stage for a number of ambitious experiments under
development or construction that target precisions in the $10^{-4}$ range.  For ultra-cold
neutrons, for example, there are lifetime experiments based on material and magnetic
trapping geometries\cite{Fomi11,Wals09} and angular correlation experiments under
development that are particularly sensitive to ($S,T$) interactions\cite{Wilb09}.  For angular
correlation measurements with cold neutron beams, the PERC\cite{Dubb08} collaboration
based in Munich have as their goal polarimetry and other systematic
errors ultimately in the low $10^{-4}$ range, and the Nab/ABba\cite{Poca09,Wilb05}
collaboration will be targeting systematic uncertainties below the
$10^{-3}$ level for their measurements as well.

Intensity frontier development should provide the opportunity to optimize existing
cold neutron beam delivery for fundamental neutron physics research, positively
impacting beta-decay experiments as well as a variety of other fundamental neutron
studies.  For ultra-cold neutron-based experiments, the intensity frontier
initiative could provide the opportunity to construct a next-generation source of
extracted ultra-cold neutrons.  Although work over the past 10 years has established
viable strategies to significantly increase ultra-cold neutron densities, experiments
remain strongly constrained by the ultra-cold neutron densities at existing sources.
A next-generation source could permit the community to capitalize on the ongoing
refinement of systematic errors in existing beta-decay, EDM and short-range interaction
searches.  In particular, for beta-decay studies, it should enable the next generation
of beta-decay experiments with ultra-cold neutrons to reach sensitivities limited by
systematic errors, and probe energy scales comparable to and in some cases above that
planned for the LHC. 


\subsection{Neutron Lifetime}
\label{sec:nlife}

The decay of a free neutron is a semi-leptonic process that neatly probes the weak mixing of the quarks in the first generation of flavors. It provides a measurement of $V_{ud}$, the first of the elements of the CKM matrix, which quantifies the relative rotation between the weak eigenstates and the flavor eigenstates of quarks.
The current best determination of $V_{ud}$ comes from the combined analysis of 20 different $0^+\rightarrow0^+$ nuclear decays, which are ultimately limited by theoretical uncertainties in the knowledge of nuclear structure and corrections associated with isospin symmetry breaking~\cite{Hardy09}.
Though experiments with free neutrons are less precise, 
improvements could yield independent information on $V_{ud}$ useful for performing unitarity tests of the CKM matrix, an important part of the program of testing the Standard Model.     
In addition, neutron $\beta$-decay includes both Fermi and Gamow-Teller transitions. In the SM framework, the neutron decay amplitude is proportional to the linear sum of the squares of the weak vector and axial vector coupling constants ($g_v$ and $g_a$). 
Calculations of the radiative corrections to the neutron lifetime have been recently revised to a precision of 0.04\%~\cite{Marciano06}.
Improving the experimental uncertainty to this level requires (1) the determination of the ratio of the vector and axial vector coupling constants ($\lambda\equiv g_a/g_v$) by measurement of one of the decay asymmetries (for example the $\beta$ asymmetry $A$) to a precision of $\Delta A/A=4\Delta\lambda/\lambda<0.16\%$, and (2) the determination of the neutron lifetime to a precision of $< 0.04\%$~\cite{Marciano06}.  

Measurements of the neutron lifetime have been approaching the 0.1\% level of precision ($\sim$ 1~s uncertainty). However, several recent neutron lifetime results~\cite{Serebrov05, Serebrov2008, Ezhov09, Pichlmaier10} are up to 7~s lower than the PDG value before 2010 ($885.70\pm 0.85$~s)~\cite{PDG06}.
This $>\mbox{6 }\sigma$ deviation has not yet been resolved.
Even though the new 2011 update of the PDG value ($881.5 \pm 1.5$~s)~\cite{PDG2011} includes all these measurements, with the uncertainty scaled up by a factor of 2.7, the PDG questions this new world average and calls upon the experimenters to clear up the current state of confusion.

The precise determination of the neutron lifetime could also have a profound impact on astrophysics and cosmology. 
In the standard big bang nucleosynthesis (SBBN) model, the abundance of the light elements can be determined with a single cosmological parameter -- the baryon-to-photon ratio $\eta_{10}$, together with the nuclear physics input of the neutron lifetime and 11 key nuclear reaction cross-sections~\cite{Burles99}.
The influence of the neutron lifetime on the abundance of the light species of primordial nuclei, in particular $^4$He, is based on two effects. First, the neutron lifetime indirectly affects the neutrino-nucleon reaction rate and the neutron-to-proton ratio when the primordial neutrons decouple from the radiation field. Second, the lifetime directly affects the number of these neutrons left to participate in nucleosynthesis, which is delayed by photo-dissociation of deuterons in the hot radiation field. 
As the primordial neutrons are protected against $\beta$-decay by fusing with protons into deuterons and then into $^4$He, a shorter neutron lifetime would result in a smaller $^4$He abundance (Y$_p$).
As a consequence, a 1\% change in the neutron lifetime leads to a 0.75\% change of Y$_p$~\cite{Mathews05}.
With a precise determination of $\eta_{10}$ from WMAP, the SBBN predicts Y$_p$ with a 0.2 -- 0.3\% precision~\cite{Steigman10}.
Of the primordial elements, Y$_p$ is particularly sensitive to the expansion rate of the universe and to a possible lepton asymmetry in the early universe~\cite{Steigman07}.
Information on Y$_p$ is attained from either direct observations of the H and He emission lines from low-metallicity extragalactic regions or from the indirect measurements using the power spectrum of the cosmic microwave background (through its effect on the electron density at recombination). The current precision of these measurements is about 1 -- 2\%~\cite{Steigman10, Izotov2010}. 
With the anticipated improvement from the Planck experiment in Europe and the James Webb space telescope in the US, the interpretation of Y$_p$ in SBBN will hinge on the accuracy of the neutron lifetime.   
In astrophysics, the weak axial coupling constant $g_A$ of neutron $\beta$-decay is a key parameter in the standard solar model, which describes the $pp$ fusion process and the CNO cycle that produce solar energy as well as generate the solar neutrinos that can be directly measured by terrestrial experiments. 
Solar neutrino fluxes thus depend indirectly on the neutron lifetime and its uncertainties.
As more measurements of the solar neutrino flux become available~\cite{Borexino08, SNO}, the precise knowledge of the neutron lifetime will have a greater impact on the expected values. 
The same concern also applies to the estimate of the anti-neutrino flux used in many neutrino oscillation experiments based at reactors~\cite{Mention11}.

It seems that the only way to settle the controversy surrounding the value of the neutron lifetime is to perform independent measurements with 0.1~s precision and rigorous control of systematic effects.
The difficulties with measuring the absolute neutron lifetime originate from the low energy of its decay products, the essential impossibility of tracking slow neutron trajectories in matter, and the fact that the lifetime is long.
The lifetime of $\beta$-decay is comparable to the timescale of many surface effects that contribute to the loss of neutrons~\cite{Lamoreaux02, Steyerl10}. To achieve a precision measurement of the $\beta$-decay lifetime, one has to control these additional sources of loss to levels better than the desired precision.
Recent lifetime measurements, which succeeded in reducing the uncertainty to a few seconds, have used ultra-cold neutrons (UCN) trapped in material bottles.
With small kinetic energies, UCN experience total reflection from material walls (with a small absorption coefficient) at any incident angle~\cite{golub1991}.
This property allows them to be trapped in material bottles for times comparable to the $\beta$-decay lifetime.
In a typical UCN bottle experiment, the UCN are loaded into a bottle (or a trap) and stored for different periods of time, and the survivors are counted by dumping them into a UCN detector outside the trap. 
There is much controversy over how to reliably correct for the loss of UCN when they interact with material on trapping walls.

The UCN$\tau$ collaboration will measure the neutron lifetime using UCN in a novel magneto-gravitational trap~\cite{Walstrom09}.  
This trap will eliminate interactions on the confining walls and the associated uncertainties by replacing the material bottle with a trap formed by magnetic fields on the sides and bottom, and closed by gravity at the top.  
Since the interactions of neutrons with magnetic fields and gravity are well understood (and can be reliably modeled in numerical simulations), the systematic uncertainties in this approach are expected to be smaller than (and independent of) those in earlier experiments using material bottles.
The novelty of the experiment originates from the use of an asymmetric trap to facilitate (1) fast draining of the quasi-bound UCN~\cite{Bowman05}, and (2) quick sampling of the entire phase space in order to suppress spurious temporal variation of the decay signals when coupled to a non-uniform detection efficiency. 
The room-temperature trap using a permanent magnet array avoids many engineering challenges of prior cryogenic experiments and allows fast turnaround time for detector prototyping. 
The design of the experiment, with an open top, provides ample room for implementing many novel detection techniques, allowing a comprehensive study of the systematic effects discussed above. 

In the preliminary report of the 2011 review on fundamental neutron physics~\footnote{http://science.energy.gov/$\sim$/media/np/nsac/pdf/mtg-63011/Kumar\_Neutron\_Interim\_Report.pdf} conducted by the National Science Advisory Committee (NSAC), the committee strongly recommends the pursuit of the planned NIST beam experiment at the 0.1\% level of accuracy, and encourages more R\&D effort on the UCN$\tau$ experiment. The UCN$\tau$ experiment uses techniques complementary to the beam experiment, and has the potential to push the precision of the neutron lifetime beyond the current state of the art, towards the 0.01\% level.
The intensity frontier of Project X brings about the unique opportunity to install a world-class UCN source, driven by the planned proton source. 
This investment will add a new facet to strengthen the scientific program aimed to probe physics beyond the SM at the TeV scale, using complementary techniques with low energy neutrons, including the search for the neutron electric dipole moment, the $n\bar{n}$ oscillation, and a comprehensive program of precision $\beta$-decay measurements.  


\subsection{Nuclear Decay Correlations}
\label{sec:nuccor}

Two tests of discrete symmetries are being considered at the National Superconducting Cyclotron Laboratory( NSCL) at Michigan 
State University via
measurements of correlation terms in nuclear beta decay. The first
project plans a differential measurement of the so-called
polarization-asymmetry correlation in the decay of $^{21}$Na as a tool
to search for deviations from maximal parity violation. The second
is the measurement of a fifth-fold correlation in $^{36}$K decay that
is sensitive to deviations from time reversal invariance. Both
measurements require spin polarized nuclei that can be produced at
the laser spectroscopy and beam polarization facility, BECOLA. The
low energy (maximum 60 keV/q) polarized nuclei will be produced by
first stopping the high energy fragments with suitable "stoppers" and
then transporting the thermal beams towards the BECOLA beamline,
where a collinearly overlapped laser light induces their polarization
by the optical pumping technique.

\subsubsection{The Differential Polarization-Asymmetry Correlation}

Measurements of pseudo-scalar quantities in beta decay can probe
possible deviations from maximal parity violation due, for instance, to
the presence of right-handed bosons with vector and axial couplings or
to the exchange of other bosons with exotic scalar and tensor couplings.

The most stringent tests of maximal parity violation in nuclear beta
decay arise from measurements of polarization-asymmetry correlations
in the decays of $^{107}$In \cite{Sev93} and $^{12}$N \cite{Tho01}.
In comparison with the beta asymmetry parameter and with the longitudinal
beta polarization, this observable offers an enhanced sensitivity to
deviations from maximal parity violation resulting from the combination
of two pseudo-scalar quantities contributing to the decay.
In previous experiments, the nuclear polarization was obtained by
either a low-temperature nuclear orientation technique or by
polarization transfer in a reaction initiated with a polarized beam.
All measurements so far have been carried out at fixed beta particle
energies by selecting a window in the beta energy spectrum with magnetic
spectrometers.

The longitudinal polarization of beta particles emitted from unpolarized
nuclei has a constant sign and also a constant sensitivity to effects
related to partial parity symmetry restoration. Beta particles emitted
from polarized nuclei can, in contrast, exhibit a longitudinal polarization
that changes sign as a function of the beta particle energy. In addition,
the sensitivity to effects associated with parity symmetry restoration
increases at lower beta particle energies. This offers a new window
for tests of maximal parity violation in beta decay, provided the SM
values can be controlled to sufficient accuracy.

A new positron polarimeter is being designed for a differential
measurement of the longitudinal polarization of beta particles emitted
from polarized $^{21}$Na nuclei.

\subsubsection{Search for Time Reversal Violation by a Fifth-Fold Correlation Measurement}

The five-fold correlation
$E_1\vec{J}\cdot(\vec{p}\times\vec{k})(\vec{J}\cdot\vec{k})$,
where $\vec{J}$ is the nuclear spin, $\vec{p}$ is the beta particle
momentum, and $\vec{k}$ is the photon momentum provides a means
to the searches for time reversal violation \cite{Mor57, Cur57, Hol72}
that is complementary to measurements of triple correlations in beta decay.
The fifth-fold correlation is $P$-odd/$T$-odd and it is generally interpreted
in terms of a possible imaginary phase between the vector and axial
couplings. The most precise result so far for such a correlation was
obtained with $^{56}$Co nuclei and yielded $E_1 = -0.011\pm0.22$ \cite{Cal77}.
This provides the weakest constraint compared with the other $T$-violating coefficients
in beta decay. The transition of interest here is an isospin-hindered
Gamow-Teller transition. The contribution from the Fermi matrix element
is finite, though small, due to a breakdown of isospin symmetry or possibly
through a contribution of second-class currents (SCC). If SCC --which
are known to be zero so far \cite{Wil00}-- were the source for the underlying
mechanism for $T$-violation, the $^{56}$Co decay provides an ideal test. However,
no $E_1$ test has been performed so far in a mirror beta decay.

It has been pointed out \cite{You95} that the superallowed decay of
$^{36}$K is a good candidate to test $T$ symmetry by a fifth-fold correlation
measurement.

The low energy, polarized $^{36}$K beam will be implanted in a host
crystal placed under a magnetic field surrounded by a set of high-resolution
germanium detectors. The decay of interest populates a $2^+$, $T = 1$ state
at 6.61 MeV in the $^{36}$Ar daughter with a branching ratio of 42\%.
The excited state in $^{36}$Ar decays to the ground state by emission of
several gamma rays with energies larger than 2 MeV, providing the conditions
needed for a $\beta\gamma$ angular correlation experiment. 

To be competitive with present limits, the new $T$-invariance measurement
will most likely require the higher beam rates like those expected at the
future FRIB facility at MSU.


\subsection{Searching for Tensor Currents in $^6$He}
\label{sec:He6}

At the University of Washington, Seattle, there is a program to search for tensor currents in the decay of $^{6}{\rm He}$. In this case the coefficients $a$ and $b$, neglecting radiative and recoil-order corrections, are given by:
\begin{eqnarray}
a = -\frac{1}{3} \frac{\left( 2 |C_A|^2 -| C_T|^2- |C_T^\prime|^2\right)}{\left( 2 |C_A|^2+| C_T|^2+ |C_T^\prime|^2\right)}~~~ \nonumber \\
b =   \frac{2 C_A \left(C_T+C_T^\prime\right)}{\left( 2 |C_A|^2+| C_T|^2+ |C_T^\prime|^2\right)}.
\nonumber
\end{eqnarray} 

\begin{figure} [htbp]
\begin{center}
\rotatebox{0.}{\resizebox{3.5in}{3.5in}{\includegraphics{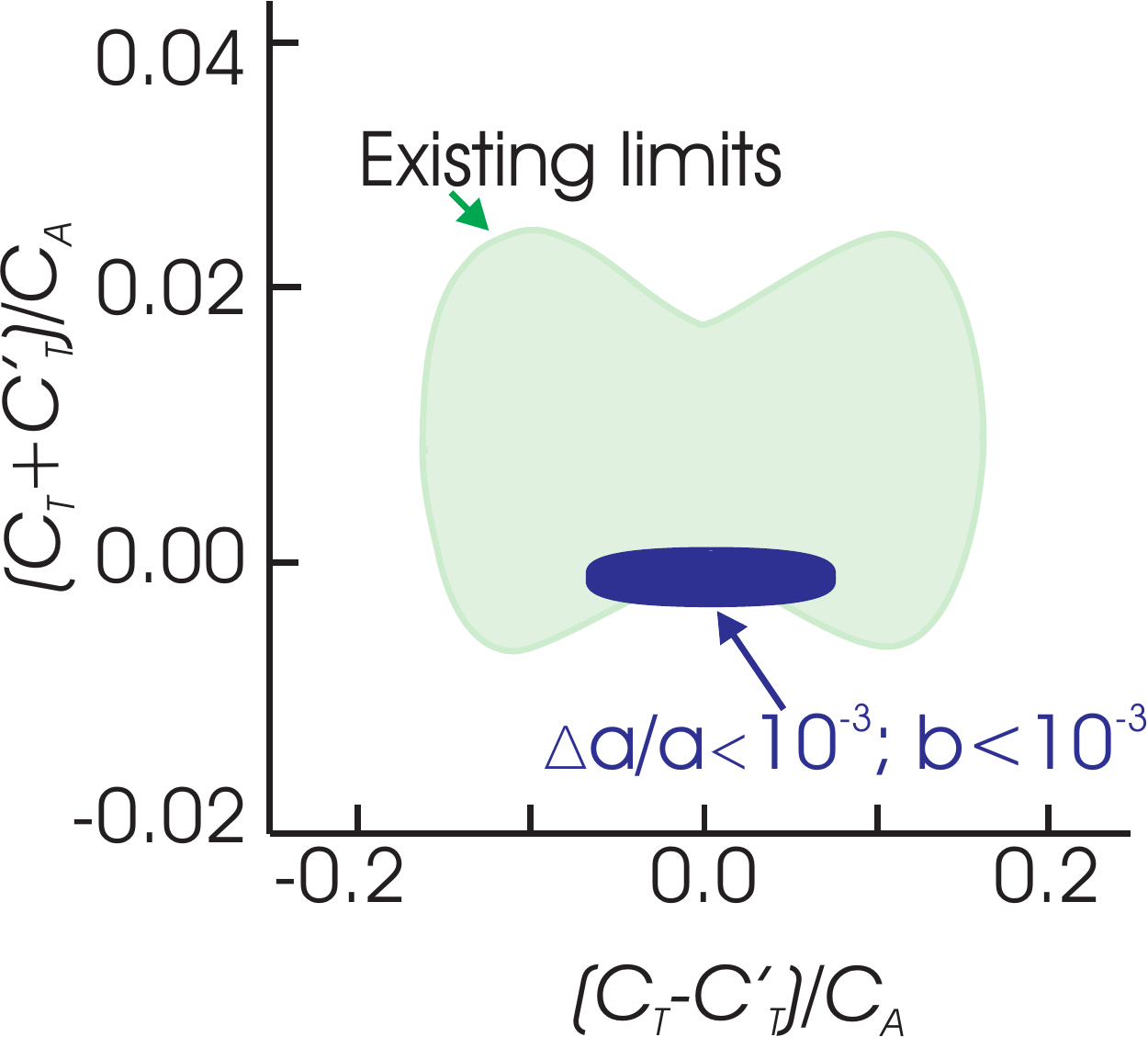}}}
\end{center}
\caption{95\% confidence intervals: presently allowed region (from ~\cite{se:06}) in green and,   in blue, the limits one would get with measurements of \protect $\Delta a/a < 10^{-3}$ and $\Delta b < 10^{-3}$. }
\label{fig:He6-limits}
\end{figure}


$a$ is sensitive {\em quadratically} to tensor currents of either chirality, while $b$ is sensitive {\em linearly} to tensor currents, but only to those with no left-handed anti-neutrinos (or right-handed neutrinos). Here we aim at detecting $a$ with a relative precision of $\sim 0.1$\% and $b$ with absolute precision of $\sim 10^{-3}$. Because of its linear dependence on the tensor couplings, $b$ is more sensitive to tensor currents. However, because it is sensitive to only one chirality, a test concentrating solely on $b$ would leave unchecked a large region of the parameter space with a particular bias. Although we do not have a concrete scenario for new physics that would show up with neutrinos with helicities opposite to that of the Standard Model, this possibility cannot be presently excluded.
Fig.~\ref{fig:He6-limits} shows the present limits and those that we aim to have from this work.

In the $^{6}{\rm He}$ case, with a $0^+ \rightarrow 1^+$ beta decay, the connection between $a,\;b$ and tensor currents is {\em direct}, as opposed to the case of mirror transitions, such as neutron decay, where $a$ and $b$ are sensitive, in addition, to the ratio of axial to vector current couplings and matrix elements plus potentially existing scalar currents. The decay of $^6{\rm He}$ presents an excellent opportunity for these searches. The large $Q$-value allows for an exploration of the shape of the spectrum over a broad energy range. By comparison, neutron beta decay presents an endpoint smaller by a factor of about 3.5, and the potential problem that capture of neutrons in the environment will generate additional backgrounds. Being the lightest radioactive nucleus with a pure GT decay, the recoiling nucleus shows high sensitivity to the momentum carried by the leptons, which is good for a determination of the correlation coefficient.
The system produces copious amounts of $^6{\rm He}$ \cite{kn:11}: more than $10^9$ atoms per second delivered to a low-background room.

Part of the program consists of using laser traps \cite{mu:11} to determine the electron-anti-neutrino correlation by coincidence
detection of the beta and $^6{\rm Li}$ recoil ion. 
There is already have a working MOT trap with $^{4}{\rm He}$ atoms and hope to trap $^{6}{\rm He}$ atoms by the end of 2011. 
The laser trapping requires exciting He to a metastable state at $\sim 20$ eV and one expects to count only about 10 e-Li coincidences 
per second in this mode.
Another part of this program is to determine the shape of the beta spectrum, for which use of  the laser-trap system is not planned.
An ISOL-type FRIB could provide up to three orders of magnitude higher production rates, allowing for further improvements in both the $e,{\overline \nu}$ correlation and the spectrum-shape measurement, which could then be done with trapped atoms.

\subsection{$\beta$-Decay with Neutral Atom Traps I}
\label{sec:trapsI}

Nuclear $\beta$ decay correlation experiments helped establish the nature of the weak interaction, a theory with spin-1 bosons coupling only to left-handed leptons. Such measurements can still assist particle physics by determining first-generation lepton-quark couplings of possible new particles, but must reach excellent accuracy of 0.1\%. Using atom trap technology, the TRIUMF Neutral Atom
Trap collaboration (TRIUMF, Texas A\&M U., U. Manitoba, Tel Aviv U., 
and U. British Columbia) is poised to complete two experimental programs 
reaching this accuracy in the next three years.

  Confined in a 1 mm-sized cloud, the nuclei undergo beta decay, producing three products: a $\beta^+$, a $\nu$, and a recoiling daughter nucleus. 
The daughter nucleus has very little kinetic energy and would stop in a nm of 
material, but it freely escapes the trap. By detecting it in coincidence with 
the $\beta^+$,  the momentum and angular distribution of the
$\nu$ can be measured~\cite{behr}. 

The atom trap methods provide many unique tools to determine 
experimental systematics.
Kinematic redundancy in most of the data set allows many
consistency tests; for example, the construction of the
$\beta$ momentum from other observables 
tests simulations of the detector response. 
One measures the cloud dimensions and
temperature by photoionizing a small fraction of the atoms and measuring their time of flight
(TOF) and position on the same MCP that measures the nuclear recoils. 
Detection of shakeoff atomic electrons in coincidence provides a high-efficiency TOF trigger and suppresses backgrounds~\cite{pitcairn}.

The process to spin-polarize nuclei and measure the polarization {\it in situ}
of the decaying species by atomic methods independent of the nuclear decays has been established. 
A measurement of the $\nu$ asymmetry $B_{\nu}$ at 3\% accuracy~\cite{melconian} has been published.
The goals of our spin correlation program include a simultaneous 
measurement of $\beta$ and recoil asymmetries 
$A_\beta$ and $A_{\rm recoil}$ at part-per-thousand
sensitivity, and eventually $B_\nu$ at 0.3\%.  
For example, a week of counting of $A_{\rm recoil}$ would have statistical sensitivity
to $C_t+C_t'$ at 0.002. 

Limits on scalar interactions coupling to the first generation of particles could be improved by measuring the $\beta$-$\nu$ correlation in the pure Fermi decay of $^{38{\rm m}}$K~\cite{gorelov}. 
The complete angular acceptance for recoils we have developed will minimize key systematic errors in our upgraded experiment, with a goal of reaching 0.1\% accuracy in $a$ and
possibly $b_{\rm Fierz}$. We determine systematic errors for this experiment from statistics-limited kinematic observables that are independent of the angular correlation.

 Concentrating on isobaric analog decays allows determination of small
recoil-order Standard Model corrections from the electromagnetic moments of 
the nuclei. The Gamow-Teller/Fermi ratio in $^{37}$K can be well-determined from measurements of the lifetime and branching ratio, as the community has achieved a reliable set of techniques for these experiments.

If successful, our correlation experiments would be complementary with constraints from radiative $\pi$ decay~\cite{bhat} 
and indirect effective field theory (EFT) dependent constraints from $\pi$ to e $\nu$~\cite{campbell}, 
and would begin
to allow sensitivity to SUSY left-right sfermion mixing~\cite{profumo}.

It is natural in the trap to measure the time-reversal violating
D $ \vec{I} \cdot \vec{p_\beta} X \vec{p_\nu}$. We have a conceptual design
for a dedicated geometry with 20\% $\beta$ efficiency (it helps that the polarized light comes
at 90$^{\rm o}$ to the $\beta$ detectors helps) so
 2x10$^{-4}$ statistical
error/week can be achieved.
The best measurement in the decay of the
neutron was recently published by
emiT, achieving 2x10$^{-4}$ sensitivity~\cite{emiT}.
However, a recent effective field theory calculation has identified a
mechanism for the electric dipole moment of the neutron 
to constrain leptoquark contributions to D (previously considered 
safe from EDM's~\cite{herczeg}), and if
there are no cancellations would imply D $<$ 3x10$^{-5}$~\cite{tulinD}. 
Since there are many possible contributions to $CP$ violation from new physics,
D could still be complementary. We will study time-reversal systematics in our present
$^{37}$K geometry.

\subsection{$\beta$-Decay with Neutral Atom Traps II}
\label{sec:trapsII}

Recent advances in the techniques of atom and ion trapping have opened up a new 
vista in precision $\beta$ decay studies due to the near-textbook source they 
provide: very cold ($\lesssim1$~mK) and localized ($\lesssim1$~cm$^3$), with 
an open geometry where the daughter particles escape with negligible 
distortions to their momenta.  

{\bf Atom traps:}\ \ 
Magneto-optical traps (MOTs) have demonstrated the ability to measure the 
angular distribution of short-lived radioactive neutral atoms. 
Experiments using MOTs have placed stringent direct limits on a possible 
fundamental scalar current in the charged weak interaction via a precise 
measurement of the $\beta-\nu$ correlation parameter, 
$a_{\beta\nu}$~\cite{gorelovPRL,scielzo}.  These experiments continue to improve, 
along with others being developed to extend to other cases (for example, see 
A.~Garc\'ia \emph{et al.}'s contribution to this workshop on searching for 
tensor interactions using trapped $^6$He).  Techniques are also being 
developed by the \trinat{} collaboration at \triumf{} to highly polarize 
laser-cooled atoms via optical pumping.  A proof-of-principle experiment 
measured the neutrino asymmetry parameter in \tsups{37}K~\cite{melconianPLB}, 
with an improved version planned to take beam at \triumf{} in the summer of 
2012.  The contributions by J.A.~Behr, R.J.~Holt and L.A.~Orozco to this 
workshop describe other physics opportunities available using MOTs.

{\bf Ion traps:}\ \ 
Penning traps of ions are best known for the incredible precision with 
which they can measure masses: relative uncertainties of $\Delta M/M\simeq
10^{-11}$ have been reported on stable species~\cite{pritchardNature}, and 
$\simeq10^{-8}$ for very short-lived ($\gtrsim 10$~ms) exotic 
ions~\cite{blaumPRL}.  These mass measurements have impacts in many fields 
of science, including fundamental physics research (CKM unitarity, testing 
nuclear models, correlation studies, etc.).  Penning traps are also used 
in other applications, including laser spectroscopy, QED effects, 
electron-capture studies and the astrophysical r-process, to name a few.

\begin{figure} [htbp]
\begin{center}
\rotatebox{0.}{\resizebox{3.5in}{3.5in}{\includegraphics{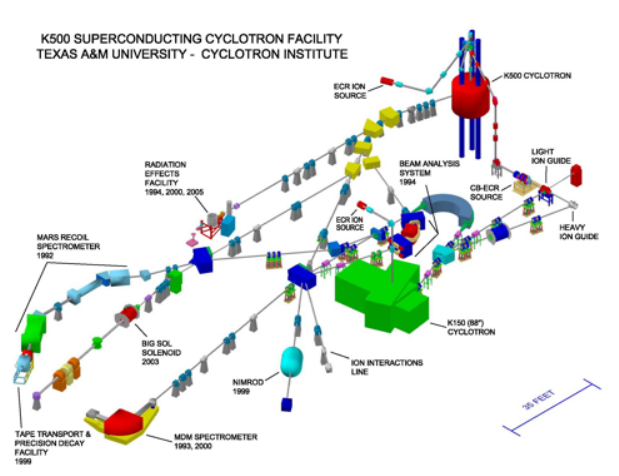}}}
\end{center}
\caption{Floorplan of the Cyclotron Institute showing the accelerators 
  and equipment, including the T{\tiny REX} upgrade components. }
\label{fig1a}
\end{figure}

{\bf Opportunities at \trex:}\ \ 
The group at Texas A\&M University is in the process of constructing a new 
double-Penning trap facility, \tamutrap, which will take advantage of the 
radioactive ion beam capabilities of the upgraded Cyclotron Institute 
facility, \trex~\cite{cyclotron-upgrade}.  The layout of the institute's 
cyclotrons and experimental equipment is shown in Fig.~\ref{fig1a}, where the 
components that are currently being built as part of the \trex{} upgrade are: 
re-commissioning the K150 cyclotron to deliver high-intensity light particle 
and heavy ion beams; the light and heavy ion guide systems; the charge-breeding 
electron cyclotron resonance (ECR) ion source and coupling of it to the K500 cyclotron, to provide high 
quality re-accelerated rare beams of both neutron and proton rich isotopes 
in the 5 to 50~MeV/u range.

\begin{figure} [htbp]
\begin{center}
\rotatebox{0.}{\resizebox{3.0in}{3.0in}{\includegraphics{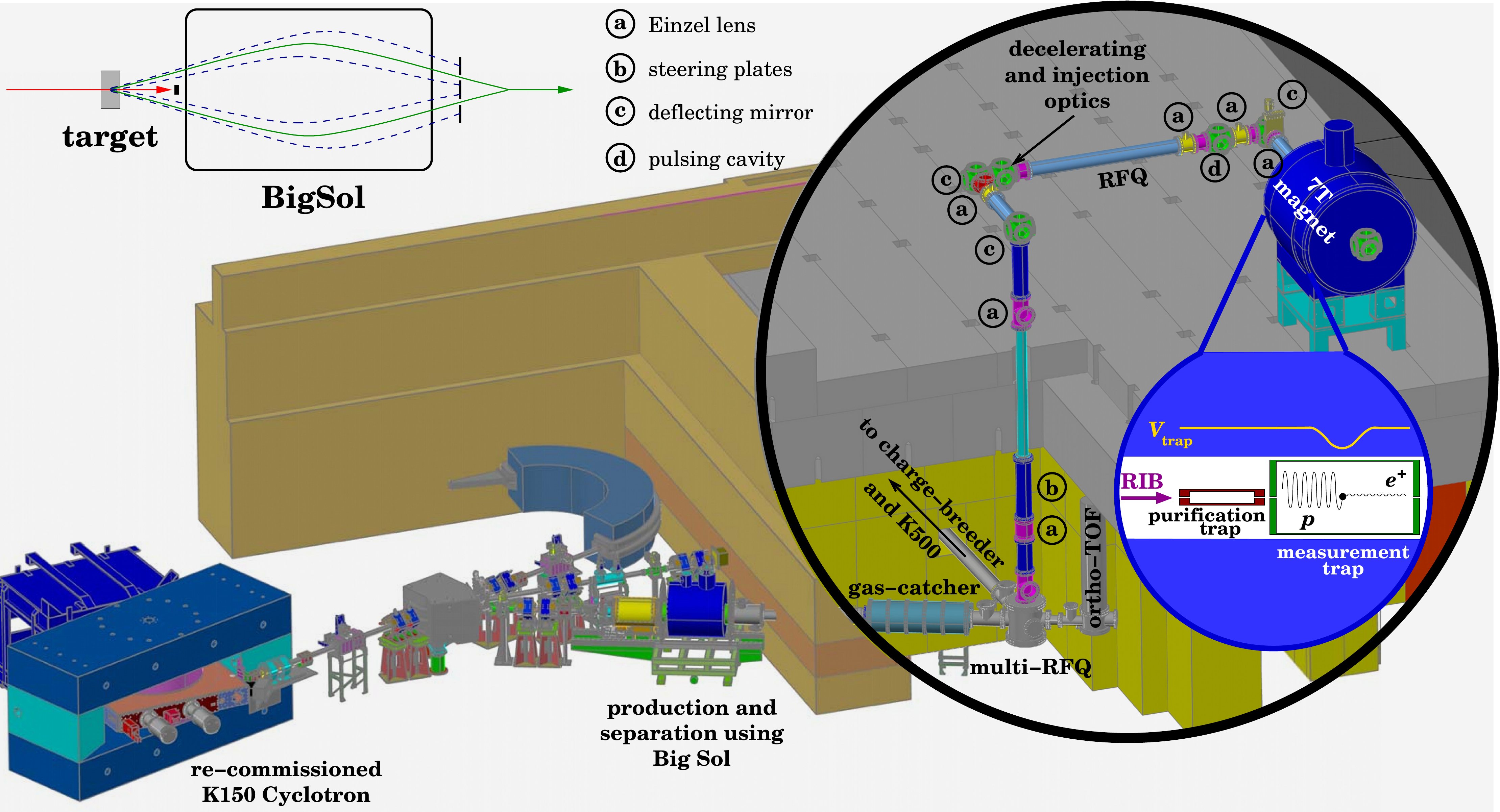}}}
\end{center}
\caption{Coupling of the K150 to the \tamutrap{} facility, including a 
  schematic of the double-trap system and principle behind the $a_{\beta\nu}$ 
  measurement. }
\label{fig1b}
\end{figure}


Figure~\ref{fig1b} shows the plans for the Penning trap facility in relation 
to the \trex{} upgrade.  Radioactive beams for \tamutrap{} will be produced 
using the K150 cyclotron which will provide a high-intensity primary beam 
that will react with a target in front of ``BigSol'', a large-acceptance 
7-Tesla solenoid which will separate the desired products from deap inelastic 
reactions.  An Argonne National Laboratory type gas-catcher~\cite{gas-catcher} 
will collect these products and transport them via a dedicated low-energy 
beamline to \tamutrap{} at $\sim10$~keV (or alternatively to the 
charge-breeding ECR for post-acceleration in the K500 cyclotron). The ions 
will be cooled and bunched using a segmented gas-filled RFQ and injected at 
80~eV into the double-Penning trap with time and energy spreads of 
$1.2-1.6~\mu$s and $6-10$~eV, respectively.

Both Penning traps will be housed in the 210-mm-bore 7-Tesla superconducting 
magnet, which has been purchased from Agilent Technologies and has  
demonstrated to have better than $4$~ppm homogeneity at full field.  The first 
trap will be a gas-filled ($\lesssim10^{-4}$~mbar) cylindrical Penning trap to 
(optionally) allow for further purification of the rare ion beam following the 
gas-catcher. The 2\tsups{nd} cylindrical ``measurement'' trap will be unique 
in that it will have the largest diameter (180~mm) of any existing Penning 
trap. The purpose of this large-diameter bore is for the flagship program 
which will perform $\beta-\nu$ correlation studies and $ft$-value measurements 
on \zerotozero{} transitions of the short-lived $\beta$-delayed proton emitters
\tsups{20}Mg, \tsups{24}Si, \tsups{28}S, \tsups{32}Ar and \tsups{36}Ca.  Measurements  
will start with \tsups{32}Ar since its decay has already been studied in 
detail~\cite{adelberger,bhattacharya}, but the others have similar decay 
schemes and so can be studied in a similar way.  The Larmour radii of the 
$\beta$s from these decays ($\lesssim5$~mm) is much less than the 
protons, allowing good spatial separation of the two when detected by 
position-sensitive Si detectors placed at either end of the measurement trap 
(see the schematic zoom-in of the magnet in the bottom-right of 
Fig.~\ref{fig1b}).  
The large bore of the magnet allows for complete radial confinement of up to 
4.3~MeV protons, which spans the energies of interest produced in these decays. 
The shape of the proton's energy spectrum depends on the value of $a_{\beta\nu}$ 
and thus the $\beta-\nu$ correlation can be investigated in a manner similar 
to that done by Adelberger \emph{et al.} at ISOLDE~\cite{adelberger} where 
a precision of 0.5\% in $a_{\beta\nu}$ was reached; one advantage  
over their experiment is the observation of the $\beta$'s energy, as well as 
separation of events where the $\beta$ and proton were emitted in the 
same/opposite hemispheres, which will help reduce systematics.  

In addition to the superallowed program, 
the system is being designed to be flexible and allow for other fundamental 
physics studies, including the ability to perform precision mass measurements.  
Although not a user facility, the Cyclotron Institute has a long history 
of collaborating with outside users and we expect that with our existing and 
planned upgrade for extending our RIB capabilities, that collaborative efforts 
will continue to be formed with groups interested in using our facilities.

{\bf The intensity frontier:}\ \ 
Both atom and ion traps provide extremely clean sources, so experiments can 
be performed with relatively small sample sizes owing to the very large 
signal-to-noise ratio.  However, high intensities with relatively long access 
times to radioactive beams would be extremely advantageous to these 
experiments:  most are systematics-limited, where the systematics 
are themselves statistics limited; \emph{i.e.}\ systematic uncertainties could 
be considerably reduced if their sources could be investigated and minimized by 
using dedicated beamtime to characterize and quantify them.  The \trex{} 
facility at Texas A\&M University is not a user facility and so can offer 
better availability of beamtimes than, for example, F{\small RIB} or \triumf; 
however, the larger facilities can provide greater intensities.  Ideally, one 
would like to have a facility that can provide both high intensities and 
long accessibility of beamtimes.  Low-energy nuclear physics programs using 
both MOTs and Penning traps would be able to capitalize on such a facility and 
meaningfully add to probes of physics beyond the Standard Model.

\subsection{High Intensity Neutron Sources}
\label{sec:nsources}

Experimentation using slow neutrons is an integral part of studies spanning fields as diverse as nuclear and particle physics, fundamental symmetries, astrophysics and cosmology, and 
gravitation.
The field possesses a coherence that derives from the unique properties of the neutron as an electrically neutral, strongly interacting, long-lived unstable particle that can be used either as a probe or as an object of study.  Experiments include measurement of neutron-decay parameters as tests of the Standard Model,  the use of parity violation to isolate the weak interaction between nucleons, and searches for a source of time reversal violation beyond the SM. These experiments provide information that is complementary to that available from accelerator-based nuclear and particle physics facilities. Neutron physics measurements also address questions in astrophysics and cosmology. The theory of Big Bang Nucleosynthesis needs the neutron lifetime and the vector and axial vector weak couplings as input, and neutron cross sections on nuclei are necessary for a quantitative understanding of element creation in the universe.

The approach to this class of experiments requires performing measurements with a high degree of statistical precision. A large fraction of the experiments using slow neutrons are statistically limited. Thus, in order for progress to be made, it is critical that more intense sources of neutrons be made available.  The experiments that probe these questions are largely performed at reactors or spallation sources that are able to generate large densities of cold or ultra-cold neutrons. For neutron decay experiments, there is a novel effort underway to create a high flux of electrons and protons from neutrons decaying upstream in a neutron guide and transport them to the measurement region.
While it is not possible to survey all the experiments that would benefit from high neutron densities, this brief note mentions just a few areas of investigation that would benefit greatly from higher fluxes of cold neutrons.  The reader is referred to the more specialized papers and reviews listed in the references for more details.

Neutron decay is an important process for the investigation of the Standard Model of electroweak interactions. As the prototypical beta decay, it is sensitive to certain SM extensions in the charged-current electroweak sector.  Neutron decay can determine the CKM matrix element $|V_\text{ud}|$ through increasingly precise measurements of the neutron lifetime and the decay correlation coefficients. The best determination of $|V_\text{ud}|$ comes from superallowed nuclear decays~\cite{PDG10}, but neutron beta decay offers a somewhat cleaner theoretical environment than the superallowed transitions due to the absence of other nucleons (although some radiative corrections are common to both systems). Presently, the experimental uncertainties from the neutron measurements are significantly larger than those from superallowed decays and arise from systematic uncertainties; more specifically, there are disagreements among experiments. Once those systematic effects are sorted out, modest improvements in the statistical precision will enable the precision of neutron experiments to be comparable to that obtained from the nuclear decays.

Searches for violations of time-reversal symmetry and/or $CP$ symmetry address issues that lie at the heart of cosmology and particle physics. The next generation of neutron EDM searches, which plan to achieve sensitivities of $10^{-27}$\,\ecm\ to $10^{-28}$\,\ecm, is the most important of a class of experiments aiming to search for new physics in the $T$-violating sector. Thus far, all observations of $CP$ violation can be entirely accounted for by a phase in the CKM matrix; however, this phase is insufficient to account for the known baryon asymmetry in the context of big bang cosmology and there is good reason to search for $CP$ and the related time-reversal violation in other systems.  Because neutron beta-decay is theoretically straightforward and final state interactions are generally small and calculable,
the neutron is a clean system in which to look for the effects of new physics.  Two $T$-odd triple correlations have recently been measured, $D\vec{\sigma}_{n}\cdot \bf{{p_e}}\times\bf{{p_\nu}}$ and $R\vec{\sigma}_{n}\cdot {\sigma_e}\times\bf{{p_e}}$, where $\vec{\sigma_n}$ is the neutron spin, $\vec{\sigma_e}$ is the neutron electron spin and $\bf{{p}}$ are the decay product momenta.  Both experiments are currently believed to be statistically limited; thus it is interesting to consider the potential of higher fluence sources.
 
The $D$ coefficient is the most sensitive probe of the phase, $\phi_{av}$, between the axial-vector and vector currents and is also sensitive to scalar and tensor interactions that could arise due to beyond Standard Model physics. The most recent result, $D = (-0.96\pm1.89 (stat)\pm1.01(sys))\times10^{-4}$~\cite{Mumm11}, represents the most sensitive test of time-reversal invariance in beta decay. Within the Standard Model, the result can be interpreted as a measure of the phase $\phi_{av} = (180.013\pm0.028)^\circ$.  The $D$ coefficient can also be related to the neutron EDM.
significantly constrains scalar couplings beyond previous limits.   The systematic uncertainty, already a factor of three below the statistical uncertainty, is dominated by background subtraction, the characterization of which should improve with improved statistics.  Any increases in available neutron fluence thus could lead to significantly improved limits on scalar couplings.

The rare decay $n \rightarrow \text{H} + \bar{\nu}$ has been discussed in several articles going back to the early sixties.
The branching ratio for this rare mode of decay is $4 \times 10^{-6}$ of the total neutron $\beta$-decay rate, making it exceedingly challenging to detect.  The hyperfine-state population of hydrogen after the bound-$\beta$-decay of the neutron directly yields the neutrino left-handedness or a possible right-handed admixture and possible small scalar and tensor contributions to the weak force.
The detection of the neutral hydrogen atoms and the analysis of the hyperfine states can be accomplished using straightforward atomic physics techniques.  The elegance and power of this approach makes it a very attractive way to test the Standard Model. A group
is attempting such a measurement and hopes to improve current limits by a factor of 10. This is an experiment for which a high-intensity neutron source would be a great asset.

The neutron decays with the emission of a soft photon. At present, experiments are only probing the inner bremsstrahlung component of this decay due, in part, to the low rate of events that one obtains at the low energy photon threshold.
With higher-intensity neutron sources, it would be interesting to pursue investigations into testing nonleading order corrections to the decay. In addition, there is recent work
studying radiative beta decay in the context of the Harvey, Hill, and Hill interaction. In particular, they are looking at the triple correlation $\bf{l_p} \cdot (l_e \times k)$, where $\bf{l_p}$ and $\bf{l_e}$ are the proton and electron momenta, respectively, and $\bf k$ is the photon momentum, as a possible source of time-reversal violation.

The last decade has also seen qualitative advances in both the quantitative understanding of nuclei, especially few-body systems, and the connection between nuclear physics and quantum chromodynamics (QCD). Low energy properties of nucleons and nuclei, such as weak interactions in n-A systems, low energy n-A scattering amplitudes, and the internal electromagnetic structure of the neutron (its electric polarizability and charge radius) are becoming calculable in the SM despite the strongly interacting nature of these systems. These theoretical developments are motivating renewed experimental activity to measure  undetermined low energy properties such as the weak interaction amplitudes between nucleons and to improve the precision of other low energy neutron measurements. The ultimate goal is to illuminate the strongly interacting ground state of QCD, the most poorly understood sector of the SM, and to clarify the connection between the strong interaction at the hadron level and the physics of nuclei.  

Independent of the theoretical model, several experimental approaches are required to narrow the range of the predictions. There are two mature experimental efforts that are able to provide input. One is an experiment to measure the parity-violating spin rotation of a polarized neutron beam in a liquid helium target,
and the other is an experiment to measure the gamma-ray asymmetry in the process $p(n,\gamma)D$.
The experiments are largely complementary in terms of their sensitivity to the effective field theory parameters (or meson exchange amplitudes). The uncertainties from initial results from both experiments are dominated by statistics and require running at more intense sources of cold neutrons to improve the ultimate precision. Improved measurements from both experiments are essential to produce parameters that can be extracted in a theoretically clean manner.

\section{Neutral Currents}
\label{sec:Neutral}

Experiments using intense beams of polarized electrons scattering from fixed targets, as well as those exploiting parity-forbidden atomic transitions, have played a vital role in developing and testing the electroweak sector of the Standard Model as well as in probing novel aspects of hadron and nuclear structure (for a review, see~\cite{Musolf:1993tb,Beck:2001dz,Beise:2004py}). Most recently, a program of parity-violating electron scattering experiments at MIT-Bates, Mainz, and Jefferson Laboratory have yielded stringent limits on contributions from the strange quark sea to the nucleon's electromagnetic properties, while an experiment with parity-violating M\o ller scattering at SLAC has provided the most stringent test to date of the running of the weak mixing angle $\theta_W$ below the weak scale. Similarly powerful measurements have been carried out with atomic cesium, yielding the most precise determination of the nuclear weak charge and intriguing indications of the so-called nuclear ``anapole moment."

These achievements have built on the steady improvements in experimental sensitivity since the pioneering measurements of parity-violating, deep inelastic electron-deuteron scattering~\cite{SMB} and atomic parity violation in the 1970s, together with refinements of the theoretical interpretation. The frontier of this field now promises a unique capability to probe possible new physics at the TeV scale, as well as explore previously inaccessible features of nucleon and nuclear structure. At the same time, the possibility to carry out a sensitive search for charged lepton flavor violation with unpolarized beams at an electron-ion collider (EIC) appears increasingly feasible. Below, we review some of the present efforts and future prospects for studies that exploit neutral current interactions with  electrons. 

To set the stage, we introduce the low-energy effective Lagrangian characterizing the parity-violating neutral weak electron-quark interaction:

\begin{equation}
\mathcal{L}^{PV}=\frac{G_F}{\sqrt{2}}
[\overline e\gamma^\mu\gamma_5e(C_{1u}\overline u\gamma_\mu u+C_{1d}\overline
  d\gamma_\mu d)
+\overline e\gamma^\mu e(C_{2u}\overline u\gamma_\mu\gamma_5 u+C_{2d}\overline
  d\gamma_\mu\gamma_5 d)].
\end{equation}

In the Standard Model, the coefficients $C_{1q}$ and $C_{2q}$ can be predicted with high precision and depend critically on 
$\sin^2\theta_W$. A similar expression obtains for the parity-violating electron-electron interaction. One goal of the parity-violation experiments is to test these predictions. As with the muon anomalous magnetic moment, any deviation from the Standard Model predictions  would indicate the presence of new physics to the extent that the theoretical predictions are sufficiently robust. Given the present level of experimental and theoretical uncertainties, the present generation of parity violation experiments are able to probe for new TeV scale physics. 

From a complementary perspective, one may assume the Standard Model values for the $C_{iq}$ are correct and use the interactions in 
Eq.~(7.5) to access properties of the nucleon and nuclei that are not readily probed by the purely electromagnetic interaction. Indeed, the determination of strange quark contributions relies on different combinations of the light quark currents that enter electromagnetic and neutral weak currents. As discussed below, a number of additional interesting aspects of nucleon and nuclear structure can be uncovered using the parity-violating electron-quark interaction. 

In much of the discussion below, the object of interest is the parity-violating asymmetry
\begin{equation}
\label{eq:apv}
A_{PV}=\frac{\sigma_{+}-\sigma_{-}}{\sigma_{+}+\sigma_{-}}\ \ \ ,
\end{equation} 
where $\sigma_{+}$ ($\sigma_{-}$) denotes the cross section for scattering by longitudinally polarized electrons with positive (negative) helicity from an unpolarized target. The corresponding parity-violating observable in atomic physics depends on the specific setup, so we defer a description to the discussion of atomic experiments below. 

\subsection{Proton Weak Charge}
\label{sec:Qweak}

\begin{figure}
\begin{center}
\rotatebox{0.}{\resizebox{3.7in}{3.3in}{\includegraphics{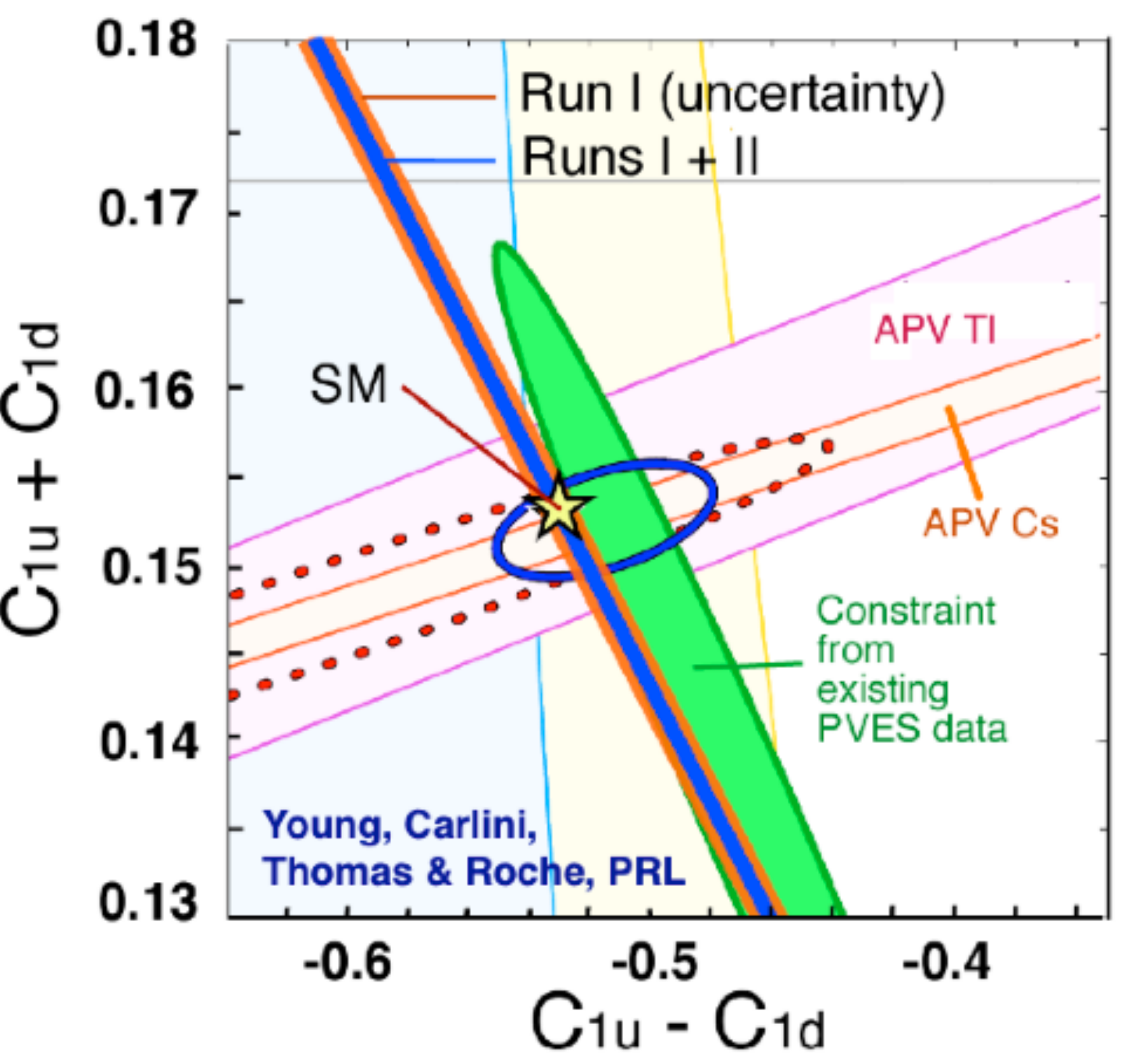}}}
\end{center}
\caption{ Knowledge of the neutral-weak effective coupling constants. The dotted contour displays the previous experimental limits (95\% CL) reported in the Particle Data Group Review\cite{Q2} together with the prediction of the Standard Model (yellow star). The filled ellipse denotes the present constraint provided by recent high precision PVES scattering measurements on hydrogen, deuterium, and helium targets (at $1\sigma$), while the solid contour (95\% CL) indicates the full constraint obtained by combining all results\cite{Q3}. All other experimental limits shown are displayed at $1\sigma$. The striking improvement possible from the future Jefferson Laboratory $Q^{p}_{W}$ measurement is shown as the blue line for Standard Model predictions.}
\label{searches-3}
\end{figure}

The Qweak collaboration\cite{Q1} is conducting the first precision measurement of the weak charge of the proton, $Q^{p}_W$, given in terms of the $C_{1q}$ as
\begin{equation}
Q_W^p = -2\left(2C_{1u}+C_{1d}\right)\ \ \ .
\end{equation} 
At leading order in the Standard Model, $Q_W^p=1-4\sin^2\theta_W$. Inclusion of higher-order electroweak  corrections leads to a Standard Model prediction with $\mathcal{O}(1\%)$ theoretical error\cite{erler,Erler:2004in}.
This experiment is being performed at Jefferson Laboratory, building on the technical advances made in the laboratory's parity violation program and using the results of earlier measurements to constrain hadronic corrections. The experiment measures the parity-violating longitudinal analyzing power in e-p elastic scattering at $Q^{2}$ = 0.026 $(GeV/c)^{2}$ employing 180~$\mu A$'s of 86\%  polarized electrons on a 0.35~m long liquid hydrogen target. The measurement will determine the weak charge of the proton with about 4.1\% combined statistical and systematic errors. This corresponds to constraints on parity-violating new physics at a mass scale of 2.3 TeV at the 95\% confidence level. This also allows $\sin^{2}\theta_{W}$ to be determined to 0.3\% accuracy, providing a competitive measurement of the running of the mixing angle.  In combination with other parity-violation measurements, a high-precision determination of the weak couplings $C_{1u}$ and $C_{1d}$ significantly improves on the present knowledge as shown in Figure~\ref{searches-3}. 
Letting $Q^2$ denote the four-momentum transfer, $\tau = Q^{2}/4M^{2}$ where $M$ is the proton mass  and $\theta$ the laboratory electron scattering angle, for forward-angle scattering where $\theta \rightarrow 0$, $\epsilon \rightarrow 1$, and $\tau << 1$, the asymmetry can be written as\cite{Musolf:1993tb}:
\begin{equation}\label{ffequation}
A_{PV} = 
\left[- G_F Q^2 \over 4 \pi \alpha \sqrt{2}\right] 
\left[  Q_{W}^{p} + F^{p}(Q^{2},\theta,E) \right] 
\rightarrow 
\left[- G_F Q^2 \over 4 \pi \alpha \sqrt{2}\right] 
\left[   Q_{W}^{p} + Q^{2} B(Q^{2})+C(E)\right]\,,
\end{equation}
where $F^{p}$ is a form factor that includes a dependence on the beam energy $E$ beyond the Born approximation. The first term, proportional to $Q^2$, is for a point-like proton. The second term $B(Q^{2})$, proportional to $Q^4$, is the leading term in the nucleon structure defined in terms of neutron and proton electromagnetic and weak form factors. Ideally one would like to measure at a low enough $Q^2$ that the proton would look like a point and hadronic corrections would be negligible.  An accurate measurement of $\sin^{2}\theta_{W}${} thus requires higher-order, yet significant, corrections for nucleon structure. Nucleon structure contributions in $B(Q^{2})$ can be suppressed by going to lower momentum transfer and energy. The numerical value of $B(Q^{2})$ has been constrained experimentally by extrapolation from existing forward angle parity-violating data at higher $Q^{2}$. The importance of the additional $E$-dependent  contributions arising from the exchange of two vector bosons between the electron and proton remains a topic of ongoing theoretical scrutiny. 


\subsubsection{Possibility of a Second Generation Measurement at the JLab FEL Accelerator Complex}

Initial simulations indicate that it is technically feasible to perform  $\sim$2\% ultra-low $Q^2$ measurement of $Q^{p}_W$ by using the existing apparatus with approximately 0.5mA of 200 MeV beam potentially available at the JLab FEL accelerator complex. The value of such a measurement would really come from its impact on a global fit. It would be another point, at a much lower $Q^2$, but with different systematic and theoretical uncertainties, rather than simply smaller ones. This would allow a better global extraction of a final result. Such a measurement would have approximately 100 times the rate of ``Qweak-1" but with an average $Q^2$ about 10 times lower, implying a smaller figure of merit (FOM). Another advantage in FOM would come from the suppression of the dilution terms (magnetic moment, plus any strange quark, etc.) and considerably smaller $E$-dependent hadronic effects. 

\begin{figure} [h]
\begin{center}
\rotatebox{0.}{\resizebox{7.0in}{2.5in}{\includegraphics{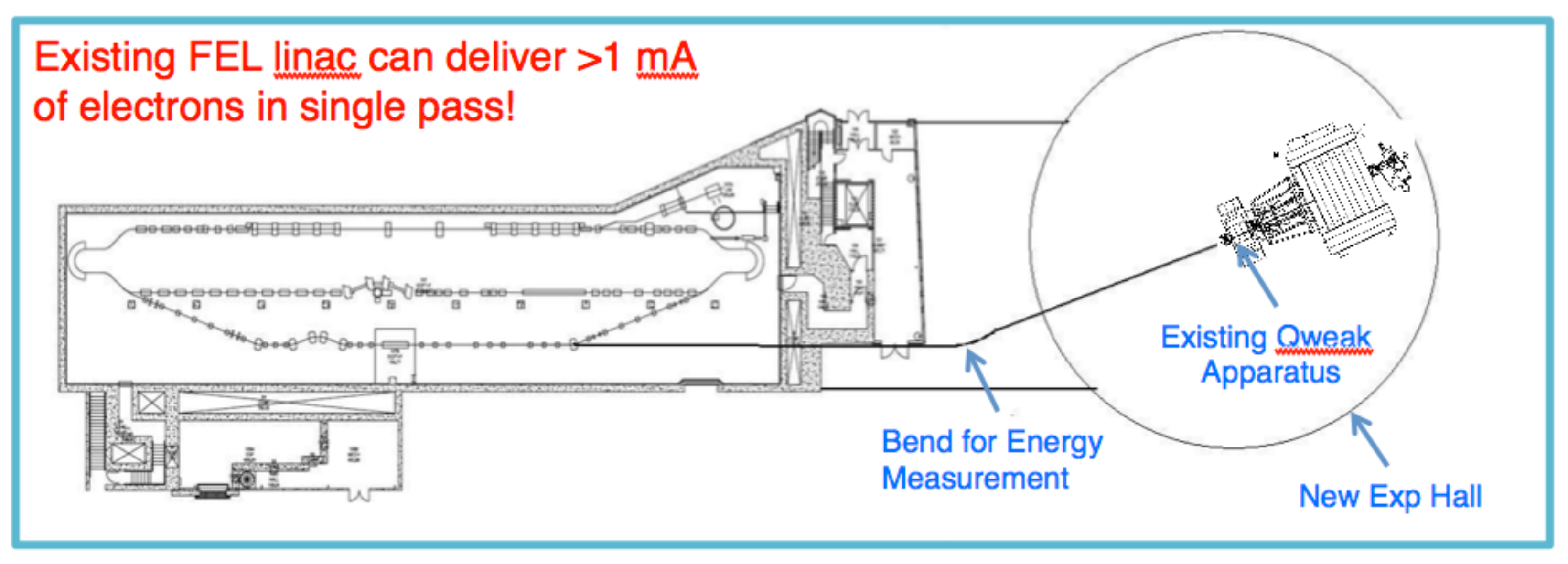}}}
\end{center}
\caption{ The basic layout of a second-generation ultra-low $Q^2$ measurement of the proton's weak charge at the existing JLab FEL accelerator complex.}
\label{FEL}
\end{figure}

A significant point to note is that the existing Qweak toroidal spectrometer design will not produce magnetic fields strong enough to polarize the proton target, which is essential for such a low $Q^2$ measurement. The existing cryogenic target system will work up to ~0.2 mA and perhaps much more. It also appears feasible to augment the experimental technique to help suppress target boiling noise at these higher beam currents by increasing the helicity flip rate to perhaps 2 KHz and by using an radio-frequency beam current cavity (BCM) downstream of the target as a transmitted beam current monitor. A possible experimental layout at the JLab FEL accelerator complex is shown in Figure~\ref{FEL} .

Realizing this capability would require the construction of a new endstation (similar in size to the existing Hall B at JLab), the addition of a polarized injector, beam polarimetry and some upgrade work on the FEL accelerator.  If the existing Qweak apparatus were simply used with 200 MeV beam it would have the following characteristics: $Q^2$ = 7 x $10^{-4}$ $(GeV/c)^2$, with an effective detector rate of 1.5~GHz/$\mu$A of beam incident on the hydrogen target. The layout shown also avoids major modifications to the installed optical beamline at the FEL. A suitable transport line would consist of 3 triplets between the existing FEL and wall, a 4 period FODO arc (using the GW dipoles now in Lab 1 for LIPSS), and then a scaled clone of, say, the transport into Hall C.





\subsection{Parity Violating Deep Inelastic Scattering}
\label{sec:pvdis}


At JLab, an experiment to measure parity violation in the deep
inelastic scattering (PVDIS) of polarized 11 GeV electrons from deuterium has been approved.
The goals of the experiment are:
(1) measure the $C_{2q}$ coefficients with high precision;
(2) search for charge symmetry violation (CSV) at the quark level;
(3) search for quark-quark correlations in the nucleon.

\subsubsection{Physics of PVDIS}

In 1978, Prescott {\it et al.}~\cite{SMB} showed that parity violation could be
observed in neutral currents by measuring the asymmetry (\ref{eq:apv})
in the deep inelastic scattering (DIS) of polarized electrons from deuterium.
By extending the kinematic range, the same group later published results that
were able to exclude alternative theories to the Standard Model that were
considered reasonable at the time.  The PVDIS asymmetry 
is sensitive to both the axial-vector (vector) electron couplings 
$C_{1q}$ and $C_{2q}$, whereas the emphasis for the Q-Weak experiment is on the $C_{1q}$.
For deuterium, one has
\begin{equation}
A^{PV}=
-\left(\frac{G_FQ^2}{4\sqrt{2}\pi\alpha}\right)
\left[a_1+\frac{1-(1-y)^2}{1+(1-y)^2}a_3\right];\ \ 
a^D_1(x)=
-\frac{6}{5}(2C_{1u}-C_{1d});\ \ 
a^D_3(x)
=-\frac{6}{5}(2C_{2u}-C_{2d}),
\label{eq:Afull}
\end{equation}
where $y=\nu/E$.  By observing the dependence of $A_{PV}$ on $y$, the Prescott
experiment was sensitive to both the $C_{1i}$'s and the $C_{2i}$'s.

As discussed above, there has been great progress more recently in the precision of the
measurements of the $C_{1q}$'s with the advent of the Qweak experiment and
precision atomic physics measurements.  On the other hand, progress with the $C_{2i}$'s has been
slower because at low energies, uncertain radiative corrections involving
the long-distance behavior of hadrons are large.
However, in DIS, these corrections are tractable since the relevant energy scale implies that all corrections can be computed perturbatively.  Hence the motivation for a
new precise experiment in PVDIS from deuterium.  The sensitivity of
the JLab experiment to new physics is given by 
Kurylov {\it et al.}~\cite{Kurylov:2003xa}.

\begin{figure} [htbp]
\begin{center}
\rotatebox{0.}{\resizebox{7.0in}{7.0in}{\includegraphics{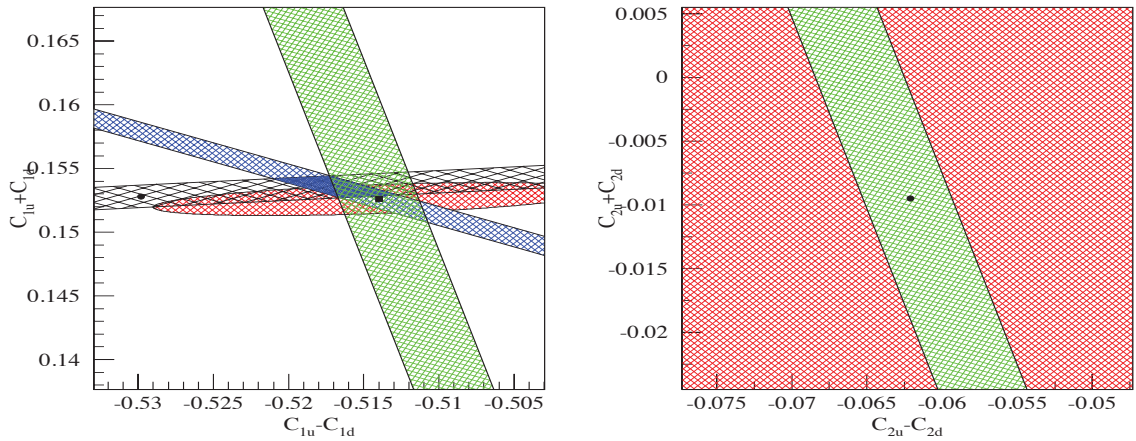}}}
\end{center}
\caption{ Left plot: Constraints on the Standard Model from
         parity violation experiments.
         The black crossed band presents the
         results from atomic parity violation in Cs atoms, the blue band 
         is the expected result from Qweak, the red ellipse
         is a PDG fit, 
         while the green band shows the anticipated JLab limit.
         Right plot: The anticipated error band from the JLab experiment.
         All limits are 1 standard deviation.}
\label{fig:pvdis}
\end{figure}

The collaboration at JLab plans  to build a new solenoid called SoLID, that
will enable one to obtain statistical precision of $<0.5$\% for a number of bins 
with $ 0.3<x<0.75$, $4<Q^2<10$ (GeV/c)$^2$, and $y\sim 1$~\cite{SOLID}.  
By designing the
apparatus for large $y$, one can maximize the sensitivity to the $C_{2i}$'s.
The  expected sensitivity  is given in Figure~\ref{fig:pvdis}.

\subsubsection{Hadronic Corrections}

The NuTeV experiment published a test of the Standard Model in neutrino-nucleus DIS .  The result differed from the prediction of the Standard Model
by about $3\sigma$~\cite{Zeller:2001hh},
generating considerable controversy and  causing a serious re-evaluation of the 
work.  Corrections, including
changes in the value of $V_\mathrm{us}$, strange sea-quarks, and improved radiative corrections
have been made.  Including these effects has not substantially mitigated the discrepancy.
Another possible explanation of the NuTeV result is charge symmetry violation 
(CSV) in the parton distribution functions (PDFs)\cite{Londergan:2003pq,Londergan:2003ij}.
Various authors have presented the case that this is a reasonable
explanation, citing the effects of PDFs \cite{Martin:2003sk}, QCD 
effects \cite{Sather:1991je,Rodionov:1994cg,Cloet:2009qs,Bentz:2009yy},
and QED effects \cite{Martin:2004dh,Gluck:2005xh}.

The JLab PVDIS  experiment is also sensitive to CSV.  If the $x$ dependence of the
CSV falls more slowly than that of the PDFs, the asymmetry should display a clear
$x$ dependence.  Moreover, these results will provide an important
test of the CSV explanation for NuTeV.  More details are given in the 
proposal~\cite{SOLID}.

There are additional important corrections to 
Equation~\ref{eq:Afull}~\cite{Hobbs:2008mm}. 
In particular, the cross sections
in this kinematic angle have significant scaling violations due to
higher-twist effects~\cite{Blumlein:2008kz}.  
However, as pointed out by Bjorken~\cite{Bj} and 
more recently by Mantry {\it et al.}~\cite{Mantry:2010ki}, the higher twist 
terms cancel in the $a_1$ term unless they are
due to quark-quark correlations.  The observation of diquarks in the nucleon,
if found, would be an exciting discovery.  The higher twist contribution of the
$a_3$ term can be subtracted from the asymmetry by using data on charged 
neutrino scattering~\cite{Bj}. 

\subsubsection{Apparatus}

The apparatus will use a large solenoid; presently the CLEOII magnet is being
considered.  Polarized electrons with an energy of 11 GeV will strike a liquid
deuterium target.  Scattered electrons with energies above 2 GeV will pass 
through a series of baffles and
then strike a detector package consisting of tracking chambers, a Cherenkov
counter to reject pions, and a calorimeter to serve as a trigger and provide
additional pion rejection.  The data will be taken with a deadtime-less flash 
ADC system with 30 independent sectors.  The tracking detectors will be
GEM with a total area of 25 m$^2$~\cite{Ketzer:2004jk}.


\subsection{Parity Violating M\o ller Scattering}
\label{sec:moller}


A collaboration at JLab has proposed to measure $A_{PV}$ in M\o ller scattering with longitudinally polarized electrons off unpolarized electrons, using the upgraded 11~GeV beam in Hall~A at JLab to an overall fractional accuracy of 2.3\%. Such a measurement would constitute more than a factor of five improvement in fractional precision over the only other
measurement of the same quantity by the E158 experiment at SLAC~\cite{e158}. 

The electron beam energy, luminosity, and stability at JLab are uniquely suited to carry out such a measurement. 
After the energy upgrade, a 11 GeV JLab beam provides a compelling new opportunity  to achieve a new benchmark in sensitivity. The physics motivation has two important aspects: 
\begin{enumerate}
\item New neutral current interactions are best parameterized model independently at low energies by effective four-fermion interactions via the quantity $\Lambda/g$, where $g$ characterizes the strength and $\Lambda$ is the scale of the new dynamics. 
The proposed $A_{PV}$ measurement is sensitive to interaction amplitudes as small as $1.5\times 10^{-3}$ times the Fermi constant, $G_F$, which corresponds to a sensitivity of $\Lambda/g =  7.5$~TeV. This would be the most sensitive probe of new flavor and $CP$-conserving 
neutral current interactions in the leptonic sector until the advent of a linear collider or a neutrino factory.
\item As noted above, within the Standard Model, weak neutral current amplitudes are functions of the weak mixing angle $\sin^2\theta_W$. The two most precise independent determinations of $\sin^2\theta_W$
differ by 3$\sigma$. While the world average is consistent with other electroweak measurements and constraints on the Higgs boson mass $M_H$, choosing one or the other central value ruins this consistency and implies very different new high energy dynamics. 
The proposed $A_{PV}$ measurement, which would achieve a sensitivity of 
$\delta(\sin^2\theta_W) = \pm 0.00029$, is the only method available in the next decade to directly address this issue at the same level of precision and interpretability.
\end{enumerate}

$A_{PV}$ in M\o ller scattering measures the weak charge of the electron
$Q^e_W$, which is proportional to the product of the electron's vector and axial-vector couplings to the $Z^0$ boson. 
The electroweak theory prediction at tree level in terms of the weak mixing angle is $Q^e_W = 1 - 4\sin^2\theta_W$; this is modified at the one-loop level~\cite{Czarnecki:1995fw, Czarnecki:2000ic, Erler:2004in} and becomes dependent on the energy scale at which the measurement is carried out, {\em i.e.} $\sin^2\theta_W$ ``runs". At low energy, $Q^e_W$  is predicted to be 
$0.0469\pm 0.0006$, a $\sim 40$\%\ change of its tree level value of $\sim 0.075$ (when evaluated at $M_Z$).

The prediction for $A_{PV}$ for the proposed experimental design is 
$\approx 35$~parts per billion (ppb) and the goal is to measure this quantity with a statistical precision of 0.73 ppb
and thus achieve a 2.3\%\ measurement of $Q^e_W$. The reduction in the numerical value of $Q^e_W$ 
due to radiative corrections leads to increased fractional accuracy in the determination of the weak mixing
 angle, $\sim 0.1$\%, comparable to the two best such determinations from measurements of asymmetries in
$Z^0$ decays in the $\mathrm{e}^+\mathrm{e}^-$ colliders LEP and SLC. 

\begin{figure} [htbp]
\begin{center}
\rotatebox{0.}{\resizebox{4.0in}{4.0in}{\includegraphics{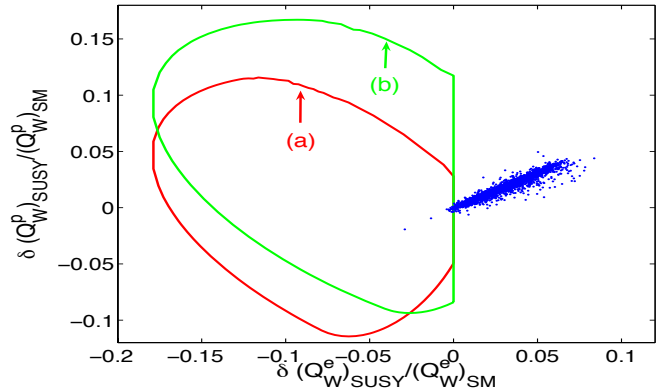}}}
\end{center}
 \caption{{\it Relative shifts in the electron and proton weak charges due to SUSY effects. Dots indicate the range of allowed MSSM-loop corrections. The interior of the truncated elliptical regions gives possible shifts due to R-parity violating (RPV) SUSY interactions, where (a) and (b) correspond to different assumptions on limits derived from first-row CKM unitarity constraints.}}
\label{qweqwpsusy}
\end{figure}

At the level of sensitivity probed, the proposed measurement could be influenced by radiative loop effects of new particles predicted by the Minimal Supersymmetric Standard Model (MSSM)~\cite{Kurylov:2003zh,nna:RamseyMusolf:2006vr}. In Fig.~\ref{qweqwpsusy}, the dots on the right-hand side show the results of a random scan over an allowed set of MSSM parameters. Deviations from the Standard Model as large  as $+8$\% are possible, corresponding to a potential deviation as large as 3.5~$\sigma$. 
If the assumption of R-parity conservation is relaxed (RPV), tree-level interactions could generate even larger deviations in $Q_W^e$. The left-hand side of Fig.~\ref{qweqwpsusy} shows the allowed region after constraints from low energy precision data have been taken into account. In this case, relative deviations of up to $-18$\% are allowed, a shift of almost 8~$\sigma$.

\begin{figure} [htbp]
\begin{center}
\rotatebox{0.}{\resizebox{3.0in}{3.0in}{\includegraphics{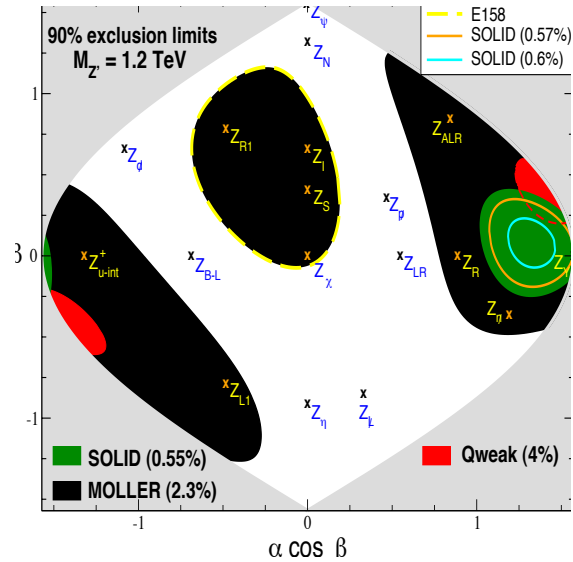}}}
\end{center}
\caption{{\it 90\% C.L.\ exclusion regions for a 1.2 TeV $Z^\prime$ ($E_6$ gauge group) for MOLLER, Qweak and SOLID, assuming they obtain exactly the SM predictions. Also shown is the contour from the E158 result.}}
\label{zprimemollerreach}
\end{figure}

A comprehensive analysis of the MOLLER experiment's sensitivity to 
TeV-scale $Z^\prime$s has recently been carried out~\cite{nna:Erler:2011iw} 
for a fairly large class of family-universal models
contained in the $E_6$ gauge group.  While models with full $E_6$ unification are already
excluded by existing precision electroweak data, the Z' bosons in these models with the same electroweak charges to SM particles are still motivated because 
they also arise in many superstring models as well as from a bottom-up approach~\cite{Erler:2000wu}. 
The MOLLER reach for a 1.2 TeV $Z^\prime$ from this model class, 
assuming the value predicted by the SM is measured, along with the current region excluded by E158, is
shown in Fig.~\ref{zprimemollerreach}. 

The measurement would be carried out in Hall A at JLab, where a 11 GeV longitudinally polarized electron beam would be incident on a 1.5 m liquid hydrogen target. M\o ller electrons (beam electrons scattering off target electrons) in the full range of the azimuth and spanning the polar angular range 5 mrad $<\theta_{lab}<$ 17 mrad, would be separated from background and brought to a ring focus $\sim 30$~m downstream of the target by a ???spectrometer system consisting of a pair of toroidal magnet assemblies and precision collimators. 
The M\o ller ring would be intercepted by a system of quartz detectors; the resulting Cherenkov light would provide a relative measure of the scattered flux. The experimental techniques for producing an ultra-stable polarized electron beam, systematic
control at the part per billion level, and calibration techniques to control normalization errors, including the degree of electron 
beam polarization at the 1\%\ level, have been continuously improved over 15 years of development at JLab.




\subsection{Atomic Parity Violation }
\label{sec:eic}

 Atomic parity violation (APV) comes from the mixing of opposite parity states by the weak interaction that allows forbidden electromagnetic (EM) transitions. The smallness of the weak interaction is enhanced experimentally by interference of an allowed and a forbidden transition, that then changes as the parity of the apparatus where it is measured changes. It has a rich history that spans more than 30 years \cite{khriplovich91,bouchiat00}, but continues to attract attention as it has been identified as a low energy area where it is possible to search for physics beyond the Standard Model
(SM) \cite{marciano90}. 

APV arises from the parity-violating exchange of $Z^{0}$-bosons between electrons and the
quarks in the nucleus \cite{bouchiat74}. The most precise measurement to date of APV was completed during the 1990s in $^{133}$Cs 
by the 
the group of  Wieman in Boulder  \cite{wood97}.  
The  extraction of weak interaction physics requires a theoretical input that has been improving significantly during the last decade ({\it e.g.},~\cite{ginges04,porsev09}).

APV measures the strength of the weak neutral current at 
very low momentum transfer. 
There are three types of such ``low energy" weak neutral current measurements
with complementary sensitivity. The atomic weak charge is predominantly 
sensitive to the weak charge of the neutron, as the proton weak charge is 
proportional to $(1-4 \sin^2{\theta_W})$, which accidentally is near zero. 
The Qweak electron scattering experiment on hydrogen 
will be sensitive to the weak 
charge of the proton (see \cite{erler,Erler:2004in,blunden11} for recent results on weak charge calculations). The SLAC E158 M\o ller scattering is sensitive to the electron's weak 
charge. 
Different Standard Model extensions then contribute 
differently~\cite{ramsey-musolf08}. 
The atomic weak charge is relatively insensitive to one-loop order
corrections from all SUSY particles, so its measurement provides a benchmark
for possible departures by the other low energy observables. 
M\o ller scattering is purely leptonic and has no
sensitivity to leptoquarks, so APV can then
provide the sensitivity to those.

{\bf Current efforts:} 
APV scales with the nuclear charge roughly as Z$^{3}$, favoring
experiments in heavy atoms (see the recent results in Yb in~\cite{tsigutkin09}). 
Ongoing efforts in APV include, but are not limited to: Yb (Berkeley \cite{tsigutkin09}), Fr (FrPNC Collaboration at TRIUMF, and the Ferrara, Legnaro, Pisa, Siena collaboration), Ra$^+$ (KVI), and Dy (Berkeley). There is an overall plan to work with different isotopes as a way to understand possible neutron distribution effects~\cite{brown09}, but also for studies of the weak interaction inside the nucleus, through anapole moment measurements (see for example \cite{gomez06} and references therein).

{\bf Exploration of new methods to measure APV:} 
All successful experiments that have measured APV have used an interference between an EM-allowed transition and the weak, parity mixing, amplitude. The result then comes from the extraction of a rate of transitions with one or the other parities. This applies also to the experiments that measure optical rotation. The measurements are subject to many complications, which have been overcome. Many atomic physics and precision metrologists have developed great tools to measure frequency and frequency shifts. However, there is no frequency shift associated with a transition dipole. An electric dipole P-odd and T-even cannot give rise to a frequency shift in a stationary atomic state perturbed by homogeneous E and B dc fields. Important proposals have appeared, and are beginning to be followed, that change the measurement of APV from a rate to a frequency shift, with the use of quadrupole transitions in a single ion \cite{fortson93} or by using light shifts that can create a linear Stark shift measurable with matter-wave interferometry  \cite{bouchiat07,bouchiat08}.

{\bf Intensity frontier:} The successful APV experiments require as many atoms as possible ($N$). Current efforts with rare isotopes incorporate laser trapping and cooling to ensure large samples for interrogation \cite{gomez06}, ensuring the large $N$ regime. Methodologies for the measurement are exploring many new avenues $-$ for example, two-photon transitions \cite{dounas-frazer11} and the proposals of Fortson and Bouchiat.  For radioactive ions, large numbers may be more complicated, given the effects of space charge on a cloud of ions; but care can be taken to pursue this avenue. The planned energy and current for nuclear studies in Project X at FermiLab would allow to produce many orders of magnitude more rare isotopes than in any other facility in the world. If the design allows for parallel operations. To achieve this will require development of appropriate targets and handling facilities. The flux, together with the multiuser parallel operation mode, has the potential to become a fantastic tool for APV research in the future.


\subsection{Electroweak Physics at an EIC}
\label{sec:eic}

Over the last decade, the Electron Ion Collider (EIC) has been considered in the US nuclear science community as a 
possible future experimental facility (beyond 12 GeV upgrade of the CEBAF at Jefferson Laboratory, and the FRIB at MSU) \cite{lrp07}
for the study of QCD. The EIC will help us explore and understand some of the most fundamental and universal aspects of
QCD \cite{lrp07, intreport}. 
This physics program requires the EIC 
to have a variable center-of-mass energy from about $\sqrt{s}=30-140$ GeV, luminosities of 
$\sim 10^{33-34} {\rm cm}^{-2} {\rm sec}^{-1}$ 
for ep collisions (100-1000 times that achieved at HERA),  polarization in both beams, and a wider range in 
nuclear species. The planned precision studies of QCD and the partonic dynamics also require the construction of 
a comprehensive detector system capable of excellent particle identification over a large momentum range, high 
momentum and energy resolutions and almost full (4$\pi$) acceptance. Currently there are two designs under consideration 
for the EIC in the US:
\begin{enumerate} 
\item  eRHIC \cite{eRHIC} at Brookhaven which will use the hadron and nuclear 
beams of the existing Relativistic Heavy Ion Collider (RHIC). The plan is to build an ERL-based electron beam facility 
of variable energy 5-30 GeV in the existing RHIC tunnel to collide with one of the RHIC beams.
\item ELIC at JLab \cite{elic} 
which will use the electron beam from the 12 GeV upgraded CEBAF under construction now. This will require construction
of a hadron/nuclear beam facility to be built next to the upgraded CEBAF complex to enable such collisions.
\end{enumerate}

With the experimental conditions available at the EIC $-$  a) center of mass energy ($\sim 100-140$ GeV),  
b) $\sim 100-1000$ times  larger luminosity in e-p collisions than HERA,  c) polarization in electron and 
proton/deuteron/helium beams, and  d) a comprehensive detector system $-$ 
it is only natural to explore what measurements would be possible at the EIC in the electroweak physics sector 
and of possible physics beyond the Standard Model. Three possible physics topics were considered so far: 

\begin{enumerate}
\item Precision measurement $\sin^{2}{\theta}_{W}$ as a function of Q, the momentum transfer, in e-p collisions \cite{sin2theta}. 
This would be the next-generation experiment beyond the SLAC-E158, 6 GeV Q-Weak, 6 GeV PVDIS and the currently planned 
12 GeV experiments (M\o ller \cite{moller} \& SoLID-PVDIS \cite{solid}) and would be complementary to atomic parity 
violation searches planned in the future.  The Q range of the EIC would be between the fixed-target experiments and
the measurements from LEP at the Z-pole. Any deviation from the expected running of the $\sin^{2}{\theta}_{W}$ would be 
a hint of physics beyond the SM.

\item Possible exploration of charged lepton flavor violation, particularly transitions between the 1$^{st}$ and the 
3$^{rd}$ generation leptons ($e-\tau$) \cite{etau} at the highest energy and highest luminosity e-p collisions \cite{lpq} at the EIC. 
This would extend searches made at HERA, in different ranges of the leptoquark couplings and masses, and will be 
complementary to future searches at the LHC, LHeC and the Super-B factories. The $\sim$100-1000 times more 
luminous collisions compared to HERA will be key to success in these searches. The angular correlations in the final
state decay particles in the known Standard Model interactions involving taus in the final state and those in which
taus are created in leptoquark interactions will form the tell-tale signs of the existence of leptoquarks interactions. 
Such studies were performed at HERA in the last decade, so the requirements of detector acceptance and the
methods of analyses are fairly well defined\cite{lpqhera}.  

\item Exploration of nucleon spin structure using the electroweak probes {\it i.e.}, the $Z$ boson (and its interference with $\gamma$) 
and $W^{+/-}$\cite{ewsf}. Due to their different couplings to the quarks and anti-quarks, these measurements will enable 
us to explore different combinations of partons, and hence allow a determination of parton distribution functions different 
from those accessible through conventional deep inelastic scattering with virtual photons.  There is ample
experience with W production in e-p scattering from HERA, where detailed studies of the structure function $xF_{3}$ have
been performed. Since the HERA proton beam was unpolarized, only unpolarized heavy quark distributions were ever
measured. With the polarized proton and neutron (via either deuteron or helium) beams at the EIC, these studies can
be extended to include not only the quark anti-quark polarization, but also details of their possible charge 
symmetry relations.
\end{enumerate}

The above studies and considerations are preliminary. Detailed detector simulations are needed to confirm the feasibility of 
these, and are under way. 



\section{Other Symmetry Tests}
\label{sec:Other}

Numerous tests of fundamental symmetries of nucleons, nuclei, and atoms  -- outside the main thrusts described above -- are being pursued.
 Among those for which descriptions were discussed at this workshop are studies of antimatter and possible novel tests of time-reversal invariance.


\subsection{Antiprotons and Antihydrogen}
\label{sec:antip}

\noindent{\bf Matter-antimatter Symmetry and $CPT$}

The discrepancy between the deviation from matter-antimatter symmetry on the cosmic scale, on one hand, and the so-far-observed perfect symmetry between particle and antiparticle properties on a microscopic scale, on the other hand, is one of the big mysteries that has not yet been satisfactorily explained by the SM. The observed baryon asymmetry in the cosmos of $(N_{B}-N_{\overline{B}})/N_{\gamma} \sim 10 \times10^{-6}$ \cite{Dolgov:2009fk} is in the SM thought to originate from the three Sakharov conditions: 
(1) baryon number violation, 
(2) $C$ and $CP$ symmetry violation, and
(3) deviations from thermal equilibrium during the expansion of the universe.
The so-far-observed violations of $CP$ symmetry in the $K$ and $B$ meson sector are, however, too small to quantitatively explain the observed baryon asymmetry, and thus other sources of matter-antimatter asymmetry may be explored. Notably, if $CPT$ symmetry is violated, then Sakharov's third criterion need not be invoked. As a concrete illustration, one may consider the Standard Model extension by Kostelecky \emph{et al.} \cite{Colladay:1997vn}, wherein it is possible to generate a large baryon asymmetry through violations of $CPT$ \cite{Bertolami:1997uq}.

 $CPT$ symmetry ensures that particles and antiparticles have perfectly equal properties. It is the result of a proof based on mathematical properties of the quantum field theories used in the SM, but certain of these properties, like point-like particles, are not valid in extensions of the SM such as string theory. Thus tests of $CPT$ by precisely comparing particle and antiparticle properties constitute a sensitive test of physics beyond the SM. Antiprotonic atoms and especially antihydrogen offer the most sensitive tests of $CPT$ in the baryon sector.


\noindent{\bf $CPT$ Tests with Antiprotonic Atoms and Antihydrogen}

For more than 20 years, low energy antiprotons have provided the most sensitive tests of $CPT$  in the baryon sector. The TRAP collaboration at LEAR of CERN has determined the maximal deviation of the charge-to-mass ratio of protons and antiprotons to 
$(Q_{\overline{p}}/M_{\overline{p}})/(Q_{p}/M_{p})+1 = 1.6(9)\times 10^{-10}$
 \cite{Gabrielse:99,Thompson:2004fk} using a Penning trap. The ASACUSA collaboration at CERN's Antiproton Decelerator has been performing precision laser and microwave spectroscopy of antiprotonic helium, an exotic three-body system containing a helium nucleus, an antiproton, and an electron exhibiting highly excited metastable states \cite{Hayano:2007}. By comparing the experimentally observed transition frequencies between energy levels of the antiproton with state-of-the-art three-body QED calculations, the most precise values for the equality of proton and antiproton mass and charge ( $  (Q_{p}+Q_{\overline{p}})/Q_{p} = (M_{p}-M_{\overline{p}})/M_{p}< 7\times10^{-10}$ \cite{Hori:2011gk}) and the antiproton magnetic moment (($\mu_{p}-\mu_{\overline{p}})/\mu_{p} < 2.9\times10^{-3}$ \cite{Pask:2009lq}) have been obtained.
 
Antihydrogen ($\overline{\mathrm{H}}\equiv \overline{p}e^{+}$), the simplest atom consisting only of antimatter, promises the highest sensitivity because its $CPT$ conjugate system, hydrogen, is one of the best-studied atoms in physics. Currently three collaborations at CERN aim at forming antihydrogen and performing precision spectroscopy of its structure. ATRAP and ALPHA have the goal of measuring the two-photon 1S--2S laser transition in antihydrogen, 
which has been determined to a relative precision of $10^{-14}$ in hydrogen, and ASACUSA aims at measuring the 1.4 GHz ground-state hyperfine transition that is known to $10^{-12}$ relative precision from the hydrogen maser. 

The progress -- as is typical for precision experiments -- is slow: the first formation of antihydrogen was reported in 2002 by ATHENA \cite{ATHENA-Hbar:02} (the predecessor of ALPHA)  and ATRAP \cite{ATRAP-Hbar:02a} using a nested Penning trap technique \cite{Gabrielse:2005sl}, the next major step happened in 2010 when ALPHA reported the first trapping of neutral antihydrogen atoms in a Joffe-Pritchard trap  \cite{Andresen:2010jba} (and later announced trapping times of 1000 seconds 
\cite{Andresen:2011jg}), and ASACUSA announced the first formation of antihydrogen in a different trap named ``cusp trap'' \cite{Enomoto:2010uq} which is expected to provide a polarized $\overline{\mathrm{H}}$ beam \cite{Mohri:2003wu} useful for measuring the ground-state hyperfine structure in an atomic beam \cite{Juhasz:2006uq}. Given these major achievements, first spectroscopy results can be expected within the next few years. 

To reach the full potential of the measurements, \emph{i.e.}, a similar precision to that obtained in hydrogen, a much longer time will be needed. The 
only facility in the world now providing low energy antiprotons, the AD at CERN, will be upgraded by an additional storage ring ELENA to decelerate antiprotons further and thus increase the number of trapped antiprotons. By the end of this decade another facility called FLAIR may go into operation at the FAIR facility under construction at Darmstadt.

\noindent{\bf Gravity of Antimatter}

Using neutral antihydrogen, the gravitation between matter and antimatter can be experimentally investigated for the first time. Scenarios exist where a difference in the gravitational interaction of matter and antimatter can arise, which are part of a general search for non-Newtonian gravity \cite{Nieto:1991uq}. The AEgIS experiment has been approved at CERN-AD and is about to start, aiming initially at a percent-level measurement of the gravitational acceleration of the antiproton. A second collaboration, GBAR, has submitted a letter of intent and is preparing a proposal, so that enhanced activities in this field can be expected in the future.


\subsection{Antihydrogen: ALPHA}
\label{sec:alpha}

The goal of the Antihydrogen Laser Physics Apparatus (ALPHA) collaboration is
to conduct fundamental studies of matter-antimatter asymmetry. 
Until recent successes~\cite{1,2} with the current ALPHA apparatus, no group had
ever trapped neutral antimatter. In 2009, the collaboration first had a hint of success,
reporting~\cite{3} 6 `'candidate'' antihydrogen atoms.  In 2010, they 
demonstrated trapped antihydrogen, reporting  38~\cite{1} antiatoms with trapping times of  0.17 s, 
subsequently optimizing performance and increasing the trapping numbers to  309~\cite{2} trapped antiatoms during the 2010 antiproton
season, and  hundreds more during 2011, while simultaneously extending the
confinement time of these antiatoms from 0.17 s to 1000 s. These results received significant coverage in the scientific press and were enthusiastically received by the general public~\cite{3a}. The ALPHA apparatus was primarily designed to
demonstrate that cold antihydrogen could be trapped, and this has been accomplished.  Motivated by this rapid
progress towards achieving conditions required for physics studies of antihydrogen, the collaboration is rebuilding the experiment. The new apparatus, 
ALPHA-II,  is  specifically designed to be a production machine on which physics studies of antimatter are conducted. 



The success to date is only a
first step on a long road of scientific exploration. Lessons from the five years
of operation of the ALPHA apparatus are  being incorporated into the new apparatus.
ALPHA-II will be  designed, constructed, and commissioned
during the next three years. The  ALPHA-II design
will allow one to perform  microwave and laser spectroscopy,
antihydrogen charge neutrality measurements, and antimatter gravity studies. Though unlikely,  any
discovery of a fundamental difference between matter and antimatter in $CPT$ or
gravity experiments may shed light on baryogenesis or dark energy.

Key to these future physics studies with ALPHA-II is improving its performance over that of ALPHA. The ALPHA-II design goals are  (a) a significant increase in trapping
rates, from a peak rate of about 4 antihydrogen atoms per hour to a peak of order 40
antihydrogen atoms per hour,   (b) improved trap geometry, allowing for enhanced
diagnostics and laser and microwave spectroscopic access,  and (c)  improved shot-to-shot reliability.
ALPHA-II will  incorporate  the techniques that the collaboration has developed on the original ALPHA apparatus,
including  evaporative cooling~\cite{evaporative} of antiprotons and positrons, autoresonant
injection~\cite{AR} of antiprotons into the positron plasma, rotating wall plasma compression~\cite{compression} and an octupole-mirror magnetic-well~\cite{Andresen2010141} neutral confinement scheme.
The experiment will increase the production rate by separating catching and mixing functionalities,  enhanced antiproton capture, creating and maintaining lower positron, antiproton, and electron
plasmas, and developing new simulation tools for fast optimization of a wide range of system parameters. The collaboration is  working on improved  plasma diagnostics to reliably measure temperature at sub-Kelvin resolution, and on new, preferably non-invasive, temperature diagnostics. 

The AD at CERN is the only facility that can provide low energy antiprotons for trapping experiments. The AD currently provides $\sim$5 MeV antiprotons to the experiments. ALPHA sends them through a degrader and traps about one in $10^3$. CERN is building a new decelerator ring that will take the AD beam and lower its energy while providing the needed additional cooling, to 100kV. The construction of ELENA should begin in 2013 and the first physics injection should follow about three years later.  This will significantly enhance antiproton capture rates, and, consequently,  antihydrogen production. New  positron-antiproton mixing schemes may be developed to best utilize increased antiproton numbers. 

 Antimatter experiments have one advantage over experiments with normal matter: One can detect, with more than 60\% efficiency, the interaction of each antihydrogen on the trap wall. Thus, with only a few antiatoms, one can perform experiments that would be impossible with normal atoms.  For example, a microwave-induced spin flip  to a high-field
state can be detected on almost a per atom basis. If the flip is successful, the antihydrogen hits the trap wall and the resulting shower is detected. The detector has both spatial and temporal resolution and serves as a critical diagnostic for inferring the temperature of the trapped antiatoms.


\noindent{\bf Exploration of $CPT$ Invariance in Antihydrogen Studies}: Antihydrogen
studies  open up $CPT$ physics in a new, previously unexplored sector. 
Projected resolution achievable on ALPHA-II should  reach that achieved in the kaon
measurements. The $CPT$ experiments can be conducted in two ways: (i) by measuring the microwave
transitions of the hyperfine interaction and (ii)  $1S \rightarrow 2S$ atomic transitions.   We  are currently testing microwave equipment on ALPHA, and will be operational with microwaves and lasers on ALPHA-II. 
The collaboration believes it can reach the level of kaon-sector precision in five years of physics operation, with ultimately higher precision depending on the
level of laser cooling that can be reached. These spectroscopic measurements,
as well as the neutrality measurements and gravity measurements discussed below,
are direct measurements on antimatter. They are not model dependent. 

\noindent{\bf Charge Neutrality Measurements}: Measurements of charge neutrality
constitute a novel method to test for $CPT$ violation. The charge neutrality of
antihydrogen can be tested, without spectroscopic techniques, using ideas from accelerator physics.  Currently, the charge of the positron and
antiproton are known only~\cite{pdg} to a factor of 4x10$^{-8}$ and 2x10$^{-9}$, respectively.  Experiments on ALPHA-II should reach a higher level of precision.
For example, with a 300s lifetime and roughly 40 antiatoms, the expected
precision is ~1x10$^{-10}$. With a longer Hbar lifetime (300s $\rightarrow$ 2700 s)  the precision is ~3x10$^{-11}$, and with a much longer Hbar lifetime
(36,000 s) the precision is ~1x10$^{-11}$.

\noindent{\bf Gravity Measurements}:    Direct measurement of the
gravitational interaction with antihydrogen is a unique and important
experiment.  Antigravity studies on ALPHA-II  do not
require high-precision laser or microwave spectroscopy, though laser  cooling will
enhance the sensitivity of the gravity  measurements. With laser cooling, the collaboration believes ALPHA-II can
reach a precision of unity or better in the
ratio of the inertial to gravitational mass of antihydrogen. (This would rule out,     for
example, the possibility that the sign of the matter-antimatter gravitational force is
opposite to the matter-matter gravity).


\subsection{Neutron Beam Tests of Time Reversal}
\label{sec:ntrv}

One advantage of the search for time-reversal invariance violation (TRIV) in neutron nuclei interactions is the possibility of an enhancement of $T$-violating observables  by many orders of  magnitude  due to the complex nuclear structure  (see, {\em i.e.}  \cite{Gudkov:1991qg} and references therein). Moreover, the variety of  nuclear systems for measuring $T$-violating parameters provides assurance  that a possible ``accidental'' cancelation of T-violating effects due to unknown structural factors related to the strong interaction in the particular system would be avoided.  Taking into account that different models of the $CP$ violation may contribute differently to a particular $T$-odd or $CP$-odd observable, which  may have  unknown theoretical uncertainties, TRIV nuclear effects could be considered  valuable complementary experiments  to  electric dipole moment (EDM) measurements.

One of the  promising approaches to the search for TRIV in nuclear reactions is the measurement of TRIV effects in  transmission of polarized neutrons through a polarized target.
For the observation of TRIV and parity violating effects,  one can consider  effects related to the $\vec{\sigma}_n\cdot({ \vec{p}}\times{\vec{I}})$ correlation, where  $\vec{\sigma}_n$ is the neutron spin, ${\vec{I}}$ is the target spin,
and $\vec{p}$ is the neutron momentum, which can be observed in the transmission of polarized neutrons through a target with  polarized nuclei.
 This correlation leads to a
difference  between the total neutron cross sections $\Delta\sigma_{\not{T}\not{P}}$  for $\vec{\sigma}_n$
parallel and anti-parallel to ${\vec{p}}\times{\vec{I}}$
and to neutron spin rotation angle  $\phi_{\not{T}\not{P}}$  around the axis
$\vec{p}\times{\vec{I}}$
\begin{equation}
\label{cc}
\Delta\sigma_{\not{T}\not{P}}=\frac{4\pi}{p}{\rm Im}(f_{+}-f_{-}),\qquad \frac{d\phi_{\not{T}\not{P}}}{dz}=-\frac{2\pi N}{p}{\rm Re}(f_{+}-f_{-}).
\end{equation}
Here, $f_{+,-}$ are the zero-angle scattering amplitudes for neutrons polarized
parallel and anti-parallel to the $\vec{p}\times{\vec{I}}$ axis, respectively;
 $z$ is the target length and $N$ is the number of target nuclei per
unit volume.
The unique feature of these TRIV effects (as well as the similar effects  related to TRIV and parity conserving correlation $\vec{\sigma}_n\cdot({ \vec{p}}\times{\vec{I}})\cdot({ \vec{p}}\cdot{\vec{I}})$) is  the absence of false TRIV effects due to the final state interactions (FSI) (see, {\em e.g.}, \cite{Gudkov:1991qg} and references therein), because these effects  are related to   elastic scattering at a zero angle. The general theorem about the absence of FSI for TRIV effects in elastic scattering was first proved  by R. M. Ryndin \cite{Ryndin:fsi} (see, also \cite{Ryndin:1965,Ryndin:1969,Gudkov:1991qg,VG-Song:2011}).  Therefore, an observation of a non-zero value of  TRIV effects in neutron transmission directly indicates  TRIV, exactly as in the case of neutron EDM \cite{Landau:1957}.

Moreover,  these TRIV effects  are  enhanced \cite{Bunakov:1982is} by a factor of about $10^6$ in neutron-induced  nuclear reactions (a similar  enhancement was observed for PV effects related to  $(\vec{\sigma}_n\cdot{ \vec{p}})$ correlation in neutron transmission through nuclear targets). Since TRIV and PV effects have  similar enhancement factors,  it is convenient to consider  the ratio $\lambda$ of TRIV to PV effect at the same nuclei and at the same neutron energy as the measure of TRIV effect, because for this ratio, most nuclear structure effects  cancel each other out.  As a result, one can  estimate $\lambda \sim g_{T}/g_{P}$, where $g_{T}$ and $g_{P}$ are TRIV and PV nucleon-nucleon coupling constants.  Theoretical predictions for $\lambda$ are varying from $10^{-2}$ to $10^{-10}$ for different models of $CP$ violation (see, for example, \cite{Gudkov:1995tp} and references therein). Therefore,  one can estimate a range of  possible values of the TRIV observable and relate a particular mechanism of $CP$ violation to their values. These estimates show that these effects could be measured at the new  spallation neutron facilities, such as the SNS at Oak Ridge National Laboratory or the J-SNS at J-PARC in Japan.

\section{Strong Interaction Issues}
\label{sec:Strong}

An important consideration for the interpretation of the aforementioned studies is the robustness of theoretical Standard Model predictions against which one makes comparisons. Obtaining sufficiently reliable computations in strongly interacting systems, such as the neutron or nuclei, as well as robust atomic many-body computations, remains an important thrust for the field. In some cases -- such as electric dipole moment searches --  theoretical uncertainties are unlikely to substantially affect the discovery potential, but may influence the interpretation of the results in terms of the specific underlying mechanism of $CP$ violation. In others, such as the tests of CKM unitarity, tiny deviations from Standard Model expectations can only be interpreted as arising from new physics to the extent that uncertainties in the Standard Model predictions are sufficiently small. 

Achieving these robust Standard Model computations requires progress on a number of fronts. Lattice QCD computations are now being used to compute the ``$\theta$"-term contribution to the neutron EDM as well as various scalar and tensor form factors associated with possible non-Standard Model contributions to neutron decay. Similarly, there is continued scrutiny of nuclear wavefunction uncertainties  entering extractions of $V_\mathrm{ud}$ from superallowed nuclear Fermi transitions, while work is ongoing with refinements of nuclear Schiff moment computations needed for the interpretation of atomic EDM searches. And novel aspects of nucleon structure associated with heavy quarks may affect expectations for dark matter direct detection experiments. 

The vast subject of strong interaction issues related to probes for new physics -- not to mention those of intrinsic interest to the QCD community -- exceeds the scope of what we are able to provide here. Several important contributions were received from the community, including discussion of possibilities for new probes of nucleon spin structure, the role of ``intrinsic" heavy quarks in the nucleon wavefunction, and efforts to sharpen lattice QCD input.
We refer the reader to the working group website for detailed write-ups and links to other material\cite{ifwref}.

\def\Discussion{\setlength{\parskip}{0.3cm}\setlength{\parindent}{0.0cm}
     \bigskip\bigskip      {\Large {\bf Discussion}} \bigskip}\def\speaker#1{{\bf #1:}\ }
\def\endDiscussion{}

\def\BABAR{\mbox{\slshape B\kern-0.1em{\footnotesize A}\kern-0.08em
  B\kern-0.1em{\footnotesize A\kern-0.12em R}}\xspace}

\chapter{Summary}

The intensity frontier explores fundamental physics with intense sources and ultra-sensitive, sometimes massive detectors.  
It encompasses searches for extremely rare processes and for tiny deviations from Standard Model expectations.  Intensity frontier
experiments use precision 
measurements to probe quantum effects of new particles and interactions.   
They typically investigate very large energy scales, even higher than the kinematic reach of 
high energy particle accelerators. The science addresses basic questions, such as: Are there new sources of $CP$ violation?  Is there 
$CP$ violation in the leptonic sector? Are neutrinos their own antiparticles?  Do the forces unify? Is there a weakly coupled hidden 
sector that is related to dark matter?  Do new symmetries exist at very high energy scales?

To identify the most compelling science opportunities in this area, the workshop {\it Fundamental Physics at the Intensity Frontier} 
was held in December 2011,  sponsored by the Office of High Energy Physics in  the US Department of Energy Office of Science.    
Participants investigated the most promising experiments to exploit these opportunities and described the knowledge that can be gained 
from such a program. Experiments that can be performed this decade with current or planned facilities were identified, as well as those 
proposed for the next decade that require new facilities or technology to reach their full potential.  Some emphasis was placed on 
experiments that could produce data by the end of this decade.  Both international and domestic facilities were considered.  The 
workshop generated much interest in the community, as witnessed by the large and energetic participation by a broad spectrum of scientists.

For the purposes of this workshop, the intensity frontier program was defined in terms of six areas that formed the basis of working 
groups:  experiments that probe ($i$) heavy quarks, ($ii$) charged leptons, ($iii$) neutrinos, ($iv$) proton decay, ($v$) light, weakly 
coupled particles, and ($vi$) nucleons, nuclei, and atoms.  Each working group was convened by a theorist and an experimenter, as well 
as an observer who was charged with providing input from the broader community.  The working groups began their tasks well in advance 
of the workshop, holding meetings and soliciting white papers.

The workshop program began with a half-day plenary session, where each working group presented an introduction to 
their field.  This was followed by 1.5 days of intense parallel sessions. The final day was devoted to plenary sessions, including talks 
that placed progress in high energy physics in the context of other areas of science,  placed the intensity frontier in the context of 
the three frontiers of high energy physics, and gave  an overview of intensity frontier science. Each working group then summarized their 
results. The workshop closed with a summary and an outline of the next steps for realizing  a world-leading intensity frontier program 
in the US. These talks and the workshop agenda are available in \cite{sum:IFW}

This report constitutes the proceedings of the workshop.  The chapters containing the working group reports provide a clear overview 
of the science program within each area of the intensity frontier.  The discovery opportunities are laid out for present, planned, and 
proposed facilities that will be available this decade or come on line during the next decade.  Here, we briefly summarize the findings 
from each working group.  

\section{Heavy Quarks}

The study of strange, charm, and bottom quark systems has a long and rich history in particle physics, and measurements of rare processes 
in the flavor sector have led to startling revelations. Examples include the discovery of $CP$ violation in the decay $K_L^0\to\pi^+\pi^-$, 
the predicted existence of the charm quark and its mass scale from the suppression of $K_L^0\to\mu^+\mu^-$ and from the
measurement of $K_L-K_S$ mixing, the 
development of the theory of flavor changing neutral currents and the  Glashow-Iliopoulos-Maiani mechanism, the formulation of the 
three-generation CKM picture, the first indication of the top-quark mass scale from $B_d^0$ meson mixing, and, recently, the establishment 
of the CKM phase as the leading source of $CP$ violation in $B$ meson decays.  These observations played a critical role in the development 
of the Standard Model.  In addition, some of the most powerful constraints on theories beyond the SM arise from flavor physics experiments. 
 While the current quark flavor data set is in agreement with SM expectations, new corrections to the SM at the level of tens of percent 
are still allowed.  Contributions to flavor processes from many theories beyond the SM arise at this level, and thus more precise 
measurements may observe new physics.  In particular, if new massive states are observed at the CERN Large Hadron Collider (LHC), 
then detailed measurements of the quark flavor sector will be necessary to determine the underlying theory and its flavor structure.  
If such states are not discovered in high energy collisions, then precision quark flavor experiments, with their ability to probe mass 
scales far beyond the reach of the LHC, provide the best opportunity to set the next energy scale to explore. Depending on the strength 
of new physics interactions, this program already indicates that the new physics scale is above $1-10^5$ TeV, and proposed experiments 
can probe even further.  Continued investigations of the quark flavor sector are thus strongly motivated.

A well-planned program of flavor physics experiments has the potential to continue this history of paradigm-changing advances.  Such a 
program exists worldwide with the LHCb experiment at the LHC, an upgraded SuperKEKB facility in Japan, a planned Super$B$ factory in Italy, 
BESIII in China, and future kaon experiments at CERN, J-PARC and possibly Fermilab. This currently envisioned program will develop over 
this decade and continue into the next decade.   It is an essential component of a balanced particle physics program.  These facilities 
have the potential to carry out a rich multi-purpose program in the strange, charm, and bottom sector and perform numerous crucial 
measurements of rare decays and $CP$-violating observables.  The expected sensitivities are given in this report and are not predicated 
on future theoretical progress, although theoretical advancements will strengthen the program by increasing the set of meaningful 
observables.  The US should have a strong involvement in these experiments in order to exploit existing expertise and share in the discoveries.

\section{Charged Leptons}

Precision measurements of charged lepton interactions provide an excellent opportunity to probe the existence of new physics.  Historically, 
the investigation of this sector has played an integral role in our understanding of particle physics.  In particular, it was the discovery 
of the muon (and tau) that first heralded the existence of the second (and third) generation of fermions.   Charged lepton reactions have 
traditionally played a strong role in constraining theories beyond the SM. Ultra-sensitive measurements are possible, since charged leptons 
are easy to produce and detect and their interactions are relatively free of theoretical uncertainties.  They provide probes of leptonic 
couplings of new particles, complementing the new-physics searches at the Large Hadron Collider.  Any indication of charged Lepton Flavor 
Violation (LFV) would be an indisputable discovery of new physics.  The experimental program consists of a large and diverse set of 
opportunities and includes multi-purpose experiments that utilize the large tau production rates at high-luminosity $B$ factories, as 
well as highly optimized experiments that explore muon transitions. LFV experiments search for the rare or SM-forbidden 
interactions of $\mu\to e\gamma$, $\mu^\pm\to e^\pm e^+e^-$ decays and $\mu N\to eN$ transitions in the presence of nucleons as well 
as lepton flavor violating tau decays.  Lepton Flavor Conserving (LFC) processes are also valuable probes for new physics and most notably 
include measurements of the anomalous magnetic moment, as well as the electric dipole moment, of the $\mu$ and $\tau$ leptons. This report 
documents how precision measurements of these transitions can be used to verify predictions of the SM and look for signs of new physics.

Significant advances in studying LFV in the muon sector can be achieved this decade.  For the rare decay $\mu\to e\gamma$, the only running 
experiment, MEG at the Paul Scherrer Institute, set a limit on the branching fraction of $2.4\times 10^{-12}$ in 2011.  In 2012 this is 
expected to decrease by roughly a factor of two, and with possible improvements the experiment can reach a sensitivity of $10^{-13}$ in 2016.   
For $\mu^\pm\to e^\pm e^+e^-$, the present limit from the SINDRUM experiment ($<10^{-12}$) is a decade old.  Two new experiments are currently 
at the proposal stage, mu3e at PSI and MuSIC at Osaka University, and have plans to improve this sensitivity by approximately four orders 
of magnitude.  Current limits for $\mu N\to e N$ conversion are placed by experiments at PSI (with Ti and Au nuclei) at the level of 
$<4.3\times 10^{-12}$ and $7.0\times 10^{-13}$, respectively.  DeeMe at J-PARC plans to improve these bounds by two orders of magnitude.  
Later in this decade, COMET at J-PARC and Mu2e at Fermilab could improve the existing constraints by four orders of magnitude.  This 
would bring the sensitivity to such conversions to $\sim 2 \times 10^{-17}$ by the end of this decade.  Signals from supersymmetric 
grand unified models are expected to lie in this range.  If no signal is observed, this would set constraints on LFV physics at the scale of 
$10^4$ TeV.  Future experiments beyond these are being considered in conjunction with more intense muon beams, which in return require 
more intense proton beams that could be provided by upgraded facilities at J-PARC or Project-X at Fermilab.

The classic example for studying LFC processes with muons is the measurement of  the muon's magnetic moment, which is predicted very 
precisely (five-loop level) in the SM. New physics contributes via radiative corrections, and would yield an anomalous value of $g\neq 2$.  
The present level of sensitivity was obtained by the E821 experiment at Brookhaven National Laboratory, with a difference between the 
measurement and the SM theoretical prediction of $(287\pm 80) \times 10^{-11}$,  a $3.6\sigma$ deviation from the SM value.  A new 
experiment, E989, plans to move the E821 muon storage ring to Fermilab and use the same experimental techniques. It aims to increase 
the statistics by a factor of 20 while keeping the other uncertainties at the same level of 0.1ppm, resulting in an overall reduction in 
the experimental error by a factor of roughly four. An alternate approach, being developed at J-PARC using lower-energy muons,  is 
expected to have the same statistical uncertainty as E989 at Fermilab but with very different systematics.  This decade will see a big 
improvement over the current limits, possibly with two approaches. 

The important observables in $\tau$ decays are: LFV and $CP$-violating decays, the electric dipole moment, and an anomalous magnetic 
dipole moment.  New physics effects usually scale as a function of the lepton mass, and hence $\tau$ observables can be very sensitive 
to new contributions.  The large $\tau$ production rates possible at the future SuperKEKB facility in Japan and Super$B$ in Italy could 
reach the level of $10^{-9-10}$ in LFV branching fractions, which represents an order of magnitude improvement over current results from 
\BABAR\ and Belle.

The charged lepton sector has significant potential to reveal crucial information on the fundamental principles of Nature, and the US has 
the opportunity to play a leading role with facilities planned for this decade.

\section{Neutrinos}

Neutrinos are the most elusive of the known fundamental particles.  They carry no electric or color charge and participate only in the 
weak interactions within the SM. This makes neutrinos difficult to detect, and exploration of the neutrino sector requires a variety of 
very intense sources accompanied by very large and sensitive detectors.  Useful sources of neutrinos can be obtained from reactions in the 
sun, cosmic ray interactions in the atmosphere, nuclear reactors, $\beta$-decay of nuclei, and man-made beams from the decay of mesons and 
muons produced at an accelerator.  These sources span a wide energy range and each is best suited to probe different aspects of the 
neutrino sector.  An assortment of neutrino experiments is essential to fully probe the properties of neutrinos.

We know amazingly little about the nature of neutrinos.  They were  thought to be massless until the discovery of neutrino flavor 
oscillations demanded a non-zero neutrino mass. Experiments with solar, atmospheric, reactor, and accelerator generated neutrinos have 
established these oscillations beyond any doubt.  The leptonic mixing matrix, similar to the CKM matrix in the quark sector, relates the 
weak eigenstate neutrinos ($\nu_{e,\nu\tau}$) to their mass eigenstates ($\nu_{1,2,3}$).  Much effort in the experimental neutrino program 
over the last decade has been geared towards determining the parameters contained in this matrix and the hierarchy of neutrino masses.  
Neutrino oscillations are  sensitive only to the difference in the squares of the masses.  Two such mass differences are presently 
measured, but the ordering of the masses is ambiguous.  This leaves room for one of two choices: either a normal or an inverted mass 
hierarchy.  This is a crucial point, as the relation of the leptonic mixing matrix parameters to the experimental observables depends 
on this ordering.  The matrix parameters consist of three mixing angles and a complex phase that is related to $CP$ violation in the 
leptonic sector. Two of these mixing angles are well measured and the third, so-called $\theta_{13}$, has recently been determined and 
takes on a surprisingly large value.  The phase responsible for $CP$ violation has not yet been measured.  $CP$-violating and mass 
hierarchy effects depend on the value of $\theta_{13}$, so this new result aids in the quest to discover $CP$ violation in the leptonic sector. 

In the last decade, some neutrino oscillation experiments reported anomalies at the $\sim 2.5\sigma$ level that point towards the 
existence of additional neutrino states known as sterile neutrinos.  Their name derives from the fact that they do not experience SM 
interactions.  These results arise from short-baseline accelerator and reactor experiments.  The potential existence of sterile neutrinos 
is an outstanding question that needs to be addressed, and plans are under way for new experiments to be constructed this decade.

One of the most fundamental open questions in the neutrino sector is whether neutrinos are their own antiparticles (Majorana type) or not 
(Dirac type).  Either choice is possible and has consequences for theories of 
neutrino mass generation.  Majorana neutrinos would violate 
lepton number, and their mass would imply the existence of new physics at high mass scales.  Dirac neutrinos would indicate 
that an additional field, the right-handed neutrino, must be included in the SM.  This forms the basis of the claim that the observation of 
neutrino mass is a discovery of new physics.  This issue can  be addressed only by looking for neutrinoless double beta decay
($\beta\beta_{0\nu}$) $-$ a process 
that also holds  the promise of measuring the absolute neutrino mass scale. Several approaches for more sensitive $\beta\beta_{0\nu}$ 
experiments are under way, and a decision on moving forward with a kiloton detector is expected in the middle of the decade.  This 
detector size guarantees observation of Majorana neutrinos if the mass hierarchy is inverted.

Neutrino physics promises to be a very rich field with 
new discoveries and insights being possible during this coming decade via experiments that are 
either currently running, being constructed, or proposed.  This field is just beginning to precisely determine the properties of 
neutrinos and their mixing. It is a global program, with an experimental emphasis on accelerator-based neutrino sources in the US and 
Japan. 2012 is the year of $\theta_{13}$, due to the turn on of the Daya Bay and RENO reactor experiments and resumed running of T2K.  
NOvA will turn on in 2014 and will have sensitivity to the mass hierarchy for roughly half of the possible parameter space.  The recent 
measurement of $\theta_{13}$ allows for better and more effective planning of future neutrino experiments; this effort is currently 
under way.  There are several proposals for large future facilities in Japan, Europe, and the US based on deep underground detectors, which 
would have sensitivity to $CP$ violation and the mass hierarchy.  These underground detectors also have the capability to observe proton 
decay (as discussed below) and to track the evolution of neutrinos escaping from a supernova burst.  This program will require intense beams 
and new underground facilities.  The US currently has a strong neutrino program and has the opportunity to take the lead in this field.

\section{Proton Decay}

We know that the lifetime of protons is long, given the existence of baryons and Earth and life on it. Within the Standard Model protons 
are stable, as baryon number is assumed to be conserved.  However, baryon number is not a fundamental symmetry of the SM, and is not 
conserved in many extensions of the SM.  In particular, Grand Unified Theories (GUTs) predict that the proton decays, 
with the decay being mediated at scales of order $10^{16}$ GeV.  Two important decay channels in GUTs are 
$p\to e^+\pi^0$ and $p\to \bar\nu K^+$, with several other modes also being possible.  The current limits on the proton lifetime in these 
two channels are $1.4\times 10^{34}$ years and $4\times 10^{33}$ years, respectively, which is a factor of five to 10 below predictions 
in well-motivated GUTs.

The search for proton decay is carried out in large underground detectors $-$ large to amass enough protons and underground to reduce 
backgrounds.  Large neutrino oscillation detectors are ideal for this task, and proton decay is an important piece of their physics 
portfolio.  The current largest underground neutrino experiment is a tens of ktons water Cherenkov detector. Future underground neutrino 
experiments with projected target masses of 200 to 500 ktons can measure lifetimes on the order of GUT expectations with exposures of roughly 
10 years. Typically, a 5.6 Megaton-year exposure could reach a sensitivity of $\tau(p\to  e^+ \pi^0) <1.3\times 10^{35}$ years and 
$\tau(p\to\bar\nu K^+)<2.5\times 10^{34}$ years.  This translates to roughly a factor of 10 improvement in the limit or to a potential 
discovery if, {\it e.g.}, the lifetime for the $e^+\pi^0$ mode is of order $6 \times 10^{34}$ years. Observation of proton decay would 
probe energy scales not accessible to any other measurement and test fundamental symmetries of Nature.

\section{New Light, Weakly Coupled Particles}

New light particles that couple very weakly to the SM fields are a common feature of extensions beyond the SM.  Their existence is 
motivated by both theoretical and observational considerations, including the $CP$ problem in strong interactions 
and the nature of dark matter and dark 
energy.  Examples of such particles include axions, hidden-sector photons, milli-charged particles, and chameleons.  These hidden-sector 
particles typically couple to the photon via a mixing effect.   Because they only interact very weakly with the photon, intense sources 
are required to produce them at rates sufficient to enable their discovery.

The parameters relevant for searches of such hidden-sector particles are their mass and coupling strength to the photon.  A variety of 
observations constrain part of this parameter space, but much territory is still open for exploration.  In particular, regions that are 
consistent with dark matter observations, and areas that offer an explanation for the present result on the anomalous magnetic moment of 
the muon, have yet to be probed.  The current constraints arise from astronomical observations, cosmological arguments, and a variety of 
laser, heavy flavor, and fixed-target experiments, and are detailed in this report.

There is much activity on the experimental front in searches for light, weakly coupled particles, with several laboratory experiments being 
either in progress or proposed.  Two ADMX-type microwave cavity searches for axions will be under way soon in the US, but require  
developments to further 
increase their mass reach.  The light shining through walls technique, where photons are injected against an opaque barrier, 
was pioneered in the early 1990s and continues to explore open regions of parameter space.  More advanced technology to create
strong electromagnetic fields is needed to make 
progress in the mid-term.  Axion helioscope searches were first carried out at Brookhaven using borrowed magnets, and now require a 
custom-built magnet to improve sensitivity.  Collider searches for hidden-sector particles can be performed via the reaction 
$e^+e^-\to\gamma\ell^+\ell^-$ at high luminosity $B$ factories or in decays of gauge bosons at the LHC.  Fixed-target experiments 
using both electron and proton beams are a promising place to search for hidden-sector particles.  The electron beam experiments APEX at 
Jefferson Laboratory  and A1 at the University of Mainz have recently performed short test runs and have plans for more extensive runs 
this decade.  HPS has been approved by JLab and will run after the 12 GeV upgrade is installed at CEBAF.  DarkLight proposes to use the 
free-electron laser beam at JLab.  Proton fixed-target experiments have the potential to explore regions of parameter space that cannot 
be probed by any other technique.  Experiments at Fermilab, such as MiniBooNe or MINOS, have yet to explore their sensitivity.  The 
intense proton source at Project X could provide a powerful extension to the search reach.
   
Impressively large regions of parameter space are currently unexplored and are ripe for the discovery of light, weakly coupled particles
that are related to the dark sector of the universe.

\section{Nucleons, Nuclei, and Atoms}

Observables involving nucleons, nuclei, and atoms have sensitivity to physics beyond the Standard Model, although they represent more 
traditional measurements in the area of nuclear physics.  This report has not attempted to cover the full and rich program of nuclear 
physics, but rather focuses on the overlap with particle physics in the quest to discover new interactions.  Measurements of electric 
dipole moments, weak decays of light hadrons, weak neutral currents, and atomic parity violation are particularly well-suited in this regard.

The existence of a particle electric dipole moment (EDM) would be a direct signature of both parity and time-reversal violation, and thus 
probe $CP$ violation. EDM searches provide an excellent tool to look for new physics and have historically placed strong model constraints, 
most notably on $CP$-violating effects in supersymmetry.  The SM prediction (via multi-loop contributions) for the EDM of the electron, 
neutron, and nucleus is $10^{-38}\,, 10^{-31}$, and $10^{-33}$ {\it e}$\cdot$cm, respectively.  EDM measurements are challenging, and the 
present experimental sensitivity is at the level of $10^{-27}\,, 2.9\times 10^{-26}$, and $10^{-27}$ {\it e}$\cdot$cm for the electron, 
nucleon, and $^{199}$Hg nucleus, respectively. Experiments searching for the electron EDM typically use the polar molecules YbF and ThO 
and ultimately expect to reach a level of $3 \times 10^{-31}$ {\it e}$\cdot$cm.  Several experiments searching for the neutron EDM are 
planned or under way and are expected to reach a sensitivity of $5\times 10^{-28}$ {\it e}$\cdot$cm, which constitutes a factor of 100 
improvement over current limits.  For atoms, future experiments using Hg, radon and radium expect sensitivity at the level 
$10^{-32}$ {\it e}$\cdot$cm.  This would require upgraded facilities such as FRIB at Michigan State University or Project X at Fermilab.  
The expected sensitivity of these future programs is at the level where signals are predicted to appear in several theories beyond the SM.

The weak decays of light hadrons provide precision input to the SM and are a sensitive test of new interactions. The ratio of decay 
channels $e\nu/\mu\nu$ for pions and kaons provides an accurate test of lepton universality and can probe energy scales up to 1000 TeV.  For 
pions, experiments under way at TRIUMF and PSI will improve the measurement error by a factor of five over the current result and probe 
the SM to the level of 0.1\%.  Na62 is expected to reach a similar level of sensitivity in kaon decays.  Neutron beta decay is the classic 
weak interaction process and it provides the most accurate determination of the CKM element $V_{ud}$, thus enabling strong tests of CKM 
unitarity.  This unitarity bound constrains new physics up to scales of $\sim 10$ TeV.  The neutron lifetime and decay asymmetries provide 
other precision tests of the weak interactions, and several programs are under way to measure observables with improved precision.

Measurements of parity-violating asymmetries in fixed-target scattering with polarized electrons yield valuable information on neutral 
currents in the weak interaction.  This allows for a precision determination of the weak mixing angle at low values of $Q^2$, which in 
turn constrains new parity-violating effects up to $2-3$ TeV.  Specifically, improved polarized M\o ller scattering experiments expect to 
determine the weak mixing angle to an accuracy of $\pm 0.00029$.  This program will take place at the upgraded JLab facilities.  In 
addition, parity violation in atomic transitions yields valuable measurements of the weak mixing angle.  New techniques that require 
intense sources are being developed and will improve on previous measurements performed during the 1990s. Project X at Fermilab would 
provide more rare isotopes for this program than any other facility.

These overviews exhibit the broad spectrum of science opportunities attainable at the intensity frontier.  While each subfield is at 
a different stage of maturity in terms of testing the SM, the proposed experiments in each area are poised to have major impact.  The 
programs involving transitions of heavy quarks, charged leptons, and nucleons, nuclei, and atoms are advanced, with the most precise SM 
predictions and a well-developed experimental effort that has spanned decades.  In this case, the next level of experimental precision 
would reach the point where effects of new TeV-scale interactions are expected to be observable.  More sensitive searches for proton decay 
and new light, weakly coupled particles can cover a large range of parameter space that is consistent with Grand Unified Theories and 
cosmological observations, respectively.  
Neutrino physics is just reaching the level of sensitivity where it is possible to probe basic neutrino properties 
and answer principal questions, and therefore holds great promise for discovery.

This workshop marked the first instance where discussion of these diverse programs was held under one roof.   As a result, it was realized that 
this broad effort has many connections; a large degree of synergy exists between the different areas as they address similar questions.  For 
example, all the programs probe new symmetries at very high energy scales.  Experiments with heavy quarks, charged leptons, neutrinos, nucleons, 
nuclei, and atoms all search for new sources of $CP$ violation.  Light, weakly coupled particles could be the source of the current anomalous 
measurement of the magnetic moment of the muon.  Charged lepton interactions can test theories beyond the so-called neutrino SM.  Measurement 
of the neutrino mass hierarchy has implications for the interpretation of results from neutrinoless double beta decay.  One conclusion from 
the workshop is that there are strong ties between the various elements that comprise the intensity frontier.

The science of the intensity frontier is also strongly linked to the energy and cosmic frontiers, where different approaches are again 
necessary to gain the most knowledge.  For example, new physics searches at the LHC are closely related to flavor physics.  As the LHC 
pushes ever higher on limits for the scale of new interactions,  the flavor sector must become less SM-like.  Eventually, one is forced to 
choose between Minimal Flavor Violation and naturalness.  Since the flavor sector can probe interactions far above the reach of a
high energy collider, 
it can point the way to the appropriate energy scale that would be ripe for exploration.  If discoveries are made at the LHC, then precision 
measurements of heavy quark and charged lepton interactions will probe the flavor structure of the new physics. Connections to the cosmic 
frontier are similarly striking.  Light, weakly coupled particles are motivated by the presence of dark matter and dark
energy in our universe and may be a 
part of a rich dark matter phenomenology.  The large underground detectors constructed to observe neutrino oscillations are ideal for 
measuring the neutrino spectrum from supernovas, and could yield invaluable information on the evolution of the supernova mechanism. 

The vibrancy of the workshop reflected the enthusiasm of the community for intensity frontier physics.  Since the workshop addressed all 
aspects of the intensity frontier, participants were forced to look across traditional boundaries and become aware of the intensity science 
program as a whole.  It was realized that there is a strong intensity frontier community with common physics interests that is addressing 
fundamental questions.  It is important for the future of the field that broad workshops such as this continue.  In this way members of the 
intensity frontier community can continue to learn from each other and see how their program contributes to our understanding of the universe.

The working group reports and the above summary show very clearly that the intensity frontier physics program is diverse and rich in opportunities 
for discovery.  Such an extensive and multi-pronged 
program is necessary to address the unresolved fundamental questions about Nature.  We are at a stage in 
our understanding where the Standard Model provides sharp predictive capabilities; however, we know it is not the final theory and there is 
more to learn.  The knowledge we seek cannot be gained by a single experiment, or on a single frontier, but rather from 
the combination of results from many distinct approaches working together in concert.

This document provides a reference for the captivating science that can be carried out at the intensity frontier this decade and next.  
The broad program has the potential to make discoveries that change paradigms.  We hope that this report assists in the effort to make 
this rich program a reality and help the US  realize its potential to become a world leader at the intensity frontier.

\newpage

\chapter{Acknowledgements}

We thank the OHEP of the US Deparment of Energy Office of Science for their support and 
participation in the planning of this workshop.  We thank Christie Aston and the DOE OHEP 
staff, Donna Nevels and the ORISE staff, and Jennifer Seivwright for their indispensible and 
cheerful assistance during the workshop and its planning stages.  We thank 
Glennda Chui and Maria Herraez for their invaluable assistance in the preparation of this 
document and Katie Yurkewicz for assistance in communicating the workshop results.

\newpage

\chapter{DOE Charge}

\begin{figure}[b!]
\centerline{\includegraphics*[height=26cm]{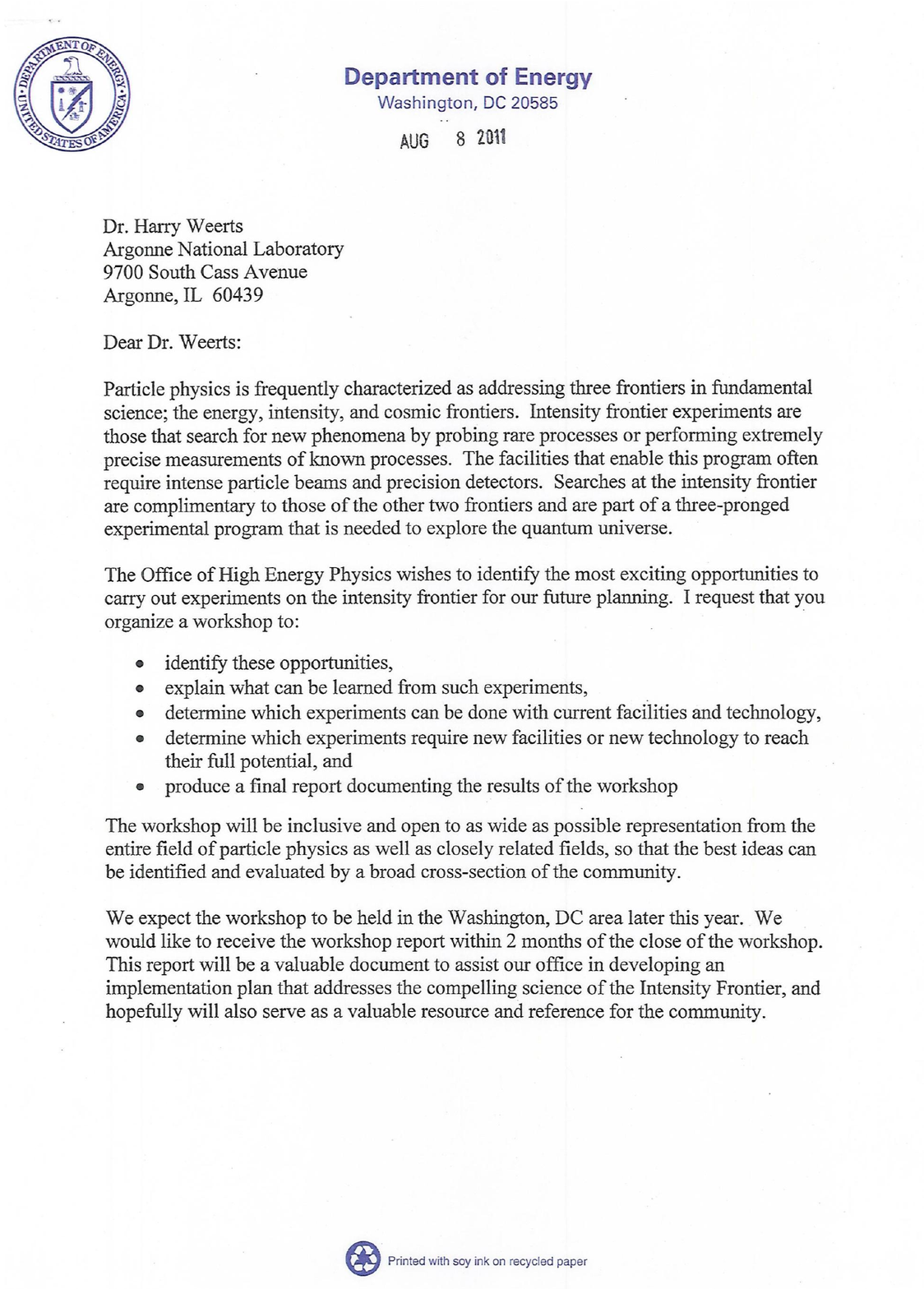}} 
\end{figure}

\begin{figure}[b!]
\centerline{\includegraphics*[height=26cm]{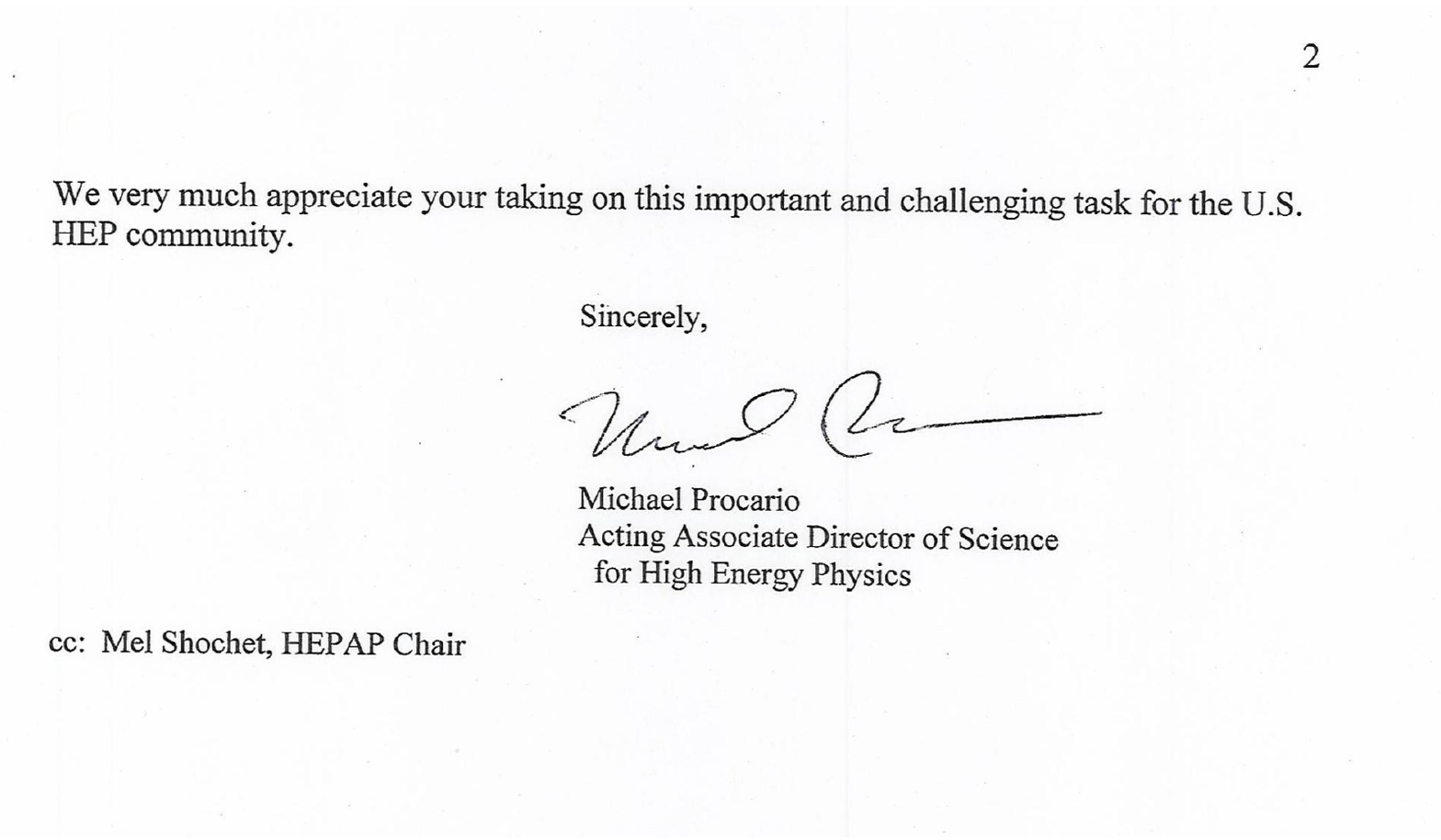}} 
\end{figure}

\end{document}